%% file: thesis.tex
\documentclass[a4paper,titlepage,oneside,fleqn]{report}
\usepackage[final]{graphicx} %% does what it says on the tin
\usepackage{amssymb} %% lots of maths symbol goodies
\usepackage{amsmath} %% do nice stuff to equations
\usepackage{setspace} %% throwback thesis typesetting requirement
\usepackage[round]{natbib} %% get biblio commands
\usepackage{xspace} %% add flexible spaces after definitions
\usepackage{subfig} %% allow easy multiple panels in figures
\usepackage{fullpage} %% Make the margins smaller
\usepackage{color} %% allow coloured output

\font\crestfont=oxcrest40 scaled\magstep4
\def\crest{{\crestfont \char1}}

\newcommand{\sun}{\odot}
\newcommand{\rsun}{R_0}

\newcommand{\vect}[1]{{\bf #1}}
\newcommand{\vecthat}[1]{\hat{\vect{#1}}}

\newcommand{\ttp}[1]{\times 10^{#1}}

\renewcommand{\eqref}[1]{equation~(\ref{#1})}
\newcommand{\bracketeqref}[1]{(equation~\ref{#1})}
\newcommand{\eqsref}[2]{equations~(\ref{#1}) and (\ref{#2})}
\newcommand{\blankeqref}[1]{(\ref{#1})}
\newcommand{\Eqref}[1]{Equation~(\ref{#1})}
\newcommand{\secref}[1]{\S\ref{#1}}
\newcommand{\tabref}[1]{Table~\ref{#1}}
\def\figref#1{Fig.~\ref{#1}}
\newcommand{\figstworef}[2]{Figs.~\ref{#1} and~\ref{#2}}

\newcommand{\fighardref}[1]{Fig.~{#1}}
\newcommand{\chapref}[1]{Chapter~\ref{#1}}
\newcommand{\appref}[1]{Appendix~\ref{#1}}

\long\def\symbolfootnote[#1]#2{\begingroup%
\def\thefootnote{\fnsymbol{footnote}}\footnote[#1]{#2}\endgroup}

\def\vot{\vect{v}_{0\perp}}
\def\near{\sim\!}

\def\b#1{{\bf#1}}

\def\kms{\,{\rm km}\,{\rm s}^{-1}}
\def\Myr{\,{\rm Myr}}

\def\Gyr{\,{\rm Gyr}}
 
\def\pc{\,{\rm pc}}
\def\kpc{\,{\rm kpc}}

\def\e{{\rm e}}
\def\d{{\rm d}}
\def\msun{M_\odot}
\def\pone{\phi_1}
\def\ptwo{\phi_2}
\def\masyr{\,{\rm mas}\,{\rm yr}^{-1}}
\def\deg{^{\circ}}

\def\percent{\,\,\text{per cent}}

\newcommand{\krh}{K10\xspace}
\def\vo{\vect{v}_0}
\def\near{\sim\!}
\def\tx{t_\parallel}
\def\ty{t_\perp}
\def\rperi{R_{\rm p}}
\def\angwid{\Delta \psi}
\def\anglen{\Delta \theta}
\def\lendeproj{\Delta \Theta}
\def\fvfps{{\sc fvfps}}

\def\vot{\vect{v}_{s}}
\def\vt{v_t}
\def\rh{\hat{\vect{r}}}
\def\vee{\vect{v}}
\def\vr{v_r}
\def\fr{F_r}
\def\arcmin{\,{\rm arcmin}}
\newcommand{\smt}{\sigma_\mu}
\newcommand{\svot}{\sigma_{\vot}}
\newcommand{\cov}{{\rm cov}}
\def\mastereqs{\eqsref{radvs:eq:master}{radvs:eq:master2}\xspace} % got bored of typing this
\def\blankmastereqs{\blankeqref{radvs:eq:master} and~\blankeqref{radvs:eq:master2}\xspace} 
\def\diffeqs{differential equations~\blankeqref{radvs:eq:master} and~\blankeqref{radvs:eq:master2}}
\def\pressetal{Press~et~al.}

\newcommand{\votx}{v_{s, x}}
\newcommand{\voty}{v_{s, y}}
\newcommand{\vpar}{v_{s\parallel}}
\newcommand{\vper}{v_{s\perp}}
\def\svper{\sigma_{\vper}}
\def\svpar{\sigma_{\vpar}}
\def\vsun{\vect{v}_0}

\def\b08{B08\xspace}
\def\vJ{\vect{J}}
\def\DvJ{\delta\vJ}
\def\vT{\mbox{\boldmath$\theta$}}
\def\vO{\vect{\Omega}}

\def\eigens{\vecthat{e}_n}
\def\eigen{\vecthat{e}}
\def\hessian{\vect{D}}

\newcommand{\basefreq}[1]{\Omega_{0,#1}}
\def\pone{\partial_1}
\def\ptwo{\partial_2}
\def\jr{J_r}
\def\rlim{r'_\text{lim}}
\def\tdyn{t_\text{dyn}}

\def\eff{\text{eff}}
\def\vw{\vect{w}}
\def\tide{\text{tide}}
\def\peri{p}
\def\apo{a}
\def\min{\text{min}}
\def\max{\text{max}}
\def\twidr{\tilde{r}}

\def\stackel{St\"{a}ckel}

\def\figureshrink{0.4}
\def\doublefigshrink{0.4}
\def\figplaceopts{}

\def\mnras{Mon. Not. R. Astron. Soc.}

\title{
\sc
On the dynamics of tidal streams in the Milky Way galaxy}
\author{Andrew M Eyre\\
Rudolf Peierls Centre for Theoretical Physics
}
\date{
\vspace{1.5cm}
\crest\\
\vspace{2cm}
Merton College\\
University of Oxford\\
\vspace{1cm}
\vspace{1cm}
A thesis submitted for the degree of Doctor of Philosophy\\
at the University of Oxford\\
\vspace{2cm}
Hilary Term 2010\\
}

\begin{document}

\onehalfspacing % this spacing is what I configured it with
\maketitle

%\singlespacing % don't print abstract -etc- with doublespacing
\pagestyle{empty} % don't print page numbers on the abstract

{\centering
{\Large\sc
On the dynamics of tidal streams in the Milky Way galaxy}\\
\vspace{7mm}
Andrew M Eyre\\
Rudolf Peierls Centre for Theoretical Physics\\
\vspace{0.5cm}
Merton College\\
University of Oxford\\
}

\vspace{1cm}

{\bf \centering Abstract\\}
\vspace{5mm}
\noindent
\input{abstract}

\vspace{2cm}
{\centering
A thesis submitted for the degree of Doctor of Philosophy\\
at the University of Oxford\\
\vspace{0.5cm}
Hilary Term 2010\\
}

\newpage
\pagestyle{plain} % no headers for now
\pagenumbering{roman} % front matter

\section*{Foreword}

%A personal foreword and dedication is to be inserted here later.
\input{foreword}

\newpage

\section*{Acknowledgements}

I would like to extend my thanks to the following, each of whose aid
has directly contributed in some way to this work: Prof.~James Binney,
for his knowledge, patience, advice and support, and for copious
quantities of his time; the members of the Oxford Dynamics group, for
comradeship and helpful critique; the anonymous referees of my papers,
for their helpful remarks, many of which advanced the work of
this thesis; Prof.~Andy Gould, whose remarks on the uncertainties in
Galactic parallax calculations prompted the analysis of
\secref{galplx:sec:uncertainty}; Sergey Koposov for his provision of
the data for \figref{galplx:fig:gd1-udot}; Michal Molcho for
her eagle-eyed proofreading of this manuscript; and the examiners
of this thesis, Vasily Belokurov and John Magorrian, for their
comments and suggestions.
\\ \\
The following people kindly provided code that aided my
calculations. I would like to acknowledge their contributions and
thank them: Prof.~James Binney, whose contributions included code for:
the calculation of action-angle variables in the isochrone potential,
the calculation of actions in \stackel\ potentials, the sampling of King
model distribution functions, the reconstruction of orbital trajectories using
radial velocity data, and many other miscellaneous routines; Carlo
Nipoti, who provided the \fvfps\ tree code, and updates thereof; and
Paul McMillan, who provided a shake-and-bake version of Prof. Walter Dehnen's
{\sc falPot} potential calculation routines.
\\ \\
I acknowledge the receipt of a PPARC/STFC award during the preparation
of this work.

\section*{Derivative publications}

The following publications have arisen out of the work of this thesis.
\\ \\
Parts of \chapref{chap:radvs} appeared in:\\
Eyre, A., \& Binney, J. 2009, \mnras, 400, 528\\
\\ 
Parts of \chapref{chap:pms} appeared in:\\
Eyre, A., \& Binney, J. 2009, \mnras, 399, L160\\
\\
Parts of \chapref{chap:galplx} appeared in:\\
Eyre, A. 2010, \mnras, 403, 1999

\tableofcontents 
\listoftables
\listoffigures
\newpage
\doublespacing % throwback thesis requirement
\pagenumbering{arabic} % normal numbers

\input{intro/intro}

\input{radial_vs/radial_vs}

\input{gal_plx/gal_plx}

\input{streamdirs/streamdirs}

\input{concs/concs}

\singlespacing

\bibliographystyle{astronat/apj/apj}
\bibliography{combined}

\appendix
\onehalfspacing

\input{appendix_stackel/ap_stackel.tex}

\end{document}

%% file: abstract.tex
We present a brief history of Galactic astrophysics, and explain the
origin of halo substructure in the Milky Way Galaxy.
We motivate our study of the dynamics of tidal streams in our Galaxy by
highlighting the tight constraints
that analysis of the trajectories of tidal streams can place
on the form of the Galactic potential.

We address the reconstruction of orbits from observations of tidal
streams. We upgrade the geometrodynamical scheme reported by
\cite{binney08} and \cite{jin-reconstruction}, which reconstructs orbits
from streams using radial-velocity measurements, to allow it to work
with erroneous input data. The upgraded algorithm can correct for both
statistical error on observations, and systematic error due to streams
not delineating individual orbits, and given high-quality but
realistic input data, it can diagnose the potential with considerable accuracy.

We complement the work of \cite{binney08} by deriving a new algorithm,
which reconstructs orbits from streams using proper-motion data
rather than radial-velocity data. We demonstrate that the new algorithm
has a similar potency for diagnosing the Galactic potential.

We explore the concept of Galactic parallax, which arises in connection
with our proper-motion study. Galactic parallax allows trigonometric
distance calculation to stars at 40 times the range
of conventional parallax, although its applicability
is limited to only those stars in tidal streams.

We examine from first principles the mechanics of tidal stream
formation and propagation. We find that the mechanics of tidal
streams has a natural expression in terms of action-angle variables.
We find that tidal streams in realistic galaxy potentials 
will generally not delineate orbits precisely, and that attempting
to constrain the Galactic potential by assuming that they do
can lead to large systematic error. We show that we can
accurately predict the real-space trajectories of streams, even when
they differ significantly from orbits.

%% file: foreword.tex
The journey of this thesis has been a long one. Although I present
thanks overleaf to those who have been of direct assistance
in my work, I would like to use this page to speak of those who
made the journey pleasurable as well as possible.

Firstly, I will mention my guide along the way, James Binney,
without whom this endeavour would never have been started, let alone
finished. While I was contemplating where to go for my PhD, a very
esteemed astrophysicist, upon learning of my list of prospective
supervisors, said: ``James Binney is virtually in a class of his own
... in terms of the breadth and profundity of his physics
knowledge. If you have the chance to work with him, I would advise you
to take it.''  It does not take a prolonged conversation with James to
confirm this opinion as abundantly true.  It has indeed been a rare
honour and a privilege to work with him.

Without my fellow travellers from Room 2.9 of the Peierls Centre,
office life would have been so much duller. They are, in order of
appearance: Ralf Donner, whose Teutonic expressivity is surely
unmatched by anyone; Sarah S\"{a}gesser, who was always in
the office, no matter how early I arrived; Rachel Koncewicz, whose
delicious coffee gave life to the office, both metaphorically and
literally; Mimi Zhang, whose inscrutable Confucianism made our chats
so challenging and yet so interesting; Mike Williams, whose track
record of involvement in creative projects---including but not limited
to: a movie, a record label, and at least three PhD projects---is
utterly bewildering; Ben Burnett, whose repartee is eclipsed only by
his skill in physics; Calum Brown, who always had to hand a bolus
of healthy Scottish realism; Francesco Fermani, whose fiery Italian spirit
was a welcome new broom; and Alfred Mallet, who finally relieved me of
the burden of being the only turncoat around.

From elsewhere in the Physics Department, I wish to mention: John
Magorrian, whose ever-welcome humour is lost on many of the Peierls
Centre residents, I am sure; Carole Jordan, who can be relied upon to
find the silver lining in any cloud; Paul McMillan and Christophe
Pichon, who both helped enspiritualize our Monday lunchtime meetings;
Will Newton, whose unerring knack for attracting the unusual led to a
high-speed bus chase along the M40, a well-timed escape from an Eton
man waving a fistful of \pounds20 notes, and numerous other close
shaves; Garrett Cotter, who has an inexhaustible supply of tales of
tutorial-room woe; and Tom Mauch, for the much-vaunted musical
education that is hopefully soon to materialize!

External to the department, I have counted amongst my friends: Maja
Starcevic and Kreso Petrinec, Claire Labrousse, Daniel Rotenberg and
Merav Haklai-Rotenberg, Jonathan Riley, Jim Naughton, and countless
others that I apologize for unfairly forgetting to mention. Whilst at
home, my enduring memory is of the abundant worldly wisdom of Charlie
being punctuated by the incessant banter of Clo. I would also like to
thank my family, who have always given me their unconditional love and
support.

Lastly, I wish to speak of Michal. She is wise and beautiful. If
Michal has taken of me but a fraction of that which I have taken of
her, she must indeed be the richest woman in the world. For without a
shadow of a doubt in my mind, there are none that are richer than me
for having shared in her life for these past few years.

\begin{flushright}
\vspace{5mm}
Andy Eyre\\
Oxford, July 2010
\end{flushright}

\newpage
\hbox{}
\begin{center}
\vspace{8cm}
To my mother and father\\
\vspace{4mm}
and to Michal
\end{center}

%% file: intro/intro.tex
\chapter{Introduction}
\label{chap:intro}

The study of astronomy has long held an ennobled position amongst the
fields of natural enquiry, of which it is undoubtedly one of the
oldest: the first written records of astronomical measurement were
made by the ancients in the city-state of Babylon.\footnote{ Perhaps
  competing for the title of oldest is the field of medicine, for which the
  {\em Edwin Smith Papyrus}, dated to the 16th century BC, contains
  rational descriptions of injury and prognosis.  By comparison, the
  earliest records of astronomical observation are the Babylonian
  tablets {\em Enuma Anu Enlil}, also dating from sometime in the 2nd
  millennium BC, but which are unfortunately far from
  rational, since they clearly claim to have been made for the purposes
  of conducting magic.  The earliest known rational attempt to explain
  astronomical phenomena is attributed to Plato's student, Eudoxus of
  Cnidus, who lived in the 4th century BC.  }
The study almost certainly goes back further, however, since many
of the Babylonian constellations were named in a Sumerian dialect,
and the Sumerian civilization had already crumbled by the 3rd
millennium BC. The origins of Sumer predate that by some thousands of years,
and are lost is the mists of prehistory:
one can perhaps imagine
our Mesopotamian ancestors staring at the heavens at night, and being
amazed by both the regularity and spectacular beauty of the slow
procession of the bodies held therein.

Even within the irrational world-views held by the ancients, it was
recognized that the processes governing the movement of the heavenly
bodies must be mechanical. This insight lead to the earliest recorded
attempts to impart mechanical descriptions on natural phenomena. For
instance, the regularity of the diurnal motion of the planets and
stars about the Earth's axis led to the quantization of the passage of
time: it is no coincidence that, even today, our measure of time is
fundamentally a measure of angle, and indeed was without alternate
physical basis until as recently as 1967.\footnote{The Thirteenth General
  Conference on Weights and Measures, 1967.}

It is therefore unsurprising that history credits astronomy
with provoking a most extraordinary series of physical
discoveries. The motions of the planets, as observed by Tycho Brahe
and resolved into orbits by Johannes Kepler, both inspired Isaac Newton and provided
him with the necessary data to inform his deduction of the laws
of classical mechanics and of universal gravitation. In the process of
formulating his theories of planetary motion, Newton discovered the
differential calculus---a necessary tool for his task---and consequently
founded physics as a mathematical discipline. Newton then went
on to write his second great treatise {\em Opticks}, which made
great advances in the geometric description of light and 
directly derived from experiments he began in order to construct
a better telescope.

Newton's theories were a well-spring of progress in physics, and much
of this progress came in pursuit of astronomy. Although
fully-formulated in {\em Principia} by Newton, classical mechanics
later evolved under the genius of the likes of Joseph-Louis Lagrange,
Pierre-Simon Laplace and William Hamilton. Lagrange and Hamilton
each reformulated Newtonian mechanics into their respective, eponymous
dynamics. These new dynamics readily admit solutions to problems
that are computationally taxing using Newtonian mechanics, and both
were directly inspired by their namesake's desire to solve problems
in celestial mechanics. As too was Laplace. Intrigued by celestial
mechanics, he developed potential theory,
culminating in his eponymous equation: he then invented spherical
harmonics in order to solve it. Laplace also laid the modern foundations
of probability theory and statistics, in order to better interpret
incomplete astronomical observations.

The list could go on. Astronomy has indeed inspired many of the
greatest scientists in their greatest work. It is yet more remarkable,
therefore, that it was only with the middle-half of the 20th
century that confirmation was finally made that there existed galaxies
other than our own.

%not until 1904aside from advances in mathematical machinery for
%describing celestial mechanics, the state of knowledge of astronomy
%advanced relatively little between Newton's day and the
%1920s. Certainly, though, that mathematical machinery would become
%vitally useful in the golden age of observational astronomy that began
%with the middle-half of the 20th century.

\section{A brief history of galactic astronomy}

The idea of the Milky Way---easily visible on a dark night away from
the glare of artificial city lighting---as a collection of stars
similar to our own Sun is very old, dating from antiquity.\footnote{
The Greek philosopher Democritus, a contemporary of Aristotle and Plato,
was the first to be recorded espousing this view, which was perhaps
informed by his strong atomist tendencies. Little of Democritus' own
work survives, but we know of his views on galactic astrophysics
by means of Plutarch, in {\em De placitis philosophorum}.}
The confirmation of this supposition came by way of the persecuted
genius Galileo Galilei whose self-made optical
telescope allowed him to observe that the Milky Way, which 
appears nebulous to the naked eye, is actually comprised
of countless distinct and individual stars.

The first mention of objects that we now know to be galaxies external
to our own came somewhat later than did those of the Milky Way.  The
Magellanic clouds were first recorded by 10th century Arabian
astronomers, and the knowledge of their existence was eventually
brought to Europe following the global circumnavigation of the 16th
century explorer Ferdinand Magellan.

It would require Newton's law of universal gravitation in order to
begin to understand the true nature of galaxies, although Newton
himself apparently failed to make any significant progress in the task.\footnote{
  Newton was traumatized by his inability to show that the Solar
  system was stable, since his limited calculations of the
  Sun-Jupiter-Earth three-body problem led him to predict the rapid
  ejection of the Earth from the system.  Newton repaired this problem by
  hypothesizing the intervention of God to reset any Jovian anomalies
  induced in Earth's orbit.  Newton's cosmology was one of an infinite
  constellation of static, but mutually gravitating stars: this
  configuration is actually unstable, and this fact was known to
  Newton, but he had no qualms in invoking divine intervention to
  maintain it indefinitely, just as he had for the Solar
  system. Newton's unwarranted belief in a
  static universe puts him good company. Albert Einstein's
  similarly-unwarranted belief in the same led him to insert an otherwise
  unprovoked constant of integration---the cosmological
  constant---into his general-relativistic field equations. Einstein
  later regretted this action, calling it ``the biggest blunder of
  (my) life'' \citep{gamow-autobio}.  The modern understanding of the
  Newtonian Solar system shows that it is indeed subject to rapid
  disintegration on account of Jovian perturbations: it is of some
  irony that Einstein's own general-relativistic corrections to
  Mercury's orbit are required in order to detune the resonance
  between Jupiter's orbit and Mercury's and thus stabilize the system,
  which would otherwise result in the ejection of Mercury, followed by
  the other inner planets in a few million years
  \citep{laughlin-stability}. If only Newton had known general
  relativity! }
The English astronomer Thomas Wright \citeyearpar{thom-wright}
%\footnote{ Wright,
%  T. 1750, {\em An Original Theory or New Hypothesis of the Universe}
%  (London) }
was the first to understand that the Milky Way might be a
rotationally-supported, flattened disk of stars, which we view as a
tract across the sky by nature of our position within it. His
speculation was made with little evidence to back it up. Wright also
speculated that the mysterious ``nebulae,'' such as the Magellanic
clouds, might well be separate galaxies in their own right. His ideas
were taken up and promulgated by Immanuel Kant, who called the
structures ``island universes.''

Remarkably, it was not until the 1920s that observational technology
improved to the point where either of these hypotheses could be
conclusively confirmed.  In 1924, while working at the Mount Wilson
Observatory, the American astronomer Edwin Hubble used the
brand-new Hooker Telescope 
to resolve individual stars in several nearby galaxies, including
M31. Several of the stars he observed were Cepheid variables, for
which the Harvard astronomer Henrietta Leavitt had some years earlier
deduced a tight period-luminosity relation.  With their intrinsic
brightness determined by this relation, the faint apparent magnitude
of the stars put them far beyond the most generous estimates for the
extent of the Milky Way galaxy at that time \citep{hubble-extra}.  M31 and its companion
nebulae could be nothing other than external galaxies. Thus, the Wrightian
conjecture of ``island universes'' made almost 200 years earlier was
proved and the study of extragalactic astronomy was born.

As has often been the case in the history of our field, fate conspired
that the mathematical tools and the experimental machinery to probe a
new area of science became available at the same time. Einstein's
general theory of relativity (\citeyear{gr}) had provided the physical
framework upon which consistent cosmological theories could be
constructed. Hubble was again instrumental in advancing the field, and
in 1929 he made the discovery of a linear relationship between the
distance to, and the measured line-of-sight velocity of, far-away
galaxies \citep{hubble-expansion}.  This was the first observational
evidence for the expansion of the Universe, and the modern study of
cosmology was born, with Hubble having launched his second new field
of natural enquiry in about as many years.\footnote{ Infamously,
  Hubble never won the Nobel Prize in Physics for his groundbreaking
  contributions to our understanding of the Universe.  At the time,
  astronomy was not considered in the remit of the physics Prize, and
  despite a growing clamour in the scientific community for the
  Prize to be awarded to Hubble, he died suddenly in 1953 having not
  received it. Astronomy and astrophysics were admitted to the list of
  eligible fields of study that very year. Indeed, at the time of his
  death---but obviously unknown to Hubble himself---the 1953 Nobel
  Prize committee, amongst whose august members were counted Enrico
  Fermi and Subrahmanyan Chandrasekhar, had already unanimously voted
  Hubble to receive that year's physics Prize \citep{soares-hubble}.
  Nobel Prizes cannot be awarded posthumously: the first Nobel Prize
  awarded for an astrophysical discovery eventually went to Hans Bethe
  in 1967, for his explanation of the nuclear fusion processes that
  power stars \citep{bethe-nobel}.}

The consequences of Hubble's two great discoveries were enormous.
The expansion of the Universe requires that at some time in the
past all matter was coincident, and hence the Universe is of
finite age. Although it would take many decades before this
age was known with any kind of precision, it was apparent
from Hubble's first observations that the age of the entire Universe
was not incomparable to the geological age of the Earth---some
billions of years.

One conclusion to be drawn from this is that
galaxies are not and were never steady-state objects. Since they
cannot be substantially older than the stars within them,
and given the great distance scales that they span, galaxies cannot
 be dynamically very old. Indeed, our own Milky Way
cannot have completed more than 70 complete revolutions
since the big bang, assuming that it formed shortly
thereafter.

It thus became---and remains---a core problem in modern astrophysics
to explain the structure, formation and evolution of galaxies.  It was
immediately possible for Hubble's contemporaries, such as James Jeans
and Arthur Eddington, to begin this difficult task because the
evocation of statistical mechanics by James Clerk Maxwell and Ludwig
Boltzmann in the late 19th century had provided the tools necessary to
begin to understand the bulk motion of stars under the influence of
gravity. In combination with orbital mechanics, this laid the
framework for the study of stellar dynamics, which deals with
systems comprised of many-fold more bodies than does celestial
mechanics, its direct intellectual predecessor.

Here we will leave our whistle-stop history tour of galactic astronomy,
but for one small diversion of direct relevance to our work.
Of the many astounding scientific discoveries of 20th century,
one of the least expected, and certainly one of the least
well-understood, has been the growing---and by now, colossal---body of evidence that
there is simply insufficient baryonic matter in most galaxies
to explain the observed kinematics.

The first observation to this effect was reported by the Swiss
astronomer Fritz~\cite{zwicky33}, who combined the virial theorem with
measurements of velocity dispersion in galaxy clusters to argue that
the bulk of their mass must be in non-luminous matter.
Further progress awaited technological advances. Up until the late
1950s, galactic astronomy consisted mostly of observations of stars at
optical or near-optical wavelengths.  The discovery of the so-called 21-cm neutral
hydrogen transition, which arises because of spin-spin coupling between
electron and proton in the
ground state of atomic hydrogen, heralded yet another revolution in galactic
astronomy. Radio astronomy observations of 21-cm emission allowed the
optically-transparent, non-ionized gaseous content of galaxies to be mapped for the first
time, and since then with ever increasing sensitivity.

The first 21-cm maps of gas in external galaxies showed that
the circular speed of matter was independent of radius.
 One corollary of this surprising observation is that the
mass interior to a given radius must grow linearly with
radius: thus, the dynamical mass in these galaxies was not concentrated
in those regions of high luminous mass, as had previously been supposed.
Even more significantly, the 21-cm
maps were used to quantify the mass of gas in these galaxies, which was then
added to star counts to estimate the total content of baryonic
matter. The resulting estimate for the mass content of these galaxies was
insufficient to explain the observed kinematics of the gas and stars,
with this so-called ``missing mass'' problem becoming most acute at
large radii where the density of baryonic matter falls rapidly, but
where the kinematics indicate that most of the matter is
concentrated \citep[\S8.2.4]{bm98}.

The description of this ``dark matter'' is one of the foremost
problems in modern astrophysics, and neither the formation nor the
evolution of galaxies can be properly studied without addressing
it. \mbox{N-body} simulation of cosmological structure formation has
provided clues as to what the dark matter distribution should look
like \citep{nfw}, and the suggested profiles have some corollary in
observations of external galaxies \citep[e.g.][]{rix-ellipticals}.
However, the universal applicability of the simulation results is far
from proved, and apart from its general presence, the distribution of
dark matter in the Milky Way in particular is still not
well-determined \citep{rave-escape-v}.

Observations are needed. Unfortunately, all attempts to directly
detect particle annihilation signatures from concentrations of dark
matter have as yet been unsuccessful \citep{direct-d-failed}, and in
any case, the detection rate of such signatures is unlikely to ever be
high enough for useful imaging to take place.  Our only option to
examine the dark matter content of the Milky Way is to utilize that
very mechanism by which we hypothesize its existence in the first
place, namely, the effects of its gravity on the dynamics of luminous,
observable matter.

It will therefore be the topic of this thesis to address
certain indirect methods of probing the
mass distribution of our Galaxy. To do this, we will examine
the mechanics of tidal streams.

\section{Galactic cannibalism in action: tidal streams}

``Galactic cannibalism''---to use the words of
\citet{ostriker-cannibalism}---refers to the merging of individual
galaxies to form a greater whole and has been increasingly recognized
over the last 30 years as playing a significant role in determining
the structure of galaxies in general, and the Milky Way in particular
\citep[\S8]{bt08}.  Indeed, it is a fundamental tenet of the
\cite{white-rees-merging} model of galaxy formation, which has
dominated our understanding since its inception. In this picture,
baryonic components are embedded in massive dark haloes, which cluster
and then merge purely under the force of gravity. Radiative processes
then cause the gaseous phase to cool and contract and eventually
form stars at the bottom of the potential well.

In support of this model, we note that collisions between external
galaxies have been known about for a long time. The M51 ``Whirlpool''
and NGC 4038/9 ``Antennae'' galaxies are all undergoing obvious
merging events, and all were discovered in the 18th century, although
their nature was not understood at the time.  Violent mergers, such as
those of the Antennae, are often associated with tidal tails:
high-energy ejecta made up of stars and gas, catapulted from the
edges of the colliding objects by immense tidal forces, and only
marginally bound, if bound at all, to the resulting combined host mass
\citep{toomre2tails}. Many more such merging systems are now known,
and these interactions are believed to be commonplace.\footnote{ Such
  a merger is forecast between the Milky Way and M31, to take place in
  about 3 billion years time \citep{mw-m31-merger}. The structure of
  both galaxies will be destroyed, and a massive elliptical galaxy
  will emerge in their stead, although it is most unlikely that
  individual solar systems will be directly affected by the
  merger. The elderly Sun will still be in the main sequence at this
  time: for whoever or whatever life inhabits the Earth, the spectacle
  in the night sky will be extraordinary.  }

The hierarchical galaxy formation model is particularly successful in
explaining the presence of the Milky Way's halo: a spheroid of old,
metal-poor stars and globular clusters, extending some tens of \kpc\
from the Galactic centre \cite[\S10.5]{bm98}. The halo contains very
little gas and dust and is hard to see how the halo stars could form
in situ. The \cite{white-rees-merging} mechanism answers this, by explaining
the halo as the phase-mixed stellar remnant of long-ago mergers.

However, up until the 1990s, very little evidence of cannibalization of its satellites by
the Milky Way had been seen at all: the Large Magellanic Cloud was
identified by \cite{mag-stream} as losing mass to the Milky Way halo,
but the mass lost is in gas and not stars, and rather than being
evidence of a gravity-driven merging event, it is most likely that the
observed `tail' is simply the streamlined wake of the Cloud's gaseous
envelope being stripped due to ram pressure from the Milky Way's own
halo gas \citep{lmc-ram-pressure}.

The advent of multi-million particle N-body simulations of
cosmological structure formation allowed quantitative predictions for
the expected number of Milky Way satellite galaxies and merger
remnants to be made \citep{nfw,moore-etal,bullock-etal}.  The outcome
was problematic: simulations showed that, although the numbers of
high-mass satellites was predicted almost perfectly, the Milky Way
ought to have accreted an order of magnitude more low-mass satellite
galaxies than were actually known at the time
\citep{klypin-missing-satellites}.  Attempts were made to repair the
``missing satellite'' problem by proposing mechanisms to shut off star
formation in low-mass haloes, thus rendering them invisible, but
the situation still remained highly unsatisfactory
\citep{klypin-missing-satellites,bullock-missing-clusters,
  moore-2006-missing-satellites}.

The predictions of the simulations went further. Long-lived stellar
substructure in the Milky Way's halo was shown to be a consequence on
ongoing merger activity, and this substructure was associated with
kinematic and chemical signatures that ought to be observable
\citep{bullock-johnston}.  Indeed, the substructure was reckoned to be
permeate the Galaxy with sufficient density to leave kinematic traces
in the Solar neighbourhood \citep{helmi-white-halo}.  In this way, some of
the first direct evidence for substructure resulting from past mergers
was identified by \cite{helmi-white-nature} using data from the
Hipparcos satellite \citep{newhipparcos}.

The discovery of the Sagittarius Dwarf galaxy by \cite{ibata-sag}, on the
far side of the Milky Way, provided dramatic evidence for ongoing
merger activity. This galaxy stands out amongst the other known
companion galaxies on account of both its high mass and close
proximity to the centre of the Milky Way.  Dynamics requires the Milky
Way to impart strong tidal forces across the Sagittarius Dwarf galaxy,
with it being so large and so close, and thus the existence of a tidal
stream of stripped dark matter and stars was predicted soon after its
discovery
\citep{velazquez-tail-prediction,johnston-spergel-hernquist-sag}.
It took nearly 10 years before the stars of
the massive Sagittarius stream were conclusively observed by
\cite{majewski-sag} in infra-red data from the Two Micron All-Sky
Survey \citep[2MASS,][]{2mass}.

It is difficult to observe merger substructure in external galaxies,
because it phase-mixes rapidly, and is only detectable thereafter in
the form of kinematic and chemical signatures, which require
observations of such precision that they can only be performed in the
Milky Way.  It was therefore with some great anticipation that data
from the automated Sloan Digital Sky Survey \citep[SDSS,][]{sdss}
arrived. This project used a purpose-built 2.5-m telescope to survey
tens of millions of stars, as faint as magnitude $r\sim 24$, away from
the Galactic plane.  Nonetheless, despite its impressive
performance, the very faint substructure of the Galactic halo still
required careful data processing in order to expose its signal above
the noise of the halo stars.

\begin{figure}
%\centerline{
%\includegraphics[width=0.75\hsize]{thesisfigs/pretty/field_of_streams_900.eps}
%} 
\centerline{
\includegraphics[width=0.75\hsize]{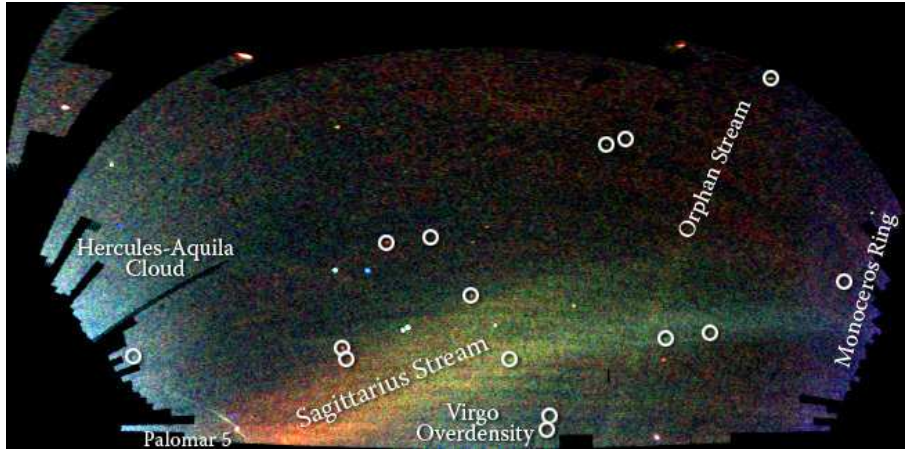}
} 
\caption
[The ``field of streams'' of  \cite{field-of-streams}]
{The ``field of streams'' of  \cite{field-of-streams}.
The observation of tens of millions of stars with SDSS, combined with
appropriate cuts to the data, expose the dramatic merger-history substructure
present in the Milky Way's halo. The colours are for $r$ magnitude, which
proxies for heliocentric distance in this image. The circles highlight
clusters/galaxies newly discovered in this image.
{\em Credit: V. Belokurov and the Sloan Digital Sky Survey. Image source:
sdss.org.}}
\label{intro:fig:fos}
\end{figure}

\figref{intro:fig:fos} shows one result of the effort: an on-sky
map along the north Galactic pole, which was reported by \cite{field-of-streams}.
This image shows the halo to be criss-crossed by extant tidal streams,
and dotted with hitherto undetected low-mass halo-dwelling galaxies.
Indeed, the observational evidence unearthed by the SDSS, and its follow-on
extension programme \citep[SEGUE,][]{segue}, revealed a dramatic
number of tidal streams in the Milky Way halo, often with no obvious
associated progenitor object
\citep{odenkirchen-delineate,majewski-sag,yanny-stream,field-of-streams,orphan-discovery,
  grillmair-orphan,gd1-discovery,ngc5466,
  grillmair-2009,newberg-streams-2009}.

These streams are almost certainly the remnants of objects of
extragalactic origin. Short of a major merger, it is hard to envisage
a scattering event that could launch an existing Milky Way globular
cluster onto a lower-energy orbit on which it cannot survive.
Hence, it is likely that these streams originated either from
former members of the cluster
system of a larger cannibalized galaxy, which could not then survive on their
new orbit around the Milky Way, or tiny galaxies whose streams are
the fading echo of their cannibalization by the Milky Way in their own
right.

The nature of the progenitors of these streams may be important for
tracing the merger history of our Galaxy \citep{helmi-review-halo}. However, in this thesis, we
will be concerned with the use of tidal streams as probes of the matter
distribution of the Milky Way itself. As such, the details of their
origin do not concern us greatly: we simply accept that at some
point in their history, these objects have found their way onto orbits
around our Galaxy, along which they experience tidal forces strong enough
to promote their disintegration.

\section{The use of tidal streams as probes of the potential}

%This thesis is concerned with the mechanics of Milky Way tidal streams,
%with a particular view to using them as probes of the Galactic potential.

The possibility that the morphology of tidal tails from colliding
galaxies could act as a probe of the potential has been recognized for
almost as long as the physics of the tails themselves has been
understood \citep{faber-tidal-tails}, although it is only much more
recently \citep{dubinksi-antennae-tails-probe} that simulation
technology has been sufficient to make a serious attempt at
constraining the potential with models of galaxies undergoing a major
merger.

However, on account of the great
disparity in mass between the Milky Way and its victims, the mergers
observed are less violent than those seen between objects of similar mass.
We should therefore be careful to distinguish the tidal {\em streams} resulting
from the simple disintegration of clusters and small galaxies in the
gravity field of a massive host---as is typically seen in the Milky Way---from those tidal {\em tails}, which result from major mergers.
The stars of the former remain in low-energy
orbits around the massive host, while the stars in the latter
are unbound, or nearly so. 

Furthermore, although the same fundamental physical principles
underlie the creation of both species, the tidal streams that
result from cannibalization of low-mass satellites, such as we see in
the Milky Way halo, are almost certainly more useful than tails from
major mergers in probing the potential of an individual galaxy. The reason is that
major mergers must involve the catastrophic agglomeration of the dark
matter haloes, wiping out all precursory substructure in favour of the
newly virialized dark halo blob.  Indeed, in the
\cite{white-rees-merging} model of galaxy formation it is the very
occurrence of this dark-matter agglomeration which seals the fate of
the merger.  Hence, any probe of the dark matter distribution
resulting from such a merger would be of comparatively little use in
constraining the dark matter distribution of the precursor objects.
Lastly, the unbound or marginally-bound tail stars from a major merger
feel the gravity of their former hosts only weakly, which therefore
has little effect on their motion, making them less sensitive probes
overall.

The use of Milky Way streams as probes of the Galaxy potential stems
from a recognition of the similarity between the trajectory of a
stream, and the orbital trajectory
of the progenitor object \citep{mcglynn-streams-are-orbits,
  johnston-delineate}.  Observing the trajectory of an orbit places
strong constraints on the gravitational force-field that gave rise to
that trajectory, and hence also constrains the distribution of matter
that generates that field \citep{binney08}. Indeed, it is precisely
the knowledge of orbital trajectories in the Solar system that allows
the mass of the Sun and the planets to be so well constrained.
Similarly tight constraints can be placed on the mass of the Milky
Way's black hole, from the orbital trajectories of S-stars near the
Galactic centre \citep{s-stars}.\footnote{Both of these latter
  problems are somewhat better conditioned than ours, because the
  shape of the applicable potential---i.e.~that of Kepler---is already
  known.}

In recognition of the diagnostic power of orbits,
many recent papers have put some considerable effort into
the attempt to locate the orbits delineated by tidal streams 
\citep{law-modelling,fellhauer-ngc5466,fellhauer-orphan,oden-2009,
willett,koposov}.
Unfortunately, success in this endeavour, to date, has been somewhat limited.
The traditional methods of finding orbits consistent with the data,
namely, to search over a range of initial conditions from which one
integrates the equations of motion, have resulted in fewer
convincing fits to the data than might be expected \citep{eb09b}.

Part of the problem is the enormous space of initial conditions
that must be searched over. Even if one is lucky enough
to stray upon an initial condition which reproduces an orbit
roughly consistent with the data, it is not clear that
a better match could not be found, with a different
initial condition, and perhaps in a different potential.
Ockham's razor requires that we be able to convincingly
falsify any theory of equal or lesser complexity to our own:
to make a strong statement about the Galactic potential,
we must be able to show that {\em no} other orbits in a
given potential are compatible with the data.

``Geometrodynamical'' methods---in the parlance of
\cite{geometrodynamics}---are one answer to this difficulty. Such techniques
utilize additional measurements of the stream, such as line-of-sight
velocity measurements, to place additional constraints upon its
trajectory \citep{jin-reconstruction,geometrodynamics,
  binney08}. These additional constraints substantially reduce the scope
of the problem: solutions are now parameterized
by the form of the potential and perhaps one other measurement.
This substantial reduction in the space that parameterizes solutions
makes an automated search over it feasible. Hence, the orbit consistent
with a stream in a given potential can be isolated; or it can be shown
that the data are incompatible with a given potential.
This first half of the this thesis will advance the work of
\cite{jin-reconstruction} and \cite{binney08} by 
specifying procedures to make such methods robust against errors in input data.
We will also derive a new method, similar to the \cite{binney08} procedure,
but one that uses proper-motion
measurement data as input instead of radial-velocity measurements.

Another issue that affects attempts to constrain the Galactic
potential using streams is the degree to which the latter truly
represent orbits. The belief that they do seems to originate in
empiricist observations, made from the results of N-body simulations
\citep[e.g.][]{mcglynn-streams-are-orbits}.  Latterly, evidence has
come to light, again from N-body simulations, that this belief may not
be strictly true \citep{choi-etal,eb09a}. Indeed, it turns out that the
belief has no basis in classical mechanics, and quite the
opposite is true: generally streams do not delineate orbits.
In the second half of this thesis, we will demonstrate this from first principles.
We will also examine techniques that
attempt to ameliorate the use of such non-orbital stream data
to constrain the Galactic potential.

\section{Overview of this thesis}

This thesis comprises several related studies
in the dynamics of tidal streams around our Galaxy.
The common thread that links these studies is the
desire to exploit observations of tidal streams to
place constraints upon the form of the Galactic potential.

The thesis is laid out in four substantive chapters according to the
descriptions that follow. In addition to those, \chapref{chap:concs}
reviews our findings in the context of astrophysics as a whole.
We also provide some ancillary results to the calculations
of \chapref{chap:mech} in \appref{appendix:stackel}.

% \chapref{chap:radvs} advances the prior
% work of \cite{binney08} by setting out procedures with which the
% orbits underlying tidal streams can be identified using only
% line-of-sight velocity measurements for stars in streams.
% \chapref{chap:pms} complements the work of \chapref{chap:radvs} with a
% procedure for identifying the orbits of streams using only
% proper-motion measurements for stars in streams.  \chap{chap:galplx}
% explores the practicality of utilizing Galactic parallax, a concept
% that arises out of the work of \chapref{chap:pms},

\subsection{\chapref{chap:radvs}: Finding the orbits delineated by tidal streams}

\cite{binney08} and \cite{jin-reconstruction} independently
reported an algorithm for reconstructing full phase-space
trajectories for tidal streams, given only the projection
of a stream's trajectory onto the plane of the sky, and measurements
for the line-of-sight velocities everywhere along the stream.
In this chapter, we demonstrate that the applicability of this
algorithm is limited by errors in the input tracks from the following
sources: likely statistical errors from observations,
and systematic errors due to the fact that streams do not precisely delineate
individual orbits.

We offer a procedure to overcome these difficulties
by specifying a parameter space, which describes
modifications to the baseline input in a way that is likely
to correct for the above-mentioned errors while still remaining
consistent with specified uncertainty in the baseline input.
We then describe procedures to search over this parameter space,
while applying the \cite{binney08} algorithm to isolate those modifications
that correspond to orbits.
In this way, we are able to find orbits consistent with stream
observations, without being hamstrung by errors in input data.

The ultimate goal of our work is to diagnose the Galactic
potential. \cite{binney08} showed that precise measurements
of streams could place stunningly tight constraints on the
potential. We illustrate the extent to which this is possible using
realistic data.

\subsection{\chapref{chap:pms}: Fitting orbits to streams using proper motions}

The major limitation on the work of \chapref{chap:radvs} is the lack
of line-of-sight velocity measurements to distant main-sequence stream stars. Although
obtaining such measurements is within the capability of the technology of
the day, it does require the commitment of 8-m class telescope
time, which is unlikely to be forthcoming soon for more than a few streams.

In this chapter, we present a possible alternative: the probing
of the potential using proper-motion measurements along streams.
Following the same logical schema as is used in \cite{binney08},
we develop an algorithm to reconstruct the orbits of streams
using such measurements.
We find that it is equally efficacious to use proper-motion
measurements of streams to reconstruct orbits and constrain the
Galactic potential.

\subsection{\chapref{chap:galplx}: Galactic parallax}

Measuring distances in our Galaxy is critical to
understanding its structure. However, line-of-sight
distances can typically be measured with only relatively
poor precision, and this lack of precision is manifest
in the most basic of Galactic parameters, for instance,
the distance to the Galactic centre \citep{mb09}.

The gold standard of
Galactic distance estimation is trigonometric parallax.
However, its applicability is effectively limited to nearby
stars. For more distant objects, alternative techniques
such as photometric distance estimation must be used.
Unfortunately, such non-geometric techniques necessarily rely on
assumptions about stellar chemistry and composition that add complexity
and uncertainty to measurements made using them.

The work of this chapter shows that by making use of the known
trajectory of stream stars on the sky, and given accurate enough
proper-motion measurements, it is possible to calculate trigonometric
distances to stars in distant streams.  This effect, which we call
``Galactic parallax'', has a range some 40 times greater than that of
conventional trigonometric parallax, given similarly accurate
measurements of motion on the sky.  We examine in detail the
practicality and the limitations of distance estimation using the
effect, and we demonstrate its utility by accurately computing the distance
to the tidal stream GD-1 \citep{gd1-discovery}.

\subsection{\chapref{chap:mech}: The mechanics of streams}

There has been some noise recently in the literature as to whether
tidal streams can be taken to precisely delineate orbits
\citep{choi-etal,eb09b,eb09a}.  Standard techniques for constraining
the potential by fitting orbits to streams rely upon the assumption that
they can \citep[e.g.][]{newberg-orphan}. In this chapter, we
demonstrate by use of analytical mechanics that streams, in general,
do not delineate orbits. We further show that constraining the
potential by assuming that streams make good proxies for orbits
can lead to serious systematic error.

However, we also show that with relatively simple models of the
phase-space distribution of disrupted clusters, it is possible to
predict perfectly the trajectory of a stream, even when this differs
significantly from the trajectory of a valid orbit. Thus, we
conjecture it may be possible to repair the fitting algorithms, by
having them utilize such stream trajectories instead.

%% file: radial_vs/radial_vs.tex
% preamble stuff for this chapter

\chapter{Finding the orbits delineated by tidal streams}
\label{chap:radvs}

\section{Introduction}

In this chapter, we first recount a procedure, independently
described by \cite{jin-reconstruction} and \citet[herein \b08]{binney08},
that permits the reconstruction of full phase-space information
for an orbital trajectory, given only the assumption of the
host Galaxy potential, and knowledge of both the track of the trajectory
across the sky, and the rest-frame line-of-sight velocity down
the track.

In \b08 it was shown that, in addition to reconstructing phase-space
information, by further examining which of
the reconstructed trajectories (if any) constitute dynamical orbits, it is
possible to utilize such tracks to diagnose the host potential. Given
good enough input data, the precision with which both the trajectory
can be reconstructed, and the potential diagnosed, was shown to be
exquisite: distances and potential parameters are predicted to better
than one per cent, which is far better than anything that has hitherto
been possible with conventional techniques.

This advent of this procedure is particularly exciting in the context of
Galactic astrophysics, since tidal streams from disrupted
satellite galaxies have been held for some time to effectively
delineate the orbit of their progenitor \citep{johnston-delineate,
odenkirchen-delineate}. Thus,
the detailed examination of the kinematics of these structures
may well prove an important method for determining the nature
of the Galactic potential.

The major limitations with the procedure as detailed in
\b08 are two-fold. Firstly, input data derived
from observations will be subject to error, which limits the
degree to which they can accurately represent an orbital track.
Secondly, it has been noted \citep[e.g.][]{choi-etal} that
tidal streams do {\em not} delineate individual orbits.
This compromises a core requirement of the \b08 procedure,
since the input tracks now no longer represent dynamical orbits,
and hence dynamics cannot be used to select the physical reconstructions
from the unphysical ones.

The work of this chapter, much of which has been published
in an article by \cite{eb09a}, is devoted to examining
these limitations, and exploring procedures by which they
may be overcome.

The chapter is laid out as follows. \secref{radvs:sec:recon} recounts
some of the work of \b08, upon which this chapter
draws heavily, and demonstrates its limitations when applied to data for a
simulated tidal stream.  \secref{radvs:sec:specifying} briefly
examines, with the aid of a simulated example, the constraints that
observations of a stream's track place on the orbits of the stream's constituent stars.
\secref{radvs:sec:identify} describes procedures by which we 
identify those
orbits (if any exist) that are consistent with such
constraints. \secref{radvs:sec:test} tests the method, with
the aid of a simulated example. \secref{radvs:sec:potential}
examines our ability to diagnose the host potential.
In \secref{radvs:sec:conclusions} we present our concluding
remarks.

For the remainder of this chapter, the reference frame used is the
inertial frame in which the Galactic centre is at rest; consequently
line-of-sight velocities are obtained by subtracting the projection of
the Sun's motion from the measured heliocentric velocities.  We assume
complete knowledge of the velocity of the Sun with respect to the
Galactic centre throughout most of this chapter.

Except where stated otherwise, orbits and reconstructions are
calculated using the Galactic potential of Model II from
\citet[Table~2.3]{bt08},
which is a slightly modified version of a halo-dominated potential
described by \cite{db98-pot}. We take the distance to the Galactic
centre to be $8\kpc$ and from \cite{reid-brunthaler} (for $V$ and $W$) and
\cite{db98-kinematics} we take the velocity of the Sun in the Galactic rest frame
to be $(U,V,W)=(10.0,241.0,7.6)\kms$.

\section{Reconstructing orbits from tracks on the sky}
\label{radvs:sec:recon}

The following formulation for reconstructing complete phase-space
information from a single point in space, an on-sky track, and the
line-of-sight velocity measurements along that track, was discovered
independently by \cite{jin-reconstruction} and \b08.
Our work draws heavily on the latter formulation, so it is that
formulation which we recount here.

Consider a tidal stream around the Galaxy, which we assume to
delineate an orbit.  Now let $\vect{r}$ be the position vector from
the Sun to a star in this stream, let $\vect{v}$ be the rest-frame
velocity of that star, and let $\Phi(\vect{r})$ describe the potential
of the Galaxy. At this point we assume the track to 
delineate an orbit perfectly, so its trajectory obeys the equations of motion
\begin{equation}
\ddot{\vect{r}} = -\nabla \Phi(\vect{r}) = \vect{F}(\vect{r}),
\end{equation}
where $\vect{F}$ is the acceleration due to the gravity of the Galaxy.
We define $v_r$ as the radial component of velocity,
\begin{equation}
v_r = \rh\cdot\vee = \dot{r},
\end{equation}
and we note that its derivative
\begin{equation}
\dot{\vr} = \fr + \vee\cdot{\d \rh \over \d t},
\label{radvs:eq:vrdot}
\end{equation}
where we have defined $\fr = \vect{F}\cdot\rh$ as that
component of the acceleration $\vect{F}$ along the line of sight.
From the definition of $\vee$ we have
\begin{equation}
\vee = {\d (r \rh) \over \d t} = \vr\rh + r{\d \rh \over \d t},
\end{equation}
which rearranges to
\begin{equation}
{\d \rh \over \d t} = {\left( \vee - v_r\rh\right) \over r}.
\end{equation}
Combining the above expression with \eqref{radvs:eq:vrdot}, we find
\begin{equation}
\dot{\vr} = \fr + {\left(v^2 - \vr^2\right) \over r}= \fr + {\vt^2 \over r},
\label{radvs:eq:dotvr2}
\end{equation}
where $\vect{v}_t$ is the component of the star's velocity
in the plane of the sky. $\vect{v}_t$ must satisfy the relation
\begin{equation}
\vt^2 = \left(r {\d \rh \over \d t}\right)^2 = \ (r\dot{u})^2 \equiv r^2\left(\dot{b}^2 + \dot{l}^2\cos^2b\right),
\label{radvs:eq:vt}
\end{equation}
where $(l,b)$ are the on-sky Galactic coordinates and where we now fix the meaning
of the parameter $u$ to be the angular distance along the track.
Combining \eqref{radvs:eq:dotvr2} and \eqref{radvs:eq:vt}, we obtain
the non-linear ODE
\begin{equation}
{\d v_r \over \d t} = \fr + r\left({\d u \over \d t}\right)^2.
\end{equation}
We can rearrange this equation for $\d t / \d u$ as follows.
Utilizing the chain rule and multiplying through by $(\d t / \d u)^2$,
\begin{equation}
{\d t \over \d u}{\d v_r \over \d u} = F_r\left({\d t \over \d u}\right)^2 + r.
\end{equation}
This equation is quadratic in $\d t / \d u$, and it can be solved for that quantity
\begin{equation}
{\d t \over \d u} = {1 \over 2 \fr}\left(
{\d \vr \over \d u} -
\sqrt{\left(\d \vr \over \d u\right)^2 - 4 \,r \fr}
\right),
\label{radvs:eq:master}
\end{equation}
where the choice of the negative root is made by requiring that $\d t / \d u$
is always positive. 
\Eqref{radvs:eq:master} forms a system of coupled ODEs along with
\begin{equation}
{\d r \over \d u} = \vr {\d t \over \d u},
\label{radvs:eq:master2}
\end{equation}
which follows from the definition of $v_r$ and the chain rule.

Momentarily assume that the host potential $\Phi(\vect{r})$ is known,
and that the line-of-sight velocity $\vr(u)$ is known everywhere along
an on-sky track $[l(u),b(u)]$.
Given a single initial distance $r_0$ to some fiducial point on that
track, \mastereqs can be integrated. The resulting solution
describes a trajectory for which full phase-space information is defined.

% Given a
% host potential $\Phi(\vect{r})$, and a single initial distance
% $r_0$ at some fiducial point on an on-sky track $[l(u),b(u)]$, along which the
% line-of-sight velocity $\vr(u)$ is known, the
% \eqsref{radvs:eq:master}{radvs:eq:master2} have a solution for which full
% phase-space information is defined.

This result was reached independently by \b08 and
\cite{jin-reconstruction}, although the latter did not realize that
only a subset of the solutions to
\eqsref{radvs:eq:master}{radvs:eq:master2} could be dynamical orbits.
If one is to unlock the full diagnostic power of streams, it is
important to isolate those solutions that are dynamical
orbits. Further, if one can show that {\em no} solution of
\eqsref{radvs:eq:master}{radvs:eq:master2} with {\em any} initial
distance $r_0$ is dynamical, then it follows that the assumed form for
the host potential $\Phi(\vect{r})$ must be wrong.  \b08 identified as
dynamical those solutions with minimal rms orbital energy variation down the
track, and for a given set of input data, isolated them all by means
of a comprehensive search over $r_0$.

\subsection{The problem with erroneous data}

\b08 showed that \mastereqs can locate dynamical orbits
with exquisite precision, if given perfect input data in the form
of $[l(u),b(u)]$ and $v_r(u)$. The example in Fig.~2 of \b08
showed this for an orbit in the Miyamoto-Nagai potential
\citep{miyamoto}. However, \b08 also showed that the
ability of the technique to identify dynamical orbits quickly
degrades when the input data are convolved with small random errors.

\begin{figure}
\centerline{
\includegraphics[width=0.5\hsize]{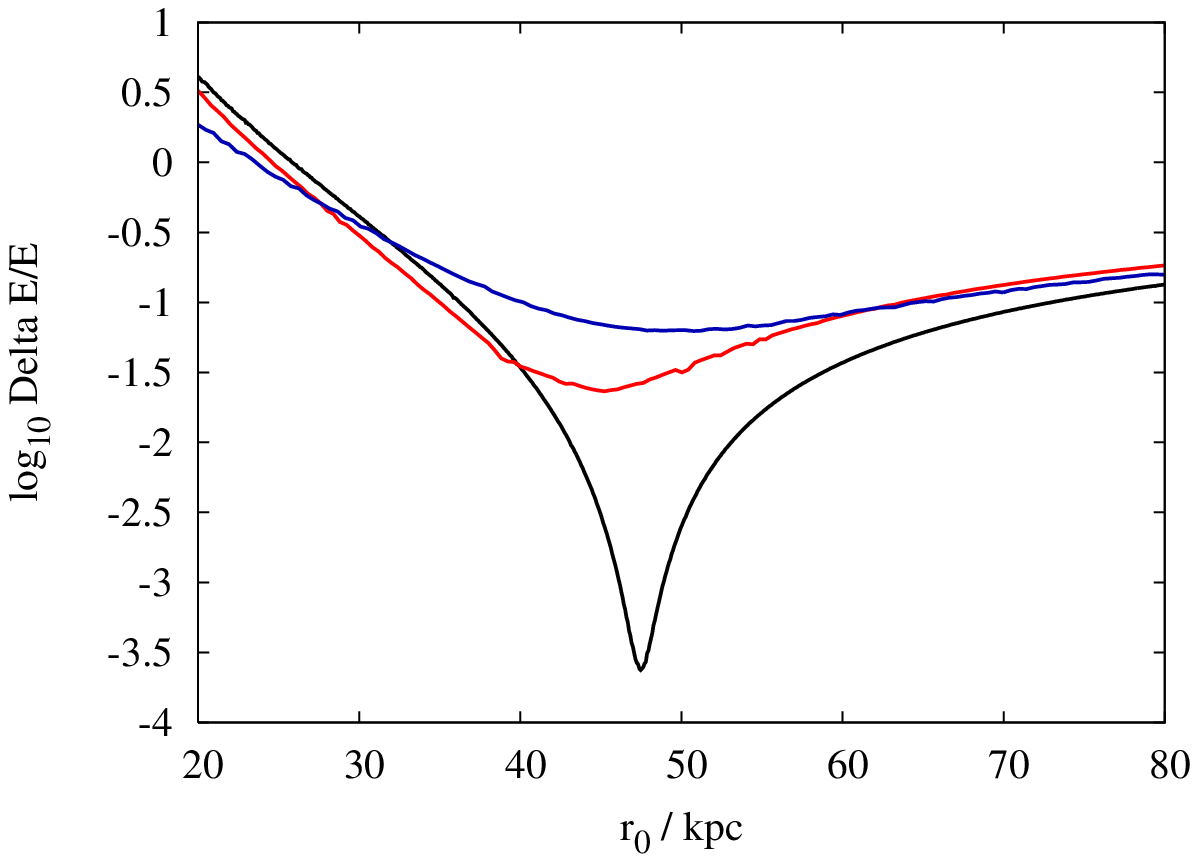}
\includegraphics[width=0.5\hsize]{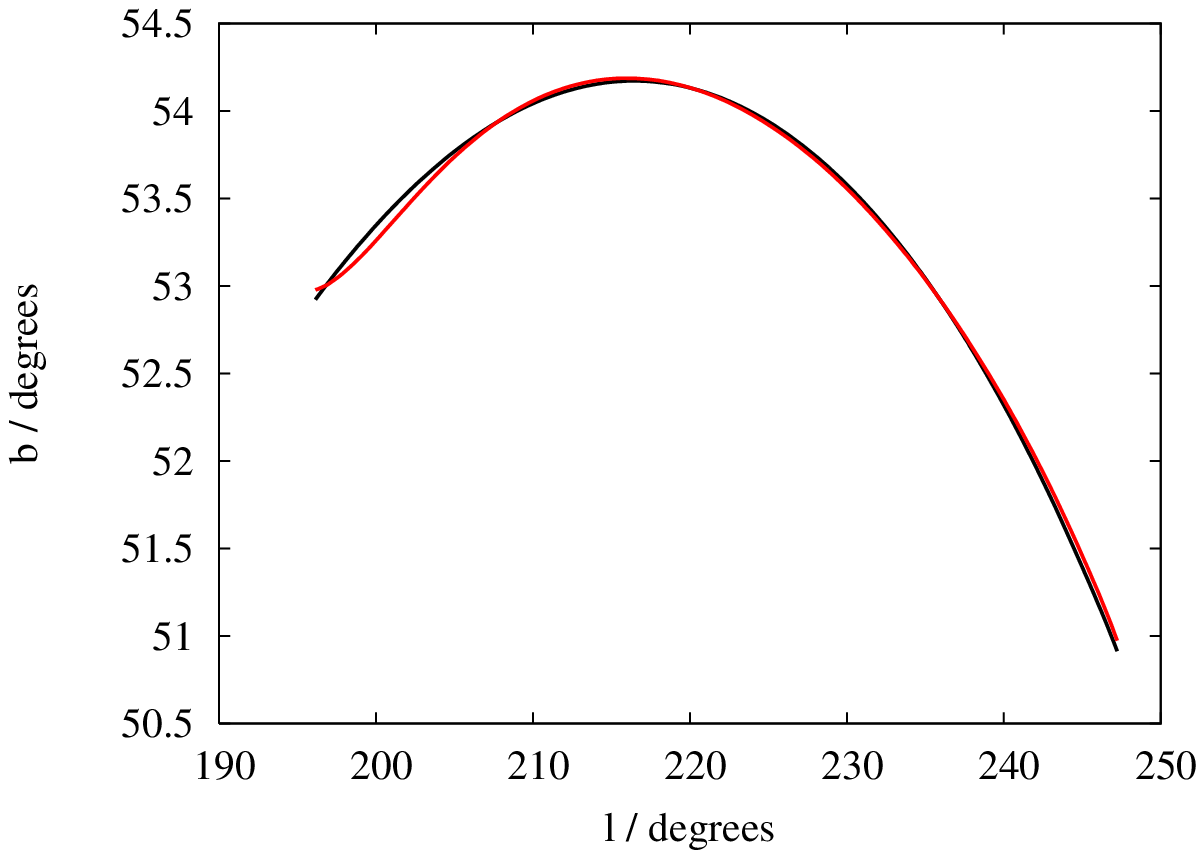}
} 
\caption[Orbital energy conservation for reconstructions using perfect
and erroneous data]
{Left panel: normalized rms orbital energy, vs. initial distance to the
  stream $r_0$, for three input tracks. Right panel: on-sky plot of
  two tracks, used as the [l,b] input data for two of the
  curves in the left panel.  The black curve in both panels is for
  the PD1 Test Orbit described in \tabref{radvs:tab:orbits}. The red
  curve in both panels is as the black curve, but with a small amount
  of random noise added to the input data. The blue curve is for a
  track derived from the simulated Orphan stream, shown in
  \figref{radvs:fig:nbody}. }
\label{radvs:fig:binneytrack}
\end{figure}

\figref{radvs:fig:binneytrack} demonstrates both these effects in the
more realistic Model II potential used throughout this chapter. The
right panel of \figref{radvs:fig:binneytrack} shows two on-sky
tracks. The black track is the projection of a segment of the PD1 Test
Orbit, described in \tabref{radvs:tab:orbits}.  The red track is
derived from the black track, but the input data $[l(u),b(u)]$ and
$v_r(u)$ have been modified by the addition of random fluctuations,
to $b(u)$ and $v_r(u)$, with the
approximate\footnote{ The detail of the random noise is as
  follows.  The input data for the black track were used as the
  baseline input $b_b(l)$ and $v_{rb}(u)$ in the Chebyshev series of
  equations~\blankeqref{radvs:eq:paramtrack}.  The coefficients $a_n$ and $b_n$, with
  $n=(1,10)$, were each randomly sampled from a uniform distribution,
  with maximum permitted values of $0.35\kms$ for the $a_n$ and $3
  \arcmin$ for the $b_n$, respectively.  The noisy red track is
  described by the $b(l)$ and $v_r(l)$ that result from
  equations~\blankeqref{radvs:eq:paramtrack}.}  
amplitude of $3 \arcmin$ and $0.35\kms$, respectively.  The
magnitude of these fluctuations is, respectively, smaller than the
on-sky width of the narrowest known stream, and smaller than the
velocity-measuring precision of an 8m-class spectrograph-equipped
telescope when observing distant stream stars.
Thus, such fluctuations are plausible estimates
for likely errors introduced by the observational process. We note that the resulting
red track is only marginally distinguishable from the black track,
even at the augmented scale of the right panel of \figref{radvs:fig:binneytrack}.

Using each of the tracks in the right panel, trajectories were
reconstructed using \mastereqs for a range of initial distances
$r_0$. Along each such reconstructed trajectory, the normalized
rms orbital energy variation
\begin{equation}
{\Delta E \over E} \equiv \sqrt{{\langle E^2\rangle \over {\langle E\rangle}^2} - 1},
\end{equation}
was computed, where $E_i = v_i^2/2 + \Phi(\vect{r}_i)$ at a point $\vect{r}_i$ along the track.
The numerical scheme used in the solution of \mastereqs was identical
to that of \b08, save for a minor upgrade to the endpoints
of the splines, detailed in \secref{radvs:sec:identify} below.

The left panel of \figref{radvs:fig:binneytrack} shows $\Delta E/E$
versus initial distance $r_0$, when each of the black and red tracks is
used as input. In the case of the perfect input of the black track, the
correct initial distance $r_0 \simeq 47\kpc$ is identified with
little scope for error, since $\Delta E/E$ reaches a deep and
sharp minimum at that point.  However, the effect of the random
fluctuations in the input data on the ability to identify dynamical
orbits has been very damaging.  The depth of the minimum has decreased by
two orders of magnitude, and the minimum has changed location to $r_0
\simeq 45\kpc$.  Although perhaps the extreme distances of $r_0 \sim
20\kpc$ and $r_0 \sim 80\kpc$ could still be ruled out, there is now
only a marginal basis on which to select a distance at $r_0 \simeq
45\kpc$ over another distance, since a
different random fluctuation could put the easily-moved
minimum elsewhere.  Most disappointingly, all
power to identify dynamical orbits from amongst the reconstructed
trajectories has been lost, since the best trajectory 
only conserves energy to one part in 40, which would wash out
most of the exquisite detail shown in Fig.~3 of \b08, which
was key to the ability of that work to diagnose the potential.

\begin{table}[t]
 \centering
% \begin{minipage}{110mm}
% \centering
   \caption[Highlighted orbits from \chapref{chap:radvs}]
{Parameters of highlighted orbits from this chapter. The
     coordinate system used is right-handed with $\hat{x}$ pointing
     away from the Galactic centre and $\hat{y}$ opposite the sense of
     Galactic rotation.
}
  \begin{tabular}{l|lll}
  \hline
  & position (x,y,z)$\rm / kpc$&
  velocity (x,y,z)$/\!\kms$\\
 \hline\hline
 N-body Orphan & $(28.1, -10.0, 34.0)$ & $(-89.2, -37.1, -76.2)$\\
 PD1 Test Orbit & $(35.5, 7.80, 37.8)$ & $(-7.97, 56.3, 47.8)$\\
\hline
\end{tabular}
  \label{radvs:tab:orbits}
%\end{minipage}
\end{table}

We therefore conclude that the \b08 procedure does
not cope well with likely observational error.

\subsection{The problem with real streams}
\label{radvs:sec:problemstreams}

The previous section, and the work of \b08 that it recaps,
are predicated on the assumption that streams precisely delineate
orbits. Recent studies involving N-body simulations of such streams
\citep[e.g.][]{dehnen-pal5,choi-etal} have made it clear that they
do not. \chapref{chap:mech} of this thesis investigates the mechanics
of tidal stream formation in some detail, and concludes with the
ability to predict the tracks of tidal streams with high precision.
However, to motivate the work of this chapter, which was reported by
\cite{eb09a} before the work of \chapref{chap:mech} was undertaken, we
content ourselves with the examination of an N-body simulation of a
stream superficially similar to the Orphan stream of
\cite{orphan-discovery}.

\begin{figure}
\centerline{
\includegraphics[width=0.5\hsize]{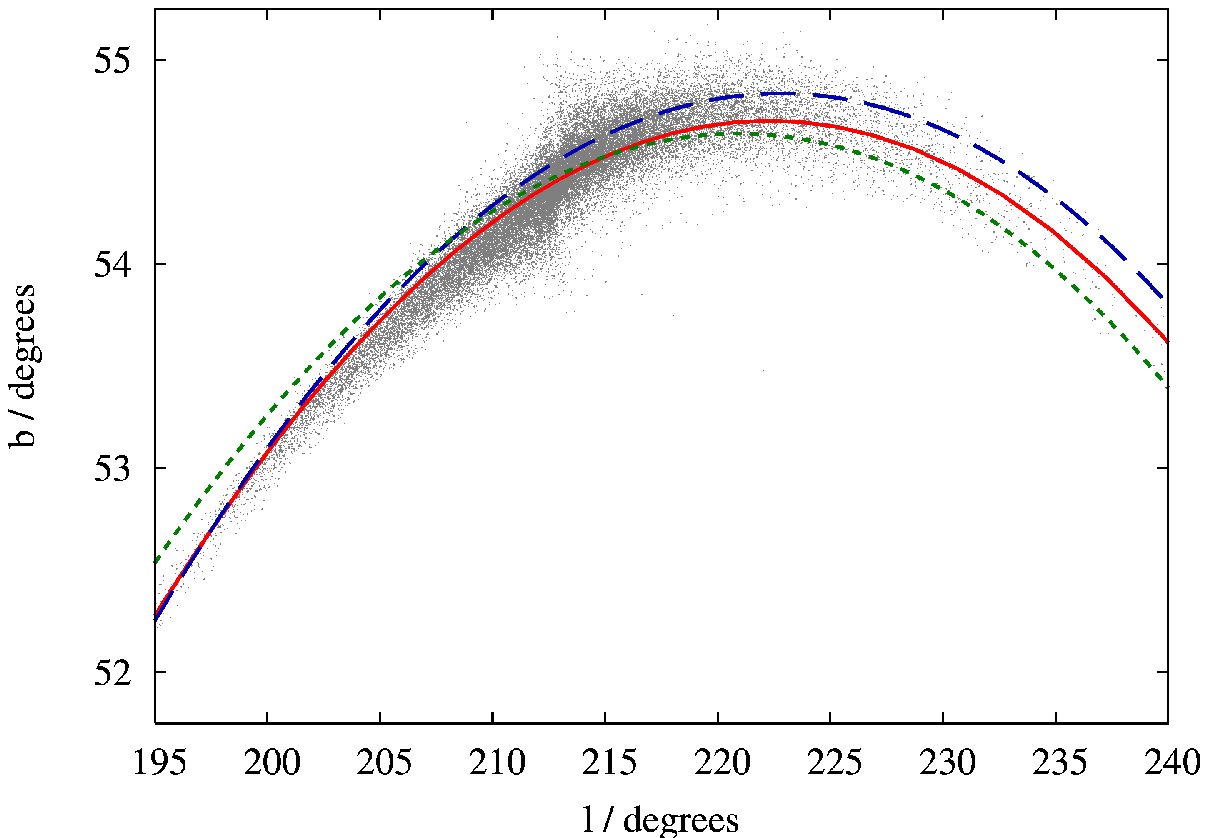}\quad
\includegraphics[width=0.5\hsize]{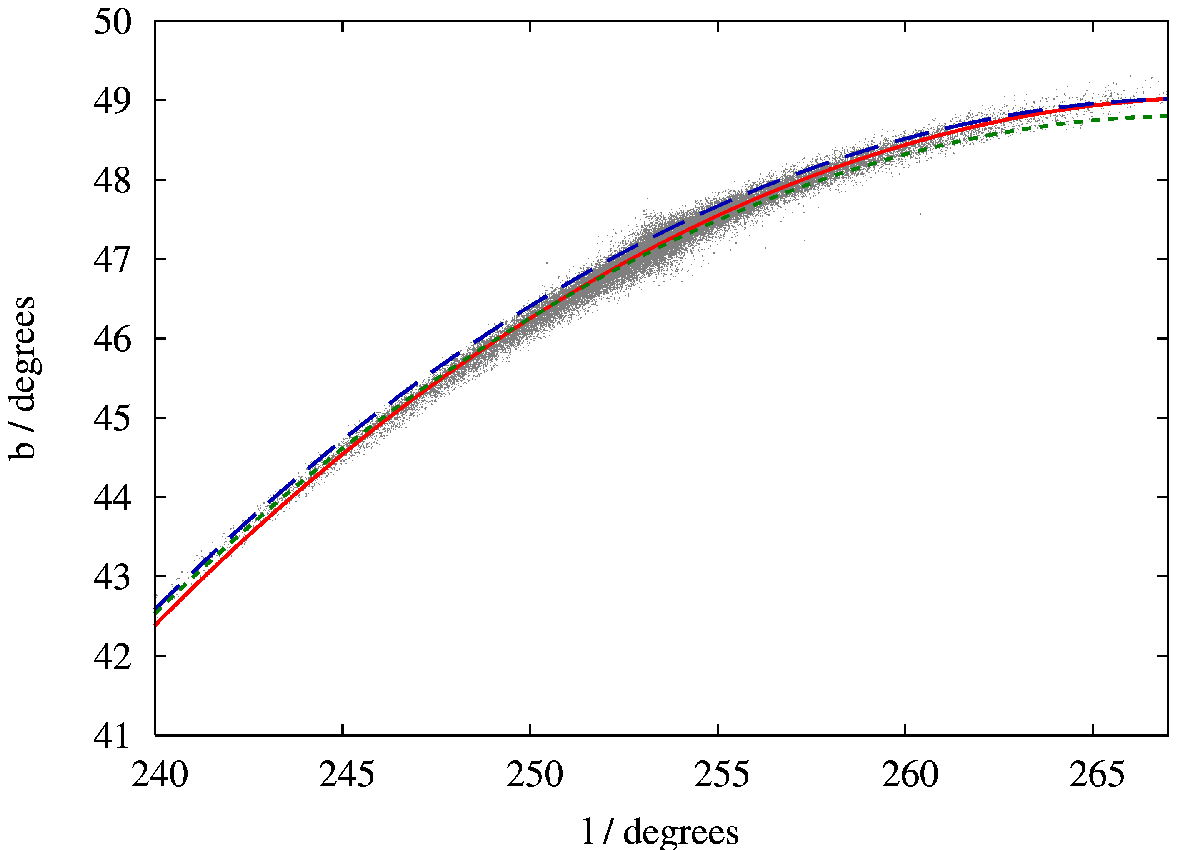}
}
\centerline{
\includegraphics[width=0.5\hsize]{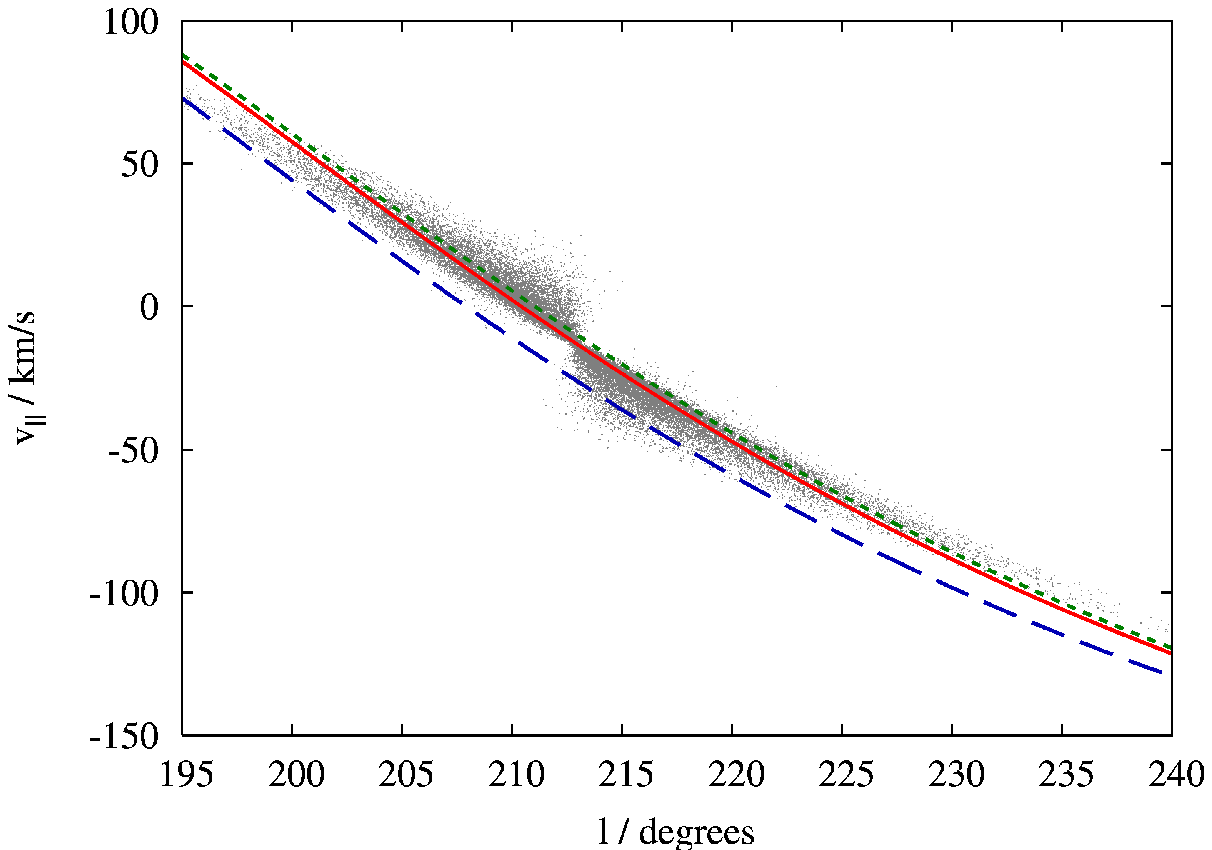}\quad
\includegraphics[width=0.5\hsize]{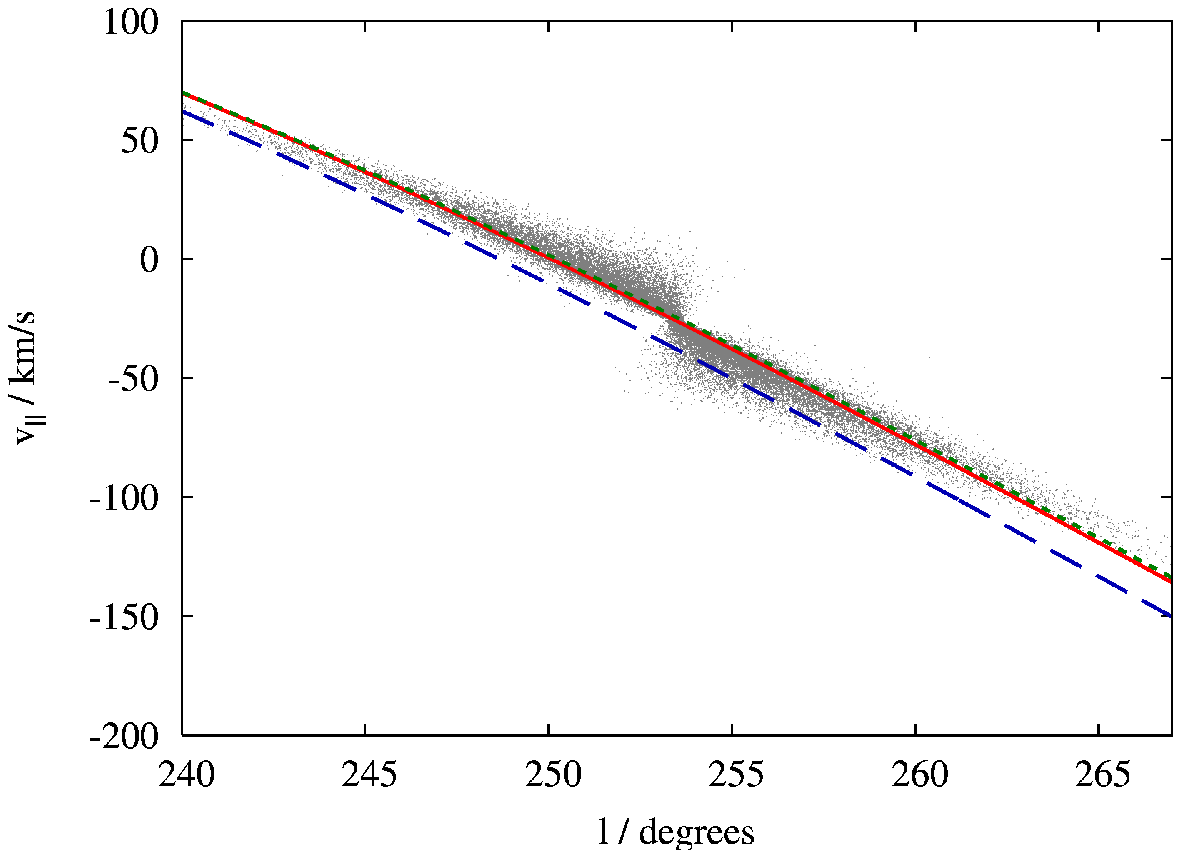}
}
\caption[On-sky projection and line-of-sight velocities for an N-body
simulation of the Orphan stream, from two perspectives]
{Full (red) lines: the orbit of a progenitor of an Orphan-like
  stream.  Broken (green/blue) lines: orbits of a star now seen
  at either end of the tidal tail.  Points: particles tidally stripped
  from an N-body model of the Orphan-like progenitor.  Upper panels:
  distributions on the sky; lower panels: line-of-sight
  velocities. The N-body model had 60,000 particles set up as a
  King model with $W=2$, $r_0 = 13.66 \, {\rm pc}$ and $ M_0 = 9381
  M_{\sun}$, on the orbit detailed in \tabref{radvs:tab:orbits}, and
  evolved for $9.43\Gyr$. The particles were advanced in time by the
  \fvfps\ tree code of \cite{fvfps}.}
\label{radvs:fig:nbody}
\end{figure}

\begin{figure}
\centerline{\includegraphics[width=0.5\hsize]{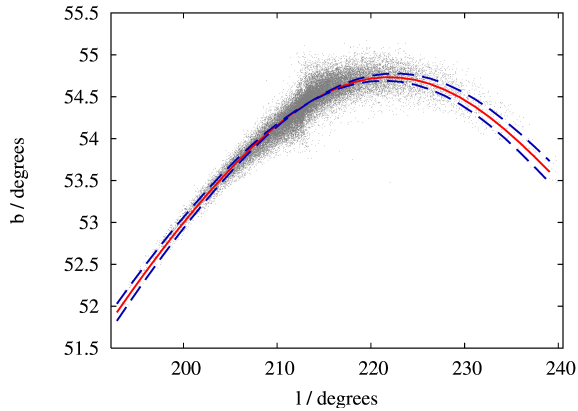}}
\caption[Baseline input for the on-sky stream track of pseudo-data sets PD2--PD7]
{ The dots are the projection of the simulated Orphan stream
  particles onto the sky.  The red curve is a smooth polynomial fit to
  a set of points selected by eye to lie down the stream track.  An
  input track $b(l)$ was defined by sampling 30 points from this
  curve, which was then used as input for the blue line in
  \figref{radvs:fig:binneytrack}, and as baseline input for the data
  sets PD2--PD7.  The blue dashed lines are the upper and lower bounds
  for PD2--PD7, as set by the penalty function $p_{\rm pos}$
  (equation~\ref{radvs:eq:ppos}), and therefore represent the
  bow-tie region for these data sets.  }
\label{radvs:fig:sky-input}
\end{figure}

The full red curves in \figref{radvs:fig:nbody} show projections of an
orbit (described in \tabref{radvs:tab:orbits}) outwardly similar to
that underlying the Orphan Stream, from two viewing locations: the
position of the Sun and a position $120^\circ$ further round the Solar
circle. Also shown in each projection are the locations of particles
tidally stripped from a self-gravitating N-body model of a cluster
launched on to that orbit. Clearly the particles provide a useful
guide to the orbit of the cluster, but they do not precisely delineate
it.  Moreover, the relationship of the projected orbit to the stream
depends on viewing angle. The line-of-sight velocities of the
particles have a similar relationship to the orbit's line-of-sight
velocity.  Hence even with perfectly error-free observations, the
track of a stream will not coincide with the progenitor orbit.

What would be the result of attempting to utilize the track of this
stream in the procedure of \b08? A set of points $\{l,b\}$ and
$\{l,v_r\}$ was selected from the simulation output, by eye, to lie
down the middle of the stream. These sets were each fitted with a
low-order polynomial to ensure smoothness, and these polynomials were
sampled at 30 points to produce a set of input data, $b(l)$ and
$v_r(l)$. The polynomial curves can be seen, along with the simulation
data from which they derive, in \figstworef
{radvs:fig:sky-input}{radvs:fig:vr-input}.

\begin{figure}
\centerline{
  \includegraphics[width=0.5\hsize]{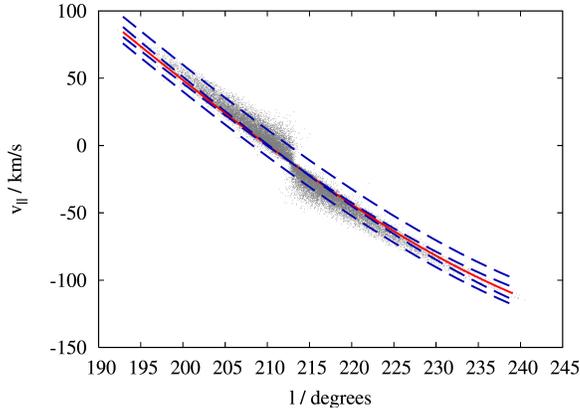}
}
\caption[Baseline input for the line-of-sight velocities of pseudo-data sets
PD2--PD4]
{ The dots are line-of-sight velocities for the particles of the
  simulated Orphan stream. The red curve is a smooth
  polynomial fit to a set of points selected by eye to lie down the
  stream track. An input track $v_r(l)$ was defined by sampling 30 points
from this curve, and was used as input for the blue line in
\figref{radvs:fig:binneytrack}, and as baseline input for the data sets
  PD2--PD4. Also shown, as dashed blue lines, are the bow-tie regions for
  PD2 (narrow) and PD4 (wide). The bounds of these regions are enforced by the penalty
  function $p_{\rm vel}$ (equation~\ref{radvs:eq:pvel}). }
\label{radvs:fig:vr-input}
\end{figure}

The input data were used to solve the reconstruction equations
\blankmastereqs for a range of initial distances $r_0$.
The blue curve in the left panel of \figref{radvs:fig:binneytrack}
shows $\Delta E/E$ for the reconstructed trajectories.
Unlike with the black curve, the minimum in the blue curve is
very shallow, with energy conservation being half an order of magnitude worse
than that of the erroneous input data, and $2 {1\over2}$ orders
of magnitude worse than that of the perfect input data. It is not
possible to make any statement about the true distance
to the stream, or the validity of the potential, from the blue
curve in \figref{radvs:fig:binneytrack}.

Quite apart then from uncertainties in the observational
data, the failure of tidal streams to properly delineate orbits has been
shown to limit the diagnostic power of the \b08 procedure.
Therefore, in order to fully exploit the dynamical potential of streams, we
have to understand how to infer the location of an underlying orbit
from measurements of the stream.

For the moment we assume that any
errors in measured quantities are negligible, which in practice means
that they are small compared to the intrinsic width of the
stream. This condition will certainly be satisfied by the on-sky
coordinates. It will not always be satisfied by the velocity data, so
we address the issue of velocity errors below.

\section{Specifying stream tracks}
\label{radvs:sec:specifying}

With what precision could the track of an orbit be specified from the
positions of the particles in \figref{radvs:fig:nbody}? First we need
to be clear that {\em any} orbit will do. Generally it will be
convenient to use the orbit that passes through some point that lies
near the centre of the observed stream, both on the sky and in
line-of-sight velocity. In some circumstances this orbit will closely
approximate the orbit of the centroid of the stream's
progenitor, but there is no requirement for it to do so. Since a
single fiducial point on the orbit can be chosen at will, this point is associated
with vanishing uncertainty. As one moves away from this point, either
up or down the stream, it becomes more uncertain where the chosen
orbit lies, and the size of the uncertainties must increase. Hence the
region of $(l,b,v_r)$ space to which the orbit is confined by
the observations is widest at its extremities and shrinks to a
point at its centre.\footnote{
Observational error in $v_r$ may lead us to demand that the orbit runs through
a fiducial point that is not justified by the positional data. In this case,
by making such a demand we may unfairly exclude from consideration
the very orbit that we seek. In extreme cases, we may exclude all orbits,
which would result in the erroneous rejection of the correct host potential.
To avoid this, in practical use, the uncertainty
associated with $v_r$ for the orbit at the fiducial point will
not be vanishing, but instead will be at a minimum and equal to the
observational uncertainty on the $v_r$ measurements themselves.
}
We call this the ``bow-tie region.''

In the top panels of \figref{radvs:fig:nbody} the leading and trailing streams are
not offset for much of the span, so our first guess may be that the orbit of the
centroid runs right down the centre of the stream.  In the lower
panels the stream has a kink at the progenitor and our best guess is that the
orbit through the point where the two halves of the tail touch runs near the
lower edge of the left-hand half and near the upper edge of the right-hand
half. In every case the uncertainty in the location of the orbit grows from
zero at the centre to roughly half the width of the stream at
its ends. Quantitatively, the largest uncertainty is then $0.15\deg$ for
the top-left panel, $\sim0.25\deg$ for the top-right panel, and  $\sim5\kms$
in the bottom panels.

\section{Identifying dynamical orbits}
\label{radvs:sec:identify}

\b08 showed that given an orbit's projection onto the sky
$[l(u),b(u)]$ and the corresponding line-of-sight velocities
$v_r(u)$, the remaining phase space coordinates can be recovered by
solving the \diffeqs.

If the input data used to solve those equations are not
derived from an orbit in the same force field as is used to derive
$F_r$, the reconstructed phase-space coordinates will not
satisfy the equations of motion. \b08 observed that violation of
the equations of motion might cause the reconstructed solution to
violate energy conservation, and therefore used rms energy
variation down the track as a diagnostic for the quality of a
solution. However, energy conservation is necessary but not sufficient
to qualify a track as being an orbit. Here we construct a diagnostic
quantity from residual errors in the equations of motion themselves,
since orbits are defined to be solutions of these equations.

We first derive the equations of motion. In the Galactic rest-frame, the
canonically conjugate momenta to the Galactic coordinates $(r, b, l)$ are
 \begin{align}\label{radvs:eq:momenta}
p_r &= \dot{r},\nonumber\\
p_b &= r^2 \dot{b},\\
p_l &= r^2 \cos^2b \, \dot{l},\nonumber
\end{align}
and the Hamiltonian is
\begin{equation}
{\rm H} = \frac{1}{2}p_r^2 + \frac{1}{2}\frac{p_b^2}{r^2}
+ \frac{1}{2}\frac{p_l^2}{r^2 \cos^2 b} + \Phi(r, b, l).
\end{equation}
The equations of motion are therefore
\begin{align}\label{radvs:eq:motion}
\dot{p_r} &= \ddot{r} = \frac{p_b^2}{r^3} + \frac{p_l^2}{r^3 \cos^2 b} - 
\frac{\partial \Phi}{\partial r},\nonumber\\
\dot{p_b} &= \ddot{b}r^2 + 2 \dot{b} \dot{r} r =
- \frac{p_l^2}{r^2}\frac{\sin b}{\cos^3 b} - \frac{\partial \Phi}{\partial b},\\
\dot{p_l} &= r^2 \ddot{l} \cos^2 b  - 2 r^2 \dot{l} \dot {b} \, \sin b \, \cos b +
2 r \dot{r} \dot{l} \cos^2 b = -\frac{\partial\Phi}{\partial l}.\nonumber
\end{align}

As in \b08, when solving \mastereqs we make extensive use of
cubic-spline fits to the data. In the examples presented in \b08
natural splines were used in order to avoid specifying the gradient of
the data at its end points. Significantly improved numerical accuracy
can be achieved by taking the trouble to specify these gradients
explicitly.  Given input data, we estimate
the quantity ${\d l}/{\d b}$ at the end points by fitting a quadratic
curve through the first three and last three points.  ${\d l}/{\d b}$
is then computed at the location of the middle point of each set, and
the very first and very last points are considered `used' and thrown
away. This quantity is then used in the geometric relation
 \begin{equation}
 \frac{\d u}{\d b}= \pm \sqrt{ 1 + \cos^2 b \left(\frac{\d l}{\d b}\right)^2},
 \end{equation}
 to compute
${\d b}/{\d u}$ at the ends of the track. The sign ambiguity is resolved by
inspection of the directionality of the input data. We then use the geometric
relation
 \begin{equation}
 \frac{\d l}{\d u} = \pm \sec b \sqrt{ 1 -
  \left(\frac{\d b}{\d u}\right)^2},
\end{equation}
to obtain ${\d l}/{\d u}$ at the ends of the track, where the sign ambiguity is
resolved in the same way. We are now able to fit cubic splines through the
input tracks, with the slopes at the ends of the $l(u)$ and $b(u)$ tracks
given as above, but at this stage the track of $v_r(u)$ is fitted
with a natural spline. The reconstruction \mastereqs are
now solved for $t(u)$, which is then fitted with a cubic spline, with the
slopes at the ends given explicitly by \mastereqs themselves.

We can now compute $l(t)$ and  $b(t)$ and fit splines to them, with the
slopes at the ends computed from ${\d l}/{\d u}$ and ${\d b}/{\d u}$
by the chain rule. The momenta \blankeqref{radvs:eq:momenta} are now calculated
explicitly, using the derivatives of the splines for $l(t)$ and $b(t)$ in place
of $\dot{l}$ and $\dot{b}$. The slopes at the endpoints,
${\d v_r}/{\d u}$, can now be calculated from \eqref{radvs:eq:motion}
and ${\d t}/{\d u}$; the $v_r(u)$ spline is refitted
using these boundary conditions, the reconstruction repeated, and
the momenta recalculated.

To compute a diagnostic quantity, the left- and right-hand sides of
the equations of motion \blankeqref{radvs:eq:motion} are now evaluated
explicitly. For each equation of motion we define a residual
 \begin{equation}
R(t) = \dot{p}_{\rm lhs}(t) - \dot{p}_{\rm rhs}(t).
\label{radvs:eq:residuals}
\end{equation}
These residuals are used to compute, for each equation of motion,
 the diagnostic quantity
 \begin{equation}
D = \log_{10} \left(\frac{\int_{t_1}^{t_2}\d t\, R(t)^2}
{\int_{t_1}^{t_2}\d t\, \dot{p}_{\rm lhs}^2}\right), \label{radvs:eq:diag}
\end{equation}
 where the residuals have been normalized by the mean-square acceleration and
the times
$t_1$ and $t_2$  correspond to the fifth and
fifth-from-last input data points: numerical artefacts from the end regions,
$0 < t < t_1$ and $t_2 < t < t_{\rm max}$, tend to dominate the integrated
quantity and are not easily reduced by modifying the input; these regions are
therefore excluded.  The largest of the three values for $D$ is used as
the diagnostic quantity for that particular input.

\subsection{Parameterizing tracks}
\label{radvs:sec:paramtracks}

Our strategy for identifying a stream's underlying orbit is to compute the
diagnostic $D$ \bracketeqref{radvs:eq:diag} for a large number of candidate tracks,
and to find which candidates yield values of $D$ consistent with their being
dynamical orbits.

We start by specifying a baseline track across the sky $[l_{\rm b}(u'),b_{\rm
b}(u')]$, where $u'$ is a parameter that increases monotonically down the
track from $-1$ to $1$.  Similarly, we specify associated baseline line-of-sight velocities
$v_{r\rm b}(u')$. The baseline track is required to come closer to
every data point than the
given uncertainty at that point. 

All candidate tracks should be smooth because orbits are. We satisfy this
condition by expressing the difference between the baseline track and a
candidate track as a low-order polynomial in $u'$.
For  streams that cover a wide range of longitudes, 
the parameterization of candidate tracks is achieved
by slightly changing the values of $b$ and $v_r$ associated with a
given value of $l$ from the values specified by the  baseline functions.
That is we write
 \begin{align}\label{radvs:eq:paramtrack}
b(u')& = b_{\rm b}(u') + \sum_{n = 0}^N b_n T_n(u'),\nonumber\\
v_r(u')& = v_{r\rm b}(u') + \sum_{n = 0}^N a_n T_n(u'),
\end{align}
where $T_n$ is the $n^{\rm{th}}$-order Chebyshev polynomial of the
first kind and $a_n$ and $b_n$ are free parameters.  These $2N$
parameters are coordinates for the space of tracks that we have to
search for orbits. When a stream does not stray far from the Galactic
plane, candidate tracks are best parameterized by adjusting the
baseline values of $b$ and $v_r$ at given longitude.  In all examples
in this chapter, the series in equations
\blankeqref{radvs:eq:paramtrack} are truncated after $N=10$. A larger
number of terms allows the correction function to produce tracks that
represent orbits better, while using insufficient terms will fail to afford
sufficient flexibility for the search procedure to find any orbits at all.
However, using more terms makes the search procedure
computationally more expensive, because the volume of the parameter space in
which the search takes place grows exponentially with the number of
terms, and locating solutions in this enlarged space requires commensurately
more effort. The number of terms used is therefore a compromise between
computational expense and the ultimate efficacy of the algorithm.

The space of tracks is defined by the $a_n$ and $b_n$ and
one extra parameter, the distance to the stream, $r_0$, at the
starting point $u=0$ for the integration of
\mastereqs.

We shall henceforth denote a point in the ($2N+1$)-dimensional space of
parameters by $\chi$.  Each $\chi$ is associated with a complete
specification of all six phase-space coordinates for every point on the
candidate orbit: $l$, $b$ and $v_r$ follow from the parameterization and the
remaining coordinates are obtained by solution of the \diffeqs.
Consequently, each $\chi$ corresponds to a value of
$D$ \bracketeqref{radvs:eq:diag} that quantifies the extent to which the
phase-space coordinates deviate from a dynamical orbit in the given potential.

\subsection{Searching parameter space}

Dynamical orbits are found by minimizing the sum 
\begin{equation}\label{radvs:eq:Dtot}
D' (\chi)=D(\chi)+p(\chi),
\end{equation}
 where $p(\chi)$ is the sum of the penalty functions:
\begin{equation}
p(\chi) = \sum_i p_{i,\rm pos} + \sum_i p_{i,\rm vel} + p_{s},
\label{radvs:eq:penaltyfn}
\end{equation}
where
\begin{equation}\label{radvs:eq:ppos}
p_{i,\rm pos} =
\begin{cases}
\Delta_{i,\rm pos}&\text{if $\Delta_{i,\rm pos}>1$},\\
0&\text{otherwise},
\end{cases}
\end{equation}
 with
 \begin{equation}
\Delta_{i,\rm pos}={\left| b(l_i) - b_{\rm b}(l_i) \right|\over\delta b
(l_i)}.
\end{equation}
 Here $\delta b_i$ is the width in $b$ of the bow-tie region at $l_i$. Similarly
\begin{equation}\label{radvs:eq:pvel}
p_{i,\rm vel}(l) =
\begin{cases}
\Delta_{i,\rm vel}&\text{if $\Delta_{i,\rm vel}>1$},\\
0&\text{otherwise},
\end{cases}
\end{equation}
 with 
\begin{equation}
\Delta_{i,\rm vel}={\left| v_r(l_i) - v_{r\rm b}(l_i) \right|
\over\delta v_r (l_i)}.
\end{equation}
 Prior information about the
distance to the stream is used by specifying the penalty function
$p_{\rm s}$ to be
\begin{equation}\label{radvs:eq:defsps}
p_{\rm s} =
\left\{
\begin{array}{ll}
\beta \left| r_0 - r_{0\rm b}\right|/\delta r &
\quad\mbox{if $\left| r_0 - r_{0\rm b} \right| > \delta r$}, \\
0 &\quad\mbox{otherwise},
\end{array} \right. 
\end{equation}
 where $\delta r$ is the half-width of the allowed range in the distance
$r_0$ to the starting point of the integrations and $r_{0\rm b}$ is the
baseline value of $r_0$. These definitions are such that $p(\chi)=0$ so long
as the track lies within the region that is expected to contain the orbit,
and rises to unity, or in the case of $p_{\rm r}$ to $\beta$, on the boundary
and then increases continuously as the orbit leaves the expected region.

In practical cases the prior uncertainty in distance is large, and the
obvious way to search for orbits is to set $\delta r$ to the large value that
reflects this uncertainty and then set the algorithm described below to work.
It will find candidate orbits for certain distances. However, we shall see
below that it is more instructive to search the range of possible distances
by setting $\delta r$ to a small value such as $0.5\kpc$ and searching for
orbits at  each of a grid of values of $r_{0\rm b}$. In this
way we not only find possible orbits, but we show that no acceptable orbits
exist outside a certain range of distances. In  this procedure the  logic
underlying $\delta r$ is very different from that underlying $\delta b$ and
$\delta v_r$.

Since $p(\chi)$ is added to the logarithm of the rms errors in the equations
of motion and since it increases by of order unity at the edge of the bow-tie region,
the algorithm effectively confines its search to the bow-tie region, where
$p=0$. Thus at this stage we do not discriminate against orbits that graze
the edge of the bow-tie region in favour of ones that run along its centre.
Our focus at this stage is on determining for which distances dynamical
orbits can be constructed that are compatible with the data.  Once this has
been established, distances that lie outside some range can be excluded from
further consideration.

The space of candidate tracks $\chi$ is 21-dimensional, so an
exhaustive search for minima of $D'$ \bracketeqref{radvs:eq:Dtot} is
impractical. Furthermore, the landscape specified by $D'$ is
complex. Some of this complexity is physical; the space should contain
continua of related orbits, and ideally $D' \to-\infty$ at
orbits. Hence deep trenches should criss-cross the space.
Superimposed on this physical complexity is a level of numerical noise
arising from algorithmic limitations in the computation of $D'
(\chi)$. The limitations include the use of finite step sizes in the
solution of \mastereqs and the subsequent evaluation of $D' (\chi)$,
the inability of a finite series of Chebyshev polynomials to precisely
describe a true orbital track, as well as the difficulty in
representing this series with a collection of sparse input points
interpolated with splines. Reconstructed tracks are therefore never
perfect orbital trajectories, even when the method is initially
presented with error-free input data, and this is manifest as a
non-zero minimum residual for each equation of motion. In
practice, this minimum residual sets a lower limit on the returned values of
$D' (\chi)$, which we refer to as the ``numerical-noise floor''.

On account of the complexity of the landscape that $D' $ defines, ``greedy''
optimization methods, which typically follow the path of steepest descent,
are not effective in locating minima. The task effectively becomes one of
global minimization, which is a well studied problem in optimization.

We have used the variant of the Metropolis ``simulated annealing'' algorithm
described in \cite{press-etal}, which uses a modified form of the downhill simplex
algorithm.  In the standard simplex algorithm, the mean of the values of the
objective function over the vertices decreases every time the simplex
deforms. In the \pressetal\ algorithm the simplex has a non-vanishing
probability of deforming to a configuration in which this mean is higher than
before.  Consequently, the simplex has a chance of crawling uphill out of a
local minimum. The probability that the simplex crawls uphill is controlled
by a ``temperature'' variable $T$: when $T$ is large, uphill moves are likely,
and they become vanishingly rare as $T\to0$. During annealing the value of
$T$ is gradually lowered from an initially high value towards zero.

One vertex of the initial simplex is some point $\chi_{\rm guess}$, and
the remaining $2N+1$ vertices are obtained by incrementing each
coordinate of $\chi_{\rm guess}$ in turn by a small amount. For the
coefficients of the Chebyshev polynomial $T_0$ this increment is
approximately the size of the allowed half widths, $\delta r, \delta
v$ and $\delta b$. Increments for coordinates representing
coefficients of higher-order polynomials $T_n$ are scaled as
$1/n$. The overall size of these increments is therefore set by the
size of the region within which we believe the global minimum to
lie. It is important to note that in each generation of a simplex, the
increments should independently have equal chance of being added to or
subtracted from the values of $\chi_{\rm guess}$ so that no part of
the parameter space is unfairly undersampled. The algorithm makes tens
of thousands of deformations of the simplex while the temperature $T$
is linearly reduced to zero.
This entire process is repeated some tens of times, after which we have a
sample of local minima that are all obtained from $\chi_{\rm guess}$.

We now update $\chi_{\rm guess}$ to the location of the lowest of the
minima just found and initiate a new search.  The entire process is
repeated until the value of the diagnostic function $D' (\chi)$ hits a
floor. When this floor lies higher than the numerical-noise floor, the
attempt to find an orbit that is consistent with the assumed inputs
has been a failure and we infer that no such orbit exists. When the
floor coincides with the numerical-noise floor, we conclude that the
corresponding $\chi$ specifies an orbit that is compatible with the
inputs. An approximate value for the numerical-noise floor for a given
problem may be obtained as follows: given input that perfectly
delineates an orbit in the potential in use, the value of $D'$
returned at the correct distance is approximately the numerical-noise
floor. Conclusive proof that a candidate track with a particular value of
$D'$ is an orbit can be obtained by integrating the equations of
motion from the position and velocity of any  point on the
track and ensuring that the time integration essentially recovers the track.

\begin{figure}
\centerline{\includegraphics[width=0.5\hsize]{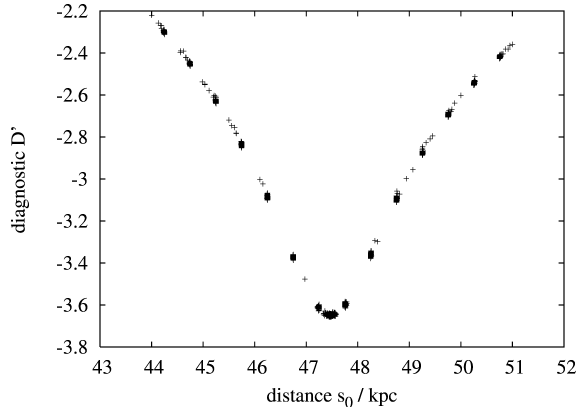}}
\caption[Values of the diagnostic $D'$ for reconstructions using PD1]
{Values of the diagnostic function $D'$
for candidate orbits reconstructed from the pseudo-data set PD1. Each group of crosses
is associated with one of 15 ranges within which the starting distance $r_0$ was
constrained to lie. For each such range the candidate orbit was reconstructed from 280
trial tracks, and each track yields a cross. In this figure the crosses are
largely superimposed because the uncertainties are small and there
is little scope for tweaking the track. This figure can be compared directly
with the black curve of \figref{radvs:fig:binneytrack}, the structure
of which is clearly reflected in these results.}
\label{radvs:fig:pd1}
\end{figure}

On account of the stochastic nature of the algorithm, an attempt to
find a solution at a particular distance occasionally sticks at a
higher value of $D'$ than the underlying problem allows. This
condition is identified by scatter in the values of $D'$ reached on
successive attempts and by inconsistency of these values with the
values of $D'$ achieved for nearby distances---we see from
\figref{radvs:fig:pd1} that the function underlying the minima is smooth. When
the magnitude of this scatter is significant, one can only confidently
declare an attempt to find an orbit a failure if the $D'$
achieved is consistently higher than the noise floor by more than the
scatter; since the diagnostic measure $D'$ quantifies the extent to
which a candidate track satisfies the equations of motion, by
definition, tracks with higher $D'$ than the noise floor plus scatter
cannot represent orbits.

When the observational constraints are weak, we expect several orbits to be
compatible with them. In particular, we will be able to find acceptable
orbits for a range of initial distances $r_0$.  It is therefore important,
for any given input, to run the algorithm starting from many different
values of $r_{0\rm b}$ with $\delta r$ set to prevent the algorithm straying
far from the specified $r_{0\rm b}$. In this way, the full range of allowable
distances can be mapped out, and dynamical orbits found for each distance in
that range. In the case of significant scatter about the noise floor, the
range of distances at which valid orbits are found is the range
within which solutions yield values of $D'$ smaller than the noise floor plus scatter.
Similar degeneracies in the parameters controlling the astrometry and
line-of-sight velocities are less of a  concern because if we have
orbits that differ in these observables, we simply concentrate on the orbit
that lies closest to the baseline track.

\section{Testing the method}
\label{radvs:sec:test}

To test this method, we used the N-body approximation to the
Orphan Stream shown in \figref{radvs:fig:nbody} as our raw data.
The baseline input data $b_b(l)$ and $v_{rb}(l)$ were chosen to be the same
30-point samples of the smooth red curves from
\figstworef{radvs:fig:sky-input} {radvs:fig:vr-input} as was used in
\secref{radvs:sec:problemstreams} to test the \b08 procedure.  To the
baseline data we attached uncertainties $\delta b$ and $\delta v_r$, which
through the penalty functions $p_{\rm pos}$ and $p_{\rm vel}$
(equations~\ref{radvs:eq:ppos} and~\ref{radvs:eq:pvel}) constrain the tracks that the
Metropolis algorithm can try.  Details of the resulting pseudo-data
sets are given below, and are summarised in \tabref{radvs:tab:testtab}.

\begin{table}
 \centering
 %\begin{minipage}{85mm}
  \caption[Configurations of test pseudo-data sets]{Configurations of pseudo-data sets used to test the method.}
  \label{radvs:tab:testtab}
  \begin{tabular}{l|lll}
  \hline
  & $\delta v_{r,{\rm max}}/\!\kms$ & $\delta v_{r,{\rm min}}/\!\kms$ &
  offset $v_r/\!\kms$ \\
 \hline\hline
  PD1 & $2\times 10^{-3}$ & $2\times 10^{-3}$ & 0 \\
  PD2 & 4 & 0 & 0\\
  PD3 & 6 & 2 & 0\\
  PD4 & 10 & 10 & 0\\
  PD5 & 6 & 2 & 2\\
  PD6 & 15 & 10.5 & 10\\
  PD7 & 4 & 0 & 0\\
\hline
\end{tabular}
%\end{minipage}
\end{table}

In one case, PD1, the above baseline input data were replaced by those
of a perfect orbit and $\delta b$ and $\delta v_r$ set very
narrow ($6\,$arcsec and $2\times10^{-3}\kms$) in order to validate the
reconstruction algorithm.

The uncertainty $\delta b(l)$ takes the same value for all the remaining
pseudo-data sets because we assume that the astrometry is sufficiently
precise for the uncertainty in position to be dominated by the offset of  the
stream from an orbit.  In all cases, $\delta b$ has a maximum value of
$0.15\deg$ at the ends of the stream, falling linearly to zero at the
position of the progenitor, consistent with the orbit of the
progenitor seen in \figref{radvs:fig:nbody}. \figref{radvs:fig:sky-input} shows this input
alongside the N-body data from which it was derived.

For the pseudo-data sets PD2 and PD3, $\delta v_r$ is set to a
maximum at the ends of the stream, and falls linearly to a minimum of zero
and $2\kms$ respectively, at the position of the progenitor. These examples
represent a case (PD2) in which the uncertainty in radial velocity is
dominated by stream width, and a case (PD3) in which there is a
significant contribution from measurement error at a level that is easily
obtainable with a spectrograph. The pseudo-data set PD4 represents the case
in which the uncertainty in $v_r$ is dominated by measurement errors:
$\delta v_r$ is held fixed at $10\kms$, which is typical for the
measurement errors in the line-of-sight velocities of SDSS stars in distant
streams.  \figref{radvs:fig:vr-input} shows the input for these data sets.

\begin{figure}
\centerline{
\includegraphics[width=0.5\hsize]{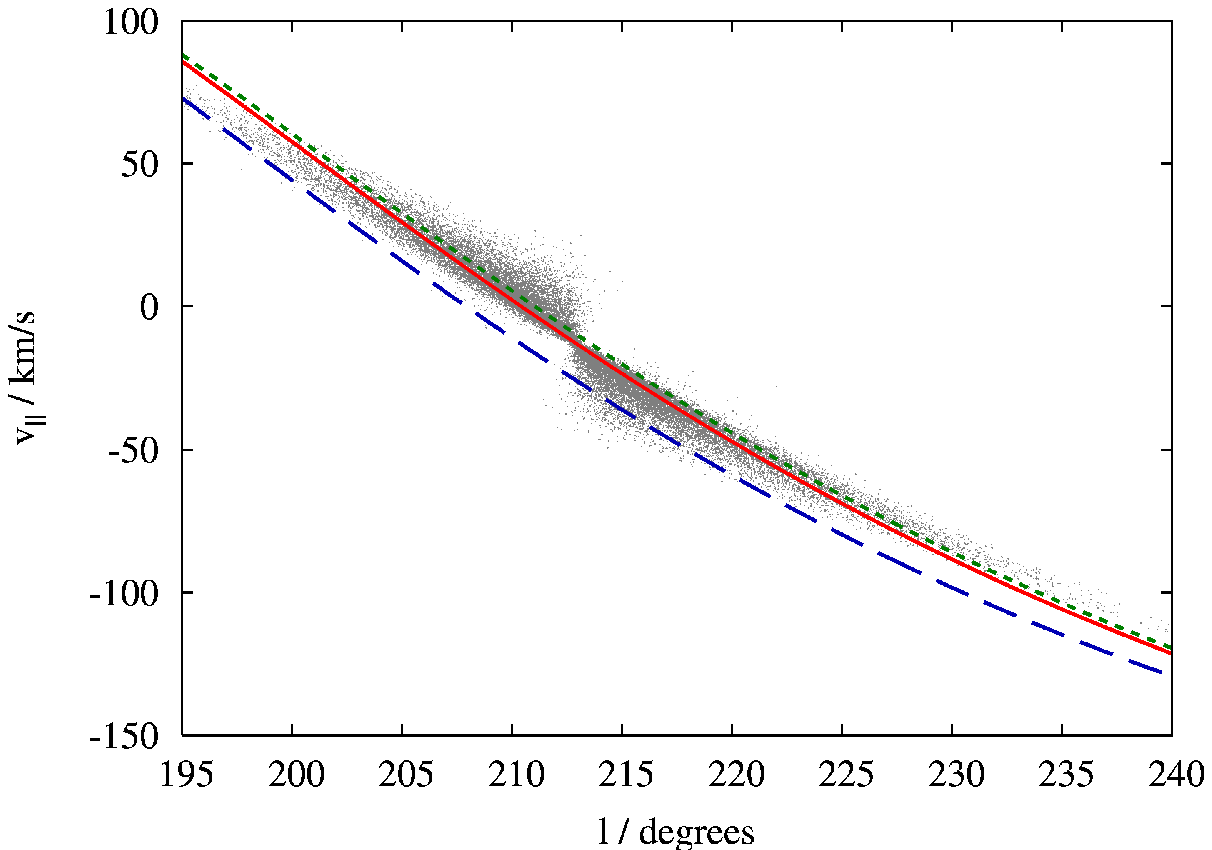}
}
\caption[Baseline input for the line-of-sight velocities of pseudo-data sets PD5--PD6]
{The broken lines show input tracks $v_r(l)$, as used for data sets PD5
  (dotted green) and PD6 (dashed blue). The unmodified input of PD2
  (full red curve) is shown for comparison.  The dots are line-of-sight velocities
  for the particles of the simulated Orphan stream, from which the
  input was derived.}
\label{radvs:fig:vr-wrong}
\end{figure}

For the pseudo-data sets PD5 and PD6, we added to the baseline data
systematic offsets in $v_r(l)$ to mimic systematic errors in radial
velocity. $\delta v_r$ varies between a maximum and a minimum as in
PD2 and PD3, with the values set to encompass the (assumed known) systematic
bias.  \figref{radvs:fig:vr-wrong} shows the input for these data sets.

The pseudo-data set PD7 is identical to that of PD2, except that the
number of raw $(l, v_r)$ points was reduced to just three: one
at either end of the N-body stream, and one at the location of the
progenitor. A quadratic curve was perfectly fitted through these three
points, and sampled at 30 locations to produce the baseline
$v_r(l)$ input. $\delta v_r$ is set to the same
maximum value as PD2 at the outermost points; $\delta
v_r$ is set to zero at the centre point. Only these three points
are allowed to contribute to the penalty function \blankeqref{radvs:eq:pvel} in this
pseudo-data set.

\begin{figure}
\centerline{
\includegraphics[width=0.5\hsize]{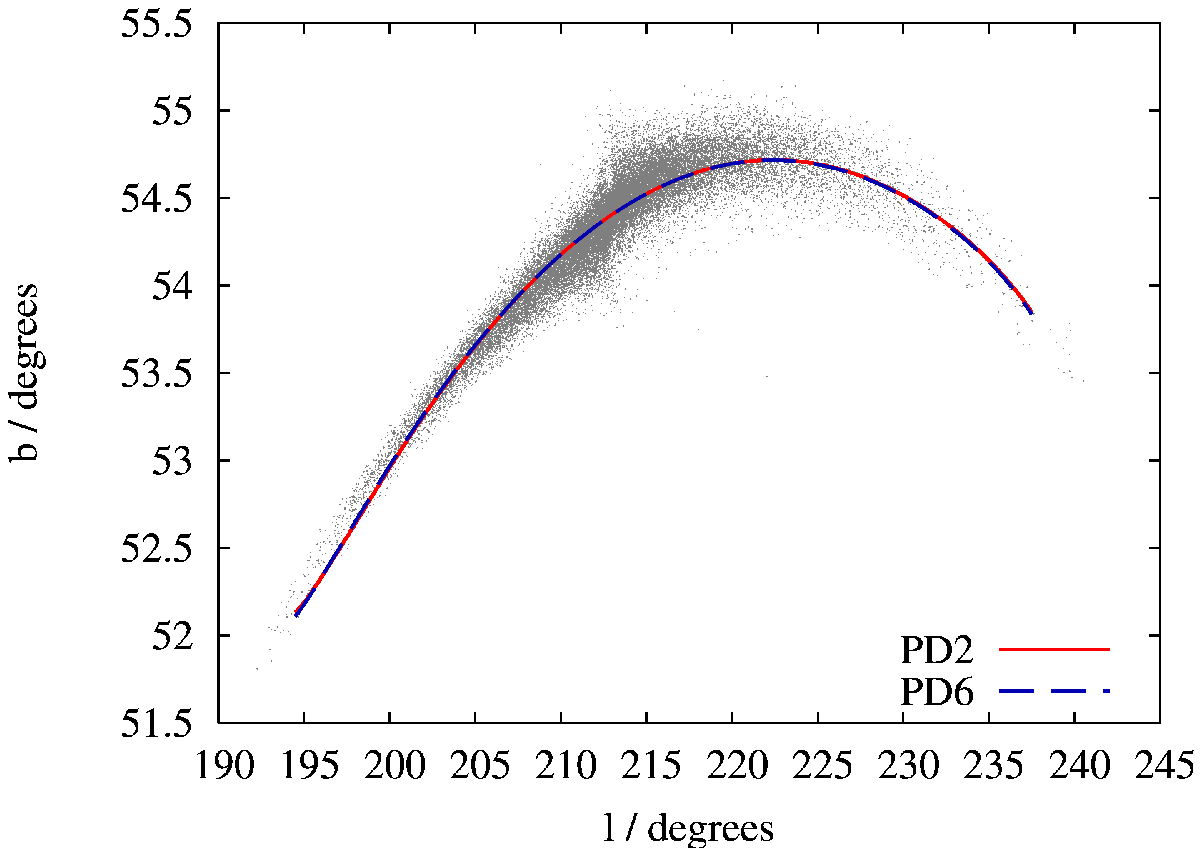}
\includegraphics[width=0.5\hsize]{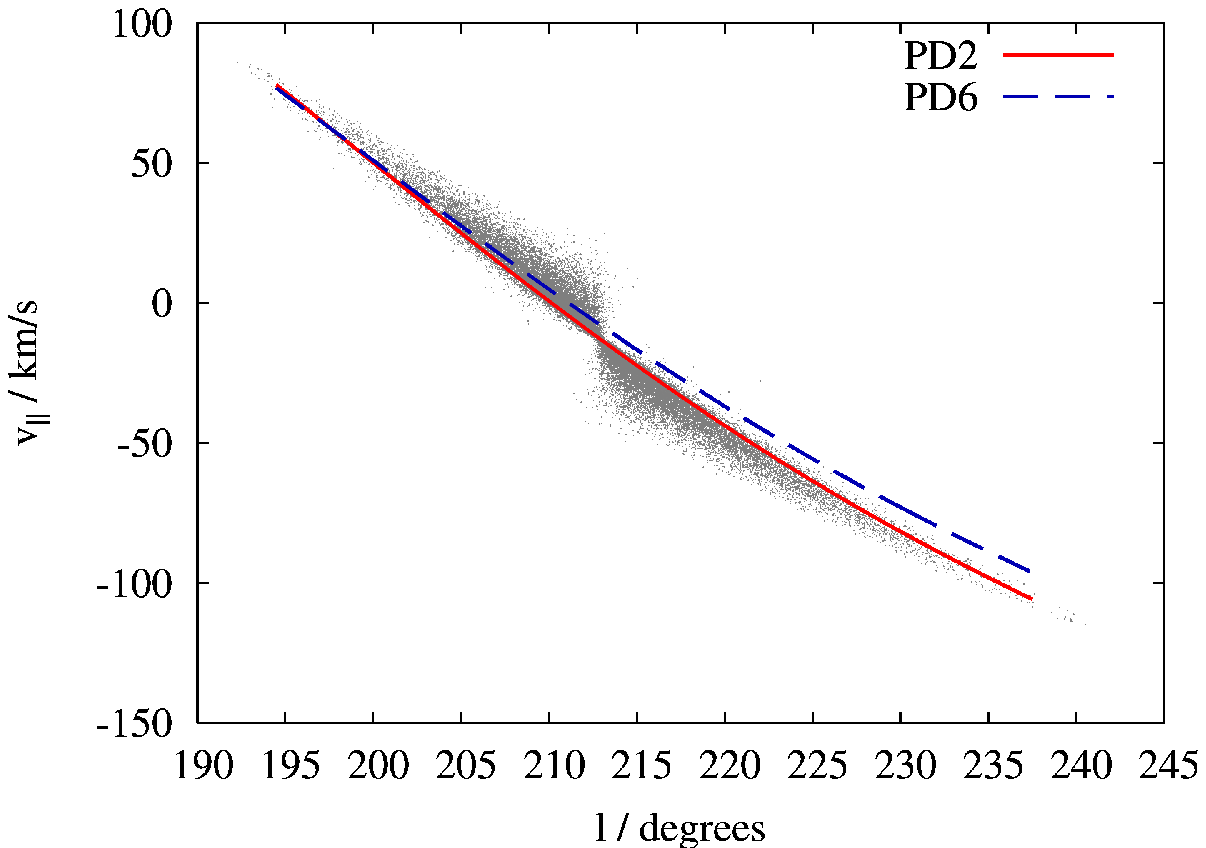}
}
 \caption[On-sky projection and line-of-sight velocities of selected orbits
reconstructed from data sets PD2 and PD6]
{Left panel: projections onto the sky of two candidate orbits and
the N-body data from which the input was derived. The full curves show the
candidate orbit at $46\kpc$ from PD2, and the dashed curves show the
candidate orbit at $43\kpc$ from PD6.  Right panel: line-of-sight velocity down the
track for the same candidate orbits. The penalty function \blankeqref{radvs:eq:penaltyfn}
 forces the sky projection and radial velocity curves of
candidate orbits to be consistent with the data; the equivalent plots for all
other tracks are very similar to these examples.  }
\label{radvs:fig:sky-vr-examples}
\end{figure}

In all of the examples, the penalty function \blankeqref{radvs:eq:penaltyfn} acts
to constrain the candidate tracks to be consistent with the data. The
contribution of the penalty function to $D'$ is therefore zero in all examples.
All candidate tracks are guaranteed to be consistent with the data, even
if they do not represent dynamical orbits. \figref{radvs:fig:sky-vr-examples} provides
example plots of $(l,b)$ and $(l,v_r)$ for candidate tracks from PD2 and
PD6 along with the raw N-body data from which the baseline data are derived.
Equivalent plots for all candidate tracks are similar to these.

For PD1, PD2 and PD3, and each of 15 values of the baseline distance
$r_{0\rm b}$, 280 optimization attempts were made, each involving
Metropolis annealing for 24,000 simplex deformations, from an initial
temperature of 0.5$\,$dex. The starting distances were constrained by
the penalty function $p_s$ \bracketeqref{radvs:eq:defsps} with $\beta = 10^6$
and $\delta r = 0.5\kpc$, so the Metropolis algorithm could only
explore a narrow band in $r_0$. After 40 optimization attempts from a
given starting point $\chi_{\rm guess}$, the starting point was
updated to the end point of the most successful of these
optimizations, and annealing recommenced from a high temperature. In
total 6 of these updates were performed. With these parameters, a
search at a single distance completes in 12 CPU-hours on 3GHz
Xeon-class Intel hardware. PD4, PD5 and PD6 follow almost the same
schema, except that 48,000 simplex deformations were made for each
of 60 attempts at the same $\chi$, which was updated 6 times. A single
distance in the latter case took 36 CPU-hours to search: the
computational load scales linearly with the number of deformations
considered.

\begin{figure}
\centerline{
\includegraphics[width=0.5\hsize]{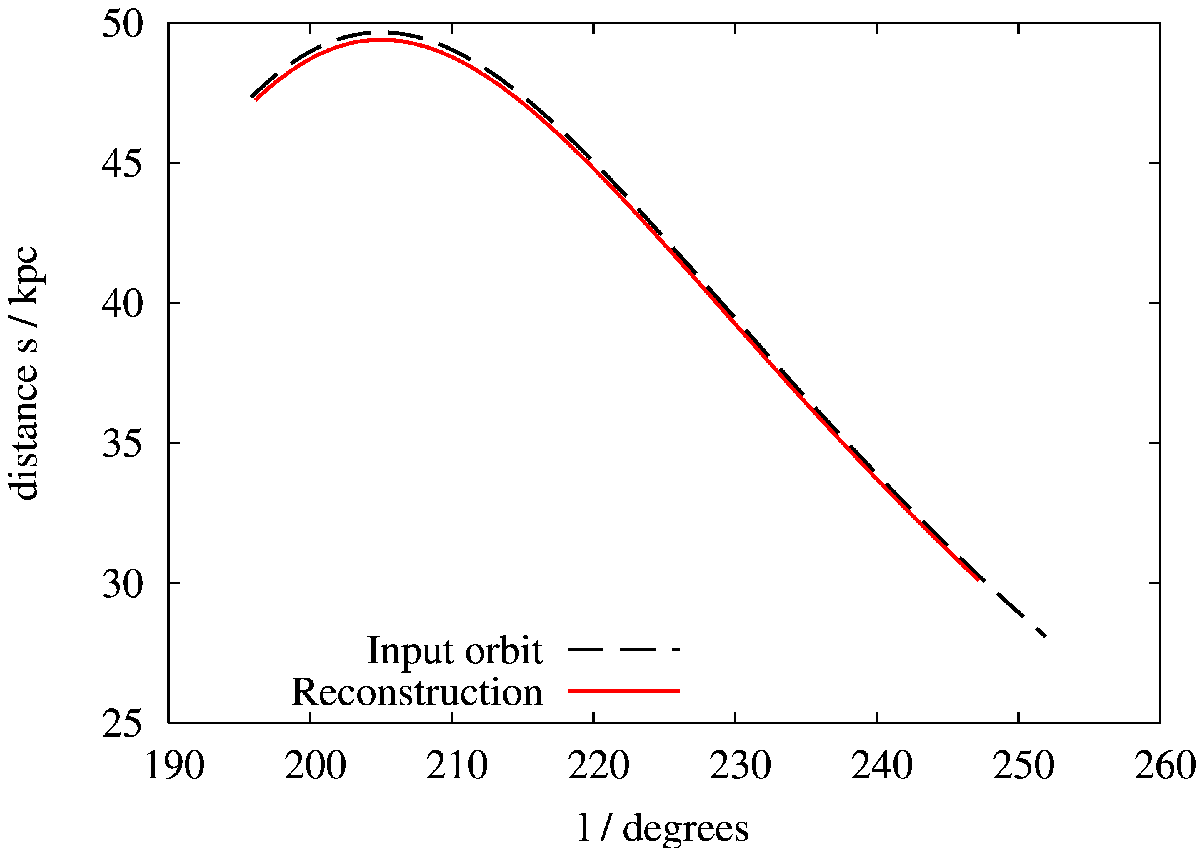}
\includegraphics[width=0.5\hsize]{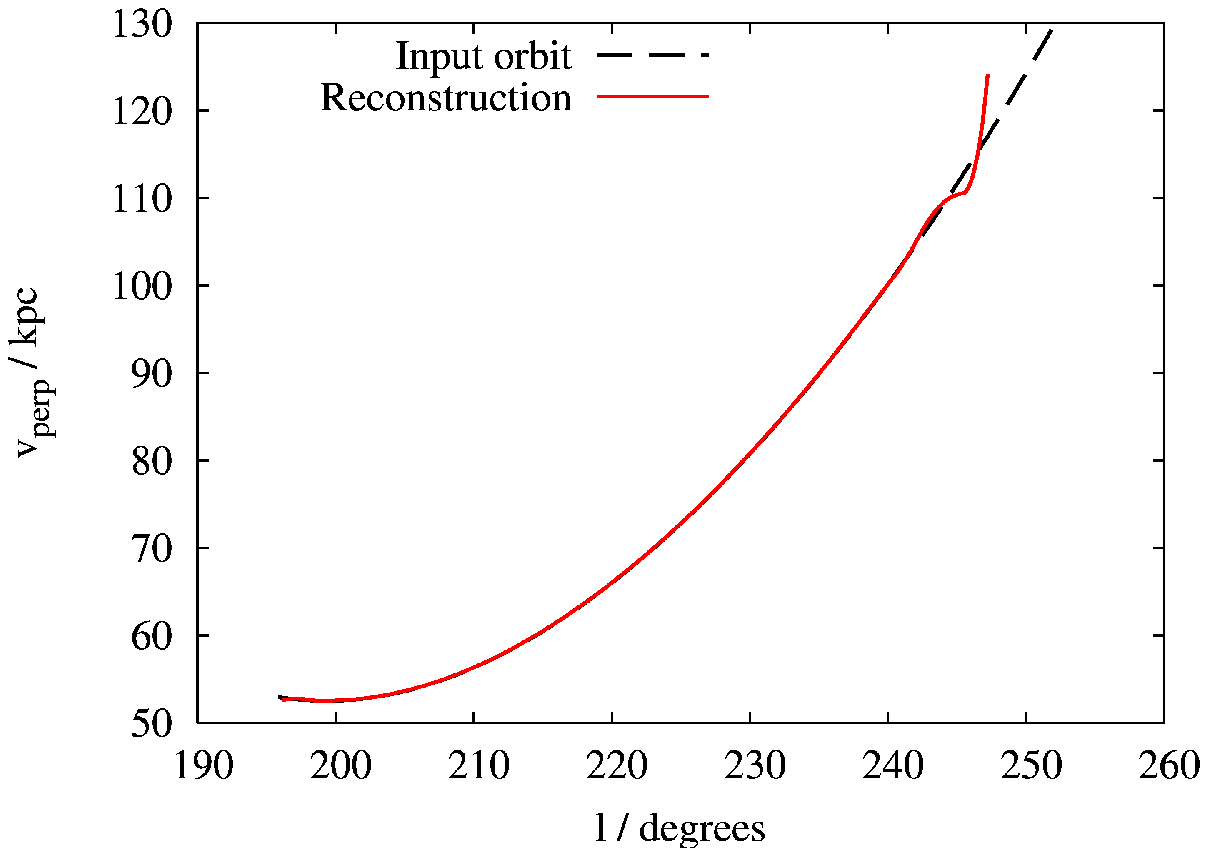}
}
\caption[Heliocentric distance and tangential velocity of the best reconstruction
from \figref{radvs:fig:pd1}]
{The left panel shows heliocentric distance, $r$,
versus galactic longitude, $l$, for the best reconstruction from \figref{radvs:fig:pd1}
and for the true orbit from which the input was generated. The two curves
are close to overlying. The right panel shows tangential velocity, $v_t$,
for the best reconstruction and for the true orbit: the 
departure of the reconstruction from the orbit near the endpoints
is symptomatic of the problems near the endpoints that necessitate their
excision from the diagnostic.} 
\label{radvs:fig:pd1-vs-truth}
\end{figure}

\figref{radvs:fig:pd1} shows the results obtained with PD1, which has very small
error bars. The diagnostic function $D'$ has a smooth minimum that
pinpoints the distance $r_0=47.4\kpc$ to the starting point with an uncertainty of
$\sim 0.2\kpc$. The
Metropolis optimization can significantly reduce $D'$ by tweaking the
input track only when $r_0$ is close to the truth. The depth of the
minimum indicates the numerical-noise floor for this particular
problem, and no significant scatter is seen between successive runs.
 We do expect the noise floor to be slightly different for
PD2--PD6 because both the underlying orbits and the input are somewhat
different. \figref{radvs:fig:pd1-vs-truth} shows that the reconstructed
solution at the minimum overlies the input orbit almost exactly. We
conclude that when the error bars are as small as in PD1, only one
orbit is consistent with the data.

\begin{figure}
\centerline{
\includegraphics[width=0.5\hsize]{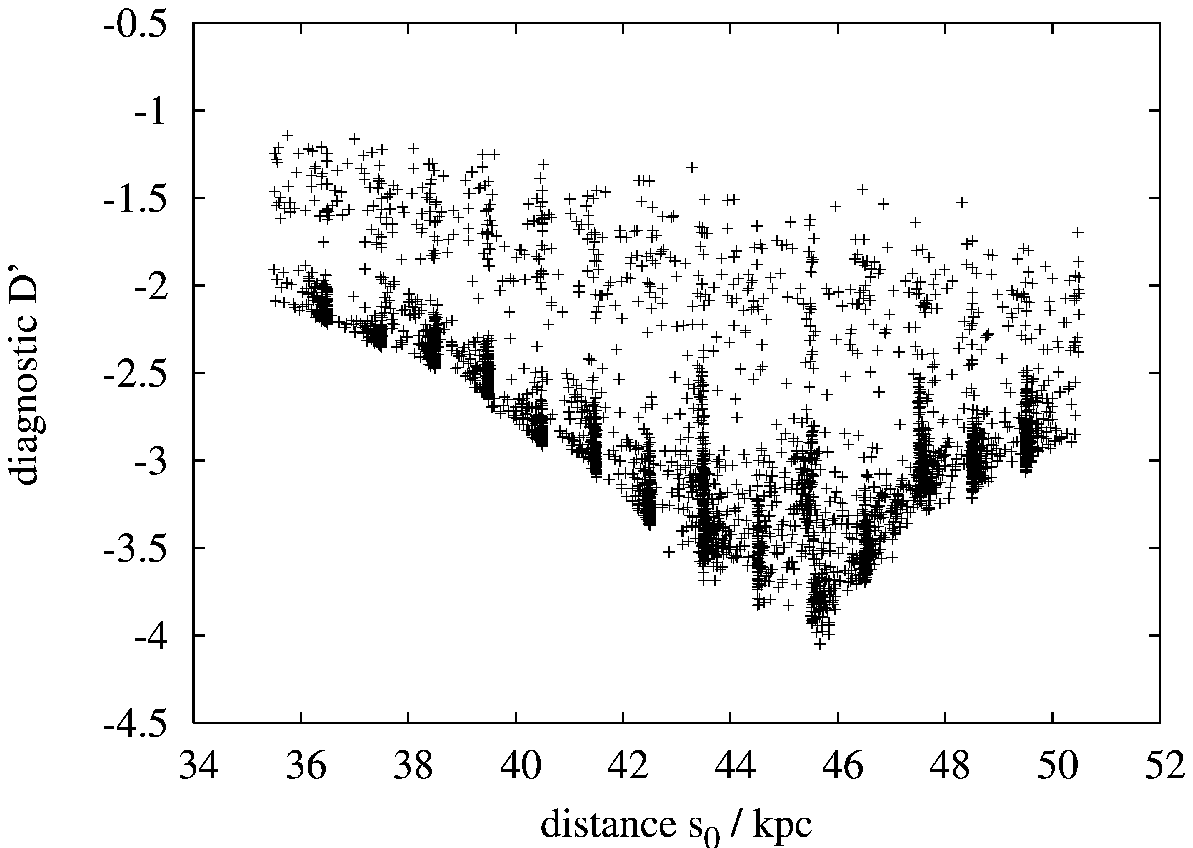}
\includegraphics[width=0.5\hsize]{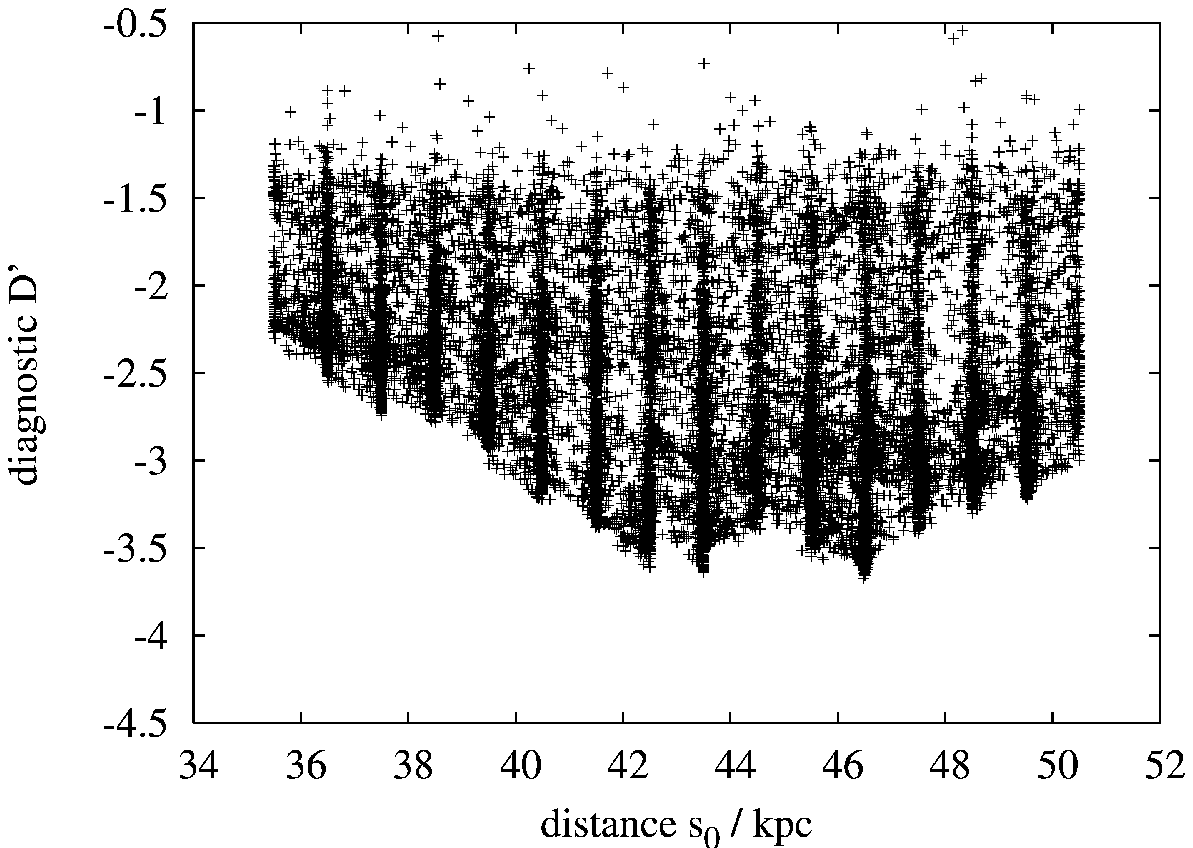}
}
\centerline{
\includegraphics[width=0.5\hsize]{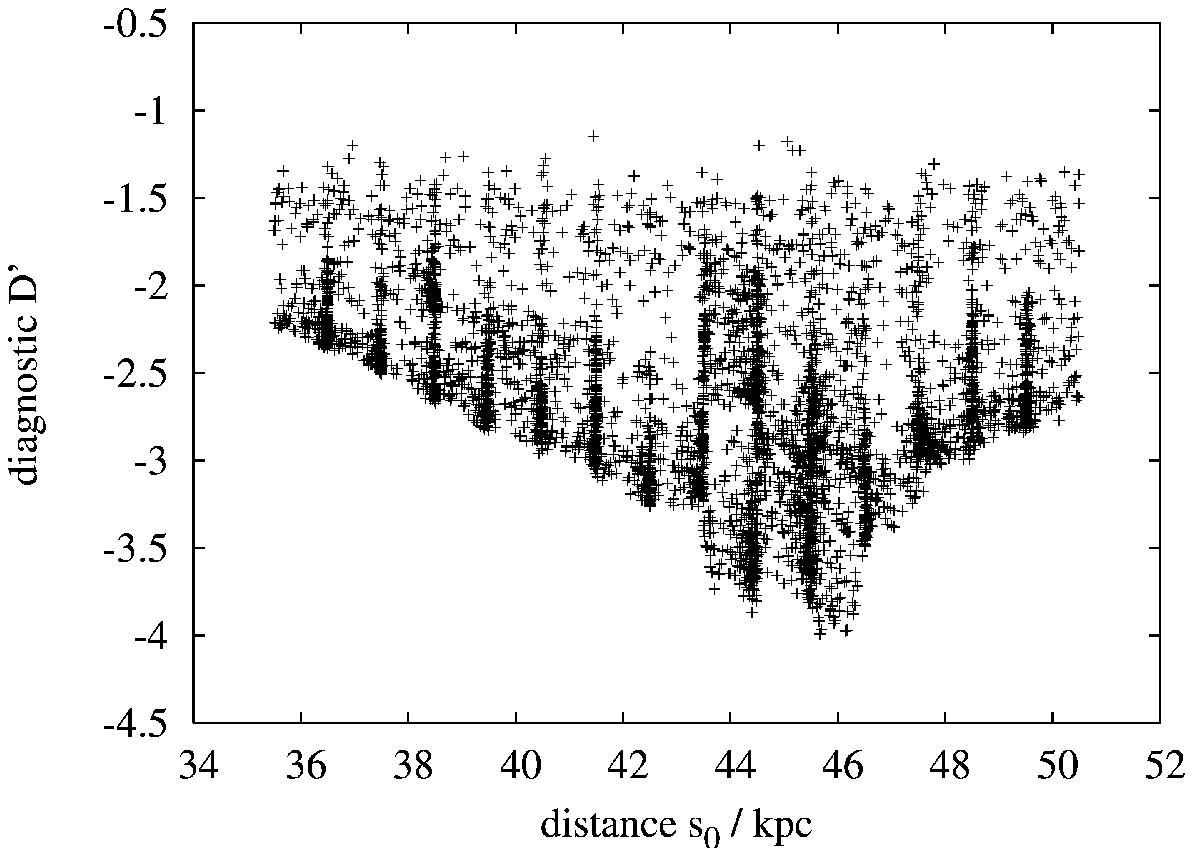}
\includegraphics[width=0.5\hsize]{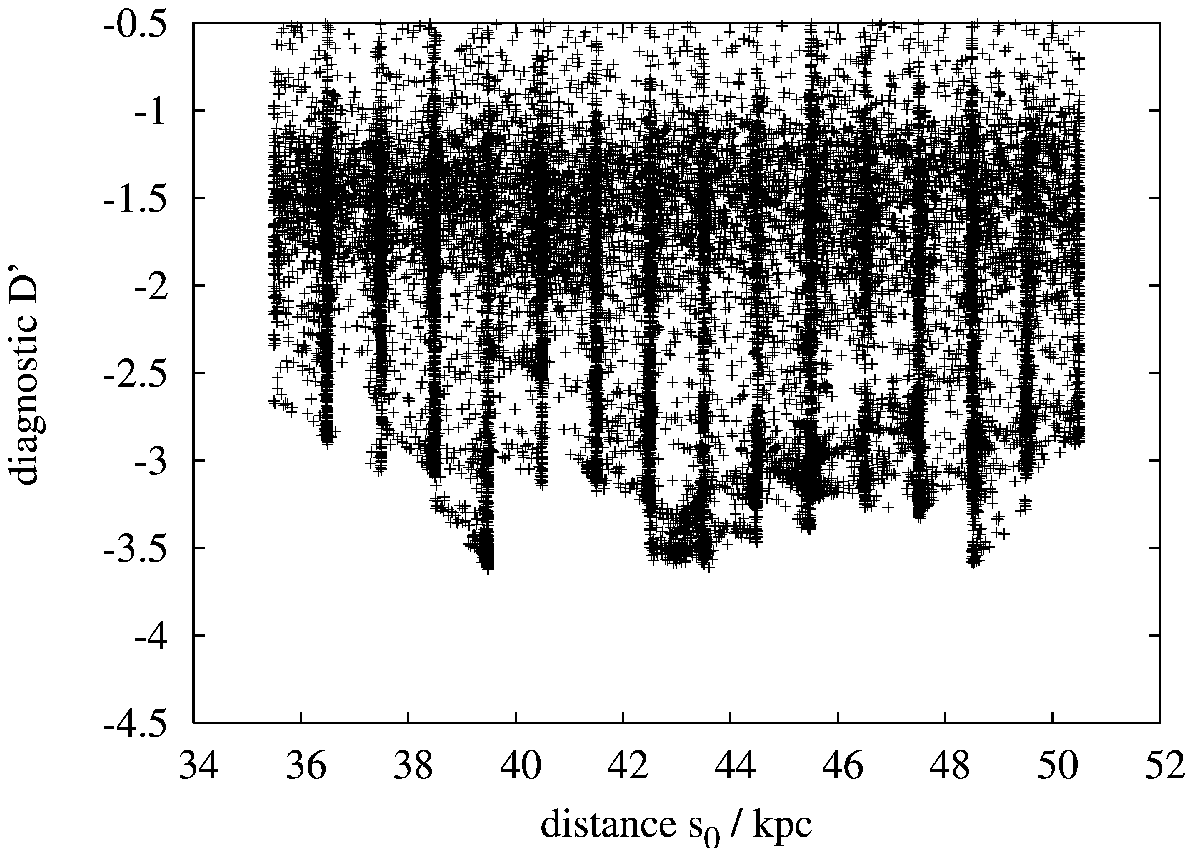}
}
\centerline{
\includegraphics[width=0.5\hsize]{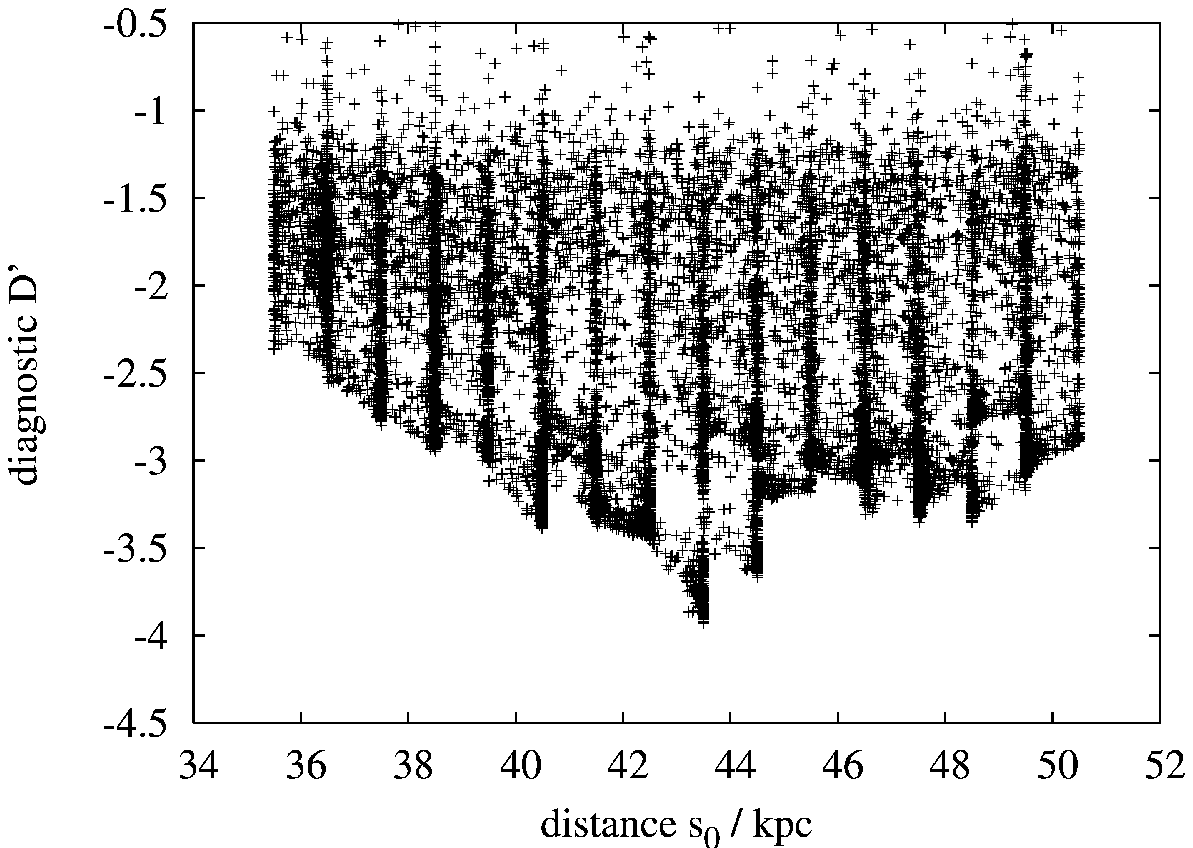}
\includegraphics[width=0.5\hsize]{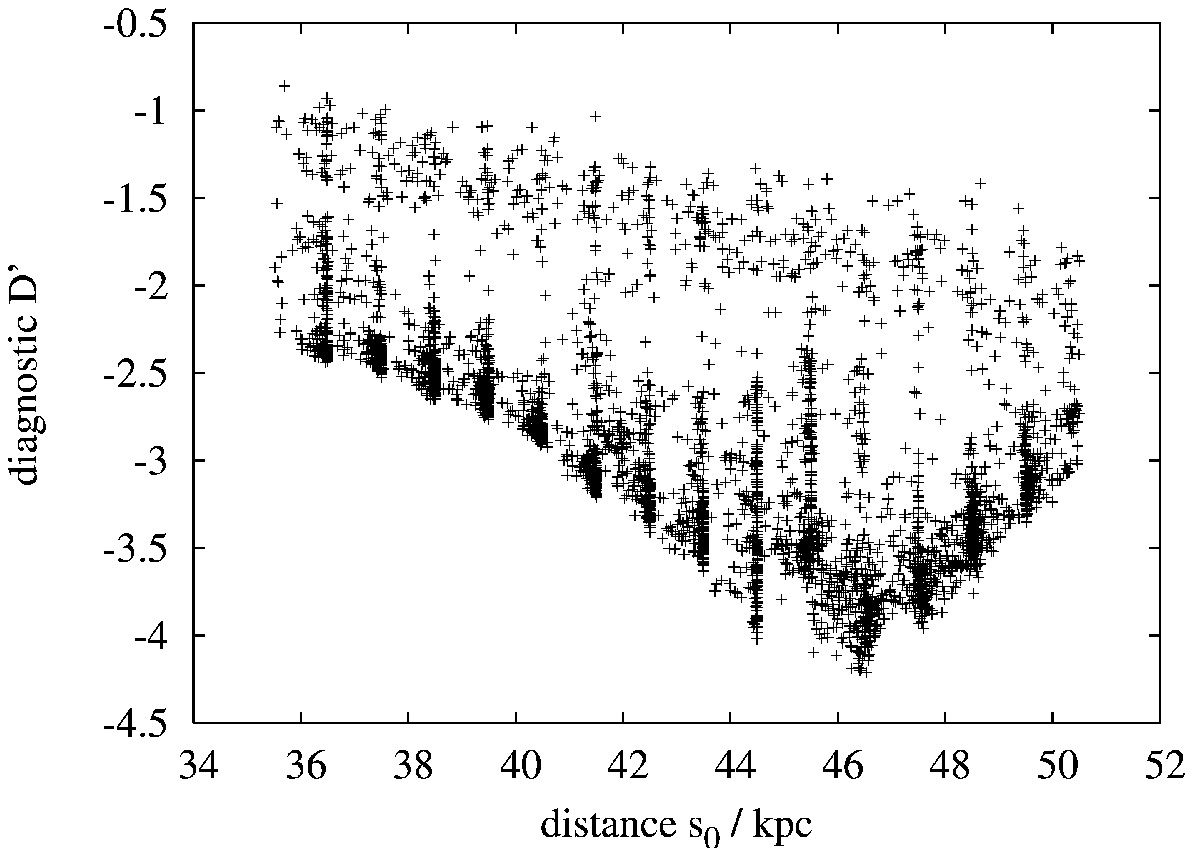}
}
\caption[As \figref{radvs:fig:pd1} but for input data sets PD2--PD7]
{The same as \figref{radvs:fig:pd1} but for input data sets:
  PD2 (upper-left), PD3 (middle-left), PD4 (lower-left), PD5
  (upper-right), PD6 (middle-right) and PD7 (lower-right).  In PD2 and
  PD3, the noise floor $D' \sim -4$ is reached only for starting
  distances in the range $44-46\kpc$. In PD4 we find an isolated good
  solution at $43\kpc$, but in contrast to what happens with data sets
  PD2 and PD3, as we change distance the value of $D'$ oscillates
  around $\sim -3.25$ for most of the range, rather than varying
  smoothly.  This behaviour arises because the volume of parameter
  space that has to be searched is large on account of the largeness
  of the velocity errors. Such an extensive volume of parameter space
  cannot be exhaustively searched with the allocated computational
  resource. Only orbits with $r_0 < 40\kpc$ could be excluded with
  confidence.  In PD5 and PD6, the noise floor $D \sim -3.5$ is now
  approached over a wider range in $r_0$: $42-47\kpc$ in PD5 with some
  confidence, and $38-48\kpc$ in PD6 with little confidence.  As the
  errors increase, the search becomes a more arduous task, and patchy
  performance of this task is reflected in the rough bottoms to the
  graphs.  In PD7, the results are very similar to those of PD2 and
  PD3, with a noise floor $D' \sim -4$ and solutions acceptable only
  in the range $44-46\kpc$.  }
 \label{radvs:fig:pd1-7}
\end{figure}

\begin{figure}
\centerline{
\includegraphics[width=0.5\hsize]{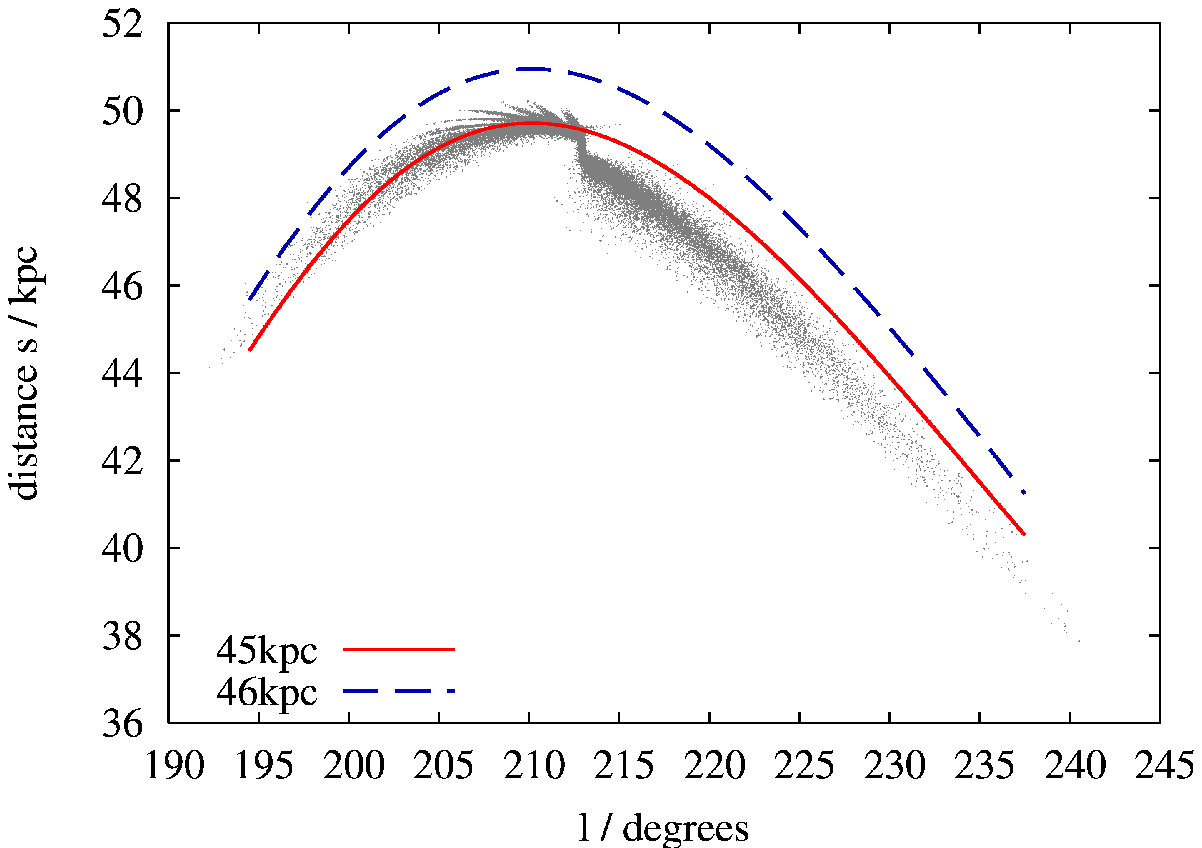}
\includegraphics[width=0.5\hsize]{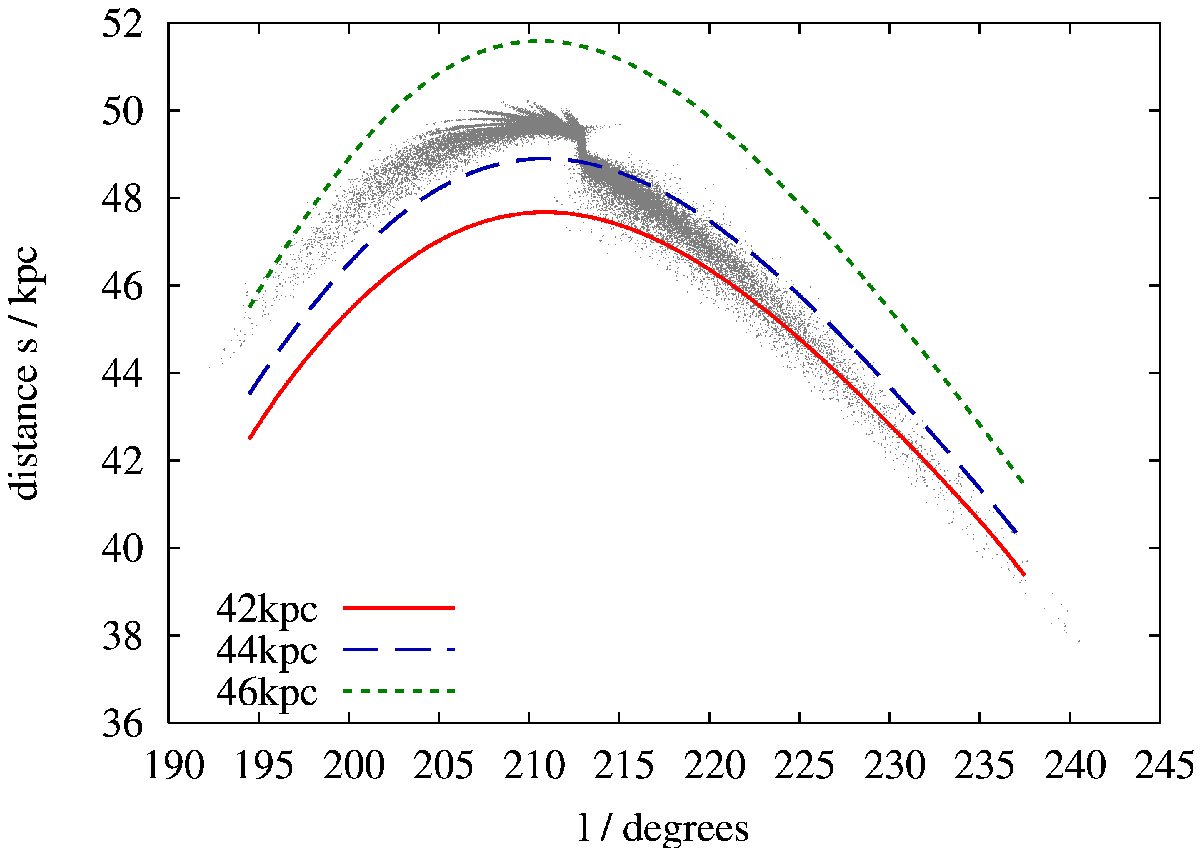}
}
\centerline{
\includegraphics[width=0.5\hsize]{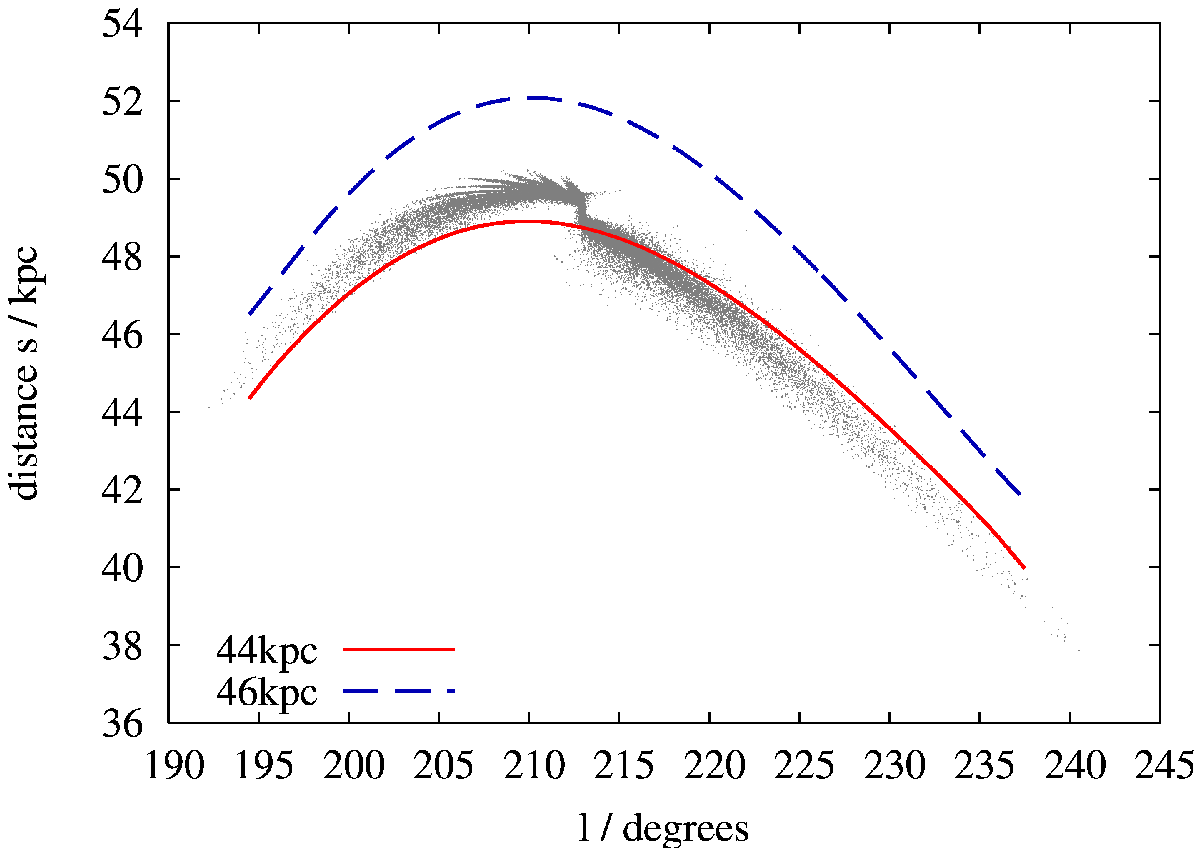}
\includegraphics[width=0.5\hsize]{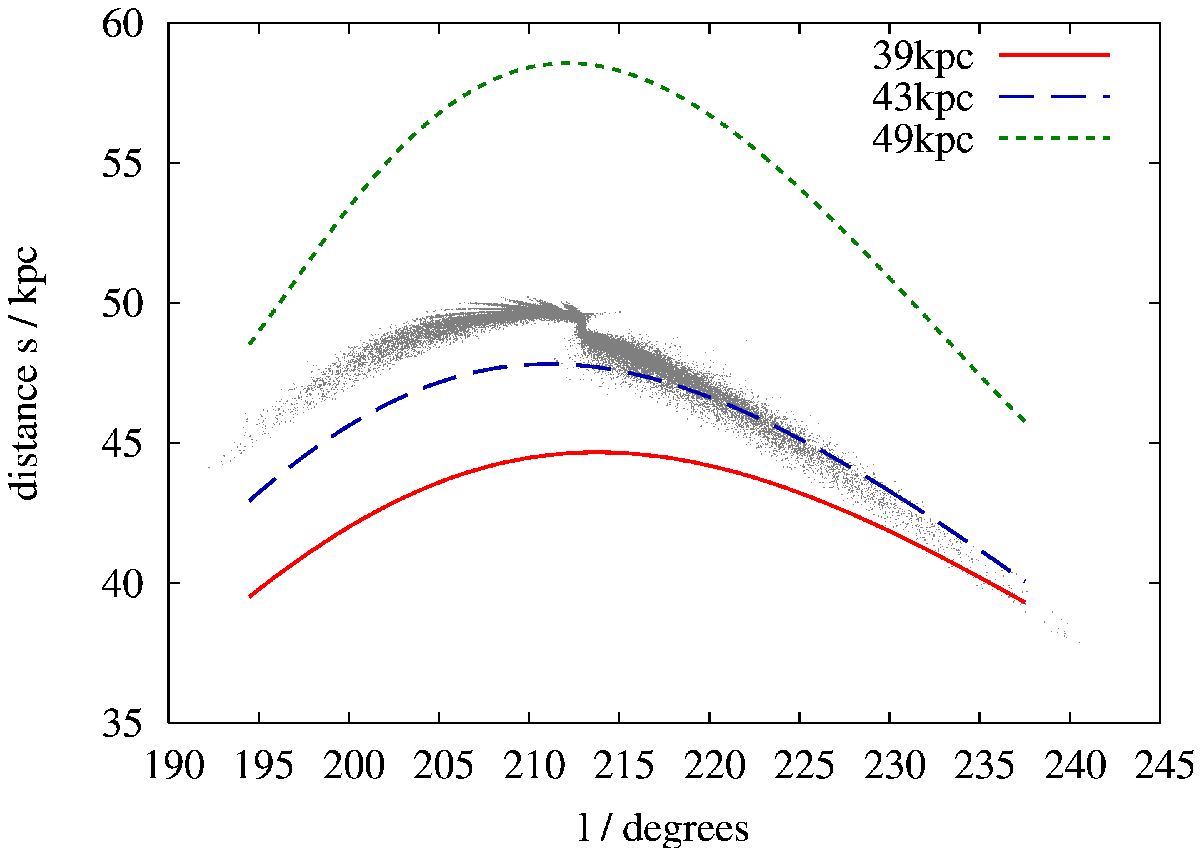}
}
\centerline{
\includegraphics[width=0.5\hsize]{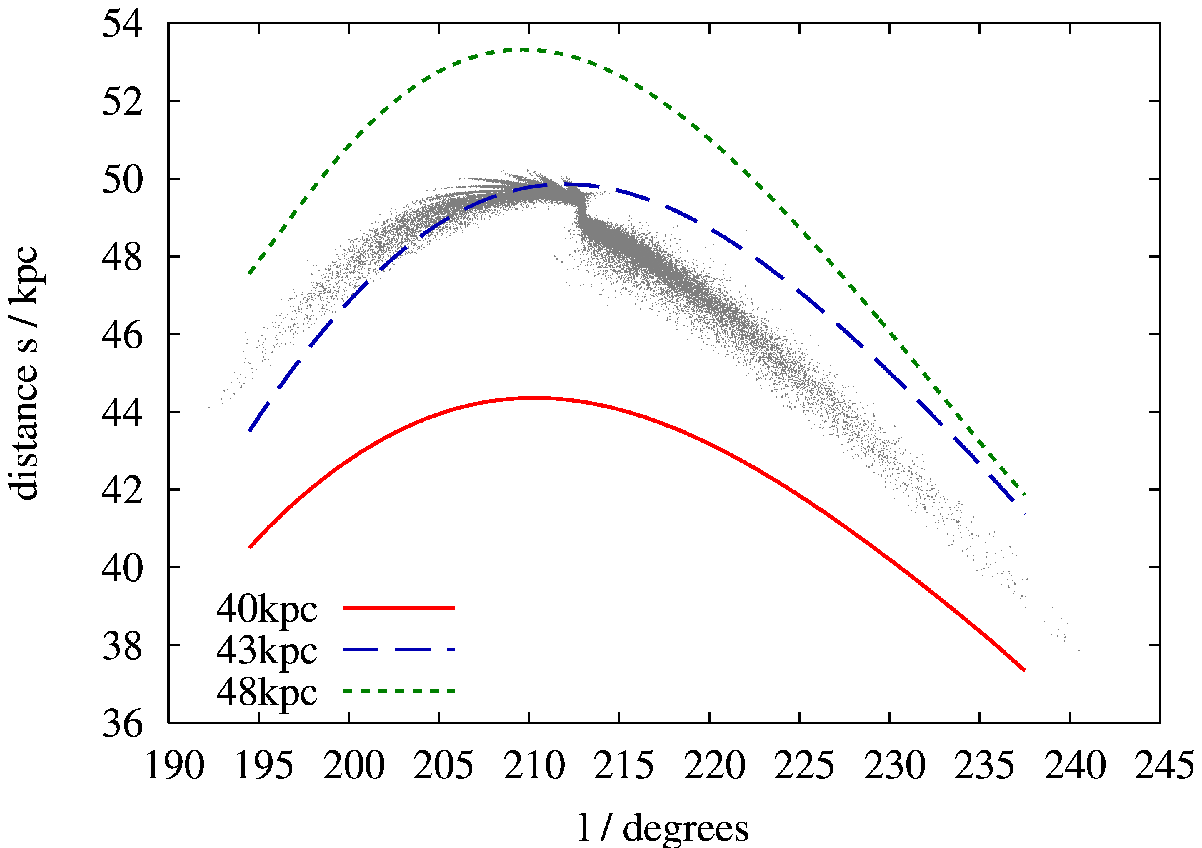}
\includegraphics[width=0.5\hsize]{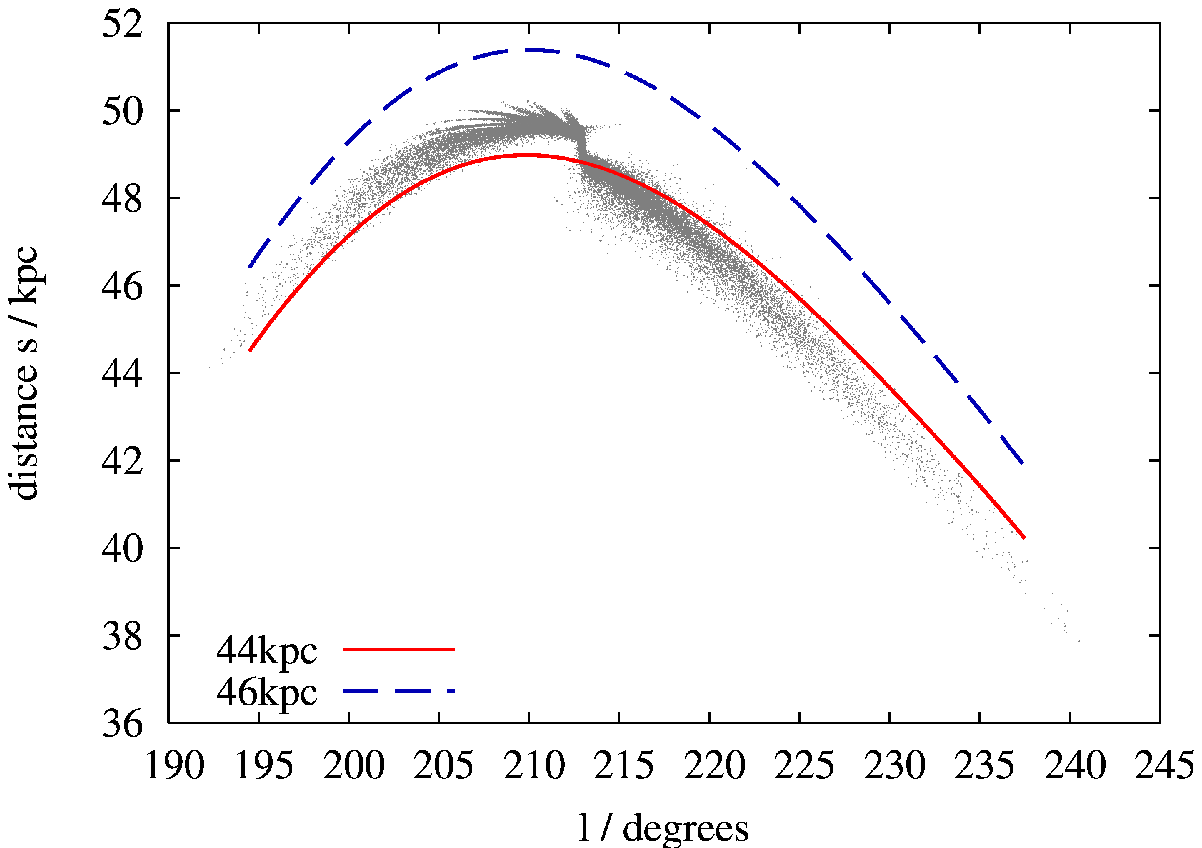}
}
\caption[Heliocentric distance for selected reconstructed orbits
from the data sets PD2--PD7]
{Heliocentric distance for selected reconstructed orbits
from the data sets:
PD2 (upper-left), PD3 (middle-left), PD4 (lower-left), PD5
(upper-right), PD6 (middle-right) and PD7 (lower-right).
The tracks selected are those with lowest $D'$ at
distances for which $D'$ approaches the noise floor in \figref{radvs:fig:pd1-7}.
}
\label{radvs:fig:pd1-7-dist}
\end{figure}

The left panels of \figref{radvs:fig:pd1-7} show the results obtained with PD2, PD3 and PD4. For
PD2 and PD3, which have small to moderate error bars, the Metropolis
algorithm reduces $D'$ to the noise floor only for $r_0$ in the range
$(44,46)\kpc$. The scatter between runs is small, $\sigma_{D'} \sim 0.1$.
The left panels of \figstworef{radvs:fig:pd1-7-dist}{radvs:fig:pd1-7-vt} show
the reconstructed distances and tangential velocities associated with
the best two solutions found at $44$ and $46\kpc$. With PD2, these reconstructions
provide a distance estimate to the stream that is, at worst, $2\kpc$ in
error, and a $v_t$ estimate that is at worst $5\kms$ in error. With PD3, the
reconstruction is, at worst, $3\kpc$ and $10\kms$ in error. For many sections of
the orbits, the errors are less than stated. Thus the method can
identify orbits consistent with the stream, and reject those that are
inconsistent with it. 

\begin{figure}
\centerline{
\includegraphics[width=0.5\hsize]{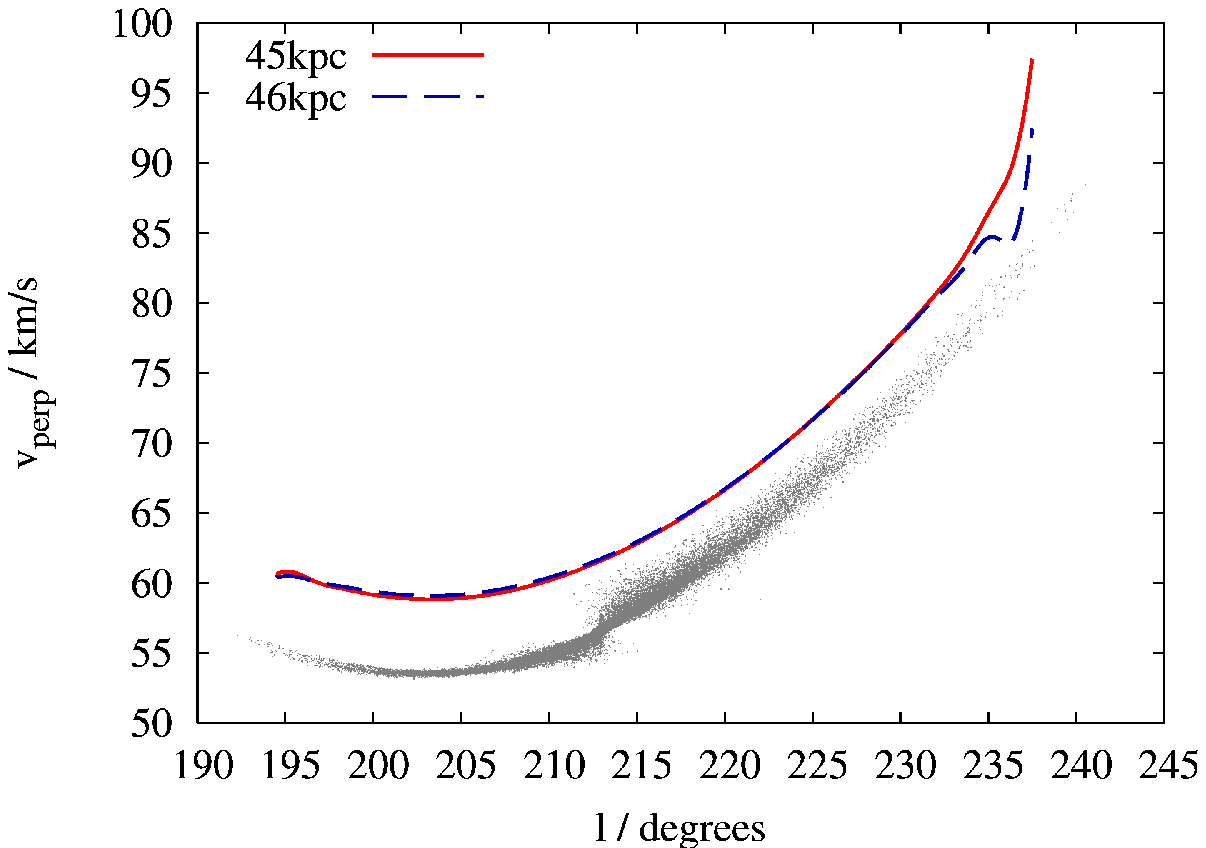}
\includegraphics[width=0.5\hsize]{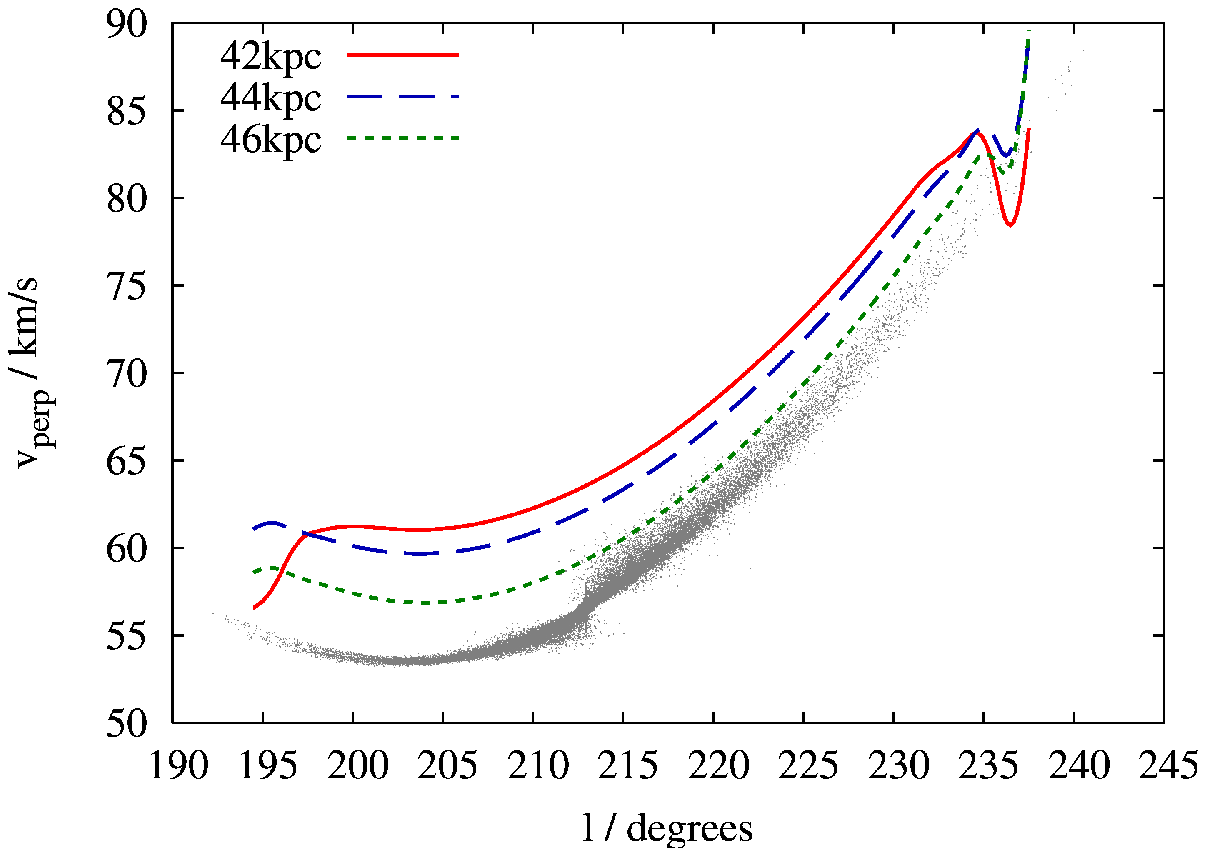}
}
\centerline{
\includegraphics[width=0.5\hsize]{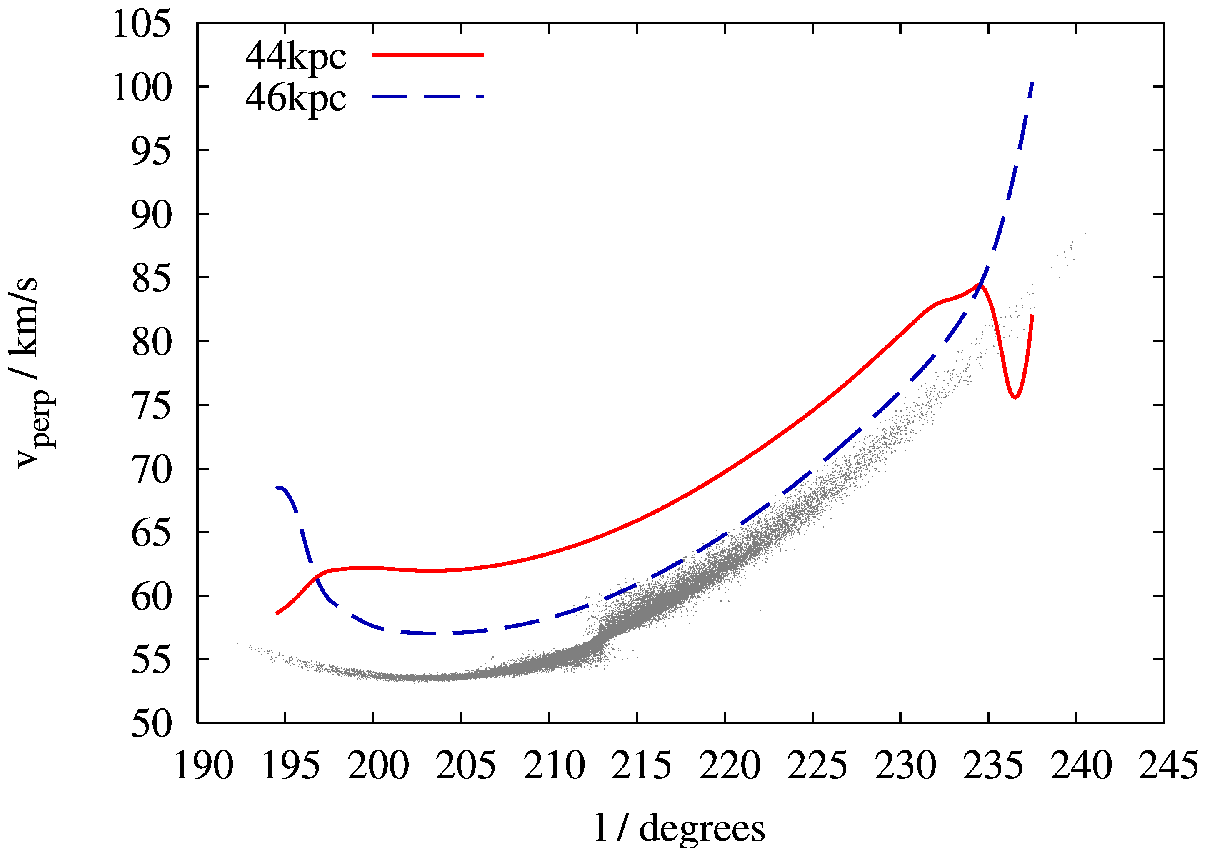}
\includegraphics[width=0.5\hsize]{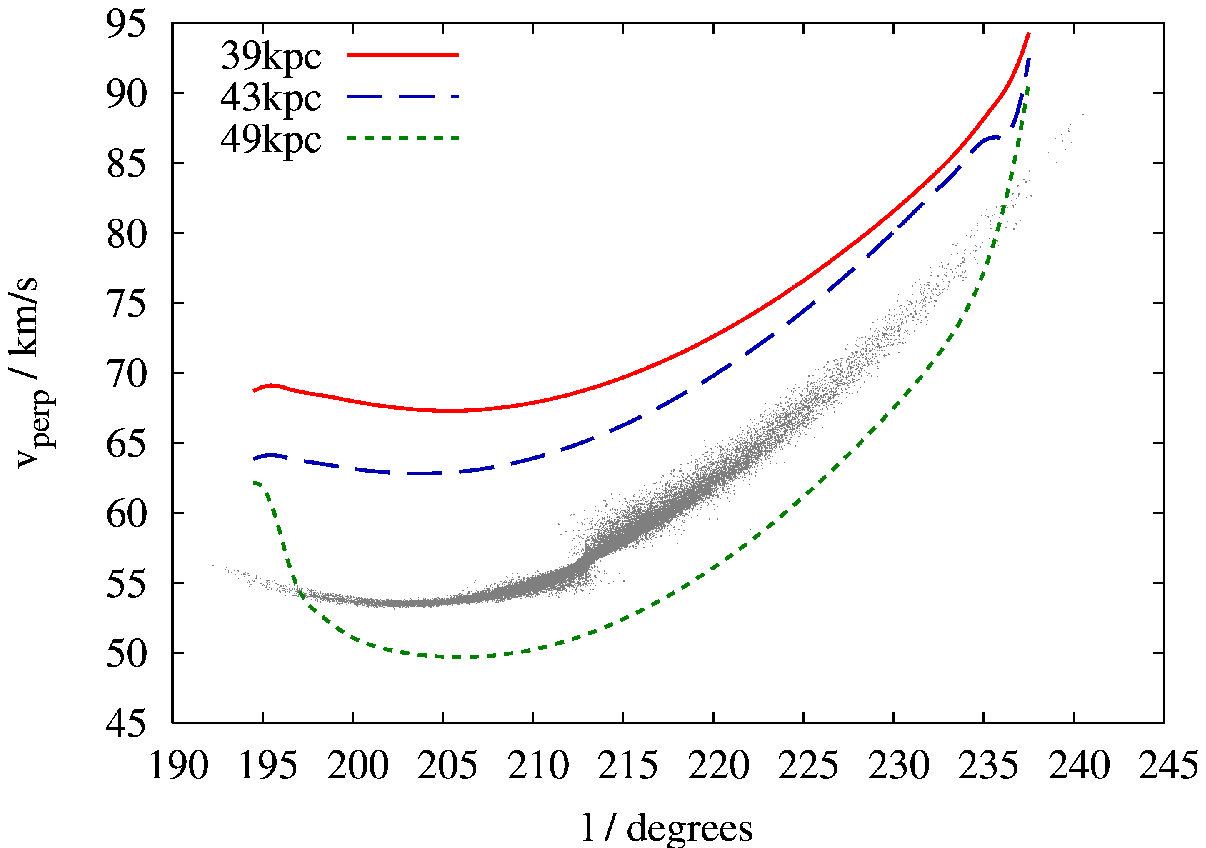}
}
\centerline{
\includegraphics[width=0.5\hsize]{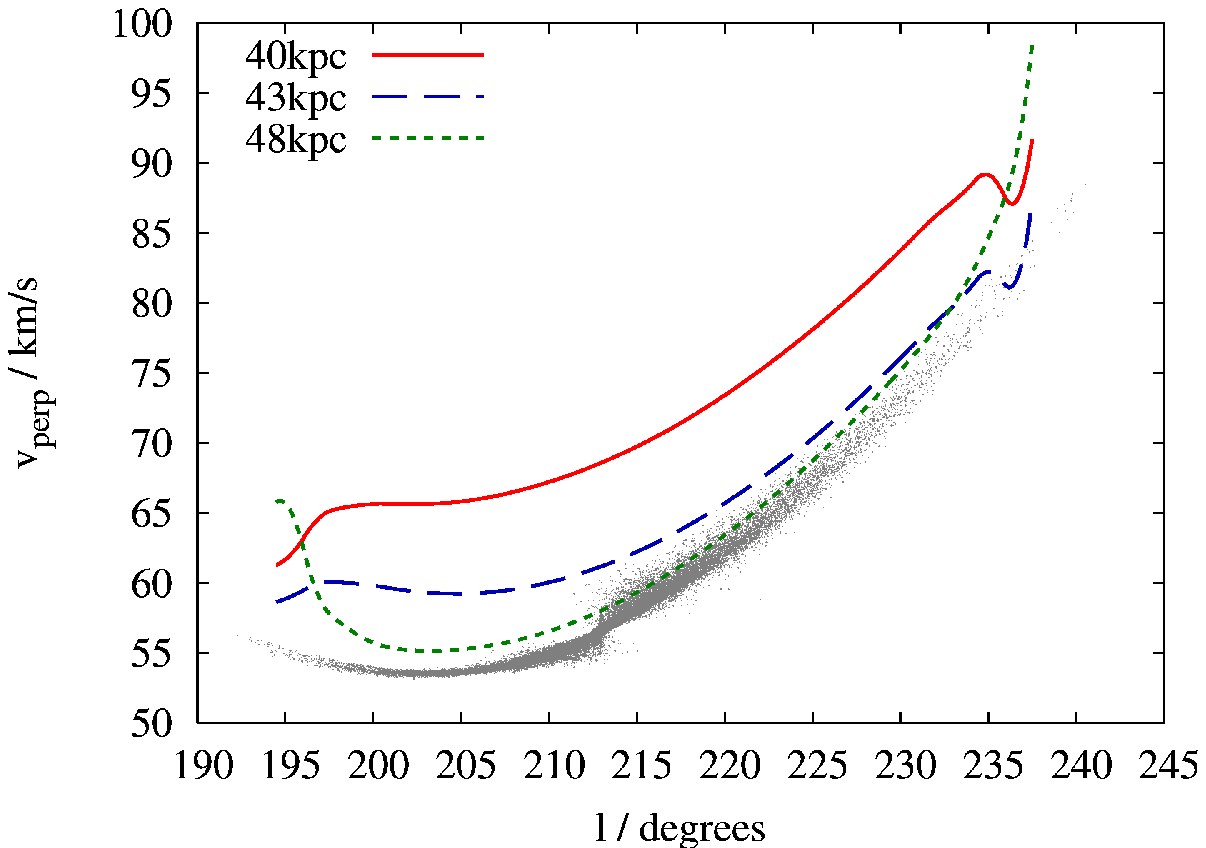}
\includegraphics[width=0.5\hsize]{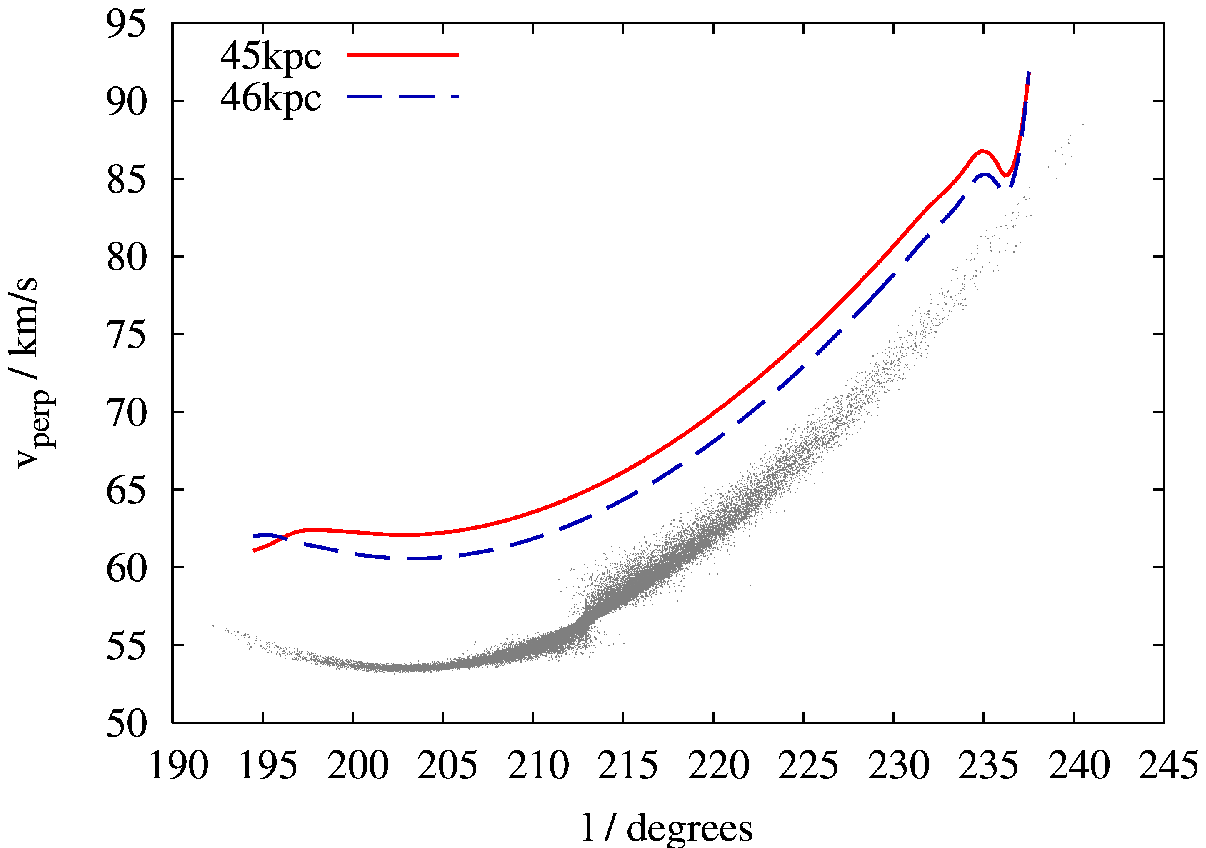}
}
\caption[Tangential velocities for selected reconstructed orbits
from the data sets PD2--PD7]
{Tangential velocities for selected reconstructed orbits
from the data sets:
PD2 (upper-left), PD3 (middle-left), PD4 (lower-left), PD5
(upper-right), PD6 (middle-right) and PD7 (lower-right).
The tracks selected are the same as in \figref{radvs:fig:pd1-7-dist}.
}
\label{radvs:fig:pd1-7-vt}
\end{figure}

For PD4, which has large error bars comparable to those for velocity data
from the SDSS, scatter between repeat runs is much larger,
$\sigma_{D'} \sim 0.5$. Only distances $r_0 < 40\kpc$ can be
confidently excluded, although a high-quality solution near $r_0 \sim
43\kpc$ provides a reconstruction in error by at most $2\kpc$ in
distance and $5\kms$ in $v_t$. \figstworef{radvs:fig:pd1-7-dist}
{radvs:fig:pd1-7-vt} show us that ignoring this high-quality solution
would give reconstructed quantities in error by up to $5\kpc$ in
distance and $12\kms$ in $v_t$. Increasing the errors associated
with the input clearly permits less satisfactory solutions to be
returned, and also decreases the ability of the search procedure to
find true orbits at particular distances, should they exist, by
increasing the volume of parameter space it has to search. The former
complaint is a physical statement about the limitations of the input data. The
latter complaint is an algorithmic one, which may be remedied by
providing the search with more computational cycles, or improving its
efficiency.

The upper-right and middle-right panels of \figref{radvs:fig:pd1-7}
shows the results obtained from input sets PD5 and PD6,
in which offsets were applied to the input velocities.  The
reconstructed distances and tangential velocities for interesting
tracks are shown in the upper-right and middle-right panels of
\figstworef{radvs:fig:pd1-7-dist}{radvs:fig:pd1-7-vt}. For PD5,
the scatter between runs is $\sigma_{D'} \sim 0.5$, and the distance
to the stream $r_0$ can be said to lie in the range $42-47\kpc$
with some confidence.  \figref{radvs:fig:pd1-7-dist} shows that the stream
does indeed lie in this range, so the algorithm has successfully
corrected the small offset. The error in reconstructed distance and
$v_t$ would be $3\kpc$ and $8\kms$ at worst.

PD6 demonstrates the limits of the method. On account of the large
scatter in $D'$, $\sigma_{D'} \sim 1$, we cannot identify the correct
distance from the middle-right panel of \figref{radvs:fig:pd1-7}. We
expect the distance range for permitted orbits to be wide with this
input, because the velocity error bars are
large. PD5 and PD6 illustrate another problem of using
large error bars: the search becomes harder because the parameter
space to be searched is much larger.  Consequently the plots of the
results for PD5 and PD6 have rough bottoms, where the algorithm has
failed to reach consistent minima for searches at adjacent distances.
This can be remedied by re-running the search with more deformations
and more iterations, and may be addressed in future upgrades to the
search procedure.

PD7 demonstrates that the accuracy of the method is not necessarily
significantly degraded when the number of velocity data points is
substantially lower than the number of positional data points, as might be
expected from a stream for which the only available radial velocities are
those of giant stars.  The results for PD7 are shown in the bottom-right panel
of \figref{radvs:fig:pd1-7} and are
directly comparable to those of PD2 and PD3 (the upper-left and middle-left panels).
In particular,
the range of allowed distances, $44-46\kpc$, is the same and the
reconstructed orbits show comparable accuracy (cf.~the upper-left and bottom-right panels
of \figstworef{radvs:fig:pd1-7-dist}{radvs:fig:pd1-7-vt}).

\section{The effect of changing the potential}
\label{radvs:sec:potential}

\b08 demonstrated the ability of orbit reconstruction to diagnose the
Galactic potential with astonishing precision when the track of an orbit on the sky
is precisely known. Here we investigate the ability of  orbit reconstruction
to identify the correct potential when the track has to
be inferred from realistic stream data. 

\begin{figure}
\centerline{
\includegraphics[width=0.5\hsize]{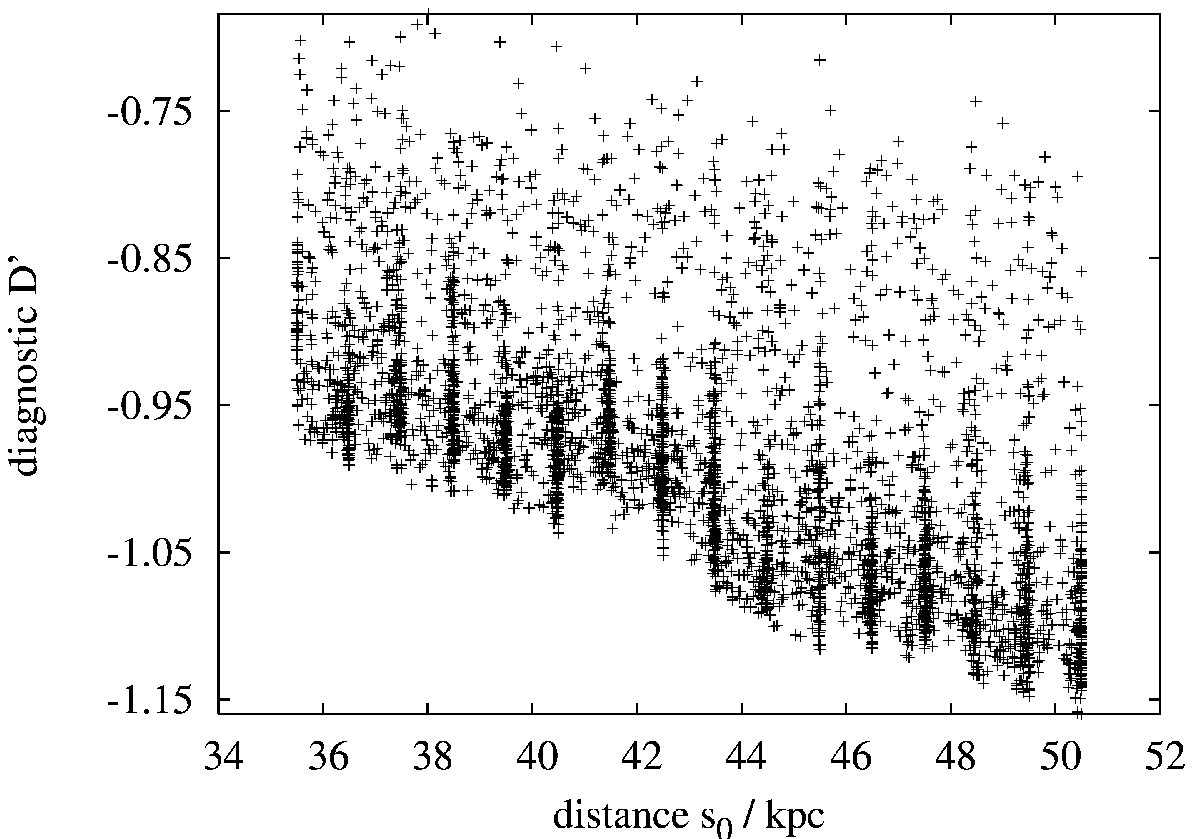}
\includegraphics[width=0.5\hsize]{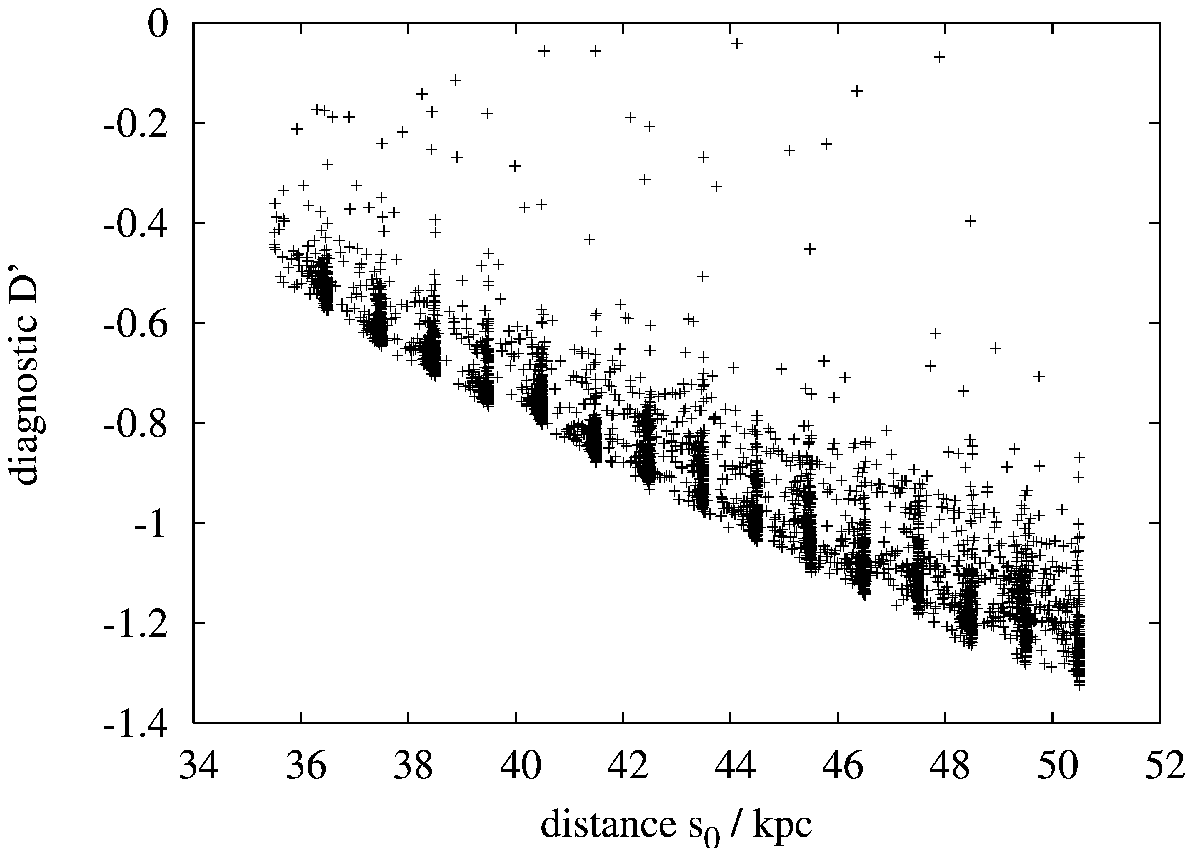}
}
\caption[As \figref{radvs:fig:pd1} but for data set PD2 analyzed in
a Kepler potential and a rising rotation-curve potential] {As
  \figref{radvs:fig:pd1}, except the reconstruction (using PD2) takes
  place in a Kepler potential (left panel), and (right panel) a
  potential with $\Phi(r) \propto r$. In both cases, the potential
  parameters are set to generate approximately the same passage time
  along the stream as in Model II. Comparing with
  \figref{radvs:fig:pd1-7} shows the values of $D'$ achieved are very
  poor, demonstrating that no orbits can be found in these potentials,
  which can therefore be excluded.  }
\label{radvs:fig:kepler}
\end{figure}

We use as our input the PD2 data set from \secref{radvs:sec:test}.  The
left panel of \figref{radvs:fig:kepler} shows the results of asking the algorithm
to find orbits in a Kepler potential with mass $M = 4.18 \times
10^{11} M_{\sun}$, which produces roughly the same passage time along
the stream as does Model II. The right panel shows the results obtained
using a potential of the form $\Phi(r) = r f_r$, which gives a rotation
curve of the form $v_c(r) = \sqrt{r f_r}$ with $f_r = 6.86 \times 10^2 (\kms)^2/\kpc$, again
chosen to produce the same passage time along the stream as does Model II.
These two potentials represent, respectively, relatively extreme falling
and rising rotation curve models. We do not offer them as realistic candidates
for the Galactic potential, but we intend to demonstrate that model potentials with
approximately correct radial force, but incorrect shape, can be excluded using
this method.

\begin{figure}
\centerline{
\includegraphics[width=0.5\hsize]{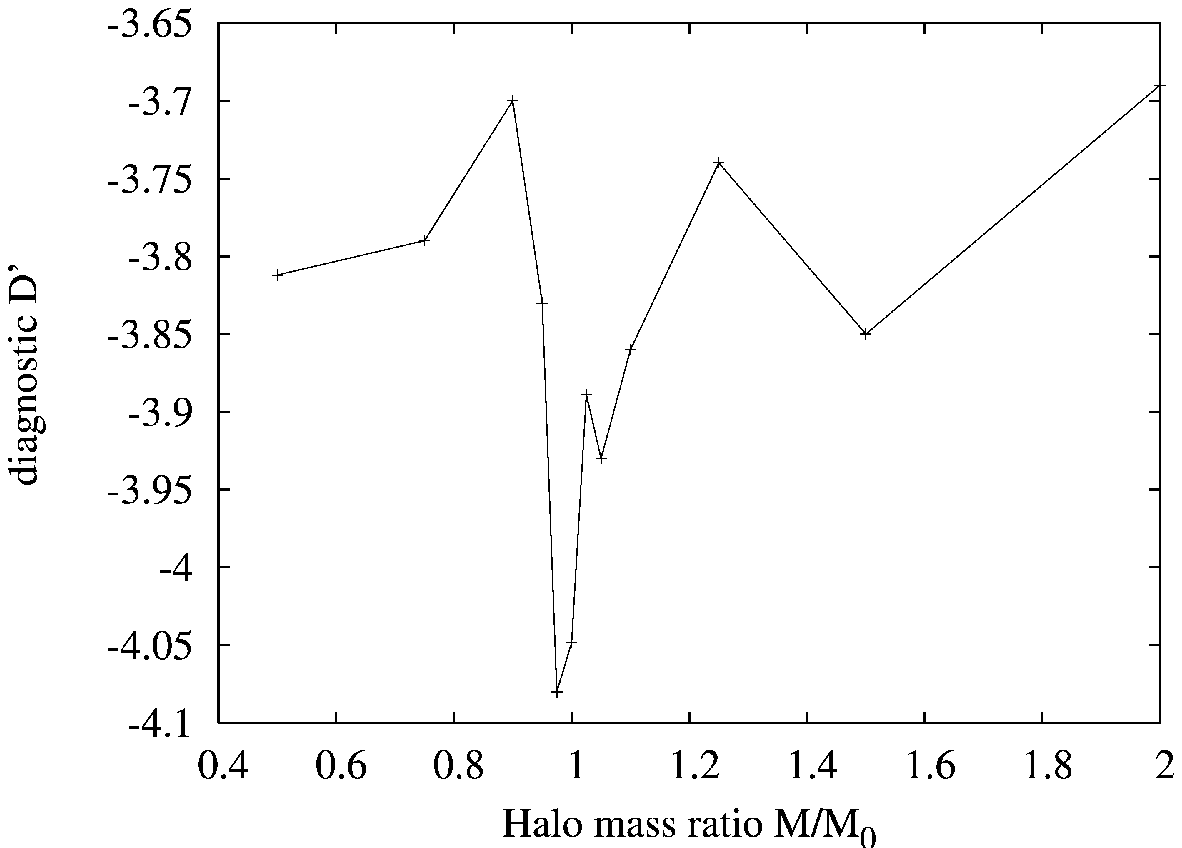}
\includegraphics[width=0.5\hsize]{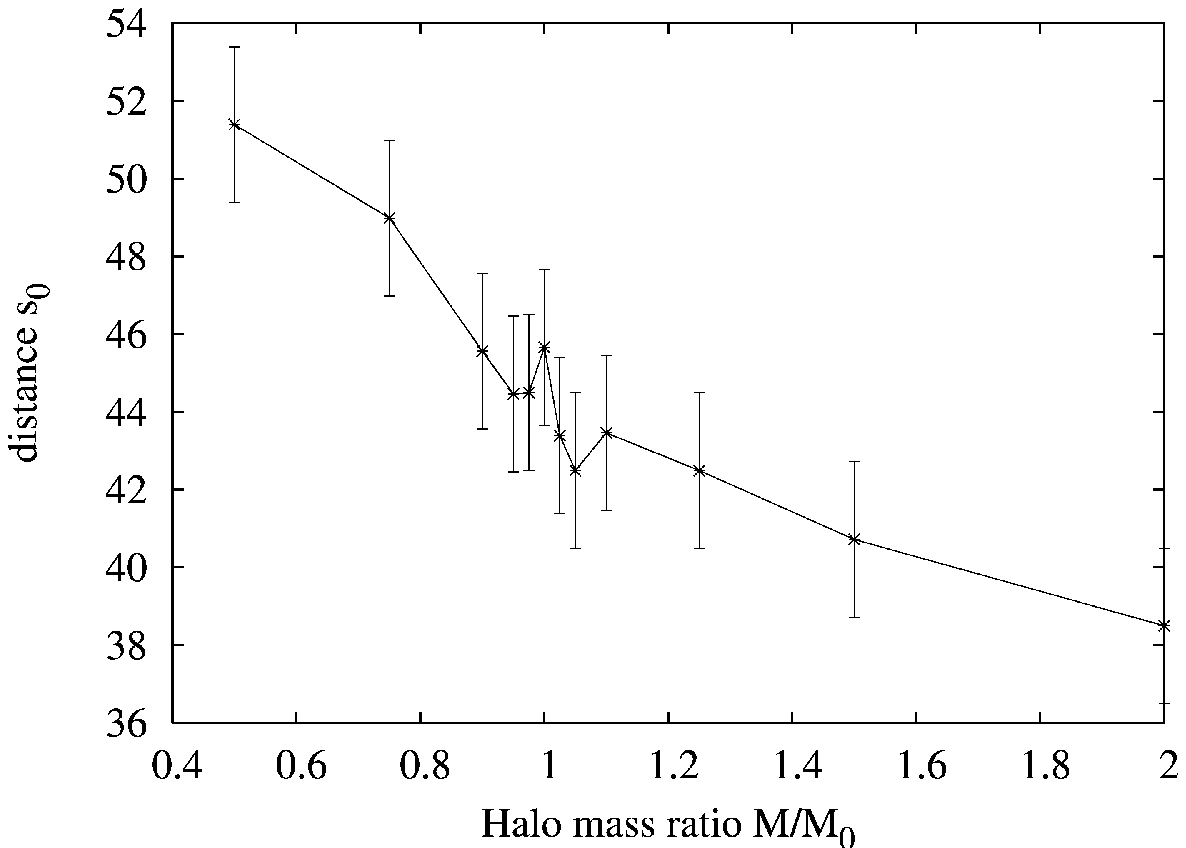}
}
\caption[Behaviour of the diagnostic $D'$ with variation in halo mass parameter]
{The left panel shows the minimum value of the diagnostic for
attempted reconstruction of the PD2 input, versus the halo mass ratio of the Dehnen-Binney
potential in which the reconstruction has been attempted. $M_0$ here is the value of the
halo mass in the Model II potential. The right panel shows the characteristic distance,
$r_0$, of the best solution, versus halo mass ratio. The error bars represent the
approximate uncertainty in the recovered result.}
\label{radvs:fig:changepot}
\end{figure}

The distance range considered in both cases spans
approximately $\pm 15$ percent of the true distance, which is the
uncertainty one might expect in distances obtained by photometry. Since all
values of $D'$ are several orders of magnitude larger than the
values one obtains with the correct potential, it is clear that no
orbits can be found at the distances considered, and both potentials
can be excluded.

\figref{radvs:fig:changepot} shows the results of
reconstructing an orbit from the PD2 data set in a potential that
differs from the Model II potential used to define the tidal stream
only in that the halo mass has been varied by the specified ratio.
This high-quality data set yields a sharp minimum in $D'$ as a
function of $r_0$ (cf.~the upper-left panel of
\figref{radvs:fig:pd1-7}) and the left panel of
\figref{radvs:fig:changepot} shows this minimum value of $D'$ as a
function of assumed halo mass. The minimum of $D'$ lies at $D'<-4$
when the mass used lies within $\sim5$ percent of the true value and
$D'>-3.85$ otherwise. Hence, with high-quality data it is possible to
constrain with remarkable precision the parameters of a model
potential that is otherwise of the correct shape.

Uncertainty estimates for potential parameters deduced in this way
present some difficulty. Any parameters for which the
minimum value of $D'$ returned is greater than the noise floor are inconsistent
with the data, and should be rejected. The uncertainty associated with
each parameter is therefore that range over which solutions can be found
that are consistent with the noise floor. Ambiguity arises, however,
because the value of the noise floor itself must be estimated from the data,
and is therefore uncertain. Furthermore, it is not clear with what probability
an otherwise consistent set of parameters would return a value of $D'$ higher
than the noise floor, on purely stochastic grounds. In practice, potential
parameters can only be confidently excluded if the scatter in returned value
of $D'$ is substantially smaller that the difference between $D'$
and the noise floor, and yet the precise level of this confidence remains difficult to
formally quantify.

The right panel of \figref{radvs:fig:changepot} shows the value of
$r_0$ at which $D'$ attains its minimum, as a function of assumed halo
mass.  We see that this value decreases systematically as the
halo mass increases.  Insight into this behaviour can be obtained by
considering the discretized equation of angular-momentum conservation
\begin{equation}
  \Delta(r\,v_t)=-r \Delta t \, \nabla_t\Phi
  =-r {\Delta v_r\over F_r} \, \nabla_t\Phi, 
\end{equation}
%  \begin{equation}
%  \Delta(sv_t)=-s{\d\Phi\over\d
%   r_t}\Delta t =-s{\d\Phi\over\d r_t}{\Delta v_r\over
%   F_r}, 
% \end{equation}
where $-\nabla_t\Phi = (\vecthat{r}\cdot\nabla\Phi\,\vecthat{r} -\nabla \Phi)$
is the component of the force in the plane of the sky, and
where $\Delta$ implies the change in a quantity between successive
data points. By expanding the left side to first order in small
quantities we can obtain an expression for $\Delta v_t$. Summing the
changes in $v_t$ along the track we have
 \begin{equation}\label{radvs:eq:vpfirst}
v_t(\hbox{end})-v_t(\hbox{start})=-\sum\left(
  \nabla_t \Phi \,{\Delta v_r\over F_r}+v_t{\Delta
    r\over r}\right).
\end{equation}
 An independent equation for $v_t$ is
\begin{equation}\label{radvs:eq:vpsecond}
 v_t=r{\Delta u\over\Delta t}=r F_r
{\Delta u\over \Delta v_r},
\end{equation}
 where we have used the radial component of Newton's second law.
Equations~\blankeqref{radvs:eq:vpfirst}
and~\blankeqref{radvs:eq:vpsecond} yield independent estimates of
$v_t(\hbox{end})-v_t(\hbox{start})$. The right side of 
\eqref{radvs:eq:vpfirst} yields an estimate that is independent of the scaling of
$\Phi$ but systematically decreases with increasing $r$, while the right side
of \eqref{radvs:eq:vpsecond} yields an estimate that is proportional to the
scaling of $\Phi$, but is almost independent of $r$, since $F_r \propto
1/r$. When the machine is asked to reconstruct the orbit with $F_r$
taken too small, it can change the right side of \eqref{radvs:eq:vpfirst} to
match the new value of the right side of \eqref{radvs:eq:vpsecond} by
increasing $r$. In this way, the discrepancies between the left sides of
\eqsref{radvs:eq:vpfirst}{radvs:eq:vpsecond}, which contribute
substantially to the diagnostic $D'$, can be largely eliminated by increasing
$r$ as $\Phi$ is scaled down.

In principle this variation in reconstructed distance with the scaling of the
potential could be combined with photometric distances to constrain the
potential. Unfortunately, \figref{radvs:fig:changepot} shows that
in the particular geometry under consideration,
even a $10$ percent distance error would produce a $\sim50$ percent error in
the estimate of the halo mass. Further work is required to discover what
effect the geometry of a particular stream has on its sensitivity to the
potential. Also of interest is whether simultaneously using multiple streams,
or streams with multiple wraps (such as the Sagittarius Dwarf stream), can
provide yet tighter constraints on the potential---\b08 indicated that much could
be gained from this approach, but more work is needed to extend the
present schema to handle multiple independent segments of streams.

\section{Conclusions}
\label{radvs:sec:conclusions}

\b08 demonstrated the tremendous diagnostic power that is available if one
knows the track of an orbit on the sky and the associated line-of-sight
velocities. Tidal streams are made up of stars that are on similar
orbits. In particular, they roughly delineate the underlying orbit, but they
do not do so exactly. We have presented a technique for identifying an
underlying orbit and thus predicting the dynamical quantities that have not
been observed, such as the distances to the stream and the proper motions of
its particles.

A complete summary of our findings of this chapter is presented below.
Following the summary is a discussion of the limitations of the work,
accompanied by suggestions on how to overcome those limitations.
\chapref{chap:concs} further reviews our findings in the context
of contemporary astrophysics.

\subsection{Summary}

The technique presented involves defining a space of tracks on the sky and sequences of
line-of-sight velocities that are consistent with the observational data,
given the observational errors and the extent to which streams deviate from
orbits.  The equations of \b08 are used to determine a candidate orbit for
each track, and then the extent to which the candidate satisfies the
equations of motion is quantified. This represents an enhancement over \b08,
in which only violations of energy
conservation along a candidate were quantified.  The resulting diagnostic quantity is
used to search for tracks that could be projections of orbits
in the Galactic potential. In practice the search is conducted for several
possible distances to a fiducial point on the stream. If constraints on this
distance are available from photometry, the computational effort of the
search can be reduced by narrowing the range of distances for which searches
need to be conducted.
Our {\em a priori} assumptions are knowledge of the Galaxy's
gravitational potential, and the Sun's velocity with respect to the
Galactic centre.

The technique is moderately computationally expensive, in practice
taking several CPU-core hours to process a given set of input for a
single fiducial distance and potential combination. This computational
expense arises from the large number of dimensions of the parameter
space that the technique must search over: the parameter space must be
sufficiently large to afford the technique the flexibility to derive a
dynamical track from the input data, but the computational time that
must be devoted to the search to ensure that the parameter space is
appropriately explored grows exponentially with the dimensionality of
the parameter space. The number of dimensions selected in practice is a
compromise between these considerations. In mitigation, the search
does lend itself to simple parallelization, and with a few dozen
CPU-cores it is feasible to fully analyze a stream in a given
potential in less than a day.

Our tests revolve around an N-body model of the Orphan Stream
\citep{orphan-discovery} on the assumption that it formed by the
disruption of a globular cluster. We show that for this stream, which
is $\sim40^\circ$ long and $0.3^\circ$ wide at its ends, distances to
and tangential velocities of points on the stream can be recovered to
within $\sim 2 \kpc$ and $\sim5 \kms$, respectively, if line-of-sight
velocities accurate to $\sim 1\kms$ are measured. As the errors in the
measured radial velocities increase, the space of tracks that must be
considered grows bigger and the search for acceptable orbits becomes
more laborious. Moreover, the range of distances for which acceptable
orbits can be found broadens. However, even with errors in radial
velocities as large as $\pm 10\kms$, the uncertainties in the
recovered distances are no greater than $\sim 10$ percent and the
recovered tangential velocities are accurate to better than $\sim 20$
percent. Zero-point errors in the input velocities that are reflected
in appropriately wide error bars broaden the range of acceptable
orbits but do not skew the results.

We have shown that the method maintains its accuracy even when very few
radial velocity points are used to define the input, as might be necessary
when radial velocities can only be measured for giant stars. In our tests,
comparable results were obtained from pseudo-data based on only three velocity
measurements and from pseudo-data based on fifteen velocity measurements.
Indeed, the results obtained with three accurate velocity measurements were
significantly superior to those obtained with fifteen lower-quality
measurements. Naturally, exactly how many points are required to provide
well-determined input will depend on the shape of the radial velocity curve
along the stream in question.

We expect the accuracy of reconstructions to depend on the geometry
of the problem in hand. In particular, we expect streams at apocentre, 
where families of orbits are compressed both on the sky and
in radial velocity, to yield poorer results than streams away from apocentre.
Unfortunately, streams are most likely to be discovered at apocentre
because both trajectory compression and low proper motions around
apocentre lead to a high density of stars at apocentre. We expect streams
that are relatively narrow to produce more accurate results, because the permitted
deviation of the orbit from the stream is then low. We also expect
to have more difficulty reconstructing orbits from streams that contain
a visible progenitor, since the potential of the progenitor will cause orbits
in the progenitor's vicinity to differ materially from orbits in the Galaxy's
underlying potential.

\b08 suggested that it should be possible, if sufficiently accurate
input is provided, to constrain the Galactic potential, since the
wrong potential will not admit an acceptable orbit. We have tested
this possibility for input with realistic errors. We find that two
potentials of significantly different shape, the Kepler potential and
$\Phi(r) \propto r$, are clearly excluded. We have also tested for
changes in scaling of an otherwise correctly-shaped potential, by
varying the mass of the assumed dark halo around the value used to
make the pseudo-data. In this case, we find the correct potential is
identified, with the diagnostic quantity generally worsening as the
halo mass moves away from its correct value by more than $\sim5$
percent. We further find a consistent relationship between the
reported stream distance and the halo mass with which the
reconstruction takes place. Although the reported distance is only
weakly dependent upon halo mass, this does open the possibility of
using alternative distance measurements, such as photometric
distances, in conjunction with these techniques to constrain the
Galactic potential.

Further work is necessary to determine a full
scheme to recover parameters of the potential from stream data.  Also
in question is the extent to which simultaneous reconstruction of
multiple streams, and reconstruction of streams with multiple wraps
around the Galaxy, might provide stronger constraints on the Galactic
potential than the short section of a single wrap that we have
considered.  It may also prove possible to refine the other main
assumption of our scheme, the location and velocity of the Sun.

\subsection{Discussion}

It is interesting to compare our method of finding orbits of progenitors with
the traditional N-body method. Firstly, our method explores each orbit at a tiny
fraction of the computational expense of N-body modelling, so it is feasible
to automate the search of orbit space. Moreover, the search lends itself to
easy parallelization.  Finally, whereas only a successful attempt to model a stream with N
body simulation yields an interesting conclusion, our method can show that no orbit is
consistent with a given range of distances. This latter facility is particularly
powerful, because by guaranteeing that no consistent orbit can be found,
it is possible to test our assumed form for the Galactic potential.

The most serious limitation on the applicability of this work is the
lack of high precision radial velocity measurements for streams. At
the time of preparation of the \cite{eb09a} work on which this chapter
draws, the radial velocity data available for a long, cold stream
were scant: perhaps the best available were the two data points for the
Orphan stream of \cite{orphan-discovery}, which were held by the
authors to be ``suggestive rather than conclusive''.

Since then, the situation has improved somewhat. \cite{newberg-orphan}
have recently reported radial velocity measurements for 7 fields along
the Orphan stream, deriving from recently released SDSS/SEGUE
spectra of giant stars. The measurements have associated uncertainties
of $\near 10\kms$. The GD-1 \citep{gd1-discovery} and Anticentre
streams \citep{anticentre-discovery} are closer to the Sun than is the
Orphan stream, and therefore make easier targets for observations.
\cite{grillmair-anticentre-radvs-pms} obtained radial velocity data
for two fields along the Anticentre stream
\citep{anticentre-discovery}, with an accuracy of $\near 7\kms$.
\cite{koposov} combined SDSS/SEGUE spectra and purposely-obtained
spectra of 24 main sequence and turn-off stars, and although they do
not do so, their results could perhaps be organized into four or five
fields with uncertainties of $\sim 4\kms$ in each.

The procedures in this chapter could be immediately applied to all
these streams. However, our experiments with pseudo-data indicate that
the strongest constraints on the potential are to be had from $\near 1\kms$
precision data, rather than $\near 10\kms$ precision data.  Except
for perhaps the \cite{koposov} observations of GD-1, the available
radial velocity data do not meet this requirement. 

Part of the
problem stems from the need to use only the brightest stars in order
to obtain useful spectra on the small-aperture telescopes that have so
far been used for this work. The density of such stars associated with
tidal streams is low, which are by their nature dynamically old
structures, and are often barely distinguishable from the field stars
even in the main sequence. The few giant stars that
are observed are insufficient in number to beat down the observational random errors
to the required level. The problem
is made worse by the difficultly in convincingly identifying giant
stars with streams, with the small statistics involved leading to the
possibility of a single contaminant star ruining an entire field.

Obtaining radial velocity measurements of streams with the required
accuracy will necessitate the observation of main-sequence stars.
Observations of main-sequence stars in distant streams such as the Orphan stream would
require the use of $8\,$m-class equipment for what is superficially
unglamourous work. However, the results of this chapter indicate
that the scientific rewards of such an undertaking would be
far-reaching.

%%% chopped stuff...

% A Sun-like main sequence star of absolute magnitude
% $M_\sun \sim 4$ has an apparent magnitude of $M \sim 20$
% at a distance of $16\kpc$, and only with great effort is it
% possible to obtain line-of-sight velocity measurements for the
% hundreds of stars necessary to obtain a precise measurement
% of a stream.

% Indeed, until very recently \citep[willett?,koposov,newberg-orphan]
% the only line-of-sight velocities available for streams
% were taken from giant stars \citep[orphan-discovery,recent-odenkirchen-pal5?].
% The problem with this approach is that streams are observed, on the
% sky, as overdensities of main-sequence stars. Determining whether
% a given star belongs to a particular stream is a difficult
% task. When examining main-sequence stars, we are able to consider the
% problem statistically and attempt to isolate overdensities in
% line-of-sight velocity measurements \citep[e.g.]{koposov}. This
% is simply not possible with velocities measured from giant stars,
% which are too few in number. Hence, it is never possible to be
% really sure that a given giant star belongs to the stream
%  \citep[recent-odenkirchen-pal5?].

%%% Local Variables: 
%%% mode: latex
%%% TeX-master: "../thesis"
%%% End: 

%% file: gal_plx/gal_plx.tex
\chapter{Fitting orbits to streams using proper motions}
\label{chap:pms}

\section{Introduction}

The previous chapter has demonstrated the remarkable
diagnostic power that orbit reconstruction from
observed stream tracks can have, when coupled with
radial velocity measurements down those streams.

However, it was proffered that perhaps the most serious problem
afflicting such work is the difficulty in obtaining precise radial
velocity measurements of stars in such streams, and particularly
of the faint main-sequence stars that constitute the bulk of the
prospective data.

One possible workaround to reconstruct phase-space
tracks using measured proper motions rather than radial velocities. This
is attractive because, although it is much harder to make
an impromptu proper-motion measurement than it is to take a spectrum,
it {\em is} possible to measure very many of the former simultaneously,
down to the faint apparent magnitudes of main-sequence stars in
distant streams.

This kind of work is already possible to some extent with the combined
SDSS-USNO catalogue of \citet{munn-etal}, as demonstrated by
\citet{koposov}, who obtain comparatively high precision proper-motion
measurements for the GD-1 stream \citep{gd1-discovery} from
lower-precision Munn et al.~raw data, on account of the hundreds
of main-sequence stars they are able to identify.  By comparison,
their spectroscopically-obtained radial velocity data number only in
the dozens, with even this frugal number making GD-1 the best observed
object in its class.  Furthermore, with the advent of advanced
astrometric projects such as the Pan-STARRS survey \citep[currently
commissioning]{pan-starrs}, and its follow-on project LSST
\citep{lsst}, the future availability of large amounts of high-quality
proper-motion data for main-sequence stars is assured. We note that
Pan-STARRS proper motions will be obtained for stars seven visual
magnitudes fainter than those to be observed by
Gaia \citep{gaia}.

In the work of this chapter, much of which has been published in
an article by \cite{eb09b}, we derive a set of equations that can reconstruct
full-phase space information from an on-sky track, if proper-motion
magnitudes are measured everywhere along that track. The application
of this technique is logically identical to the reconstruction of
trajectories using radial-velocity data (\citealp
{jin-reconstruction,binney08}; \chapref{chap:radvs}).
We will demonstrate numerically that the reconstructed tracks
can be tested for dynamicity, just as were those
in \chapref{chap:radvs}. We further illustrate that the ability
to use reconstructed tracks as probes of the Galactic potential
is retained when using proper-motion data.

The remainder of this chapter is arranged as follows.
\secref{pms:sec:recon} derives a set of reconstruction
equations that use proper motions to generate candidate orbit
trajectories.
\secref{pms:sec:recontests} demonstrates the efficacy of the
reconstruction equations with a numerical test.
\secref{pms:sec:conclusions} discusses our findings and sums
up.

\section{Reconstructing orbits with proper motions}
\label{pms:sec:recon}

As in \chapref{chap:radvs}, we derive a reconstruction algorithm
on the basis that a stream precisely delineates an orbit.
The probable effects of this incorrect assumption
will be noted when applications of the work are discussed.

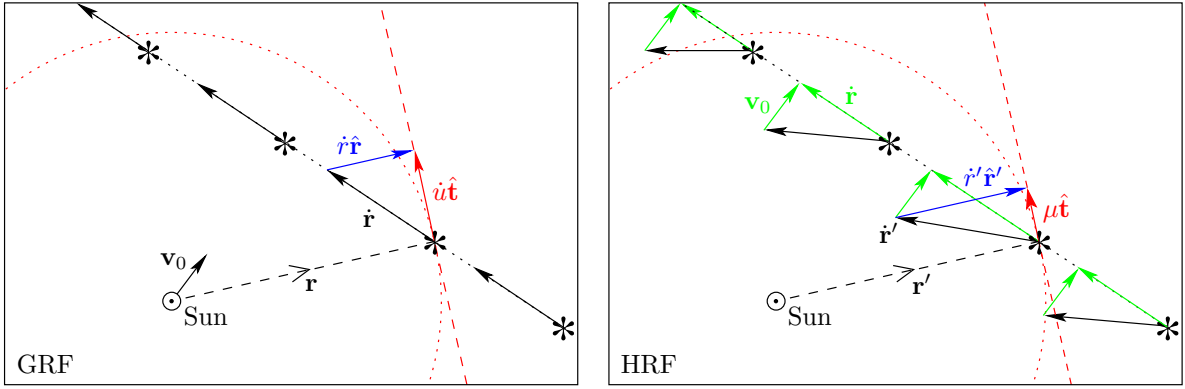
\begin{figure}
\centerline{
\input{grf.pstex_t}\quad
\input{hrf.pstex_t}
}
\caption[Diagrams of velocities in the grf and the hrf] { Diagrams
  illustrating the velocities of the Sun and of a stream of stars, as
  seen from (left) the Galactic rest-frame, and (right) the
  heliocentric rest-frame. Black arrows represent the velocities of
  objects. The components of velocity along the line-of-sight (blue
  arrows), and in the plane of the sky (red arrows), are shown for a
  star in the stream at position $\vect{r}$. A section of the plane of
  the sky at the radius $r$ is shown as a dotted red circle, while the
  dashed red line shows the tangent to that circle at the position
  $\vect{r}$.  Note that, even while the stream stars have been taken
  to follow one another in the grf, in the hrf the velocities of the
  stream stars do not point along the stream.}
\label{pms:fig:grf-hrf}
\end{figure}

Consider two inertial frames of reference: one in which the
Galactic centre is at rest (the grf), and another in which the
Sun is instantaneously at rest (the hrf).
In the grf, let $\vect{x}_0$ be the position vector of
the Sun, $\vect{x}$ the location of a star in the stream and let
$\vect{r}$ be the vector from the Sun to the star. By definition
\begin{equation}
\vect{r} = r \vecthat{r} = \vect{x} - \vect{x}_0,
\label{pms:eq:grf}
\end{equation}
where $\vecthat{r}$ is the direction from the Sun to the star.
Now let $\vect{x}'$ be the vector obtained by
transforming $\vect{x}$ from the grf
frame to the hrf. Again, by definition 
\begin{equation}
\vect{r}' = r' \vecthat{r}' = \vect{x}' - \vect{x}'_0.
\label{pms:eq:hrf}
\end{equation}
Consider the measurement of $\vecthat{r}$ and $\vecthat{r}'$, each
made by one two observers, both at the location of the Sun, but each in his
respective frame.  Instantaneously the frames must
coincide, so $\vect{r} = \vect{r}'$ at that given moment.  However, the
instantaneous derivatives $\d \vecthat{r}/ \d t$ and $\d \vecthat{r}'
/ \d t$ will not be equal.  The former derivative is tangent to the
stream, and is the proper motion that would be made by an observer
stationary with respect to the Galactic centre.
The latter derivative is the terrestrial observable proper motion of the star,
and includes a component perpendicular to the stream, due to the reflex
motion of the Sun.
Diagrams illustrating these configurations are shown in \figref{pms:fig:grf-hrf}.

Now let $u$ be the on-sky angle along the stream from some fiducial point.
We may write
\begin{equation}
\frac{\d\vecthat{r}'}{\d t} = \mu \vecthat{t}; \qquad  \frac{\d\vecthat{r}}{\d t} = \dot{u}
\vecthat{p},\label{pms:eq:derivatives}
\end{equation}
where $\mu$ and $\vecthat{t}$ are the magnitude and direction of the
measured proper motion, while $\dot{u}$ and $\vecthat{p}$ are the
on-sky magnitude and direction of the star's motion along the stream. 
Space velocities measured in the grf differ
from those measured in the hrf by the velocity of the Sun,
that is
\begin{equation}
{\d \vect{r}' \over \d t} = {\d \vect{r} \over \d t} - \vsun,
\label{pms:eq:transform}
\end{equation}
where $\vsun$ is the grf velocity of the Sun.
Differentiating equations \blankeqref{pms:eq:grf} and
\blankeqref{pms:eq:hrf}, and combining the results
with \eqref{pms:eq:derivatives} and \eqref{pms:eq:transform},
we obtain
\begin{align}
{\d \vect{r} \over \d t} &=
\dot{r}\vecthat{r} + r \mu \vecthat{t} \nonumber\\
& = {\d \vect{r}' \over \d t} - \vsun 
 = \dot{r}' \vecthat{r}' + r' \dot{u} \vect{p} - \vsun  = \dot{r}' \vecthat{r} + r \dot{u} \vect{p} - \vsun,
\end{align}
where in the last step we have made use of the instantaneous equality
$\vect{r} = \vect{r}'$. We note that $\d r' / \d t = \dot{r}'$ is the
spectroscopically measured heliocentric velocity, and that
$\d r / \d t = \dot{r} = v_r$ is the projection along the line
of sight of the star's velocity with respect to the Galactic centre.
Equating the components in the plane of the sky, we have
\begin{equation}
r \mu \vecthat{t} = r \dot{u} \vecthat{p} - (\vect{v}_0 - \vecthat{r} \cdot 
\vect{v}_0 \,\vecthat{r} ) = r \dot{u} \vecthat{p} - \vect{v}_{s},
\end{equation}
where $\vect{v}_{s}$ is the component of the Sun's velocity perpendicular to the
line of sight. This equation has just two unknowns, $\dot{u}$ and $r$, and we
can in principle solve for both through
\begin{equation}
\dot{u}\vecthat{p} = \mu \vecthat{t} + \frac{\vect{v}_{s}}{r}.
\label{pms:eq:fundamental}
\end{equation}
Specifically, since both $\vecthat{t}$ and $\vecthat{p}$ can in principle be
deduced from the observations and $\vect{v}_{s}$ may be presumed known,
we could determine $r$ such that the right side is parallel to $\vecthat{p}$, and
then read off $\dot{u}$ from the magnitude of the right side. Startlingly,
this permits the distance to the stream to be recovered
from the observables with no additional assumptions. This method
embodies the idea of Galactic parallax, and is examined further in
\chapref{chap:galplx}.
For now, we will consider that case where the uncertainty in the direction $\vecthat{t}$
is significant, prohibiting the use of the Galactic parallax procedure.
We proceed to eliminate $\vecthat{t}$ by squaring up
\begin{equation}
(\dot{u})^2 - 2\frac{\vect{v}_{s} \cdot \vecthat{p}}{r}\dot{u}
+ \frac{\left| \vect{v}_{s} \right|^2}{r^2} - \mu^2 = 0.
\end{equation}
The roots of this quadratic equation in $(r \dot{u})$ are given by,
\begin{equation}
r \dot{u} = \vect{v}_{s} \cdot \vecthat{p} \pm \sqrt{ 
   (\vect{v}_{s} \cdot \vecthat{p})^2 + (r \mu)^2 - 
   \left| \vect{v}_{s} \right|^2}.
\label{pms:eq:udot}
\end{equation}
The sign ambiguity is resolved as follows.
Using this expression to eliminate $\dot u$ from
\eqref{pms:eq:fundamental}, we find
\begin{equation}
\mu\,\vecthat{t}\cdot\vecthat{p}
=\pm\sqrt{ 
   (\vect{v}_{s} \cdot \vecthat{p})^2  - 
   \left| \vect{v}_{s} \right|^2+ (r
   \mu)^2}.
\end{equation}
 Since $\mu$ is inherently positive, we see that the sign of the radical in
\eqref{pms:eq:udot} must be chosen to agree with the sign of
$\vecthat{t}\cdot\vecthat{p}$. Even though the directions of individual
proper motions may be uncertain, it should be possible to decide whether they
are on average opposed to the direction of travel along the stream.
This ambiguity resolved, \eqref{pms:eq:udot} now makes $\dot u$ into a
function of both $r$ and quantities that can be determined from the observations.

Now let $\vect{F}(\vect{r}) = -\nabla\Phi(\vect{r})$ be the Galaxy's gravitational
acceleration. We recall that when we resolve the
star's equation of motion along the line of sight, we
obtain \bracketeqref{radvs:eq:dotvr2}
\begin{equation}
\label{pms:eq:eq-of-m}
{\d\vr\over\d t}= \dot{u} {\d\vr \over \d u} = F_r+{\vt^2\over r},
\end{equation}
where $F_r = \vecthat{r}\cdot\vect{F}$ is the radial component of the
acceleration, $v_t = \sqrt{v^2 - v_r^2}$ is the magnitude of
the on-sky component of the star's velocity, and where we have explicitly written
the time derivative in terms of $u$ to make it clear that we refer to
phase along the stream, and not a measured flux.

Since \eqref{pms:eq:udot} makes $\dot u$ a known
function of  $r$, in conjunction with \eqref{pms:eq:eq-of-m} we can now write down
a system of three coupled nonlinear ODEs for the
unknowns along the stream
\begin{align}
\label{pms:eq:diff-eqs}
{\d r\over\d u}&={\vr \over \dot{u}},\nonumber\\
{\d\vr\over\d u}&={F_r+r\dot u^2 \over \dot{u}},\\
\d t \over \d u & =  {1 \over \dot{u}}.\nonumber
\end{align}
If initial conditions on $(r,\vr,t)$
were given at some point on the stream, by
integrating these equations we can obtain complete phase-space coordinates
at every point along the stream. We can trivially set $t=0$ at a fiducial
point along the stream, and like in \chapref{chap:radvs}, we can guess
an initial distance $r = r_0$ at that point.
To obtain the initial condition on $\vr$ we write
\begin{align}\label{pms:eq:Fperp}
F_t&={\d\vect{v}\over\d t}\cdot\vecthat{p}
={\d\over\d t}(\vr\vecthat{r}+r\dot u\vecthat{p})\cdot\vecthat{p}\nonumber\\
&=\vr{\d\vecthat{r}\over\d t}\cdot\vecthat{p}+\vr\dot u+r\ddot u\\
&=2v_\parallel\dot u+s\ddot u,\nonumber
\end{align}
 where we have used \eqref{pms:eq:derivatives} to eliminate
 $\d\vecthat{r}/\d t$. Hence
\begin{equation}
\frac{\d(r \dot{u})}{\d t} = \dot{u}{\d (r \dot{u}) \over \d u}
= F_t - \vr \dot{u}. \label{pms:eq:tan-eq-of-m}
\end{equation}
The lhs of this equation is obtained by explicitly differentiating \eqref{pms:eq:udot}
\begin{equation}
\frac{\d (r\dot{u})}{\d u}  = \alpha
+ {1 \over \beta} \left( \alpha\gamma
+ \mu r^2 \frac{\d\mu}{\d u}
+ {r \mu^2 v_r \over \dot{u}} - \vot \cdot \frac{\d\vot}{\d u} \right),
\label{pms:eq:dsudot/dt}
\end{equation}
where we have defined
\begin{align}
\alpha =& \frac{\d\vot}{\d u}\cdot\vecthat{p}
+ \vot \cdot \frac{\d\vecthat{p}}{\d u},\nonumber\\
\beta =& \sqrt{(\vot \cdot \vecthat{p})^2 + (r\mu)^2 - v_{s}^2},\\
\gamma =& \vot \cdot \vecthat{p}\nonumber.
\end{align}
We can evaluate the terms in these quantities as follows.
By differentiation of $\vect{v}_s = \vsun - \vecthat{r}\cdot\vsun\,\vecthat{r}$,
we find that
\begin{align}
\frac{\d\vot}{\d u} & = -\vecthat{r}\cdot\vect{v}_0\,\frac{\d\vecthat{r}}{\d u}
- \frac{\d\vecthat{r}}{\d u}\cdot\vect{v}_0\,\vecthat{r}. \nonumber\\
& =  -\vecthat{r}\cdot\vect{v}_0\,\vecthat{p}
- \vecthat{p}\cdot\vect{v}_0\,\vecthat{r},
\end{align}
and so
\begin{equation}
\frac{\d\vot}{\d u}\cdot\vecthat{p} = -\vecthat{r}\cdot\vect{v}_0,
\end{equation}
and
\begin{equation}
\vot\cdot\frac{\d\vot}{\d u} = -(\vecthat{r}\cdot\vect{v}_0)\, \vot \cdot \vecthat{p},
\end{equation}
all of which can be evaluated given the initial condition $r_0$
and the observables.
Since \eqref{pms:eq:dsudot/dt} is linear in $\vr$, we can combine it with
\eqref{pms:eq:tan-eq-of-m} and rearrange to give
\begin{equation}
v_r = {1 \over {\beta + s \mu^2/\dot u }}{\left(
{\beta F_t \over \dot{u}} - \alpha\beta - \alpha\gamma
  + \vot \cdot\frac{\d\vot}{\d u}- \mu {\d\mu\over\d u}s^2\right)}.
\end{equation}
Once an initial distance $r=r_0$ has been chosen, the right side of this equation can be
evaluated from the observables at the fiducial point on the stream. The initial
conditions required for the integration of equations \blankeqref{pms:eq:diff-eqs}
are then known, and full phase-space information for the stream is defined.

In practice, it is clear that not all values of $r_0$ will provide
solutions to \eqref{pms:eq:fundamental}. In particular, if
$r$ is significantly too small, \eqref{pms:eq:fundamental} will become
insoluble at some point along the track. This pathology manifests
itself as $\dot{u} \rightarrow 0$ which halts the solution of
the differential equations \blankeqref{pms:eq:diff-eqs}. This event signals that the measured
value of $\mu$ is too small to be consistent with the reflex motion of the Sun
at the proposed distance, and it effectively sets a
geometrical lower bound on $r_0$. There is no similar geometrical
upper bound, since $\dot{u} \rightarrow \mu$ as $r_0 \rightarrow
\infty$ and \eqref{pms:eq:fundamental} always has a solution.
In the range above the lower bound, $r_0$ must be searched over, with
dynamics used to isolate any physical solutions, just as for 
the radial-velocity reconstruction in \chapref{chap:radvs}.

\section{Tests}
\label{pms:sec:recontests}

The implementation of the radial-velocity reconstruction algorithm
from \chapref{chap:radvs} was adapted to solve
equations~\blankeqref{pms:eq:diff-eqs}. Given an on-sky track
$[l(u),b(u)]$ and heliocentric proper motions $\mu(u)$, along with an
assumed potential $\Phi(\vect{r})$ and an initial distance $r_0$, the
algorithm returns a trajectory for which full phase-space information
is known.  Aside from the system of equations being solved, the sole
difference between the implementation in \chapref{chap:radvs} and that
here is the following minor procedural upgrade: instead of the
fiducial point from which the integration begins being located at one
of the ends of the input track, it was instead located in the
middle. This leads to a somewhat more accurate start to the
integration, since the data constrain the derivatives much more
tightly in the middle of the track than they do at the ends.

\begin{figure}
\centerline{
\includegraphics[width=0.45\hsize]{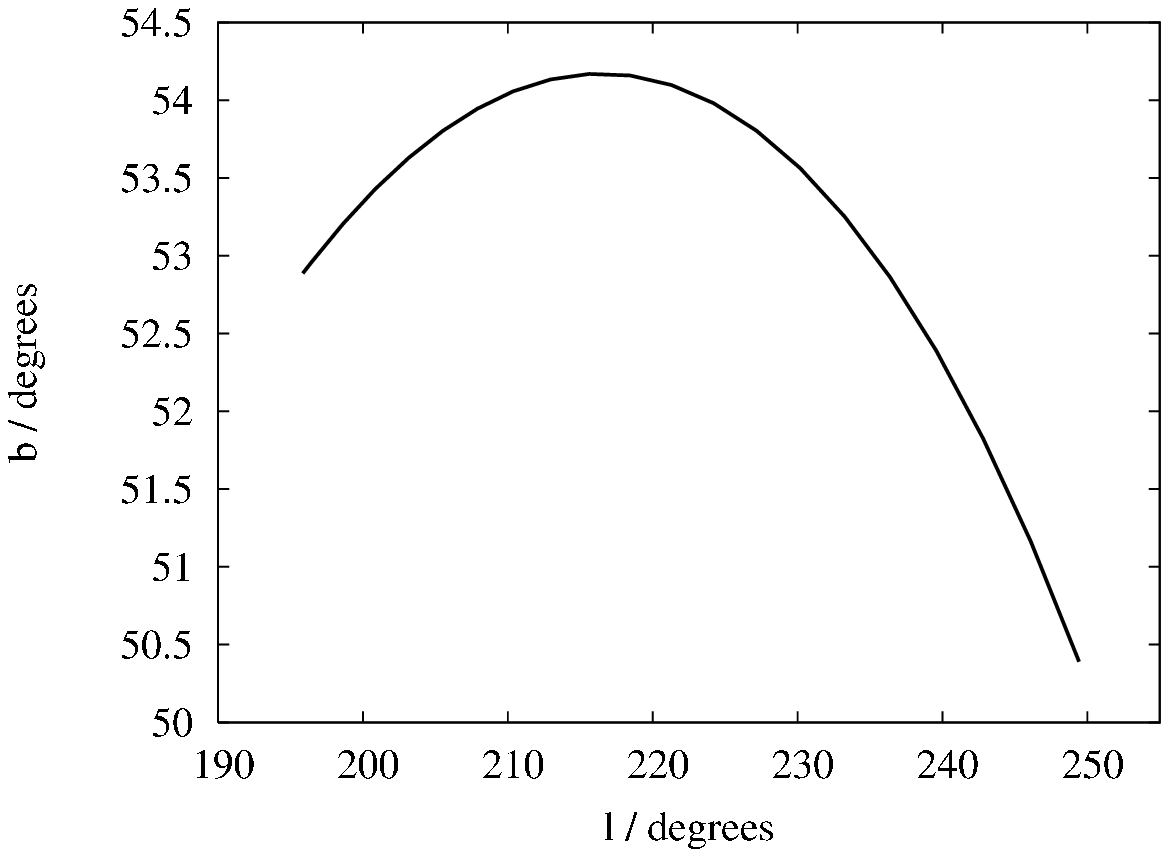}
\includegraphics[width=0.45\hsize]{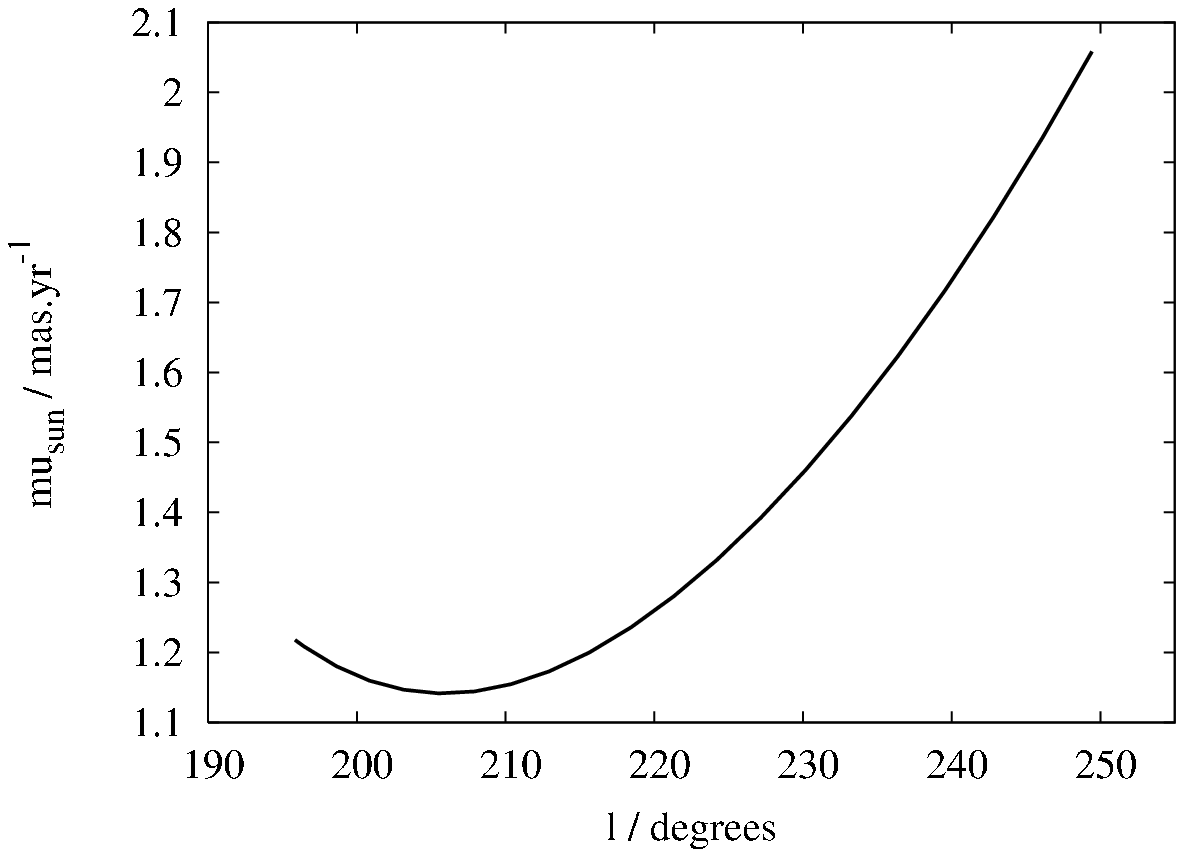}
}
\caption[Test input for the algorithm]
{
Test input for the algorithm.
The input is derived from a perfect orbit in the Model II potential, and is
described in the text.
Left panel: on-sky projection of the orbit. Right panel: heliocentric
proper-motion magnitude.
}
\label{pms:fig:test-input}
\end{figure}

To test the algorithm, a segment of the PD1 Test Orbit from
\chapref{chap:radvs} was integrated in the Model II potential
of \cite{bt08}. This segment, which is identical to that
used to test the radial velocity reconstruction algorithm
in \figref{radvs:fig:binneytrack}, was used to generate the input data sets $[l(u),b(u)]$
and $\mu(u)$, assuming the Sun to be on a circular orbit of radius $\rsun=8\kpc$, at which
Model II generates a circular velocity $v_c = v_0 = 227\kms$.
The input data sets, which represent a perfect orbit in the Model
II potential, are shown in \figref{pms:fig:test-input}.

The reconstruction equations
\blankeqref{pms:eq:diff-eqs} were then solved for this input,
and for a range of initial distances $r_0$, with each initial
distance producing a different reconstructed trajectory.
As in \cite{binney08} and \secref{radvs:sec:recon},
the trajectories were tested for dynamicity by
examining the conservation of orbital energy along them.
The normalized rms orbital energy variation
\begin{equation}
{\Delta E \over E} \equiv \sqrt{{\langle E^2\rangle \over {\langle E\rangle}^2} - 1},
\end{equation}
was computed for each trajectory,
where $E_i = v_i^2/2 + \Phi(\vect{r}_i)$ at a point $\vect{r}_i$ along the track.

\begin{figure}
\centerline{
\includegraphics[width=0.45\hsize]{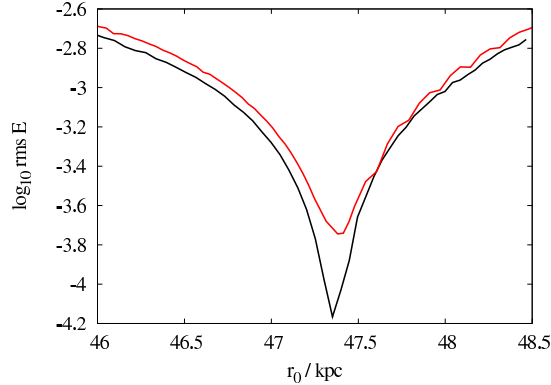}
}
\caption[Orbital energy conservation for reconstructions using proper motions]
{
Diagnostic quantity $\Delta E/E$ for a range of initial distances,
$r_0$, for input corresponding to a perfect orbit in the Model II
potential. Black line: the procedure performed in the Model II
potential. Red line: the procedure performed in a potential
with halo asymmetry parameter $q=1$, but otherwise identical
to the Model II potential.}
\label{pms:fig:test-de}
\end{figure}

The black curve of \figref{pms:fig:test-de} shows the
diagnostic quantity $\Delta E/E$ as a function of initial
distance $r_0$, where the reconstruction algorithm has been given the correct form of
the host potential $\Phi(\vect{r})$. As with the reconstruction
using radial velocities in \chapref{chap:radvs}, the 
procedure has highlighted the correct distance to the
stream with superb precision. Indeed, in this example, the
energy conservation of the best reconstructed trajectory
exceeds, by half an order of magnitude, the energy conservation
of the equivalent track when reconstructed from radial
velocity data. The reasons for this are twofold. Firstly,
the upgraded technique of beginning the integration from the
centre of the track has reduced numerical noise
slightly. Secondly, the two reconstruction techniques, while
intellectually equivalent, are mathematically distinct, and it
is understandable that one technique may fare better than another,
given otherwise identical input.

The red curve in \figref{pms:fig:test-de} is for reconstructions
using the same input data as the black curve, but in this case
we have asked the algorithm to reconstruct in a potential similar to
Model II, but for which the halo asymmetry parameter \citep[Table~2.3]
{bt08} has been changed from
its Model II value of $q=0.8$ to $q=1.0$. Thus, the algorithm is
attempting to reconstruct trajectories using a potential which
is less flattened than the truth. 

A minimum in the red curve is still very distinct, and the minimum
indeed occurs at approximately the correct distance, as can be seen by
comparison with the black curve. We understand this, because we have
not altered the halo mass, only its shape. Thus, the magnitude of the
force at the distance of the stream remains approximately correct.
The results of \secref{radvs:sec:potential} showed that the force
strongly determines the distance $r_0$ at which the optimum trajectory
is found.  However, the minimum in the red curve shows substantially
worse energy conservation than does the black curve, being half an
order of magnitude above the noise floor, as determined by the minimum
of the black curve. Thus, as in a reconstruction using radial
velocities, when given perfect input, the algorithm can successfully
diagnose an incorrect potential.

\section{Conclusions}
\label{pms:sec:conclusions}

We have complemented the work of \cite{binney08} and
\chapref{chap:radvs} by showing that, given proper motions
measured everywhere along a segment of a single orbit,
then full phase-space information can be reconstructed
for that orbit, just as it can if radial-velocity measurements
are known.

However, proper motions are two-dimensional vectors rather than
scalars like line-of-sight velocities, and this fact is rather useful.
We have found that, in principle, an orbit can be recovered from
proper-motion measurements on purely geometric grounds, with no
assumptions made about the potential, and no scanning over distances
to find dynamical tracks required. Since any orbit so recovered {\em
  must} correspond to an orbit in the correct potential, the latter
would be tightly constrained.

We explore this exciting discovery is full detail in
\chapref{chap:galplx}.  In the work of this chapter, we have chosen to
sacrifice some of the diagnostic power of the proper motions by
utilizing only their magnitude, and not their direction, which we deem
to be the most difficult quantity to measure. The resulting
scheme is logically identical to the scheme of
\cite{binney08} for reconstructing trajectories from
radial-velocity data. Just as in the latter, trajectories
reconstructed from proper motions are
not forced to be dynamical, and we must hunt for those that are,
by computing a diagnostic criterion for each candidate trajectory.

% This means that $r_0$ is not strictly determined.

Our tests have indicated that reconstructions using proper
motion data are at least as potent as are radial-velocity
reconstructions when it comes to both isolating dynamical orbits, and
diagnosing the potential.  However, like with the radial velocity
reconstruction, we expect random input errors, and the failure of
actual stream tracks to delineate individual orbits, to be destructive
of our ability to identify dynamical solutions.

In order to unlock the diagnostic power of streams using proper-motion
measurement, it is likely that techniques similar to those presented in
\chapref{chap:radvs} will be necessary, affording the algorithm the ability to detect
those dynamical solutions that can be found with constrained
modification of its input.
In principle, the adaptation of the procedures
to the new algorithm ought to be straightforward,
given the intellectual similarity between the two sets of reconstruction equations,
and given the similarity in our method of implementing them.

However, we will leave this interesting endeavour for future work.
Instead, in the following chapter we will revisit our assertion that
recovery of the distance to a stream star can proceed 
from nothing more than measurement of its on-sky coordinates and 
proper-motion vector.

%%%%%%%%%%%%%%%%%%%%%%%%%%%%%%%%%%%%%%%%%%%

\chapter{Galactic parallax for the tidal stream GD-1}
\label{chap:galplx}

\section{Introduction}

Conventional trigonometric parallax has long been used to calculate
accurate distances to nearby stars. The regular nature of the
parallactic motion of a star, caused by the Earth's orbit around the Sun, allows
this motion to be decoupled from the intrinsic proper motion of the
star in the heliocentric rest-frame. Hence the distance to the
star can be calculated. However, the maximum
 baseline generating such parallaxes is obviously limited to 2AU. For
a given level of astrometric precision, this imposes a fundamental
limit to the observable distance. Indeed, the accuracy of
parallaxes reported by the Hipparcos mission
data \citep{newhipparcos} falls to $20$--$30$ percent at best for distances
$\near 300\pc$ and only then for the brightest stars.
Upcoming astrometric projects such
as Pan-STARRS \citep{pan-starrs}, LSST \citep{lsst} and the
Gaia mission \citep{gaia} will achieve similar uncertainty
for Sun-like stars as distant as a few kpc, and at
fainter magnitudes than was possible with Hipparcos. This
extended range will encompass less than 1 percent of the total number
of such stars in our Galaxy.

It is clear that it will {\em not} soon be possible to calculate
distances to many of the stars in our Galaxy with conventional
trigonometric parallaxes. Alternative means to compute distances to
stars are therefore required. Photometry can be used to estimate the
absolute magnitude of a star which, when combined with its observed
magnitude, allows its distance to be computed. Unfortunately, all
attempts to calculate such photometric distances are hindered by the
same problems: obscuration by intervening matter alters both observed
magnitude \citep{vergely} and colour \citep{reddening1,drimmel}, and
it is difficult to model appropriate corrections without a reference
distance scale. The effects of chemical composition and age further
complicate matters \citep{juric}. It is therefore difficult to compute
photometric distances with an accuracy much better than 20 percent,
even for nearby stars, and distances to faint stars are less accurate
still \citep{juric}.

Consequently, there exists a demand for long-range distance
measuring tools that can complement the existing tools by
overcoming some of their limitations. The work of this chapter,
much of which was published in articles by \cite{eb09b} and \cite{galplx},
describes such a tool, which measures distances by utilizing
the parallactic motion between a distant star and the Sun, rather
than that between a star and the Earth as in conventional parallax. 

In the general case, it is not possible to
obtain distances this way, because the parallactic motion of 
the star and its intrinsic proper motion are
inextricably mixed up.
However, in the special case where the star can be associated
with a stellar stream, its rest-frame trajectory can be predicted
from the locations of the other associated stars.
Using this trajectory, the proper motion in the Galactic
rest-frame can indeed be decoupled from the reflex
motion of the Sun, and the component of 
its motion due to parallax can be computed.

\begin{figure}
\centerline{
\input{gpdiagram.pstex_t}
}
\caption[Diagram illustrating the Galactic parallax effect]{
An illustration of the Galactic parallax effect.
The Sun is at some distance $r$ from a star in a stream. The stream track is
oriented perpendicularly to the plane of the diagram. In the rest-frame of
the Galaxy, the Sun has a velocity $\vsun$ (marked in red), while the velocity of the stream
star points vertically out of the page, consistent with it travelling along
the stream. In the heliocentric rest-frame, however, the stream star
acquires a velocity component $-\vsun$ (marked in black) in the plane of the diagram, due to
the reflex motion of the Sun. Observationally, this reflex motion causes
the stream star to acquire
a component of proper motion, equal to $v_0 /r \equiv \Pi v_0$ projected onto the plane
of the sky, that is perpendicular to the stream track. This is the Galactic
parallax effect, and allows $\Pi$ to be calculated if $\vsun$ and the proper
motion are known.
}
\label{galplx:fig:gpdiagram}
\end{figure}
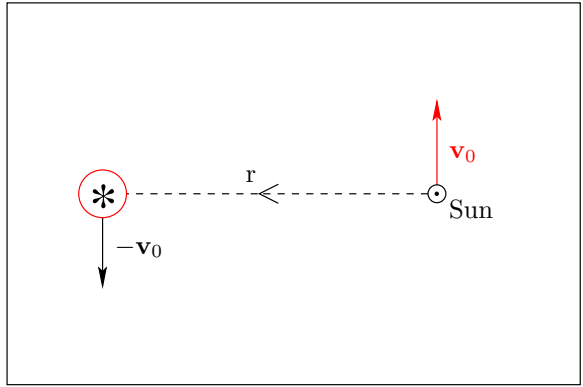

In the work of the preceding chapter it was discovered that by
measuring the proper-motion vector of a star belonging to a stream, it
is possible to directly recover the distance to that star.
This discovery is an embodiment of the Galactic parallax
effect: since we assume that the peculiar motion of the star is
tangent to its stream, {\em any} proper motion observed perpendicular
to that stream can only be due to the reflex motion of the Sun. If the
motion of the Sun in a common rest-frame is already known through
alternate means, then a parallax can be computed from this
perpendicular motion. A diagram illustrating the effect is shown
in \figref{galplx:fig:gpdiagram}.

% Most notably, no form for the host potential
% needs to be assumed, unlike with reconstruction from
% line-of-sight velocities (\chapref{chap:radvs}) or from proper-motion
% magnitudes alone (\chapref{chap:pms}), both of which require such
% an assumption to be made.

% Consider the angular motion of a star on the sky, which is
% generated by the relative real-space motion between that star and the observer.
% This motion is given by
% \begin{equation}
% \nu = {(\v - \v0) \over r},
% \end{equation}
% where $(\v - \v0)$ is the real-space relative motion, and $r$ is the distance
% from the observer to the star. Such is the basis for all parallax effects:
% for a fixed relative motion, the observed angular motion diminishes with
% the inverse of distance.

Such Galactic parallaxes have the same geometrical basis as
conventional trigonometric parallaxes, and as such are free from errors
induced by obscuration and reddening. However, the range of Galactic
parallax significantly exceeds that of conventional parallax.  This is
because the Sun orbits about the Galactic centre much faster than the
Earth orbits the Sun, and because, unlike with conventional parallax,
the Galactic parallax effect is cumulative with
continued observation. In realistic cases, for a typically oriented
stream, we can expect the Galactic parallax to be observable at nearly
40 times the distance of the equivalent trigonometric parallax, based
on 3 years of observations. For increased range, one simply observes over
a longer baseline.

This large range means that Galactic parallax might prove
a powerful tool to complement conventional parallaxes, and
validate other distance measuring tools. It is exciting to note that the
capabilities of astrometric projects such as LSST, which
will observe the conventional parallax of a G star at a distance of $\near 1\kpc$
with 20 percent uncertainty, will put much of the Galaxy in range of
Galactic parallax calculations with similar accuracy.

The main restriction on the use of Galactic parallax is the
requirement for stars to be part of a stream.  However, the continuing
discovery of significant numbers of streams
\citep{odenkirchen-delineate,majewski-sag,yanny-stream,orphan-discovery,
grillmair-orphan,gd1-discovery,ngc5466,
grillmair-2009,newberg-streams-2009}
using optical
surveys implies that they are a staple feature of the Galactic
environment, rather than a rarity.  The deep surveys of Pan-STARRS and
LSST are likely to find yet more, increasing the 
number of applications for Galactic parallax.

The remainder of this chapter explores the viability of using Galactic parallax to
estimate distances and demonstrates its practicality by applying it to
data for the GD-1 stream \citep{gd1-discovery} published by \citet[herein
\krh]{koposov}; we choose to work with the latter over the
earlier analysis of the same stream by \cite{willett} on account
of the significantly smaller proper-motion uncertainties ($1\masyr$
vs $4\masyr$) cited in the later work.
Throughout the chapter, the Solar motion is assumed to be
\begin{equation*}
(U,V,W) = (10.0,252,7.1) \pm (0.3,11,0.34)\kms,
\end{equation*}
consistent with \cite{ab09},
\cite{reid-brunthaler} and \cite{gillessen}.

The chapter is arranged as follows.
\secref{galplx:sec:galplx} outlines the calculation of Galactic
parallax from observational data.
\secref{galplx:sec:uncertainty} examines how  uncertainties
 affect Galactic parallax calculations, and
\secref{galplx:sec:practicality} discusses the practicality
of using Galactic parallax as a distance measuring tool in light
of those findings.
\secref{galplx:sec:galplxtests} demonstrates the use of Galactic
parallax on pseudo-data, and shows that our uncertainty estimates
are correct, and that the technique is practical.
\secref{galplx:sec:gd1} uses Galactic parallax to find distances
to the tidal stream GD-1.
\secref{galplx:sec:conclusions} presents our concluding remarks.

\section{Galactic parallax}
\label{galplx:sec:galplx}

Suppose that a star is part of a stellar stream, and has a
location relative to the Sun described by $(\vect{x}-\vect{x}_0)
= r\vecthat{r}$, where $r$ is the distance to the star, and
$\vect{x}_0$ is the position of the Sun.
In the plane of the sky, let the tangent to
the trajectory of the stream, near the star,
be indicated by the vector $\vecthat{p}$. Assume the
velocity of the Sun, $\vect{v}_0$, in the Galactic rest-frame (grf) 
is known, and for the time being assume that streams perfectly
delineate orbits. \chapref{chap:pms} showed that if the measured proper motion of
the star is $\mu \vecthat{t}$, then \bracketeqref{pms:eq:fundamental}
\begin{equation}
\dot{u}\vecthat{p} = \mu \vecthat{t} + \frac{\vot}{r}
= \mu \vecthat{t} + \Pi \vot,
\label{galplx:eq:fundamental}
\end{equation}
where $\Pi \equiv 1/r$ is the Galactic parallax,
$\dot{u}$ is the proper motion as would be seen from the grf,
and $\vot$ is the Sun's velocity projected onto the plane
of the sky. We note that $\dot{u} = v_t/r$, where $v_t$ is that
component of the star's grf velocity perpendicular to the line of sight,
and that
\begin{equation}
\vot = (\vect{v}_0 - \vecthat{r} \cdot \vect{v}_0\, \vecthat{r} ).
\end{equation}
\Eqref{galplx:eq:fundamental} is a vector expression and can be solved
simultaneously for both $\dot{u}$ and $\Pi$ provided that $\vecthat{p}$,
$\vecthat{t}$ and $\vot$ are not parallel.
The stream direction $\vecthat{p}$ will not typically be
known outright, but must be estimated from the positions of
stream stars on the sky. We can achieve this by fitting a low order
curve through the position data, the tangent of which is then taken
to be $\vecthat{p}$. The curve must be
chosen to reproduce the gross behaviour of the stream,
but we must avoid fitting high-frequency noise, because $\vecthat{p}$ is
a function of the derivative of this curve, which is sensitive
to such noise.

\section{Uncertainty in Galactic parallax calculations}
\label{galplx:sec:uncertainty}

We begin by taking the cross-product of \eqref{galplx:eq:fundamental}
with $\vecthat{p}$ in order to eliminate $\dot{u}$, and we render the
resulting equation into an orthogonal on-sky coordinate system, whose
components are denoted by $(x,y)$. Re-arranged for $\Pi$, we find
\begin{equation}
\Pi = { {\mu \left(t_x \sin \alpha - t_y \cos \alpha\right)} \over
{\voty \cos \alpha- \votx \sin \alpha} }, \label{galplx:eq:pi1}
\end{equation}
where the $(x,y)$ suffixes denote the corresponding components of
their respective vectors, and where we have defined the
angle $\alpha \equiv \arctan (p_y/p_x)$.

The choice of coordinates $(x,y)$ is arbitrary.
We are therefore free to choose the coordinate system in which
$\alpha = 0$, i.e.~that system in which the $x$-axis points
along the stream trajectory, $\vecthat{p}$. \Eqref{galplx:eq:pi1} becomes
\begin{equation}
\Pi = -{\mu \ty \over \vper}, \label{galplx:eq:pi2}
\end{equation}
where we now identify the $y$-component of the various vectors
as that component perpendicular ($\perp$) to the stream trajectory,
and the $x$-component as that component parallel ($\parallel$) to
the trajectory. \Eqref{galplx:eq:pi2} shows explicitly that the Galactic
parallax effect is due to the reflex motion of stream stars perpendicular
to the direction of their travel.

Uncertainties in the $(x,y)$ components of the measured quantities
$\mu \vecthat{t}$ and $\vot$, and uncertainty in $\alpha$,
can be propagated to $\Pi$ using \eqref{galplx:eq:pi1}. When we set
$\alpha = 0$, this equation becomes
\begin{equation}
{\sigma^2_\Pi \over \Pi^2} =
{\smt^2 \over \mu^2 \ty^2} + 
{\svper^2 \over \vper^2}
+ {\sigma_\alpha^2 \over \vper^2}\left(\vpar + {\mu \tx \over \Pi} \right)^2,
\label{galplx:eq:error:pi}
\end{equation}
where we anticipate the uncertainty in $\mu\vecthat{t}$ to be
isotropic, and so we have set $\sigma_{\mu t_x}= \sigma_{\mu t_y} =
\smt$.

We assume $\smt$ to be known from observations; it may contain
any combination of random and systematic error.
$\sigma_{\vper}$ is calculated directly from
the error ellipsoid on $\vect{v}_0$, which is assumed known. Any
error on $\vo$ affects all data in exactly the same way. However, the
projection of error on $\vo$ to $\vot$ varies with position on the
sky. Hence, the effect of $\svot$ is to produce a systematic error in
reported distance that varies along the stream in a problem-specific
way.

Uncertainty in $\alpha$ arises from three sources.  Firstly, a
contribution $\sigma_{\alpha,m}$ arises due to the possible
misalignment of the stream direction with the rest-frame proper
motions of the stars. The magnitude of this misalignment is dependent
upon both the geometry of the stream and the nature of the host
potential, as is detailed extensively in \chapref{chap:mech} of this
thesis. However, the misalignment is for the most part predictable,
given an initial estimate of the stream's orbit. The tangent vectors
$\vecthat{p}$ can therefore be corrected for the most part of this
misalignment, and the residual error is expected to be small.

The second contribution arises because
the on-sky trajectory $\vecthat{p}$ is chosen by fitting a smooth
curve through observational fields, so $\vecthat{p}$ need not be
exactly parallel to the underlying stream.
Further, since $\vecthat{p}$ depends on the derivative of the
fitted curve, it is likely to be much less well
constrained for the data points at the ends of the stream than for
those near the middle.

We can quantify this effect. At the endpoints, the fitted curve is likely
to depart from the stream by at most $\angwid$, the angular width of the stream
on the sky. For a low-order curve, this departure is likely to have been gradual over
approximately half the angular stream length, $\anglen$, giving a contribution to
$\sigma_\alpha$ from fitting of
\begin{equation}
\sigma^2_{\alpha,f} = {4\angwid^2 \over \anglen^2}.
\label{galplx:eq:errora:curvefitting}
\end{equation}

The third contribution to $\sigma_\alpha$ arises as follows.
Since the stream has finite width, at any point, the stars within it
have a spread of velocities, corresponding to the spread
in action of the orbits that make up the stream.
If the stars in a stream show a spread in velocity $(\sigma_{v_x}, \sigma_{v_y})$
about a mean velocity $\vect{v_t} = r\dot{u}\vecthat{p}$, this effect
contributes
\begin{equation}
\sigma^2_{\alpha,v} = {1\over v_t^2}\left( \sigma_{v_y}^2 \cos^2 \alpha +
\sigma_{v_x}^2 \sin^2 \alpha \right),
\label{galplx:eq:alpha:vspread}
\end{equation}
to the uncertainty in $\alpha$ for a single star.
Again we can choose $\alpha = 0$, such that $\sigma_{v_y}
= \sigma_{v\perp}$, the velocity dispersion perpendicular to the stream direction.
\Eqref{galplx:eq:alpha:vspread} becomes
\begin{equation}
\sigma^2_{\alpha,v} = {\sigma^2_{v\perp} \over v_t^2} = {\sigma^2_{v\perp} \over 
(v \sin \beta)^2},
\label{galplx:eq:alpha:sigw}
\end{equation}
where we have introduced $v$, the grf speed of the stream, and
$\beta$, the angle of the stream to the line of sight.
$\sigma_{v\perp}$ has its origin in the random
motions of stars that existed within the progenitor object.
In fact, if we assume the stream has not spread significantly
in width, then the width and the velocity dispersion \cite[\S 8.3.3]{bt08}
are approximately related by
\begin{equation}
{\sigma_{v\perp} \over v} \simeq {w \over \rperi} = {r\angwid \over \rperi},
\end{equation}
where $w$ is the physical width of the stream, and $\rperi$ is the
radius of the stream's perigalacticon. This gives
\begin{equation}
\sigma_{\alpha,v} = {r \angwid \over \rperi \sin \beta}.
\label{galplx:eq:alpha:equiv}
\end{equation}
If secular
spread has made the stream become wider over time, then 
this relation will over-estimate $\sigma_{\alpha,v}$, since
$\sigma_{v\perp}/v$ is roughly constant.
$\angwid$ therefore represents
an upper bound on the true value of $\sigma_{\alpha,v}$ through this relation.
 This argument also 
assumes that the stream was created from its progenitor in a single
tidal event. Real streams do not form in this way. However, repeated
tidal disruptions can be viewed as a superposition of ever younger streams,
created from a progenitor of ever smaller $\sigma_{v\perp}$.
\Eqref{galplx:eq:alpha:equiv} holds for each of these individually.
Thus, $\angwid$ remains a good upper bound for $\sigma_{\alpha,v}$
through this relation.

In reality, we do not measure the proper motion of individual
stream stars, but rather the mean motion of a field of $N$ stars. 
The contribution to $\sigma_\alpha$ is from the error on this mean.
Putting this together
with \eqref{galplx:eq:errora:curvefitting} gives our final expression
for $\sigma_\alpha$
\begin{equation}
\sigma^2_\alpha = {\sigma^2_{\alpha,v}\over N} + \sigma^2_{\alpha,f} 
+ \sigma_{\alpha,m}^2
= { r^2 \angwid^2 \over N
\rperi^2 \sin^2\beta} + {4\angwid \over \anglen^2}
+ \sigma_{\alpha,m}^2.\label{galplx:eq:errora:random}
\end{equation}
We note that the first term represents a random error, while the second
and third terms represent systematic errors that will vary with position down
the stream. In general, $\sin \beta$ and $\rperi$ are {\em a priori}
unknown. We can infer $\sin \beta$ from radial velocity information,
either directly where the measurements exist, or indirectly from
Galactic parallax distances. Guessing $\rperi$ requires assumptions
to be made about the dynamics, but in general we expect the ratio $r/\rperi \simeq 1$
or less.

Explicit evaluation of $\sin \beta$
and $\rperi$ are not necessary to evaluate the uncertainty if the
contribution from $\sigma_{\alpha,v}$
is overwhelmed by the error from fitting, $\sigma_{\alpha,f}$.
We can see this will be the case when the number of observed stars
per field
\begin{equation}
N > \Big( { r \anglen \over 2 \rperi \sin \beta} \Big)^2.
\label{galplx:eq:condition}
\end{equation}
We expect this to be true in almost all practical cases.

\subsection{Uncertainty in tangential velocity calculations}

\Eqref{galplx:eq:fundamental} can also be used to solve for $\dot{u}$
\begin{equation}
\dot{u} = {\mu (t_y + t_x) + \Pi (\voty + \votx) \over \cos \alpha
+ \sin \alpha},\label{galplx:eq:udot1}
\end{equation}
which becomes
\begin{equation}
\dot{u} = \mu (\tx + \ty)  + \Pi (\vpar + \vper) = \mu \tx + \Pi \vpar,
\label{galplx:eq:udot2}
\end{equation}
when we set $\alpha = 0$. \Eqref{galplx:eq:udot1} combined with
\eqref{galplx:eq:pi1} can be used to explicitly propagate
uncertainties in the measured quantities to $\dot{u}$. 
When $\alpha = 0$, the uncertainty in $\dot{u}$ is
\begin{eqnarray}
{\sigma^2_{\dot{u}} \over \dot{u}^2} &=&
{\vot^2 \smt^2 \over \mu^2 (\ty \vpar - \tx \vper)^2}
+ { \ty^2 (\vper^2 \svpar^2 + \vpar^2 \svper^2)
\over \vper^2 (\ty \vpar - \tx \vper)^2} \nonumber\\
&&{}-{ 2 \ty^2 \vpar \vper \cov(\vpar, \vper)
\over \vper^2 (\ty \vpar - \tx \vper)^2}
+ {\vpar^2 \sigma_\alpha^2 \over \vper^2}.
\label{galplx:eq:error:udot}
\end{eqnarray}
$\sigma_{\vpar}$ and $\cov(\vpar,\vper)$ are calculated directly from
the error ellipsoid on $\vect{v}_0$, which we have assumed known.

\section{Practicality of Galactic parallax as a distance measuring tool}
\label{galplx:sec:practicality}

Using \eqref{galplx:eq:pi2} to eliminate $\mu t_\perp$ from
\eqref{galplx:eq:error:pi}, and taking the dot product of $\vecthat{p}$ with
\eqref{galplx:eq:fundamental} to simplify the last term, we obtain
\begin{eqnarray}
{\sigma^2_\Pi \over \Pi^2} &=& {1 \over \vper^2} \Big\{
(r \smt)^2 + 
\sigma_{\vper}^2
+ (r\dot{u})^2\sigma_\alpha^2 \Big\}\nonumber\\
& =& {1 \over \vper^2} \Big\{
(r \smt)^2 + 
\sigma_{\vper}^2
+ v^2 \left({r^2 \angwid^2 \over \rperi^2 N} + {4\angwid^2 \over \lendeproj^2}
+ {\sigma_{\alpha,m}^2\over \sin^2 \beta}\right)
\Big\},
\label{galplx:eq:error3}
\end{eqnarray}
where we have noted that $r\dot{u} = v_t = v\sin\beta$, and we have related
the observed stream length, $\anglen$, to the deprojected length,
$\lendeproj = \anglen / \sin \beta$. We note that the last
term is independent of $r$, since $r \angwid = w$, $\angwid/\lendeproj$
and $\sigma_{\alpha,m}$ are all constant,  and that for a stream of given physical dimension,
the uncertainty in $\Pi$ only has dependence upon
the angle of the stream $\beta$ in the misalignment term $\sigma_{\alpha,m}$.

What level of uncertainty does \eqref{galplx:eq:error3} predict, when
realistic measurement errors are introduced? The answer to this is
dependent upon both the physical properties of the stream $(R_p, \angwid,
\lendeproj, v)$ and the geometry of the problem in question $(r, \vper, \beta)$.

We progress by assuming `typical' values for some of these quantities.
The average magnitude of $\vot$ taken over the whole sky is $v_0\, \pi
/ 4$. The average perpendicular component, for a randomly oriented
stream, is $2 / \pi$ of this value.  We therefore assume a typical
value for $\vper$ of $v_0 / 2 \sim 120 \kms$. We also assume a typical
grf velocity equal to the circular velocity, $v = v_c \sim 220\kms$,
and a value for the stream angle to the line of sight, $\sin^2 \beta \sim 0.5$,
consistent with a randomly oriented stream.

\cite{mb09} recently summarised the current state of knowledge of $\vo$. The
uncertainty quoted is typically $\near 5$ percent on each of $(U,V,W)$.
Correspondingly, we estimate a typical value for the uncertainty
$\sigma_{\vper}$ of 5 percent of $\vper$, or $6\kms$, which is dominated
by the error on V.

The GD-1 stream that we consider below is exceptionally thin and long,
with $\angwid \near 0.1\deg$ and $\anglen \near 60\deg$.  The Orphan
stream \citep{grillmair-orphan,orphan-discovery} is of similar length,
but about 10 times thicker.
%Both of these streams are near apsis, so
%$\anglen = \lendeproj$.
We therefore take $\angwid \near 1\deg$,
$\lendeproj \sim 60\deg$ as typical of the streams to which one would
apply this method.  If hundreds of stars are observed for each proper
motion datum, then \eqref{galplx:eq:condition} is true for all
realistic combinations of $(r,\rperi)$, so we can ignore the
contribution of $\sigma_{\alpha,v}$ to $\sigma_\alpha$. The
contribution from fitting error is $\sigma_{\alpha,f} \simeq 1.9\deg$.
The results of \chapref{chap:mech} will show that stream misalignment
is highly dependent upon the specifics of the problem in question, but
we find that in spherical logarithmic potentials, the typical
uncertainty due to misalignment $\sigma_{\alpha,m} \ll 1\deg$. If the
misalignment were larger, then the results of \chapref{chap:mech}
could be used to make a correction to the tangent vectors derived
from the stream.  In either case, the residual errors
$\sigma_{\alpha,m} \ll 1\deg$.  Hence, we will take $\sigma_\alpha
\simeq 2\deg$ as our typical value.

The individual SDSS-USNO proper motions \citep{munn-etal} used by
\krh have a random uncertainty $\smt\sim 4\masyr$. After averaging over
hundreds of stars and accounting for a contribution from non-stream stars,
\krh report a random uncertainty of $\smt \sim 1 \masyr$ on their GD-1 data.
For a stream $10\kpc$ distant, with these proper motions and the typical values mentioned,
 \eqref{galplx:eq:error3} reports an
uncertainty of $\sigma_\Pi / \Pi \sim 40$ percent. By far the
greatest contribution comes from the first term in \eqref{galplx:eq:error3},
hence, the error on proper-motion measurement is dominating our
uncertainty.

To obtain an uncertainty of $\sigma_\Pi / \Pi < 20$ percent with
\cite{munn-etal} proper-motion measurements, we would need to restrict
ourselves to streams less than $5\kpc$ distant. $20$ percent error is
also possible at $10\kpc$ given optimum problem geometry. This is
clearly competitive with the $\sim 20\pc$ at which one could observe a
standard trigonometric parallax, with similar accuracy, using
astrometry of this quality. However, previous work
\citep[\krh]{willett} shows that SDSS photometry combined with
population models produce distance estimates accurate to $\sim 10$
percent for stars in streams at $8\kpc$.  The accuracy of Galactic
parallax is therefore not likely to be as good as that of photometric
distances for distant streams, using data this poor, unless the
problem geometry is favourable.

Proper-motion data from the Pan-STARRS telescope
is expected to be accurate to $\near 1\masyr$ for Sun-like stars at
$10\kpc$ \citep{pan-starrs-3pi}. \krh reduce raw data with accuracy
$\near 4\masyr$ to processed data accurate to $\near 1\masyr$, even
though the expected proper motion of the stars is of the same size as
the errors. It is not unreasonable to expect a similar analysis
applied to Pan-STARRS raw data, where the relative error would be much
less than unity, to yield processed data accurate to $\sim
0.2\masyr$. In truth, the ability of Pan-STARRS to detect very
faint stars will increase the number of stars identifiable with a stream,
and thus reduce the uncertainty in the mean proper motion further than this,
but we use $0.2\masyr$ as a conservative estimate.

The same $10\kpc$ distant stream would have a parallax error of
$\sigma_\Pi / \Pi \simeq 11$ percent with data this accurate.  An
error of less than $20$ percent is possible for a typical stream less
than $\sim 23\kpc$ distant, and for a stream with favourable geometry
less than $50\kpc$ distant. \cite{juric} report that SDSS photometric
distances for dwarf stars have $\near 40$ percent error at
$20\kpc$. Thus, the accuracy of Galactic parallax derived from
Pan-STARRS data should be at least comparable to distance estimates
from photometric methods, even in the typical case.

Future projects such as LSST and Gaia will each obtain proper motions
accurate to $\sim 0.2\masyr$ for Sun-like stars $10\kpc$ distant \citep{ivezic,
gaia}. These
data would allow a distance estimate for our typical stream accurate
to $8$ percent, and a stream with favourable geometry accurate to
$4$ percent. Error in the proper motion no longer dominates the
uncertainty in these calculations.  We might expect such accurate
astrometric surveys to reduce the uncertainty in the Solar motion; in
this case, the error in Galactic parallax would be lower still.

Gaia will not observe Sun-like stars beyond $10\kpc$, but LSST will, with
accuracy of $0.4\masyr$ for dwarf stars $30\kpc$ distant
\citep{ivezic}. The accuracy of the parallax to our typical stream at
this distance would be about $14$ percent with these data, and
$6$ percent is achievable with optimum geometry. A typical stream
could be measured to 20 percent accuracy out to $40\kpc$, and a stream
with favourable geometry out to $54\kpc$; this range approaches the
limit of LSST's capability for detection of dwarf stars.  Such data
will put the Orphan stream, which is about $20-30\kpc$ distant
\citep{grillmair-orphan, orphan-discovery, sales-orphan}, in range of accurate trigonometric
distance estimation. For comparison, photometric distances from SDSS
data are hardly more accurate than 50 percent for this stream
\citep{orphan-discovery}.

\begin{figure}
\centerline{
\includegraphics[width=0.45\hsize]{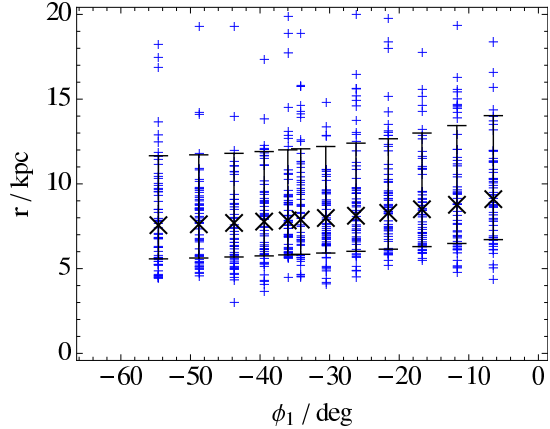}
}
\caption[Galactic parallax distances computed using erroneous pseudo-data]
{Crosses: Galactic parallax distances
computed from the pseudo-data, with no extraneous errors.
Error bars: the random scatter expected in Galactic
parallax distances, with measurement errors
 as mentioned in the text. Plus signs: Galactic parallax
distances computed from 60 Monte Carlo realisations of each
pseudo-datum convolved with the measurement errors. The analytic
uncertainty estimate and the Monte Carlo realisations are
in good agreement.}
\label{galplx:fig:pd-MC}
\end{figure}

\section{Tests of Galactic parallax}
\label{galplx:sec:galplxtests}

To test the method, pseudo-data were prepared from an orbit fitted to
data for the GD-1 stream by \krh. The orbit is described by
the initial conditions
\begin{equation*}
\vect{x}=(-3.41,13.00,9.58) \kpc, \qquad \vect{v} =
(-200.4,-162.6,13.9) \kms,
\end{equation*}
where the $x$-axis points towards the Galactic centre, and the
$y$-axis points in the direction of Galactic rotation. The orbit was
integrated in the logarithmic potential
\begin{equation}
\Phi(x,y,z) = {v_c^2 \over 2} \log \left(
x^2 + y^2 + \left( { z \over q } \right)^2 \right),
\end{equation}
where $v_c = 220\kms$ and $q = 0.9$. The resulting trajectory was
projected onto the sky, assuming a Solar radius $R_0 = 8.5 \kpc$.  The
implicit assumption is made that the GD-1 stream is closely
approximated by an orbit: the work of \chapref{chap:mech} will show
that this assumption is a fair one. Several points were sampled from
the on-sky track, and each was taken to be a separate datum in the
pseudo-data set. The proper motion for each datum was computed by
projecting the difference between its grf motion and the Solar motion
onto the sky.

The pseudo-data were transformed into the rotated coordinate system
used by \krh to facilitate comparison with their data; the
transformation rule is given in the appendix to \krh. The stream is very
flat in this coordinate system, so the dependence of $\phi_2$ on
$\phi_1$ is relatively weak. This helps to increase the quality of the fitted curve
and minimises the corresponding error in $\sigma_{\alpha,f}$.

\begin{figure}
\centerline{
\includegraphics[width=0.45\hsize]{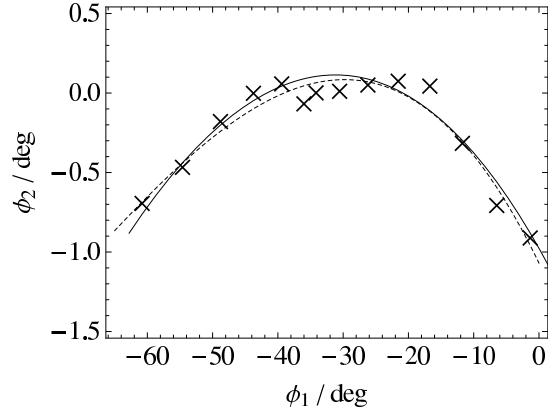}
}
\caption[The \krh orbit for the GD-1 stream, and pseudo-data derived from it]{Full line: the orbit for the GD-1 stream,
taken from \krh. Crosses: pseudo-data
derived from that orbit, but randomly scattered in $\phi_2$
according to a Gaussian distribution with a dispersion of $\sigma_{\phi_2} = 0.1\deg$.
Dotted line: a cubic polynomial fitted to the pseudo-data,
used to estimate stream direction.}
\label{galplx:fig:pseudodata}
\end{figure}

To simulate the observed scatter in the real positional data, the
pseudo-data were each scattered in the $\phi_2$ coordinate according
to a randomly-sampled Gaussian distribution with a dispersion
$\sigma_{\phi_2} = 0.1\deg$. The resulting positional pseudo-data are
plotted in \figref{galplx:fig:pseudodata}, along with the orbit from which
they were derived (full curve). A cubic polynomial representing $\phi_2(\phi_1)$ was
least-squares fitted to the pseudo-data, the tangent of which was used to estimate
$\vecthat{p}$.  In the case of the pseudo-data, uniform
weights were applied to each datum for the fitting processes. The
resulting curve is also shown in \figref{galplx:fig:pseudodata} (dotted curve).

\begin{figure}
\centerline{
\includegraphics[width=0.45\hsize]{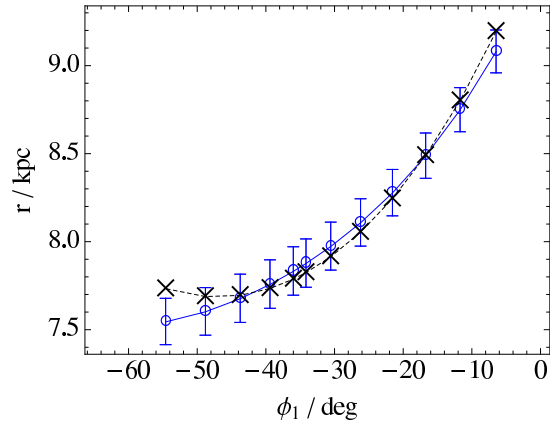}
}
\caption[Galactic parallax distances computed from the pseudo-data of
\figref{galplx:fig:pseudodata}]
{Dotted line: the orbit of the GD-1 stream,
taken from \krh. Crosses: the true distance of each
pseudo-datum. Circles: Galactic parallax
distances computed from the pseudo-data.
The error bars represent
 the distance error expected from the polynomial fitting
 procedure.
No extraneous error was added. The error bars
are shown to be a good estimate of likely error from
the fitting procedure, and the agreement of the
 distances overall is excellent.}
\label{galplx:fig:pd-distance}
\end{figure}

When the correct orbit is
used to calculate $\vecthat{p}$, and precise values for the measured
proper motion $\mu \vecthat{t}$ and Solar reflex motion $\vot$ are
used, the distance is recovered perfectly from \eqref{galplx:eq:pi2}.
\figref{galplx:fig:pd-distance} compares the recovered distance when
$\vecthat{p}$ is estimated using the polynomial fit to the
pseudo-data, but still using accurate values for $\mu \vecthat{t}$ and
$\vot$. Our pseudo-data stream is $\angwid \simeq 0.1\deg$ wide and
$\anglen \simeq 60\deg$ long. \Eqref{galplx:eq:errora:curvefitting}
therefore estimates $\sigma_{\alpha,f} \simeq 0.38\deg$. The recovered
distances in \figref{galplx:fig:pd-distance} are in error by only $\sim 2$
percent across most of the range, which is the approximate uncertainty
predicted by \eqref{galplx:eq:error:pi} for this value of
$\sigma_{\alpha,f}$. Thus, the estimation of $\vecthat{p}$
from the observed stream is good, and contributes little error to the
distance calculations.

The \krh observational data for the GD-1 stream, discussed below, have a similar
uncertainty $\sigma_\alpha \sim 0.38\deg$ due entirely to the fitting process,
and proper-motion uncertainties $\sigma_\mu \sim 1\masyr$.
\figref{galplx:fig:pd-MC} shows the recovered distances from
\figref{galplx:fig:pd-distance} with error bars for the expected uncertainty
in recovered distance, given these
measurement uncertainties and the uncertainty in $\vo$ quoted
in \secref{chap:galplx}.  Also plotted for each datum are the
distances recovered from 60 Monte Carlo realisations of the
pseudo-data input values, convolved with the errors given above.

\Eqref{galplx:eq:error:pi} is found to be a good estimator
for the uncertainty, with approximately 80 percent of
the Monte Carlo realisations falling within the error bars.
The error in parallax for the \krh data is thus predicted
to be about 50 percent, of which the greatest contribution
comes from the uncertainty in proper motion.

\section{Distance to the GD-1 stream}
\label{galplx:sec:gd1}

\begin{figure}
\centerline{
\includegraphics[width=0.45\hsize]{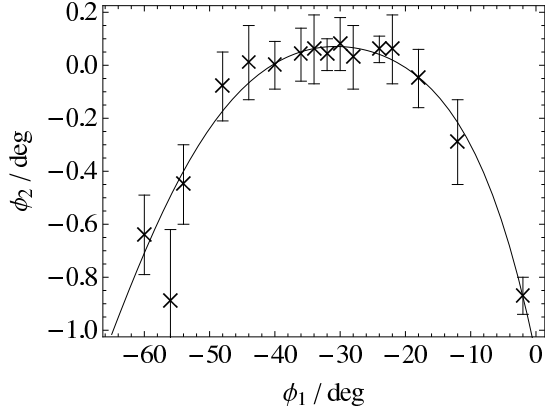}
}
\caption[On-sky position data for the GD-1 stream]
{Crosses: on-sky position data for the GD-1
stream, as published in \krh. The error bars
represent the quoted uncertainties. Full line:
linear least-squares fit of a cubic polynomial, $\phi_2(\phi_1)$,
to these data; the inverse-square of the uncertainties
was used to weight the fit.}
\label{galplx:fig:gd1-fit}
\end{figure}

\figref{galplx:fig:gd1-fit} shows the on-sky position
data for the GD-1 stream, as published in \krh.
Also shown in \figref{galplx:fig:gd1-fit} is a
linear least-squares fit of a cubic polynomial
to these data, used to estimate $\vecthat{p}$. The weights for the fit
were the inverse-square uncertainties for each position field, as given
by \krh.

\krh provide measured proper-motion data for five fields
of stars, spanning the range $\phi_1 \sim (-55, -15)\deg$,
along with uncertainties for these measurements.
The uncertainty in
$\vot$ is computed for each individual field from the
uncertainty in $\vo$ given in \secref{chap:galplx}.
Uncertainty in the stream direction is $\sigma_\alpha \sim 0.38\deg$,
which is entirely contributed by the curve
fit to the stream; since hundreds of stars contributed to the
calculation of the proper motions, the contribution from
the first term in \eqref{galplx:eq:errora:random} is negligible.
We have also assumed that the contribution to $\sigma_\alpha$ from
stream misalignment is negligible. This assumption is validated
by the results of \chapref{chap:mech}, which predicts that the
projection of GD-1
should be perfectly aligned with the orbits of its stars.

\begin{figure}
\centerline{
\includegraphics[width=0.45\hsize]{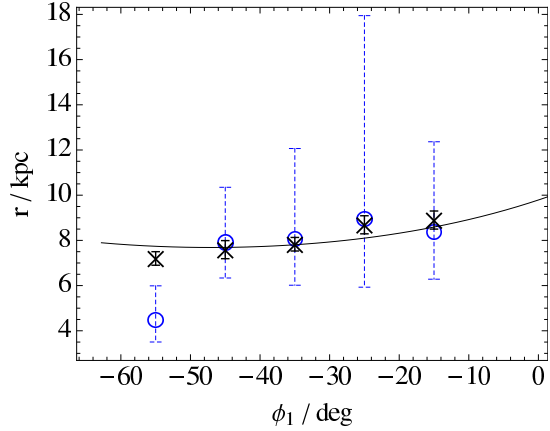}
}
\caption[Galactic parallax distances for the GD-1 stream]
{Circles: Galactic parallax distances for the GD-1 data
  presented in \krh. Dotted error bars: the uncertainty estimated by
  \eqref{galplx:eq:error:pi}, given the \krh measurement
  uncertainties. Crosses: the photometric distances reported in \krh,
  along with their error bars. Full line: the orbit for GD-1 taken
  from \krh.  With the exception of the datum at $\phi_1 \sim -55$
  deg, the Galactic parallax distances are in excellent agreement with
  the photometric distances from \krh. The dotted error bars appear to
  seriously over-estimate the true error in the distance estimates.
}\label{galplx:fig:gd1-dist}
\end{figure}

\figref{galplx:fig:gd1-dist} shows the Galactic parallax distances for each of these
data, along with the \krh photometric distances. The dotted error bars
represent the expected error in distance for
the uncertainties given. The small solid error bars are the uncertainties
reported by \krh for their photometric distances. The \krh orbit used to
compute the earlier pseudo-data is plotted for comparison.

With the exception of the datum at $\phi_1 \sim -55 \deg$, the
parallax distances and the \krh distances are in remarkable
agreement. However, the dotted error bars vastly overestimate the true
error in the results.  If we ignore the datum at $\phi_1 \sim -55
\deg$, the scatter in the distance, $\sigma_r \sim 1\kpc$, is similar
to that of the photometric distances, and consistent with a true
random error of $\sigma_\mu \sim 0.3\masyr$, and negligible systematic
offset. We cannot explain this discrepancy, except by suggesting that
the \krh proper-motion measurements are more accurate than the
published uncertainties suggest. This is corroborated by the top-right
panel of Fig.~13 from \krh in which the $\mu_{\phi_2}$ data, with the
exception of the datum at $\phi_1 \sim -55 \deg$, show remarkably
little scatter within their error bars.

\begin{figure}[t]
\centerline{
\includegraphics[width=0.5\hsize]{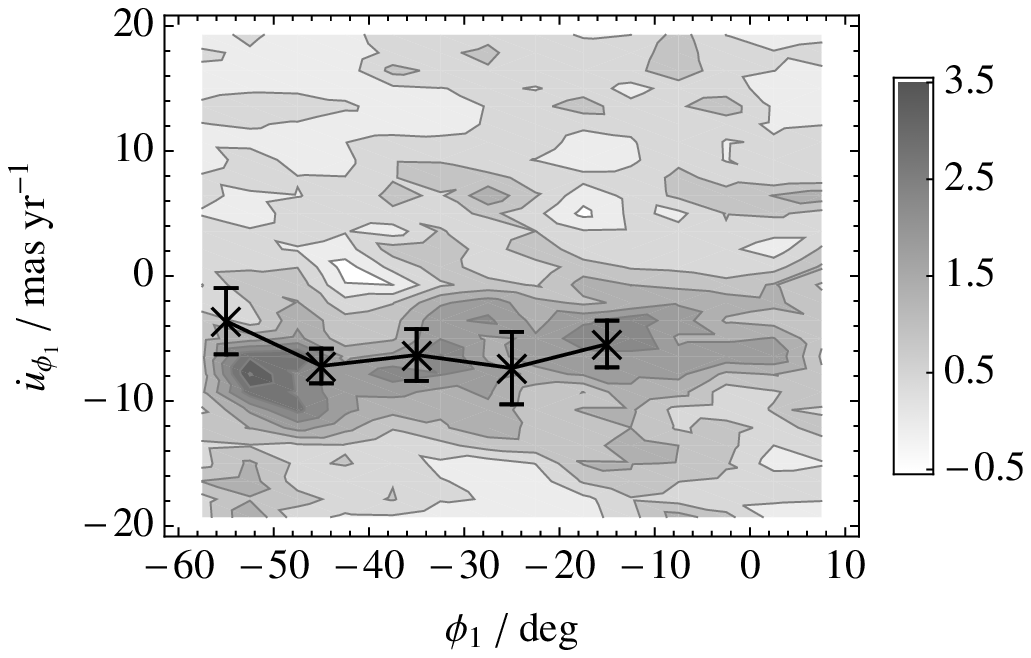}
\includegraphics[width=0.5\hsize]{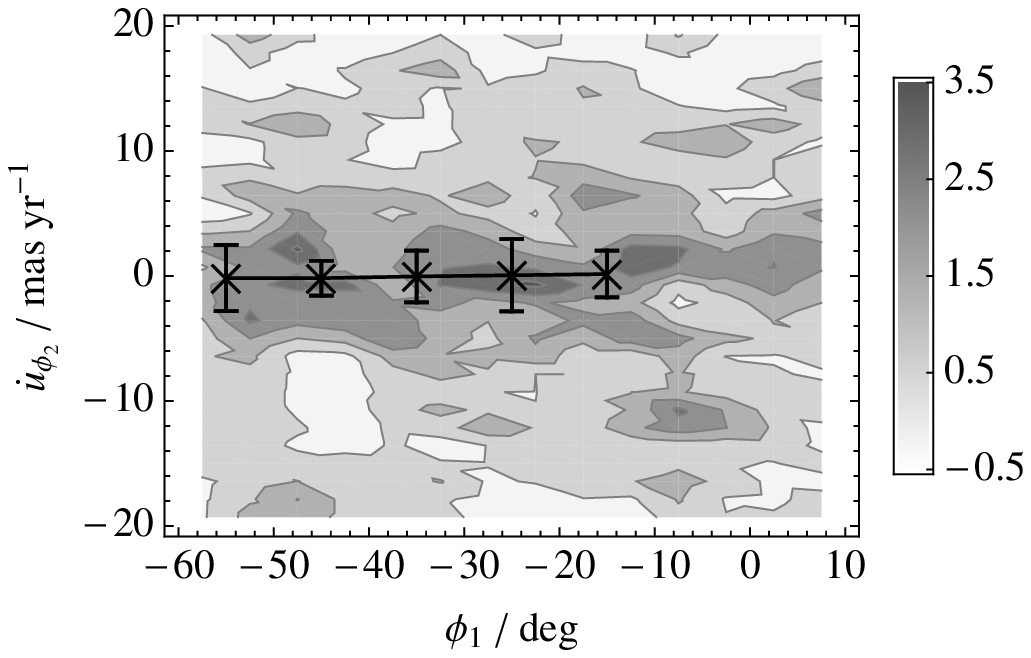}
}
\caption[Galactic rest-frame proper motions for the GD-1 stream]
{Full lines: Galactic rest-frame proper motion,
$\dot{u}$, calculated from the \krh data using
\eqref{galplx:eq:udot2}. The (left, right) panels show
the $(\phi_1, \phi_2)$ components respectively. Plotted in the background
are the observational data from Fig.~9 in \krh;
the greyscale shows the number of stream stars, per bin, with the
given motion. 
The data are broadly consistent,
except for the datum at $\phi_1 \sim -55$ deg in the left panel.
}\label{galplx:fig:gd1-udot}
\end{figure}

\figref{galplx:fig:gd1-udot} shows the Galactic rest-frame proper motions,
$\dot{u}$, calculated from \eqref{galplx:eq:udot2} along with their error bars,
from \eqref{galplx:eq:error:udot}.  In the background are plotted the data
from Fig.~9 of \krh, which show the density of stars with a given grf
proper motion in the sample of stars chosen to be candidate members of
the stream, and after subtraction of a background field.  The \krh grf
proper motions have been calculated by correcting measured proper
motion for the solar reflex motion, using an assumed distance of
$8\kpc$ (Koposov, private communication); this assumption will cause a
systematic error in the \krh proper motions, of order the distance
error, which changes with position down the stream. The apparently
large width of the stream in this plot is due to uncertainty in the
underlying \cite{munn-etal} proper-motion data.

The stream is clearly visible in this plot as the region of high
density spanning $\phi_1 \sim (0, -60) \deg$ with $\dot{u}_{\phi_2}
\simeq 0 \masyr$ and $\dot{u}_{\phi_1}$ falling slowly between
$(-6, -10 )\masyr$. Despite the
expected systematic error, the estimates of
$\dot{u}$ from the parallax calculation are consistent with these
data, with the exception of the same datum at $\phi_1 \sim -55 \deg$
that also reports an anomalous distance.

We explain this suspect datum as follows. From inspection of the
top-right panel of Fig.~13 from \krh, it is apparent that the
$\mu_{\phi_2}$ measurement for this datum is not in keeping with the
trend. Conversely, the corresponding $\mu_{\phi_1}$ measurement is not
obviously in error. If the magnitude of $\mu_{\phi_2}$ for this datum
has been over-estimated by the \krh analysis, then \eqref{galplx:eq:pi2} will
over-estimate the parallax, and hence under-report the distance.
\figref{galplx:fig:gd1-dist} indicates that the distance for this datum
is indeed under-reported.

The effect of such an error in $\mu_{\phi_2}$ on the
grf proper-motion, $\dot{u}$, can be understood by
considering \eqref{galplx:eq:udot2}. If $\Pi$ is over-estimated,
$\dot{u}$ will be either over-estimated or under-estimated, depending
on the relative sign of the two terms. In the case of GD-1,
$\mu \tx$ and $\vpar$ have opposite signs, so an over-estimated
$\Pi$ will result in an under-estimated $\dot{u}$. This 
too corresponds with the behaviour of the suspect datum in
\figref{galplx:fig:gd1-udot}.

It is unknown why this particular datum should be significantly in
error while the other data are not. There are no obvious structures in
the lower panel of \figref{galplx:fig:gd1-udot} which might cause the fitting
algorithm in \krh to mistakenly return an incorrect value for
$\mu_{\phi_2}$. Nonetheless, if the scatter in the other data are
accepted as indicative of their true statistical error, it is clear
that the datum at $\phi_1 \sim -55\deg$ cannot represent the proper
motions of GD-1 stars at that location.
%%This may be because the \krh fitting procedure has picked out
%%a dynamical structure other than the GD-1 stream, or it may be that
%%the procedure has failed and has fit noise from halo stars, rather
%%than a signal from the stream.
We therefore predict that an appropriate
re-analysis of the proper-motion data, taking care to ensure that 
a signal from GD-1 stream stars is properly detected, will return a revised
proper-motion of $\mu_{\phi_2} \sim -3 \masyr$.

In summary, it seems that Galactic parallax measurements confirm the
\krh photometric analysis, and predict that the stream is
approximately $(8 \pm 1)\kpc$ distant, where the uncertainty denotes
the scatter in the results. Since Galactic parallax and photometric
estimates are fundamentally independent, it seems unlikely that
systematic errors in either would conspire to produce the same shift
in distance; this suggests that no systematic error is present.

We also calculate a grf proper motion for the stream of
$\dot{u}_{\phi_1} = (-7 \pm 2) \masyr$, corresponding to a grf tangential
velocity of $(265 \pm 80)\kms$ in a direction $(\dot{u}_l \cos b, \dot{u}_b)
\simeq (0.8, -0.6)$. This implies that the stream is on a retrograde orbit, inclined to
the Galactic plane by $\near 37\deg$, which is in accordance with previous
results \citep{willett,koposov}.

The galactocentric radius of $\near 14.5\kpc$ does not seem to be
changing rapidly along the stream's length, which subtends $\near 12\deg$
when viewed from the Galactic centre. This implies that the observed
stream is at an apsis. The grf velocity of the stream is faster than the
circular velocity, $v_c \sim 220\kms$.  This implies that the stream
is at pericentre, although the large uncertainty prevents a firm
conclusion from being drawn. We note that the radial velocity
data in \krh would also imply that the stream is observed at
pericentre.

\section{Conclusions}
\label{galplx:sec:conclusions}

In this chapter, we have demonstrated the practical application of a
technique for computing distances using Galactic parallax. This
technique utilises the predictable trajectories of stars in a stream
to identify the contribution of the reflex motion of the Sun to the
observed proper motion.  The parallax and the Galactic rest-frame
proper motion follow from this.

Galactic parallax is a geometric phenomenon, and the distances
obtained from it are in every way as fundamental as those
obtained using conventional trigonometric parallax.
The only assumption made is knowledge of the Galactic rest-frame
velocity of the Sun. It is also a requirement that the observed stars
are part of a stream. Recent evidence
\citep{odenkirchen-delineate,majewski-sag,yanny-stream,orphan-discovery,
grillmair-orphan,gd1-discovery,ngc5466,
grillmair-2009,newberg-streams-2009}
indicates that tidal streams are a common constituent
of the Galactic halo, and so this technique should have widespread
application.

The key hurdle to the widespread application of this 
technique is the lack of high-accuracy
proper-motion data for distant stream stars. Given the
advent of next-generation astrometric projects such
as Pan-STARRS \citep{pan-starrs}, LSST \citep{lsst}
and Gaia \citep{gaia}, proper-motion catalogues
with billions of entries of the required precision
will become available in the next few years. The technique
should then become the standard method of determining
the distance to remote stream stars.

\subsection{Summary}

We have presented a method for calculating the Galactic
parallax of tidal streams. We first determine the on-sky
direction of the stream by fitting it with a curve. We then
combine the tangents of this curve with measured proper-motion
data to estimate the parallax of the stream.

We have derived an expression for the uncertainty in
Galactic parallax calculations. We include contributions
from measurement errors in proper motion and the Solar motion,
error in the estimation of stellar trajectories from the
stream direction, and algorithmic error in the estimation of
stream direction itself.

The uncertainty for calculations involving a particular stream
depends upon the size, location and orientation of the stream, as well
as upon measurement errors. We estimate that using individual proper
motions accurate to $4\masyr$, available now in published surveys
\citep{munn-etal}, the parallax of a $10\kpc$ distant stream with typical
geometry can be measured with an uncertainty of 40
percent. The parallax of a stream with optimum geometry could be 
measured with approximately half this uncertainty. 

Proper-motion data
from the forthcoming Pan-STARRS PS-1 survey \citep{pan-starrs,pan-starrs-3pi} will yield the
distance to a typical $10\kpc$ distant stream with 11 percent accuracy, or
the distance to a stream at $23\kpc$ with 20 percent accuracy; with favourable
geometry this accuracy could be achieved for a stream as distant
as $50\kpc$. With data of this
quality, the uncertainty in distances from Galactic parallaxes
will be considerably lower than those of photometric distances
to remote streams.

The LSST \citep{lsst,ivezic} and the Gaia
mission \citep{gaia} will produce proper-motion data that are more
accurate still.  Such data would allow the distance to stars in a
$10\kpc$ typical stream to be computed to an accuracy of order of 8
percent, where the limitation is now imposed by uncertainty in the
solar motion and in the stream trajectory. It is likely that LSST and
Gaia data will allow the uncertainty in the solar motion to be
significantly reduced, so in reality much better precision can be
expected at this distance.  For streams $30\kpc$ distant, LSST proper
motions will allow distance estimates as accurate as 14 percent to be
made in the typical case, and 6 percent with optimum geometry. Thus,
the high-quality astrometric data that is expected to be available in
the next decade will allow parallax estimates for very distant streams
to be made with unparalleled accuracy.

To test the method presented, we have created
pseudo-data simulating the GD-1
stream \citep{gd1-discovery}. When the
method is provided with error-free pseudo-data,
the correct parallax is computed perfectly.
When errors are introduced into the pseudo-data,
the reported parallax degrades in line with
the uncertainty estimates.

We applied the method to the astrometric data for the GD-1
stream in \cite{koposov}. With the exception of a single datum, the
Galactic parallax is remarkably consistent with the photometric
distances quoted by \cite{koposov}.
Indeed, the uncertainty in the measured proper motions quoted by
\cite{koposov} should produce significant error in the Galactic
parallax. However, the scatter in the results is consistent with
a random error of only $\sim 0.3\masyr$, and if the photometric
distances of \krh are believed, no systematic offset.
This is at odds with the typical uncertainty in
the proper motion of $\sim 1\masyr$ reported by \cite{koposov}.  We
cannot explain this discrepancy, other than to suggest that the
\cite{koposov} method for estimating error in the proper motions is
producing significant over-estimates.

The Galactic rest-frame proper motions predicted for the stream are
also consistent with observational data from \cite{koposov}, with the
exception of the same datum that also reports an inconsistent distance.  We
conclude that the proper-motion associated with this datum is erroneous,
and we predict that reanalysis of the stream stars near this datum
will reveal a reduced proper-motion measurement of $\mu_{\phi_2} \sim 3\masyr$.

Photometry and Galactic parallax produce fundamentally independent
estimates of distance. The quality of the corroboration of the
\cite{koposov} photometric distance estimates for GD-1 by the Galactic
parallax estimates presented here therefore lends weight to the
conclusion that the predicted distance, in both cases, is correct.  On
this basis, we conclude that the GD-1 stream is about $(8 \pm 1) \kpc$
distant from the Sun, on a retrograde orbit that is inclined $37\deg$
to the Galactic plane with a rest-frame velocity of $(265 \pm
75)\kms$. We also conclude that the visible portion of the stream
is probably at pericentre.

\subsection{Future directions}

The prospect of being able to map trigonometric distances in the Galaxy to high
accuracy at a range of tens of kiloparsecs is indeed exciting.  The 
distances generated using this method, although limited to stars in
streams, could be used to calibrate other distance measuring tools,
such as photometry, that would be more widely applicable. The technique is
immediately applicable to any stream for which proper-motion data are
currently available, although we anticipate limited accuracy until better
proper-motion data are available.

Given enough parallax data points along a given stream, an orbit can
be constructed by connecting those points. This orbit is predicted
independently of any assumption about the Galactic potential, which it
must strongly constrain. Constraints on the Galactic potential impose
constraints on theories of galaxy formation and cosmology. It would
seem that the combination of dynamics and Galaxy-scale precision
astrometry, such as provided by this method, could well have profound
implications for astrophysics in the future.

At present, however, it is not obvious how to combine all sources of
astrometric and dynamical information, to produce the tightest
constraints on the potential.  A significant theoretical effort is
therefore required to explore methods for combining this information, in
anticipation of the arrival of higher quality astrometric data in the
next few years.

% and the resulting trajectory is guaranteed to be a dynamical
% orbit in whatever gravitational field might actually apply.

%% file: grf.pstex_t
\begin{picture}(0,0)%
\includegraphics{grf.pstex}%
\end{picture}%
\setlength{\unitlength}{3947sp}%
\begingroup\makeatletter\ifx\SetFigFont\undefined%
\gdef\SetFigFont#1#2#3#4#5{%
  \reset@font\fontsize{#1}{#2pt}%
  \fontfamily{#3}\fontseries{#4}\fontshape{#5}%
  \selectfont}%
\fi\endgroup%
\begin{picture}(3629,2424)(-11,-1573)
\put(2686,-403){\makebox(0,0)[lb]{\smash{{\SetFigFont{10}{12.0}{\familydefault}{\mddefault}{\updefault}{\color[rgb]{1,0,0}$\dot{u}\vecthat{t}$}%
}}}}
\put(1888,-970){\makebox(0,0)[lb]{\smash{{\SetFigFont{10}{12.0}{\rmdefault}{\mddefault}{\updefault}{\color[rgb]{0,0,0}$\vect{r}$}%
}}}}
\put(2251,-571){\makebox(0,0)[lb]{\smash{{\SetFigFont{10}{12.0}{\rmdefault}{\mddefault}{\updefault}{\color[rgb]{0,0,0}$\dot{\vect{r}}$}%
}}}}
\put(2086,-110){\makebox(0,0)[lb]{\smash{{\SetFigFont{10}{12.0}{\familydefault}{\mddefault}{\updefault}{\color[rgb]{0,0,1}$\dot{r}\vecthat{r}$}%
}}}}
\put( 76,-1486){\makebox(0,0)[lb]{\smash{{\SetFigFont{10}{12.0}{\familydefault}{\mddefault}{\updefault}{\color[rgb]{0,0,0}GRF}%
}}}}
\put(1126,-1186){\makebox(0,0)[lb]{\smash{{\SetFigFont{10}{12.0}{\familydefault}{\mddefault}{\updefault}{\color[rgb]{0,0,0}Sun}%
}}}}
\put(976,-811){\makebox(0,0)[lb]{\smash{{\SetFigFont{10}{12.0}{\familydefault}{\mddefault}{\updefault}{\color[rgb]{0,0,0}$\vect{v}_0$}%
}}}}
\end{picture}%

%% file: hrf.pstex_t
\begin{picture}(0,0)%
\includegraphics{hrf.pstex}%
\end{picture}%
\setlength{\unitlength}{3947sp}%
\begingroup\makeatletter\ifx\SetFigFont\undefined%
\gdef\SetFigFont#1#2#3#4#5{%
  \reset@font\fontsize{#1}{#2pt}%
  \fontfamily{#3}\fontseries{#4}\fontshape{#5}%
  \selectfont}%
\fi\endgroup%
\begin{picture}(3629,2424)(-11,-1573)
\put(1905,-990){\makebox(0,0)[lb]{\smash{{\SetFigFont{10}{12.0}{\rmdefault}{\mddefault}{\updefault}{\color[rgb]{0,0,0}$\vect{r}'$}%
}}}}
\put(1698,-670){\makebox(0,0)[lb]{\smash{{\SetFigFont{10}{12.0}{\rmdefault}{\mddefault}{\updefault}{\color[rgb]{0,0,0}$\dot{\vect{r}}'$}%
}}}}
\put(2726,-508){\makebox(0,0)[lb]{\smash{{\SetFigFont{10}{12.0}{\familydefault}{\mddefault}{\updefault}{\color[rgb]{1,0,0}$\mu\vecthat{t}$}%
}}}}
\put(2228,-313){\makebox(0,0)[lb]{\smash{{\SetFigFont{10}{12.0}{\familydefault}{\mddefault}{\updefault}{\color[rgb]{0,0,1}$\dot{r}'\vecthat{r}'$}%
}}}}
\put(850,172){\makebox(0,0)[lb]{\smash{{\SetFigFont{10}{12.0}{\familydefault}{\mddefault}{\updefault}{\color[rgb]{0,1,0}$\vect{v}_0$}%
}}}}
\put(1126,-1186){\makebox(0,0)[lb]{\smash{{\SetFigFont{10}{12.0}{\familydefault}{\mddefault}{\updefault}{\color[rgb]{0,0,0}Sun}%
}}}}
\put(1486,203){\makebox(0,0)[lb]{\smash{{\SetFigFont{10}{12.0}{\familydefault}{\mddefault}{\updefault}{\color[rgb]{0,1,0}$\dot{\vect{r}}$}%
}}}}
\put( 76,-1486){\makebox(0,0)[lb]{\smash{{\SetFigFont{10}{12.0}{\familydefault}{\mddefault}{\updefault}{\color[rgb]{0,0,0}HRF}%
}}}}
\end{picture}%

%% file: gpdiagram.pstex_t
\begin{picture}(0,0)%
\includegraphics{gpdiagram.pstex}%
\end{picture}%
\setlength{\unitlength}{3947sp}%
\begingroup\makeatletter\ifx\SetFigFont\undefined%
\gdef\SetFigFont#1#2#3#4#5{%
  \reset@font\fontsize{#1}{#2pt}%
  \fontfamily{#3}\fontseries{#4}\fontshape{#5}%
  \selectfont}%
\fi\endgroup%
\begin{picture}(3624,2424)(-11,-1573)
\put(676,-736){\makebox(0,0)[lb]{\smash{{\SetFigFont{10}{12.0}{\familydefault}{\mddefault}{\updefault}{\color[rgb]{0,0,0}$-\vect{v}_0$}%
}}}}
\put(2776,-511){\makebox(0,0)[lb]{\smash{{\SetFigFont{10}{12.0}{\familydefault}{\mddefault}{\updefault}{\color[rgb]{0,0,0}Sun}%
}}}}
\put(2776,-136){\makebox(0,0)[lb]{\smash{{\SetFigFont{10}{12.0}{\familydefault}{\mddefault}{\updefault}{\color[rgb]{1,0,0}$\vect{v}_0$}%
}}}}
\put(1501,-286){\makebox(0,0)[lb]{\smash{{\SetFigFont{10}{12.0}{\familydefault}{\mddefault}{\updefault}{\color[rgb]{0,0,0}r}%
}}}}
\end{picture}%

%% file: streamdirs/streamdirs.tex
% any chapter-specific preamble stuff? %

\chapter{The mechanics of streams}
\label{chap:mech}

\section{Introduction}

Many of the techniques used to harness the diagnostic power
of tidal streams rely upon the assumption that such streams
delineate orbits precisely \citep{binney08,eb09b,oden-2009,willett,koposov,galplx}.

Although both the author \citep[\chapref{chap:radvs};][]{eb09a} and
others \citep{dehnen-pal5,choi-etal,montuori-nbody-streams} have shown
evidence that this is not necessarily the case, there has not to date
been an exposition of how streams form that fundamentally addresses
this issue.  In particular, it is necessary to understand under what
circumstances tidal streams delineate orbits, by what measure they are
in error when they do not, and what can be done to correct this error.

The few studies that have been made have either focused on
N-body simulations \citep[e.g.][]{choi-etal,montuori-nbody-streams},
the confusion of which makes the predominant physics hard to isolate,
or have attempted to describe the problem in terms of conventional
phase-space coordinates and classical integrals
\citep[e.g.][]{dehnen-pal5,choi-etal}, which makes the problem
intractably hard.  In particular, the work of \cite{choi-etal} made
some progress towards understanding the dynamical structure of
clusters at the point of disruption, and they provide a qualitative
picture of the evolution of tidal tails, understood in terms of
classical integrals. However, they are unable to make predictions for
stream tracks on the basis of this picture alone, and they are
ultimately forced to rely on N-body simulation.  In the work of this
chapter, we approach the problem using action-angle variables
\citep[\S3.5]{bt08}, in which the physics of stream formation turns
out to have a natural and simple expression.

We confine our investigation to the formation of
long, cold streams, such as may form from tidally disrupted
globular clusters. We do so for
two reasons. Firstly, a low mass for the progenitor
cluster simplifies the understanding of its orbital
mechanics, because of the lack of tidal friction and other feedback
effects in their interaction with the host galaxy.
Secondly, thin, long streams provide the strongest
constraints upon the Galactic potential \citep[\chapref{chap:radvs};][]
{binney08,eb09a}, because any orbit delineated
by them can be observationally identified with less ambiguity.
It is therefore long, cold streams that are of primary interest
for use in probing the potential.

We study the mechanics of stream formation immediately following the
tidal disruption of a progenitor cluster. In most of the work that
follows, the assumption is made that stream stars feel only the
potential of the Galaxy; i.e.~the stream stars do not
self-gravitate. This assumption is generally a fair one: the stars in
streams are generally spaced too widely for their self-gravity to be
of consequence \citep{dehnen-pal5}.  Indeed, we will demonstrate in
\secref{mech:sec:disruption} below that self-gravity becomes
negligible shortly after stars are stripped from the cluster.

The chapter is arranged as follows: The remainder of this introduction
discusses the action-angle variables in which we perform our analysis.
\secref{mech:sec:tremaine} discusses the basic mechanics of stream
formation and propagation. \secref{mech:sec:2d} discusses the detail
of stream formation in spherical systems, and explores
some examples. \secref{mech:sec:mapping} describes the principles and
pitfalls of mapping streams from action-angle space to real-space.
\secref{mech:sec:fitting} describes the consequences of optimizing
potential parameters by assuming that streams follow orbits.
\secref{mech:sec:actions} examines the action-space distribution of
disrupted clusters using N-body simulation.
\secref{mech:sec:nonsph} discusses stream formation in
flattened systems, using oblate axisymmetric
\stackel\ potentials as an example. Finally, \secref{mech:sec:conclusions}
presents our concluding remarks.

\subsection{Action-angle variables}
\label{mech:sec:aavars}

We will approach the analysis of stream formation and propagation
by describing the problem using action-angle variables. The usefulness
and theoretical basis of action-angle variables is extensively
discussed in \S3.5 of \cite{bt08}. Here, we merely note that
action-angle variables are a set of canonical coordinates, like
conventional phase-space coordinates,
that can be used to describe systems in Hamiltonian mechanics.

The coordinates are special because the canonical momenta,
called actions, $\vJ$, are integrals of the motion, and are
thus constant with the passage of time. The equation of motion
for $\vJ$ then reads
\begin{equation}
\dot{J_i} = 0 = -{\partial H \over \partial \theta_i},
\label{mech:eq:jmotion}
\end{equation}
where we have introduced the angle variables, $\vT$. \Eqref{mech:eq:jmotion}
requires that the Hamiltonian, $H(\vJ)$ be independent of $\vT$, and therefore
a function of $\vJ$ only. The equation of
motion for $\vT$ is then,
\begin{equation}
\dot{\theta_i} = {\partial H \over \partial J_i} \equiv
\Omega_i(\vJ),
\label{mech:eq:eqnsmotion}
\end{equation}
where the frequencies $\vT(\vJ)$, being functions of $\vJ$ only, are
constant. The solution to \eqref{mech:eq:eqnsmotion} is very simple
\begin{equation}
\vT(t) = \vT(0) + \vO\,t.
\end{equation}
Hence, the motion of a system described by action-angle variables
is very easy to predict. Moreover, because the actions for the system
are constant, any approximation that we may make to
$H(\vJ)$, if valid at one time, is automatically valid at all times.
This particular feature is of critical importance in simplifying 
the analysis of the propagation of streams in the work that follows.

\S3.5 of \cite{bt08} discusses at length the calculation of
action-angle coordinates in various systems. Here, we note that
standard methods for calculating action-angle coordinates require the
Hamilton-Jacobi equation \citep[equation 3.205,][]{bt08} to
separate. This condition is met by all spherically symmetric
potentials, but excludes any asymmetric potential that is not of
\stackel\ form (see \secref{mech:sec:stackel} below). Given
this condition, the action corresponding to a coordinate $q$ is given by
\begin{equation}
J_q = {1 \over 2\pi} \oint p_q \, dq,
\label{mech:eq:jact}
\end{equation}
where the integral is over the closed path that encloses
a single oscillation of the coordinate $q$ along an orbit.

We note that just as spherical systems are naturally
described by spherical polar coordinates $(r,\vartheta,\phi)$,
the natural actions for such systems are $(J_r, L)$,
where $L$ is the angular momentum, and we have without
loss of generality confined our motion to the $(x,y)$ plane.
We can qualitatively understand the meaning of these actions.  An
orbit with finite $L$ we understand to be circular at, or to oscillate
in epicycles around, some guiding-centre radius $r_g$. An orbit with
$J_r = 0$ always has zero radial momentum, and thus
corresponds to a circular orbit. Conversely, an orbit with
comparatively large $J_r$ must be very eccentric.
Hence, $J_r$ quantifies the radial motion of an orbit, while $L$
quantifies its azimuthal motion. We note for
completeness that orbits with $L=0$ are plunging orbits.

In the sections that follow, we utilize the equations
from \S3.5.2 of \cite{bt08} to transform between the
action-angle coordinates $(J_r,L,\theta_r,\theta_\phi)$ 
and conventional phase-space coordinates, when investigating
spherical potentials. The action-angle coordinates
used when investigating non-spherical potentials 
are discussed in \secref{mech:sec:stackel} below.

\section{The formation of streams from tidally stripped clusters}
\label{mech:sec:tremaine}

Consider a low-mass cluster, just past the point of disruption, such that the
stripped stars no longer feel the effects of its gravity. The cluster is on
a regular orbit, identified by its actions $\vJ_0$, in a fixed background potential,
which has a Hamiltonian $H$ in terms of the actions $\vJ$. Suppose further that in
the locality of $\vJ_0$, the
Hamiltonian is well described by the Taylor expansion,
\begin{equation}
H(\vJ) = H_0 + \vO_0\cdot\DvJ + {1 \over 2}\DvJ^T\cdot\hessian\cdot\DvJ,
\label{mech:eq:taylor-H}
\end{equation}
where $\delta\vJ = \vJ - \vJ_0$, and $\vect{D}$ is the Hessian of $H$
\begin{equation}
D_{ij} = \left.{\partial^2 H \over \partial J_i \, \partial J_j}\right|_\vJ,
  \label{mech:eq:hessian}
\end{equation}
and $\vO_0$ is the frequency of the cluster's orbit
\begin{equation}
\Omega_{0,i} = \left.{\partial H \over \partial J_i}\right|_\vJ.
\end{equation}
The frequency $\vO$ of a nearby orbit $\vJ$ is then
\begin{equation}
\vO(\vJ) = \vO_0 + \hessian\cdot\DvJ.\label{mech:eq:omega}
\end{equation}
If the disrupted cluster has some spread in actions $\Delta \vJ$ and
angles $\Delta \vT_0$, then the spread in angles after some time $t$ is
given by \citep[\S8.3.1]{tremaine99,bt08},
\begin{equation}
  \Delta \vT(t) = t \Delta \vO + \Delta\vT_0 \simeq t\Delta\vO, \label{mech:eq:angle_t}
\end{equation}
where the near equality is valid when $\Delta\vO \, t \gg \Delta\vT_{0}$,
and where we have introduced the spread in frequencies, $\Delta \vO$,
which are related to the spread in actions via
\begin{equation}
  \Delta \vO = \hessian \cdot \Delta \vJ.
  \label{mech:eq:d-dot-j}
\end{equation}
In the absence of self-gravity, the action-space distribution $\Delta
\vJ$ is frozen for all time.  $\hessian$ and $\vO$ are functions of
$\vJ$ only, and so are similarly frozen.  The secular evolution of a
disrupted cluster is therefore to spread out in angle-space, with its
eventual shape determined by $\Delta \vO = \hessian \cdot \Delta \vJ$,
and its size growing linearly with $t$.

%We now note the  What will be the direction and relative magnitude of this spread?

For a given $\Delta \vJ$, what does $\Delta \vO$ look like?
\Eqref{mech:eq:angle_t} and \eqref{mech:eq:d-dot-j}  show that the
function of $\hessian$ is to act as a linear map between a star's
position in action-space and its position in angle-space.  We note
that $\vect{D}$ is a Hessian and is therefore symmetric.
Associated with $\hessian$ are three orthogonal directions, $\eigen_n$
($n=1,3$), corresponding to the eigenvectors of $\hessian$ if it is
evaluated as a matrix. Each of these directions is
associated with an eigenvalue, $\lambda_n$. 

Consider a cluster whose stars are distributed isotropically 
within a unit sphere about some mean $\vJ_0$ in action-space.
The corresponding angle-space structure, resulting from the
mapping of this sphere by $\hessian$, will be an ellipsoid
with semi-axes of length $t\lambda_n$ and direction $\eigen_n$.

If the $\lambda_n$ are finite and approximately equal, such an
isotropic cluster will spread out in angle-space with no
preferred direction, with the density initially falling as $t^3$.
Eventually, the cluster will uniformly populate the whole of
angle-space; in real-space the cluster will uniformly fill the entire
volume occupied by the orbit $\vJ_0$.  For a cluster disrupting in an
actual galaxy, the structure would quickly fall below the level of
observability.

If one of the $\lambda_n$ is much smaller than the others, then
$\hessian$ will act to map the cluster into a highly flattened
ellipsoid in angle-space.  In real-space, the cluster will occupy a
2-dimensional subspace of the orbital volume of $\vJ_0$. The precise
configuration of this subspace is likely to be complex. However, the
density will initially fall as $t^2$, and in a real galaxy, such a
structure will rapidly become invisible.

If two of the $\lambda_n$ are small, then $\hessian$ acts to map
the cluster into a line in angle-space. In this case, the resulting
real-space structure will be a filament. The density of this
filament will fall linearly with $t$. In a real galaxy, such
a structure may therefore persist with a significant overdensity
for some time. It is this case that describes the formation of
tidal streams \citep[\S8.3.1]{bt08}, and it is this case
that we investigate in detail in this chapter.
Finally, if all the $\lambda_n$ are zero, there will be no spread at all,
and even an unbound cluster will remain intact indefinitely.

We note that there is no {\em a priori} reason for any of the
$\lambda_n$ to be small.
The existence of $\vect{D}$ imposes no conditions on $H$ in
general, save that it must be twice differentiable near to $\vJ_0$.
We can write a Hamiltonian for which, for particular $\vJ_0$ at least,
the $\lambda_n$ and the $\eigen_n$ are arbitrary.
It must therefore be a peculiar property of realistic Galactic
potentials that causes disrupted clusters to form streams.

Lastly, a word on validity. The Taylor expansion \eqref{mech:eq:taylor-H}
is valid when,
\begin{equation}
D_{ij} \gg {1 \over 3}{\partial D_{ij} \over \partial J_k} \delta J_k =
{1 \over 3}{\partial^3 H \over \partial J_i \partial J_j \partial J_k} \delta J_k.
\label{mech:eq:condition2}
\end{equation}
This condition approximates to
\begin{equation}
\delta J_i \ll J_i,
\label{mech:eq:condition}
\end{equation}
if $H$ is dominated by some low power of $J$. In general then, we expect
our analysis to be valid if the spread in action of the stars
in the cluster is small compared to the actions themselves, which is likely
to be true for cold clusters in high-energy orbits around massive hosts.
The condition (\ref{mech:eq:condition2}) is met in detail
for all the examples considered below.
%We will examine in detail the validity
%of \eqref{mech:eq:taylor-H} for the case of the isochrone potential in
%\secref{mech:sec:validity} below. 

\subsection{The geometry of streams in phase-space}

We have seen that the condition for a stream to form is
that one of the eigenvalues $\lambda_n$ of the Hessian, $\hessian$,
must be very much larger than the other two. Herein, we will number the
$\lambda_n$ and their corresponding $\eigen_n$ such that
\begin{equation}
\lambda_1 > \lambda_2 \ge \lambda_3.
\end{equation}
We now ask, under what conditions does the direction of this stream point down
the progenitor's orbit?

In action-angle coordinates, the trajectory in angle-space of an orbit
$\vJ$ is given by the frequency vector, $\vO(\vJ)$.
One
condition for a stream to delineate precisely the progenitor orbit\footnote{
Alternately, one can consider this orbit to be the mean orbit of the stream
stars. Although the actions $\vJ$ of stars bound to the cluster are not constant,
the mean action $\vJ_0$ does remain constant, since by energy conservation
\begin{align*}
&\d H = \vO_0 \cdot \d\vJ_a + \vO_0 \cdot \d\vJ_b = 0,\\
&\d \vJ_a  = -\d\vJ_b.
\end{align*}
This restatement of the law of conservation of momentum tells us
that any change in the action $\vJ_a$ of a single star must be matched by
an equal and opposite change in the mean action $\vJ_b$ of the other
stars.}
is that the long axis of the angle-space
distribution $\Delta\vT$ must be aligned with the progenitor frequency
$\vO_0$. We will see later that this condition is not sufficient to
ensure real-space alignment between streams and orbits, 
but it is a required condition.

For simplicity, let us restrict ourselves once more to clusters that
are isotropic in $\Delta\vJ$. The angle-space distribution $\Delta\vT$
is now an ellipsoid with semi-axes of length $\lambda_n$, and with the
semi-major axis of the distribution aligned with the principal
direction of $\hessian$, i.e.~$\eigen_1$.  The $\eigen_n$ are given by
the expression
\begin{equation}
  (\vect{\nabla_J} \vect{\Omega})\, \eigen_n = \lambda_n \eigen_n,
  \qquad (n=1,3),
\end{equation}
where subscript $n$ denotes a label, i.e.~no summation.
Since the $\eigen_n$ and $\lambda_n$ are properties of $\hessian$ which is
assumed constant with respect to $\vJ$ over the cluster,
$\lambda_n$ and $\eigen_n$ are constant with respect to $J$. 
We can therefore rearrange and integrate the above expression
\begin{align}
  \vect{\nabla_J} (\vO \cdot \eigen_n) & = \lambda_n \vect{\nabla_J}(\delta \vJ\cdot\eigen_n),\\
  \vO\cdot\eigen_n & = \lambda_n \, \delta \vJ\cdot\eigen_n + k_n, \label{mech:eq:omega.e}
\end{align}
where $k_n$ is a constant of integration. Comparing \eqref{mech:eq:omega.e}
with \eqref{mech:eq:omega} we find
\begin{equation}
  k_n = \vO_0\cdot\eigen_n.
\label{mech:eq:k_n}
\end{equation}
Since $\hessian$ is symmetric, the $\eigen_n$ are orthogonal. For
$\eigen_1$ to be perfectly aligned with $\vO_0$, both $\eigen_2$ and
$\eigen_3$ must be orthogonal to $\vO_0$.
Hence
\begin{equation}
k_n = \vO_0\cdot\eigen_n = 0, \qquad (n=2,3),
\end{equation}
is the required condition for a stream formed from an isotropic cluster
to delineate the progenitor orbit.

Nothing we have said thus far shows that this ought to be the case for
either general or specific potentials. In the section that follows,
we will determine in detail the form that a Hamiltonian must take,
if it is to satisfy this condition.

\section{Stream formation in spherical potentials}
\label{mech:sec:2d}
\subsection{The general case in systems with two actions}
\label{mech:sec:general2d}

In this section, we will explicitly solve \eqref{mech:eq:omega.e} in
the case of a general Hamiltonian described by two actions, $H(J_1,
J_2)$, such as can be used to describe regular motion in any
spherical potential. The resulting solution will be the form of the
Hamiltonian in any stream-forming system, and our goal is to relate
the terms in that solution to the geometry of stream formation
described by $\hessian$. We will then
use these relations to examine the geometry of streams that form in
some example systems.

We begin by writing out \eqref{mech:eq:omega.e} explicitly
\begin{equation}
  \Omega_1 \,e_{n,1} +  \Omega_2 \, e_{n,2} = e_{n,1}\,\partial_1 H  + e_{n,2}\,\partial_2 H  =
  \lambda_n( e_{n,1} \, \delta J_1 + \e_{n,2} \, \delta J_2 ) + k_n,
\label{mech:eq:explicit2d}
\end{equation}
where the $\{1,2\}$ suffixes denote the respective vector components, and
where we have used the shorthand $\partial_i \equiv \partial/\partial J_i$.
\Eqref{mech:eq:explicit2d} has two solutions, one corresponding to each
principal direction $\eigen_n$, where $n=(1,2)$.
We first define the constants
\begin{align}
\alpha_n &\equiv e_{n,2}/e_{n,1}, \label{mech:eq:2dalpha}\\
\beta_n &\equiv k_n/e_{n,1}. \label{mech:eq:2dbeta}
\end{align}
\Eqref{mech:eq:explicit2d} then becomes
\begin{equation}
  \partial_1 H + \alpha_n \partial_2 H = \lambda_n(\delta J_1 + \alpha_n \delta J_2) + \beta_n, \label{mech:eq:2dpde}
\end{equation}
which is a PDE for $H$ in $\delta \vect{J}$. The solution to the
homogeneous equation for \blankeqref{mech:eq:2dpde} is
\begin{equation}
  H = f(\delta J_1 - {\delta J_2 \over \alpha_n}),
\end{equation}
where $f$ is an arbitrary function of the characteristic coordinate,
$\left(\delta J_1 - {\delta J_2 \over \alpha_n}\right)$. The
general solution for the inhomogeneous equation \blankeqref{mech:eq:2dpde} is
\begin{equation}
  H = f\left(\delta J_1 - {\delta J_2 \over \alpha_n}\right) + {\lambda_n \over 2} (\delta J_1^2 + \delta J_2^2)
  + \beta_n \delta J_1 + H_0.\label{mech:eq:2dH}
\end{equation}
This solution relates the form of the Hamiltonian to the geometry of
streams that are created within it, with this geometry being explicitly
described by $(\alpha, \beta, \lambda)_n$. We must take care to note
that although $n=(1,2)$ and it appears that we have specified two forms
for the Hamiltonian, this is not strictly the case. The Hamiltonian has {\em one}
unique form $H(\vJ)$, but the Taylor expansion \eqref{mech:eq:taylor-H}
and the above \eqref{mech:eq:2dH} can {\em always} be compared in two
different ways. This can be understood by considering a second-order Taylor
expansion of \eqref{mech:eq:2dH} itself: the quadratic term that
would appear in $\alpha_n$ as a result of expanding out $f$ can always
be solved for two roots, $(\alpha_1, \alpha_2)$, for the same values
of $H$ and $\vJ$. Indeed, we will perform this calculation explicitly below.

Consider again \eqref{mech:eq:2dH}. There are two
possibilities. Firstly, the Hamiltonian of the system may globally
take a single form of this type. In this case, the geometry of stream
propagation in action-angle space will be the same for all streams of
equivalent $\Delta\vJ$ configuration, no matter what the orbit of the
progenitor. We shall see that this is the case with the Kepler
potential, discussed below.

The other possibility is that the Hamiltonian does not
take a form that can be globally described by \eqref{mech:eq:2dH} at
all. Expanding $f$ in a power series,
\begin{equation}
  f(x) = \gamma_n x + \delta_n x^2,
  \label{mech:eq:powerseries}
\end{equation}
where we have considered any constant term already subsumed into $H_0$,
and we have neglected any terms in $\delta \vJ^3$,
 gives
\begin{align}
  H &= H_0 + \gamma_n\left(\delta J_1 - {\delta J_2 \over \alpha_n}\right)
  + \delta_n \left(\delta J_1 - {\delta J_2 \over \alpha_n}\right)^2
  + {\lambda_n \over 2}\left(\delta J_1^2 + \delta J_2^2\right)
  + \beta_n \delta J_1\nonumber\\
  &= \delta J_1^2\left(\delta_n + {\lambda_n \over 2}\right) + \delta J_2^2\left({\delta_n \over
      \alpha_n^2} + {\lambda_n \over 2}\right) - 2{\delta_n \over \alpha_n}\, \delta J_1 \, \delta J_2 
  + \delta J_1 \, (\gamma_n + \beta_n) - {\gamma_n \over \alpha_n} \delta J_2 + H_0.
  \label{mech:eq:2dH-greek}
\end{align}
We rewrite the Taylor expansion \blankeqref{mech:eq:taylor-H} in our current notation,
\begin{equation}
H(\vJ) = H_0 + \Omega_{0,i}\,\delta J_i + {1 \over 2}D_{ij}\,\delta J_i \,\delta J_j.
\end{equation}
Comparing coefficients between the above expression and \eqref{mech:eq:2dH-greek} gives,
\begin{align} 
  \alpha_n & =  {-2 \delta_n \over \pone \basefreq{2}},\nonumber\\
  \beta_n & =  \basefreq{1} + \alpha_n \basefreq{2},\nonumber\\
  \gamma_n & =  - \alpha_n\basefreq{2},\nonumber\\
  \delta_n & =  {1 \over 4}\left\{ (\pone \basefreq{1} - \ptwo \basefreq{2})
    \pm \sqrt{(\pone \basefreq{1} - \ptwo \basefreq{2})^2 + 4(\pone \basefreq{2})^2}\right\},\nonumber\\
  \lambda_n & = \pone \basefreq{1} - 2 \delta_n.
%  \delta_n\left( 1 - {1 \over \alpha_n^2}
%  \right) & = 
%{1 \over 2}(\pone \basefreq{1} - \ptwo \basefreq{2})\nonumber\\
%{-2 \delta_n \over \alpha_n} & = \pone \basefreq{2}
\label{mech:eq:2d-greek}
\end{align}
where we can now see explicitly that choosing $n = (1,2)$ is
equivalent to setting the sign of the radical in the above expression
for $\delta_n$.

We are now in a position to deduce the geometry of streams for a
given potential, if we know the form of $H(\vJ)$. In particular, we
can deduce the angle between the principal direction $\eigen_1$
and $\vO_0$, since
\begin{equation}
  \vO_0 =   \left\{ \gamma_n + \beta_n , {- \gamma_n \over \alpha_n} \right\},
\end{equation}
and from \eqref{mech:eq:k_n} and the definition of $\beta_n$ \blankeqref{mech:eq:2dbeta},
\begin{equation}
  \vO_0 \cdot \eigen_1 = e_{1,1} \beta_1,
\end{equation}
and so the misalignment angle is
\begin{equation}
  \vartheta = \arccos {e_{1,1} \beta_1 \over \left|\vO_0 \right| }.
  \label{mech:eq:misalignment}
\end{equation}
where, since the $\eigens$ are normalized to unity
\begin{equation}
  e_{1,1} = \pm \sqrt{1 \over 1 + \alpha_1^2}.
\end{equation}
We will now proceed to study the formation of streams in various
example potentials.

\subsection{Kepler potential}
\label{mech:sec:kepler}

\begin{table}
  \centering
  \caption[Parameters for the spherical potentials used in \chapref{chap:mech}]
{Parameters for the spherical potentials used in this chapter.}
  \begin{tabular}{l|ll}
    \hline
    & $M / 10^{10}\msun$ & $b / \kpc$ \\
    \hline\hline
    Kepler & 10.75 & 0 \\
    Isochrone& 28.52 & 3.64 \\
    \hline
  \end{tabular}
  \label{mech:tab:potentials}
\end{table}

In the study of galactic dynamics, there are remarkably few potentials
of interest for which a Hamiltonian can be written as a
function of $\vJ$ in closed form. One such potential is the
Kepler potential
\begin{equation}
  \Phi(r) = { -GM \over r},
\end{equation}
for which the Hamiltonian is,
\begin{equation}
  H(\vect{J}) =  { -(GM)^2 \over 2 (J_r + L)},
\end{equation}
where $J_r$ is the radial action, and $L$ is the angular momentum. Comparing
the above expression with \eqref{mech:eq:2dH}, we immediately see that
$H$ is of the form required for streams to form with globally consistent
geometry. The most obvious solution for the parameters \blankeqref{mech:eq:2d-greek} is
\begin{equation}
\left(\alpha,\beta,\lambda\right)_2 = \left(-1, \, 0, \, 0\right),
\end{equation}
for $\eigen_2$, which has a null eigenvalue, and lies
perpendicular to $\vO_0$. Some algebra will also confirm the other
solution
\begin{equation}
\left(\alpha,\beta,\lambda\right)_1 = \left(1,\, 2\Omega_0,\, 
2\partial_{J_r} \Omega_0\right),
\end{equation}
for the principal direction $\eigen_1$, which points precisely along
$\vO_0$.
Streams therefore perfectly delineate progenitor orbits in Kepler potentials,
and exhibit secular spread strictly along the orbit, and do
not grow wider with time.

\begin{figure}[\figplaceopts]
  \centerline{
    \includegraphics[width=\figureshrink\hsize]{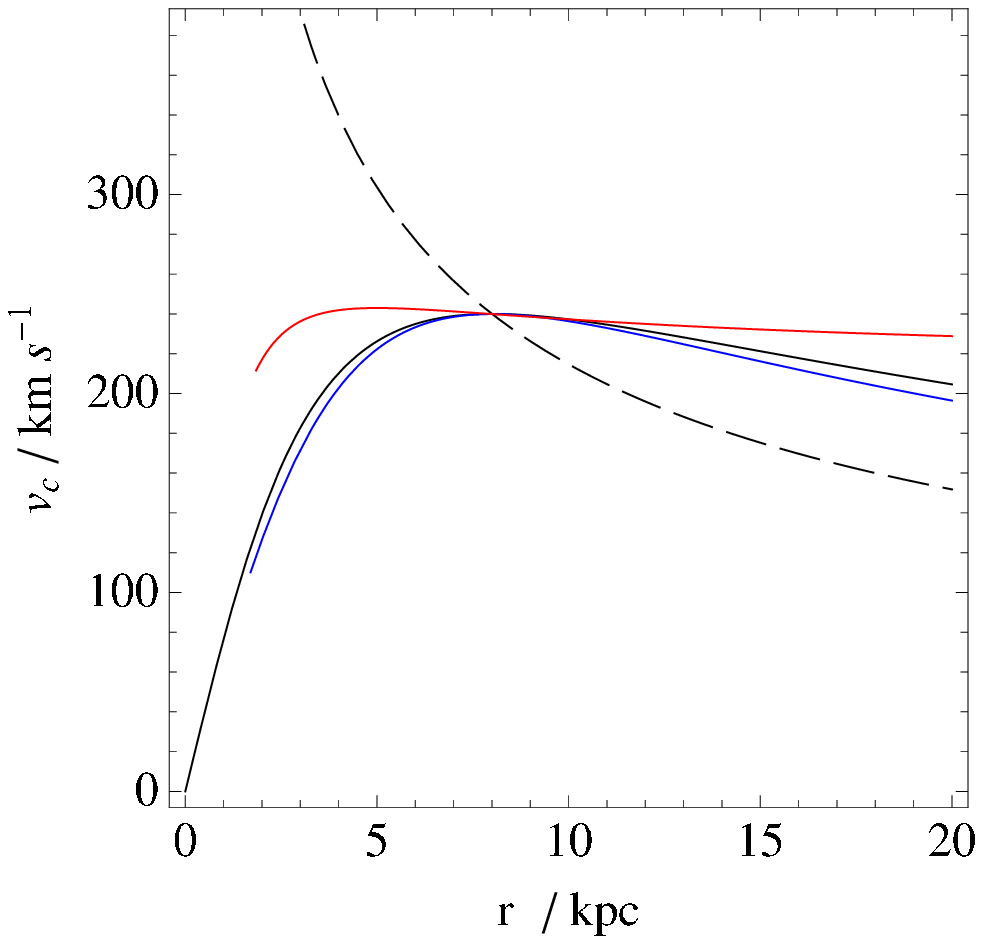}
  }
  \caption[In-plane rotation curves for selected potentials from \chapref{chap:mech}]
{The in-plane rotation curves of the potentials specified in
    \tabref{mech:tab:potentials} and
    \tabref{mech:tab:stackpots}. Solid black line: the isochrone potential.
    Dashed black line: the Kepler potential. Red line: the \stackel\
    potential SP2. Blue line: the \stackel\ potential SP1.  
    Parameters in all potentials have been tuned to give a circular
    speed of $v_c = 240\kms$ at the Solar radius $\rsun =
    8\kpc$.
%     In the case of the isochrone potential, $b$ was chosen
%     such that the maximum of the curve occured at $\rsun$, giving a
%     compromise between high circular speed interior to $\rsun$, and
%     flatness exterior to $\rsun$.
}
  \label{mech:fig:rotationcurve}
\end{figure}

\begin{figure}[\figplaceopts]
  \centerline{
    \includegraphics[width=\doublefigshrink\hsize]{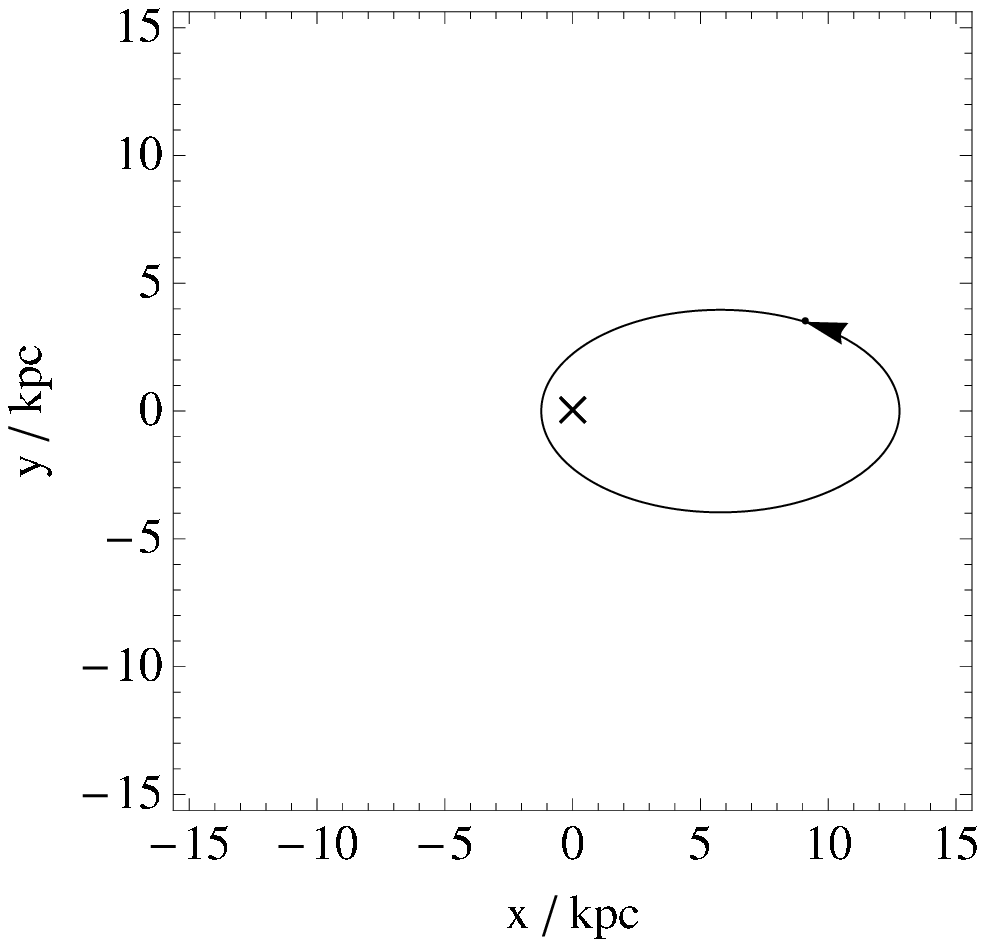}
    \includegraphics[width=\doublefigshrink\hsize]{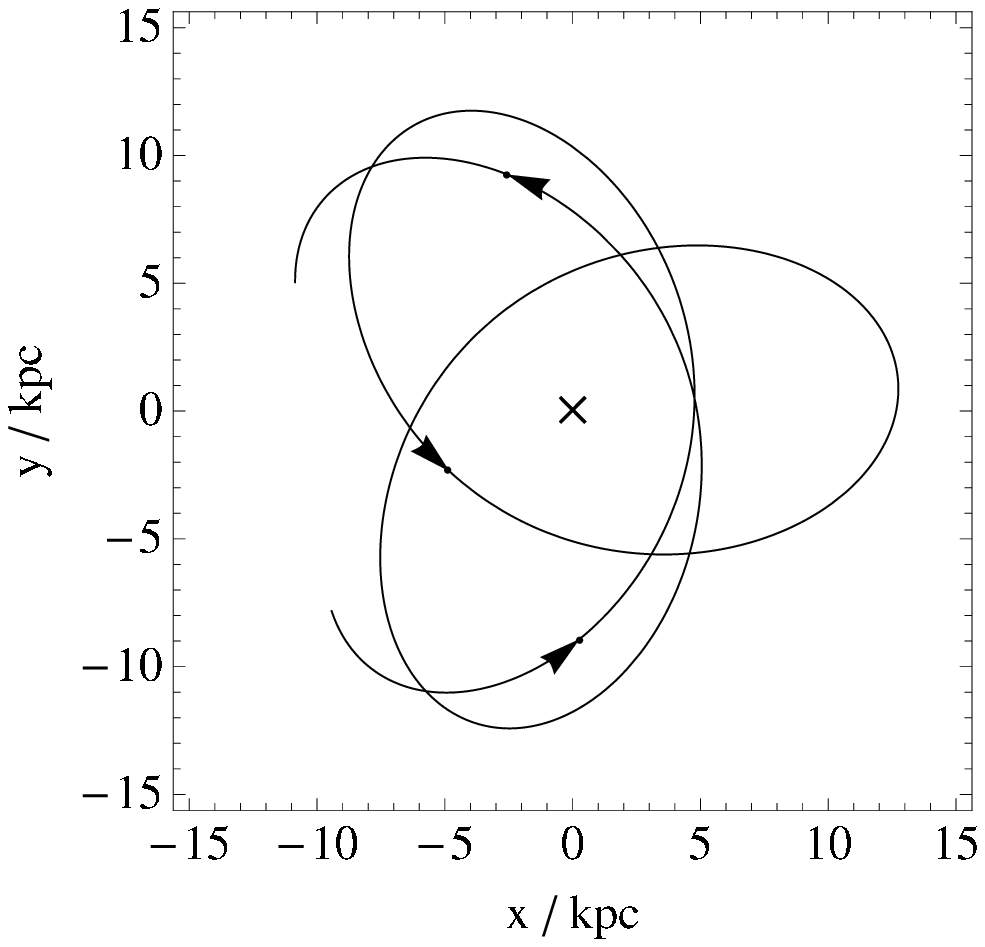}
  }
  \centerline{
    \includegraphics[width=\doublefigshrink\hsize]{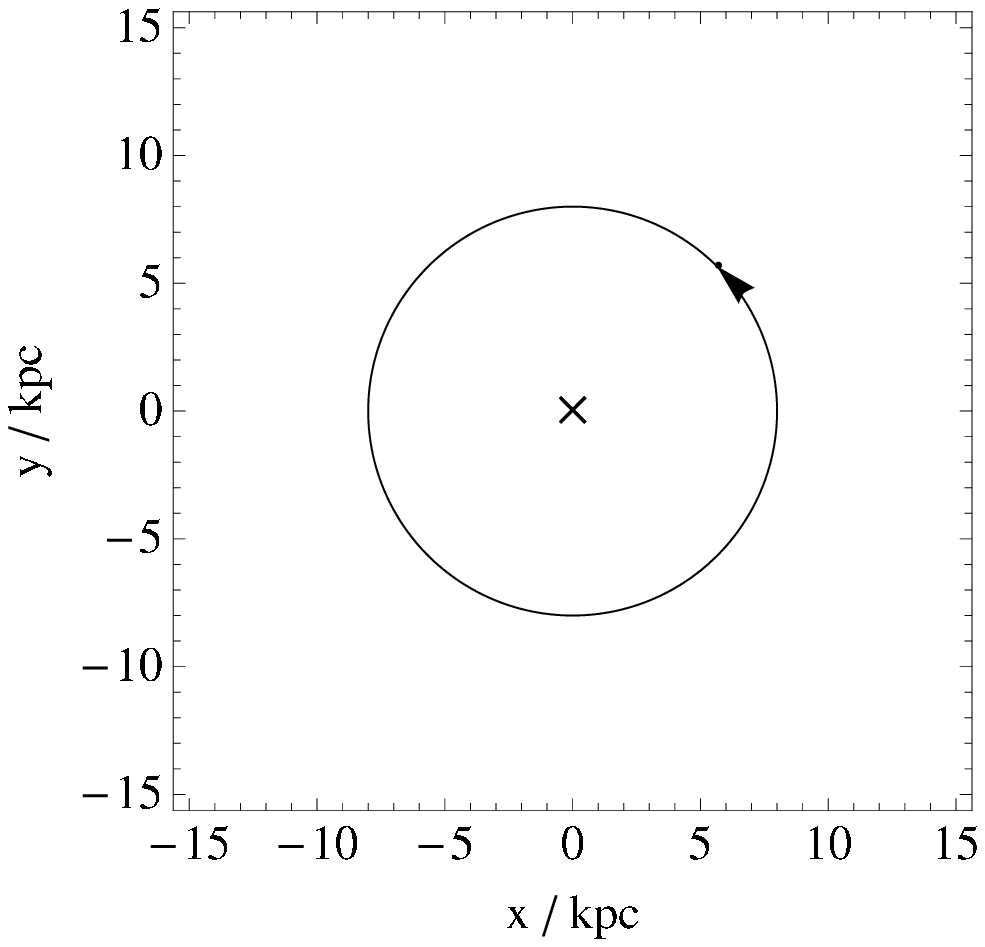}
    \includegraphics[width=\doublefigshrink\hsize]{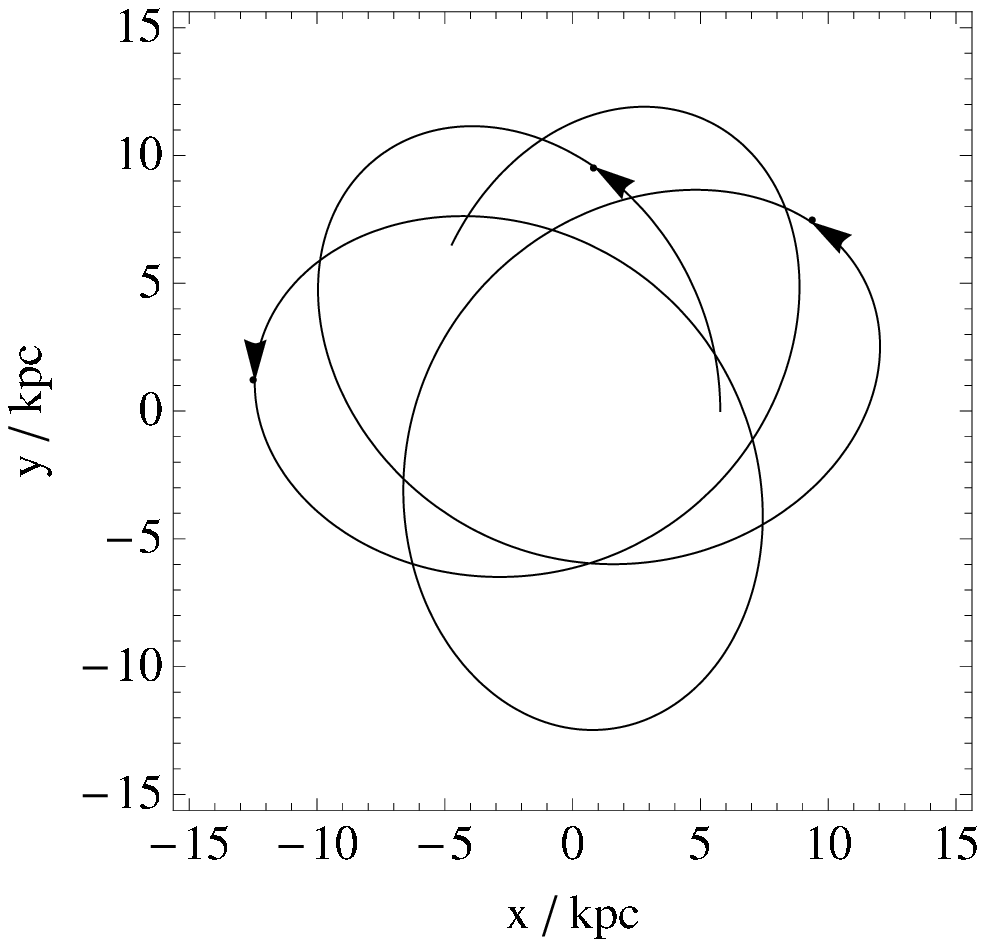}
  }
  \centerline{
    \includegraphics[width=\doublefigshrink\hsize]{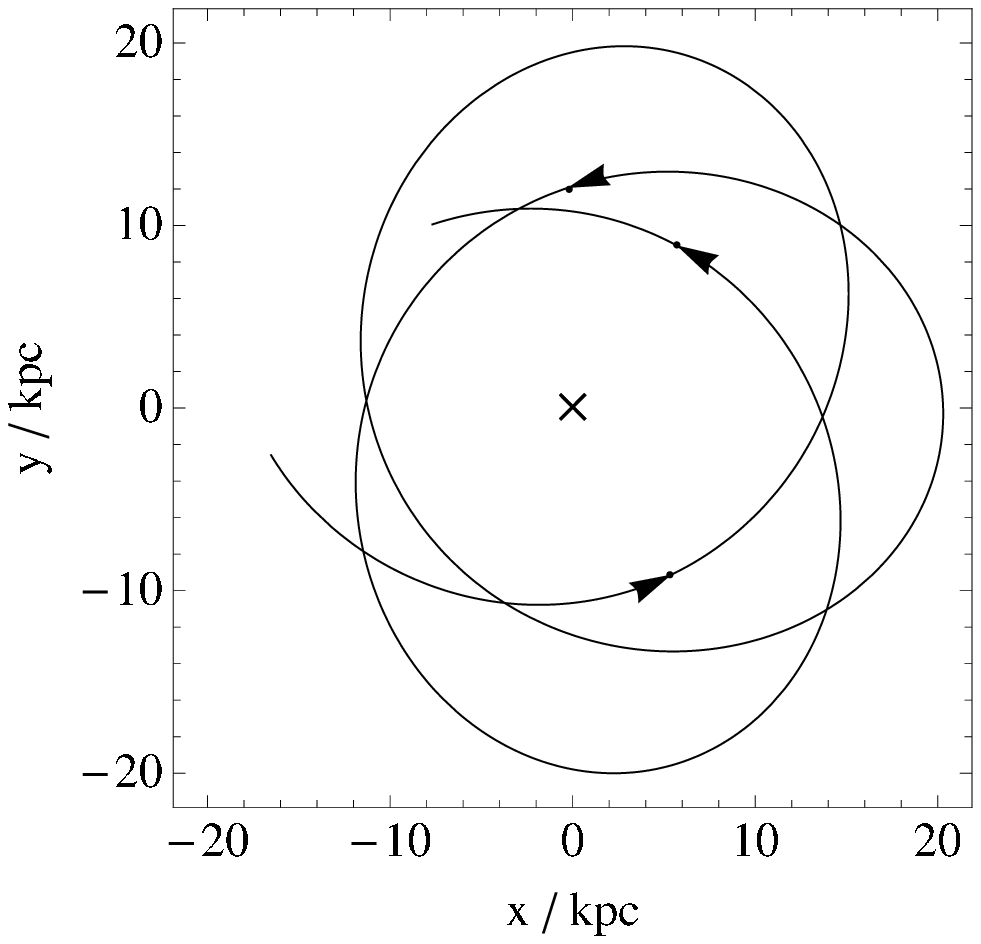}
    \includegraphics[width=\doublefigshrink\hsize]{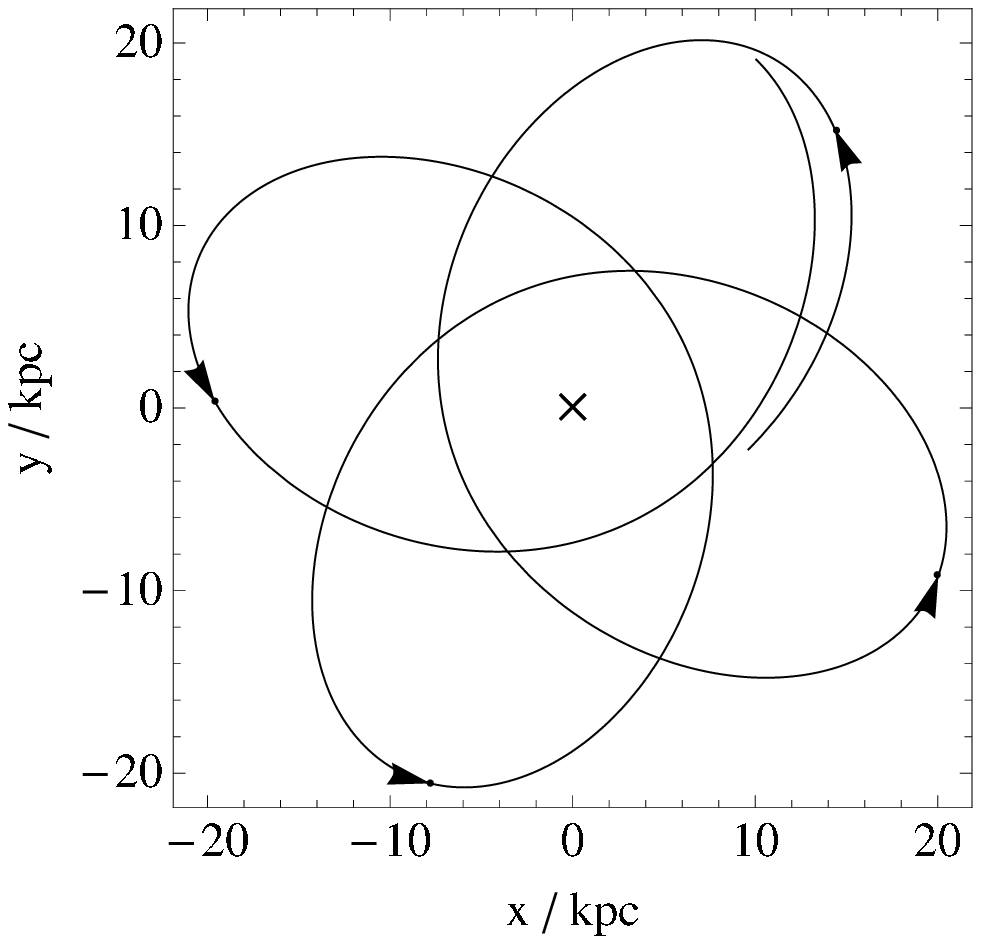}
  }
  \caption[%% spherical orbits
Real-space trajectories for selected orbits in the spherical potentials
of \tabref{mech:tab:potentials}
]
{ %% spherical orbits
The real-space trajectories of the orbits (left-to-right, top-to-bottom)
K1, I1, I2, I3, I4 and I5, as described in \tabref{mech:tab:orbits}. The
trajectories were evaluated
in the Kepler potential (for K1) and the isochrone potential (for I1--I5)
of \tabref{mech:tab:potentials}. 
}
  \label{mech:fig:sphericalorbits}
\end{figure}

\subsubsection{Numerical tests}

\tabref{mech:tab:potentials} describes a Kepler potential in which
we will now confirm this prediction numerically. The parameter $M$ of
this potential was chosen to reproduce a fiducial circular velocity
$v_c = 240\kms$ at the approximate solar radius of $\rsun=8\kpc$.
\figref{mech:fig:rotationcurve} shows the rotation curve in this
potential, along with the rotation curves for other potentials
in use in this chapter.

A cluster of 50 test particles was created, with a Gaussian distribution of particles
in action-angle space, defined by $\sigma_J = 1 \kpc\kms$ and $\sigma_\theta = 5\ttp{-3}$
radians. This cluster was placed on the orbit K1, given in
\tabref{mech:tab:orbits}. in the aforementioned Kepler
potential. The orbit has a pericentre radius of about $1.5\kpc$ and an apocentre of
about $13\kpc$, and is illustrated in \figref{mech:fig:sphericalorbits}.

\begin{table}
  \centering
  \caption[Highlighted orbits in the potentials of \tabref{mech:tab:potentials}]
  {Actions and apses for selected orbits in the spherical potentials
    used in this chapter. We generally choose $J_\phi = L$ so all orbits
    remain in the $(x,y)$ plane. The trajectories of these orbits are illustrated
    in \figref{mech:fig:sphericalorbits}
  }
  \begin{tabular}{l|ll|ll}
    \hline
    & $J_r / \kpc\kms$ & $L / \kpc\kms$ & $r_p / \kpc$ & $r_a / \kpc$ \\  %& $E$ \\
    \hline\hline
    K1 & $780.$ & $1016.$ & $1.5$ & $13$ \\ %& $-0.767$ \\
    I1 & $313.$ & $1693.$ & $5$ & $13$ \\ %& $-1.469$ \\ 
    I2 & $0.$ & $1920.$ & $8$ & $8$ \\%& ?\\
    I3 & $207.$ & $1920.$ & $6$ & $12.5$ \\ %& ?\\
    I4 & $571.7$ & $2536.$ & $11$ & $20$ \\ %& $-0.942244$\\
    I5 & $215.4$ & $3127.$ & $7$ & $20$ \\ %& $-0.898452$\\
    \hline
  \end{tabular}
  \label{mech:tab:orbits}
\end{table}

\begin{figure}[\figplaceopts]
  \centerline{
    \includegraphics[width=\doublefigshrink\hsize]{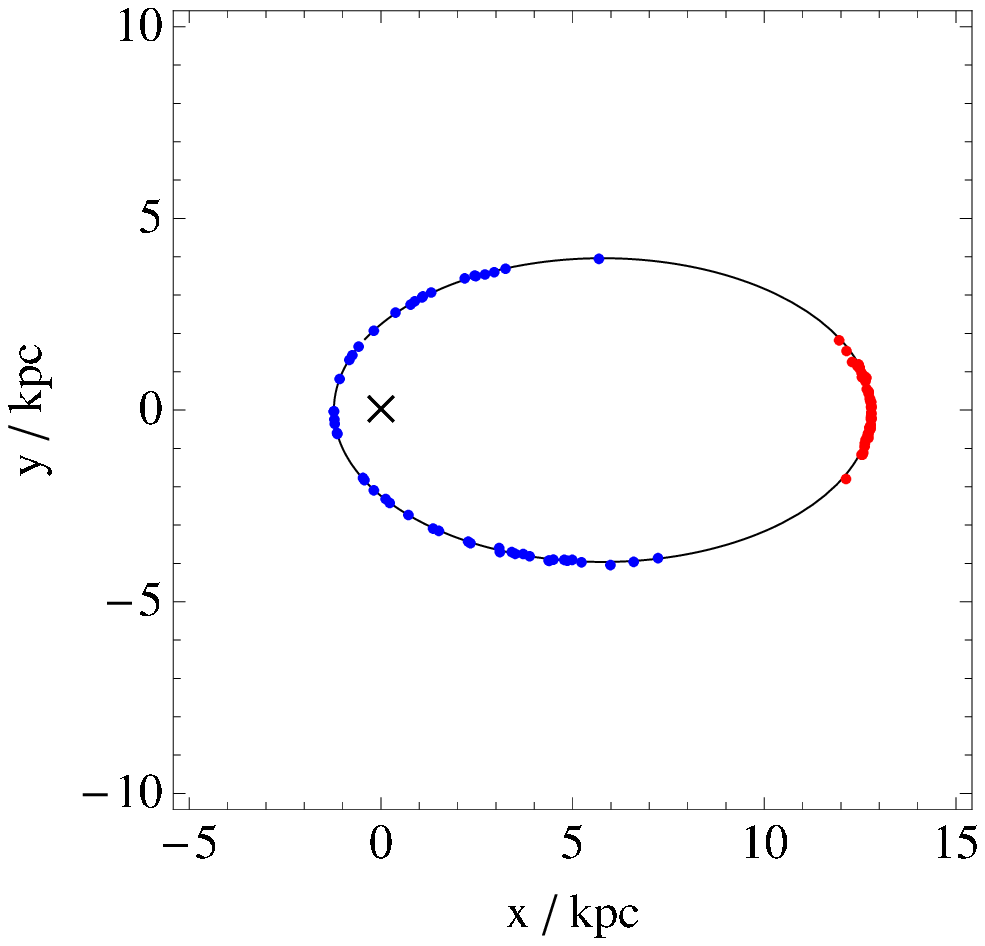}
    \includegraphics[width=\doublefigshrink\hsize]{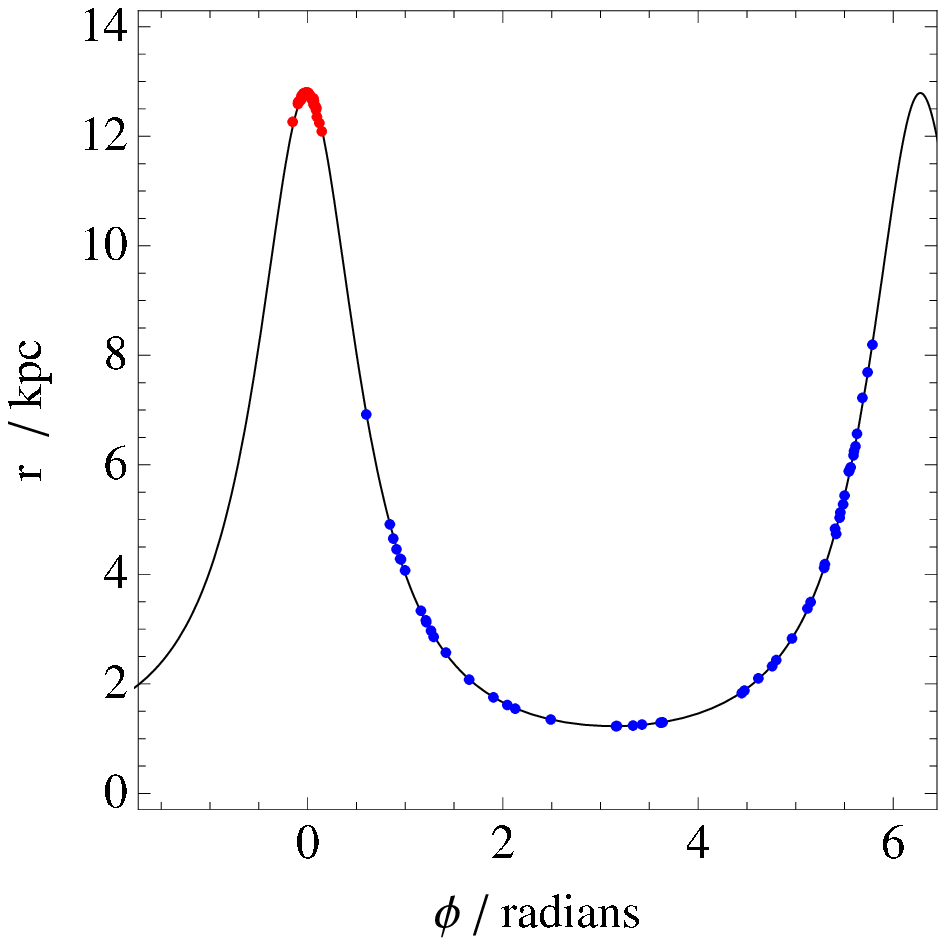}
  }
  \caption[A stream formed from the orbit K1]
{The solid line shows the orbit K1 (\tabref{mech:tab:orbits})
    in a Kepler potential (\tabref{mech:tab:potentials}), on which a cluster of 50 test
    particles (initial conditions described in the text) has been
    evolved. The particles were released at apocentre. The red dots show the position
    of the test particles near apocentre, after 24 complete orbits, at $t=4.02\Gyr$.
    The blue dots show the same test particles near pericentre, approximately half an
    orbit later. In both cases, the dots delineate the underlying orbit precisely.
  }
  \label{mech:fig:kepler}
\end{figure}

\begin{figure}[\figplaceopts]
  \centerline{
    \includegraphics[width=\doublefigshrink\hsize]{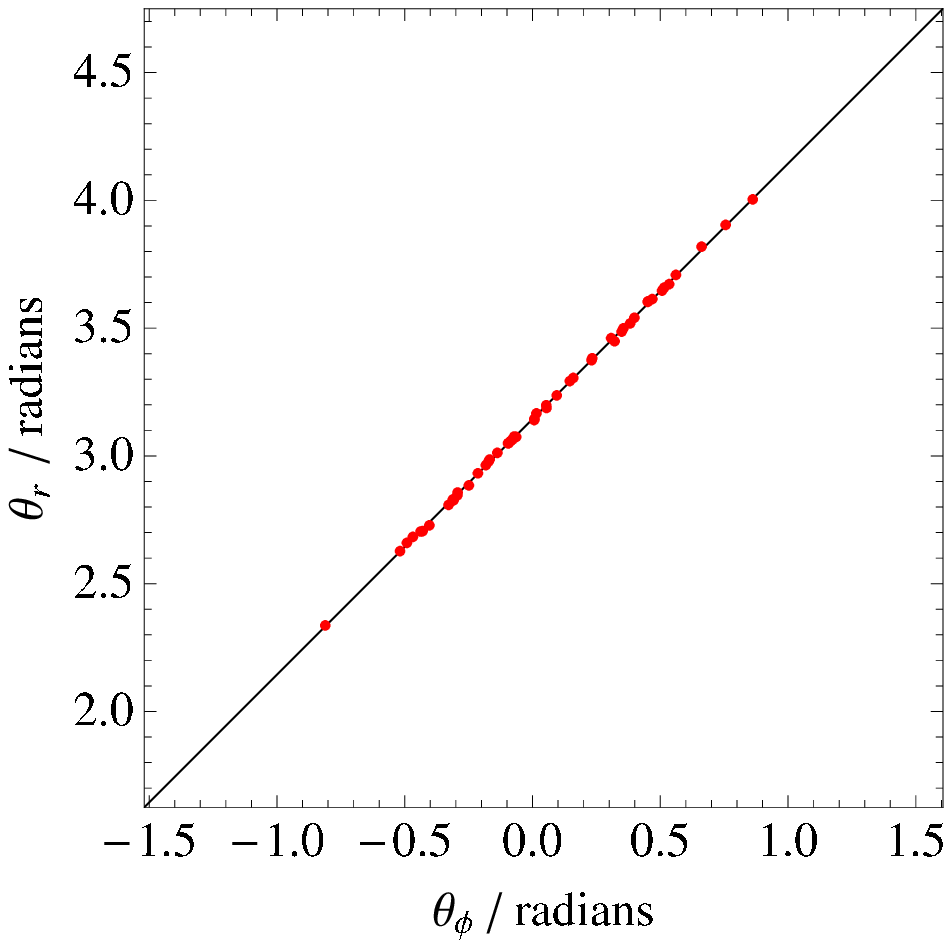}
  }
  \caption[Angle-space configuration for the particles shown at apocentre in
    \figref{mech:fig:kepler}]
{Angle-space configuration for the particles shown at apocentre in
    \figref{mech:fig:kepler}.
    The solid line shows the frequency vector $\vO_0$, with which the
    stream particles are perfectly aligned.
    There is also
    no secular spread in the direction perpendicular to the stream motion,
    as predicted in \secref{mech:sec:kepler}.
  }
  \label{mech:fig:kepler-angles}
\end{figure}

The cluster was released at apocentre, and evolved for $4.02\Gyr$,
equal to 24 complete orbits, by integrating the equations of motion
for each particle in the relevant potential. \figref{mech:fig:kepler} shows the
real-space configuration of the particles at the end of this time, and
\figref{mech:fig:kepler-angles} shows the configuration of
the same particles in angle-space.

The cluster has elongated to form a stream, that
\figref{mech:fig:kepler} shows to cover approximately half the orbit when
the centroid is at pericentre. There is no spread in width in
either real-space or angle-space. The stream delineates the
cluster's orbit in both figures perfectly. \figref{mech:fig:kepler} shows that
this is true irrespective of the real-space location of the centroid.
Thus, the prediction of the previous section is validated.

\subsection{Spherical harmonic oscillator}
\label{mech:sec:sho}

The spherical harmonic oscillator potential applies for motion within
a sphere of uniform density \citep[\S3.1a]{bt08}, and is therefore
of relevance to galaxy cores in the absence of a black hole. It has the
form
\begin{equation}
\Phi(r) = {1\over2}\Omega^2 r^2.
\end{equation}
The Hamiltonian of the harmonic oscillator
takes a particularly simple form
\begin{equation}
H(J_r,L) = \Omega \left(L + 2J_r\right),
\label{mech:eq:2dH-sho}
\end{equation}
where $\Omega$ is a constant. Comparing the above
expression with \eqref{mech:eq:2dH}, we see that like the Kepler
Hamiltonian, the form of \eqref{mech:eq:2dH-sho} admits the
same stream geometry everywhere. In this case, however, the
solutions to \eqref{mech:eq:2d-greek} are trivial, since
$\hessian$ is a null matrix, and $\beta_n$ and $\lambda_n$ are identically zero
for both $n=(1,2)$.

We conclude that clusters in harmonic potentials will always remain in the same
configuration in angle-space, and will not spread out. Hence,
streams cannot form in harmonic potentials. We further note that,
in any case, it would be difficult to tidally strip a cluster in a harmonic
potential, since the tidal force $dF_{\rm tide}$ across a cluster
\begin{equation}
dF_{\rm tide} \simeq {\partial^2 \Phi \over \partial r^2} \d r
= \Omega^2 \d r,
\end{equation}
is independent of galactocentric radius $r$. Thus, a cluster that is bound
at apocentre in such a potential will remain bound elsewhere along its orbit.
\subsection{Isochrone potential}
\label{mech:sec:isochrone}

The isochrone potential \citep[\S2.2.2d]{bt08} is a simple potential
which has several useful properties. It behaves as a harmonic
oscillator in the limit of small radius, and as a Kepler potential
at large radius, thus providing a reasonable model for a spherical
galaxy across all radii.
%Further, like its limiting cases, it has
%closed forms for its action-angle variables in terms of $E$ and
%$L$ \citep[\S3.5.2]{bt08}.
The form of the potential is
\begin{equation}
  \Phi(r) = { -GM \over b + \sqrt{b^2 + r^2}},
\end{equation}
where $b$ is a scaling constant, and the Hamiltonian is
\begin{equation}
  H(\vect{J}) = { -(GM)^2 \over 2[J_r + {1 \over 2}
    (L + \sqrt{L^2 + 4GMb})]^2}.
\end{equation}
This Hamiltonian is not of the form of \eqref{mech:eq:2dH}, and does not admit
a globally applicable stream geometry. We therefore need to calculate
the parameters \blankeqref{mech:eq:2d-greek} directly from $H$.
We require the frequencies, which are obtained by direct differentiation
\begin{align}
  \Omega_r &= { (GM)^2 \over [J_r + {1 \over 2}
    (L + \sqrt{L^2 + 4GMb})]^3},\\
  \Omega_\phi &= {1 \over 2}
  \left( 1 + {L \over \sqrt{L^2 + 4GMb}} \right) \Omega_r. \label{mech:eq:iso-freqs}
\end{align}
We also require the derivatives of the frequencies with respect to the actions.
We note that $\Omega_r$ can be written as a function of $H$ only,
\begin{equation}
  \Omega_r = {(-2H)^{3/2} \over GM},
\end{equation}
and therefore its derivatives with respect to the actions are
\begin{eqnarray}
  {\partial_{J_r}\Omega_r(H)} &=& \Omega'_r(H) \, \Omega_r,\\
  {\partial_L \Omega_r(H)}  &=& \Omega'_r(H) \, \Omega_\theta,
\end{eqnarray}
where
\begin{equation}
  \Omega'_r(H) = -{3\sqrt{-2H} \over G M}.
\end{equation}
We must calculate $\partial_L \Omega_\phi$ directly from
\eqref{mech:eq:iso-freqs}. We find
\begin{equation}
  \partial_L \Omega_\phi = {2 G M b \over \left(
L^2 + 4 G M b \right)^{3/2}} \Omega_r
+ {1 \over 2}
\left(
1+ {L \over \sqrt{L^2 + 4 G M b}} \right) \partial_L \Omega_r.
\end{equation}
From these expressions, we are now in a position to piece together
values for the parameters \blankeqref{mech:eq:2dH-greek}, although the full expressions for
$(\alpha,\beta,\gamma,\delta,\lambda)_n$ are not algebraically
neat, and are therefore not very instructive. We do not repeat them here.
We do note however that in the limit of $b \rightarrow \infty$ we recover the
harmonic oscillator case of \secref{mech:sec:sho}, and that in the
limit of $b \rightarrow 0$ we recover the Kepler case of
\secref{mech:sec:kepler}, as is required.

\begin{figure}[\figplaceopts]
  \centerline{
    \includegraphics[width=\doublefigshrink\hsize]{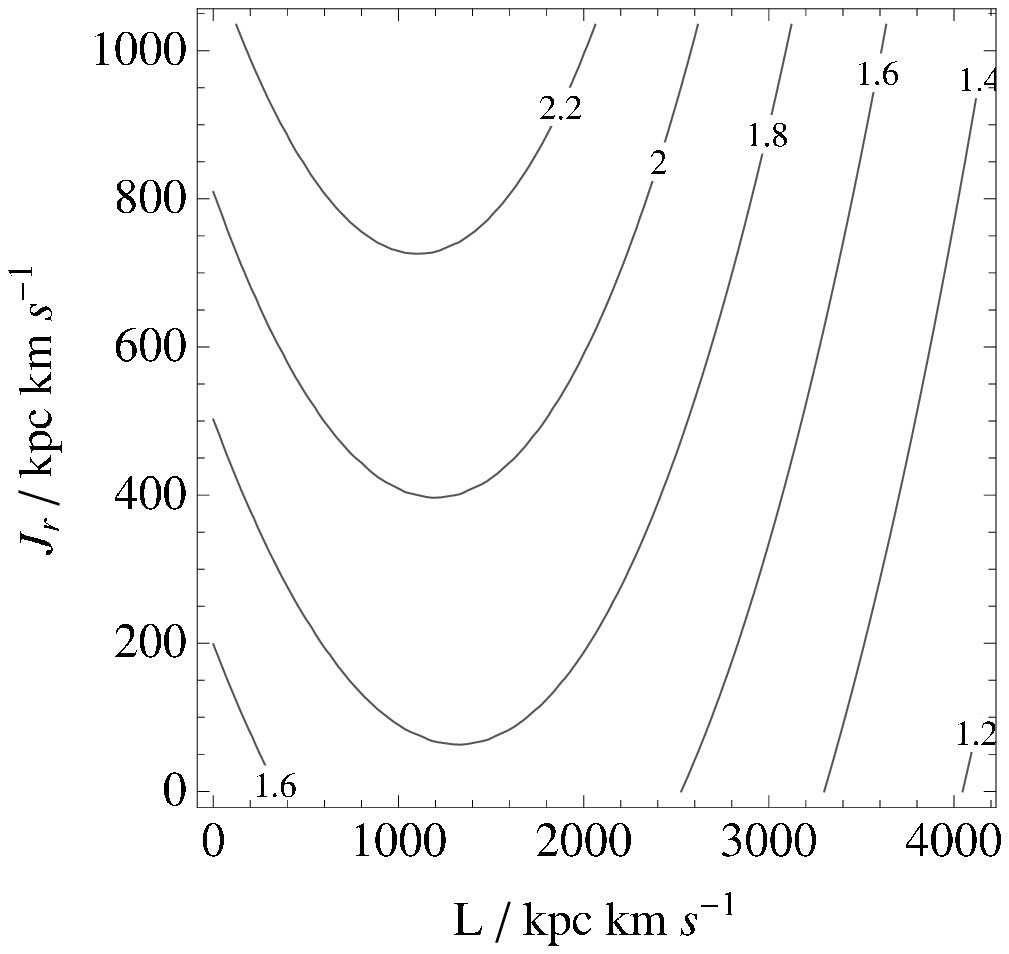}
    \includegraphics[width=\doublefigshrink\hsize]{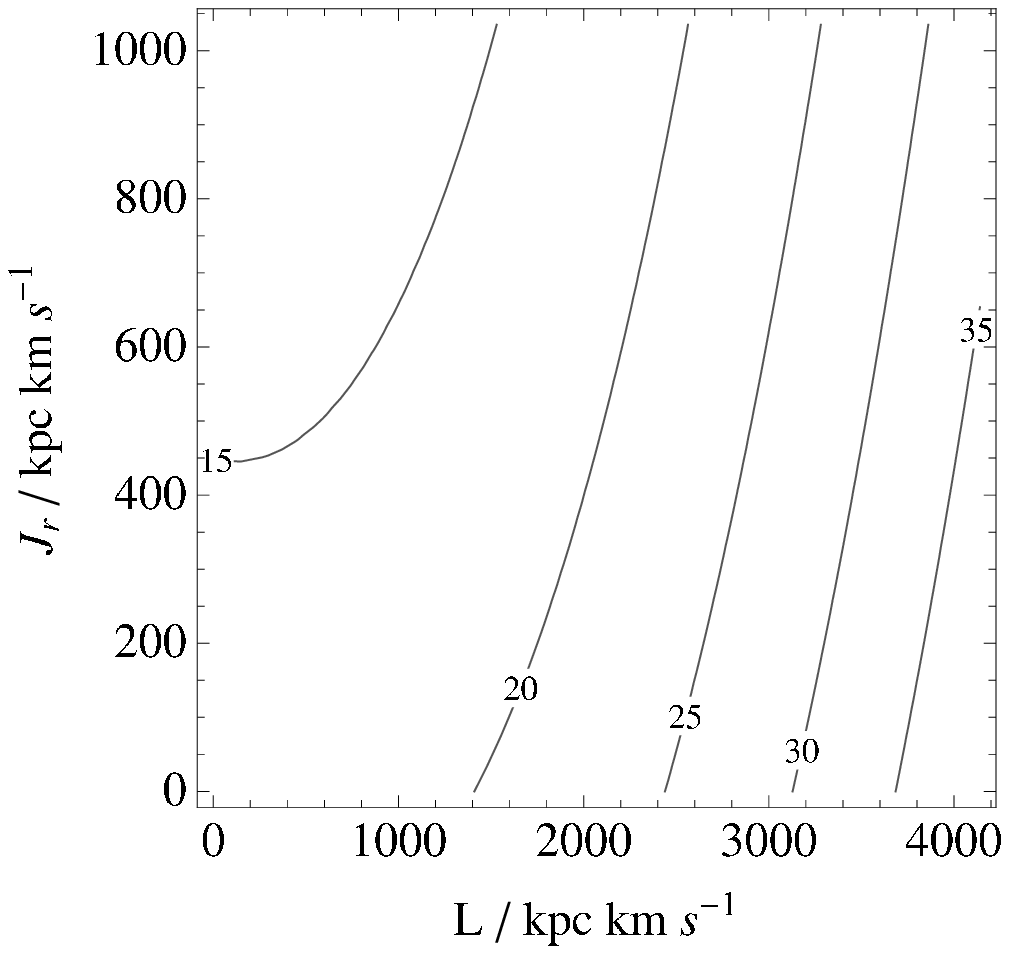}
  }
  \caption[Details of stream geometry in the isochrone potential of
    \tabref{mech:tab:potentials}]{Details of stream geometry in the isochrone potential of
    \tabref{mech:tab:potentials}. Left panel: the misalignment angle
    $\vartheta$, in degrees, between the principal direction of
    $\hessian$ and $\vO_0$, shown as a contour plot against the
    actions of the progenitor orbit.  In all cases, $\vartheta$ is
    between $1.2\deg$ and $2.2\deg$ in angle-space. Right panel: The
    ratio of the eigenvalues $\lambda_1/\lambda_2$. The
    ratio is $>10$ everywhere and rises sharply with increasing
    $L$.  The actions shown in both plots cover a range of
    interesting orbits, which are described in
    \tabref{mech:tab:isochrone-extrema}.  }
  \label{mech:fig:isochrone-hessian}
\end{figure}

\begin{table}
  \centering
  \caption[Coordinate extrema of selected orbits from \figref{mech:fig:isochrone-hessian}
  ]{
    The coordinate extrema of selected orbits from \figref{mech:fig:isochrone-hessian},
    illustrating the variety of orbits covered by that figure. The actions are expressed in $\kpc\kms$,
    while the apses are in \kpc.}
  \begin{tabular}{ll|ll}
    \hline
    $J_r$ & $L$ & $r_\peri$ & $r_\apo$ \\
    \hline\hline
    $1000$ & $4000$ & $13$ & $46$\\
    $1000$ & $\sim 0$ & $\sim 0$ & $12$\\
    $0$ & $4000$ & $20.3$ & $20.3$\\
    \hline
  \end{tabular}
  \label{mech:tab:isochrone-extrema}
\end{table}

In order to proceed, we must work with a specific example.
\tabref{mech:tab:potentials} describes an isochrone potential,
chosen to have the rotation curve maximized at $v_c = 240\kms$
at the assumed solar radius of $\rsun = 8\kpc$. The rotation
curve for this potential is plotted in \figref{mech:fig:rotationcurve}.

What then is the geometry of streams formed in this potential?
\figref{mech:fig:isochrone-hessian} shows the misalignment angle $\vartheta$,
given by \eqref{mech:eq:misalignment}, and the ratio of the eigenvalues
$\lambda_1/\lambda_2$, both as functions of $\vJ$. The range of $\vJ$ shown covers
a variety of interesting orbits, described in \tabref{mech:tab:isochrone-extrema}.

The left panel shows that the principal direction of $\hessian$ is
misaligned with the progenitor orbit for all values of $\vJ$,
by $1$--$2\deg$. The
misalignment is at a minimum for both low energy and high energy circular
orbits, and at a maximum for eccentric orbits with a guiding centre
close to $r=b$.  The right panel shows that the ratio of the eigenvalues
$\lambda_1/\lambda_2$ varies from 15 to 25 across the range, with the ratio maximized for
high energy circular orbits, and minimized for high energy plunging
orbits.

Thus, we expect an isotropic cluster of test particles in this
potential to form a stream in angle-space that is misaligned with
$\vO_0$ by $1$--$2\deg$ and is $15$--$20$ times longer than it is
wide.

\subsubsection{Numerical tests}

We have created a cluster of 150 test particles, randomly sampled from
a Gaussian distribution in action-angle space, defined by $\sigma_J =
0.2 \kpc\kms$ and $\sigma_\theta = 10^{-3}$ radians. This cluster was
placed on the orbit I1 from \tabref{mech:tab:orbits}, which has
an apocentre radius of $13\kpc$ and a pericentre radius of
$\sim 5\kpc$, and is illustrated in \figref{mech:fig:sphericalorbits}.

\begin{figure}[\figplaceopts]
  \centerline{
    \includegraphics[width=\figureshrink\hsize]{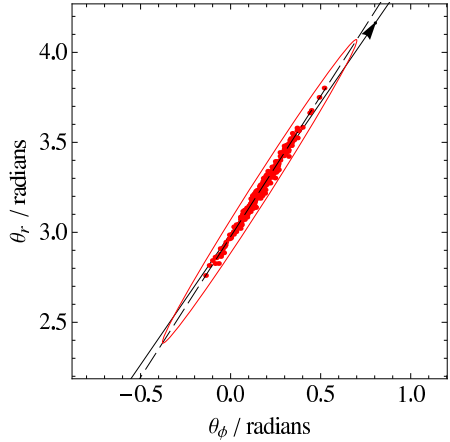}
  }
  \caption[Angle-space distribution of a stream formed from the orbit
I1]
{The angle-space distribution of a cluster of test particles, 
    evolved on orbit I1 (\tabref{mech:tab:orbits}) in the isochrone
    potential of \tabref{mech:tab:potentials} for 94 complete azimuthal
    circulations. The particles are shown at apocentre. The frequency
    vector $\vO_0$ of the progenitor orbit is shown with an arrowed black line.
    The stream is slightly
    misaligned with $\vO_0$; the dashed line is a
    straight line fit to the positions of the test particles, and
    clearly demonstrates this misalignment. Also plotted with a red
    solid line is the image in angle-space of a circle in action-space,
    mapped by $\hessian$. The shape and
    orientation of the image reflects the $\lambda_n$ and
    $\eigen_n$ of $\hessian$ for this orbit. The ellipse is
    clearly misaligned with the underlying orbit, but is perfectly
    aligned with the stream particles.
}
  \label{mech:fig:isochrone-angles}
\end{figure}

The cluster was released at apocentre, and evolved for 94 complete
azimuthal circulations, equal to a period of $t=22.75\Gyr$.
\figref{mech:fig:isochrone-angles}
shows the angle-space configuration of the particles
after this time. The arrowed, black line shows the orbit of the underlying
cluster, $\vO_0$. The dashed line shows a straight line fit to
the distribution of particles, which is clearly misaligned
with the black line. We further note that, unlike
the stream in the Kepler potential shown in 
\figref{mech:fig:kepler-angles}, this stream is clearly
also increasing in width. 

We can predict the shape of this distribution precisely. Plotted
as a red ellipse in \figref{mech:fig:isochrone-angles} is the angle-space image of a
circle in action-space, having been mapped by $\hessian$. After some time, the
angle-space distribution of an isotropic cluster of test particles
should take the form of a scaled version of this image. We see that
the image and the particle distribution are indeed comparable, and that
the dashed line is perfectly aligned with the principal axis of the
image.

\begin{figure}[\figplaceopts]
  \centerline{
    \includegraphics[width=\doublefigshrink\hsize]{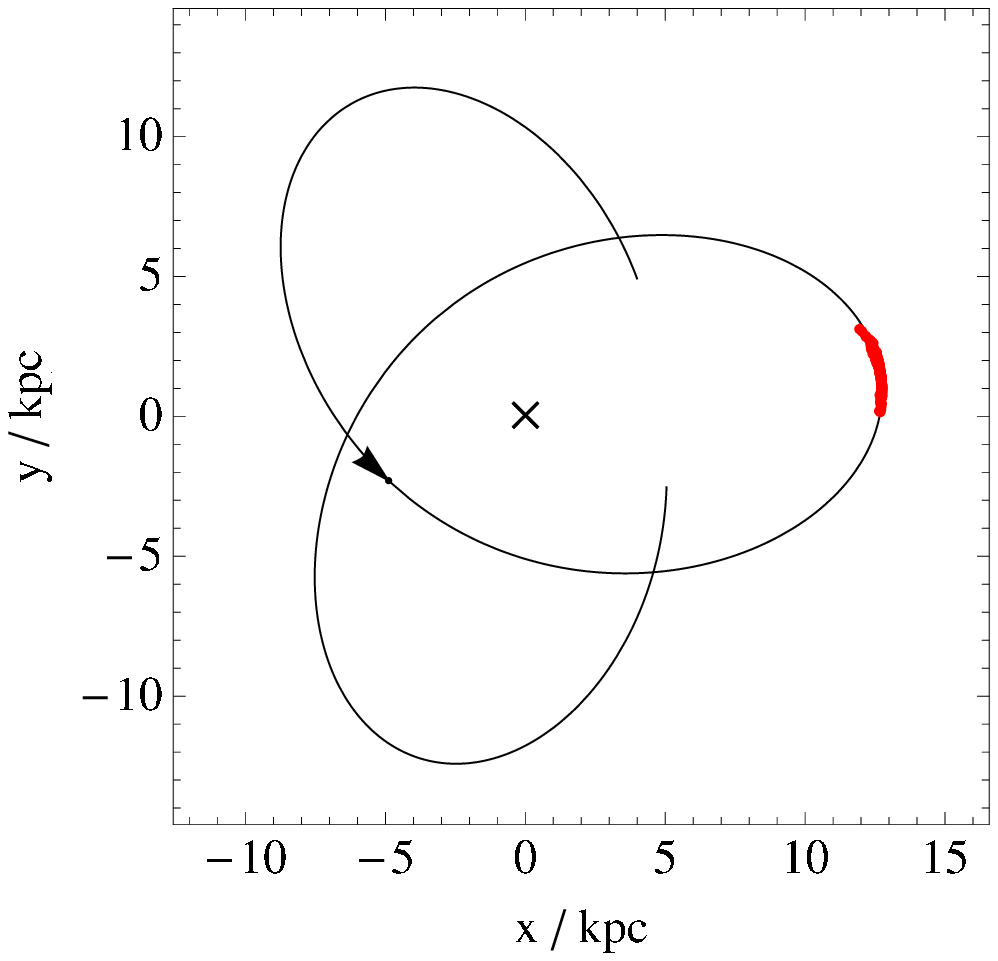}
    \qquad
    \includegraphics[width=\doublefigshrink\hsize]{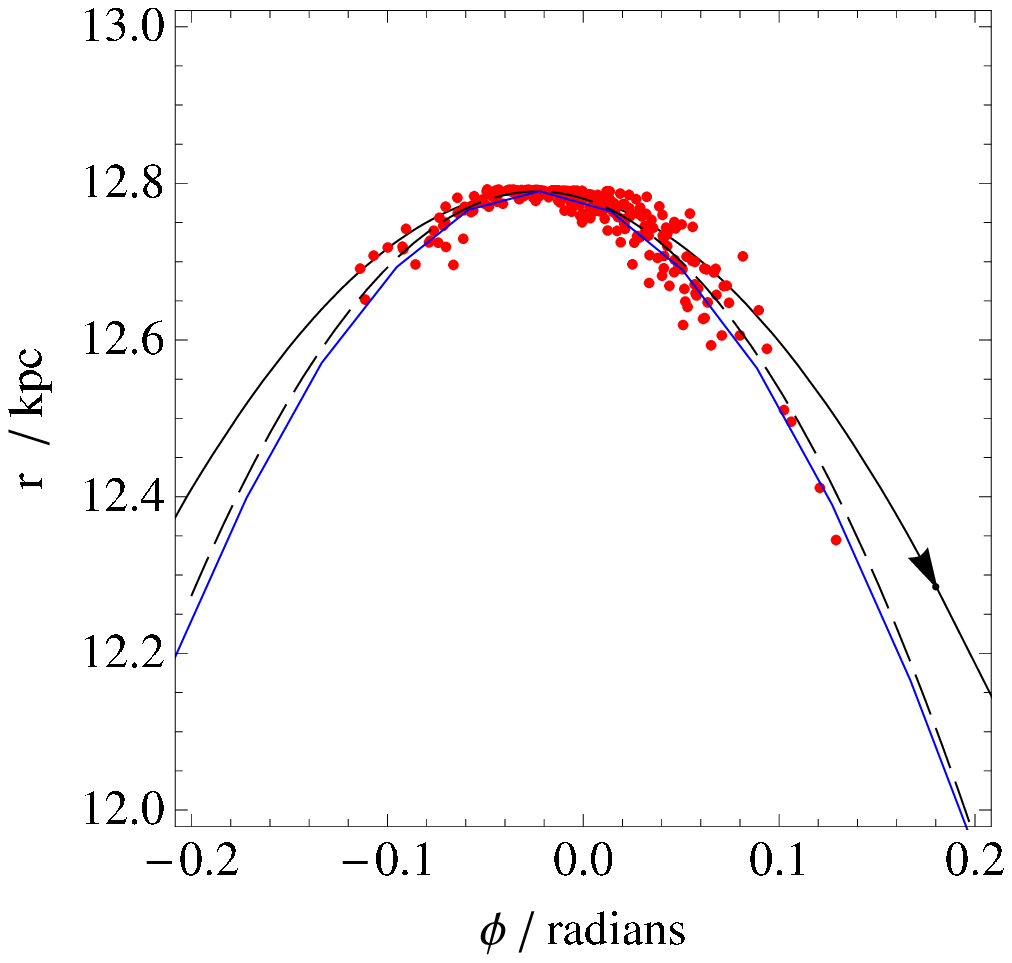}
  }
  \caption[Real-space configuration for the stream
    shown in \figref{mech:fig:isochrone-angles}]
{ Real-space configuration for the stream of test particles
    shown in \figref{mech:fig:isochrone-angles}. The arrowed black
    line shows the trajectory of the progenitor orbit.  The left panel
    shows an overview of the particles and the orbit; the cross marks
    the centre of the potential. The right panel shows a zoomed-in
    view of the particles, and is plotted in polar coordinates. The
    stream formed by the dots falls away in radius faster than does
    the orbit, in both forwards and backwards directions: the stream
    clearly does not follow the progenitor orbit. The dashed line is a
    quadratic curve least-squares fitted to the stream, which shows
    that the stream has a substantially greater curvature than does
    the underlying orbit. The solid blue line is the track predicted
    in \secref{mech:sec:mappingtests} from the dashed line in
    \figref{mech:fig:isochrone-angles}: it agrees perfectly with
    the stream.}
  \label{mech:fig:isochrone}
\end{figure}

How does this misalignment manifest itself in real-space? 
\figref{mech:fig:isochrone} shows the real-space configuration
of the cluster at the end of the simulation. The left panel shows an overview
of the cluster in the orbital plane, while the right panel
shows a close-up view of the cluster. The orbit of the cluster
is drawn with a solid black line. In the right panel, the particles
have been least-squares fitted to a quadratic curve, shown as a
dashed line.

The progenitor orbit is clearly a poor representation of the
stream. Although the orbit passes through the centroid of the stream,
as expected, the curvature of the orbit is too low to match the stream
adequately. Thus, the small misalignment in angle-space is manifest as
a significant change in stream curvature at apocentre.

We conclude that, in this case, the track of the stream makes a poor proxy
for the orbit of its stars, and that in general, streams cannot be relied
upon to delineate orbits. In \secref{mech:sec:fitting} we will
highlight the errors that can be made in attempting to optimize potential
parameters based on the assumption that they do delineate orbits.
Firstly, however, we will discuss the details involved in the mapping
of streams from action-angle space to real-space, in order that we 
may properly predict the track of a stream.

\section{Mapping streams from action-angle space to real space}
\label{mech:sec:mapping}

\subsection{Non-isotropic clusters}
\label{mech:sec:nonisotropic}

Up until this point, we have considered our streams to form from a
cluster of particles that is isotropic in $\vJ$, resulting in a stream
that is perfectly aligned in angle-space with the principal direction
of $\hessian$.

It is not obvious that this is a fair assumption. The structure in
angle-space, given by equations \blankeqref{mech:eq:angle_t}
and \blankeqref{mech:eq:d-dot-j}, is linearly dependent
upon the action-space distribution that generates it. By properly
choosing that distribution, we can create streams of arbitrary shape
in angle-space\footnote{Systems in which one of the $\lambda_i$ is
  null are the exception to this statement. Structures in such systems
  are limited to that subset of angle-space which is spanned by the
  non-null eigenvectors of $\hessian$. The Kepler potential is one
  such system; all structures are mapped to a line pointing precisely
  along the frequency vector $\vO_0$.}. Clearly, nature does not create
clusters with arbitrary action-space distributions, so
arbitrary-shaped streams do not emerge. But what kind of action-space
distribution should be considered reasonable, that we may think of how
it maps?

In \secref{mech:sec:actions} below, we will investigate the action-space
distribution of real clusters, by means of N-body simulation. Here, we
only require a qualitative understanding, in order to guess what kind
of action-angle structures we should learn to map. Since the action-space
distribution arises from the random motion of stars within the cluster,
there is unlikely to be much complex structure. We
will therefore assume that the distribution is ellipsoidal. But what
should be the axis ratio of this ellipse?

Consider a cluster with isotropic velocity dispersion, $\sigma$, that
is on an orbit with apocentre $r_\apo$ and pericentre $r_\peri$, where it is
tidally disrupted. In our spherical system, the radial action is
given by the closed integral
\begin{equation}
J_r = {1 \over 2\pi} \oint p_r \, dr,
\label{mech:eq:jr}
\end{equation} 
where the integration path is one complete radial oscillation.
Now consider a particle whose radial momentum $p_r$ differs from that of the
cluster average by $\delta p_r \sim \sigma$. 
% The difference in the particle's orbital
% energy $E$ is then $\delta E = p \, \delta p_r$. Since this $\delta E$ is
% conserved, if the particle's total momentum $p$ does not vary greatly
% along its orbit, then $\delta p_r$ will be roughly constant along the orbit.
We can take a finite difference over \eqref{mech:eq:jr} and thus obtain an expression for the
the difference in radial action between the particle and the cluster
\begin{equation}
\delta J_r \sim {1 \over \pi} \delta p_r \Delta r \sim \sigma \Delta r,
\label{mech:eq:deltajr}
\end{equation}
where $\Delta r = (r_\apo - r_\peri)$ is the amplitude of the radial oscillation.
Now consider another particle, whose azimuthal velocity differs from that of the
cluster by $\delta v_t \sim \sigma$. The difference in angular momentum
between this particle and the cluster is
\begin{equation}
\delta L \sim r_p \, \delta v_t \sim r_p \sigma,
\end{equation}
where we have performed our calculation at pericentre,
because that is where the cluster is stripped.
For the purposes of this section, we are interested in the relative size
of the spread in radial action $\Delta J_r$ and the spread in
angular momentum $\Delta L$ for a disrupting cluster. We see that is
approximately given by
\begin{equation}
{\Delta J_r \over \Delta L} \sim {\Delta r \over \pi r_p}.
\label{mech:eq:djr/dl}
\end{equation}
Although this ratio will take on every value between $(0, \infty)$ as
we move from a circular orbit to a plunging one, for the orbits likely
to be occupied by stream-forming clusters, it will typically be of order
unity.

\begin{figure}[\figplaceopts]
  \centerline{
    \includegraphics[width=\doublefigshrink\hsize]{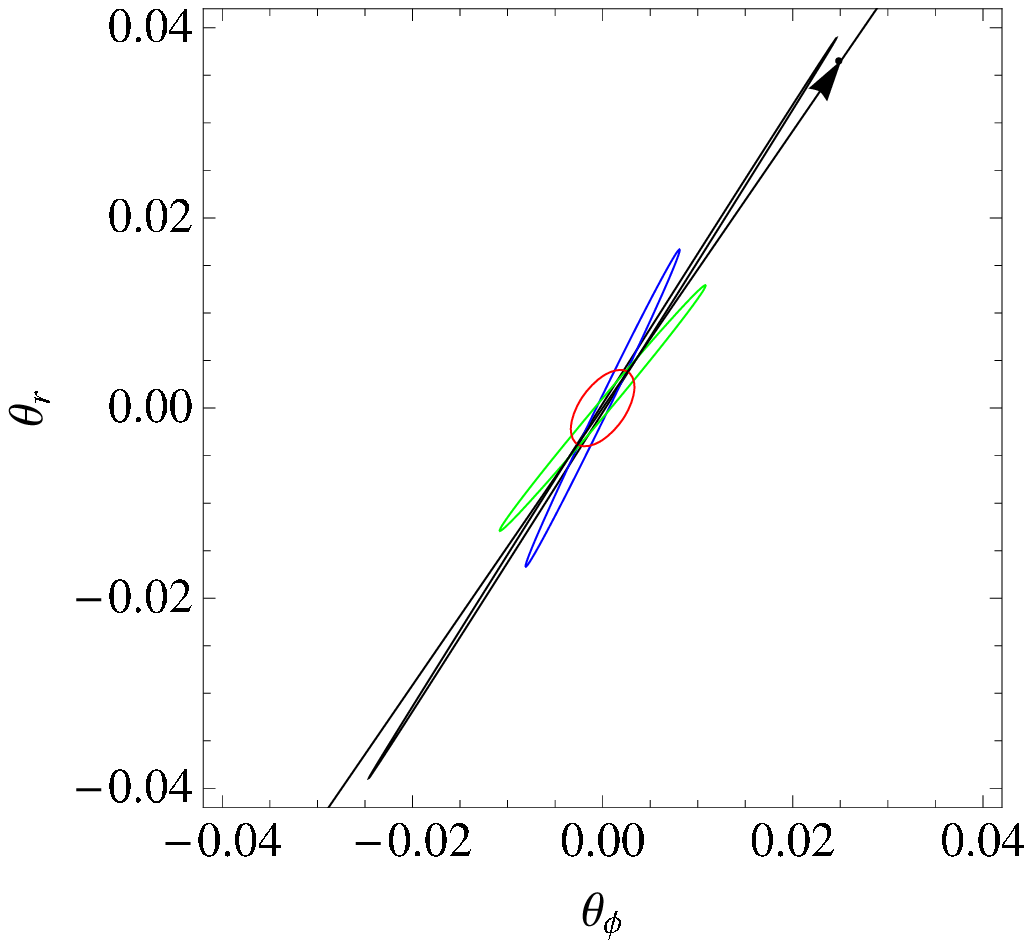}
    \qquad
    \includegraphics[width=\doublefigshrink\hsize]{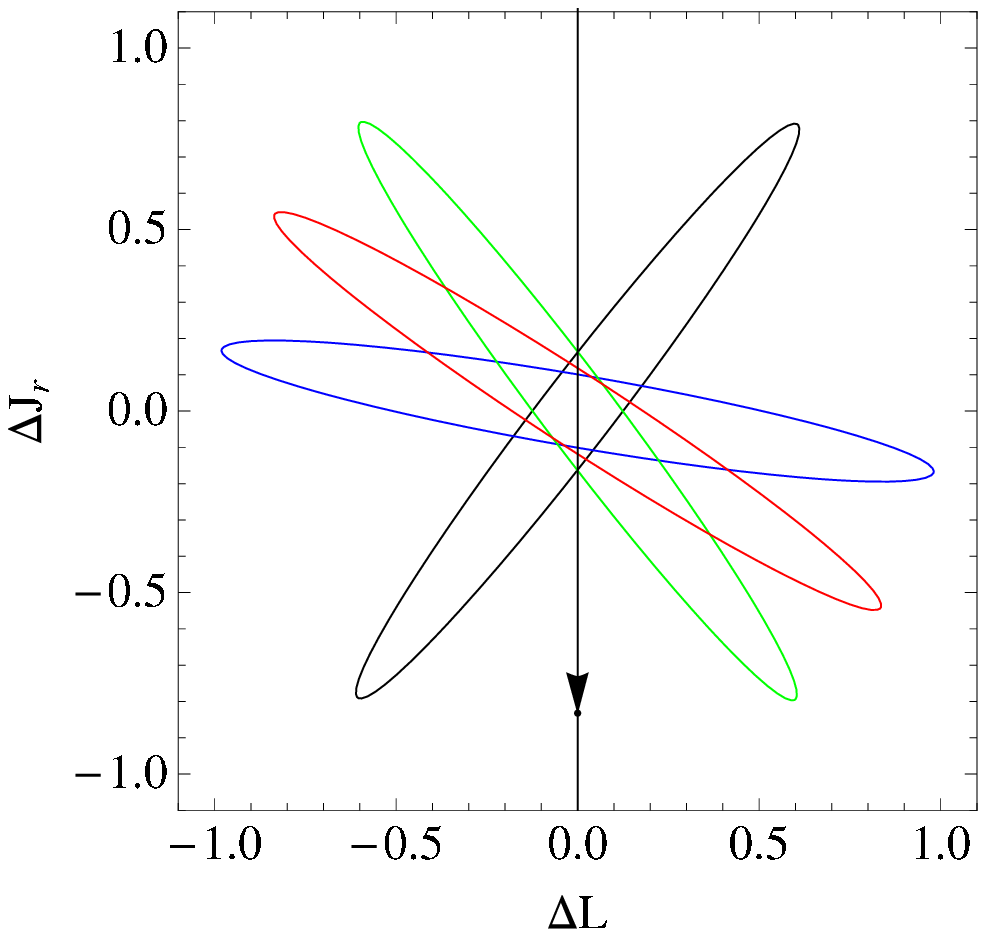}
  }
  \caption[Mapping of action-space to angle-space for the orbit I1]
{The right panel shows a selection of ellipses in action-space,
    each of axis ratio 10, but oriented in different directions.
    The left panel shows the image in angle-space that results from
    mapping each of the action-space ellipses with the Hessian $\hessian$,
    calculated for the orbit I1 in the isochrone potential of 
    \tabref{mech:tab:potentials}. In the left panel, the arrowed black
    line is the frequency vector $\vO_0$; in the right panel, the arrowed
    black line is the inverse map of the frequency vector, $\hessian^{-1}
    \vO_0$.
    We see that regardless of the shape in action-space, the 
    mapped images are all elongated and roughly aligned with the
    orbit, although the alignment is generally not perfect.
    In this example, the misalignment of the red and green images
    is about $10\deg$.
  }
  \label{mech:fig:isochrone-ellipses}
\end{figure}

The right panel of \figref{mech:fig:isochrone-ellipses} shows a set of
ellipses, with axis ratio 10, placed at various orientations in
action-space. The left panel of \figref{mech:fig:isochrone-ellipses}
shows the images of these ellipses in angle-space, following mapping
by $\hessian$ when evaluated on I1. Note that all the images are both
elongated and roughly oriented towards the principal direction.  We
conclude that the images of most action-space ellipses under this
map---and thus, most streams formed in this potential---would be highly elongated and oriented
to within a few degrees of the principal direction, which is itself
oriented to within a few degrees of the frequency vector $\vO_0$.

Since the ratio of the eigenvalues for this orbit is $\sim 17$, it is
not possible to produce an image in angle-space that is not elongated
towards the principal direction, by mapping an action-space ellipse of
axis-ratio 10.  We note from \figref{mech:fig:isochrone-hessian} that,
for this potential, the ratio of eigenvalues does not vary much, and
nor does the principal direction stray from $\vO_0$ by more than a few
degrees. We therefore conclude that reasonable action-space
distributions will always result in the formation of streams in this
potential, and that such streams will always be oriented in angle-space to
within a few degrees of $\vO_0$.

\subsection{The mapping of action-angle space to real space}
\label{mech:sec:map}

\begin{figure}[\figplaceopts]
  \centerline{
    \includegraphics[width=\figureshrink\hsize]{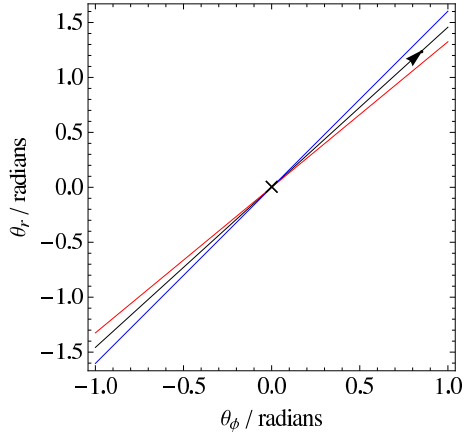}
  }
  \caption[Three trajectories in angle-space, derived from the orbit I1]
{Three trajectories in angle-space. The black line is the
    trajectory of orbit I1, in the
    isochrone potential of \tabref{mech:tab:potentials}. The red
    line has a frequency ratio $\Omega_r / \Omega_\phi$ that is $10
    \percent$ lower than I1. Conversely, the blue line has a
    frequency ratio that is $10 \percent$ higher than I1.  We
    note that the red line has retarded radial phase (relative to the
    black line) on the leading
    tail, and advanced radial phase on the trailing tail. Conversely,
    the blue line has advanced radial phase in the leading part, and
    retarded radial phase in the trailing part.
}
  \label{mech:fig:mapping-lines}
\end{figure}

\figref{mech:fig:mapping-lines} shows three trajectories in angle
space. The black line is the trajectory of I1 in the
isochrone potential given in \tabref{mech:tab:potentials}. The red
line has a frequency ratio $\Omega_r / \Omega_\phi$ that is $10 \percent$
lower than the black line, and is therefore rotated from it by approximately
$2.9\deg$.  Conversely, the blue line has a frequency ratio that is
$10 \percent$ higher than the black line, and is therefore rotated from it by
about $2.5\deg$.

The red and blue lines were chosen to represent likely streams in angle-space
that could form in the isochrone potential, given the results of the
previous section. We note that the red line has retarded radial phase (relative to the
black line) on the leading tail, and advanced radial phase on the
trailing tail. Conversely, the blue line has advanced radial phase in
the leading part, and retarded radial phase in the trailing part.

%freq: {0.178434,0.12245}
%*1.1 -> 58.04 + 2.5 deg
%*1.0 -> 55.54
%*0.9 -> 52.67 - 2.87 deg

\begin{figure}[\figplaceopts]
  \centerline{
    \includegraphics[width=\doublefigshrink\hsize]{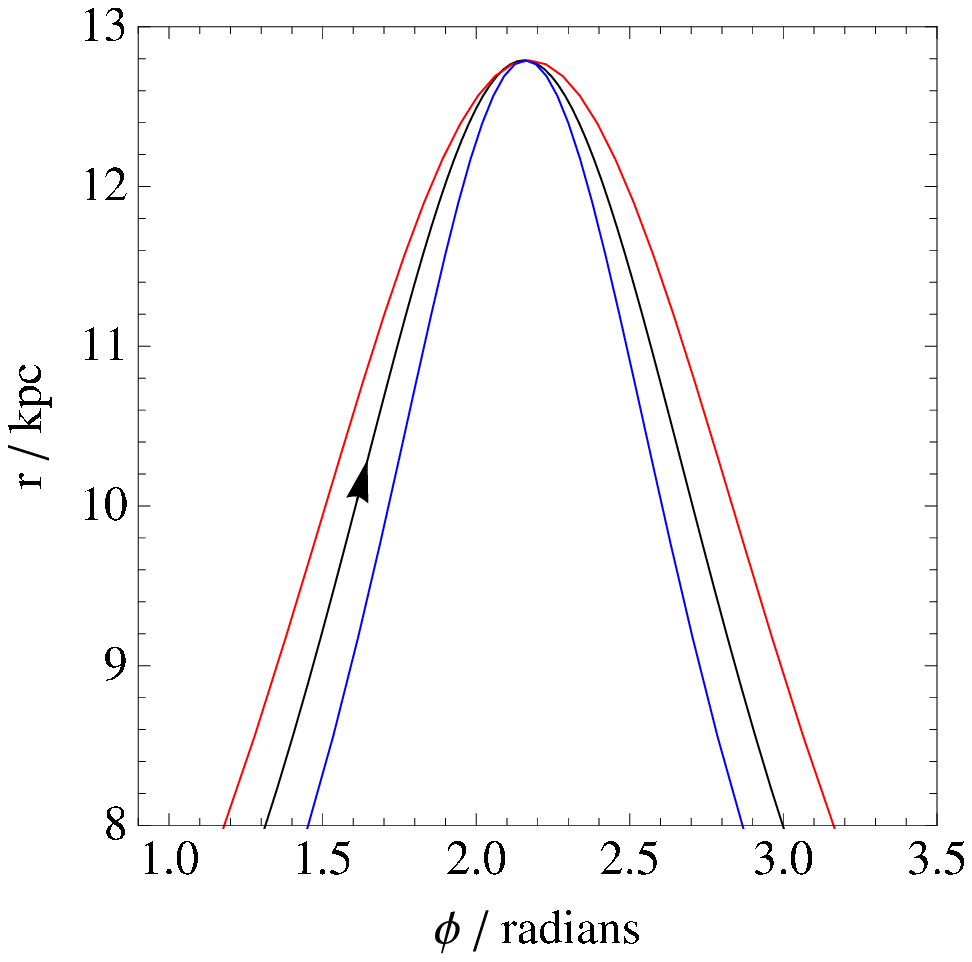}
    \qquad
    \includegraphics[width=\doublefigshrink\hsize]{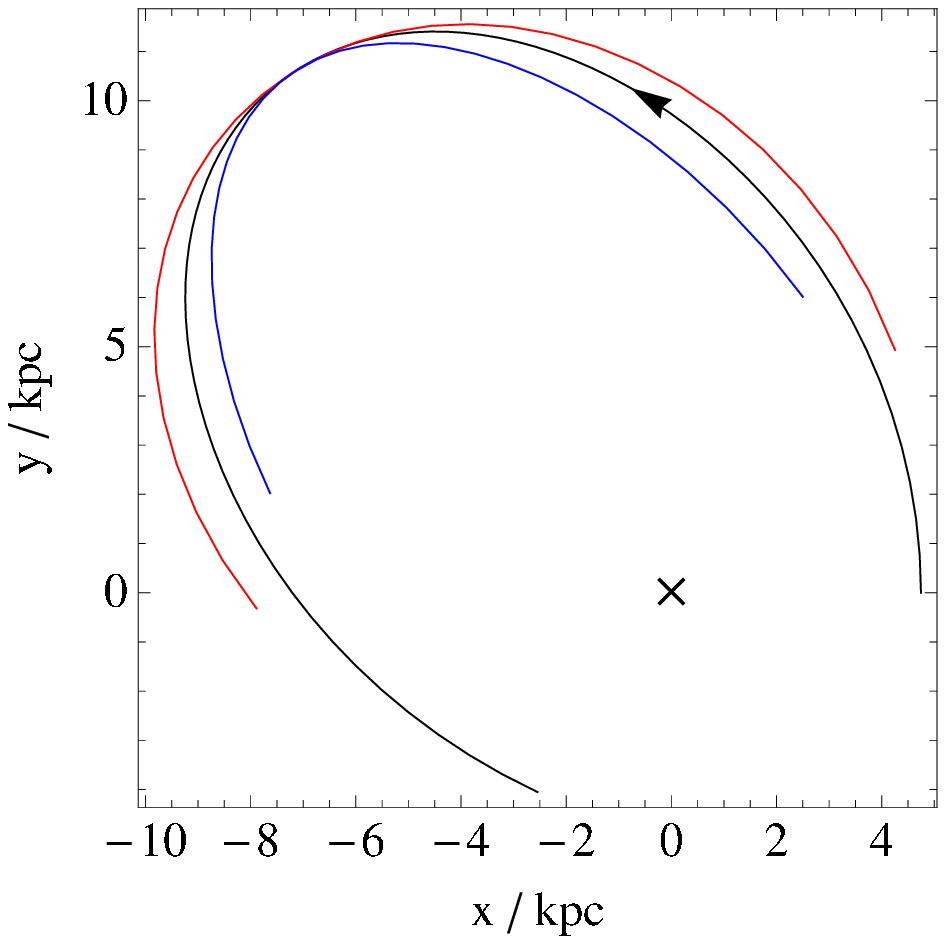}
  }
  \caption[Real-space trajectories of the lines shown in
    \figref{mech:fig:mapping-lines}, phase-matched near apocentre]
{ Plots of the real-space trajectories of the lines shown in
    \figref{mech:fig:mapping-lines}, phase-matched near apocentre.
    Left panel: the trajectories plotted in polar coordinates. Right
    panel: the trajectories plotted in Cartesian coordinates; the
    centroid of the potential is marked with a cross.
}
  \label{mech:fig:mapping-apo}
\end{figure}

How do these lines map into real-space? \figref{mech:fig:mapping-apo}
shows the real-space curve obtained from the lines in
\figref{mech:fig:mapping-lines}, having chosen the point of
intersection such that the lines are phase-matched near apocentre.
The mapping is done by solving numerically for the real-space roots of the
equations that relate action-angle variables to real-space
coordinates: for spherical potentials, the appropriate equations
are given in \S3.5.2 of \cite{bt08}.
The curves in \figref{mech:fig:mapping-apo} have been
drawn by assuming that all points along each line have the same $\vJ$.
Thus, this figure represents the real-space curves of streams oriented
in angle-space according to \figref{mech:fig:mapping-lines}, but
formed from clusters of vanishingly small $\Delta\vJ$.

In \figref{mech:fig:mapping-apo} we see that the red line has
systematically lower curvature than the black line. Conversely, the
blue line has systematically greater curvature than the black line.
We understand this, because the red curve has retarded radial phase
on the leading tail, and advanced radial phase on the trailing tail,
and thus is flattened with respect to the orbit. Similarly,
the blue line has advanced radial phase on the leading tail, and
retarded radial phase on the trailing tail, and thus appears
curved with respect to the orbit.

\begin{figure}[\figplaceopts]
  \centerline{
    \includegraphics[width=\doublefigshrink\hsize]{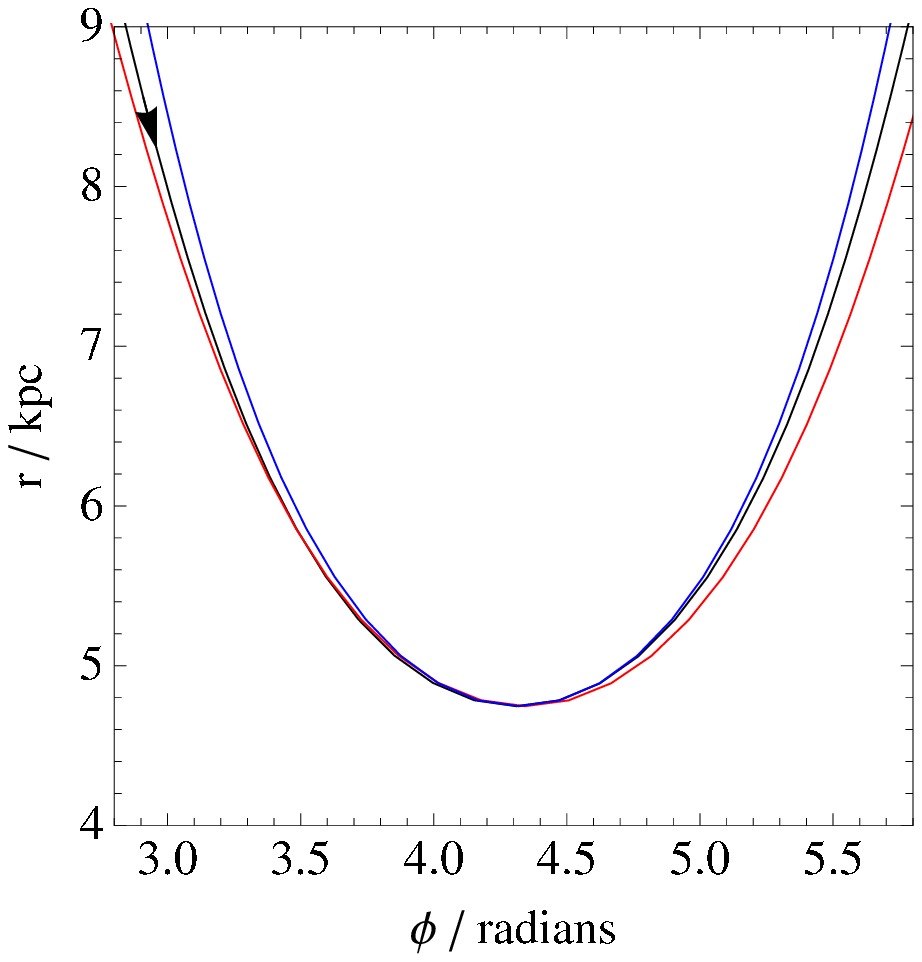}
    \qquad
    \includegraphics[width=\doublefigshrink\hsize]{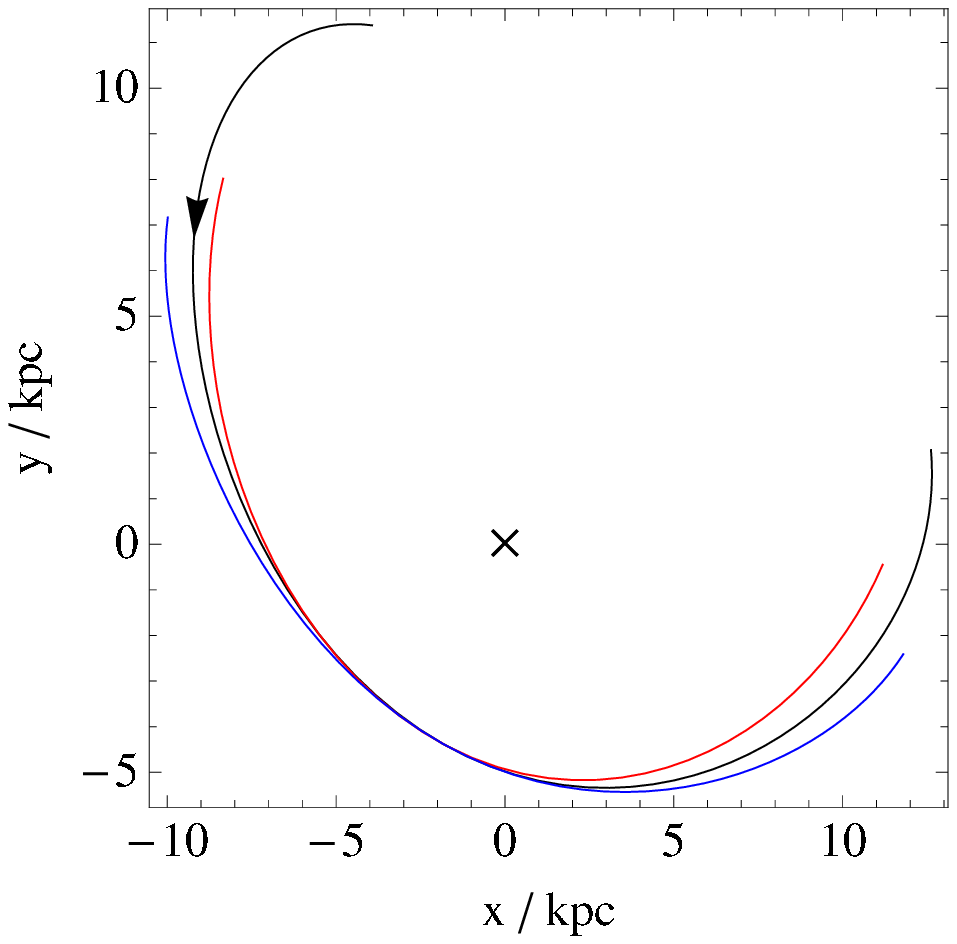}
  }
  \caption[As \figref{mech:fig:mapping-apo}, but phase-matched near pericentre]
{
    Similar to \figref{mech:fig:mapping-apo}, but with the trajectories
    phase-matched near pericentre. The red line is again observed to have
    systematically lower curvature than the black line, and the blue
    line is again observed to have systematically greater curvature than
    the black line.}
  \label{mech:fig:mapping-peri}
\end{figure}

\begin{figure}[\figplaceopts]
  \centerline{
    \includegraphics[width=\doublefigshrink\hsize]{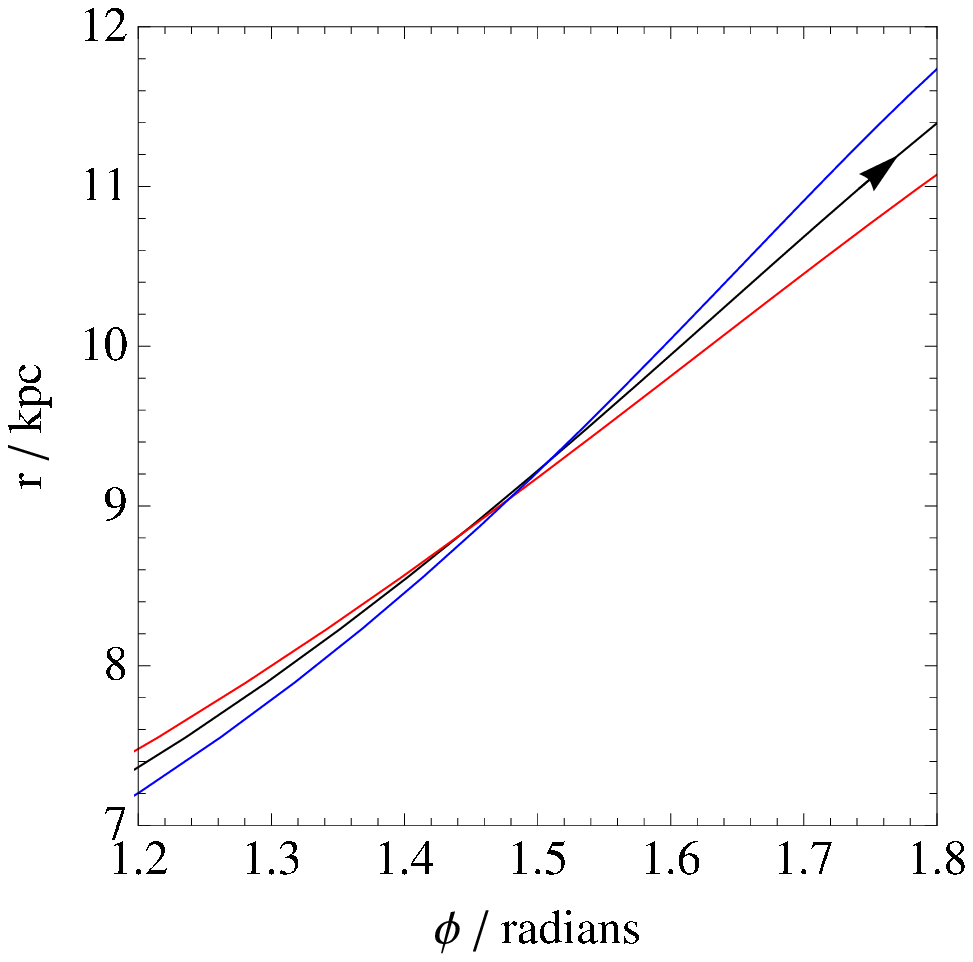}
    \qquad
    \includegraphics[width=\doublefigshrink\hsize]{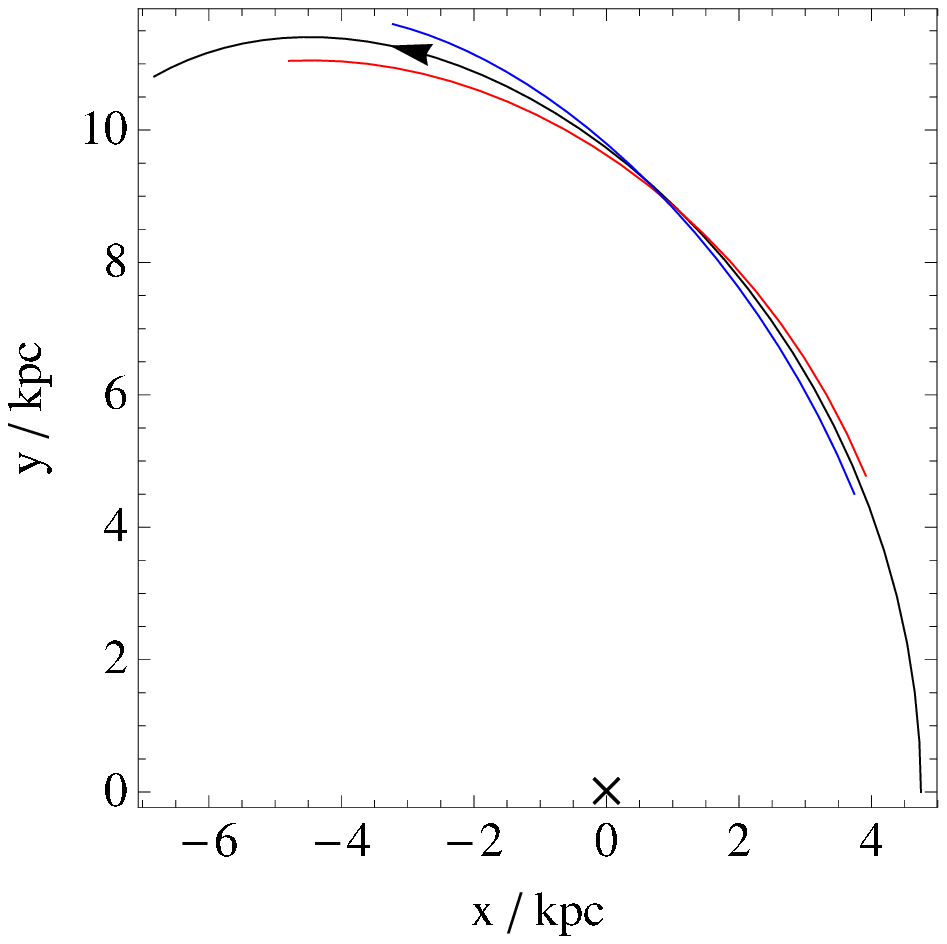}
  }
  \caption[As \figref{mech:fig:mapping-apo}, but
    phase-matched at a point well away from apsis]{
    Similar to \figref{mech:fig:mapping-apo}, but with the trajectories
    phase-matched at a point well away from apsis. In this case, a misalignment between
    the stream and $\vO_0$ in angle-space is expressed as a
    real-space misalignment, rather than a curvature error as at apsis.
    We note the similarity between the left panel and \figref{mech:fig:mapping-lines}.
}
  \label{mech:fig:mapping-mid}
\end{figure}

\figref{mech:fig:mapping-peri} shows the same lines, but now
phase-matched at pericentre. Similarly to
\figref{mech:fig:mapping-apo}, the red line again appears flattened
with respect to the orbit, and the blue line appears curved with
respect to the orbit. \figref{mech:fig:mapping-mid} also shows the same
lines phase-matched at a point well away from apsis. In this case, a
misalignment between the stream and $\vO_0$ in angle-space is
expressed as a real-space misalignment, rather than a curvature error
as at apsis.  We note the similarity between the left panel and
\figref{mech:fig:mapping-lines} which occurs because, unlike at apsis,
the mapping between angle-space and plane polar coordinates is
relatively undistorted near this point.

\subsection{Numerical tests}
\label{mech:sec:mappingtests}

\begin{figure}[\figplaceopts]
  \centering{
    \includegraphics[width=\figureshrink\hsize]{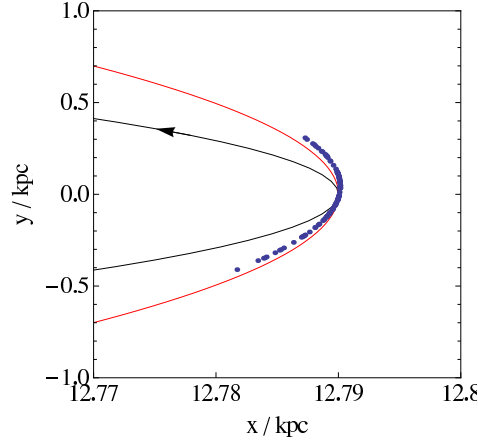}
  }
  \caption[A stream of test particles to check that the mapping from angle-space
to real-space is accurate]{Real-space configuration of a 100 particle test stream,
    intended to check that our mapping of angle-space to real-space is
    accurate. The stream has an action-space distribution chosen to
    map under $\hessian$ to a line in angle-space oriented
    perpendicularly to $\vO_0$.  The stream was released on orbit I1
    in the isochrone potential of \tabref{mech:tab:potentials} and
    integrated for $23.6\Gyr$. The black curve is the orbit I1, while
    the red curve is the predicted stream track. The red curve is
    clearly a far better match to the particles than the black curve; the slight phase
    mismatch between the apocentre of the particles and the red curve
    is due to numerical error in the simulation process.}
\label{mech:fig:dot-check}
\end{figure}

How can we be sure that our calculation of stream tracks from lines in
angle-space is correct?  We constructed a cluster of 100 test
particles by randomly sampling a segment of a line in action-space,
chosen such that its angle-space image would be a line oriented
perpendicularly to $\vO_0$.

A point on the orbit I1 was chosen by integrating backwards $23.6\Gyr$
from apocentre. The cluster was placed at this point, released, and
evolved forwards for $23.6\Gyr$.  \figref{mech:fig:dot-check} shows
the real-space configuration of the resulting stream at the end of
this time. Also plotted are the orbit I1, as a black curve, and the
stream track predicted from the line-segment in action space, as a red
curve. The stream and the prediction are seen to agree
almost perfectly\footnote{The slight phase mismatch between the apocentre
  location of the particles and the apocentre location of the
  predicted track is due to finite numerical precision in the
  translation between action-angle variables and position-velocity
  coordinates when calculating the initial conditions for the
  simulation.}.

We may further test our apparatus by predicting the track of the stream
shown in \figref{mech:fig:isochrone}. The blue line in the right
panel of that figure shows the stream track predicted from a line
in angle-space that is oriented precisely along the principal
direction of $\hessian$ when evaluated for the orbit I1. The blue
line is a much better match to both the stream data and the fitted
line than is the orbit.

\subsection{Are trajectories insensitive to small changes in $\vJ$?}
\label{mech:sec:trajectory-j}

So far, all the real-space tracks we have computed from 
streams in angle-space have assumed that the corresponding
action for all points along that stream is that of the progenitor,
$\vJ_0$. 

This assumption is only strictly valid in the case of a
vanishingly small action-space distribution, and for asymptotically
large time since disruption of the cluster. If a mapping
into real-space from a line in angle-space is made under this
assumption, then a stream generated by a sufficiently broad
action-space distribution will not be accurately represented,
even though the representation in angle-space may be exact.
This is because the small changes in action that
give rise to the small changes in frequency also cause
small changes in real-space trajectory as well.

When computing a stream track in real-space, it is possible to
correct for this effect. By inverting \eqref{mech:eq:d-dot-j}
and eliminating $\Delta\vO$ using \eqref{mech:eq:angle_t}, we find
that for a star separated from a fiducial point on the stream by angle
$\delta\vT$, the difference in action between the star and the fiducial
point, $\delta\vJ$ is given by
\begin{equation}
\delta\vJ = {1 \over t_d} \hessian^{-1} \, \delta\vT,
\label{mech:eq:correction}
\end{equation}
where $t_d$ is the time since the star and the fiducial point were
coincident.
We may therefore guess the correction $\delta \vJ$ for a star's true action
$(\vJ_0 + \delta \vJ)$ from its position in the stream, provided we know $t_d$.

We typically take the fiducial point to be the centroid of the stream,
following which we may assume $t_d$ to be the time since the first
pericentre passage of the cluster on its present orbit. Although this
assumption neglects the possibility that the star could have been torn
away during a subsequent pericentre passage, we note that during tidal
disruption, it is the fastest moving stars which become unbound. The
cluster core that remains after a pericentre passage therefore has
lower velocity dispersion. Stars subsequently torn from that cluster
will therefore have a smaller distribution in action-space.
Consequently, the stars with the largest $\delta\vT$ from the
centroid---i.e.~those for which the $\delta\vJ$ correction will be
most important---must have been torn away at the earliest time, and so
the assumption that $t_d$ equals the time since the first pericentre
passage remains good.

But just how important is this effect? For small changes in $J_r$, the
trajectory changes we discuss are expressed as changes in the radial
amplitude $\Delta r$, while the guiding centre radius $r_g$, which
is purely a function of $L$, is held constant. We
can estimate the magnitude of the effect as follows. 

Consider a cluster on an
orbit close to circular, whose radial action is given by \eqref{mech:eq:jr}.
Orbital energy $E$ is conserved, so close to apsis $r=r_0$, the
radial momentum $p_r$ is given according to
\begin{equation}
E = p_r(r)^2 + \Phi_{\rm eff}(r) = p_r(r)^2 + \Phi_{\rm eff}(r_0)
+ \left.{\d \Phi_{\rm eff} \over \d r }\right|_{r_0}(r - r_0),
\label{mech:eq:p-near-apsis}
\end{equation}
where we have defined the effective potential $\Phi_{\rm eff}(r)
= \Phi(r) + L^2/2r^2$. Since at apsis $p_r = 0$, then
\begin{equation}
E = \Phi_{\rm eff}(r_0),
\end{equation}
so from \eqref{mech:eq:p-near-apsis} we see
\begin{equation}
p_r(r) = \sqrt{-(r - r_0)\left.{\d \Phi_{\rm eff} \over \d r }\right|_{r_0}}
= \sqrt{(r-r_0)F_{\rm eff}(r_0)},
\end{equation}
where we have defined the effective force,
$F_{\rm eff}(r') = -\partial\Phi_{\rm eff}/\partial r |_{r'}$. 
If $(r_a,r_p)$ are apocentre and pericentre respectively, then we
see that
\begin{equation}
p_r(r) \simeq
\begin{cases}
\sqrt{|F_\eff(r_a)|(r_a - r)} & \text{if } r \simeq r_a, \\
\sqrt{|F_\eff(r_p)|(r - r_p)} & \text{if } r \simeq r_p. 
\end{cases}
\end{equation}
Hence, we might define the global approximation to $p_r$
\begin{equation}
\tilde{p_r}(r) = { \sqrt{\left|F_{\rm eff}\right|(r - r_p)(r_a - r)} \over
\sqrt{r_a - r_p}},
\label{mech:eq:prapprox}
\end{equation}
with $F_{\rm eff}$ a constant set equal to the value of $F_{\rm eff}(r)$ taken at apsis,
at both of which we assume it to take approximately the same value.
We note that
\begin{equation}
\int_{r_p}^{r_a} \sqrt{(r - r_p)(r_a - r)}\,\d r = {\pi \over 8} \Delta r^2,
\end{equation}
so that when combined with \eqref{mech:eq:prapprox}, \eqref{mech:eq:jr} becomes,
\begin{equation}
J_r = {1 \over 8}\sqrt{\left|F_{\rm eff}\right| \Delta r^3}.
\label{mech:eq:bodgejr}
\end{equation}
We can deduce the value of $F_{\rm eff}$ as follows. We note that
\begin{equation}
\Phi_\eff = \Phi + {L^2 \over 2r^2},
\end{equation}
and that its derivative is
\begin{equation}
{\d \Phi_\eff \over \d r} = {\d \Phi \over \d r} - {L^2 \over r^3}.
\end{equation}
If the rotation curve is relatively flat, then $F_\eff(r)$ evaluated
at $r_a \simeq r_g + \Delta r/2$ is
\begin{align}
F_\eff &= \left.{\d \Phi_\eff \over \d r}\right|_{r_a}
= {v_c^2 \over r} - {L^2 \over (r_g + {1 \over 2}\Delta r)^3}
\simeq {v_c^2 \over r_g}\left(1 - {\Delta r \over r_g}\right) - {L^2 \over r_g^3}
\left(1 - {3 \over 2}{\Delta r \over r_g}\right)\\
& = -\Delta r \left({v_c^2 \over r_g^2} + {3 L^2 \over 2 r_g^4}\right)
= -{5 v_c^4 \Delta r \over 2 L^2},
\end{align}
and similarly when evaluated at $r_p$, but with opposite sign.
\Eqref{mech:eq:bodgejr} then becomes
\begin{equation}
%J_r \simeq {1 \over 8}\Delta r^2 \, \sqrt{{v_c^2 \over r_g^2} + {3L^2 \over 2r_g^4}}.
J_r \simeq {1 \over 8}\sqrt{5 \over 2} {v_c^2 \Delta r^2 \over L}.
\label{mech:eq:jrchange}
\end{equation}
Differentiating the above expression, we find
\begin{align}
{\d J_r \over \d \Delta r} &= 
{1 \over 4}\sqrt{5 \over 2} {v_c^2 \Delta r \over L}\\
%{1 \over 4}\Delta r \, \sqrt{{v_c^2 \over r_g^2} + {3L^2 \over 2r_g^4}}.\\
%&=\sqrt{{1 \over 2} J_r \sqrt{{v_c^2 \over r_g^2} + {3L^2 \over 2r_g^4}}}.
&=\sqrt{{1 \over 2} \sqrt{5 \over 2} {J_r v_c^3 \over L}}.
\label{mech:eq:amplitude}
\end{align}
Hence, for a small change $\delta J_r$ we can estimate the corresponding change
in the radial amplitude, $\delta\Delta r$, which is likely to be a good estimate
for the positional error we would make in assuming that a star with
action $J_r + \delta J_r$ actually had action $J_r$.

\begin{figure}[\figplaceopts]
  \centering{
    \includegraphics[width=\doublefigshrink\hsize]{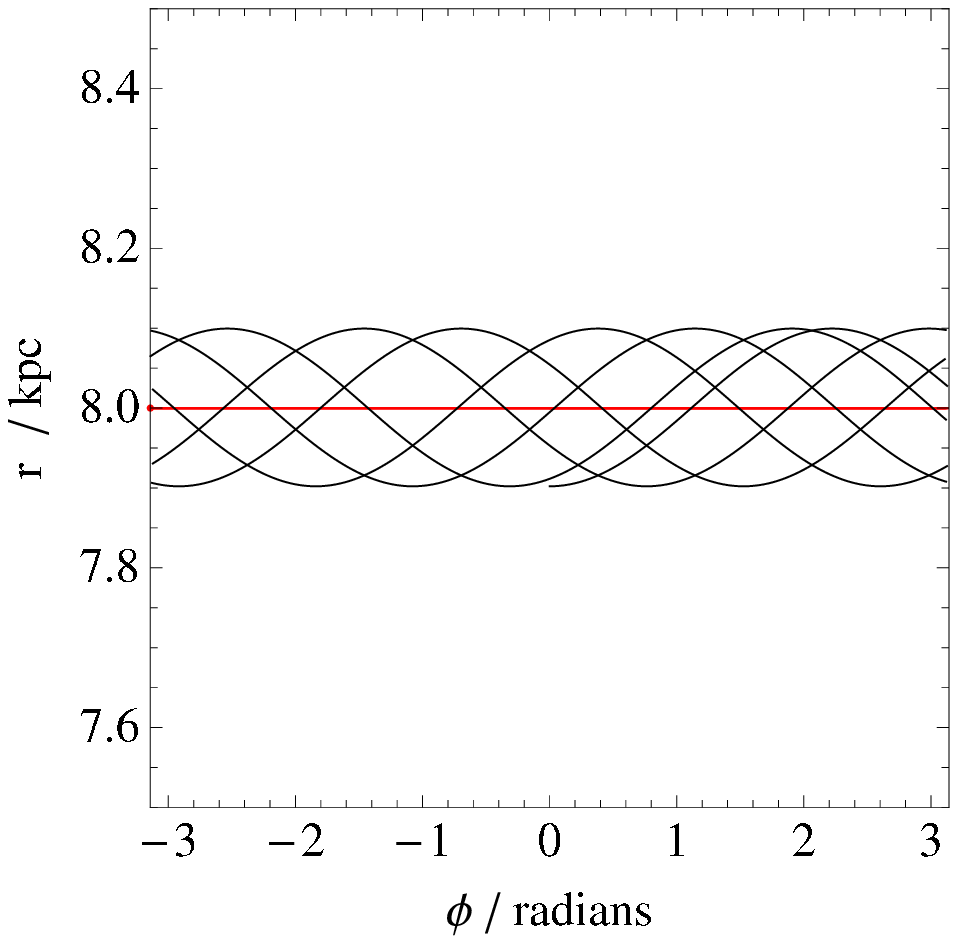}
    \includegraphics[width=\doublefigshrink\hsize]{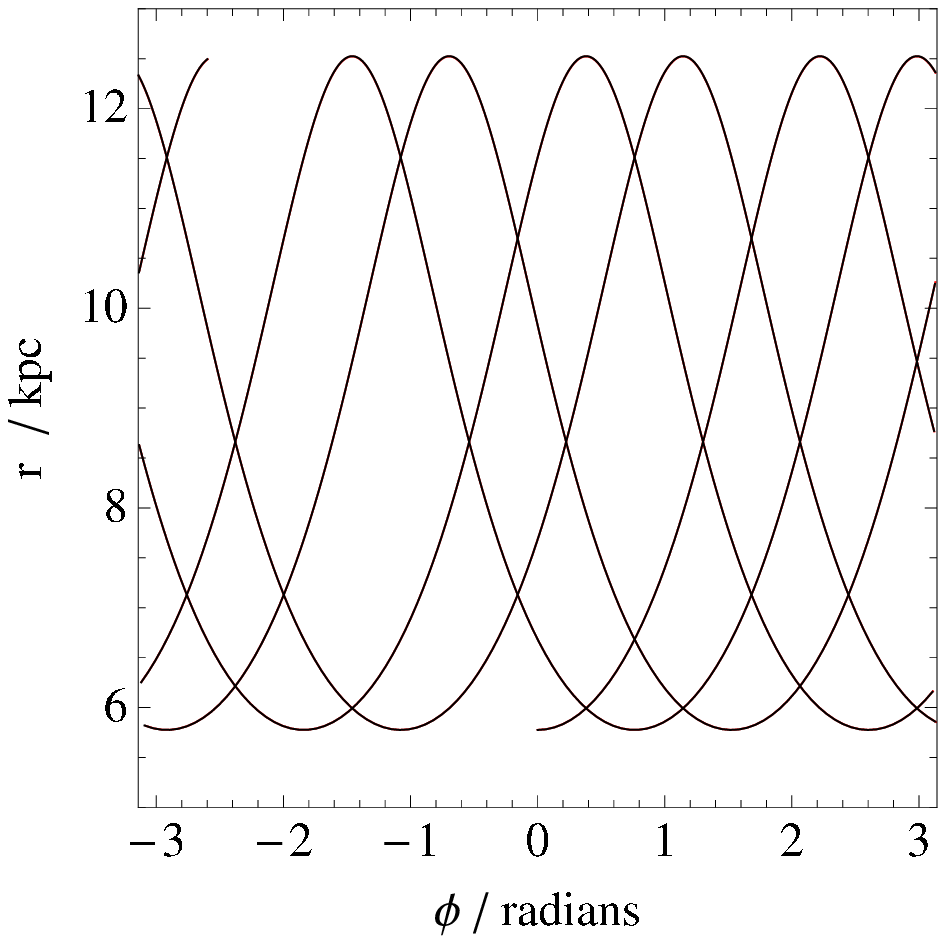}
  }
  \caption[Demonstration of the effect of changes in action $J_r$ on the real-space
    orbital trajectory of streams]
{Plots to demonstrate the effect of changes in action $J_r$ on the real-space
    orbital trajectory of streams. The left panel shows that a small change in
    $\jr$ ($\delta \jr = 0.21 \kms\kpc$) to a circular orbit
    (I2 in \tabref{mech:tab:orbits}) produces a change of $\delta\Delta r\sim 0.2\kpc$ in
    radial amplitude. The right panel shows the effect of the same
    perturbation $\delta\jr$ on an eccentric orbit (I3 in \tabref{mech:tab:orbits}) with the same angular momentum as
    the circular orbit, but with $\jr = 207\kms\kpc$. The effect on radial
    amplitude is so small as to be invisible in this plot.    
}
  \label{mech:fig:j-trajectory}
\end{figure}

We can confirm the predictions of the above equations numerically. The left panel of
\figref{mech:fig:j-trajectory} shows the real-space trajectory
of the circular orbit I2, with radius $r=8\kpc$, in the isochrone potential of
\tabref{mech:tab:potentials}. Also plotted is the trajectory
of an orbit that has identical $L$ to I2, but $J_r = 0.21\kms\kpc$.

Clearly, \eqref{mech:eq:amplitude} ceases to have meaning when faced
with orbits very close to circular, so we rely on the integral
form given by \eqref{mech:eq:jrchange} instead for our estimate.
\Eqref{mech:eq:jrchange} predicts $\Delta r = 0.19\kpc$ for this
perturbation from circular, which appears from the left panel of
\figref{mech:fig:j-trajectory} to be close to exact.

From \eqref{mech:eq:amplitude} we see that the magnitude of the effect
diminishes as $\delta\Delta r \sim 1/\Delta r \sim 1/\sqrt{J_r}$. The right panel of
\figref{mech:fig:j-trajectory} shows this to be the case. The panel
shows two trajectories in the isochrone potential: one for the orbit
I3, and one for the same orbit with $J_r$ incremented by
$0.1\percent$. \Eqref{mech:eq:amplitude} predicts a change $\delta\Delta
R \sim 8\pc$. Close inspection of the trajectories confirms an actual
$\delta\Delta r \sim 7\pc$, so the prediction is correct, but as is clear
from \figref{mech:fig:j-trajectory}, corrections of such magnitude are
negligible.

What follows is an estimate for the positional error made by incorrectly
guessing $L$. Consider again a cluster on an orbit close
to circular. The angular momentum of the cluster is related to the
guiding centre radius $r_g$ and the circular velocity $v_c$ by
\begin{equation}
L = v_c r_g.
\end{equation}
Consider now a star whose angular momentum is suddenly reduced by $\delta L$.
This star is now at apocentre, since its guiding centre radius has been 
reduced by
\begin{equation}
\delta r_g = {\delta L \over v_c},
\label{mech:eq:dLdrg}
\end{equation}
where we have assumed that the rotation curve is flat. Pericentre
radius will have been reduced by of order twice the change in
guiding centre radius, hence we may write
\begin{equation}
\delta \Delta r = {2 \delta L \over v_c}.
\label{mech:eq:l-amp}
\end{equation}
Thus, we can predict the change in radial amplitude $\delta\Delta r$ for
a small change in angular momentum $\delta L$, which is likely to be a good estimate
for the positional discrepancy we would encounter in assuming that a star with
angular momentum $L + \delta L$ actually had angular momentum $L$.

\begin{figure}[\figplaceopts]
  \centering{
    \includegraphics[width=\doublefigshrink\hsize]{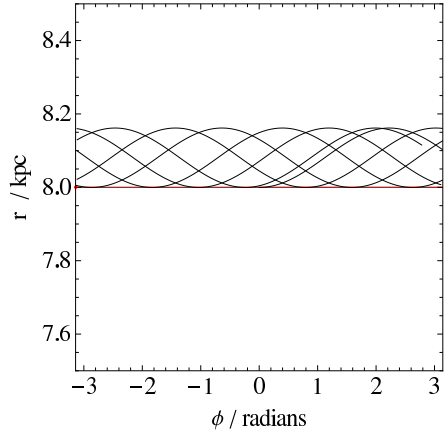}
  }
  \caption[Demonstration of the effect of changes in angular
    momentum $L$ on real-space orbital trajectory]
{Plot to demonstrate the effect of changes in angular
    momentum $L$ on real-space orbital trajectory. The trajectory of
    two orbits is plotted: the red line shows the trajectory of the
    orbit I2 (\tabref{mech:tab:orbits}), while the black line shows the same orbit, except with
    $L$ incremented by $1\percent$.  }
  \label{mech:fig:L-trajectory}
\end{figure}

Again, we can check the predictions of this expression numerically. 
\figref{mech:fig:L-trajectory} shows the trajectories of two orbits
in our isochrone potential. The red line is orbit I2, while
the black line is the same orbit, but with the angular momentum increased
by $1\percent$. \Eqref{mech:eq:l-amp} predicts $\Delta r \simeq 0.16\kpc$
for this change. \figref{mech:fig:L-trajectory} shows that this estimate
is close to exact.

The only example we have shown thus far where this effect could be of consequence
is the isochrone-potential stream shown in \figref{mech:fig:isochrone}. For this
cluster, \eqref{mech:eq:amplitude} predicts that a positional error of $\sim 3\pc$
would be accrued by assuming all stars have the same radial action. Similarly,
\eqref{mech:eq:l-amp} predicts that a positional error of $\sim 1.5\pc$ will
be accrued by assuming that all stars have the same angular momentum. These errors
are insignificant, so no corrections are required in this case.

However, we shall see in some later examples that the errors will not
be negligible. In these cases, the correction described by
\eqref{mech:eq:correction}, properly accounting for variation in
action down the stream, will be required. In such cases, we will assume that we
accurately know $t_d$, the time since the first pericentre passage. In general we would not
know $t_d$ accurately, although given $\vJ_0$ we could make
a reasonable guess as to its value. However, even a poor,
but finite, guess for the value of $t_d$ would likely produce
a more accurate real-space stream track than would assuming $\vJ=\vJ_0$
everywhere along the stream.

\section{The consequences of fitting orbits}
\label{mech:sec:fitting}

We confirmed in \secref{mech:sec:isochrone} that streams formed in the isochrone
potential do not necessarily delineate orbits. In this section, we briefly
examine the consequences should one attempt to constrain the parameters of
the potential by assuming that these streams {\em do} delineate orbits.

We construct an experiment as follows. We
first create three sets of pseudo-data, each containing a set of
phase-space coordinates corresponding to one of the three stream
tracks in \figref{mech:fig:mapping-apo}.  We choose to consider the
streams from \figref{mech:fig:mapping-apo} because they are at
apocentre, and many actual observed streams (e.g.~the Orphan stream of
\cite{orphan-discovery}) are discovered close to apocentre.

We now wish to measure the quality with which an orbit for a given set
of isochrone potential parameters can be made to fit the data.  Unlike
in Chapters \ref{chap:radvs} and \ref{chap:pms}, where orbits were
reconstructed from data for which only partial phase-space information
is available, for this exercise we have granted ourselves pseudo-data
with full and accurate positional and velocity information. This
simplifies considerably the matter of finding an orbit that is close
to the best fitting one.

We choose an orbit as follows. We first select a datum near the centroid
of the stream, and declare that our chosen orbit must pass directly through this
datum. Although it may be that some nearby orbit, one that does not pass directly through
this point, would make a better-fitting orbit, any such orbit must pass
very close to the selected datum, because it is close to the centroid. Thus, such an orbit
cannot be much better-fitting than one that passes directly through the datum.
Having chosen a datum, for a given set of potential parameters, an orbit is defined.

Having chosen our orbit, a goodness-of-fit
statistic $\chi^2$ is calculated as follows. For each datum in the stream, with
phase-space coordinate $\vect{w}_i$, a location along the orbit
$\vect{w}'_i$ is chosen that minimizes the square difference
\begin{equation}
(\vect{w}_i - \vect{w}'_i)^2.
\end{equation}
Having obtained the ${\vw'_i}$, the goodness-of-fit $\chi^2$ is defined by
\begin{equation}
\chi^2 = \sum_{i,j} {(w_{i,j} - w'_{i,j})^2 \over \sigma^2_j},
\label{mech:eq:chisq}
\end{equation}
where $j$ are the phase-space coordinates, and $\sigma_j$ is the
rms of $(w_{i,j}+w'_{i,j})/2$ over $i$. This $\chi^2$ statistic
provides a dimensionless measure of the phase-space distance
between the best-fitting orbit in a given potential, and the pseudo-data.

If the pseudo-data set were a sample of a perfect orbit in some potential,
we expect the value of $\chi^2$ to be exactly zero, when the correct
potential parameters are considered. As the potential parameters
are varied away from their true values, we expect the value of $\chi^2$
to rise, as the best-fitting orbit becomes a steadily worse
representation of the data. Hence, we expect minima in $\chi^2$ to be associated
with the potential parameters that are optimum, from the perspective
of fitting an orbit to the data. We seek such minima by plotting contours
of $\chi^2$ over a range of likely values for the potential parameters.

\subsection{Results}

Three clusters of 50 test particles were created, with small initial
angle-space distributions. The action-space distribution of each was a
segment of a line, $\Delta J = 0.2\kms\kpc$ long, and oriented such
that, after a long time, the angle-space distribution of each would
precisely match one of the three streams shown in
\figref{mech:fig:mapping-apo}.

Each cluster was placed on the orbit I1 in the isochrone potential of
\tabref{mech:tab:potentials}, and integrated forward for $77.04\Gyr$
to produce a thin stream about $10\kpc$ in length, that has its centroid at apocentre.
The streams are then similar to those shown in \figref{mech:fig:mapping-apo}.
We declare each one of these streams to be a pseudo-data set for
our experiment. 

\begin{figure}[\figplaceopts]
  \centering{
    \includegraphics[width=0.48\hsize]{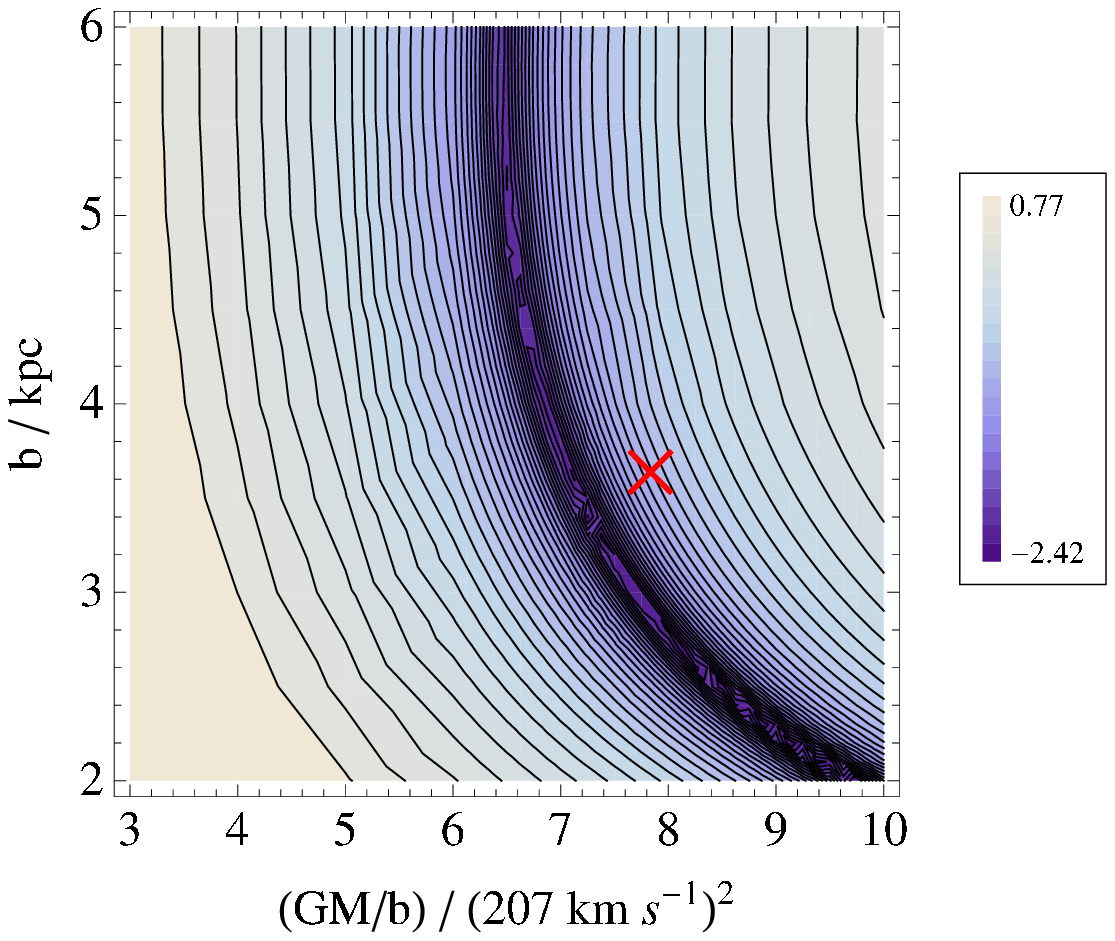}\\
    \includegraphics[width=0.48\hsize]{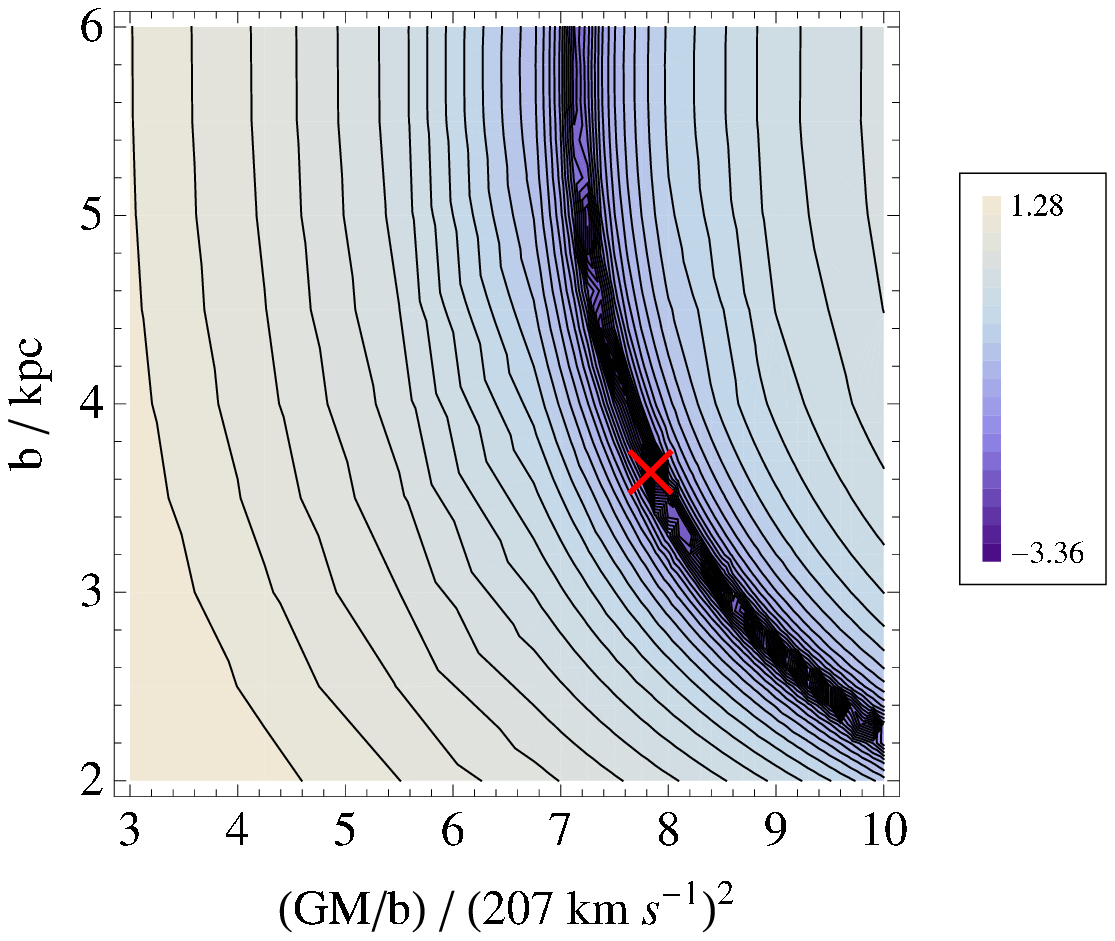}\\
    \includegraphics[width=0.48\hsize]{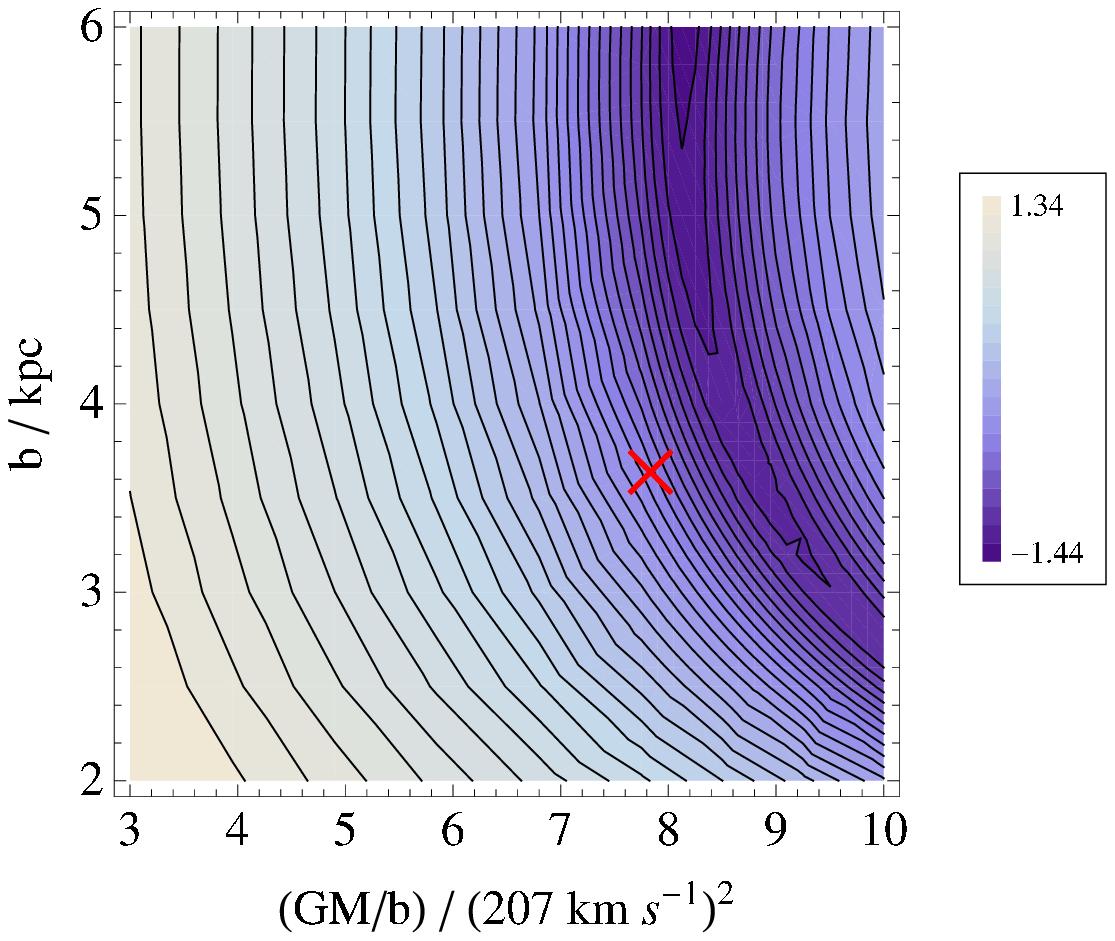}
  }
  \caption[Plots of goodness-of-fit for stream pseudo-data to the
best-fitting orbit, for a range of potential parameters]
{ Plots of goodness-of-fit $\chi^2$ for the pseudo-data to
    the best-fitting orbit in the potential, for a range of potential
    parameters $(GM/b,b)$. The pseudo-data sets are 50 particle realisations of
    those streams shown in \figref{mech:fig:mapping-apo}. Top panel:
    the red stream from \figref{mech:fig:mapping-apo};
    middle panel: the black stream from the same figure;
    bottom panel: the blue stream from the same figure.
    In each case, a valley of low values, corresponding
    to approximately degenerate solutions, can be seen. The red cross
    marks the correct parameters for the potential. In the case of the
    low-curvature stream of the top panel, the orbit fitting technique
    reports a lack of interior mass when compared with the middle
    panel.  In the middle panel, the valley of solutions passes
    directly through the correct parameter coordinate, as expected. In
    the case of the high-curvature stream of the bottom panel, the
    orbit-fitting algorithm reports an excess of mass interior to the
    stream when compared with the middle panel.  }
\label{mech:fig:orbit-fit-general}
\end{figure}

\figref{mech:fig:orbit-fit-general} shows contours for the
goodness-of-fit $\chi^2$, for the
best-fitting orbit, in an isochrone potential with parameters $GM$ and
$b$ as shown.  In each case, a valley of low values, corresponding to
a family of approximately degenerate solutions, can be seen.  This
degeneracy occurs because the trajectories (being at apocentre) are
relatively insensitive to the shape of the rotation curve, but are
very sensitive to the magnitude of the force. Hence, the family of
solutions comprises those combinations of the $(GM,b)$ parameters that
give rise to the same circular velocity at the radius of the stream.
In each plot, a red cross marks the correct parameters for the potential.

In the case of the low-curvature stream of the top panel of
\figref{mech:fig:orbit-fit-general}, for all values of $b$, the orbit
fitting technique reports a lack of interior mass when compared with
the middle panel.  In the middle panel, the valley of solutions passes
directly through the correct parameter coordinate, as expected. Thus,
the performance of the orbit fitting technique is validated. In the case of the
high-curvature stream of the bottom panel, the orbit-fitting procedure
reports an excess of mass interior to the stream for all values of
$b$, when compared with the middle panel.

Thus, the attempt to constrain the potential parameters by using
misaligned streams as proxies for orbits has led us to err. In the
case of the low-curvature stream, for any given value of $b$, we have
underestimated the mass by about $11\percent$. In the case of the
high-curvature stream we have overestimated the mass by approximately
$14\percent$.

We do note that the lowest value of $\chi^2$ seen for
the two misaligned streams is significantly higher that the lowest
value of $\chi^2$ seen for the perfectly aligned stream. This
indicates that in the case of the misaligned streams, although optimum
values for the parameters $(GM,b)$ are being found, the fit to the
orbit is still not perfect there. This is precisely the effect utilized
in \chapref{chap:radvs} to try to identify the correct potential, and it is discussed
there in further detail.

\begin{figure}[\figplaceopts]
  \centering{
    \includegraphics[width=\doublefigshrink\hsize]{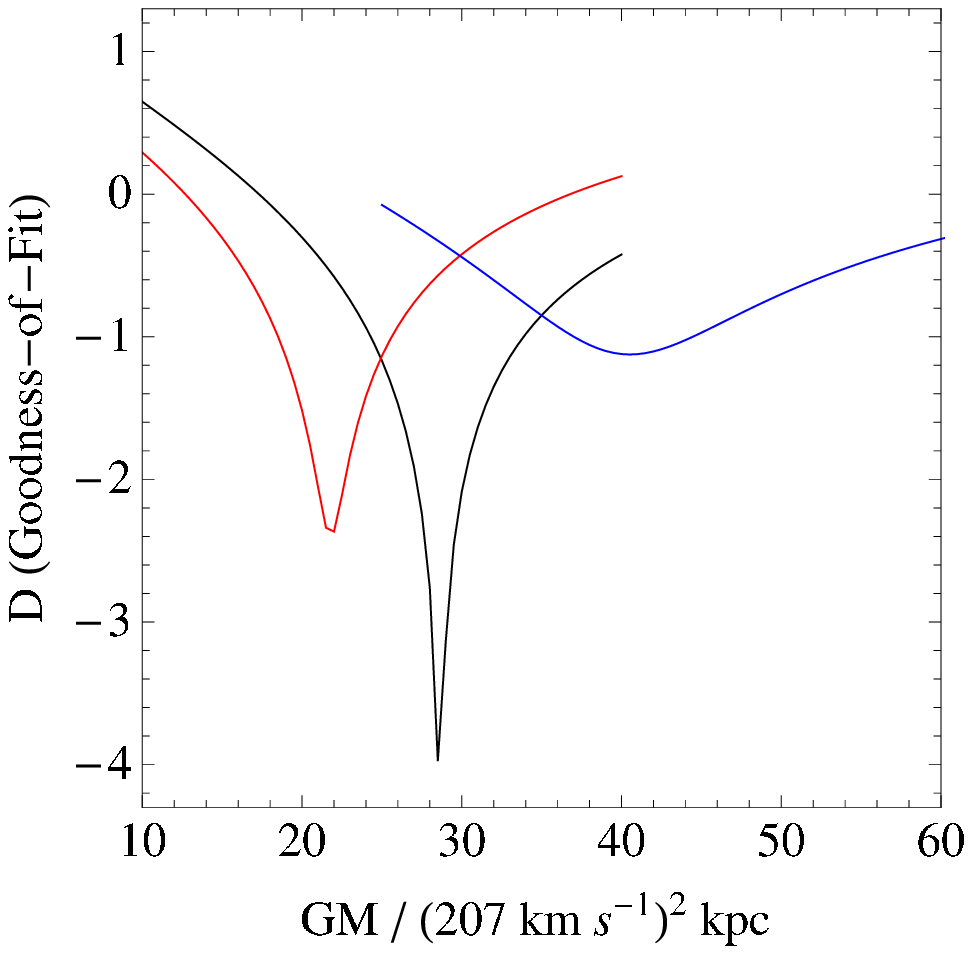}
    \includegraphics[width=\doublefigshrink\hsize]{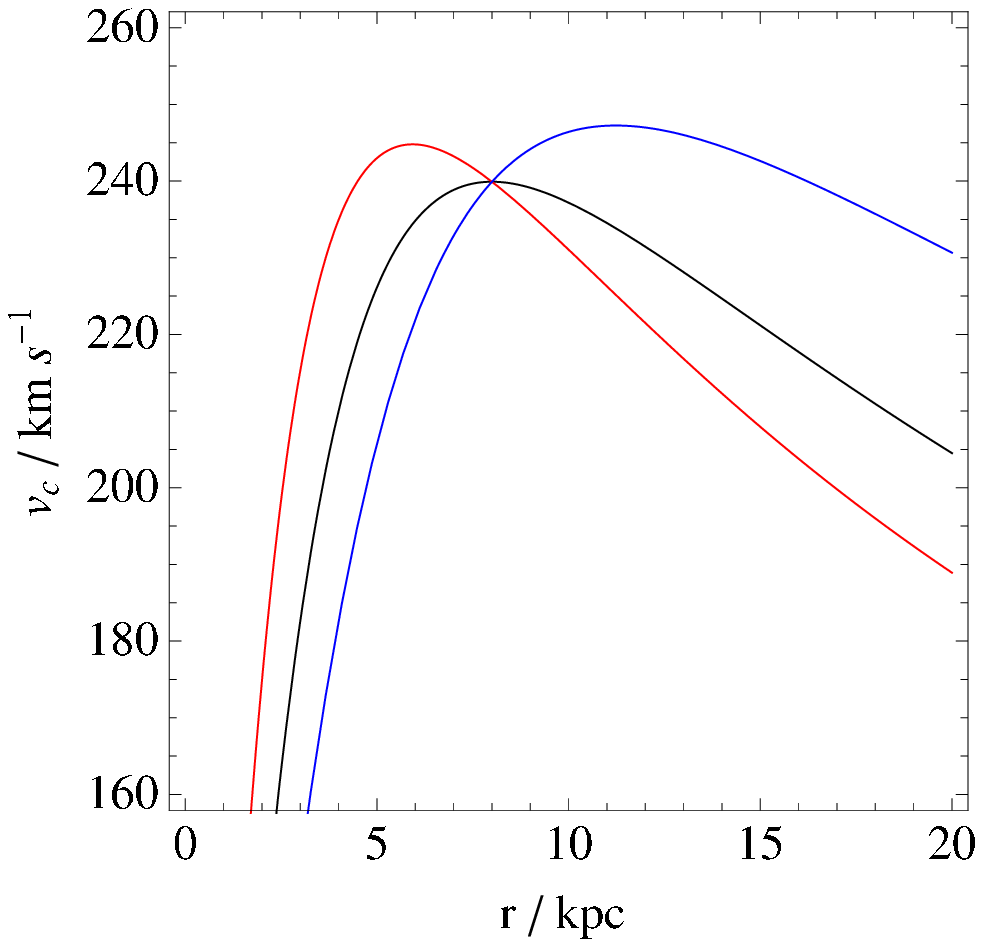}
  }
  \caption[Plots of goodness-of-fit for stream pseudo-data to the best-fitting orbit,
versus potential mass parameter]
{Left panel: goodness-of-fit $\chi^2$ for each pseudo-data
    set, to the best-fitting orbit in the potential. The colours of
    the curves for each pseudo-data set identify it with the
    corresponding stream from \figref{mech:fig:mapping-apo} from which
    it was derived. Right panel: rotation curves for the optimum
    potential found for each pseudo-data set.  Unlike in
    \figref{mech:fig:orbit-fit-general}, we have required that the
    circular velocity $v_c = 240\kms$ at $\rsun=8\kpc$.  This reduces
    the potential to one of a single parameter; in this case,
    mass. The quality of the fits to the red and blue data sets is
    significantly degraded. Similarly to
    \figref{mech:fig:orbit-fit-general}, utilizing the low-curvature,
    red data set as a proxy for an orbit causes us to underestimate
    the host mass by approximately 21\percent. Utilizing the
    high-curvature, blue data set causes us to overestimate the host
    mass by approximately 54\percent.  }
\label{mech:fig:orbit-fit-specific}
\end{figure}

In the cases presented in
\figref{mech:fig:orbit-fit-general}, we have placed no constraints on
the parameters of the potential other than those implied by the
stream. This is a somewhat unrealistic test, since in practical usage
one would generally require any acceptable potential to
reproduce other observed features of the Milky Way galaxy, such as the
circular velocity at the Solar radius.  Most of the combinations of
$(GM,b)$ in the family of solutions do not reproduce the correct
circular velocity. We therefore repeat the test, while considering
only those combinations of parameters that do.

\figref{mech:fig:orbit-fit-specific} shows the value of $\chi^2$
obtained for the best fitting orbit, versus galaxy mass parameter
$GM$, while requiring $b$ to take that value which gives the fiducial
$v_c = 240\kms$ at $\rsun=8\kpc$.
Like with \figref{mech:fig:orbit-fit-general}, the quality of the fit
at optimum $GM$ is significantly degraded for the misaligned streams
when compared to the perfectly aligned stream. The correct value of
$GM$ is obtained when fitting to the stream that perfectly delineates
its orbit, thus validating the technique. However, the
error in $GM$ when deduced using the low-curvature stream is $21\percent$,
and the error in $GM$ when deduced using the high-curvature stream is
$54\percent$. Hence, our attempt to use
misaligned stream tracks to constrain the potential, while simultaneously
requiring the potential to be consistent with other observations, has
led us to yet greater error. 

In conclusion, we find that there is a risk of substantial systematic
errors in parameter estimation being made, if one attempts to
constrain the potential using streams, and one assumes that streams
perfectly delineate orbits.

\section{The action-space distribution of disrupted clusters}
\label{mech:sec:actions}

Up until now, we have relied upon the qualitative estimate from
\secref{mech:sec:nonisotropic} for what the action-space distribution
of a disrupted cluster might be. In this section, we investigate
the action-space distribution of various cluster models using N-body
simulation. We further utilize our N-body models to
confirm the misalignment between streams and orbits, and to demonstrate
that we can accurately predict the real-space track of the stream.

The action-angle coordinates, as
defined in the host galaxy potential, have limited usefulness when
applied to the particles in a bound cluster, because the actions are
not constant with time. Nonetheless, they remain a valid set of canonical
coordinates and can be legitimately used to describe the phase-space
distribution of the cluster. Moreover,  we shall see that a disrupted
cluster gives rise to a characteristic distribution in action-space, from which
a precise track of the stream can be predicted.

\begin{figure}[\figplaceopts]
  \centering{
    \includegraphics[width=\doublefigshrink\hsize]{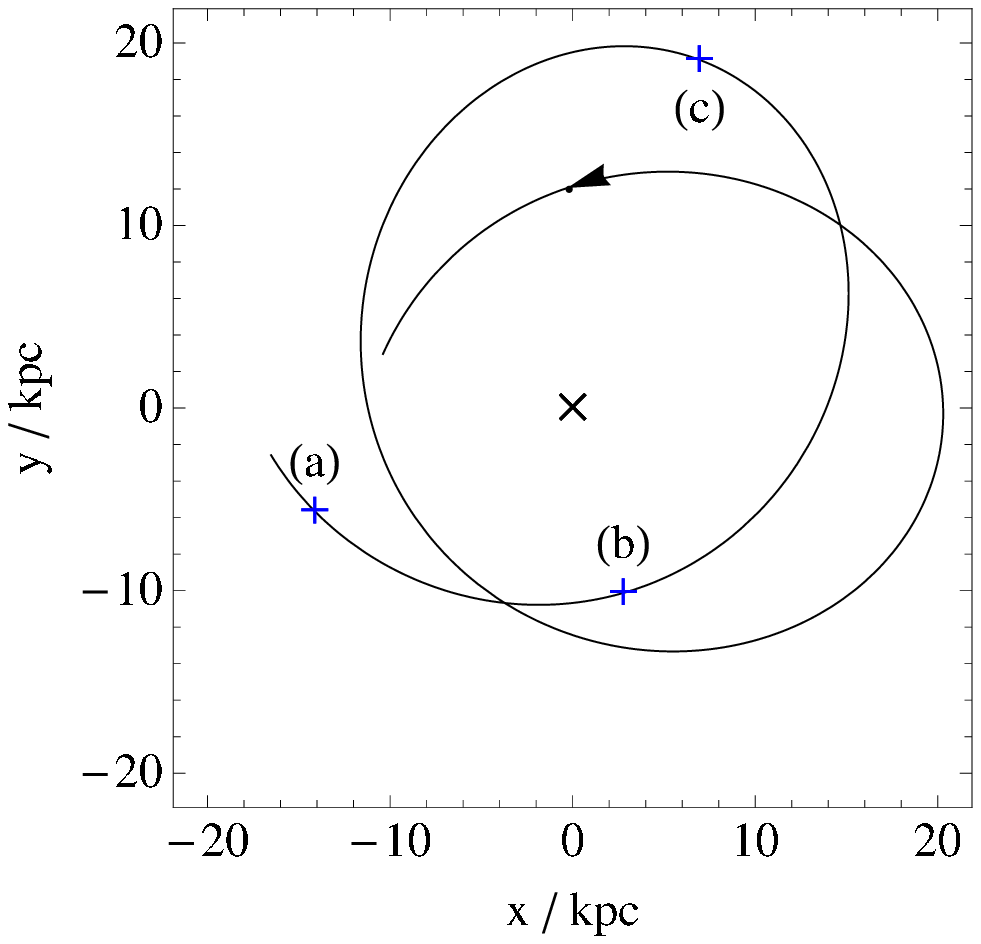}
    \includegraphics[width=\doublefigshrink\hsize]{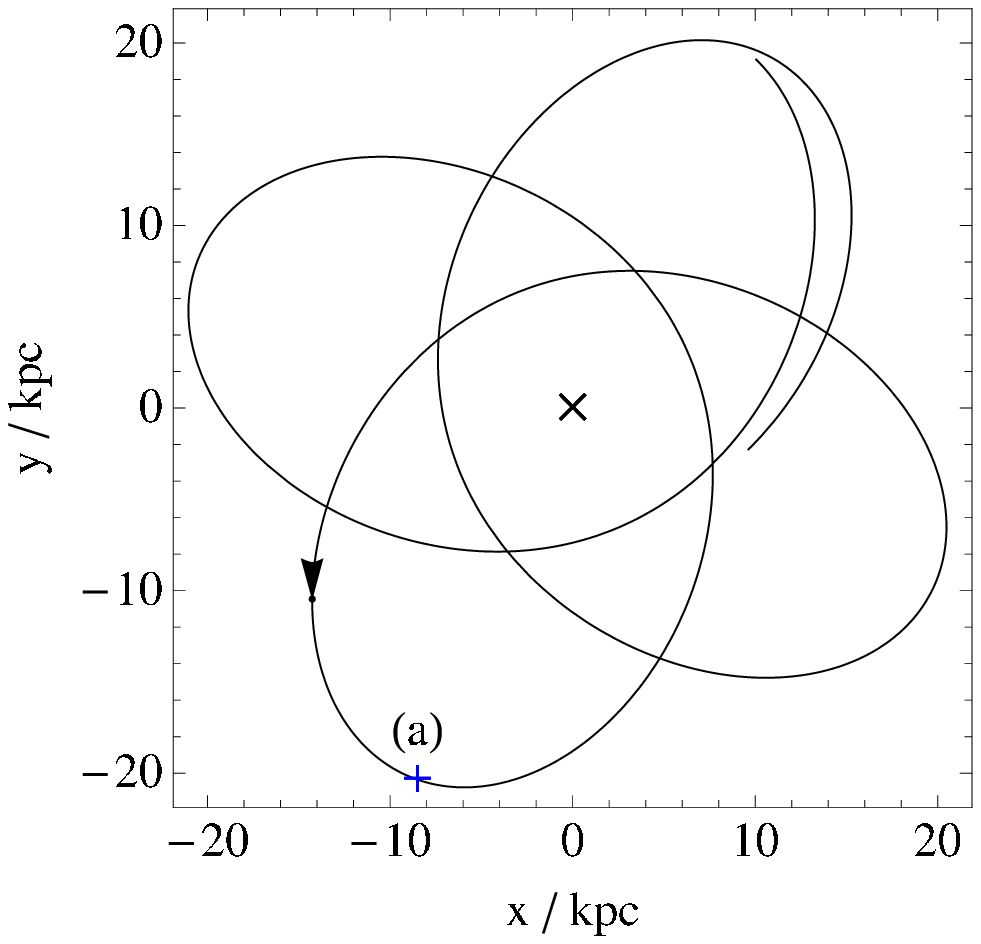}
  }
  \caption[Plan views of the orbits used in \secref{mech:sec:actions}]
{Plan views of the orbits used in this section. The left
    panel shows I4, with (a), (b) and (c) marking the positions of the
    cluster corresponding to the top-left, top-right and middle-left
    panels of \figref{mech:fig:nbody-run1} respectively. The right
    panel shows I5, with (a) marking the position of the cluster
    corresponding to the bottom-right panel of
    \figref{mech:fig:nbody-runs234}. In both panels, the potential in
    use was the isochrone potential described in
    \tabref{mech:tab:potentials}.  }
\label{mech:fig:nbody-orbit-plans}
\end{figure}

\figref{mech:fig:nbody-orbit-plans} shows a segment of each of the
orbits I4 and I5 in the isochrone potential of
\tabref{mech:tab:potentials}. These orbits were chosen to be fairly
representative of those occupied by tidal streams in our Galaxy: they
have apocentre radius $\sim 20\kpc$ and are moderately eccentric to
allow for efficient tidal stripping. For
our investigation, we wish to launch model clusters on each of these
orbits, where the otherwise-stable cluster has been chosen such that
its outermost stars will be torn away by tidal forces close to
pericentre. The process by which we choose our model clusters is
detailed in the next section.

\subsection{Cluster models}
\label{mech:sec:clusters}

We choose to work with King models \citep{kingmodel,bt08} for our
clusters, since these simple models are both easy to generate and
are fairly representative of some observed globular cluster profiles \citep[e.g.~Fig.~6.18,]
[]{bm98}. The profile of a
King model can be defined\footnote{ Equivalently, one may specify the
  concentration parameter, $c \equiv \tilde{r}_t/\tilde{r}_0$, the
  ratio of the truncation radius $\tilde{r}_t$ to the core radius
  $\tilde{r}_0$. There is a one-to-one correspondence between $c$ and
  $W$, though the form of the relation is not trivial. A plot of $c$
  versus $W$ can be seen in \citet[\S4.3.3c]{bt08}} by $W \equiv
\Phi_0/ \sigma^2$, the ratio of the central potential to the squared-velocity
parameter $\sigma^2$. The resulting family of models have similar
profiles in terms of $\rho(\tilde{r}/\tilde{r}_0)/\rho_0$, where
$\rho_0$ is the central density and $\tilde{r}_0$ is the core
radius. An exact model is specified by choosing $\rho_0$ and
$\tilde{r}_0$, in addition to $W$, either directly or through a relation with another
parameter.

For a given orbit, we specify our models as follows. Following the argument of
\cite{dehnen-pal5}, we note that a cluster of mass $M_c$ orbiting at a
galactocentric radius $r$ from the centre of a host galaxy with circular velocity
$v_c$, will be tidally pruned to the cluster radius $\tilde{r}_\tide$, where
\begin{equation}
\tilde{r}^3_\tide \simeq {GM_c \over v_c^2} r^2.
\label{mech:eq:rtide}
\end{equation}
We freely choose a profile parameter $W$ and a cluster mass $M_c$, and we
also specify a galactocentric stripping radius $r_s > r_p$, where
$r_p$ is the pericentre radius of the orbit concerned. We then
set $\tilde{r}_t$, the cluster truncation radius,
equal to $\tilde{r}_\tide$ from \eqref{mech:eq:rtide}, where 
$r\rightarrow r_s$ and $v_c \rightarrow v_c(r_s)$. The resulting cluster
will remain intact while $r\gg r_s$, but will have its outermost stars
tidally stripped when $r \sim r_p$.

\begin{table}
  \centering
  \caption[Details of the cluster models used in \secref{mech:sec:clusters}]
{Details of the cluster models used in this section. Defining parameters
  are in the central block of columns, while derived parameters occupy the right-hand block.}
  \begin{tabular}{l|lll|llll}
    \hline
    & $W$ & $M_c$ & $r_s$ & $\rlim$ & $\sigma$ & $\tdyn$ & $\epsilon$\\
    \hline\hline
    C1 & 2 & $10^4\msun$ & $12\kpc$ & $48.6\pc$ & $1.18\kms$ & $12.6\Myr$ & $1.0\pc$ \\
    C2 & 2 & $10^5\msun$ & $12\kpc$ & $104.8\pc$ & $2.54\kms$ & $12.6\Myr$ & $2.2\pc$ \\
    C3 & 6 & $10^4\msun$ & $12\kpc$ & $48.6\pc$ & $1.14\kms$ & $2.36\Myr$ & $0.32\pc$ \\
    C4 & 2 & $10^4\msun$ & $11\kpc$ & $45.5\pc$ & $1.22\kms$ & $11.79\Myr$ & $0.94\pc$ \\
    C5 & 2 & $10^4\msun$ & $11\kpc$ & $45.7\pc$ & $1.22\kms$ & $11.79\Myr$ & $0.94\pc$ \\
    C6 & 2 & $10^4\msun$ & $15\kpc$ & $56.3\pc$ & $1.10\kms$ & $15.6\Myr$ & $1.2\pc$ \\%gd1
    C7 & 2 & $10^4\msun$ & $12\kpc$ & $48.2\pc$ & $1.19\kms$ & $12.3\Myr$ & $0.99\pc$ \\%ostr
    \hline
  \end{tabular}
  \label{mech:tab:clusters}
\end{table}

\subsection{The disruption of a cluster}
\label{mech:sec:disruption}

%\tabref{mech:tab:clusters} specifies the resulting cluster models
%for this entire chapter.

The low-mass cluster model C1 (\tabref{mech:tab:clusters}) was
specified for the orbit I4 (\tabref{mech:tab:orbits}) according to the
schema in \secref{mech:sec:clusters}. We chose a low value for the
profile parameter of $W=2$ for our basic cluster model, in order to
ensure the presence of many particles near the cluster truncation
radius $\twidr_t$ during successive stripping events.

A $10^4$ particle realization of the cluster model C1 was made by
random sampling of the King model distribution function \citep[equation
4.110]{bt08}. This cluster was placed at a point shortly after apocentre on the
orbit I4 in the isochrone potential of \tabref{mech:tab:potentials}.
The cluster was evolved forward in time in the aforementioned potential
by the \fvfps\ tree code of \citet{fvfps}, using a time step of
$\d t = \tdyn/100$ and a softening length $\epsilon$ as specified
in \tabref{mech:tab:clusters}. The simulated time period was
$4.81\Gyr$, or almost 14 complete radial orbits.

\begin{figure}[\figplaceopts]
\centering{
    \centerline{
    \includegraphics[width=\doublefigshrink\hsize]{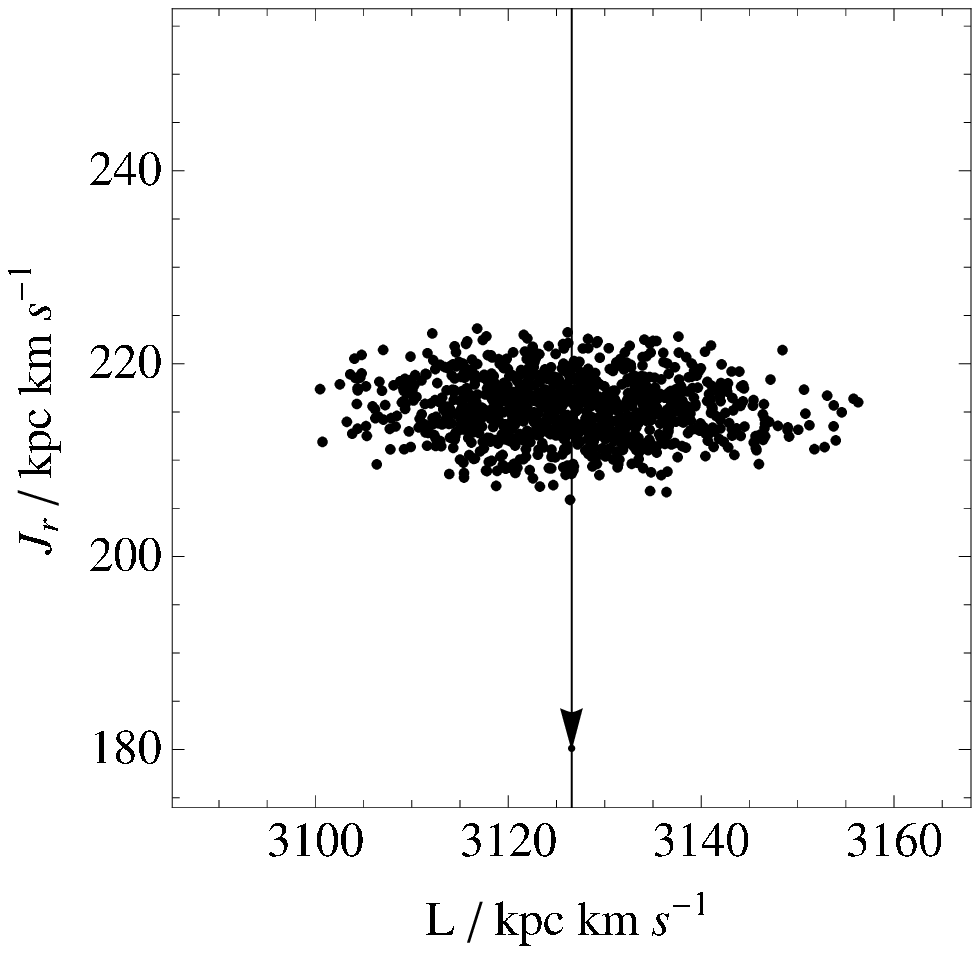}
    \qquad
    \includegraphics[width=\doublefigshrink\hsize]{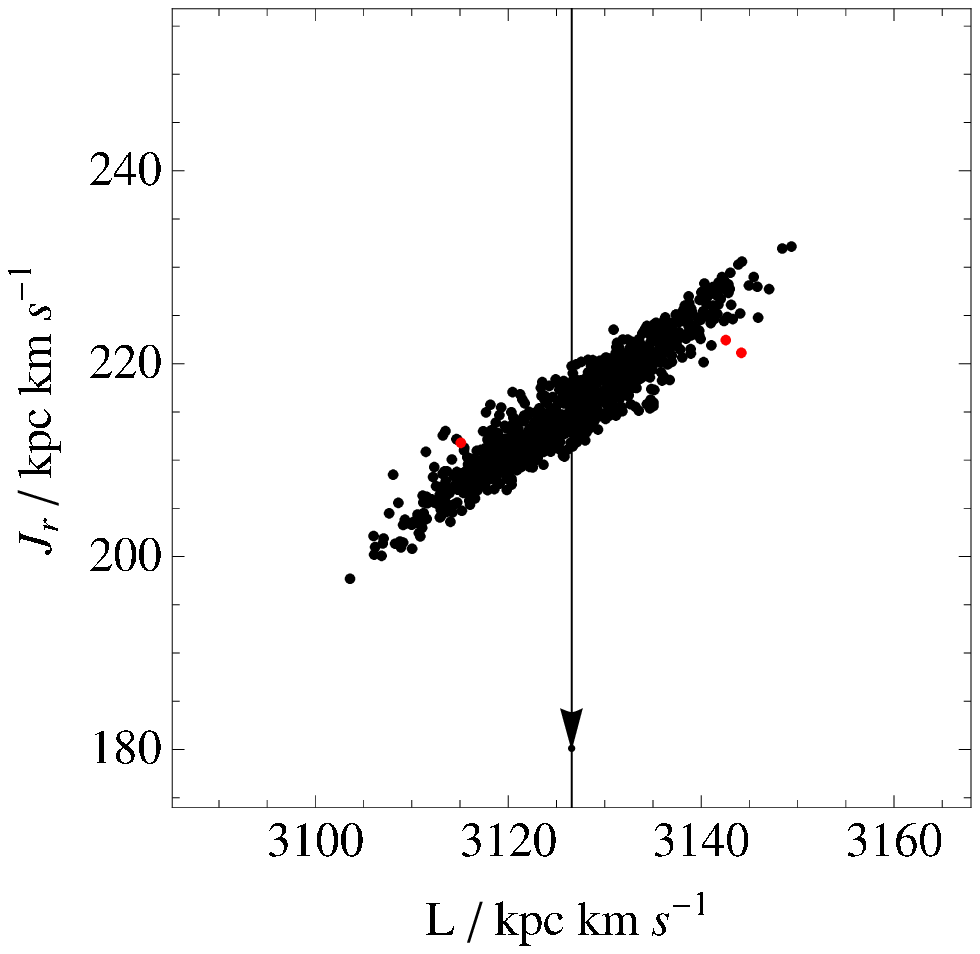}
    }
}
  \centering{
    \centerline{
    \includegraphics[width=\doublefigshrink\hsize]{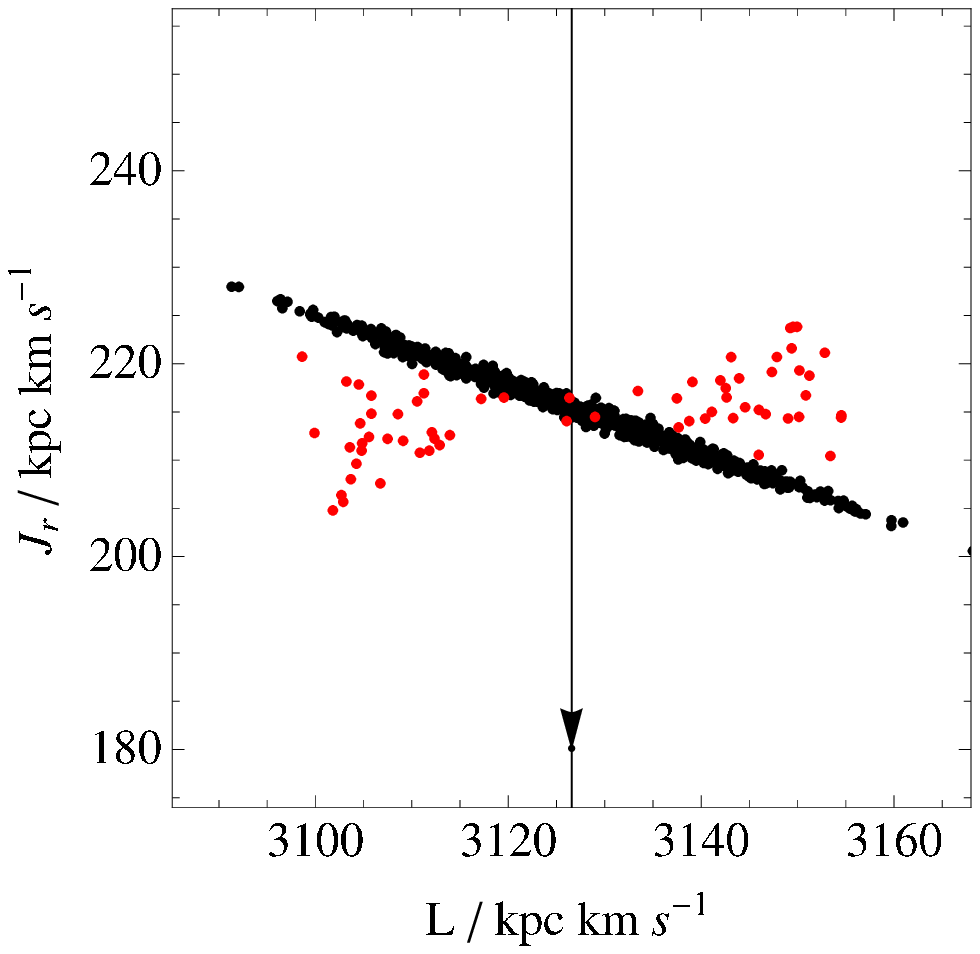}
    \qquad
    \includegraphics[width=\doublefigshrink\hsize]{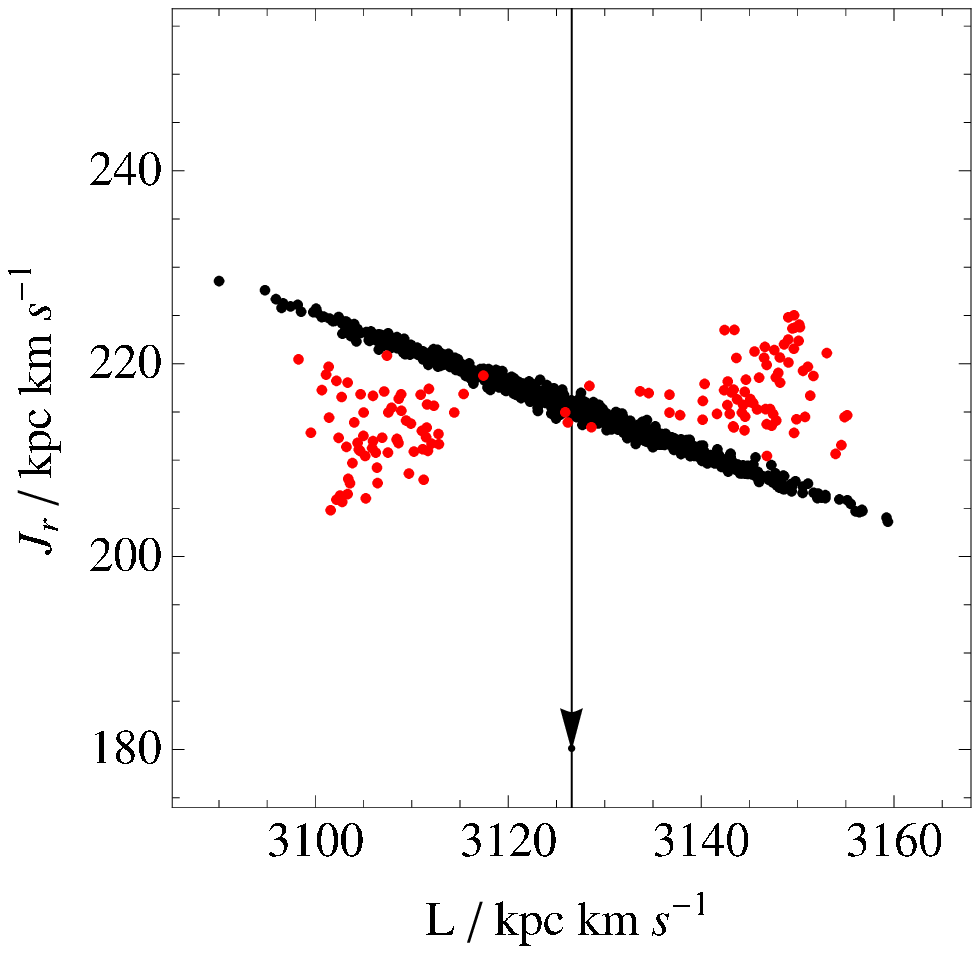}
    }
  }
  \centering{
    \centerline{
    \includegraphics[width=\doublefigshrink\hsize]{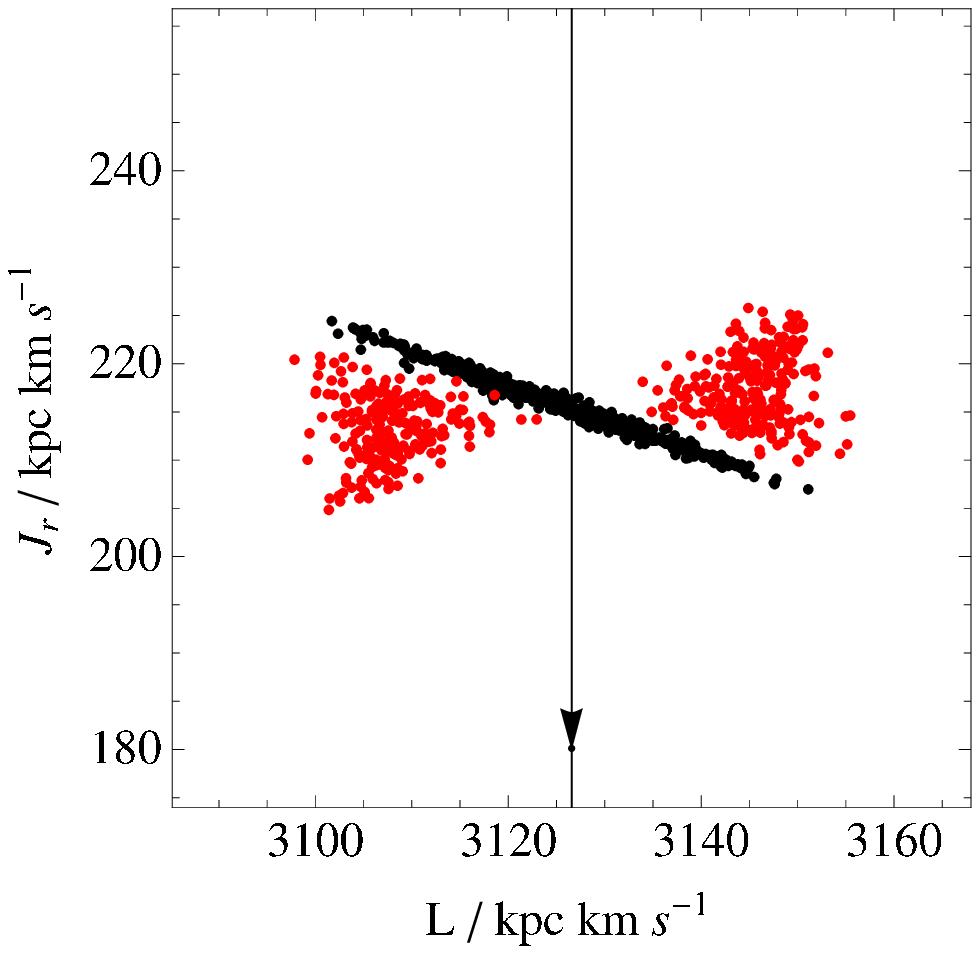}
    \qquad
    \includegraphics[width=\doublefigshrink\hsize]{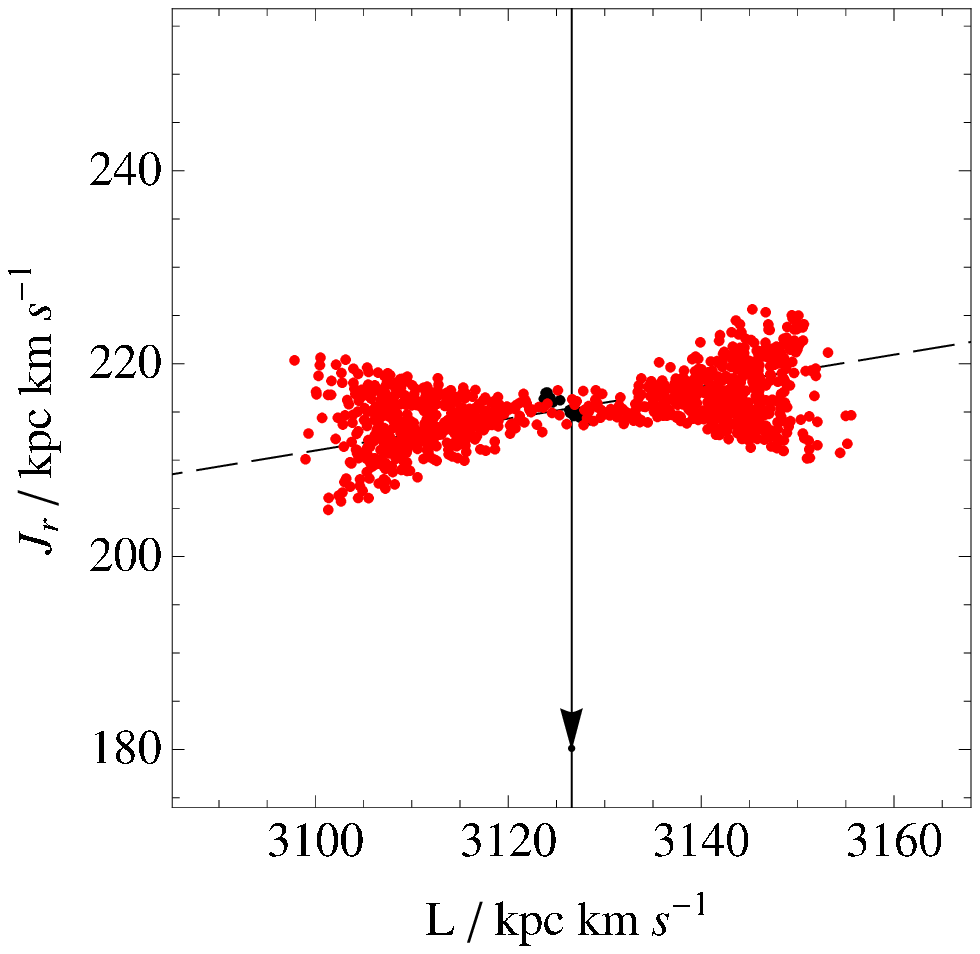}
    }
  }
  \caption[Action-space distribution of the cluster model C1, at different times along the
    orbit I4]{ Plots showing the action-space distribution of particles
    for the N-body cluster model C1, at different times along the
    orbit I4.  From left-to-right and top-to-bottom, these times are:
    shortly after release; first pericentre passage;
    first apocentre passage;
    second apocentre passage;
    7th apocentre passage;
    14th apocentre passage.
    The solid black line is the inverse map of the frequency vector,
    $\hessian^{-1}\vO_0$.  The dashed line in the bottom-right panel represents a
    least-squares linear fit to the particle distribution.  }
\label{mech:fig:nbody-run1}
\end{figure}

\figref{mech:fig:nbody-run1} shows the evolution of the action-space
distribution of the cluster model C1 as a function of time. In all the
panels of that figure, an arrowed black line shows the mapping of the
frequency vector from angle space into action space,
$\hessian^{-1}\vO_0$. This vector shows the direction that maps onto
$\vO_0$ in angle-space, so any action-space distribution that is aligned with this
vector will be aligned with $\vO_0$ in angle-space.

In all cases involving the isochrone potential,
$\hessian^{-1}\vO_0$ is oriented exactly along the $J_r$ axis. We can understand
this from \eqref{mech:eq:iso-freqs}, which shows that in the isochrone
potential, the frequency direction $\hat{\vO}_0$ is a function of $L$
only, and is independent of $J_r$. Thus, a line of constant
$\hat{\vO}_0$ must map into action-space as a line of constant $L$.
We note that this is a peculiar feature of the isochrone
potential, and is not true for a general potential.

The upper-left panel of \figref{mech:fig:nbody-run1} shows
the configuration of the cluster immediately after release, at
position (a) in \figref{mech:fig:nbody-orbit-plans}.
The distribution is ellipsoidal but without additional substructure, which we
expect since the distribution in action results entirely from the approximately
spheroidal density profile, and the approximately isotropic velocity dispersion,
of the cluster. We understand the ellipticity of the action-space distribution
from \secref{mech:sec:nonisotropic}. Indeed, the prediction of \eqref{mech:eq:djr/dl}
of $\Delta J_r/\Delta L \sim 0.3$ for this orbit can be seen to be approximately
correct.

In the top-right panel of \figref{mech:fig:nbody-run1}, the cluster
has now moved from its point of release to its first pericentre
passage, marked as position (b) in
\figref{mech:fig:nbody-orbit-plans}.  We see that the ellipse has
flattened somewhat, and has rotated anticlockwise. In the middle-left
panel, the cluster has now progressed to the subsequent apocentre passage,
marked as position (c) in \figref{mech:fig:nbody-orbit-plans}.  Here, the
cluster is again flattened, but now it has rotated clockwise.

We can qualitatively understand this behaviour as follows. Consider
a particle in a cluster near apsis. What changes to the actions
will be made by perturbations to the velocity of this particle?
A perturbation $\delta v$ to the transverse velocity will cause a
change to the angular momentum
\begin{equation}
\delta L = r \delta v.
\label{mech:eq:dL=dv}
\end{equation}
By means of the mechanism described by \eqref{mech:eq:l-amp}, this $\delta L$
will cause a change in the guiding centre radius $r_g$, which will cause a
corresponding change in the radial action, according to
\eqref{mech:eq:amplitude}. Conversely, a perturbation to the
radial velocity will cause negligible change to the radial action,
since
\begin{equation}
\delta E \simeq p_r \delta p_r = \dot{r} \delta v \sim 0,
\label{mech:eq:de0}
\end{equation}
and $J_r(E,L)$ remains unchanged. Hence, the distribution in both $J_r$
and $L$ is governed primarily by the transverse velocity, when the
cluster is at apsis, and their highly correlated distribution
reflects this. 

\begin{figure}[\figplaceopts]
  \centerline{
    \includegraphics[width=\doublefigshrink\hsize]{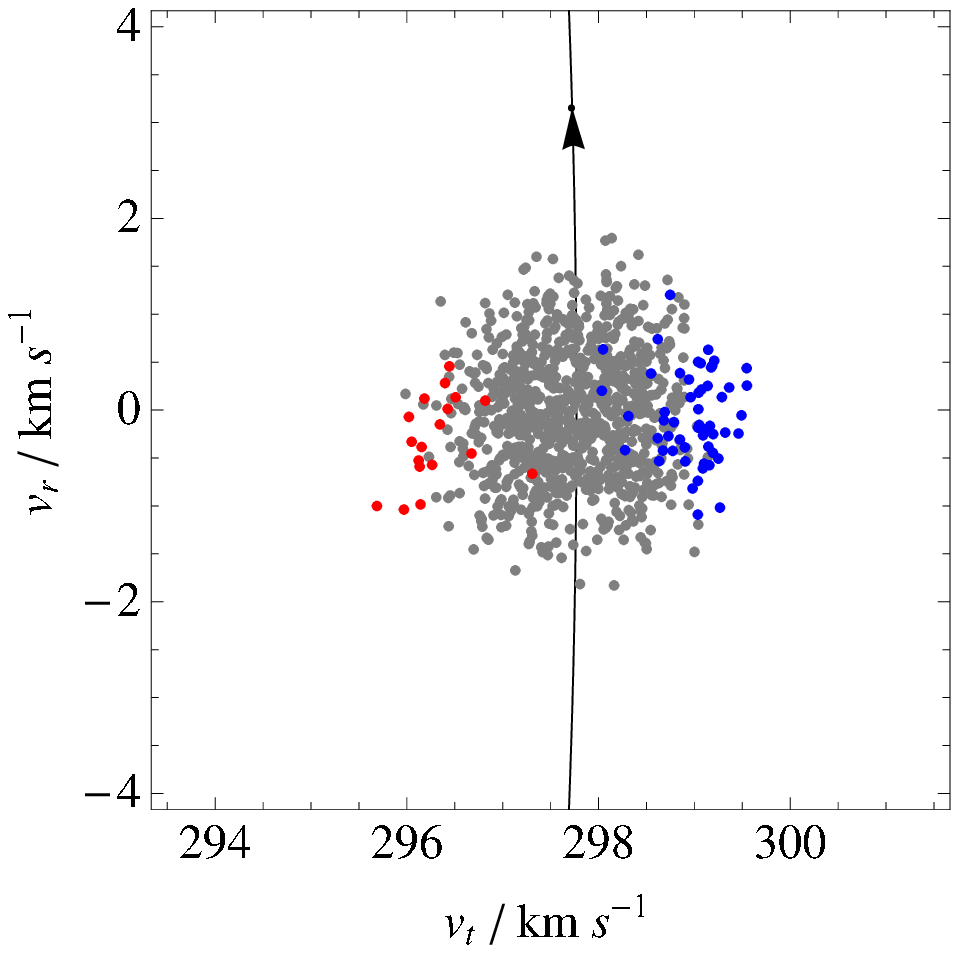}
    \qquad
    \includegraphics[width=\doublefigshrink\hsize]{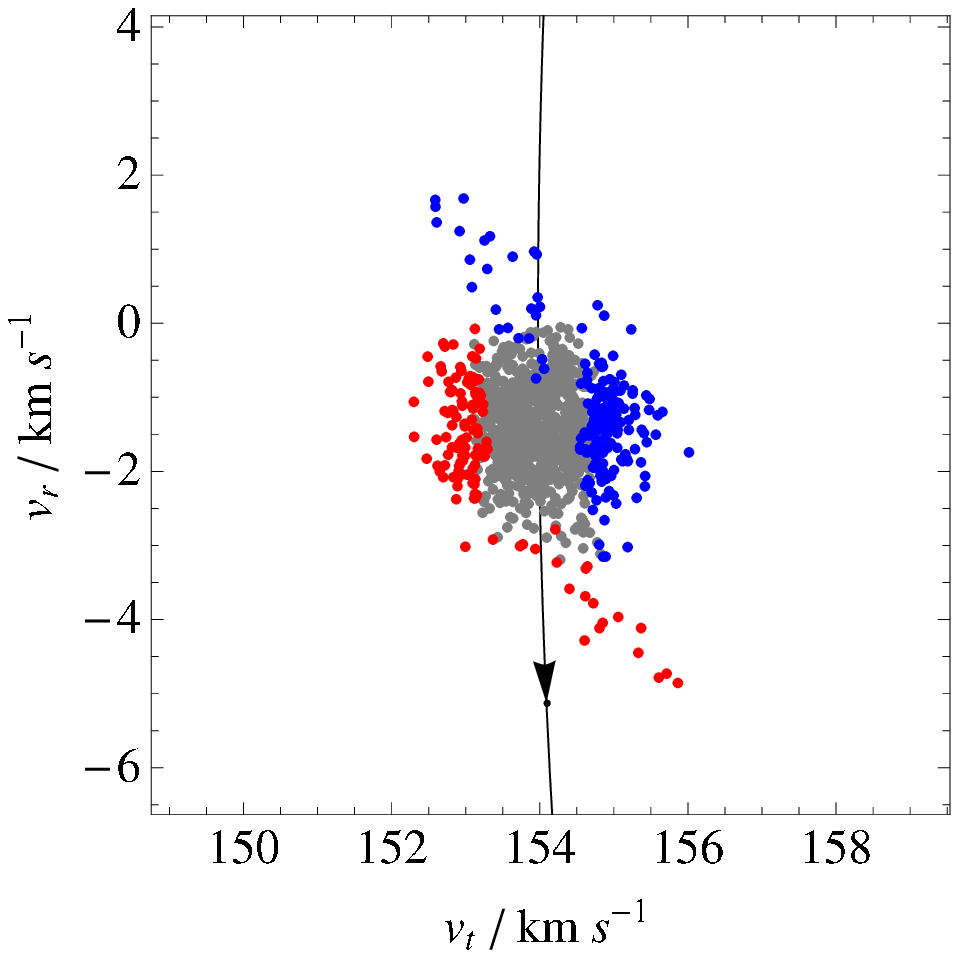}
  }
  \caption[Radial velocity versus tangential velocity for the C1 cluster model, on
the I4 orbit]{
Plots of radial velocity against tangential velocity for the C1 cluster model, on
the I4 orbit, at (left panel) the first pericentre passage, and
(right panel) the first apocentre passage.
}
\label{mech:fig:nbody-v}
\end{figure}

We can confirm this analysis by examining \figref{mech:fig:nbody-v}. The left
panel shows $(v_r,v_t)$ for the cluster near its first pericentre passage,
while the right panel shows the same for the cluster near its subsequent
apocentre passage. The particles coloured red are those with $L < 3110\kpc
\kms$ and those coloured blue have $L > 3140\kpc\kms$. The boundary
between the colours is very sharp in the $v_t$ direction, as one would
expect, since the particles have been coloured on the basis of $L$. However,
we note that particles with the full range of $v_r$ contribute to both the
blue and red regions equally, despite \figref{mech:fig:nbody-run1}
showing that $J_r$ is very different for these two regions. Hence $J_r$
must be independent of $v_r$ near apsis.

We complete our explanation of the orientation of the action-space distribution
by taking note of the sign of the change in $J_r$ near apsis. Increasing
$L$ will always increase $r_g$, and when the cluster is at pericentre,
this pushes $r_g$ further away, and so increases the radial action.
Conversely, when the cluster is at apocentre, increasing $r_g$ brings
it closer to the cluster, and so decreases the radial action. Thus, we see that
near pericentre, particles with high $L$ will have high $J_r$ and the distribution
will be rotated to have a positive gradient $\Delta J_r / \Delta L$.
Conversely, at apocentre, particles with high $L$
will have low $J_r$, and so the distribution will be rotated
to have a negative gradient $\Delta J_r / \Delta L$.

We can predict the value of the gradient of this distribution,
by combining \eqref{mech:eq:amplitude} and \eqref{mech:eq:l-amp}.
We find
\begin{equation}
{\d J_r \over d L} \sim \pm\sqrt[4]{10 J^2_r \over L^2} \sim
\pm\sqrt{\pi J_r \over L},
\label{mech:eq:gradient}
\end{equation}
where the sign of the radical depends on the apsis under consideration,
as detailed above. We can see from the upper-right panel
of \figref{mech:fig:nbody-run1} that this equation predicts approximately
the correct gradient, when evaluated for the orbit I4. Furthermore,
we note that this equation implies that the gradient will be
steeper for an orbit of greater eccentricity.

We now examine again \figref{mech:fig:nbody-run1}, and note that the
black particles are those still bound to the cluster, while the red
particles are those that are unbound, where we have defined `bound' to
mean particles that are within $r_\tide$ of the cluster barycentre. In
the middle-left plot, which shows the cluster configuration at the
first apocentre passage, the few unbound particles form an
approximately horizontal distribution.  These are particles that have
been stripped from the cluster near pericentre, i.e.~in the top-right
panel of \figref{mech:fig:nbody-run1}.

We note two things. Firstly, the bulk of the stripped particles have
large $\Delta J$ from the cluster centroid; this is simply a
consequence of the high-speed stars being most likely to be stripped.
Secondly, we note that although the red particles in the middle-left
panel span approximately the same range in $L$ as the black particles
in the top-right panel, they span a range in $J_r$ that is only about
half that spanned by the black particles.

\begin{figure}[\figplaceopts]
  \centerline{
    \includegraphics[width=\doublefigshrink\hsize]{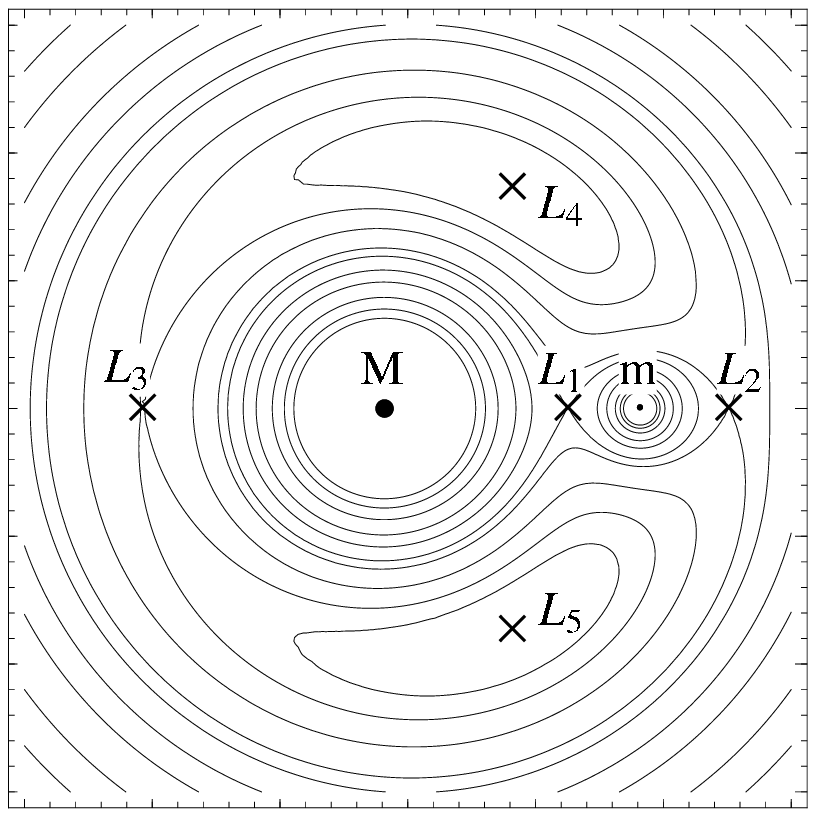}
  }
  \caption[ Diagram showing the Lagrange points in the reduced
  three-body problem ]{ Lagrange points of neutral force, for the
    reduced three-body problem of a satellite of mass $m$ on a
    circular orbit around a host galaxy of mass $M > m$. The lines are
    contours of constant effective potential $\Phi_{\rm eff}$, as
    given for this problem by equation (8.87) of \cite{bt08}, who
    also provide a diagram similar to this. In the example shown, the frame
    rotates with the orbiting masses and the entire configuration is
    static with respect to time. In the case of a cluster on an
    eccentric orbit, the Lagrange points will not be static, but the points
    $L_1$ and $L_2$ through which cluster particles escape are always
    aligned radially with respect to the cluster and the host galaxy.}
\label{mech:fig:lagrange}
\end{figure}

We explain this as a result of the cluster's self-gravity, as
follows. Particles that are stripped from the cluster mostly escape
through the Lagrange points $L_1$ and $L_2$. As illustrated in
\figref{mech:fig:lagrange}, these two points are oriented along a
radial that runs through the barycentres of the host galaxy
and the cluster. Thus, it is the particles' radial velocity which
initially carries them away from the cluster, and it is from this
velocity component that the particles pay most of the energetic
penalty for escaping, with the consequence that the radial velocity
dispersion of the escaping stars is reduced. This reduction in the
radial velocity dispersion corresponds to a compression of the
distribution in $J_r$ for escaping particles. Once unbound, the
particles are carried further away from the cluster along a complex
trajectory that ends up with the now-free particle drifting away from
the cluster according to the mapping in \eqref{mech:eq:angle_t}.
Thus, the final sum of the energetic penalty is paid from the
difference in action $\vJ - \vJ_0$ between the particle and the
cluster, with the net result that the unbound distribution uniformly
shrinks. The latter effect is minor compared to the compression in
$J_r$, because much more work is done in becoming unbound than in
escaping to infinity once already unbound. Hence, the complete effect
is to generate an unbound distribution, that looks like the high
$\Delta \vJ$ wings of the pericentre distribution, but is compressed
in $J_r$ and shrunk slightly.

Looking again at \figref{mech:fig:nbody-run1}, we note that in
the bottom-left panel, which corresponds to the 7th apocentre
passage, many particles have now escaped, and the size of the
bound distribution has visibly shrunk. We understand this as a
consequence of the most energetic particles having already
escaped the cluster, leaving behind a colder core. By the time
of the 14th apocentre passage, shown in the bottom-right panel,
almost all the particles have escaped. We note that the positions
of many of the red particles have remained static between the bottom-left
and bottom-right panels. The action-space distribution of the
unbound particles is therefore frozen in place, confirming that
self-gravity is unimportant in streams.

\begin{figure}[\figplaceopts]
\centering{
    \centerline{
    \includegraphics[width=\doublefigshrink\hsize]{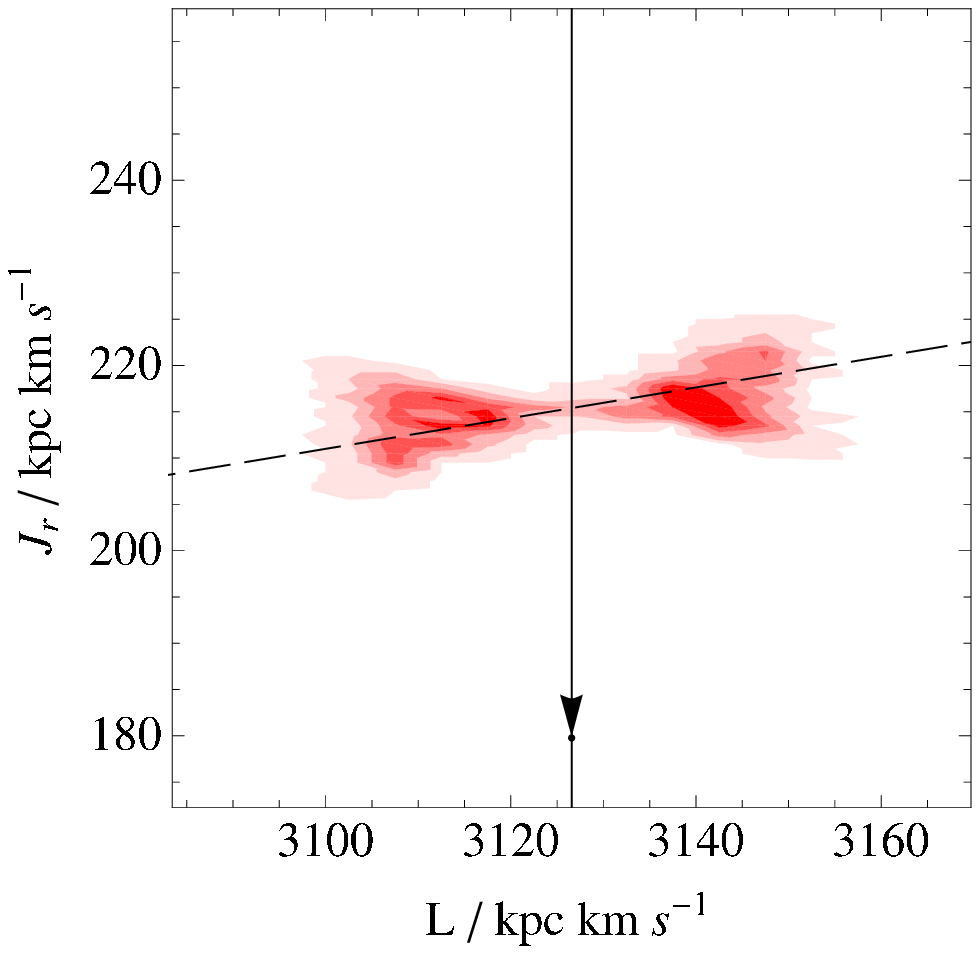}
    \qquad
    \includegraphics[width=\doublefigshrink\hsize]{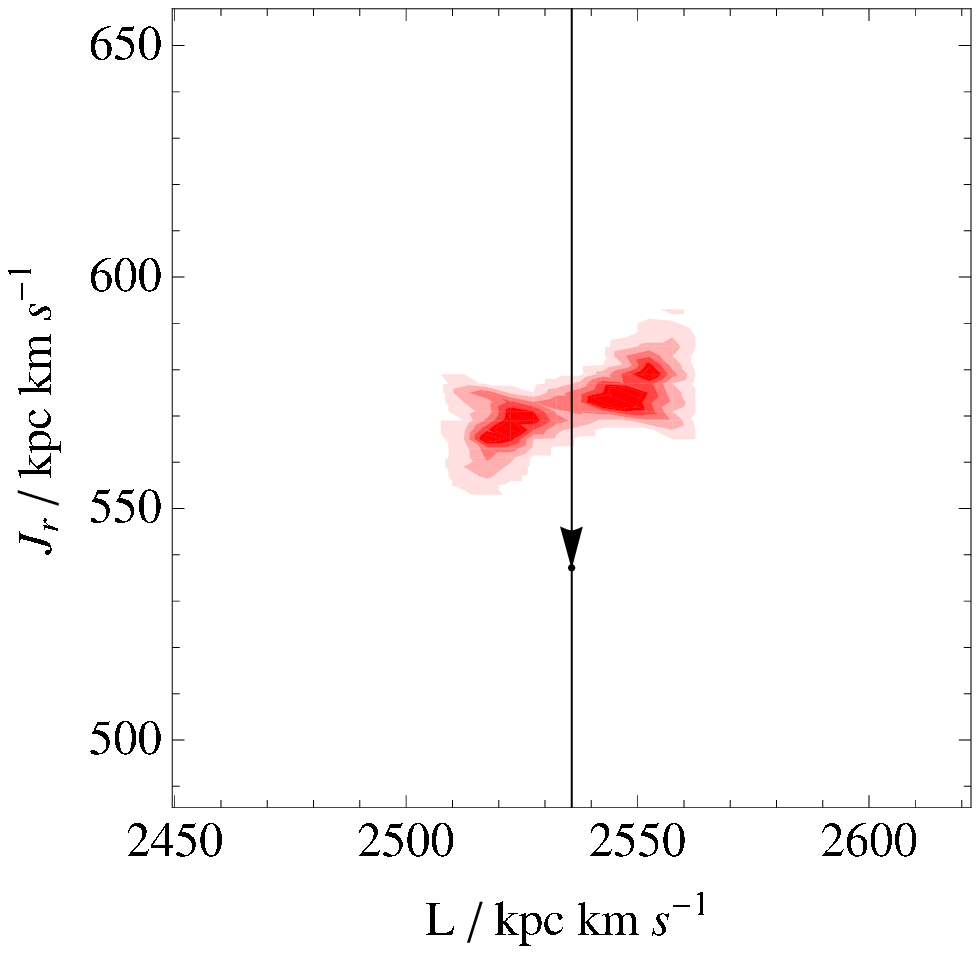}
    }
}
\caption[Particle-density plots for selected panels from Figs.~\ref{mech:fig:nbody-run1} and~\ref{mech:fig:nbody-runs234}]{ Plots showing the density of particles in
  action-space, corresponding to (left) the lower-right panel of
  \figref{mech:fig:nbody-run1}, and (right) the lower-right panel of
  \figref{mech:fig:nbody-runs234}.  The edges of the shaded areas
  represent contours of constant density, where the latter has been
  computed by placing particles into bins of width $\sim 2 \kpc \kms$.
  Darker shading represents regions of higher particle density:
  the gross ``bow-tie'' structure is clearly seen, but no
  sub-structure is visible in either plot that is obviously
  unattributable to sampling noise.}
\label{mech:fig:nbody-particle-density}
\end{figure}

The scatter plots in the bottom panels of \figref{mech:fig:nbody-run1}
are filled too densely with particles to allow proper examination of
the variation in particle density within the distribution. In order to
elucidate this variation, the left panel of
\figref{mech:fig:nbody-particle-density} shows a plot of particle
density for the bottom-right panel of
\figref{mech:fig:nbody-run1}.  The density field exhibits the same gross
``bow-tie'' structure that is visible in \figref{mech:fig:nbody-run1},
and it further shows that particles are concentrated towards the centres of
the lobes of the distribution, with particle density remaining low near
the centroid.  We explain this simply as a consequence of the
number-density distribution of particles in velocity-space: the
velocity-centroid of a cluster is populated with few particles
because, although the phase-space density of particles there is high,
the real-space volume associated with the lowest velocities is
vanishingly small as one approaches the centroid, and so the
velocity-space density is low there.\footnote{Precisely the same effect is seen
in the Maxwell-Boltzmann distribution of particle speeds in a gas.}
This aside, no significant
fluctuations in particle density can be seen in the left panel of
\figref{mech:fig:nbody-particle-density} that are obviously
unattributable to the effects of sampling noise.

In conclusion, we have qualitatively understood the distortion
of a cluster in action-space as it passes through pericentre
and apocentre along its orbit. We have found that stars
stripped at pericentre form a distribution that is
derived from the pericentre distribution of bound stars,
but is compressed in $J_r$.
We have found that the pericentre distribution will exhibit
a high correlation between $J_r$ and $L$, and that the
gradient of this correlation in $(L,J_r)$ scales as $\sim \sqrt{J_r / L}$.
We typically assume that our cluster orbit will have large $L$ and comparatively
smaller $J_r$: this would only be untrue for extreme plunging orbits which are not likely to be relevant
to the problem in hand.\footnote{
A cluster on such an orbit would encounter severe difficulties while passing
close the Galactic centre, where it would experience substantial tidal
forces from the supermassive black hole. However, it is unlikely to
even make it that far: since it must feel the maximum possible tidal
force from the Galaxy at some point along its plunging orbit, it is likely
to break-up well away from pericentre.}
Hence the gradient in $(L,J_r)$ will typically be less than unity,
and the compression will only act to shrink it still further.
Hence, the stripping mechanism always results in an action-space distribution that is both
flattened and very roughly oriented along $\hat{L}$.

\subsubsection{The effect of changing the cluster model or orbit}
\label{mech:sec:changingorbit}

We now investigate the qualitative effects of changing the cluster model parameters,
or the cluster orbit, on the action-space distribution of a disrupted cluster.
The cluster models used in this section are C1 to C4, detailed
in \tabref{mech:tab:clusters}. These clusters were created according to the
schema of \secref{mech:sec:clusters}, taking the orbit to be
I4 in the isochrone potential of \tabref{mech:tab:potentials}
for the models C1--C3, and taking the orbit to be I5 in the same potential for
the model C4.

The cluster model C1, the evolution of which along the orbit I4
was detailed in the previous section, is used as our baseline
to which we compare the distributions of the other clusters. The cluster
model C2 has the same profile parameter, $W=2$, as does C1, but is
10 times more massive. The result is a cluster that is both
heavier and proportionately larger while being stripped at the same
galactocentric radius, $r_s$.

The cluster model C3 has the same mass as does C1, but is considerably
more concentrated\footnote{ \fighardref{4.8} of \cite{bt08} shows the
  density profile for King models with a variety of values of $W$.},
with a profile parameter $W=6$. The cluster has an identical
truncation radius $\twidr_t$ and velocity scale $\sigma$, but has
significantly fewer particles near to $\twidr_t$, when
compared with C1. The particles of C3 are generally more tightly bound
to the cluster than are those of C1.

The cluster model C4 has the same mass and profile parameter as C1, but
is specified for the orbit I5, which has lower $L$ than I4, and thus
a smaller pericentre radius. The resulting cluster is slightly more
compact, allowing it to survive to a closer galactocentric radius
than can C1.

A $10^4$ particle realization of each of the models C1--C3 was placed at a
point shortly after apocentre on the orbit I4, and evolved forward in
time by the \fvfps\ tree code, using a time step of $\d t = \tdyn/100$
and a softening length $\epsilon$ as specified in
\tabref{mech:tab:clusters}. The total period of the simulation
was $2.36\Gyr$, or almost 7 complete radial orbits.
Additionally, a $10^4$ particle realization of C4 was placed at a
point shortly after apocentre on the orbit I5, and evolved forward in
time by the \fvfps\ tree code, for a total period of $2.21\Gyr$, or
almost 7 complete radial orbits.

\begin{figure}[\figplaceopts]
\centering{
  \centerline{
  \includegraphics[width=\doublefigshrink\hsize]{thesisfigs/mech/nbody_j_run1_apo7.eps}
  \qquad
  \includegraphics[width=\doublefigshrink\hsize]{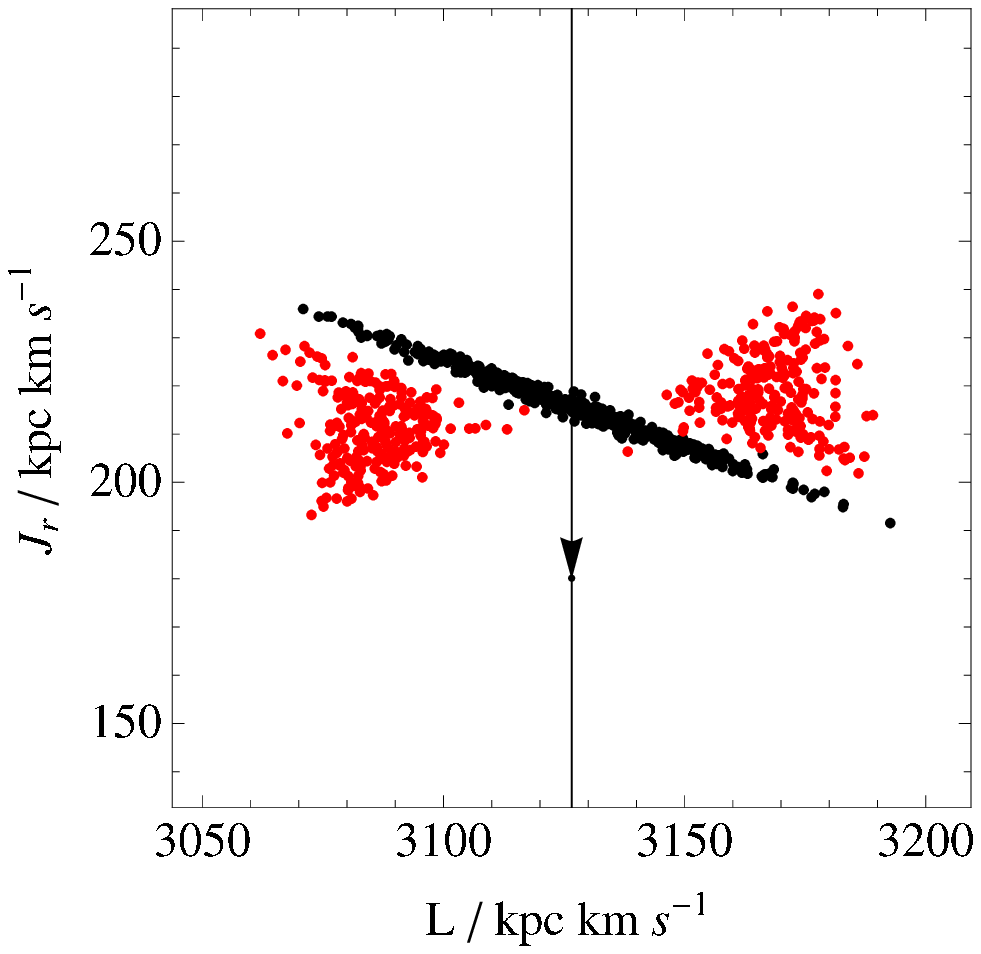}
}
 \centerline{
\includegraphics[width=\doublefigshrink\hsize]{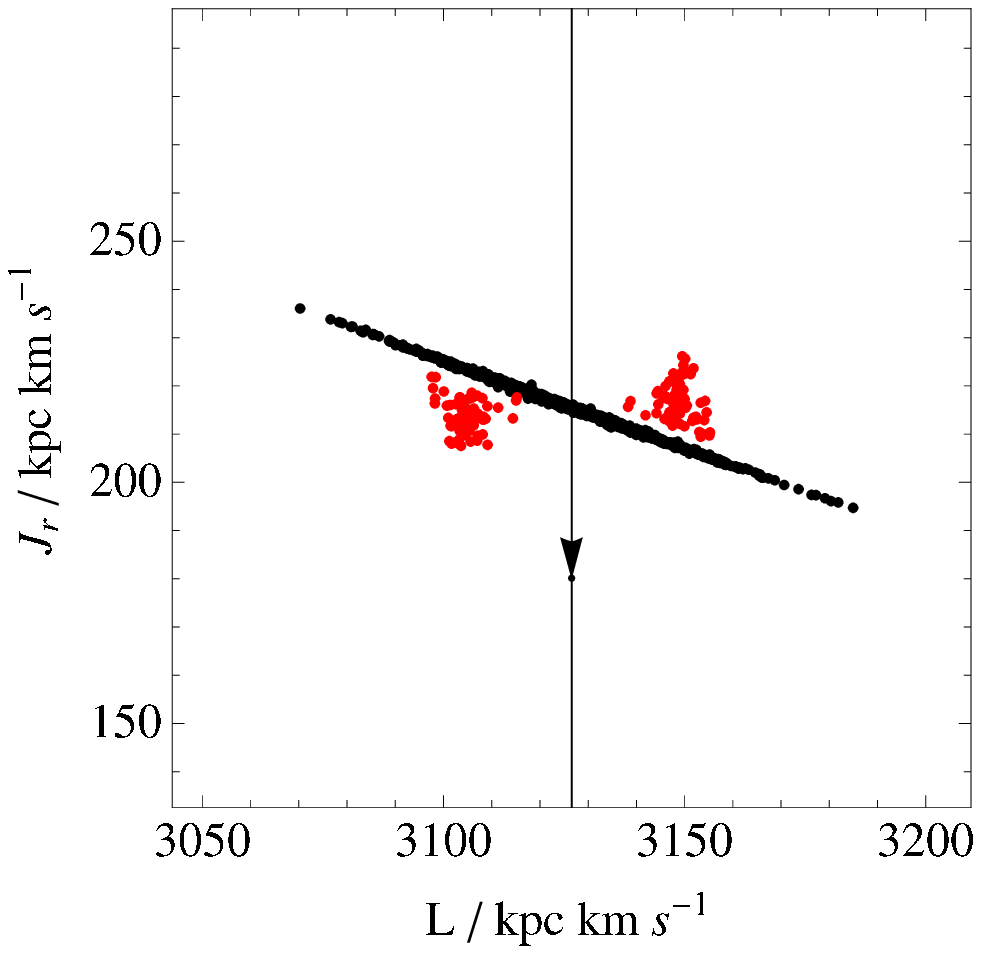}
\qquad
\includegraphics[width=\doublefigshrink\hsize]{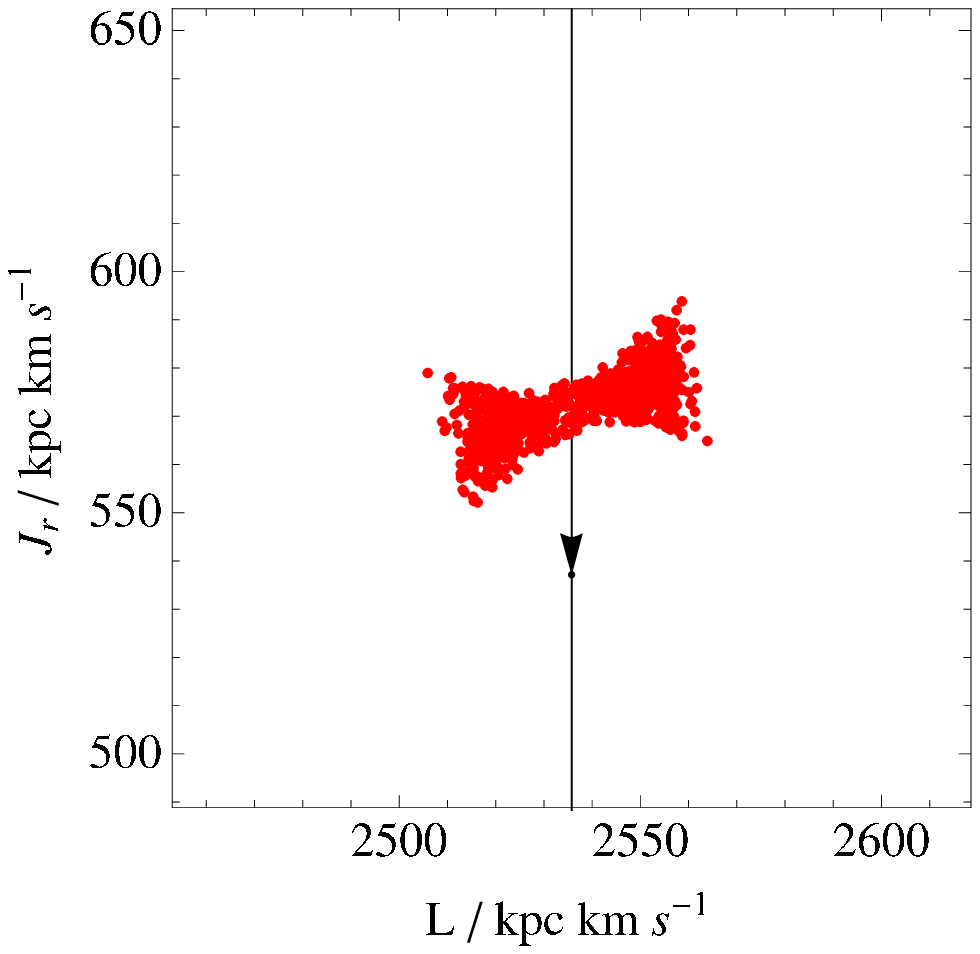}
}
}
\caption[Action-space distribution of particles for different cluster models
on various orbits]
{Plots showing the action-space distribution of particles for cluster models
on various orbits. Each plot shows the distribution at the 7th apocentre passage.
Black points are bound to the cluster, red points orbit free in the host
potential. From left-to-right and top-to-bottom, the panels show: the C1 cluster
on the orbit I4; the C2 cluster on the orbit I4; the C3 cluster
on the orbit I4; and the C4 cluster on the orbit I5.
}
\label{mech:fig:nbody-runs234}
\end{figure}

\figref{mech:fig:nbody-runs234} shows the action-space distribution for 
the cluster at the seventh apocentre passage, for each of these simulations.
The top-left panel shows C1 on the orbit I4, and is identical to the
bottom-left panel in \figref{mech:fig:nbody-run1}.
The top-right panel shows the cluster C2 on the same orbit; the bottom-left
panel shows the cluster C3 on the same orbit; and the bottom-right panel
shows the cluster C1 on the orbit I5. 

In the top-right panel, we see that the action-space distribution of the
more massive cluster is qualitatively identical to that of C1, except
that the distribution is approximately twice the scale. We conclude
that cluster mass plays little role in determining the structure of the
action-space distribution of disrupting clusters, but can determine the scale.

The bottom-left panel shows the distribution from the more centrally concentrated
cluster C3. In this case, the shape of the distribution of unbound particles
is approximately the same as for C1. However, the scale is much smaller: with 
more particles being near the core of the cluster, they have to work harder to escape,
resulting in a colder action-space distribution. However, the distribution
is still qualitatively similar to that of C1. Thus we conclude that cluster
concentration can determine the scale of the action-space distribution, but
not the general shape.

The bottom-right panel shows the distribution from the cluster C4 but
on the orbit I5, which has the same apocentre radius as I4 but a
pericentre radius about $33\percent$ smaller. Unlike in the other panels,
 the cluster has become completely unbound by the 7th apocentre
passage on I5, which is likely to be a result of $(r_s - r_p)$ being
slightly larger for C4 on I5, when compared to the other clusters on I4,
resulting in more efficient stripping at pericentre.

The distribution shown in this plot has approximately the same scale
in $\Delta L$ as does the distribution from I4, but has approximately
twice the scale in $\Delta J_r$. We can understand this, on account of
the $\Delta J_r / \Delta L$ described by \eqref{mech:eq:gradient} being steeper
for I5 than for I4. Further, since we have already noted that this
cluster was stripped faster than was C1, the energetic penalty for escaping
must be lower,  and so we expect less compression in $J_r$.
The resulting distribution is similar to
that of the top-left panel, but less compressed in the $\hat{J_r}$
direction. Thus we conclude that changing the cluster orbit can
distort the shape of the action-space distribution, but does not
affect its basic structure.

The bottom-right panel is filled too densely with particles to gauge
properly the density variation within it. We therefore provide a
particle-density plot for the bottom-right panel, which is shown in
the right panel of \figref{mech:fig:nbody-particle-density}.  This
density plot correctly reproduces the gross structure observed in the scatter
plot and reveals that the particles are concentrated in the lobes of
the distribution, as was also true for the density plot in the left
panel of \figref{mech:fig:nbody-particle-density}.  Aside from this,
the plot reveals no further density fluctuations in the action-space
distribution that cannot be attributed to sampling noise.

\subsection{Predicting the stream from the action-space distribution}

\begin{figure}[\figplaceopts]
  \centerline{
    \includegraphics[width=\doublefigshrink\hsize]{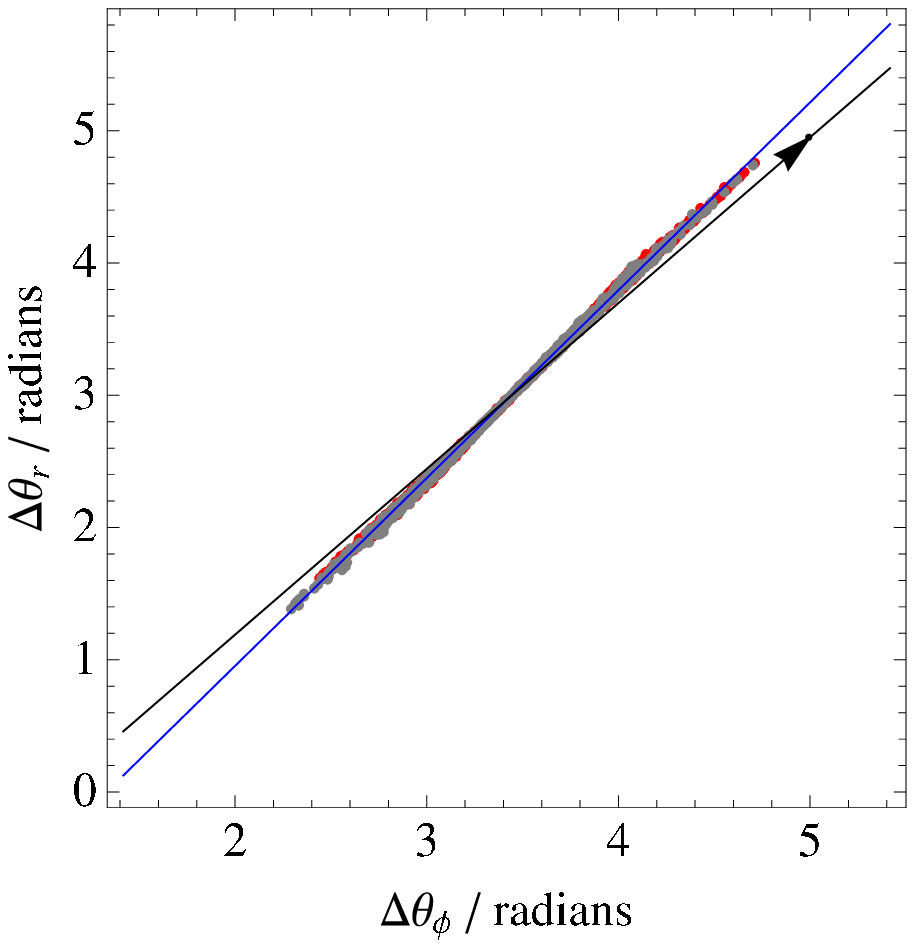}
    \qquad
    \includegraphics[width=\doublefigshrink\hsize]{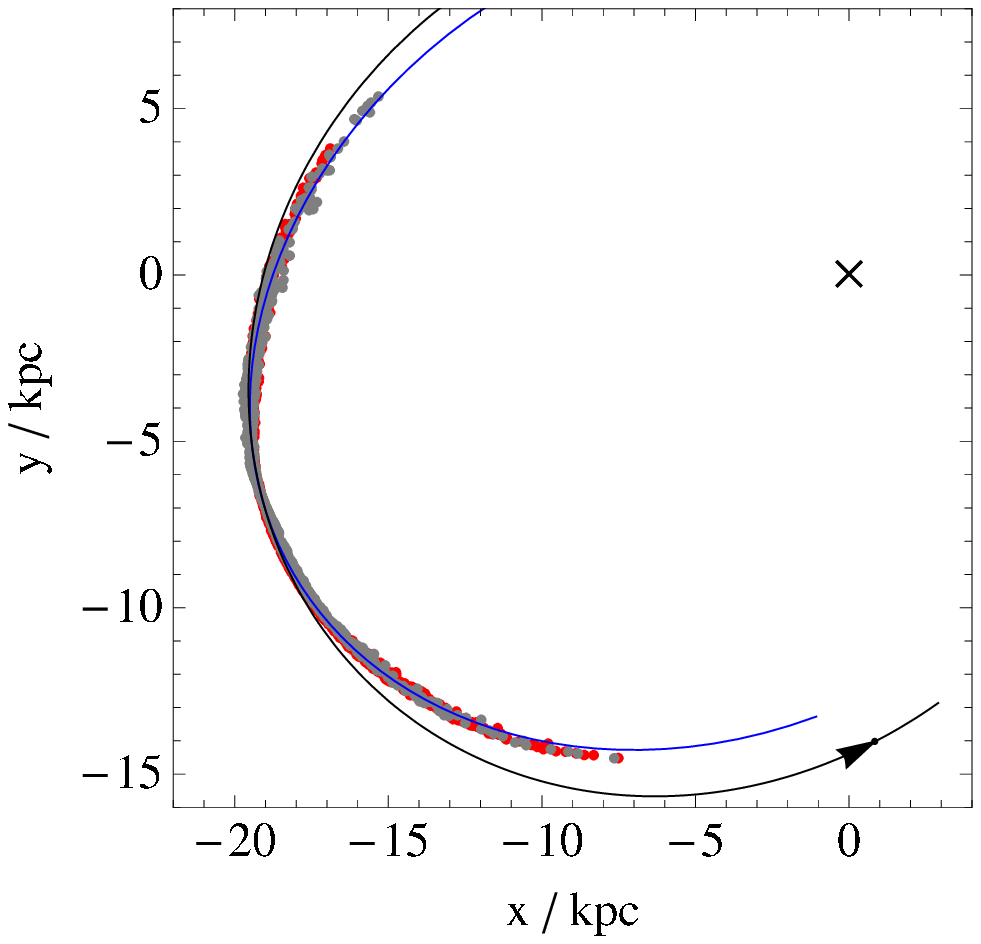}
  }
  \caption[Angle-space and real-space distribution of particles for cluster C1 on orbit I4]
{Plots of the distribution of particles for cluster C1, near
    its 14th apocentre passage on orbit I4. The left panel shows the
    angle-space distribution, while the right panel shows the
    configuration in real-space.  The grey particles show positions
    directly computed from the N-body simulation, while the red
    particles show those positions predicted from mapping the
    action-space distribution in the bottom-right panel of
    \figref{mech:fig:nbody-run1}. The two distributions almost
    precisely overlap.  In both plots, the black arrowed curves show
    the trajectory of the progenitor orbit, while the blue curves show
    the mapping of the dashed line from \figref{mech:fig:nbody-run1}.
    The blue curve is everywhere a much closer match to the stream particles than is the
    progenitor orbit.}
\label{mech:fig:nbody:isoch}
\end{figure}

We have determined the action-space distribution for several disrupted cluster models
by means of N-body simulation. We now ask whether we can accurately predict the
real-space path of the stream, given the action-space distribution of
one of those models.

%the bottom-right panel
%of \figref{mech:fig:nbody-run1}

Suppose that we know the time $t_\peri$ since a cluster's first
pericentre passage. The angle-space distribution is predicted by
\eqref{mech:eq:angle_t}, where $\hessian$ is evaluated on the progenitor orbit
$\vJ_0$ and $t\rightarrow t_\peri$. We will use as an example the action-space distribution
shown in the bottom-right panel of \figref{mech:fig:nbody-run1}, which corresponds
to the 14th pericentre passage of the simulated cluster C1 on the orbit I4.
The left panel of \figref{mech:fig:nbody:isoch} shows the angle-space configuration
corresponding to this panel: the grey
particles are for angles directly computed from the results of the N-body simulation,
while the red particles are for those predicted by \eqref{mech:eq:angle_t}, assuming
$t_\peri$ is known. Also plotted is the frequency vector $\vO_0$, shown as a
black arrowed line.
The distributions of black and grey particles in this panel agree perfectly.
Furthermore, both distributions are obviously
misaligned with the progenitor orbit. 

The right panel of \figref{mech:fig:nbody:isoch} shows the real-space
configuration of particles for the same scenario. The grey particles
are again plotted directly from the results of the N-body simulation,
while the red particles result from the mapping into real-space
of the red particles from the left panel.  As in angle-space, the two
distributions agree perfectly. Furthermore, the real-space
manifestation of the misalignment of the stream with the orbit can be
seen: the stream delineates a track that has substantially lower
curvature than the orbit.

Our attempt to predict the real-space stream configuration
from the action-space distribution has been completely successful.
However, any
complete model of the bottom-right panel of
\figref{mech:fig:nbody-run1} must necessarily be rather
complicated. It might not be possible to guess the form of this
distribution without full N-body modelling. The dashed line in the
bottom-right panel of \figref{mech:fig:nbody-run1} is a least-squares
fit of the action-space distribution to a line. This represents a rather more simple
model of the action-space distribution, which it might well be
possible to guess {\em ab initio} for a cluster on a given orbit.

How good a prediction for the stream track can we get from this line? The blue
lines in \figref{mech:fig:nbody:isoch} show the results of mapping this line
into both angle-space and then real-space. It is clearly an excellent fit to the
stream, in marked contrast to the orbit, which represents a very poor model
of the stream by comparison. Thus, even a very simple model of the
cluster in action-space---albeit one deduced from an accurate knowledge of the
distribution---allows
us to predict stream tracks accurately.

Finally, we note that during the mapping of this line into real-space,
we need to make a correction to the cluster's action, as described by
\eqref{mech:eq:correction}, to account for the variation in action
down the stream.  Evaluating \eqref{mech:eq:amplitude} and
\eqref{mech:eq:l-amp} for the orbit I4, and taking $\d L \sim
25\kpc\kms$ and $\d J_r \sim 10\kpc\kms$ from the bottom-right panel
of \figref{mech:fig:nbody-run1}, we predict errors in the real-space
of up to $\sim 0.2\kpc$ on account of the finite $L$ distribution,
and up to $\sim 0.15\kpc$ on account of the finite $J_r$ distribution.
These errors would be serious enough to be seen in
\figref{mech:fig:nbody:isoch}, and thus the correction is required. We note,
however, that even these substantial errors are insignificant compared
to the several-$\kpc$ discrepancy between the stream and the orbit.

\section{Non-spherical systems}
\label{mech:sec:nonsph}
We have investigated the formation of streams in spherical potentials,
and demonstrated that they do not necessarily delineate streams, but
that we can accurately predict their paths.
Unfortunately, many real stellar systems in the Universe are not spherical.
In particular, our own Galaxy, whose potential we are interested
in probing with streams, is probably significantly flattened.
In this section, we investigate the formation of streams in flattened
potentials, and in particular we ask by how much they are misaligned
with orbits. We also demonstrate that our apparatus is capable of correctly
predicting stream tracks in flattened potentials, just as it is for spherical
ones.

\subsection{The general case with three actions}

By analogy with \secref{mech:sec:general2d}, we now consider the
case of a general axisymmetric stream-forming system, which
is described by a Hamiltonian in three actions, $H(J_1,J_2,J_3)$.
\Eqref{mech:eq:omega.e} can be written explicitly as
\begin{align}
  \Omega_1 e_{n,1} + \Omega_2 e_{n,2} + \Omega_3 e_{n,3} 
&  = e_{n,1} \, \partial_1 H +  e_{n,2}\, \partial_2 H  +  e_{n,3}\, \partial_3 H \nonumber\\
&  = \lambda(e_{n,1}\, \delta J_1  + e_{n,2}\,\delta J_2  +  e_{n,3}\,\delta J_3) + k_n,
\qquad (n=1,3).
\end{align}
Defining the constants $\alpha_n = e_{n,2}/e_{n,1}$, $\epsilon_n = e_{n,3} / e_{n,1}$,
$\beta_n = k_n/e_{n,1}$ we have
\begin{equation}
  \partial_1 H + \alpha_n  \partial_2 H + \epsilon_n \, \partial_3 H
  = \lambda_n(\delta J_1 + \alpha_n \delta J_2 + \epsilon_n \, \delta J_3) + \beta_n.
  \label{mech:eq:3d:inhom}
\end{equation}
As in the spherical case, we can find a general solution to the homogeneous
form of this equation by considering an arbitrary function $f$ of the
characteristic coordinates $\left(\delta J_1 - {\delta J_2 / \alpha_n}\right)$ and 
$\left(\delta J_1 - {\delta J_3 / \epsilon_n}\right)$. We add to this a particular
solution for the inhomogeneous equation \blankeqref{mech:eq:3d:inhom}, to give
the general solution
\begin{equation}
  H = f\left(\delta J_1 - {\delta J_2 \over \alpha_n}
  , \delta J_1 - {\delta J_3 \over \epsilon_n}\right)
  + \beta_n \delta J_1 +{\lambda_n \over 2}\left(\delta J_1^2 +
  \delta J_2^2 + \delta J_3^2\right).\label{mech:eq:3d:hamil}
\end{equation}
A system with a Hamiltonian of the form
\eqref{mech:eq:3d:hamil} is expected to exhibit globally consistent
stream-forming geometry, described by the quantities
$(\alpha,\beta,\lambda,\epsilon)_n$. Unfortunately, apart from trivial extensions
of two-action systems such as the Kepler potential, there is no known
Hamiltonian of the form \eqref{mech:eq:3d:hamil} that is of relevance
to galactic dynamics. Indeed, there is no known closed form for {\em any}
non-trivial Hamiltonian in three actions that is relevant to galactic
dynamics.

In principle, one would expand $f$ as a power series in its two variables,
and relate the resulting coefficients to a Taylor expansion of the 
Hamiltonian, as was done for the two-action case in \secref{mech:sec:general2d}. However, in the
absence of a suitable analytic case to study, we will proceed no further.
In the remainder of the section, the values of $\beta_n$ and $\lambda_n$ will be computed
directly from $\hessian$ by numerical means, which itself will be constructed
numerically from non-algebraic expressions for the frequencies and their derivatives. 

\subsection{\stackel\ potentials}
\label{mech:sec:stackel}

The only known integrable systems of three actions are those described
by \stackel\ potentials \citep[\S3.5.3]{bt08}. An exhaustive treatment
in the context of galaxy modelling is given in \cite{de-z-1}
and \cite{de-z-2}.

\stackel\ potentials are of particular interest to us because, in the
appropriate coordinate system, the Hamilton-Jacobi equation is
separable. Action-angle variables can therefore be defined for these
systems in terms of integrals over a finite path. Paul \stackel\ first
showed that the appropriate coordinate system is that of confocal
ellipsoidal coordinates, and that indeed this is the only coordinate
system in which the Hamilton-Jacobi equation separates
\citep[p.228]{de-z-1,bt08}. It is in keeping with this that we note that
Cartesian, spherical polar and cylindrical polar coordinates are
themselves limiting cases of these coordinates.

We refer the reader to \cite{de-z-1} for a detailed treatment of
confocal ellipsoidal coordinates. We merely note here that we
are adopting the \cite{de-z-1} notation convention, and that
we are restricting ourselves to considering only oblate
axisymmetric potentials. In this case, the coordinate system
becomes the prolate spheroidal coordinates
$(\lambda, \phi, \nu)$, where $(\lambda, \nu)$ are the roots
for $\tau$ of
\begin{equation}
{R^2 \over \tau + \alpha}
+ {z^2 \over \tau + \gamma} = 1,
\end{equation}
where $(\alpha, \gamma)$ are scaling constants, and 
$q = \sqrt{\gamma/\alpha}$ is the potential shape parameter,
and $(R, \phi, z)$ are the familiar cylindrical
polar coordinates.  The permitted range for $(\lambda, \nu)$ is given
by the inequality
\begin{equation}
-\gamma \le \nu \le -\alpha \le \lambda.
\end{equation}
In the meridonal plane $(R, z)$, the coordinates $(\lambda, \nu)$ define a set of
elliptic coordinates, while in the equatorial plane, $(\lambda, \phi)$ define
a set of polar coordinates. A plot of the curves of constant $(\lambda, \nu)$
in the meridonal plane is shown in Figure 24 of \cite{de-z-1}. Of note is that curves of
constant $\lambda$ are a family of confocal ellipses, and curves of constant
$\nu$ are a family of confocal hyperbolas, and that everywhere the two sets
of curves are orthogonal.

In limit of $\lambda \rightarrow -\alpha$, we note that the direction $\hat{\lambda}$
of increasing $\lambda$ is tangent with the radial vector
$\hat{R}$ in cylindrical polar coordinates, while the direction $\hat{\nu}$
is tangent with the axial vector $\hat{z}$. We also note that in the limit of $\lambda \gg
-\alpha$, the direction $\hat{\lambda}$ is tangent with the radial vector
$\hat{r}$ in spherical polar coordinates, while the direction $\hat{\nu}$
is tangent with the polar vector $\hat{\theta}$.

The character of orbits in these potentials is as follows.  The orbits
circulate in $\phi$. The orbits nutate in $\lambda$ between two apses,
where $-\alpha \le \lambda_\min \le \lambda \le \lambda_\max$. The
orbits bounce in $\nu$, with a floor of $\nu_\min=-\gamma$
corresponding to $z=0$, and an apex of $\nu = \nu_\max$ corresponding
to $z=\pm z_\max$.  Consecutive excursions in $\nu$ take place
sequentially above and then below the equatorial plane: this is a
consequence of the degenerate ellipsoidal coordinate system.

The limit of $\lambda \gg -\alpha$ will almost always apply for the
example orbits that we examine below.
We can qualitatively understand the meanings of the actions
$(J_\lambda, J_\phi, J_\nu)$ in this limit as follows.  $J_\lambda$ is a
generalization of the radial action $J_r$ in spherical systems. Thus,
an orbit with large $J_\lambda$ will be eccentric, while an orbit with
null $J_\lambda$ will be confined to an ellipsoidal shell centred on the origin.
An orbit with large $J_\nu$ will make excursions above and below the
equatorial plane, while an orbit with null $J_\nu$ is confined to the
equatorial plane. Since we are only considering axisymmetric potentials,
then $J_\phi = L_z$, the $z$-component of angular momentum, always.
Our procedure for calculating these actions, and their corresponding
angles, from conventional phase-space coordinates is discussed
further in \secref{mech:sec:stackmisalignment} below.

Finally, we mention the form of the potential. In prolate ellipsoidal coordinates, an oblate axisymmetric \stackel\ potential
takes the form \citep{de-z-2},
\begin{equation}
\Phi(\lambda, \nu) = -{(\lambda + \gamma)G(\lambda) - (\nu + \gamma)G(\nu)
\over \lambda - \nu},
\label{mech:eq:stackelpotential}
\end{equation}
where de Zeeuw's function $G(\tau)$ is determined once the density
profile $\rho(z)$ along the $z$-axis has been chosen
\citep[see e.g.~equation 23,][]{de-z-2}. Thus, the model is
completely specified upon choosing $\rho(z)$ and the scaling
parameters $(\alpha, \gamma)$.
%separability shown in \cite{de-z-3} and refs

\subsection{Galaxy models with \stackel\ potentials}

\cite{de-z-2} shows that if one requires the density everywhere to be
non-negative it is not possible to write down a \stackel\ model in
which the density $\rho(r)$ falls off with distance from  he $z$-axis more
rapidly than $r^{-4}$ as $r\rightarrow\infty$. This is because an
elementary density on the $z$-axis $\rho(z)=\delta(z - z_0)$ provides
an off-axis density term that falls as $r^{-4}$. This behaviour 
rules out many classes of galaxy models, including disks with
exponentially falling density profiles.
However, we can construct
%by means of an integral over the elementary
%densities $\delta(z-z_0)$,
models in which the density falls more
slowly than $r^{-4}$ as $r\rightarrow\infty$. In particular,
models with asymptotically flat rotation curves,
i.e.~those in which $\rho(r) \sim r^{-2}$, are allowed.

In the models used in this section, we specify the $z$-axis density profile
\begin{equation}
\rho_z(z) = {-\gamma \rho_0 \over (z^2 - \gamma)}
= {-\gamma \rho_0 \over \tau},
\label{mech:eq:rhozed}
\end{equation}
where we have made use of $z^2 = \tau + \gamma$. In this case, de Zeeuw's
function $G(\tau)$ can be written in closed form
\citep[equation 49,][]{de-z-2}. Models specified by \eqref{mech:eq:rhozed}
become spherical at large radii and have a rotation curve that
is asymptotically flat, with
\begin{equation}
\lim_{r\rightarrow\infty}v^2_c = -4\pi G \rho_0\gamma.
\end{equation}
In the core of these models, the surfaces of constant density are
approximately ellipsoidal, with axis ratio\footnote{
Equation 46 in \cite{de-z-2} presents a formula for this quantity, but
on inspection it must be wrong: it permits the density axis ratio $a_z/a_R$ to take
all values between $(0,\infty)$ for $q=(0,1)$,
and yet the formula is derived under the requirement
that the central density profile is always oblate. The correct form for the
expression is presented here, calculated directly from equation 35 in \cite{de-z-2}.
}
\begin{equation}
{a_z^2 \over a_R^2} = {2 q^2 \over (1 - q^2)^2} \left(1 - q^2 + q^2 \log q^2\right),
\end{equation}
where the central potential axis ratio $q = \sqrt{\gamma/\alpha}$. The models
are completely specified by choosing a shape with $q$, a mass scale with $\rho_0$,
and a distance scale with $\gamma$.
The combination of an asymptotic logarithmic `halo' and a flattened
`disk' in these models allow them to make a fair representation of the
observed properties of disk galaxies, although the lack of freedom in the
models severely restricts the shape of the flattened density profile
that can be achieved.

\begin{table}
  \centering
  \caption[Parameters for the \stackel\ potentials used in \chapref{chap:mech}]
{Parameters for the example \stackel\ potentials used in this chapter.}
  \begin{tabular}{l|lll}
    \hline
    & $\rho_0 / 10^{10} \msun / \kpc^3$ & $-\alpha / \kpc^2$ & $-\gamma / \kpc^2$ \\
    \hline\hline
    SP1 & $0.361$ & $29.64$ & $8.89\ttp{-3}$ \\
    SP2 & $0.266$ & $1.893$ & $0.322$ \\
    \hline
  \end{tabular}
  \label{mech:tab:stackpots}
\end{table}

\tabref{mech:tab:stackpots} describes two such models. SP1 was chosen
to simulate the highly flattened potential that may be felt in
proximity to a heavy disk. The density axis ratio is fixed to be
10 near the solar radius of $\rsun=8\kpc$, which is approximately
the same ratio as for the (exponential) thin disk profile of the Milky
Way \citep[see][Table 2.3]{bt08}. The specification is completed by
requiring the rotation curve to peak at $\rsun=8\kpc$ with a
circular speed $v_c = 240\kms$. 

\begin{figure}[\figplaceopts]
  \centerline{
    \includegraphics[width=\doublefigshrink\hsize]{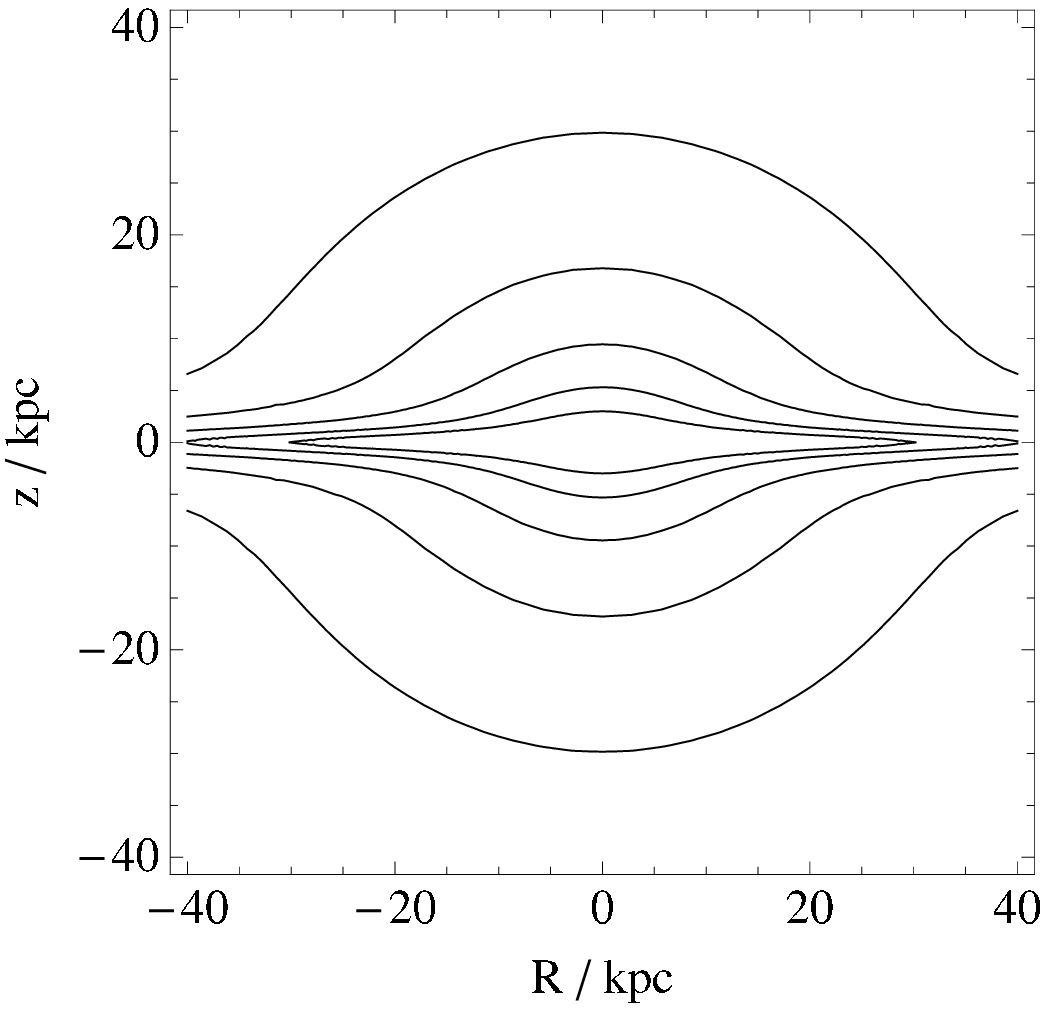}
    \qquad
    \includegraphics[width=\doublefigshrink\hsize]{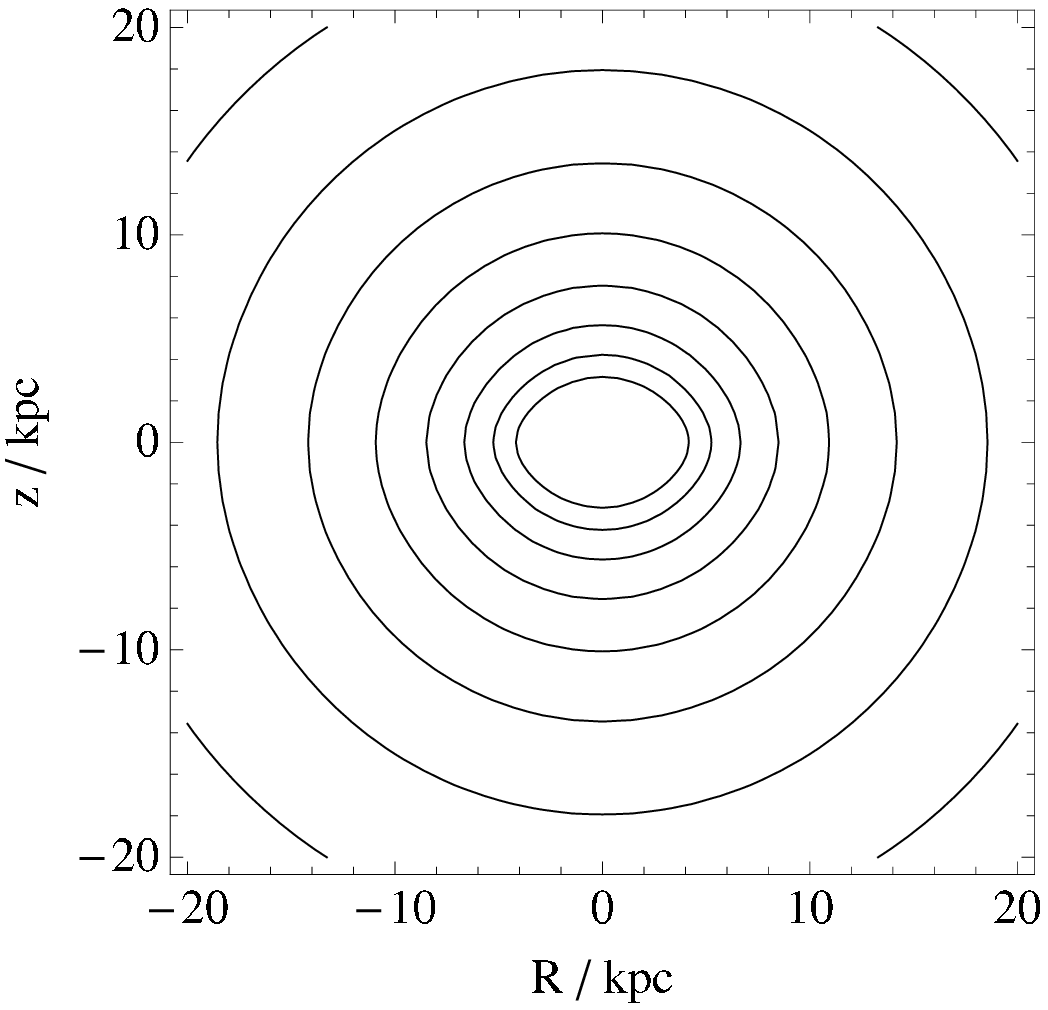}
  }
  \centerline{
    \includegraphics[width=\doublefigshrink\hsize]{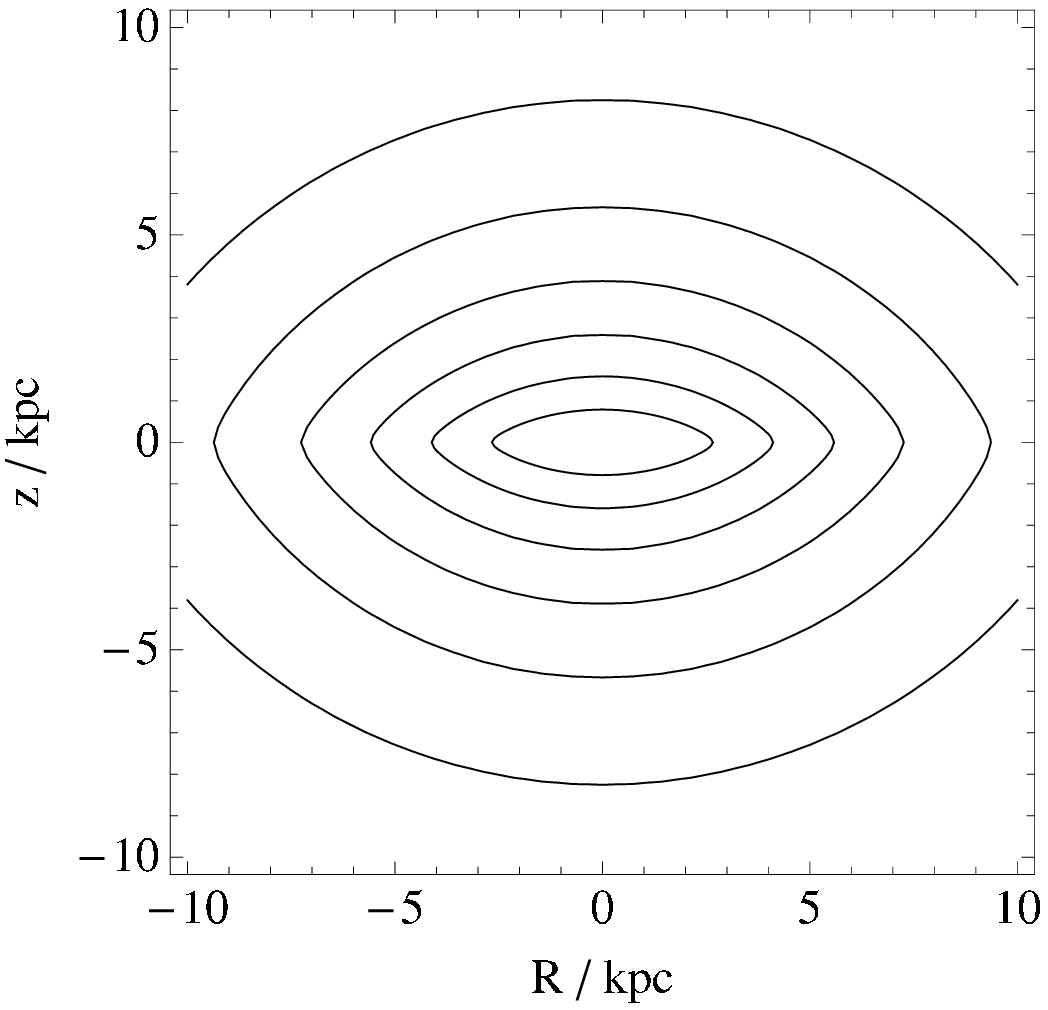}
    \qquad
    \includegraphics[width=\doublefigshrink\hsize]{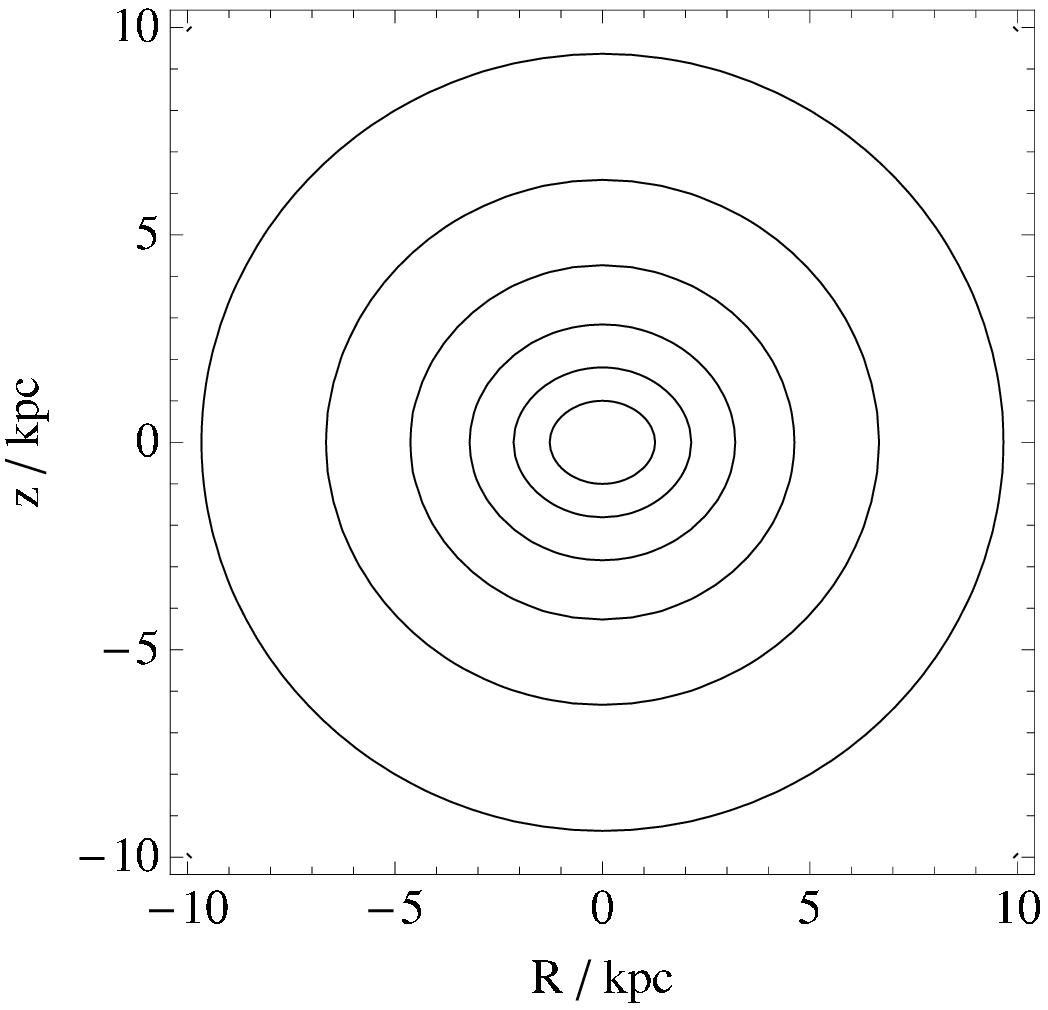}
  }
  \caption[Contours of equal density and potential, for the \stackel\ models
used in \chapref{chap:mech}]
{
\stackel\ models in use in this chapter. The left panels show the
flattened model SP1, while the right panels show the rounder
model SP2. The top panels show contours of $\log \rho/\rho_0$, while
the bottom panels show contours of the potential $\Phi$.
}
\label{mech:fig:stackel-pots}
\end{figure}

We note that the asymptotic circular velocity in this model is
$v_c = 42\kms$, which can be regarded as the halo contribution to the
circular speed. The model is too centrally concentrated, and the halo
contribution is too weak, to realistically model the Milky
Way. However, it is highly flattened, and so makes an interesting
example in which to study stream geometry. The rotation curve for
this model is plotted as the blue curve in
\figref{mech:fig:rotationcurve}, while contours of constant density
and potential for this model are shown in the left panels of
\figref{mech:fig:stackel-pots}.

The SP2 model was chosen to provide a force field better matched to
that of the Milky Way at those radii where streams are typically
observed, while still being somewhat flattened near the plane. The
model was required to have $v_c = 240\kms$ at $\rsun=8\kpc$, and $v_c \ge
235\kms$ at $R=20\kpc$. The rotation curve was also required to peak
at a radius not larger than $R = 5\kpc$. 
Finally, the model was required to be as flat as possible, subject
to satisfying these constraints. These requirement completely
specify the model, which has an asymptotic circular velocity of $v_c =
215\kms$. The disk contribution to this model is therefore arguably
rather weak, but it is in most respects a more reasonable model for
the Milky Way galaxy than is SP1. The rotation curve for SP2 is plotted as
the red curve in \figref{mech:fig:rotationcurve}, while the right panels of
\figref{mech:fig:stackel-pots} show contours of constant density and
constant potential for this model.

\subsection{Stream misalignment in \stackel\ potentials}
\label{mech:sec:stackmisalignment}

{\begin{figure}[\figplaceopts]
  \centerline{
    \includegraphics[width=\doublefigshrink\hsize]{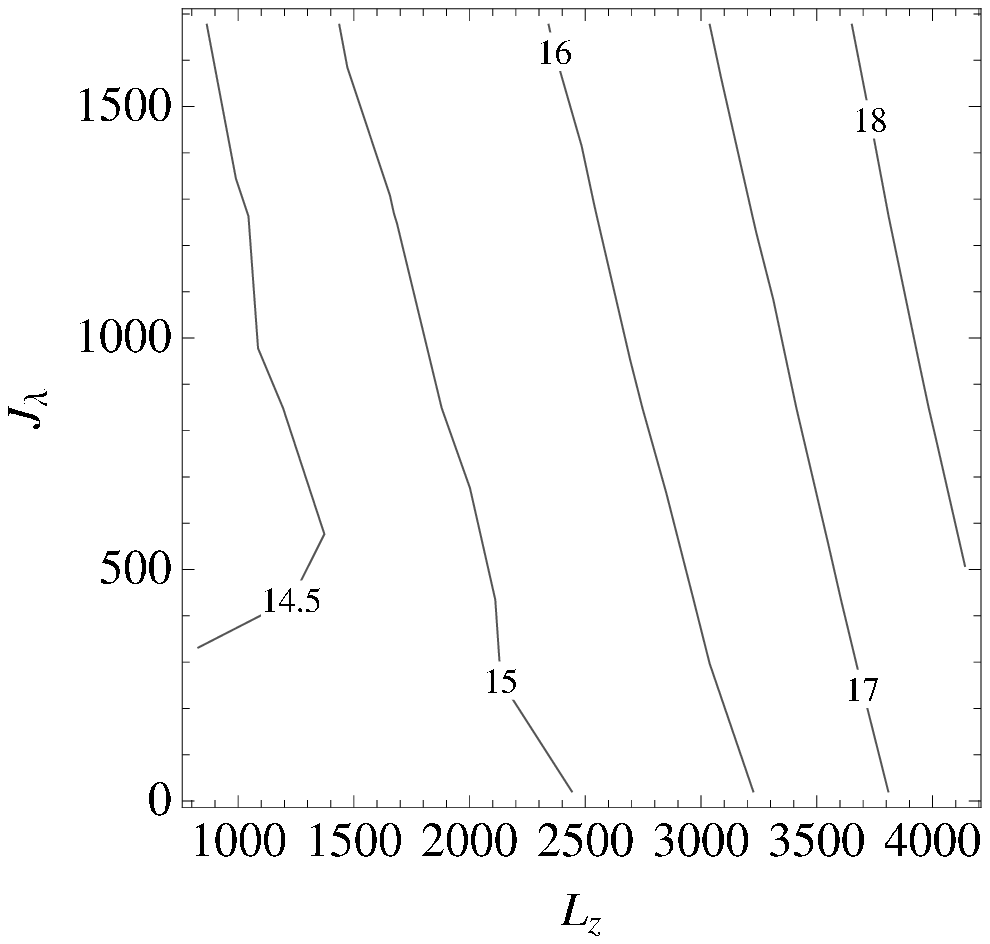}
    \includegraphics[width=\doublefigshrink\hsize]{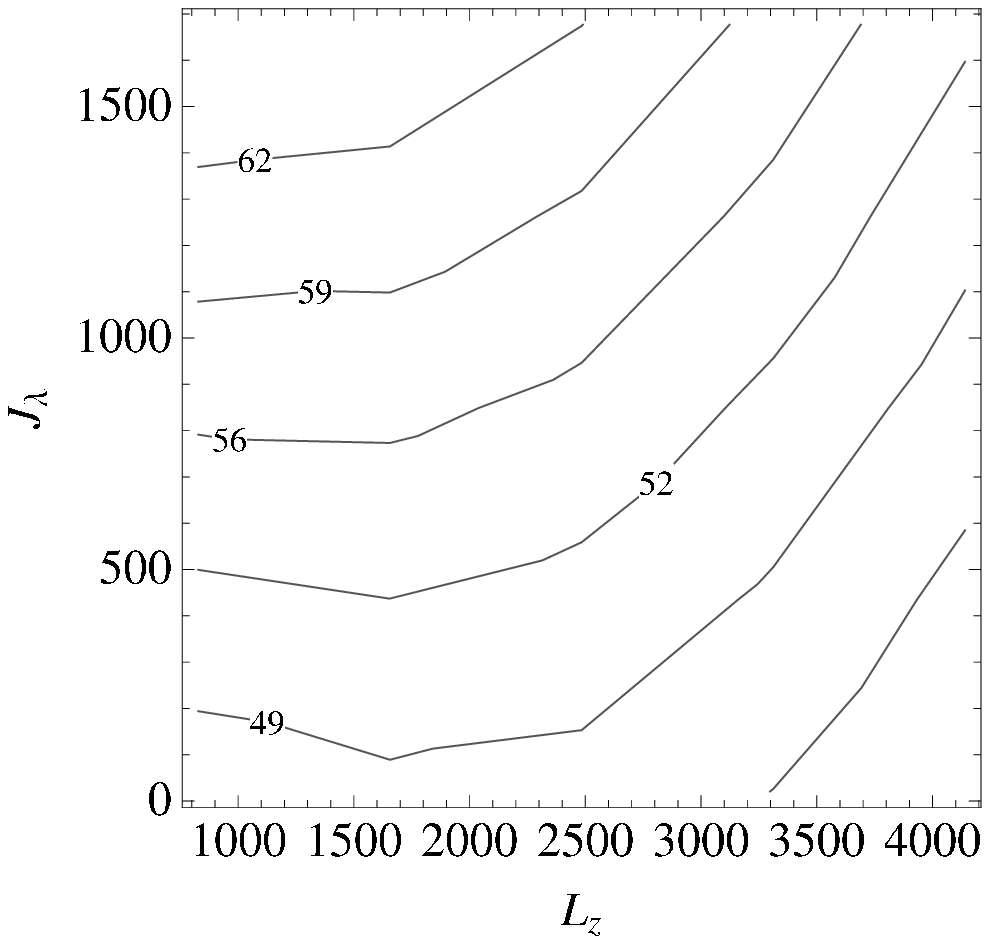}
  }
  \centerline{
    \includegraphics[width=\doublefigshrink\hsize]{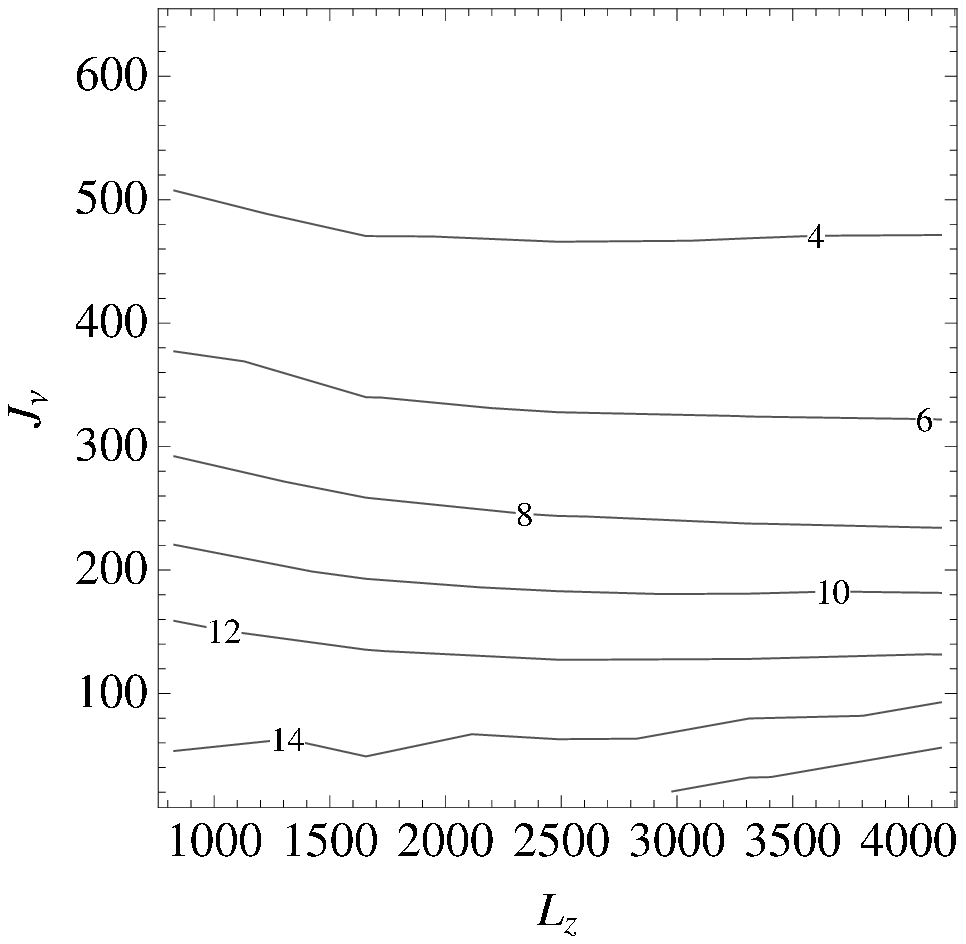}
    \includegraphics[width=\doublefigshrink\hsize]{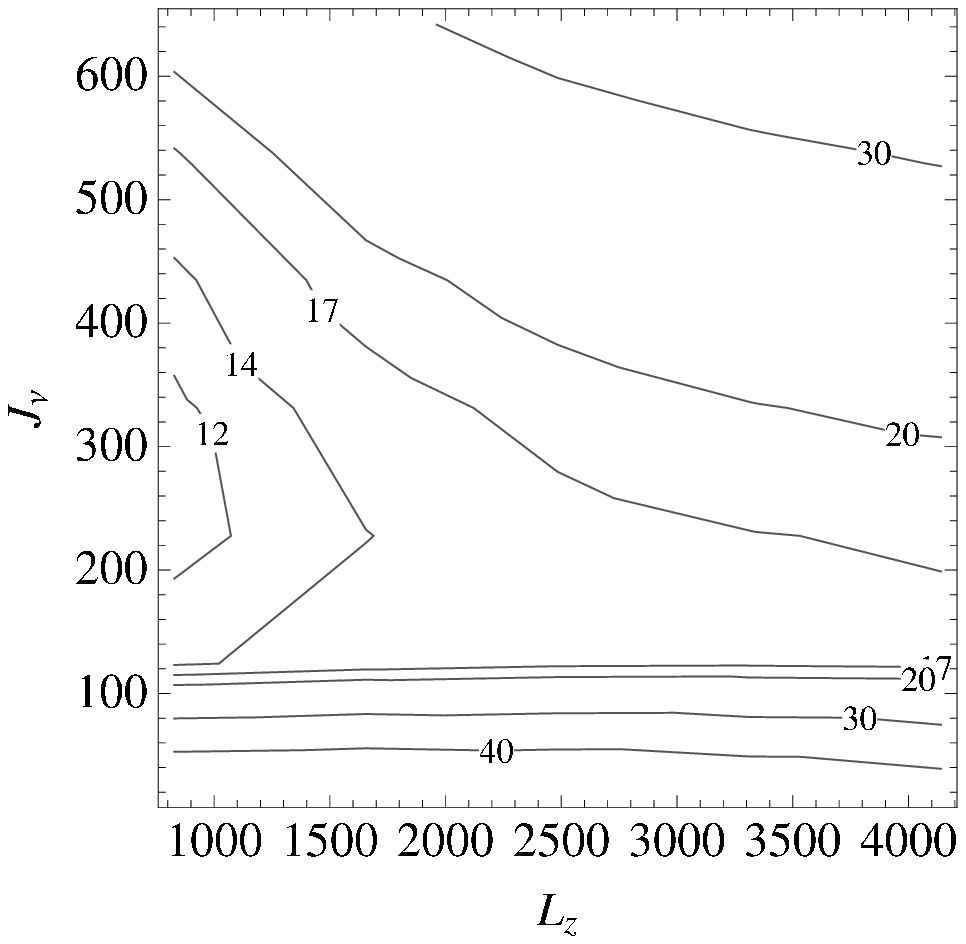}
  }
  \caption[Details of stream geometry for the \stackel\ potential SP1]
{Details of stream geometry for the \stackel\ potential SP1.
Left panels: contours for the misalignment angle $\vartheta$, in degrees, between the principal
eigenvector of $\hessian$ and $\vO_0$, shown as a function of $\vJ$. Right panels:
contours for the eigenvalue ratio $\lambda_1/\lambda_2$.
The top panels show the plane in action-space with $J_\nu = 20.7\kpc\kms$, while
the bottom panels show the plane in action-space with $L_z=414\kpc\kms$.
The range of actions covers a variety of interesting orbits: details of 
orbits at the extremes of the range are given in \tabref{mech:tab:sp1-orbit-extrema}
  }
  \label{mech:fig:sp1-hessian}
\end{figure}

We now consider the geometry of streams formed in the \stackel\
potentials SP1 and SP2. Although the form of the \stackel\ potential allows the
Hamilton-Jacobi equation to separate,
and thus allows the actions $\vJ$ to be
defined in terms of an integral over a single coordinate,
expressions for $\vJ$ do not exist in closed form. Instead, the integrals
in the expressions for $\vJ$ have to be evaluated numerically. Similarly,
expressions for both the frequencies $\vO$ and their derivatives
$\nabla_\vJ \vO$ can be written down, but not in closed form, and the
integrals that they contain must too be evaluated numerically.

The details of the expressions for $\vJ$, $\vO$ and $\nabla_\vJ \vO$,
and how to evaluate them, appear in \appref{appendix:stackel}. Here,
we simply note that having evaluated these quantities for a particular
orbit, the eigenvectors $\eigen_n$ and the eigenvalues $\lambda_n$ are
computed directly from the matrix $\hessian(\vJ_0) =
\left.\nabla_\vJ\Omega\right|_{\vJ_0}$ by standard methods.

\begin{table}
  \centering
  \caption[Coordinate extrema of selected orbits from \figref{mech:fig:sp1-hessian}]
  {The coordinate extrema of selected orbits from \figref{mech:fig:sp1-hessian},
    illustrating the variety of orbits covered by that figure.
    The actions are expressed in $\kpc\kms$, while the apses are in \kpc.}
  \begin{tabular}{lll|lll}
    \hline
    $J_\lambda$ & $L_z$ & $J_\nu$ & $R_p $ & $R_a$ & $|z|_{\text{max}}$ \\
    \hline\hline
    $20$ & $828$ & $20$ & $3.5$ & $5$ & $0.74$ \\
    $20$ & $4140$ & $20$ & $20$ & $26$ & $2.5$ \\
    $1680$ & $4140$ & $20$ & $14$ & $70$ & $7$ \\
    $1680$ & $828$ & $20$ & $1.75$ & $27$ & $3$ \\
    $414$ & $828$ & $20$ & $2$ & $10$ & $1.25$ \\
    $414$ & $4140$ & $20$ & $16$ & $38$ & $3.5$ \\
    $414$ & $4140$ & $640$ & $22$ & $56$ & $34$ \\
    $414$ & $828$ & $640$ & $3$ & $20$ & $17$ \\
    \hline
  \end{tabular}
  \label{mech:tab:sp1-orbit-extrema}
\end{table}}

We now examine the geometry of streams in the SP1 potential.  The left
panels of \figref{mech:fig:sp1-hessian} show contour plots of the
misalignment $\vartheta$ in angle-space between the principal
direction of $\hessian$ and the frequency vector $\vO_0$, where
$\vartheta$ is calculated from \eqref{mech:eq:misalignment}, as was
the case for systems of two actions. The right panels of the same
figure show contours of the ratio
$\lambda_1/\lambda_2$.  The range of actions shown in
these plots covers a variety of interesting orbits; the apses of the
orbits at the extremes of the range are described in \tabref{mech:tab:sp1-orbit-extrema}

As with the equivalent plots for the isochrone potential
(\figref{mech:fig:isochrone-hessian}), we see that the principal
direction of $\hessian$ is never perfectly aligned with $\vO_0$.  We
see that in this very flattened potential, those streams with low
$J_\nu$ have the greatest degree of misalignment, at about $\sim
15\deg$.  These orbits spend much of their time near the disk, and
never get very far from it.  The misalignment diminishes with
increasing $J_\nu$, falling to $\sim 4\deg$ for orbits with apses in
$z$ of some tens of \kpc.  Hence, in this very flattened potential,
there is much more prospect for dramatic misalignment than with the
isochrone potential, which \figref{mech:fig:isochrone-hessian} shows
to cause only comparatively smaller
misalignments. \figref{mech:fig:sp1-hessian} also shows that the
eigenvalue ratio $\lambda_1/\lambda_2 > 10$ everywhere for the SP1 potential; thus, we
conclude that highly elongated streams will form on all orbits which
permit a cluster to be disrupted.

\begin{figure}[\figplaceopts]
  \centerline{
    \includegraphics[width=\doublefigshrink\hsize]{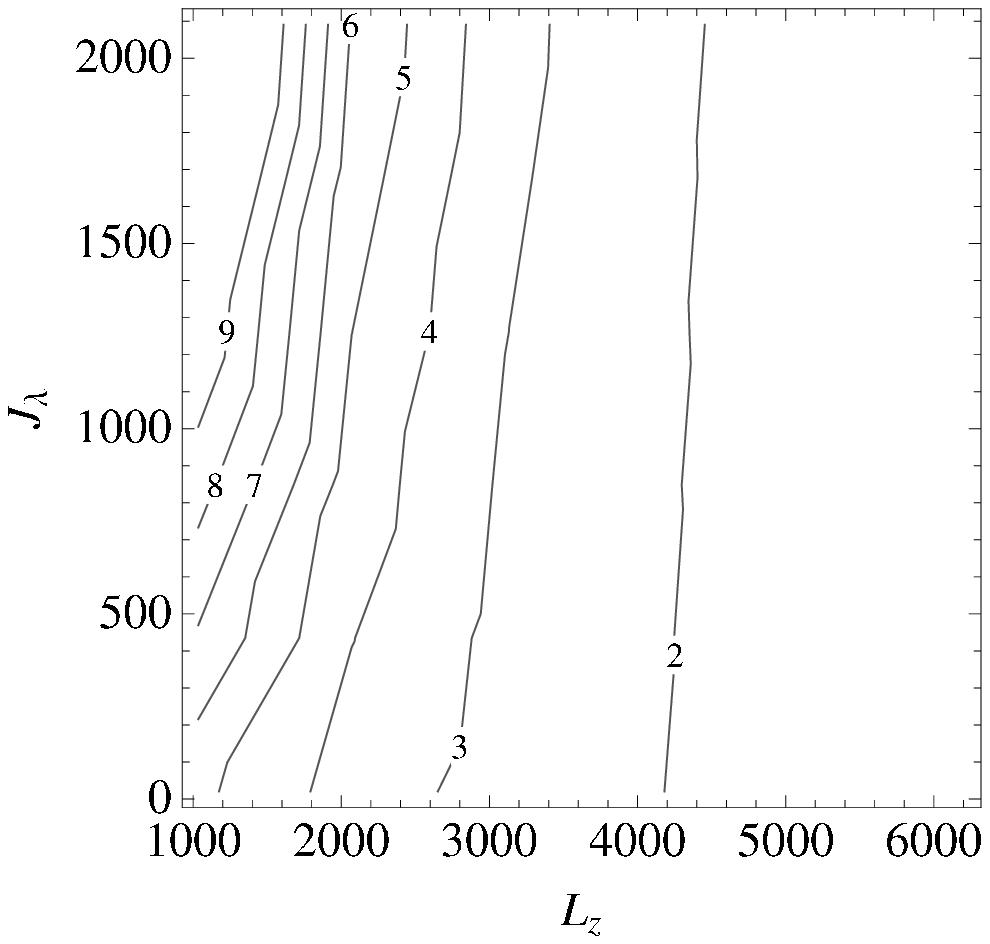}
    \includegraphics[width=\doublefigshrink\hsize]{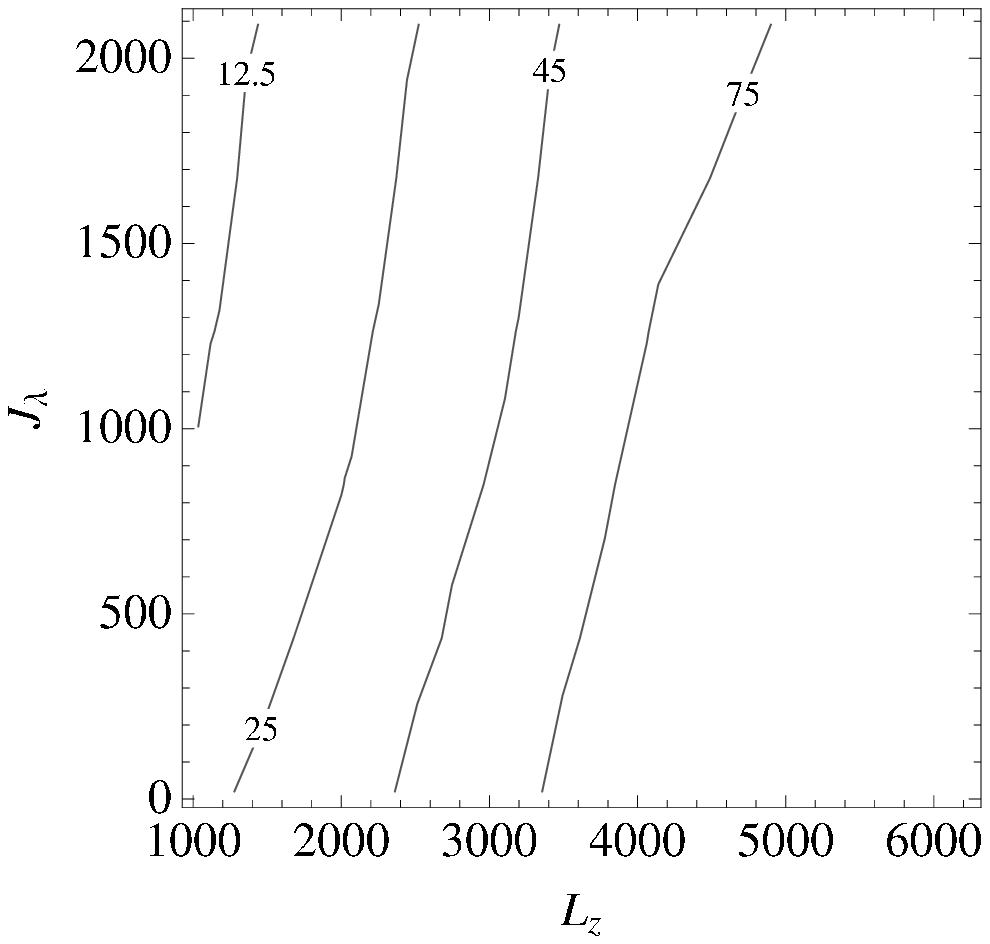}
  }
  \centerline{
    \includegraphics[width=\doublefigshrink\hsize]{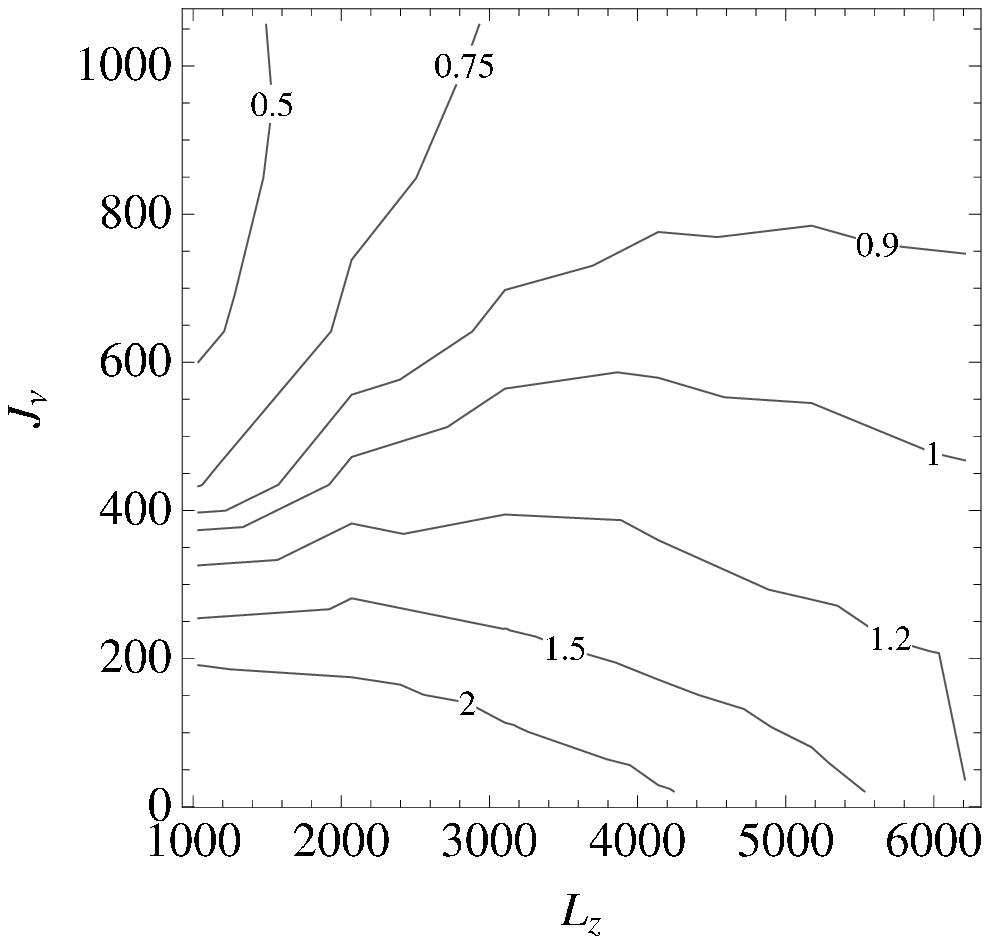}
    \includegraphics[width=\doublefigshrink\hsize]{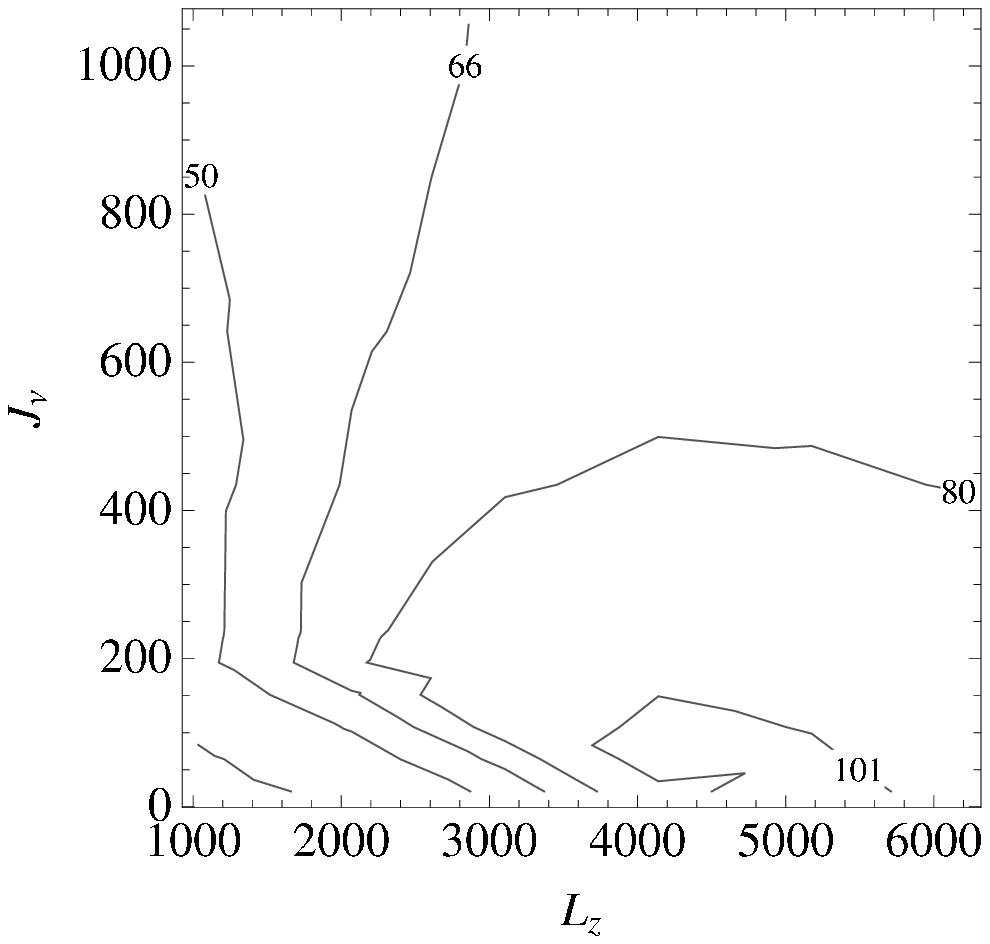}
  }
  \caption[Similar to \figref{mech:fig:sp1-hessian}, but for the \stackel\
    potential SP2]
{Similar to \figref{mech:fig:sp1-hessian}, but for the \stackel\
    potential SP2.  The top panels show the plane in action-space with
    $J_\nu = 20.7\kpc\kms$, while the bottom panels show the plane in
    action-space with $L_z=414\kpc\kms$.  The range of actions covers
    a variety of interesting orbits: the behaviour of orbits at the
    extremes of the range is shown in
    \tabref{mech:tab:sp2-orbit-extrema} }
  \label{mech:fig:sp2-hessian}
\end{figure}

\figref{mech:fig:sp2-hessian} shows the equivalent plot to
\figref{mech:fig:sp1-hessian}, but for the SP2 potential. Like with
the SP1 potential, the principal eigenvector is everywhere misaligned
with $\vO_0$. The misalignment is maximized for plunging orbits, and
minimized for highly inclined orbits.  However, the magnitude of the
misalignment is everywhere much smaller than is observed in the SP1
potential. For orbits that remain close to the plane, the
magnitude of the misalignment is comparable to or larger than that
seen in the isochrone potential (\figref{mech:fig:isochrone-hessian}).
For highly inclined orbits, the misalignment is slightly less than is seen
in the isochrone potential, for orbits with similar apses.

\figref{mech:fig:sp2-hessian} also shows that, like in the
SP1 potential, the ratio of the eigenvalues of $\hessian$ is
everywhere large. Hence, disrupted clusters should always
form elongated streams in this potential.

\begin{table}
  \centering
  \caption[Coordinate extrema of selected orbits from \figref{mech:fig:sp2-hessian}]
  {The coordinate extrema of selected orbits from \figref{mech:fig:sp2-hessian},
    illustrating the variety of orbits covered by that figure.
    The actions are expressed in $\kpc\kms$, while the apses are in \kpc.}
  \begin{tabular}{lll|lll}
    \hline
    $J_\lambda$ & $L_z$ & $J_\nu$ & $R_p $ & $R_a$ & $|z|_{\text{max}}$ \\
    \hline\hline
    $20$ & $1030$ & $20$ & $3.5$ & $5$ & $1.5$ \\
    $2100$ & $1030$ & $20$ & $2$ & $30$ & $7$ \\
    $20$ & $6200$ & $20$ & $27$ & $31$ & $3$ \\
    $2100$ & $6200$ & $20$ & $15$ & $65$ & $10$ \\
    $410$ & $1030$ & $20$ & $2.5$ & $10$ & $5$ \\
    $410$ & $6200$ & $20$ & $21$ & $40$ & $7$ \\
    $410$ & $1030$ & $1050$ & $3$ & $22$ & $20$ \\
    $410$ & $6200$ & $1050$ & $22$ & $50$ & $32$ \\
    \hline
  \end{tabular}
  \label{mech:tab:sp2-orbit-extrema}
\end{table}

In conclusion, we have found that in flattened \stackel\ potentials
with asymptotic logarithmic behaviour, streams will form
from disrupted clusters on all realistic orbits, and that
such streams will be generally misaligned with the orbits of
the stars that compose them. If we take such potentials
to be representative of the potential of our own Galaxy,
we must conclude that, generally, streams observed in and around the
Milky Way galaxy will not be perfectly aligned with orbits.

The precise behaviour of any given stream depends on both the
potential and the action-space distribution of its stars.
To proceed further we must again consider specific
examples, by means of N-body simulation.

\subsection{A stream in the \stackel\ potential SP1}

%SO1 ints {2.74965, 79.9857, 8.52861}
%SO1 acts freqs {{1.21887, 0.0989435}, {0.0942653, 0.0739833, 0.335126}}

%OS ints {4.77579,66.7292,74.9929}
%OS acts freqs {{7.25627,2.57663},{0.0635126,0.0439677,0.088977}}

%GD1 ints {4.30316,96.1624,55.2421}
%GD1 acts {{0.693134,1.72598},{0.0940358,0.0676554,0.137194}}

\begin{figure}[\figplaceopts]
  \centerline{
    \includegraphics[width=\doublefigshrink\hsize]{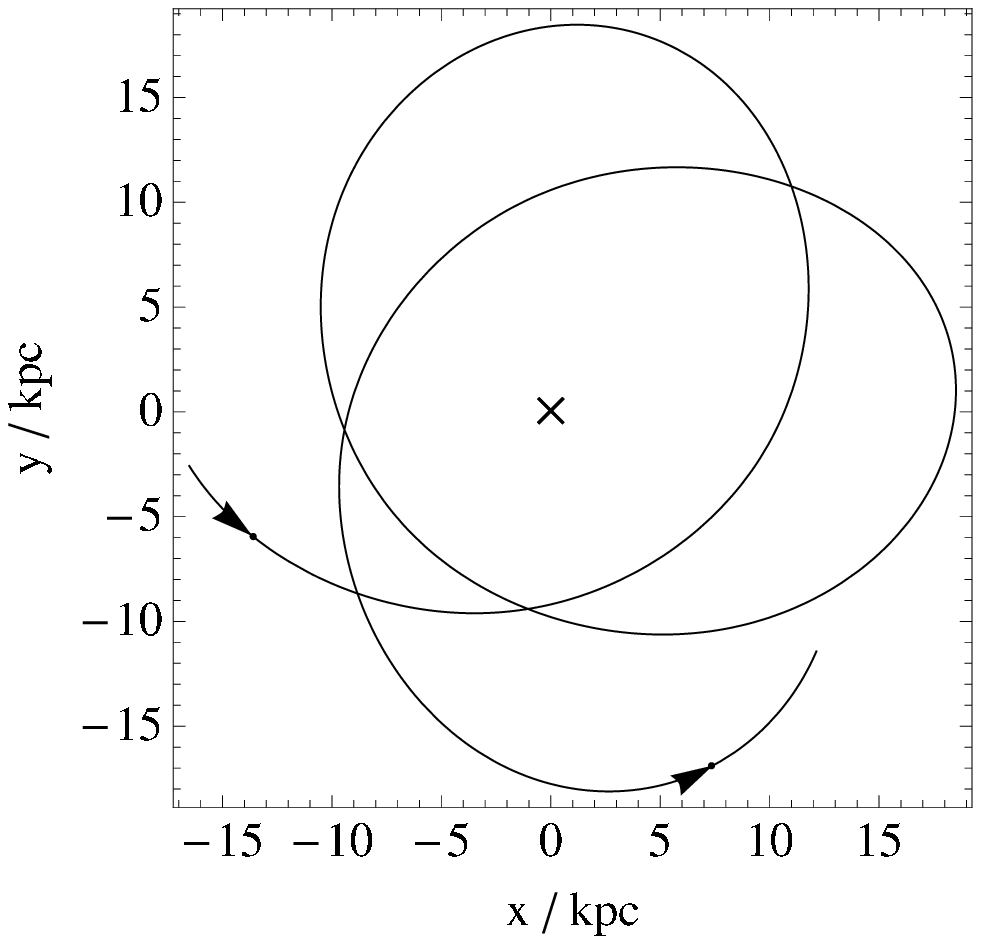}
    \includegraphics[width=\doublefigshrink\hsize]{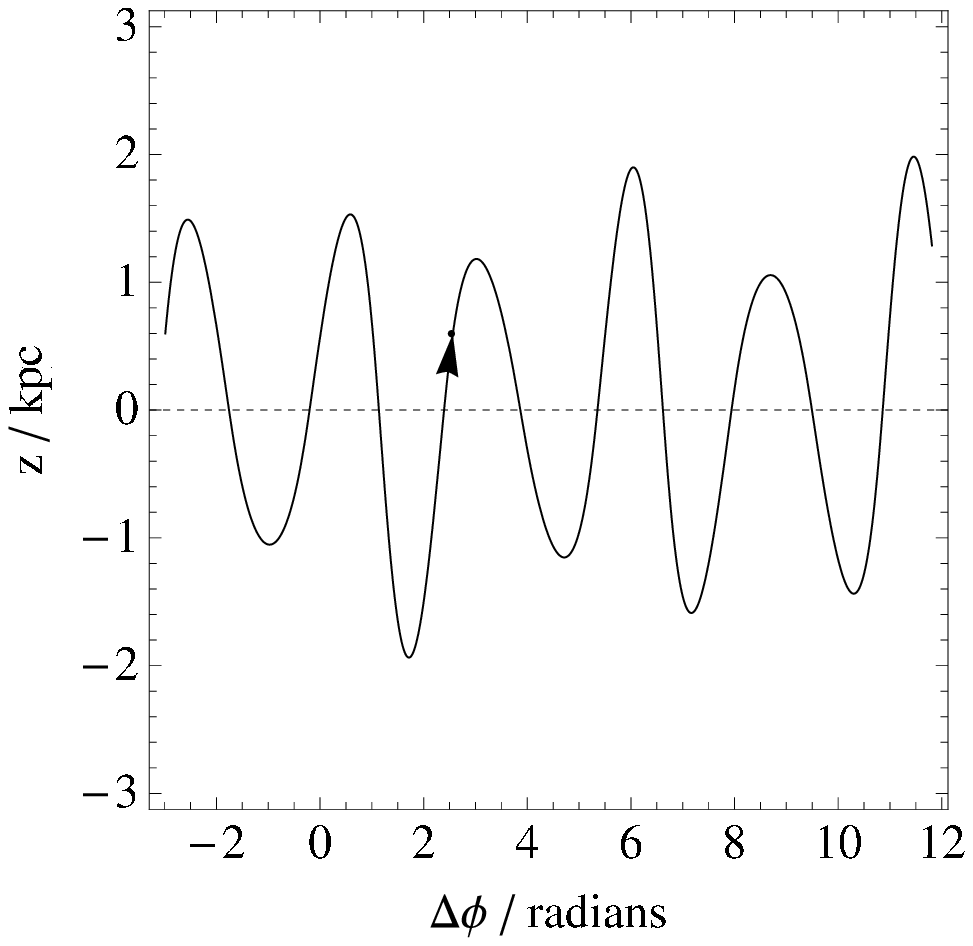}
  }
\centerline{
    \includegraphics[width=\doublefigshrink\hsize]{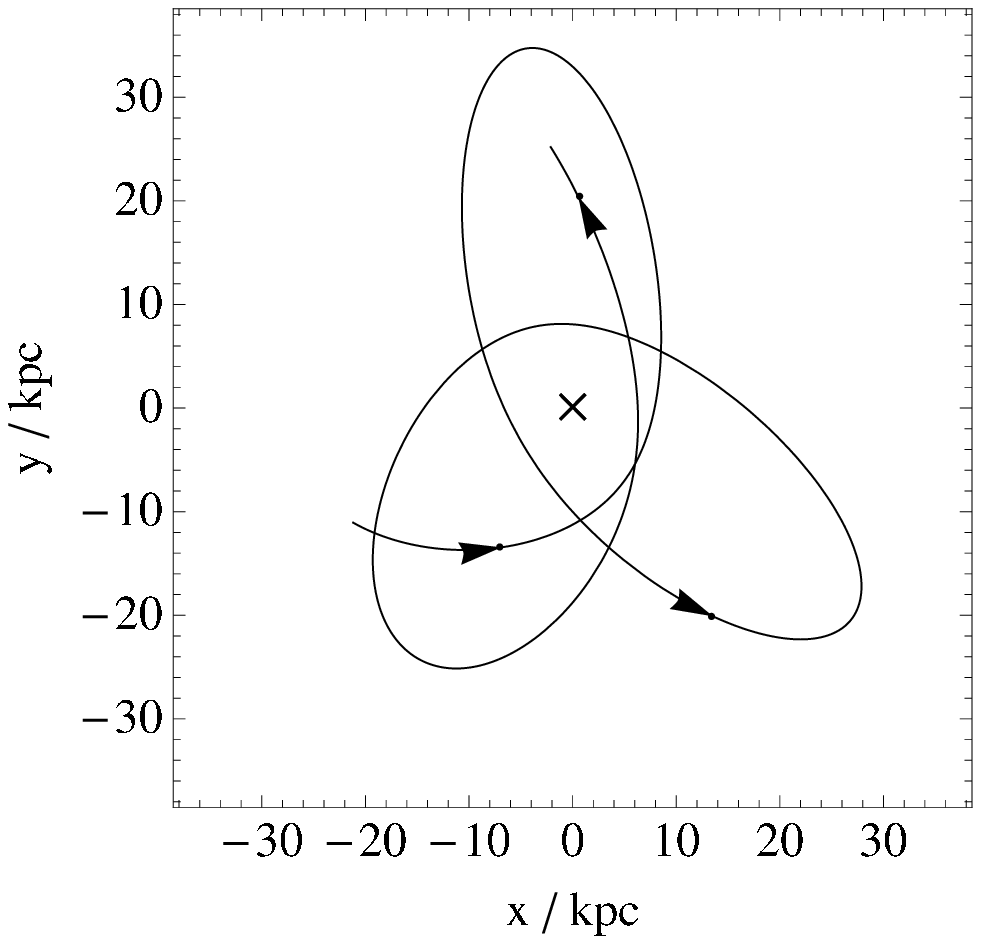}
    \includegraphics[width=\doublefigshrink\hsize]{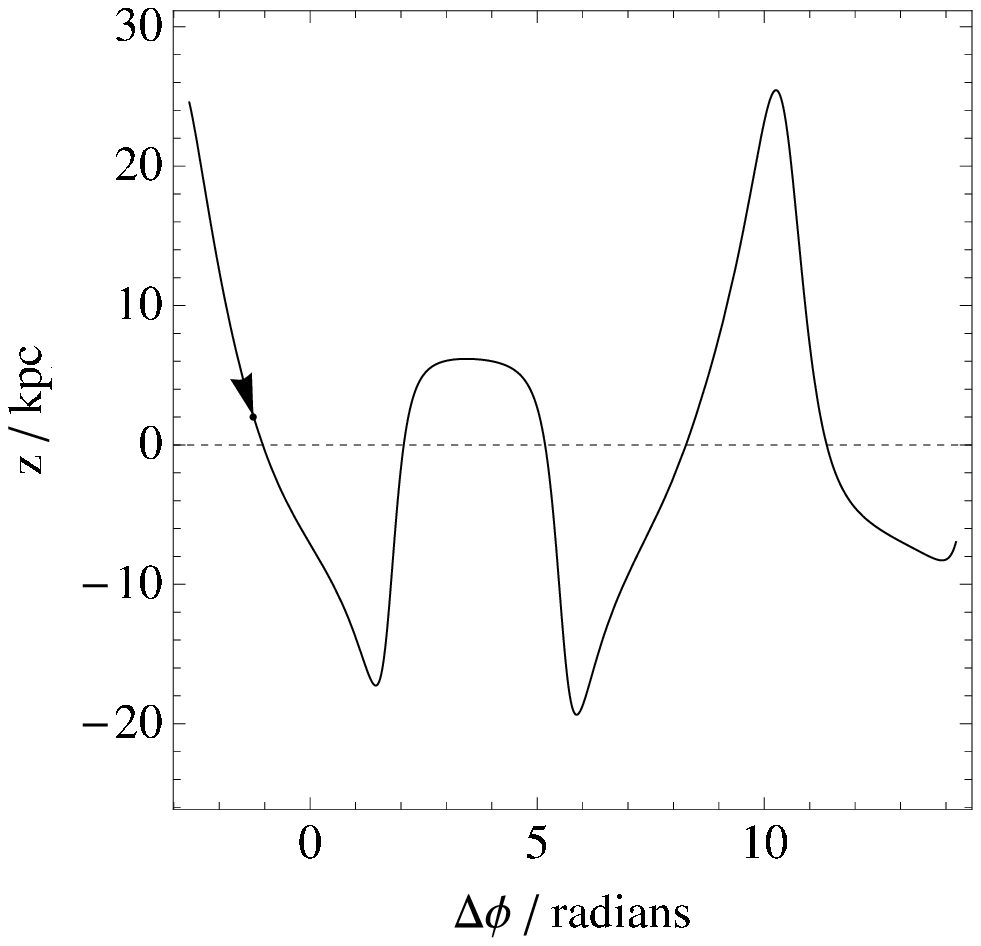}
  }
\centerline{
    \includegraphics[width=\doublefigshrink\hsize]{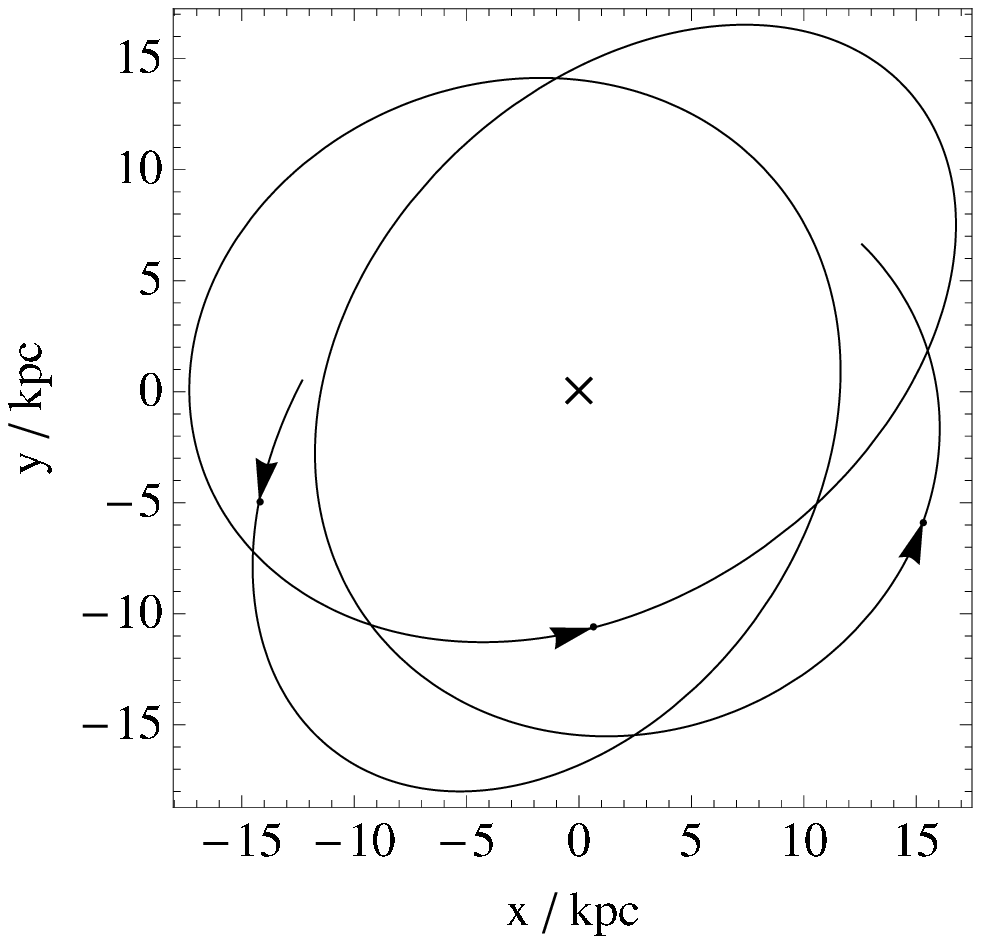}
    \includegraphics[width=\doublefigshrink\hsize]{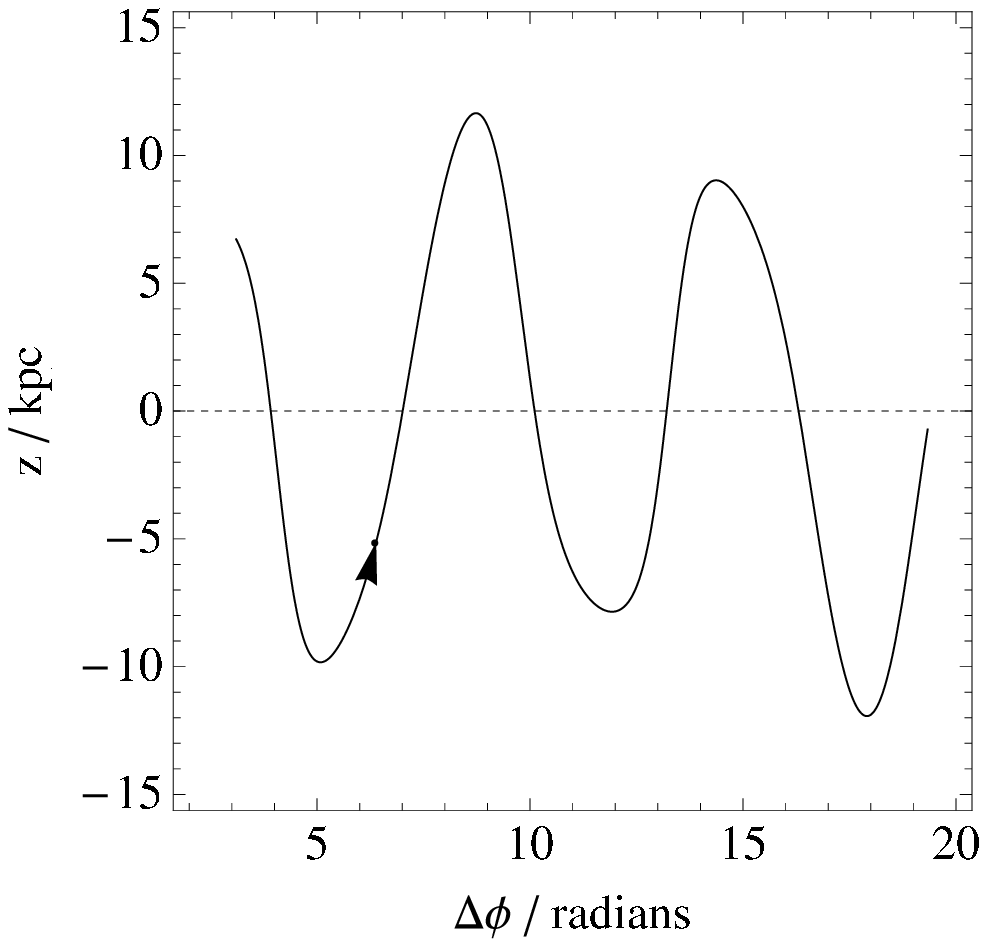}
  }
  \caption[Real-space trajectories for selected orbits in the \stackel\
potentials of \tabref{mech:tab:stackpots}]
{
The real-space trajectories of the orbits SO1 (top panels),
GD1 (middle panels) and OS1 (bottom panels), as described in
\tabref{mech:tab:stackorbs}. The trajectories shown were evaluated
in the \stackel\ potentials SP1 (for SO1) and SP2 (for GD1 and OS1),
which are described in \tabref{mech:tab:stackpots}.
}
\label{mech:fig:stack-orbit-examples}
\end{figure}

The top panels of \figref{mech:fig:stack-orbit-examples} show the
real-space trajectory of the orbit SO1 (\tabref{mech:tab:stackorbs})
in the \stackel\ potential SP1 (\tabref{mech:tab:stackpots}). This
orbit has apses of approximately $R=(8,18)\kpc$ in the galactic plane,
and $z=(-2,2)$ above and below the plane. It is thus fairly
representative of an eccentric orbit that might be occupied by a
globular cluster embedded in a galactic disk.

\begin{table}
  \centering
  \caption[Highlighted orbits in the potentials of \tabref{mech:tab:stackpots}]
  {Actions and apses for selected orbits in the \stackel\ potentials
    used in this chapter. Since all
    examples are in axisymmetric potentials, $J_\phi = L_z$ generally.
    The trajectories of these orbits are illustrated in \figref{mech:fig:stack-orbit-examples}.
}
  \begin{tabular}{l|lll|lll}
    \hline
    & $J_\lambda / \kpc\kms$ & $L_z/\kpc\kms$ & $J_\nu/\kpc\kms$ &
    $R_p$ & $R_a$ & $|z|_{\text{max}}$\\
    \hline\hline
    SO1 & $252.3$ & $2618.$ & $20.5$ & $8$ & $18$  & $2$ \\
    GD1 & $143.4$ & $2871.$ & $357.3$ & $10$ & $19.5$ & $12$ \\
    OS1  & $1502.$ & $2391.$ & $533.4$ & $6$ & $37$ & $26$ \\
    \hline
  \end{tabular}
  \label{mech:tab:stackorbs}
\end{table}

The cluster model C5 (\tabref{mech:tab:clusters}) describes a King
model specified for the orbit SO1 according to the schema of
\secref{mech:sec:clusters}. The model has the same mass and
profile parameter as does C1, and is very similar in all other
attributes, because the orbit SO1 is not
entirely dissimilar to the orbit I4 for which C1 was specified.
A $10^4$ particle realization of the C5 was made by
random sampling of the King model distribution function.
This cluster was placed close to apocentre on the orbit
SO1 and evolved forward in time by the \fvfps\ tree
code, with time step $\d t=\tdyn/100$ and softening
parameter $\epsilon$ as specified in \tabref{mech:tab:clusters}.
The total period of the simulation was $2.15\Gyr$, or 7
complete radial oscillations.

\subsubsection{Action-space distribution}
\label{mech:sec:stackeldistribution}

\begin{figure}[\figplaceopts]
  \centerline{
    \includegraphics[width=\doublefigshrink\hsize]{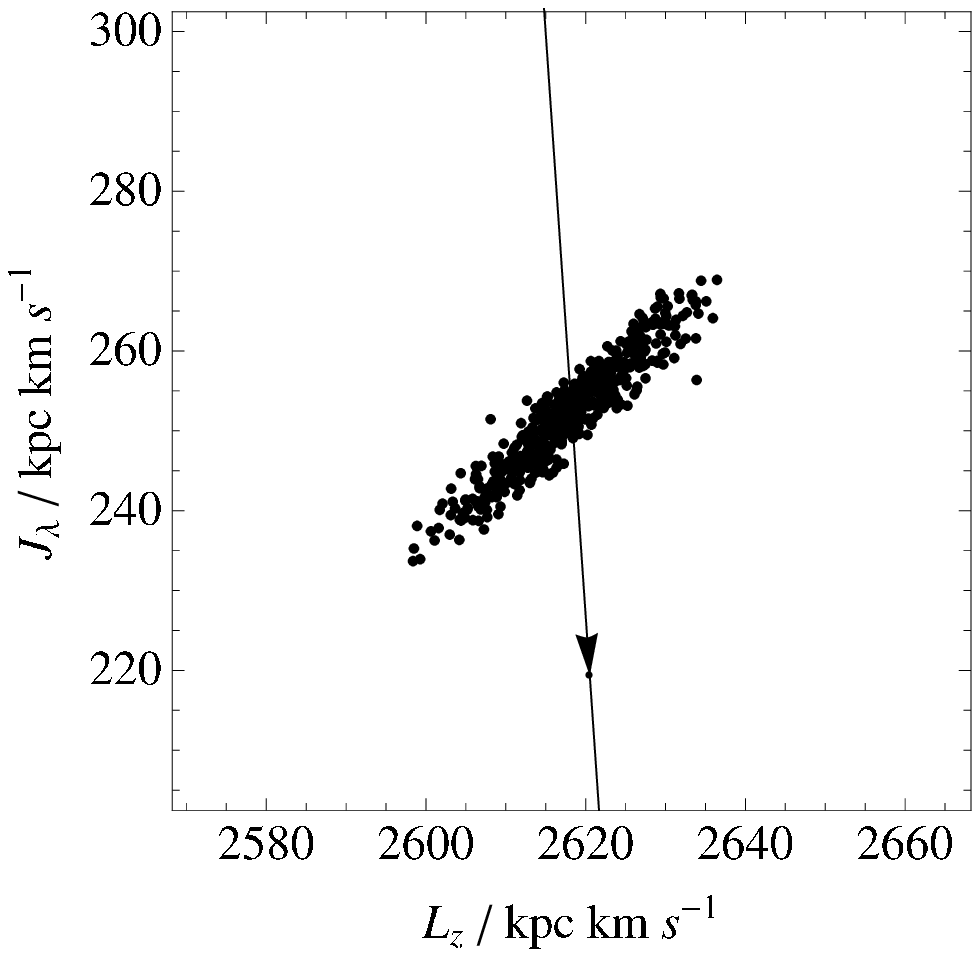}
    \includegraphics[width=\doublefigshrink\hsize]{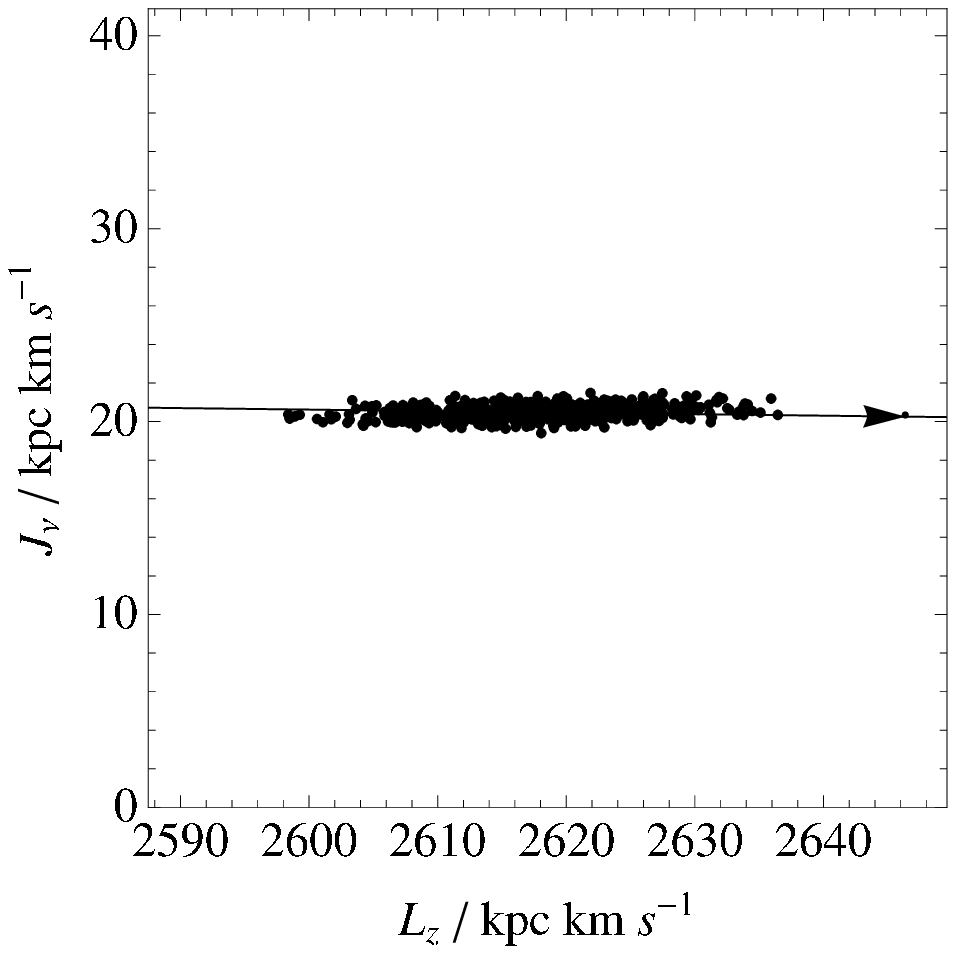}
  }
  \centerline{
    \includegraphics[width=\doublefigshrink\hsize]{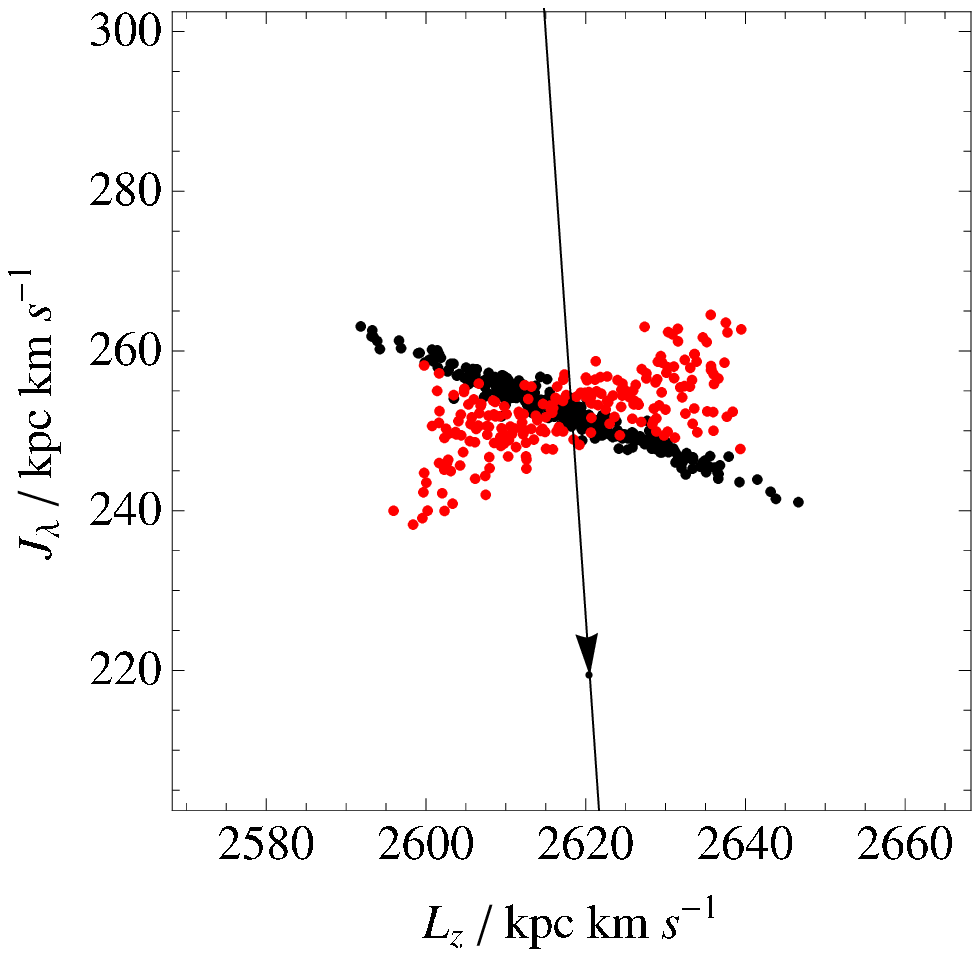}
    \includegraphics[width=\doublefigshrink\hsize]{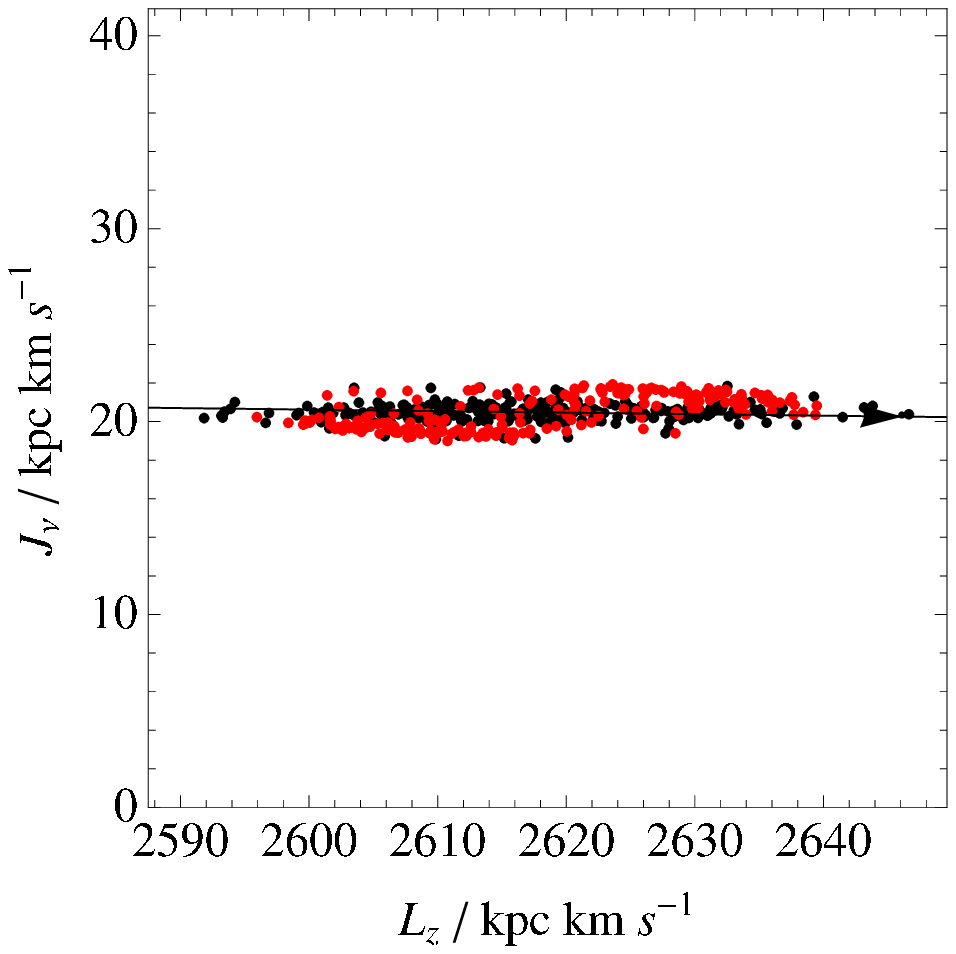}
  }
  \centerline{
    \includegraphics[width=\doublefigshrink\hsize]{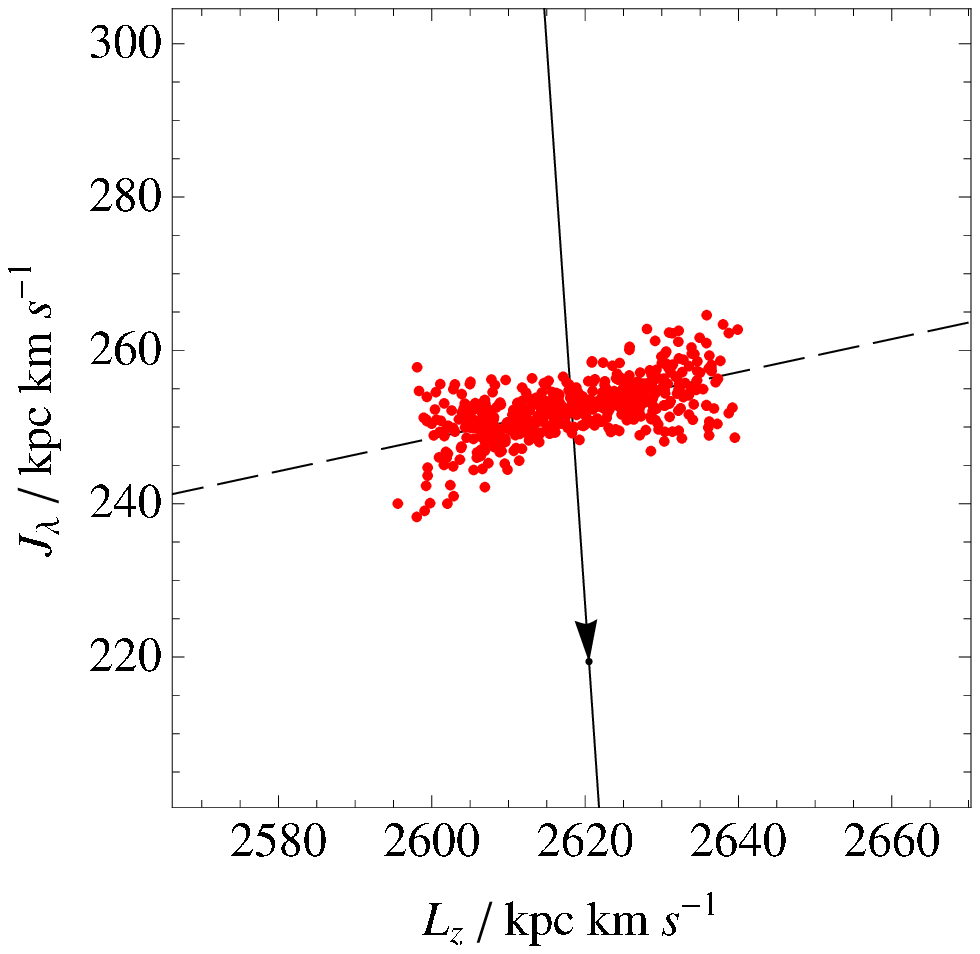}
    \includegraphics[width=\doublefigshrink\hsize]{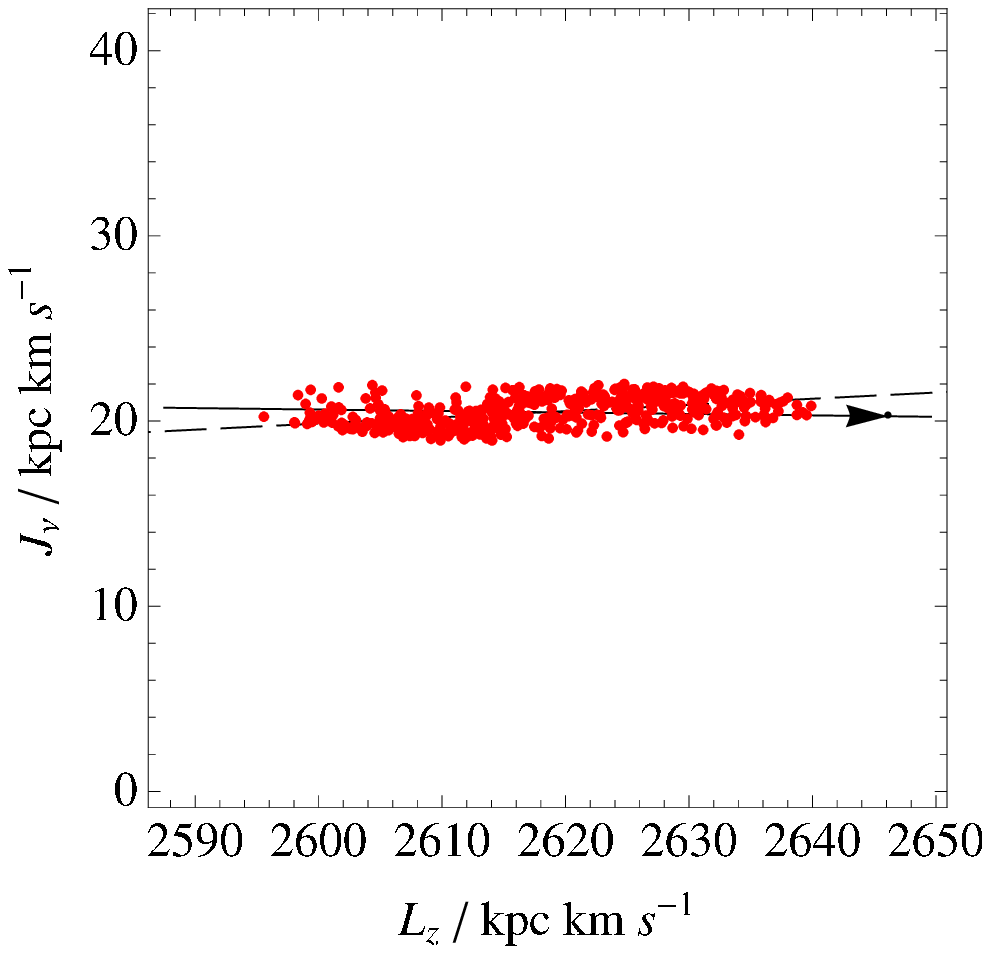}
  }
  \caption[Action-space distribution for the simulated cluster
C5, on the orbit SO1 in the \stackel\ potential SP1,
at various points in time]{Action-space distribution for the simulated cluster
C5, on the orbit SO1 in the \stackel\ potential SP1,
at various points in time. The left panels show the orthographic
projection of distribution onto the $(L_z,J_\lambda)$ plane, while
the right panels show a similar projection onto the
$(L_z,J_\nu)$ plane. Distributions are shown at the following times:
top panels, the first pericentre passage;
middle panels, the subsequent apocentre passage;
bottom panels, the 7th apocentre passage.
Black particles are bound to the cluster, while
red particles orbit free in the host potential.
The dashed lines in the bottom panels are lines that have
been least-squares fitted to the particles.
%%The red ellipses in the bottom panels are the projected
%%images of circles mapped by $\hessian$: the image in
%%the bottom-right panel is plotted with 1/3 the scale of the
%%image in bottom-left panel.
}
\label{mech:fig:stack-xt-jays}
\end{figure}

\figref{mech:fig:stack-xt-jays} shows the evolution of the action-space
distribution of this simulated cluster with time. Each row of panels
shows the distribution at a different point in time. The left panel
in each row shows the orthographic
projection of the actions onto the $(L_z,J_\lambda)$ plane, while
the right panel of each row shows a similar projection onto the
$(L_z,J_\nu)$ plane. In all panels, the appropriate projection
of the mapped frequency vector, $\hessian^{-1}\vO_0$, is shown
as an arrowed black line.

The top row shows the actions when the cluster is near to its first
pericentre passage. In the left panel, the distribution is somewhat
flattened, and oriented with positive gradient in $\Delta J_\lambda/
\Delta L$. This behaviour is analogous with that seen in the top-right
panel of \figref{mech:fig:nbody-run1}: the motion of the cluster
is predominantly in $(\lambda, \phi)$, thus $(J_\lambda, L_z)$ are good
proxies for the radial action $J_r$ and angular momentum $L$ respectively.
$J_\lambda$ and $L_z$ are therefore highly correlated for a cluster near apsis on this orbit,
in analogy with the mechanism described by \eqref{mech:eq:gradient} in \secref{mech:sec:disruption}.

Conversely, the distribution in the right panel, while being narrow,
is oriented almost exactly along $\hat{L}_z$. We can understand the
shape as follows. For this orbit, which is confined to be close to the
plane, $J_\nu \sim J_z/2$, where the factor of 2 appears because
$J_\nu$ is defined on a path restricted to only one side of the
plane. $J_z$ can be estimated by close analogy with
\eqref{mech:eq:deltajr}. Hence, the spread in $J_\nu$ for a cluster of
velocity dispersion $\sigma$ is approximately
\begin{equation}
\Delta J_\nu \simeq {1\over 2} \Delta J_z \sim {1 \over 2\pi} \delta p_z \Delta z
\simeq {1 \over 2\pi} \sigma \Delta z.
\end{equation}
By analogy with \eqref{mech:eq:djr/dl} we find
\begin{equation}
{\Delta J_\nu \over \Delta L_z} \sim {\Delta z \over 2\pi R_p},
\label{mech:eq:djnu/dl}
\end{equation}
where $R_p$ is the galactocentric pericentre radius in cylindrical
coordinates. Evaluating this expression for the orbit SO1 gives
$\Delta J_\nu/\Delta L_z \sim 0.04$, which we see from the top-right
panel of \figref{mech:fig:stack-xt-jays} is close to exact.

The flat orientation of the top-right panel we explain by pointing out
that, as the top panels of \figref{mech:fig:stack-orbit-examples}
show, the motion in $\nu$ in this example is almost decoupled from the
radial motion. This means that the $\nu$ coordinate need not be at
apsis when the cluster is at pericentre, and thus the arguments of
\secref{mech:sec:disruption}, which force a correlation between
$J_\lambda \sim J_r$ and $L_z \sim L$ near pericentre, do not
apply. For an orbit in which $J_\nu$ is more strongly coupled to the
radial motion, we would expect to see the characteristic tilting of
the $(L_z, J_\nu)$ distribution near pericentre and apocentre, as a
correlation between $J_\nu$ and $L$ is forced.

The middle panels of \figref{mech:fig:stack-xt-jays}
show the action-space distribution at the subsequent apocentre passage.
Bound and unbound particles are shown in black and red, respectively.
The action-space structure of the unbound particles in the left panel bears
striking similarity to that shown in \figref{mech:fig:nbody-run1}, as might
be expected when $(J_\lambda, L_z)$ make good proxies for $(J_r,L)$.
The same physical principle for the disruption of the cluster applies
here as it does in the spherical case; that is, the particles will
escape the cluster through the Lagrange points $L_1$ and $L_2$ by first
travelling radially, so the action-space distribution will be
compressed in this direction. Thus, the range of $\Delta L_z$ and $\Delta J_\nu$
for the unbound stars is about the same as in the top panels, but
the range of $\Delta J_\lambda$ is markedly less.

The bottom panels of \figref{mech:fig:stack-xt-jays} show action-space
distribution at the 7th apocentre passage. The structure is
essentially the same as that of the middle panels, except that all
particles are now unbound.  Also plotted in each of the bottom panels
is a dashed line, which has been least-squares fitted to the unbound
particles.
%, and a red ellipse, which is the orthographic projection of
%the image of a circle, in the plane of that panel, mapped under
%$\hessian$.
We note that the image of the frequency vector and the dashed line are
highly misaligned in the bottom-left plot;
%and that the red ellipse
%does not predict a strong distortion in this plane;
hence, we expect the stream to be significantly misaligned with $\vO_0$ in
%the $(\theta_\phi, \theta_\lambda)$ plane in
angle-space.

\begin{figure}[\figplaceopts]
\centering{
    \centerline{
    \includegraphics[width=\doublefigshrink\hsize]{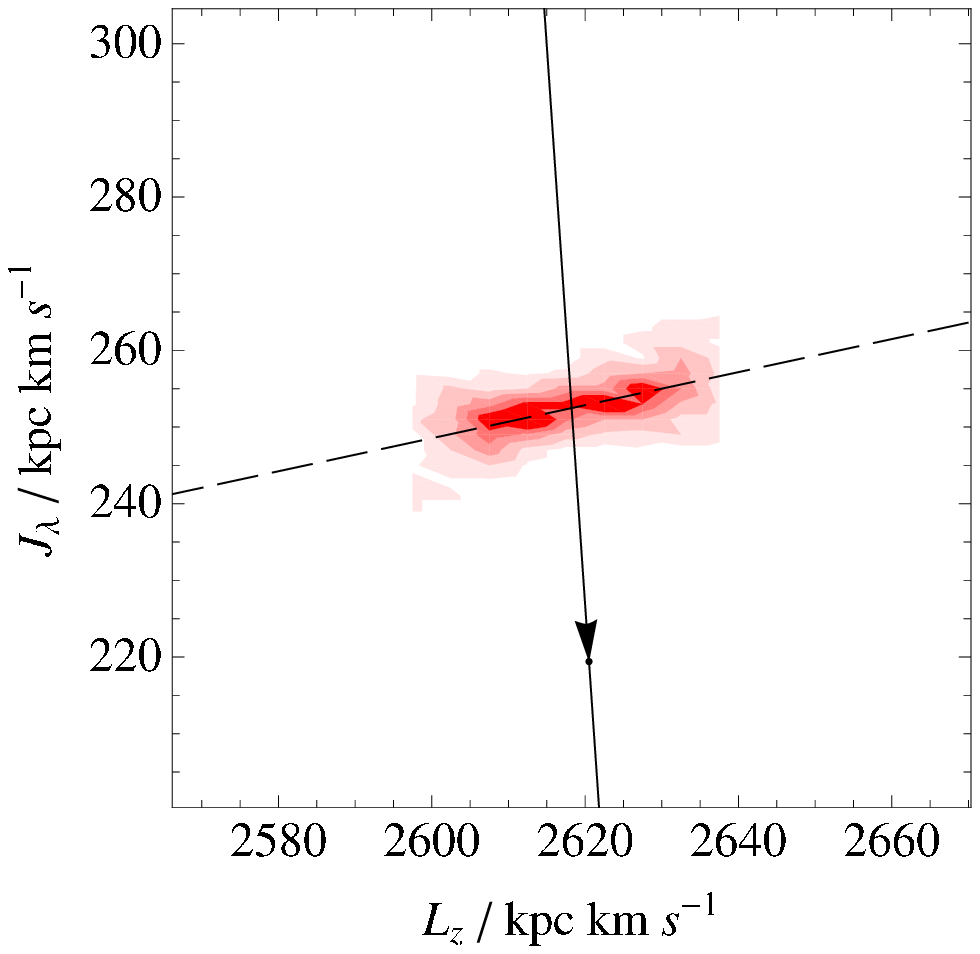}
    \qquad
    \includegraphics[width=\doublefigshrink\hsize]{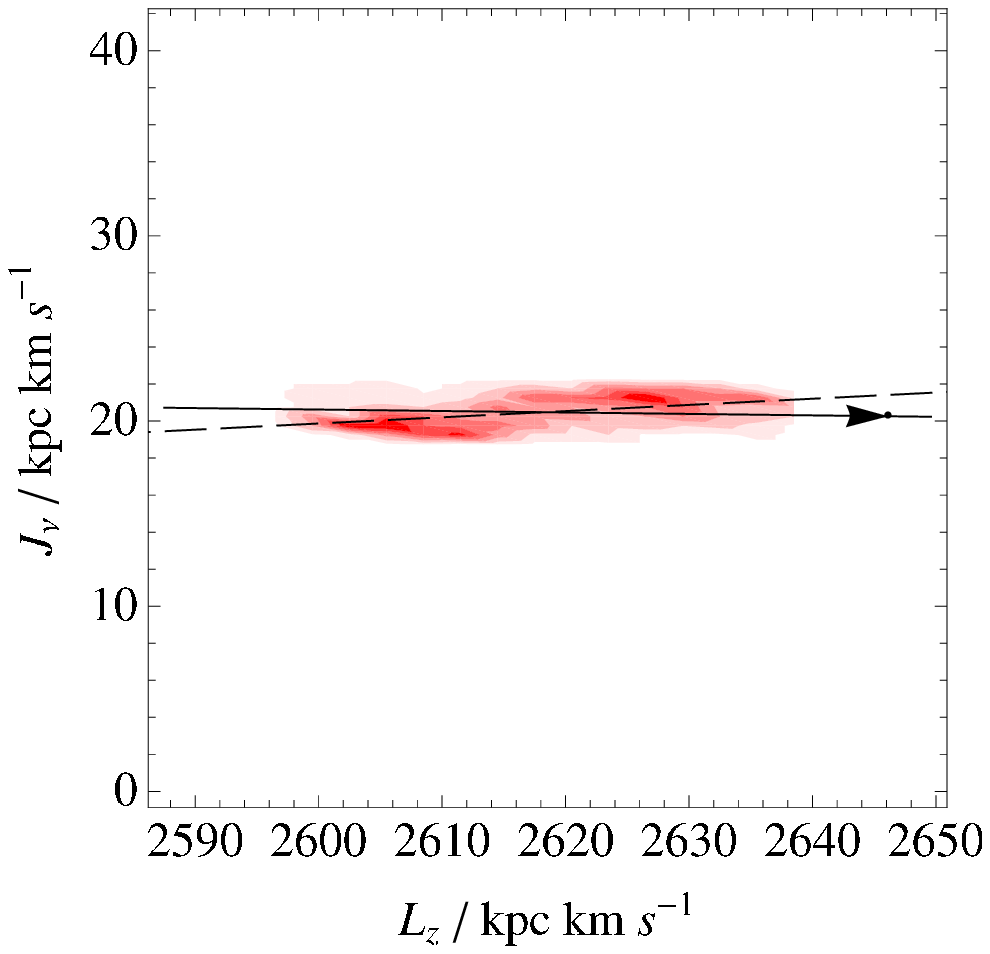}
    }
}
\caption[Particle-density plots for selected panels from \figref{mech:fig:stack-xt-jays}]
{
  Plots showing the (column) density of particles in
  action-space, corresponding to (left)
  the bottom-left and (right) the bottom-right panels of
  \figref{mech:fig:stack-xt-jays}.
  The density was estimated 
  by placing particles into bins of width $\sim 2 \kpc \kms$.
  Darker shading represents regions of higher particle density, with the
  edges of the shaded regions
  representing contours of constant density.
  As was true of \figref{mech:fig:nbody-particle-density},
  no significant sub-structure in the density field is visible
  that is not attributable to sampling noise.
}
\label{mech:fig:stack-xt-particle-density}
\end{figure}

In order to better examine the variation in particle density within
the bottom panels of \figref{mech:fig:stack-xt-jays},
\figref{mech:fig:stack-xt-particle-density} shows equivalent plots of
particle density. As was seen in \figref{mech:fig:nbody-particle-density},
the plots reveal the particles to be primarily concentrated in the
lobes of the distribution, but they fail to reveal any significant
additional structure in the particle-density field. 

% We also note
% that, although the image of the frequency vector and the stream are
% more closely aligned in the bottom-right plot, the almost
% perpendicular misalignment between the image of the frequency vector
% and the principal direction of $\hessian$ implies that the frequency
% vector has a substantial component that is not aligned with the
% principal direction in the $(\theta_\phi,\theta_\nu)$
% plane. Conversely, we expect the strongly-distorting $\hessian$ to map
% the particle distribution to a shape that {\em does} point along the
% principal direction in that plane. Hence, we expect the stream to be
% misaligned with $\vO_0$ in the $(\theta_\phi,\theta_\nu)$ plane also.

In conclusion, we find that in very flattened potentials, disrupted clusters form
an action-space distribution that is wholly analogous with that
found for disrupted clusters in spherical potentials. In the following
section, we test the ability of a simple straight-line model of the
action-space distribution to predict the track of the stream.

\subsubsection{The effects of disk shocks}

Unlike with a spherical potential, an axisymmetric potential allows
for tidal forces other than those felt during pericentre passage to
act upon an orbiting cluster.
In particular, the passage of a cluster through a massive galactic disk
will subject a cluster to a tidal force that is of comparable magnitude to
that felt when close to pericentre.\footnote{To see
this, we note that the tidal force in the direction $\vecthat{x}$
at any given point,
\begin{equation}
{\d \vect{F} \over \d x} = (\vecthat{x} \cdot \nabla) \, \vect{F}
= - (\vecthat{x} \cdot \nabla) \, \nabla\phi \sim 4 \pi G \rho,
\end{equation}
is proportional to $\rho$, which we understand to be the mean density
of tide-inducing matter interior to the point under consideration.
Consider then the example of a Milky Way cluster undergoing both pericentre
passage, and encountering disk shocks, at a galactocentric radius of
$R_0 \sim 8\kpc$.  The disk density in the solar neighbourhood is
$\rho_d \sim 0.1 \msun \pc^{-3}$ \citep[Table~1.1]{bt08}. Meanwhile,
the mean density interior to the Sun required to give a circular speed
of $v_c \sim 240\kms$ at $R_0$ is $\rho_g \sim 0.06 \msun
\pc^{-3}$. Hence, the magnitude of the tidal effects felt during disk
passage is indeed comparable to those felt during pericentre passage.
}

The tidal stress imposed on a cluster at pericentre has a tensile
component, which acts to strip stars from
the cluster. Conversely, the tidal stress imposed by a disk passage is entirely
compressive in nature. Hence, stars are not actively
stripped from a cluster during a disk passage. Instead, the action of
such `disk shocks' is to heat the cluster, perhaps repopulating the
outer edges of the cluster, the stars from which were stripped during
a previous pericentre encounter \citep[\S5.2a]{spitzer87}.

The net effect of disk shocks on the stripping process is a faster and
more complete disruption of the cluster than would take place for an
unshocked cluster exposed to equivalent pericentric tidal
stress. Since the vast majority of stars continue to be stripped at
pericentre even when the effect of disk shocks is significant, the
gross action-space distribution resulting from the stripping of a
shocked cluster will remain as previously described.  However, since the
disk shocks act to increase the velocity dispersion of the cluster
between stripping events, it is likely that the wings of the
resulting action-space distribution will be populated with more stars
than would otherwise have been the case.

\subsubsection{Predicting the stream track}

\begin{figure}[\figplaceopts]
  \centerline{
    \includegraphics[width=\doublefigshrink\hsize]{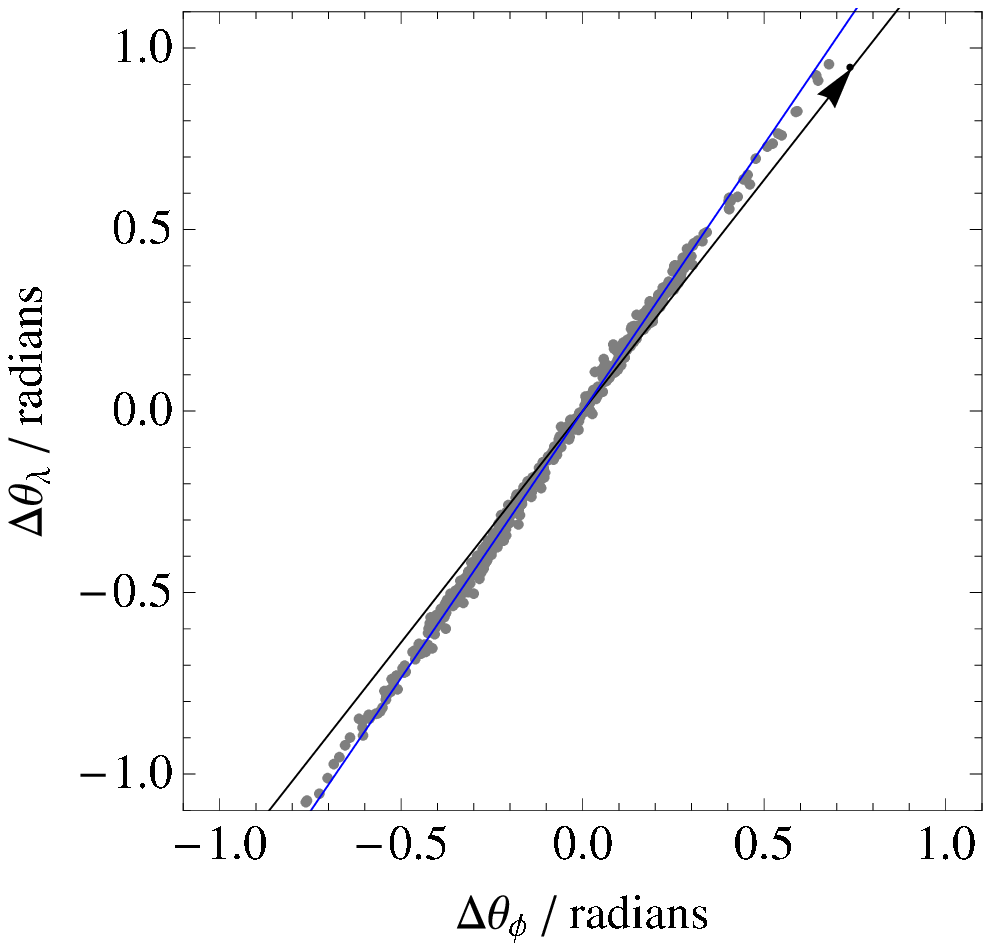}
    \qquad
    \includegraphics[width=\doublefigshrink\hsize]{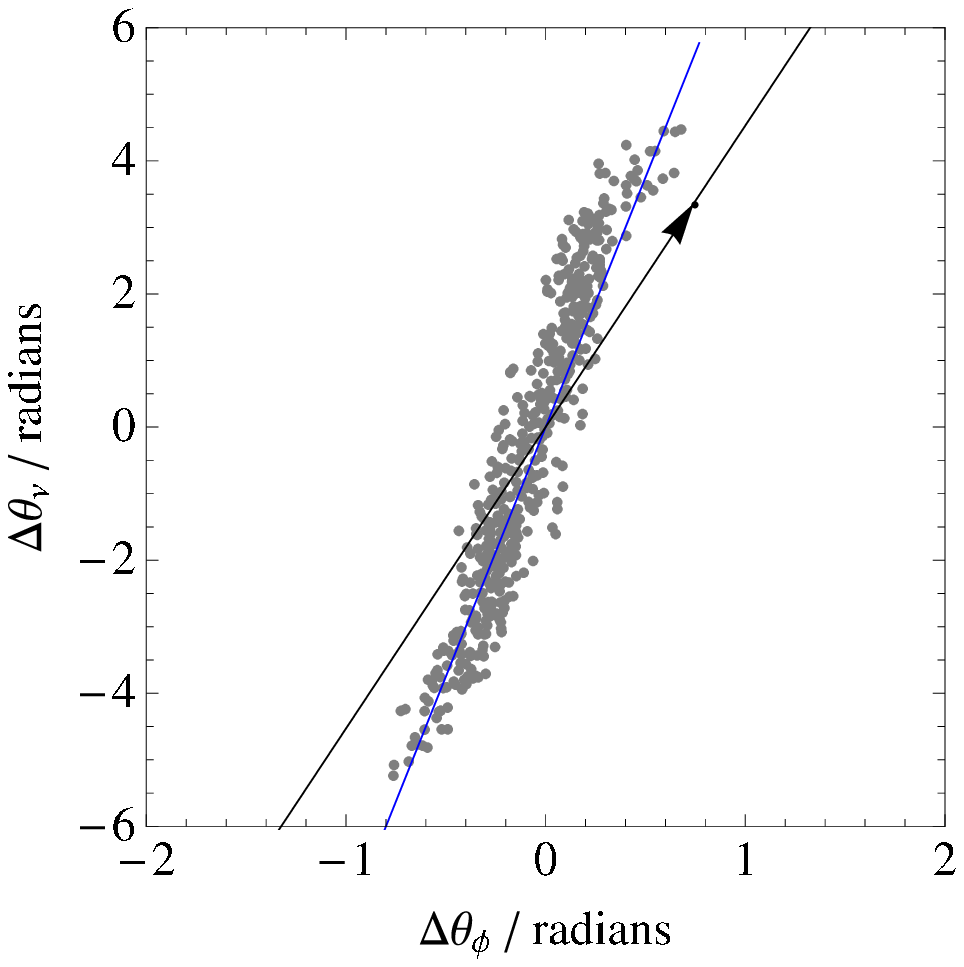}
  }
  \caption[
Angles for the N-body cluster shown in the bottom panels
    of \figref{mech:fig:stack-xt-jays}]
{ Angles for the N-body cluster shown in the bottom panels
    of \figref{mech:fig:stack-xt-jays}. The blue line shows the
    predicted stream, resulting from the mapping of the dashed
    line in bottom panels of \figref{mech:fig:stack-xt-jays}. The blue
    line is clearly a much better representation of the stream than is
    $\vO_0$, represented by an arrowed black line.  }
\label{mech:fig:stack-xt-angles}
\end{figure}

\figref{mech:fig:stack-xt-angles} shows the angle-space configuration
for the simulated cluster near its 7th apocentre passage. The grey
particles are for angles that have been computed directly from the
output of the N-body simulation. The arrowed black line is $\vO_0$,
while the blue line is the map of the dashed-line from
\figref{mech:fig:stack-xt-jays}. In both projections, the blue line is
clearly a much superior match to the data than is the orbit.

\begin{figure}[\figplaceopts]
  \centerline{
    \includegraphics[width=\doublefigshrink\hsize]{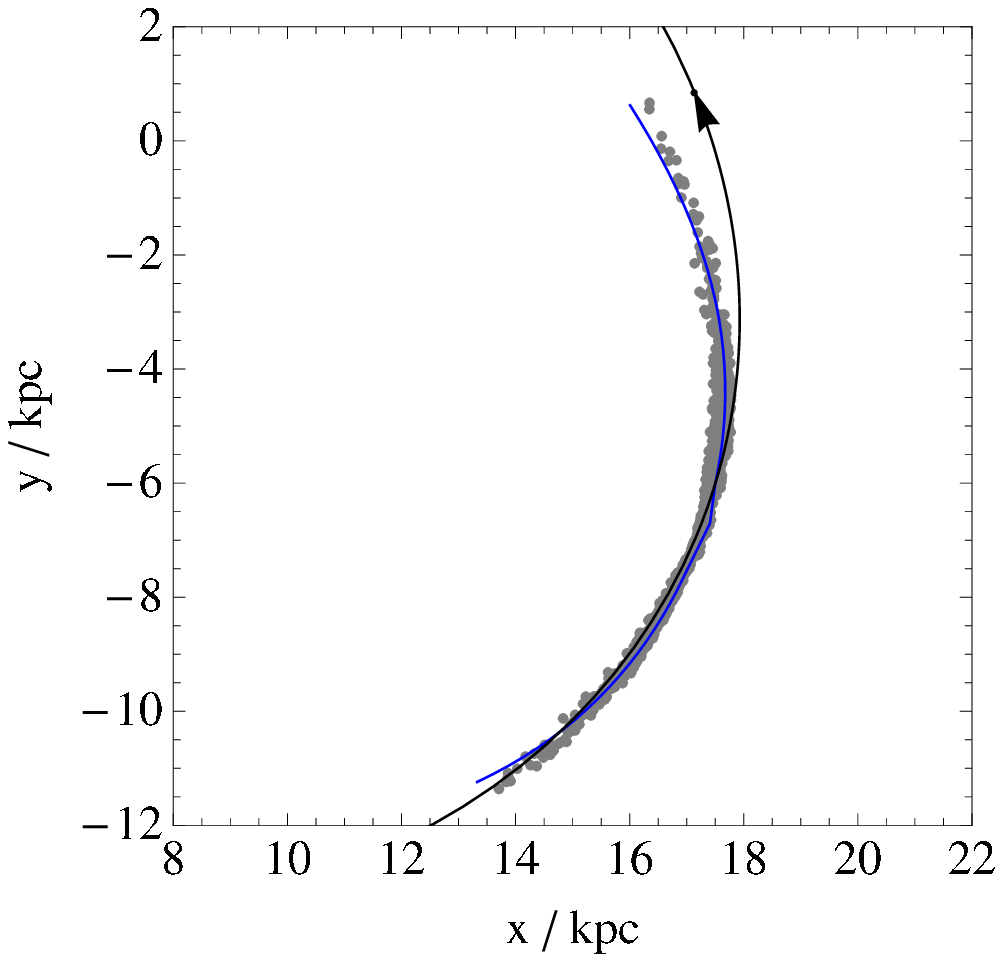}
    \qquad
    \includegraphics[width=\doublefigshrink\hsize]{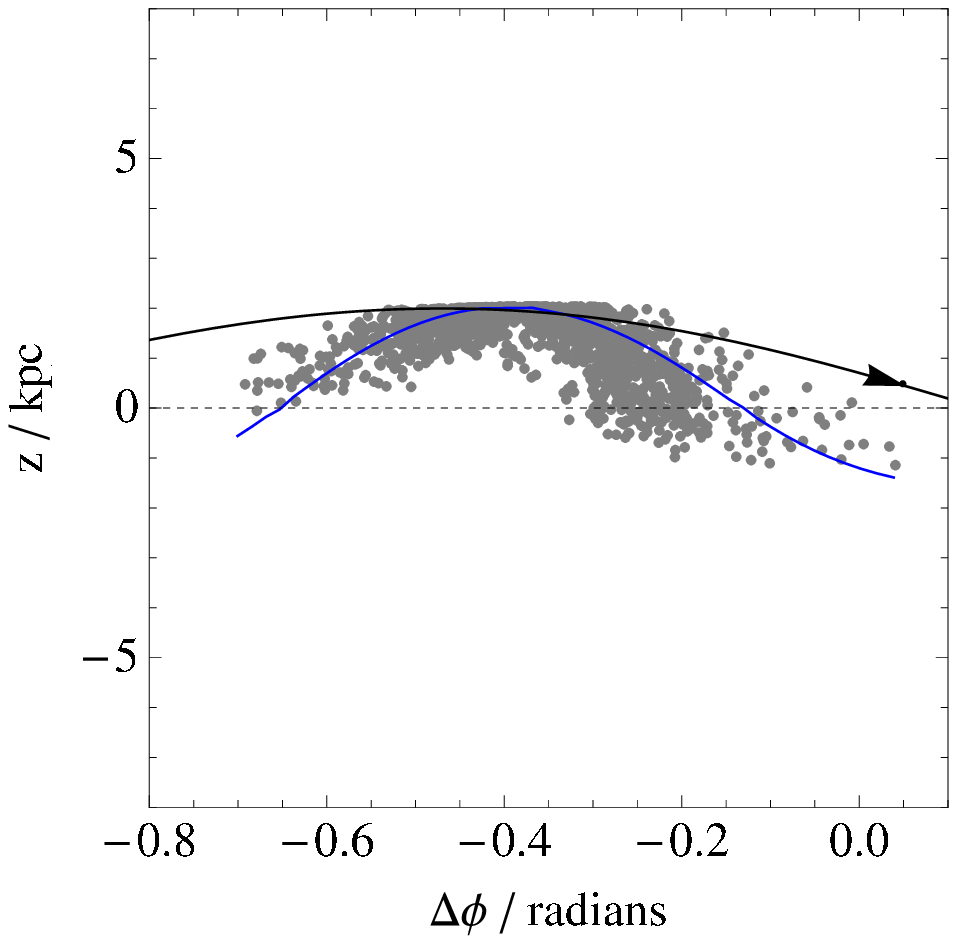}
  }
  \caption[
Real-space configuration of the cluster model C6 evolved on the orbit SO1]
{ Real-space representation of an N-body simulation of
    cluster model C6 evolved on the orbit SO1.  The
    potential is the flattened \stackel\ potential of
    \tabref{mech:tab:potentials}.  The points are shown near the 7th
    apocentre passage following release. The black line shows the
    trajectory of an individual orbit; the blue line shows the
    predicted stream path.  }
\label{mech:fig:stack-xt-nbody}
\end{figure}

\figref{mech:fig:stack-xt-nbody} shows equivalent plot to
\figref{mech:fig:stack-xt-angles}, but in real-space.
The misalignment between the stream and the progenitor orbit in angle-space
is seen to map into a large misalignment in real-space. Attempting to
constrain halo parameters by fitting orbits to the stream shown
in \figref{mech:fig:stack-xt-nbody} would not produce sensible results.
Conversely, the map of the dashed-line model for the action-space distribution,
shown in \figref{mech:fig:stack-xt-jays}, clearly provides an excellent
proxy for the track of the stream.

We therefore conclude that, in flattened systems, we are able to 
accurately predict the track a of stream using a simple, but well informed,
action-space model, while the corresponding progenitor orbit makes a substantially
less accurate proxy for the track of a stream.

\subsection{Realistic examples in the \stackel\ potential SP2}
\label{mech:sec:realistic}

In this section, we draw on the work of the previous sections to construct
N-body simulations for two actual observed Milky Way streams, in order
to examine to what extent these streams can act as proxies for progenitor orbits.

In the following sections, we take as our model of the Milky Way the
\stackel\ potential SP2, described in \tabref{mech:tab:stackpots}.
This model does reproduce an approximately correct in-plane
rotation curve outside of the solar circle, precisely where our
example streams reside. However, this model is arguably
insufficient in representing the flattening of the potential
in proximity to the Milky Way's massive disk.
\secref{mech:sec:stackmisalignment} showed that stream misalignment
is likely to worsen in the presence of a flatter potential. Thus,
if our results below are in error, it is likely that streams make
poorer proxies for orbits, not better ones.

Unfortunately, there is insufficient flexibility in \stackel\ models
to allow for an accurate representation of both halo and disk across a
reasonable range of radii. Correcting for any error due to
insufficient flattening, or even properly assessing the magnitude of its
effects, is therefore beyond the scope of this chapter. It is with
this caveat in mind that we now proceed.

\subsubsection{Tidal stream GD-1}

The tidal stream GD-1 \citep{gd1-discovery} is associated with one
of the most complete sets of full phase-space observations for any stream in
the Milky Way, and has been utilized in several recent attempts
\citep{willett,koposov} to constrain the Galactic potential by
fitting an orbit to it. Furthermore, it is the example used by this
author \citep[\chapref{chap:galplx};][]{galplx} to demonstrate the
applicability of Galactic parallax, the analysis of which would be subject to
systematic error if the track of the stream were not tangent with the
orbits of the stars. It is therefore of primary importance that
we understand the behaviour of this particular stream.

The orbit GD1, described in \tabref{mech:tab:stackorbs} and
illustrated in the middle panels of
\figref{mech:fig:stack-orbit-examples}, was chosen to approximate the
observed features of the GD-1 stream. In particular, the orbit was
required to follow the conclusions of \chapref{chap:galplx}: i.e.~it
should be at pericentre, about $7\kpc$ distant from our fiducial Solar
location; it should have an orbital plane inclined to the Galactic
plane by $37\deg$, and be on a retrograde orbit; and it should
reproduce approximately the track on the sky as reported in
\cite{koposov}.

Having selected an orbit in the SP2 potential that roughly meets
these requirements, the cluster model C6 was specified according to
the schema of \secref{mech:sec:clusters}. In the absence of any
evidence as to the true nature of the GD-1 progenitor, the model was
chosen to have the same mass and profile parameters as does C1.
The stripping radius $r_s = 15\kpc$ was chosen to be slightly larger
than the pericentre radius of the GD1 orbit.

A $10^4$ particle realization of the C6 model was made by random
sampling of the King model distribution function.  This cluster was
then placed close to apocentre on the orbit GD1, at a point $4.57\Gyr$
prior to the present pericentre location, equivalent to $14 {1\over2}$ radial
orbits. The cluster was then evolved forward in time by the \fvfps\
tree code, with time step $\d t=\tdyn/100$ and softening parameter
$\epsilon$ as specified in \tabref{mech:tab:clusters}.

\begin{figure}[\figplaceopts]
  \centerline{
    \includegraphics[width=\doublefigshrink\hsize]{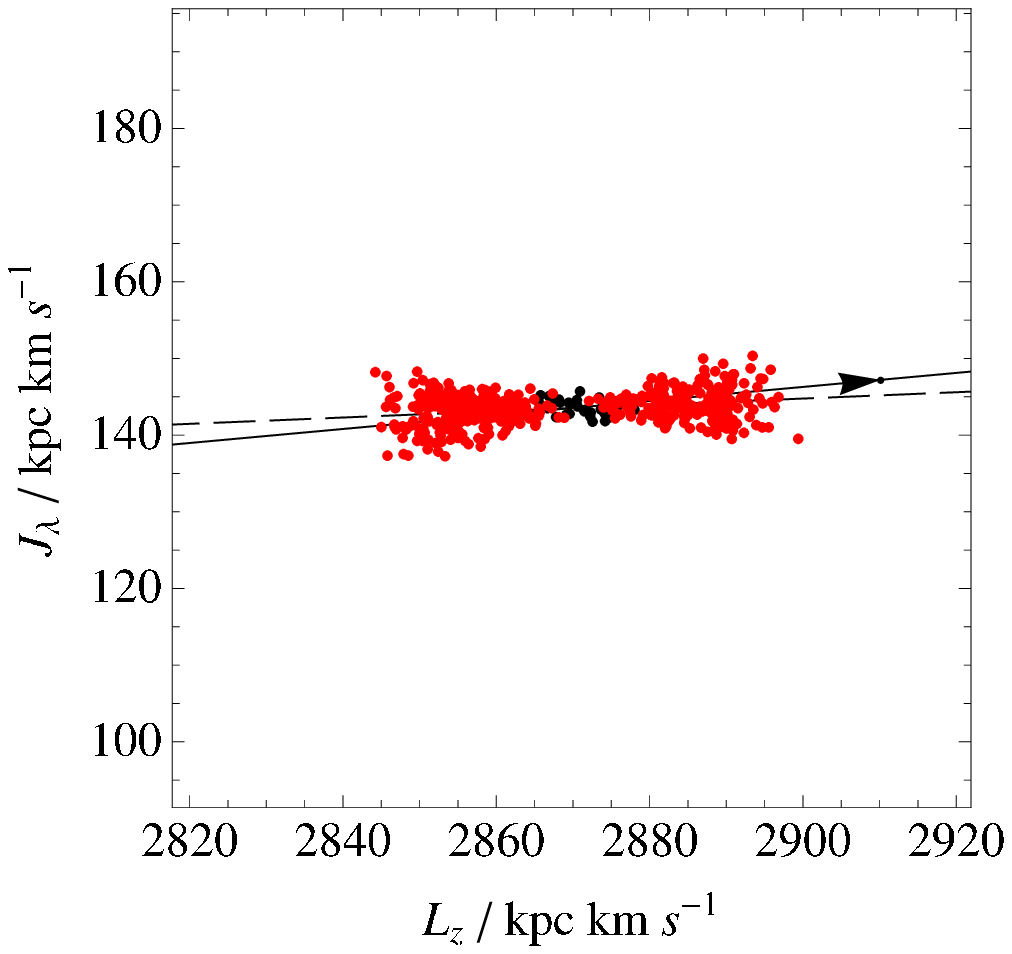}
    \qquad
    \includegraphics[width=\doublefigshrink\hsize]{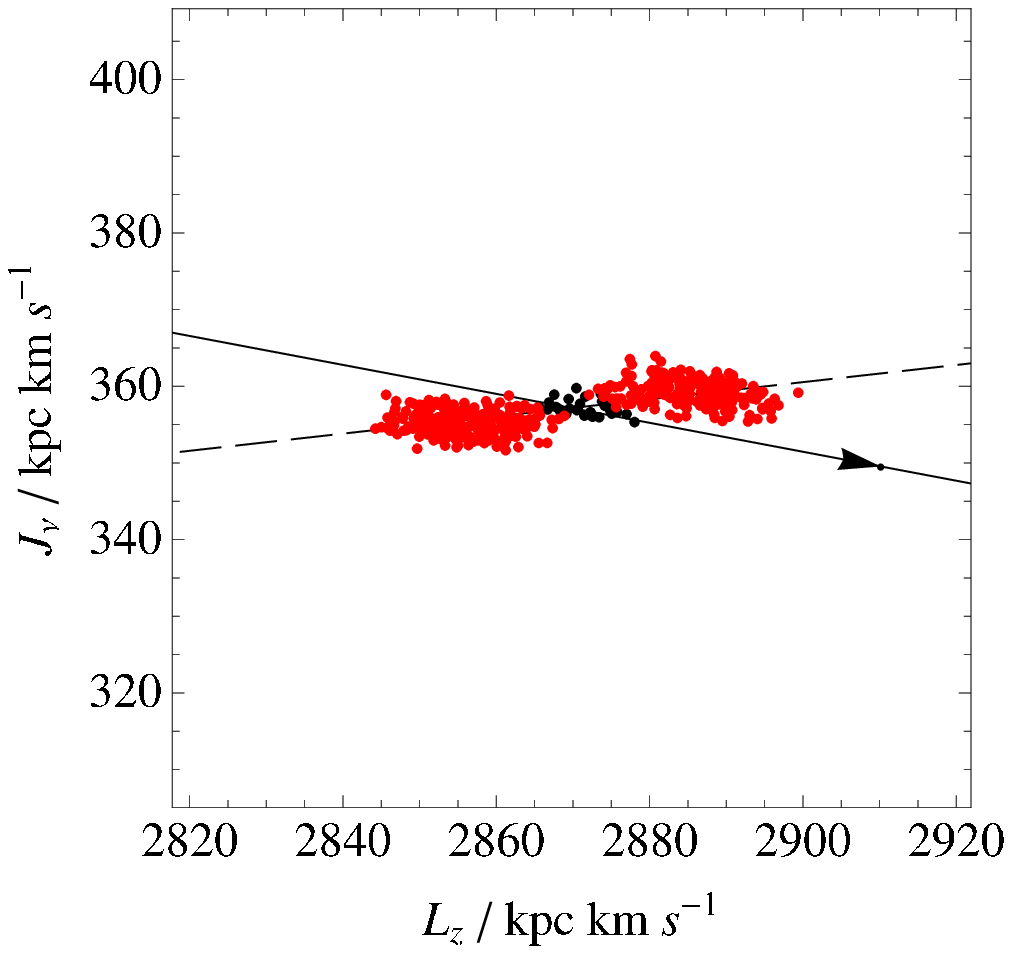}
  }
  \caption[
Actions for the cluster C6 on orbit GD1 in the \stackel\
    potential SP2]
{ Actions for the cluster C6 on orbit GD1 in the \stackel\
    potential SP2. The result is intended to simulate a possible
    configuration of the GD-1 Milky Way stream. The cluster is shown
    near pericentre after $14 {1\over2}$ radial orbits. The arrowed black line is
    the image of the frequency vector, $\hessian^{-1}\vO_0$; the
    dashed line is a straight-line fit to the data.
}
\label{mech:fig:stack-gd1-actions}
\end{figure}

\begin{figure}[\figplaceopts]
\centering{
    \centerline{
    \includegraphics[width=\doublefigshrink\hsize]{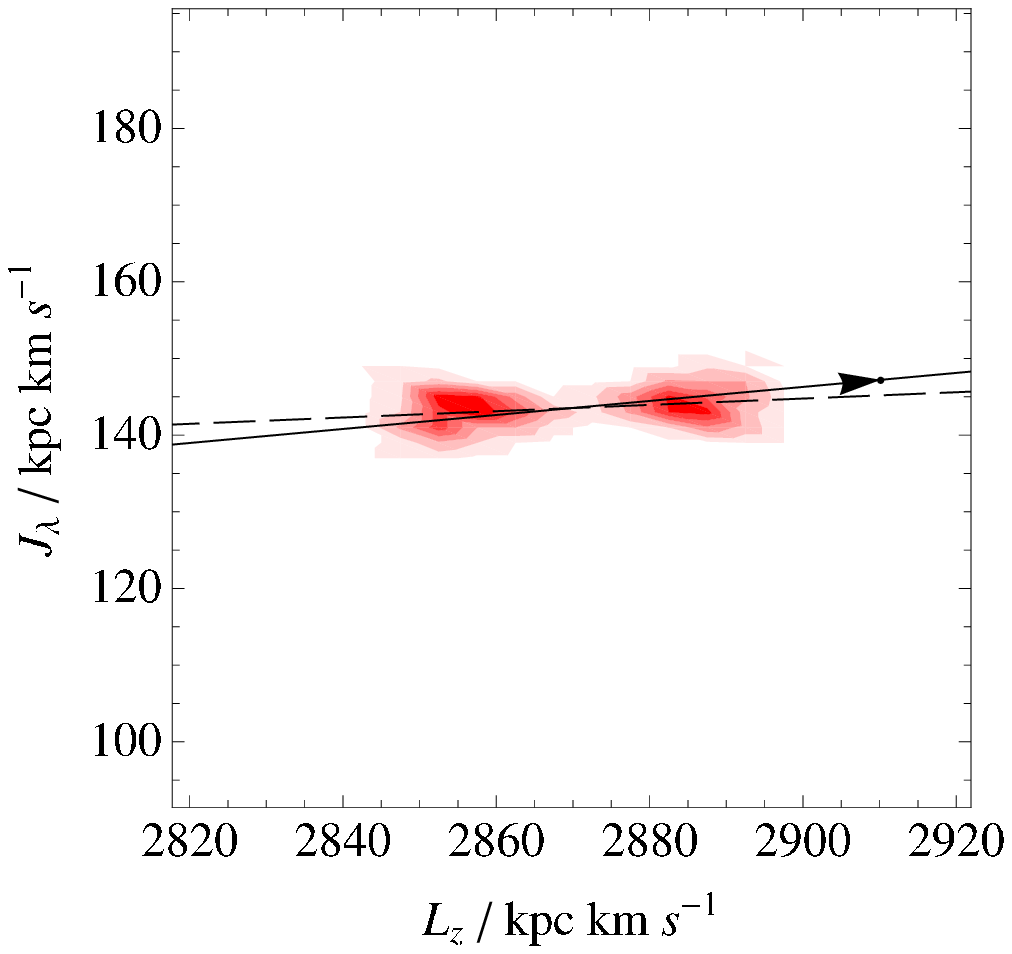}
    \qquad
    \includegraphics[width=\doublefigshrink\hsize]{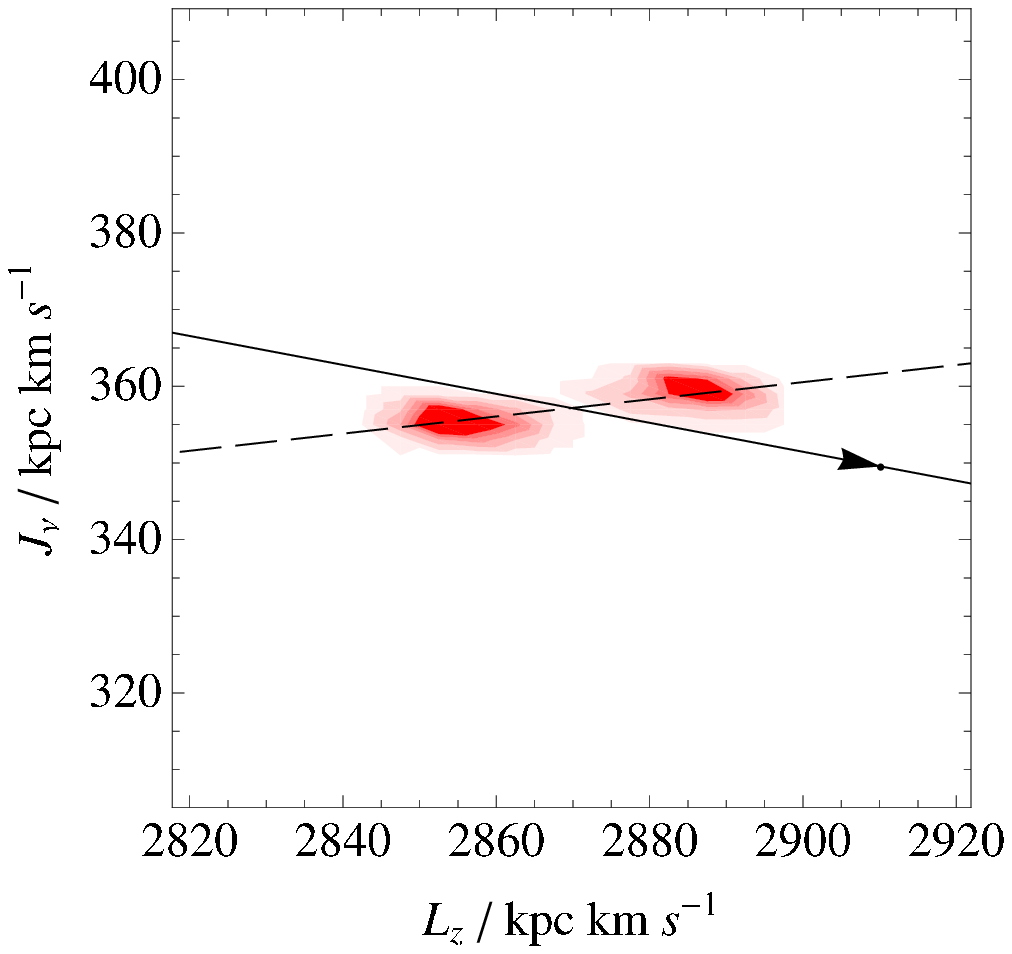}
    }
}
\caption[Particle-density plots for the scatter plots of \figref{mech:fig:stack-gd1-actions}]
{
  Plots showing the (column) density of particles in
  action-space, corresponding to the scatter plots of
  \figref{mech:fig:stack-gd1-actions}.
  The density was estimated 
  by placing particles into bins of width $\sim 2 \kpc \kms$.
  Darker shading represents regions of higher particle density, with the
  edges of the shaded regions
  representing contours of constant density.
}
\label{mech:fig:stack-gd1-particle-density}
\end{figure}

\figref{mech:fig:stack-gd1-actions} shows the action-space
configuration of the simulated GD-1 cluster at the end of the
simulation period, with the cluster centroid at pericentre,
and the stream near its present location. Bound and unbound particles are plotted in
black and red, respectively; we note that the cluster has been
almost completely disrupted at this time. The image of the frequency
vector, $\hessian^{-1}\vO_0$, is shown in this figure as an arrowed
black line, while the dashed line is a least-squares fit
to the unbound particles. In addition, \figref{mech:fig:stack-gd1-particle-density} shows
particle-density plots corresponding to the scatter plots of \figref{mech:fig:stack-gd1-actions}.
%Each panel also shows the image of
%a circle, in the plane of that panel, mapped under $\hessian$
%and projected back onto the plane; these images are plotted
%as red ellipses.

The action-space distribution does not look dissimilar to those in
either \figref{mech:fig:nbody-runs234} or
\figref{mech:fig:stack-xt-jays}.  In this eccentric, highly inclined
orbit, the ellipsoidal coordinates $(\lambda$,$ \nu$) effectively
parametrize the spherical polar coordinates $(r,\theta)$.  The
action $J_\lambda$ can therefore be compared to the radial action
$J_r$, with $J_\nu$ becoming some function of the total angular
momentum $L$ and $L_z$.  In this example, $J_\nu$ is comparatively
small, so $L \sim L_z$ and we can
understand the left panel of \figref{mech:fig:stack-gd1-actions} by
direct analogy with the results of
\secref{mech:sec:disruption} and \secref{mech:sec:stackeldistribution}.

We will not attempt to disentangle the angular momentum $L$, in order to make
sense of the right panel.  Instead, we merely note that in comparison
to the previous examples, the action-space image of the frequency
vector $\hessian^{-1}\vO_0$ is aligned much more closely with the long
axis of the action-space distribution.
% Further, unlike in \figref{mech:fig:stack-xt-jays},
%the strongly-distorting image of $\hessian$ in the right panel is
%{\em not} aligned perpendicularly to the image of the frequency
%vector.
We therefore expect the misalignment between the
angle-space stream and the frequency vector to be less here than in
previous examples.

We can actually understand this alignment as a property of spherical,
logarithmic potentials, which is approximately what this distant orbit
feels in the \stackel\ potential SP2.  In such
a potential, where
\begin{equation}
\Phi(r) = v_c^2 \log {r / r_0},
\end{equation}
the circular frequency $\Omega_\phi$ is given by
\begin{equation}
\Omega_\phi = {v_c^2 \over L}.
\end{equation}
To first order, the radial frequency $\Omega_r$ is given by \citep[\S3.2.3]{bt08}
\begin{equation}
\Omega^2_r = \left.{\partial^2 \Phi_\eff \over \partial r^2}\right|_{r_g}
= \left.{\partial^2 \Phi \over \partial r^2}\right|_{r_g} + {3 L^2 \over r_g^4}
= -{v_c^2 \over r^2_g} + {3 v_c^4 \over L^2}
= {2 v_c^4 \over L^2} = 2 \Omega^2_L.
\end{equation}
Although the above equation is only true to first order---and indeed, it is
the higher-order terms that contribute to $\hessian$ and give rise
to the stream geometry seen in \figref{mech:fig:sp2-hessian}---we
see that
\begin{equation}
\vO_0 = (\Omega_r, \Omega_\phi) \sim \Omega_\phi(\sqrt{2}, 1)
\simeq {v_c^2 \over L} (\sqrt{2},1).
\end{equation}
The direction of $\vO_0$ is approximately constant, and the magnitude
is a function of $L$ only. Hence, the image of $\hat{\vO}_0$ in action-space
points almost exactly along $\hat{L}$.
% Since cluster orbits are likely to have high $L$ when compared
% to $J_r$, then \eqref{mech:eq:djr/dl} requires the action-space distribution
% of a cluster to be elongated along $\hat{L}$. The compression
% of the radial actions during particle escape only serves to
% augment this elongation, since the radial action is always
% perpendicular to $\hat{L}$.

We have seen from the arguments in \secref{mech:sec:disruption} that
the action-space distributions of clusters formed in spherical
potentials are likely to be flattened and oriented towards $\hat{L}$.
Thus, in spherical logarithmic potentials, we expect the long axis of
such distributions to be aligned closely with the action-space image
of the frequency vector. Correspondingly, we expect streams in such
potentials to be more closely aligned with $\vO_0$ in angle-space than was
seen in the isochrone potential or in the highly-flattened SP1
potential.

\begin{figure}[\figplaceopts]
  \centerline{
    \includegraphics[width=\doublefigshrink\hsize]{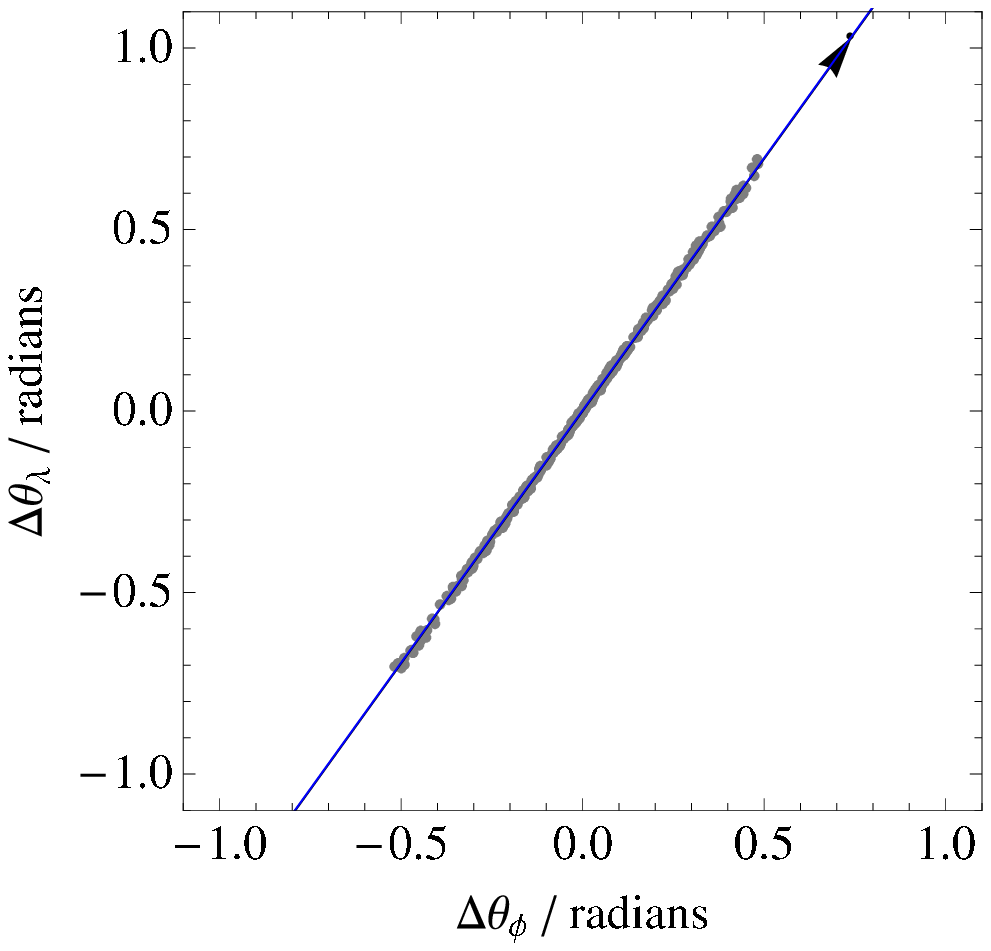}
    \includegraphics[width=\doublefigshrink\hsize]{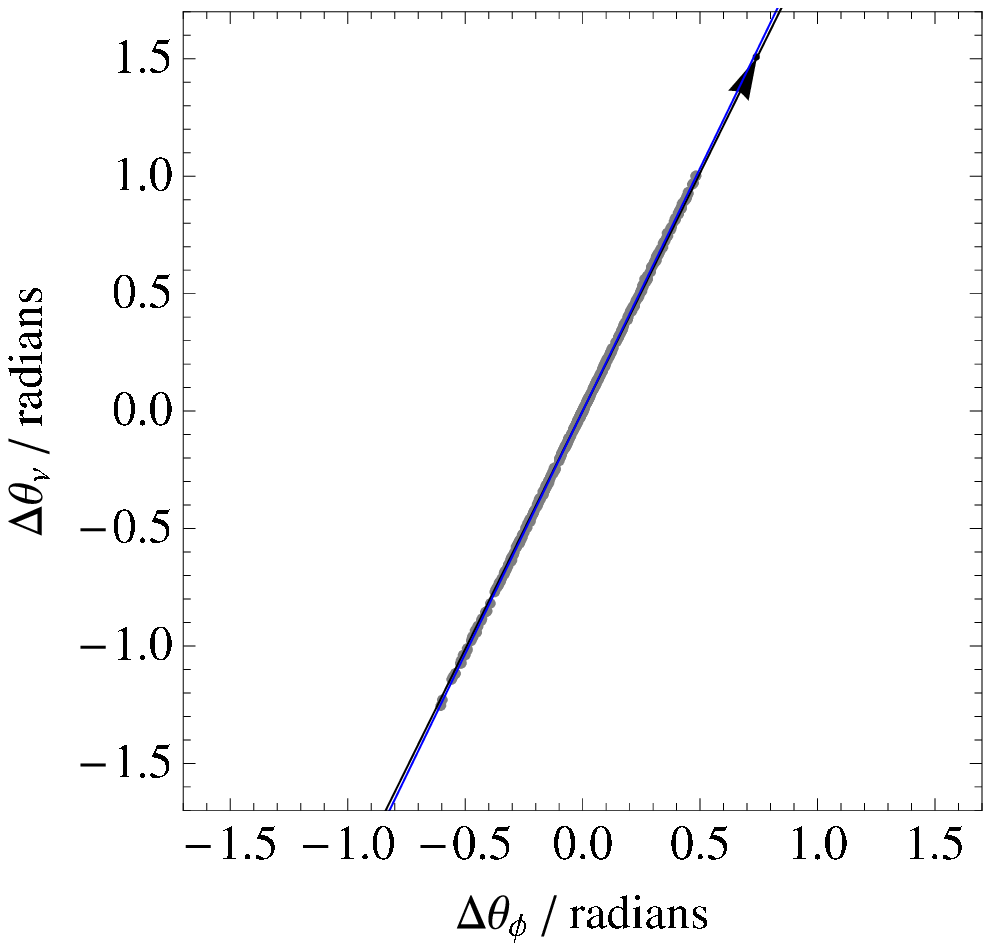}
  }
  \caption
[Angle-space configuration of the simulated GD-1 stream,
    shown in \figref{mech:fig:stack-gd1-actions}]
{ Angle-space configuration of the simulated GD-1 stream,
    shown in \figref{mech:fig:stack-gd1-actions}. The grey particles
    are computed directly from the results of the N-body simulation,
    while the blue line is the angle-space map of the dashed line from
    \figref{mech:fig:stack-gd1-actions}.  The arrowed black like shows
    the frequency vector, $\vO_0$.}
\label{mech:fig:stack-gd1-angles}
\end{figure}

Returning to our example, \figref{mech:fig:stack-gd1-angles} shows the
angle-space configuration of the simulated GD-1 stream at the end of
the simulation period. The grey particles are for angles that were computed
directly from the output of the N-body simulation. The arrowed black
line is $\vO_0$, while the blue line is the angle-space map of the
dashed line from \figref{mech:fig:stack-gd1-actions}. As expected
given the close alignment of the action-space distribution with the
image of $\vO_0$, the stream, the frequency vector and the predicted track are
all seen to agree almost perfectly.

\begin{figure}[\figplaceopts]
  \centerline{
    \includegraphics[width=\doublefigshrink\hsize]{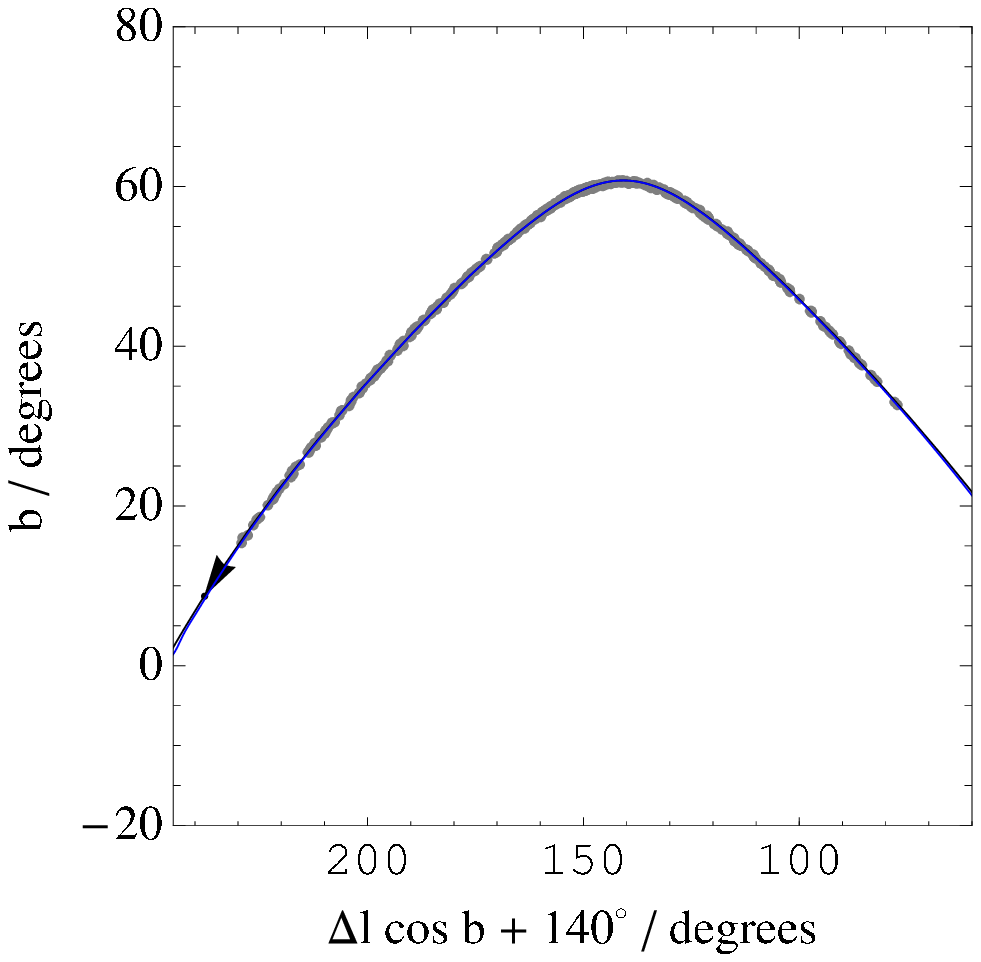}
    \includegraphics[width=\doublefigshrink\hsize]{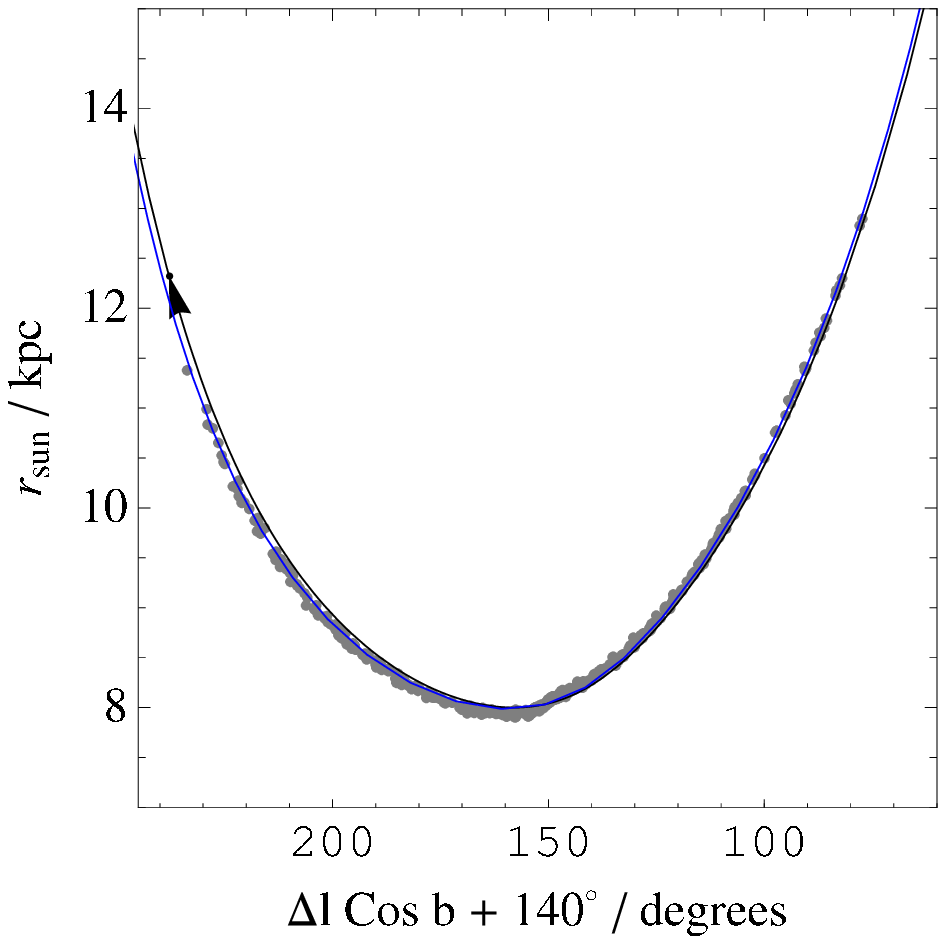}
  }
  \caption[
Real-space configuration of the simulated GD-1 stream,
    shown in \figref{mech:fig:stack-gd1-actions}]
{ Real-space configuration of the simulated GD-1 stream,
    shown in \figref{mech:fig:stack-gd1-actions}. Plots are shown in
    Galactic latitude/longitude and heliocentric radius, to aid
    comparison with observations. Left panel: the on-sky projection of the stream,
    which is seen to agree perfectly with the progenitor orbit. 
    Right panel: heliocentric distance, versus Galactic longitude.
    In the right panel there exists a tiny anomaly between the
    heliocentric distance of the stream and the trajectory of the orbit
    on the far left of the plot, but the agreement between stream and
    orbit is otherwise excellent.}
\label{mech:fig:stack-gd1-space}
\end{figure}

\figref{mech:fig:stack-gd1-space} shows the real-space configuration
of the simulated GD-1 stream and the end of the simulation period.
The plots have been rendered in Galactic
coordinates $(l,b)$ and heliocentric distance $r_{\rm sun}$ to aid
with comparison to observations.  The GD-1 stream is seen to be an
excellent proxy for the progenitor orbit.  This is expected, given the
perfect alignment of the stream with $\vO_0$ in
\figref{mech:fig:stack-gd1-angles}. A small discrepancy of $\sim
0.3\kpc$ does exist between the heliocentric distance of the stream and
the orbit at the extreme end of the leading tail, but the match is
otherwise perfect. In the case of the on-sky projection, the stream
perfectly delineates the orbit for all longitudes. In both panels, the
blue line, which is the track predicted from the dashed line in
\figref{mech:fig:stack-gd1-actions}, is a perfect match to the stream
everywhere.

In conclusion, we find that the simulated GD-1 stream delineates its orbit
perfectly. This is fortunate for our analysis of the Galactic parallax
of the GD-1 stream in \chapref{chap:galplx}, which does not now need to
be revisited. We have understood the result in terms of a peculiar property
of spherical logarithmic potentials, which causes the $\hat{L}$ direction
in action-space to be mapped close to $\vO_0$ in angle-space.
Since action-space distributions are naturally oriented close to
$\hat{L}$, the result is an angle-space stream closely aligned with $\vO_0$.

\subsubsection{Orphan stream}

Our investigation into the utilization of tidal streams began with the
Orphan stream \citep{orphan-discovery}, and the observation that
standard techniques \citep[e.g.][]{fellhauer-orphan} had difficulty
in finding an orbit that described it exactly.  It is thus fitting
that we end with the Orphan stream as our final case-study in this
thesis.

The orbit OS1, described in \tabref{mech:tab:stackorbs} and illustrated
in the bottom panels of \figref{mech:fig:stack-orbit-examples}, is based on
the simulated Orphan stream shown in \chapref{chap:radvs}, which in
turn is based on the tentative velocity data presented in
\cite{orphan-discovery}. The orbit OS1 roughly reproduces the
on-sky track, distances, and velocities from \cite{orphan-discovery}
when integrated in the SP2 potential.
Recently, a newer analysis of the Orphan
stream by \cite{newberg-orphan} utilizing SEGUE spectra \citep{segue} has produced
radial velocity data that are considerably more certain, and
unfortunately inconsistent with, the \cite{orphan-discovery} radial
velocity data points. Nonetheless, we still present our simulation
of the Orphan stream, in order to dispel any notion based on the
previous section, that tidal streams in approximately spherical,
logarithmic potential must always perfectly delineate orbits.

The cluster model C7 was specified according to
the schema of \secref{mech:sec:clusters}, for the orbit OS1 in
the SP2 potential. Like with GD-1, evidence as to the true properties
of the Orphan stream progenitor is lacking. Hence, the model was
chosen to have the same mass and profile parameters as does C1;
the stripping radius $r_s = 12\kpc$ was chosen to be slightly larger
than the pericentre radius of the orbit.

A $10^4$ particle realization of the C7 model was made.  This cluster
was then placed close to apocentre on the orbit OS1, at a point
$6.52\Gyr$ prior to the present apocentre location, equivalent to
14 complete radial orbits. The cluster was then evolved forward in time by
the \fvfps\ tree code, with time step $\d t=\tdyn/100$ and softening
parameter $\epsilon$ as specified in \tabref{mech:tab:clusters}.

\begin{figure}[\figplaceopts]
  \centerline{
    \includegraphics[width=\doublefigshrink\hsize]{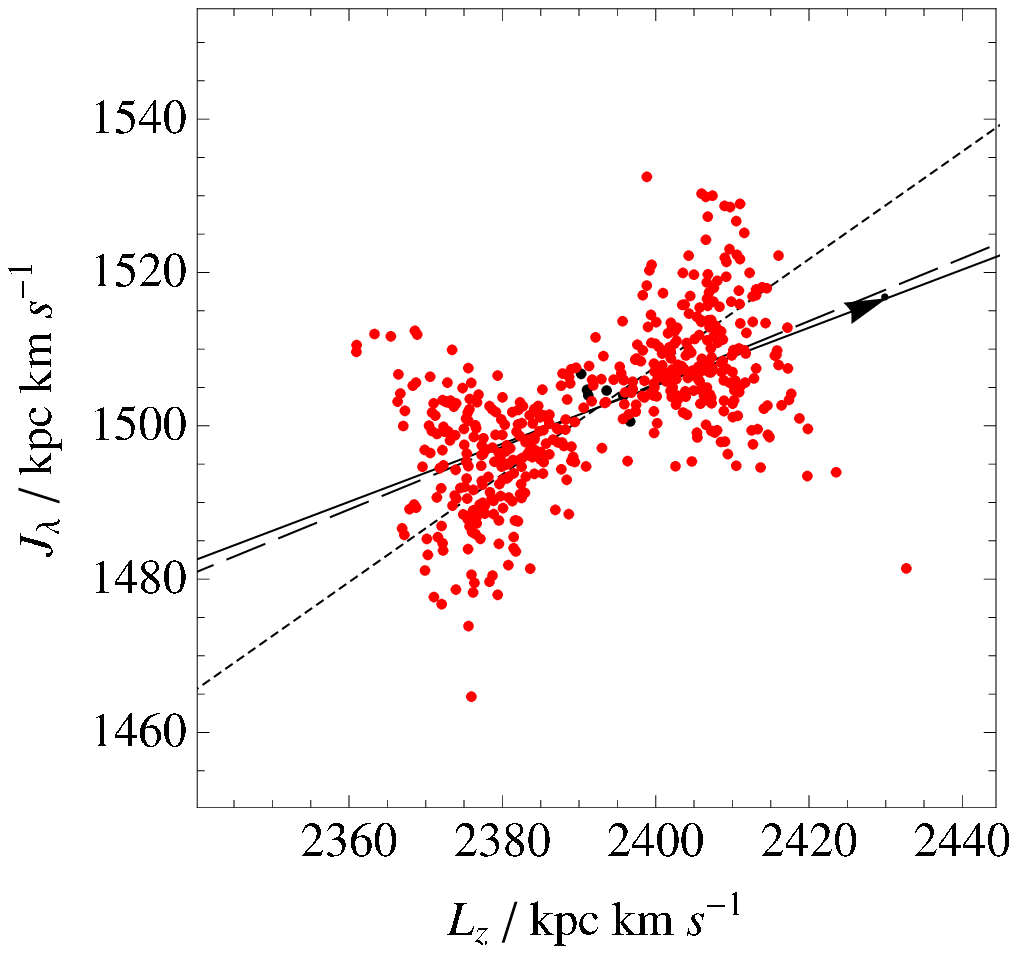}
    \qquad
    \includegraphics[width=\doublefigshrink\hsize]{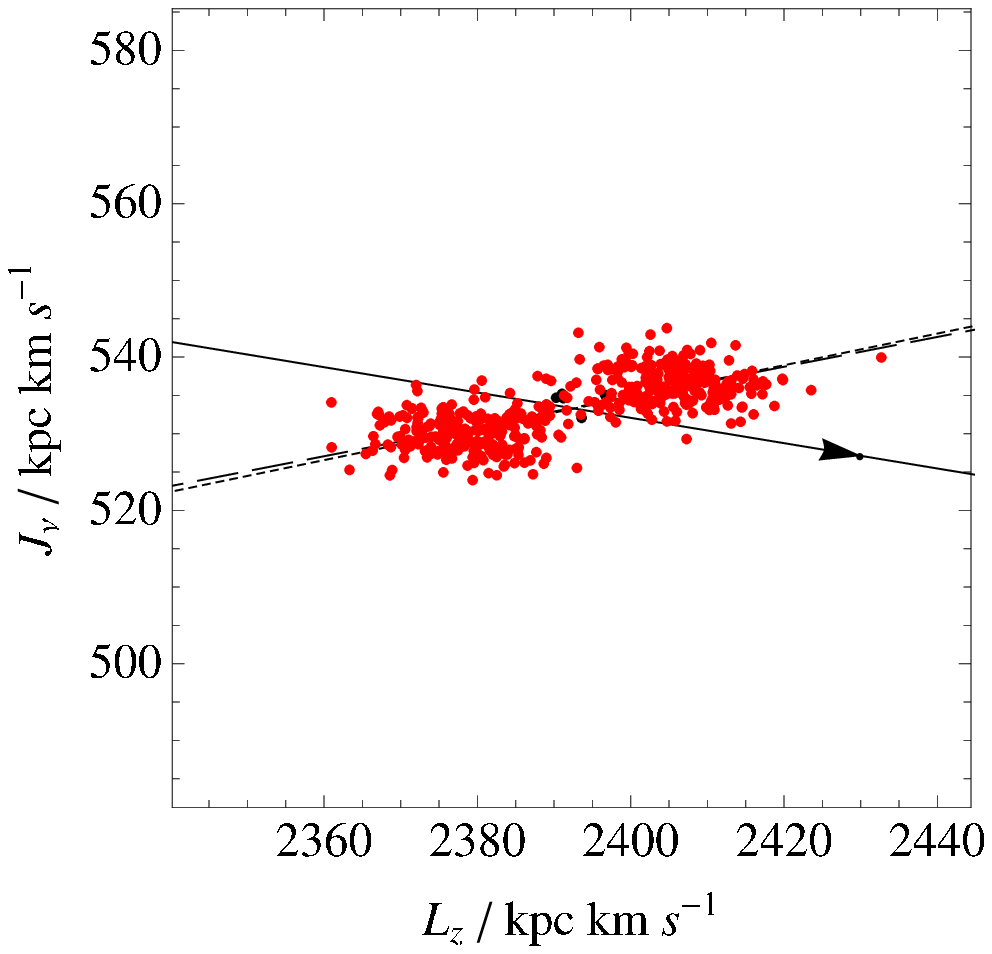}
  }
  \caption[Actions for the N-body cluster C7 on orbit OS1 in the
    \stackel\ potential SP2]{ Actions for the N-body cluster C7 on orbit OS1 in the
    \stackel\ potential SP2. The simulation is one possible
    configuration for the Orphan stream.  The cluster is shown near
    apocentre after 14 complete radial orbits. The arrowed black line is the
    image of the frequency vector, $\hessian^{-1}\vO_0$.  The dashed
    line is a least-squares fit to the data, and the dotted line is a
    weighted least-squares fit to the data, discussed in the text.
%    The red ellipses are the images of circles, in the plane of 
%    each panel, transformed by $\hessian$ and projected back into
%    the plane. The ellipse in the right panel is plotted at 1/3 the
%    scale of that in the left panel.
}
\label{mech:fig:stack-orp-actions}
\end{figure}

\begin{figure}[\figplaceopts]
\centering{
    \centerline{
    \includegraphics[width=\doublefigshrink\hsize]{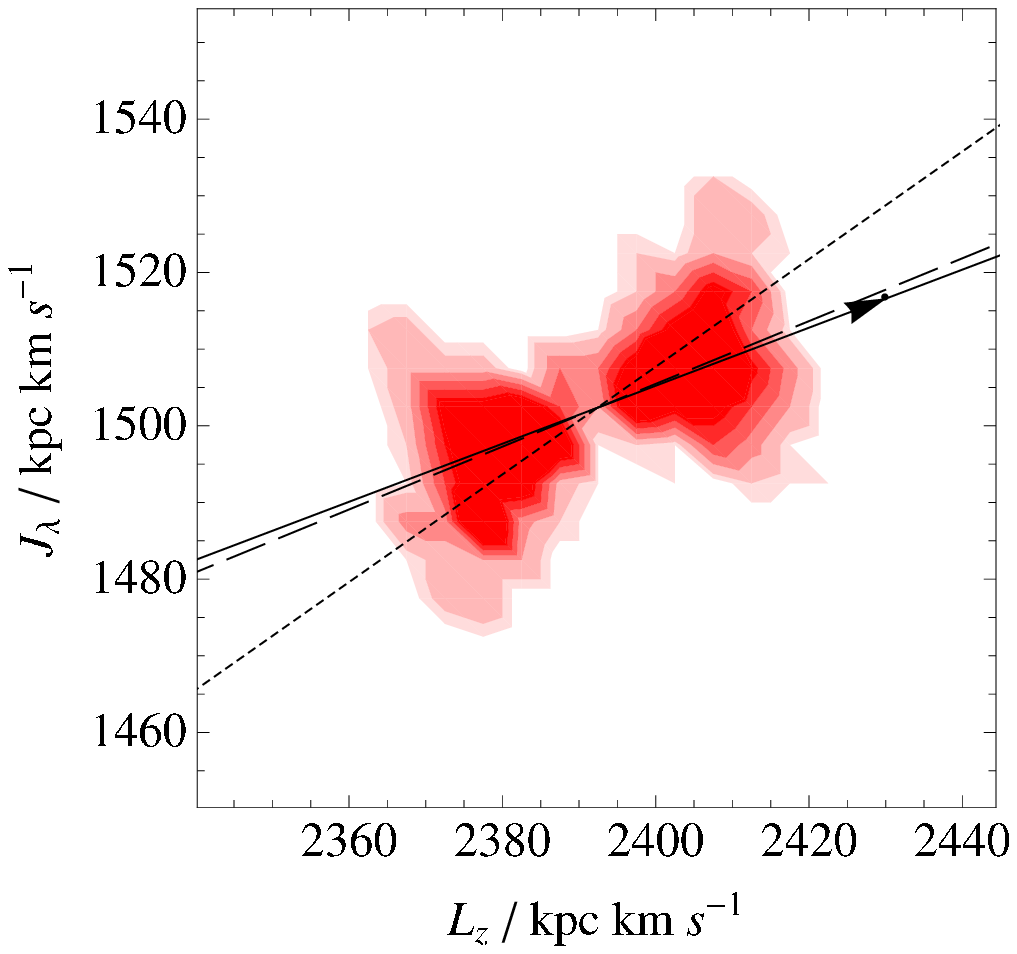}
    \qquad
    \includegraphics[width=\doublefigshrink\hsize]{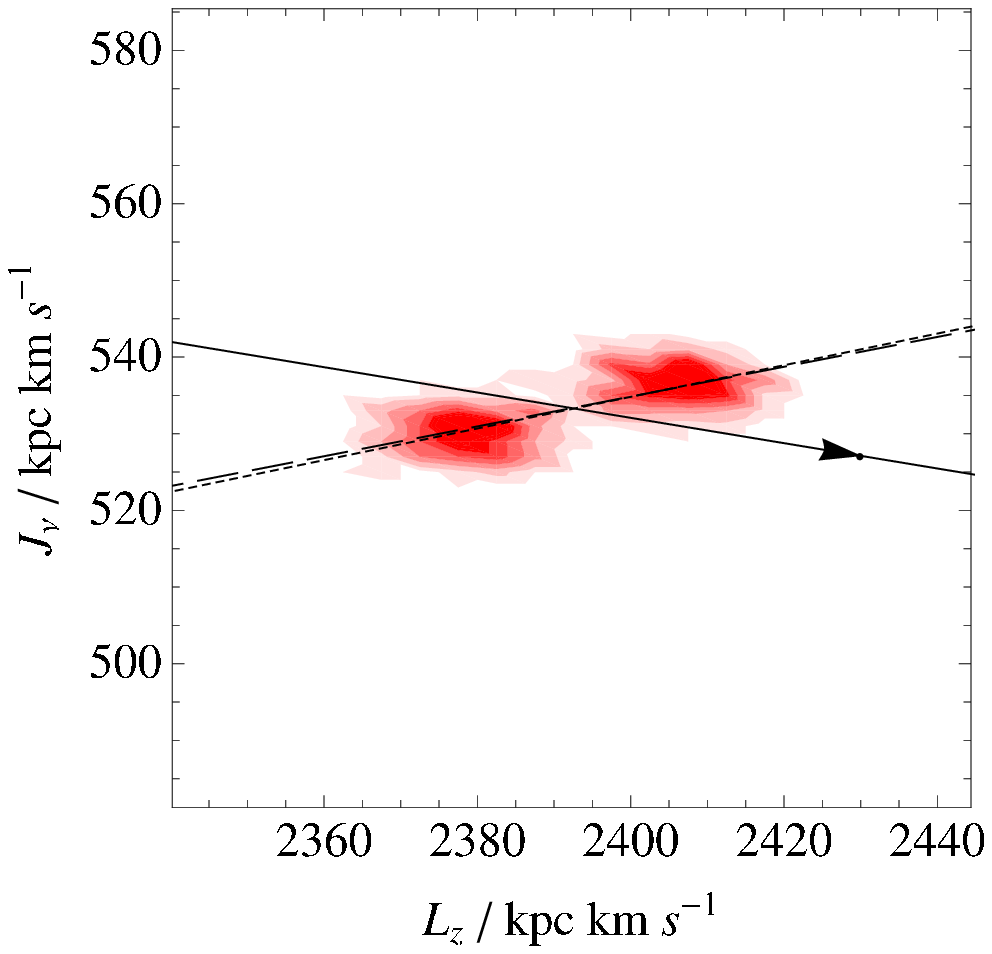}
    }
}
\caption[Particle-density plots for the scatter plots of \figref{mech:fig:stack-orp-actions}]
{
  Plots showing the (column) density of particles in
  action-space, corresponding to the scatter plots of
  \figref{mech:fig:stack-orp-actions}.
  The density was estimated 
  by placing particles into bins of width $\sim 2 \kpc \kms$.
  Darker shading represents regions of higher particle density, with the
  edges of the shaded regions
  representing contours of constant density.
}
\label{mech:fig:stack-orp-particle-density}
\end{figure}

\figref{mech:fig:stack-orp-actions} shows scatter plots for the
action-space distribution of the model Orphan stream at the end of the
simulated period, while \figref{mech:fig:stack-orp-particle-density}
shows the corresponding particle-density plots. The distribution is
immediately comparable to that of \figref{mech:fig:stack-gd1-actions},
except that the structure in the left panel appears tilted when
compared to that of GD-1.  We understand this to result from a
combination of two factors. First, the orbit OS1 is more eccentric
than is the orbit GD1. From \eqref{mech:eq:gradient} we expect this to
result in a distribution that is steeper when viewed in $(L,
J_r)$. Secondly, unlike in the two previous examples, a large
component of $L$ is in other than the $z$-direction.  Hence, a
distribution that is approximately aligned with $\hat{L}$ will not be
aligned with $\hat{L}_z$. Again, we will not attempt to disentangle
$L$ in order to precisely understand the mechanics of the plots, but
we note once more that the image of the frequency vector
$\hessian^{-1}\vO_0$ and the long axis of the action-space
distribution are closely aligned. We thus expect the angle-space
stream to be well represented by $\vO_0$.

Also shown in \figref{mech:fig:stack-orp-actions}
%are the projected
%images of a circle, in the plane of each panel, mapped under
%$\hessian$;
are two least-squares fitted lines: the dashed curve
is an unweighted fit to the particles, while the dotted line
is a weighted fit to the particles, described in detail below.

\begin{figure}[\figplaceopts]
  \centerline{
    \includegraphics[width=\doublefigshrink\hsize]{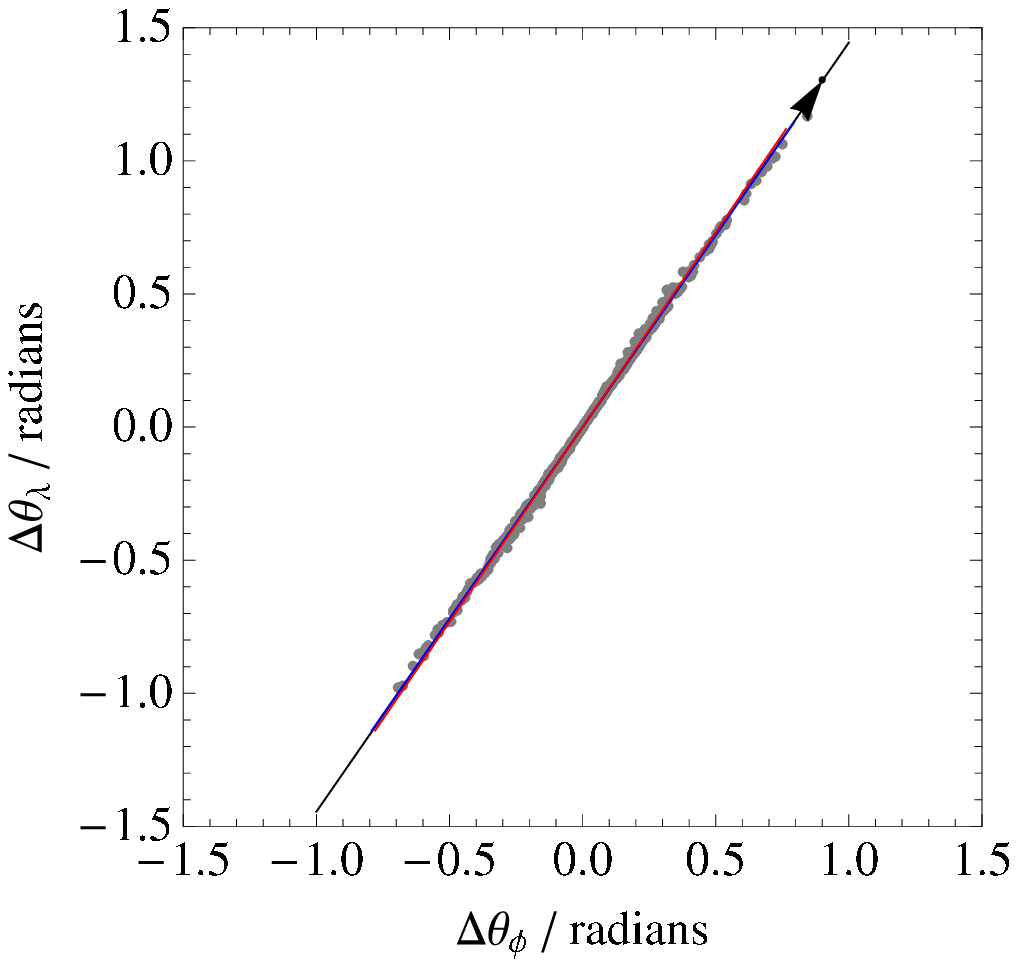}
    \includegraphics[width=\doublefigshrink\hsize]{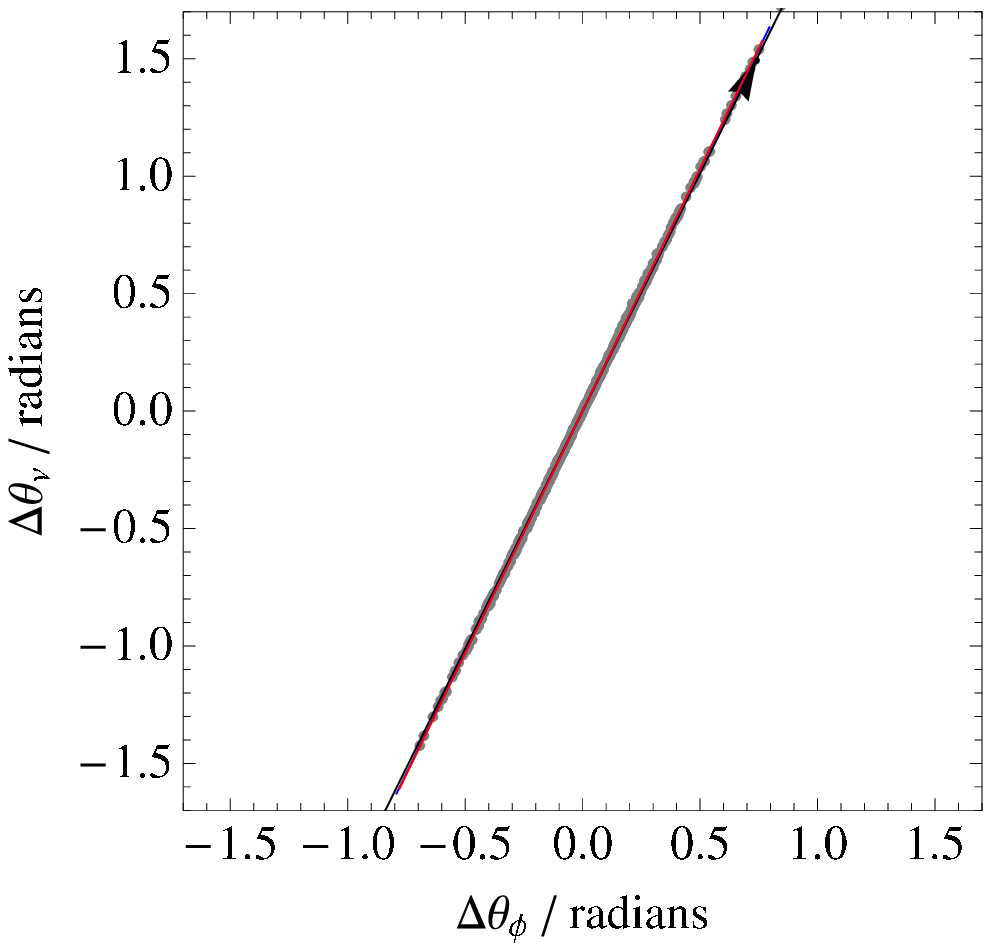}
  }
  \caption[Angle-space configuration of the simulated Orphan stream,
    corresponding to the actions shown in \figref{mech:fig:stack-orp-actions}]
{ Angle-space configuration of the simulated Orphan stream,
    corresponding to the actions shown in \figref{mech:fig:stack-orp-actions}.
    The angles of the grey particles have been computed directly from the
    N-body simulation. The arrowed black line is $\vO_0$, the
    blue line is the image of the dotted line from \figref{mech:fig:stack-orp-actions},
    and the red line is the image of the dashed line from \figref{mech:fig:stack-orp-actions}.}
\label{mech:fig:stack-orp-angles}
\end{figure}

\figref{mech:fig:stack-orp-angles} shows the angle-space configuration
of the simulated Orphan stream at the end of the simulation run. The
grey particles are for angles that were computed directly from the
output of the N-body simulation. In both panels $\vO_0$, shown as a
black arrowed line, is almost indistinguishable from the red and blue
lines, which are the angle-space images of the dashed and dotted lines,
respectively, from \figref{mech:fig:stack-orp-actions}. All three
lines almost perfectly delineate the stream, although close inspection
shows the red line to pass on either side of the particles at the
extremes of the tail, while $\vO_0$ and the blue line pass directly
through.

\begin{figure}[\figplaceopts]
  \centerline{
    \includegraphics[width=\doublefigshrink\hsize]{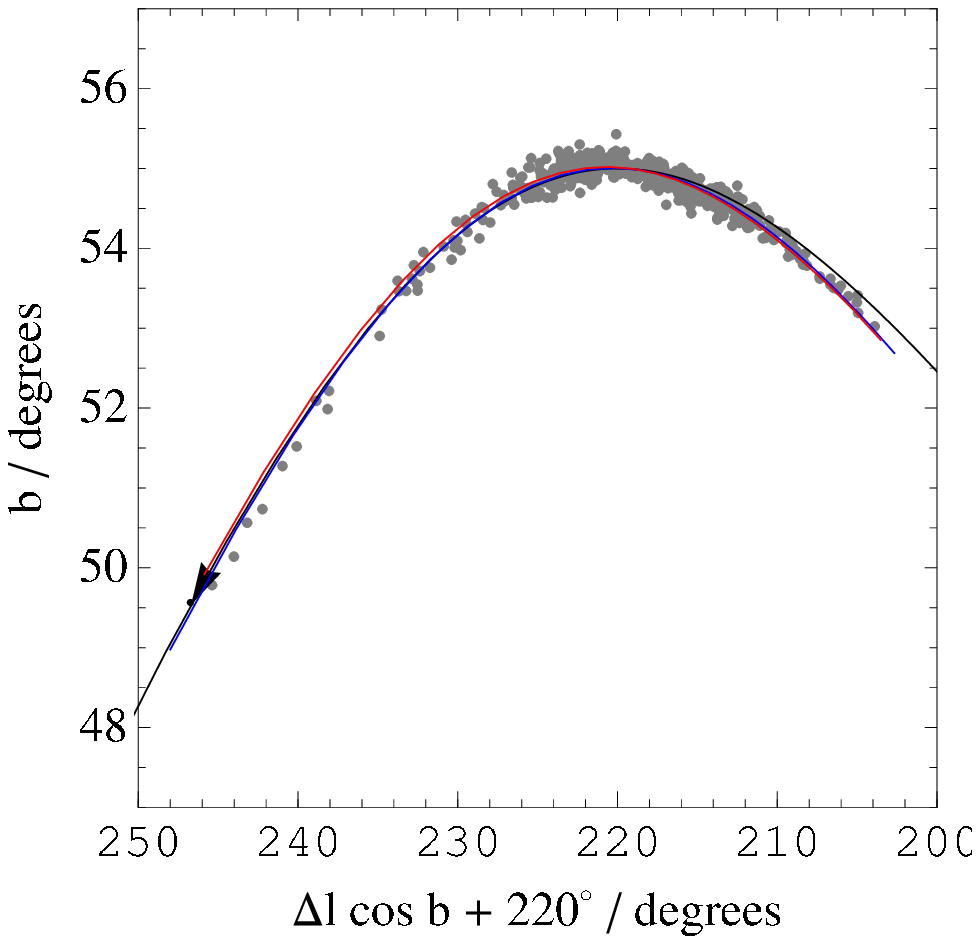}
    \qquad
    \includegraphics[width=\doublefigshrink\hsize]{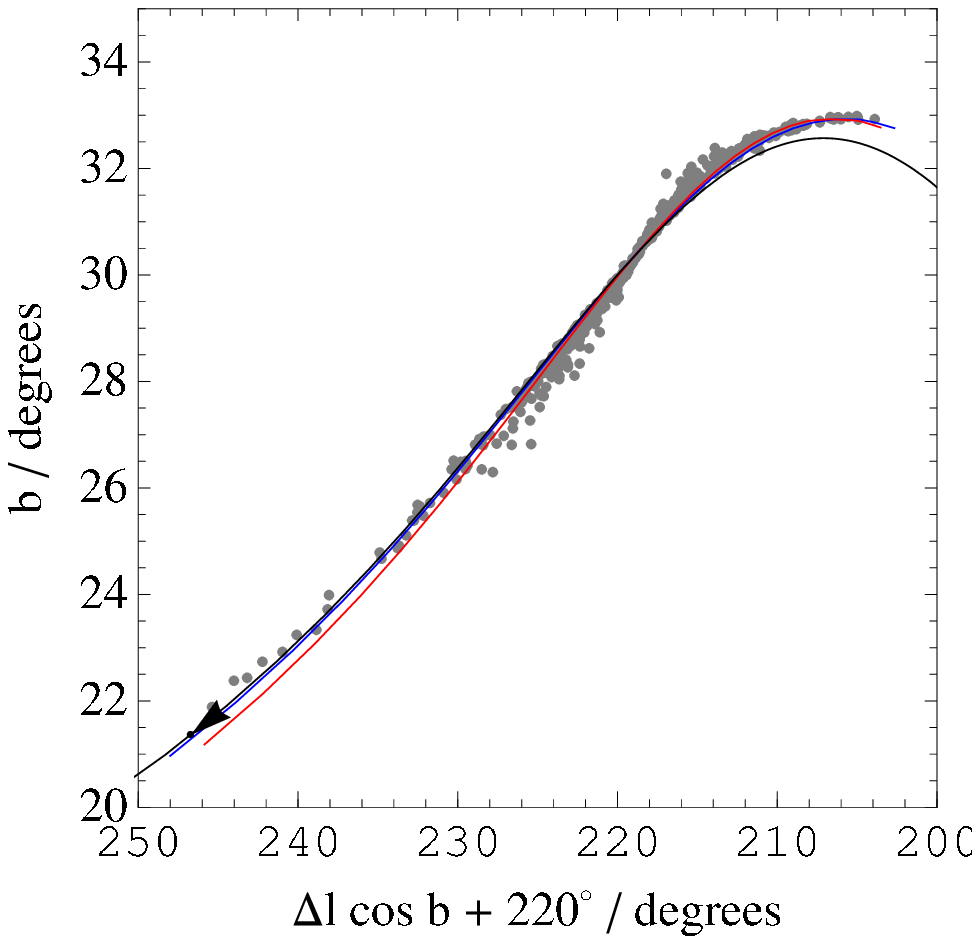}
  }
  \caption[Real-space configuration of the simulated Orphan stream]
{ Real-space configuration of the simulated Orphan stream.
    The N-body particles are plotted in grey. The black line is the
    trajectory of the progenitor's orbit, while the blue and red
    lines are the real-space mappings of the blue and red lines
    from \figref{mech:fig:stack-orp-angles}. There is a
    clear anomaly between the orbit and the stream on the right-hand
    side of both plots. On the left-hand side of the right panel,
    there is an anomaly between the red curve and the stream. }
\label{mech:fig:stack-orp-space}
\end{figure}

\figref{mech:fig:stack-orp-space} shows the real-space configuration
of the simulated stream, at the end of the simulation run.  The grey
particles are plotted directly from the output of the simulation,
while the lines are the real-space equivalents of those shown in
\figref{mech:fig:stack-orp-angles}.

We immediately note that the stream and the orbit trajectory do not
coincide on the right side of each plot. Thus, the stream is not
well represented by an orbit in this region, despite the stream
and $\vO_0$ coinciding everywhere in \figref{mech:fig:stack-orp-angles}.
The explanation is that, although there is no misalignment between
the stream and the orbit in angle-space, the changes in trajectory
induced by the variation in action down the stream have caused the stream and the orbit
to become misaligned in real-space. The detail of this mechanism
was described in \secref{mech:sec:trajectory-j}. In mapping from
angle-space to real-space, the blue and red curves include the
correction for this effect, as specified in \eqref{mech:eq:correction}. Thus, the
predicted tracks from the red and blue curves match the stream
much better than does the orbit, on the right-hand side of each
plot, where the magnitude of this effect is maximized.

Looking again at \figref{mech:fig:stack-orp-space} we see that on the
left side of the right panel, the red curve is a slightly poorer
representation of the stream than is the orbit or the blue curve. This
occurs because the dashed-line model to the action-space distribution,
shown in \figref{mech:fig:stack-orp-actions}, is generating the wrong
$\delta \vJ$ correction when utilized in \eqref{mech:eq:correction}.
The wrong correction is generated because the stars at the ends of the
tidal tails, rather than corresponding to the middling regions of the
action-space distribution which are well modelled by the dashed line,
actually correspond to the upper-right and lower-left extremities
of the action-space distribution, which are well removed from the
dashed-line model.
Hence in this example, the dashed-line model does not provide an appropriate $\delta
\vJ$ correction for the ends of the stream, and
correspondingly the red line in \figref{mech:fig:stack-orp-space}
fails to properly predict the stream track at the end of the leading
tail.

% The unweighted dashed-line model effectively assumes that the
% `average' of the action-space distribution will dominate the
% angle-space structure, while in this example the angle-space structure
% is dominated by the stars from the upper-right and lower-left
% extremities of the action-space distribution.  In this particular
% example, the stars at these extremities are boosted significantly by
% the mapping under $\hessian$, and it is these stars that make up the
% ends of the tidal tails.  However, these same stars are at the
% uppermost and lowermost wings of the action-space distribution, and
% the dashed line is a poor model for them.  Because of this, the red
% line in \figref{mech:fig:stack-orp-space} fails to properly predict
% the stream track at the end of the leading tail.

The remedy is a slightly more sophisticated model of the action-space
distribution. The dotted line in \figref{mech:fig:stack-orp-actions}
is a least-squares fit to the data, like the dashed line, but with
each data point weighted by $\sigma_i = \hessian \cdot \Delta
\vJ_i$. That is, we grant greater weight to those data points which map
to the fastest diverging stars in the stream. The resulting model,
though still simple, better predicts the variation in action down the
stream, when utilized in \eqref{mech:eq:correction}, than does the
dashed line. The results of mapping the dotted line into real-space
are shown as the blue line in \figref{mech:fig:stack-orp-space}.
Unlike the red line or the orbit, the blue line matches the stream
track everywhere.

In conclusion, we find that the action-space distribution of a
simulated Orphan stream exhibits the same characteristics as those
observed in both the isochrone potential, and for other orbits in
\stackel\ potentials. We find that, although the principal direction
of $\hessian$ is misaligned with $\vO_0$, the stream in angle-space
is nonetheless aligned with $\vO_0$, because it is a property of
spherical, logarithmic potentials that the $\hat{L}$ axis in
action-space maps to $\vO_0$ in angle-space, and it is a natural
feature of disrupted clusters that their action-space distribution is
also aligned with $\hat{L}$. Conversely, we find that the stream is
not well represented by the progenitor orbit at all points in
real-space, because the variation in action of the stars as we move
down the stream causes sufficient change to the trajectory of those
stars, that the stream track and the progenitor orbit diverge.

We find that we can predict the path of the stream perfectly, but that
the simple model of a line least-squares fitted to the action-space
distribution is not sufficient to do so. An improved model, in which
the data points are weighted by the quantity
$\hessian\cdot\Delta\vJ$ for the least-squares fit, results in
an action-space model that accurately predicts the stream
track everywhere.

\section{Conclusions}
\label{mech:sec:conclusions}

In this chapter, we have studied in detail the mechanics of the
disruption of star clusters in various model potentials.
We have been interested in learning of the conditions required
for the formation of tidal streams from such clusters, and in particular,
to what extent such streams delineate the orbits of their stars.
With regard to the latter, we have been motivated by the catalogue of techniques
which attempt to utilize tidal streams to place constraints on the Galactic
potential, many of which assume that streams perfectly delineate orbits.

We utilize action-angle variables extensively in our approach to the
problem. It is found that these coordinates allow a convenient and
natural description of the physical processes that occur in cluster
disruption and stream formation. In the absence of more advanced
techniques \citep[e.g.][]{mcmillan-torus}, this approach has
restricted the galaxy models under consideration to those for which
action-angle variables are readily computed. Nonetheless, we do expect
our findings to generalize to more sophisticated models of the
Galactic potential.

In the broadest strokes, we have found that tidal streams will always
form when a cluster in a realistic orbit around a realistic host
galaxy is subject to tidal stripping. We show that these streams need
not be well representative of an orbit in the host galaxy, and that in
general they will not be, although clusters on particular orbits in
particular potentials can well make an accurate
representation. Particularly alarming is our finding that serious
systematic errors can be made by attempting to constrain the Galactic
potential using a stream that does not delineate an orbit. It is of
some relief, however, that we find ourselves able to accurately predict the
tracks of streams, even when they are not representative of an orbit,
and so it may be possible to repair the potential-constraining procedures, by
having them fit such stream tracks instead.

A complete summary of our findings in this chapter is presented below.
The section that follows discusses some of the consequences of our findings,
and presents some directions for future work.
In \chapref{chap:concs} of this thesis we further review our findings in the
context of contemporary Galactic astrophysics.

\subsection{Summary}

\subsubsection{The physics of streams formation}

In \secref{mech:sec:tremaine}, we studied the formation of streams by
considering the motion of small clusters of test particles, when
described by action-angle coordinates of the host potential. We
recalled from \S8.3.1 of \cite{bt08} that streams form in
angle-space, when an initial, small structure in action-space results
in a corresponding structure in frequency-space. This structure in
frequency-space gives rise to a structure in angle space, which grows
secularly and quickly dominates the original angle-space form of the
cluster. Since the particles only feel the gravity of the host,
the action-space structure remains frozen.

We show that the frequency-space structure is given by the 
transformation of the action-space structure under the linear
map $\hessian = \nabla_\vJ H(\vJ)$, which is the Hessian of the Hamiltonian
in the actions $\vJ$. This linear map can be
characterized by the image produced from a unit sphere in action-space
under its transformation. The sphere will be transformed into an
ellipsoid, in which directions of the semi-axes correspond to the
eigenvectors of the map, and the lengths of the semi-axes correspond
to the eigenvalues of the map.  Streams form when one of the
eigenvalues of the map---and hence, the magnitude of the stretch in
the principal direction---is very much larger than the other two.

\subsubsection{Stream formation in systems of two actions}

In \secref{mech:sec:2d} we investigate the form of this linear map for
the general case of a system described by two actions.  We find
explicit relations between the geometry of the map and the form of the
Hamiltonian, when written as a function of actions. We find that the
Hamiltonians of the Kepler potential and of the spherical harmonic
oscillator take a form in which the geometry of the linear map is
globally consistent.

In the former case, one of the eigenvalues of the
map is null, so the angle-space image of any action-space
structure, under this map, will always be a perfect filament. We show
that this filament will always be aligned precisely with the frequency
vector of the progenitor orbit. Hence, disrupting clusters always form streams in the Kepler
potential, and those streams always perfectly delineate the progenitor
orbit. We go on to confirm this prediction numerically.

In the case of the spherical harmonic oscillator we show that the
eigenvalues of the map are both null. Hence, the image of an action-space
structure under this map is trivially null. The interpretation of this
result is that streams cannot form in spherical harmonic oscillator
potentials; all angle-space structure are perfectly preserved.

We also consider the case of the isochrone potential, for which the
Kepler potential and the spherical harmonic oscillator are limiting
cases. We find that in general the eigenvalues of the map are non-zero,
but that the ratio between them is everywhere large. Hence, disrupting
clusters will always form streams in this potential. Unlike with the
Kepler potential, we find that the principal direction of the map is
not generally aligned with the frequency vector of the progenitor
orbit. Hence, the angle-space stream formed from a spherical
cluster in action-space will also not be aligned with that frequency
vector.  We find that the misalignment in angle-space will typically
be of order a few degrees. We go on to confirm this prediction numerically.
We also show that this angle-space misalignment manifests
itself in real-space as a failure of the progenitor orbit to delineate
the track of the stream.

\subsubsection{Considerations for the mapping of streams from action-angle space into real-space}

In \secref{mech:sec:mapping} we deduce general limits on the
anisotropy of the action-space structure formed from a disrupting
cluster.  In doing so, we address the concern that a stream of
arbitrary angle-space orientation could be created from a sufficiently
non-isotropic action-space structure. We find that any realistic
cluster, if it forms a stream at all, will form a stream that is not
misaligned from the frequency vector, in angle space, by more than a
few degrees.

We then take this result and examine the real-space tracks
of lines in angle-space that are misaligned from the frequency
vector by a few degrees. We find that the nature of the anomaly
between the real-space track of such a line, and the orbit to
which the frequency vector corresponds, depends on the phase
of the orbit at which the observation is made. Specifically,
we find that if the observation is made with the orbit
close to apsis, then the angle-space misalignment 
manifests itself as a change in curvature of the stream
track, with respect to the orbit. We find that if the
observation is made at a point far from apsis, then
the angle-space misalignment corresponds to a real-space
misalignment, of similar magnitude.

The real-space position of a star is a function of both its action and
its angle. Since it is a finite structure in action-space that gives
rise to an angle-space stream, it follows that the action of the stars in
a stream must vary down the track.
In order to predict the real-space track of a stream, it is not then
sufficient to know only the angle-space structure; one must also account
for this variation in action along the angle-space stream.
We investigate this variation,
and deduce a method to detect when it will be of consequence for
predicting the real-space track of the stream. We also show that
the variation can be predicted, provided one can guess the time
since the cluster's first tidal stripping event, and provided
one has an appropriate model of the action-space structure
of the stream.

\subsubsection{The consequences of fitting orbits using misaligned streams}

In \secref{mech:sec:fitting}, we examine the
consequences of attempting to utilize a misaligned stream to constrain the
parameters of an isochrone potential, while assuming that the track
does properly delineate an orbit.

When the stream perfectly delineates its progenitor orbit,
we find that the potential parameters are isolated perfectly. Thus, we demonstrate
the efficacy of our optimization technique. When the
real-space curvature of the stream track is less than that of the
progenitor orbit, we find that the mass parameter of the potential is consistently under-reported,
by 11\percent\ in our example case. When the curvature
of the stream track is greater than that of the progenitor orbit,
we find that the mass parameter is consistently over-reported, by 14\percent\ in our example case.

However, in deducing these results, we have
allowed ourselves to consider unrealistic combinations of parameters, which would be
ruled out by independent observations.
We therefore repeat the
exercise, but we now consider only that family of potential
parameters which correctly reproduces a fiducial circular velocity
at the Solar radius.

When the stream perfectly delineates the progenitor orbit, we
again find that the potential parameters are correctly identified.
However, when the stream track has less curvature than the orbit, the best-fitting
potential has a mass that is 21\percent\ smaller than the truth. When the
stream track has greater curvature than the orbit, the best-fitting
potential has a mass parameter that is 54\percent\ larger than the truth.
Hence, adding extra constraints to the fitting process has resulted in the
systematic errors becoming worse.

In summary, we find that large systematic errors can be made when attempting to optimize
potential parameter by assuming that streams act as proxies for orbits,
when they do not.

\subsubsection{The action-space distribution of N-body clusters}

In \secref{mech:sec:actions} we utilized N-body simulation
of King model clusters to examine the action-space distribution
resulting from the disruption of a live cluster.

We found that the action-space distribution of a disrupted cluster
takes a characteristic shape, which is flattened, and is oriented with
a small positive gradient when viewed in in the $(L,J_r)$ plane. We
find that the greater the eccentricity of the cluster orbit, the
larger this positive gradient will be, but that in general, the
distribution will be roughly oriented in the $\hat{L}$ direction in
action-space.  We are able to explain all features of this
distribution in terms of the basic physical processes that apply to
clusters undergoing tidal disruption. Hence, we predict that the
action-space distribution of all disrupting clusters will take the
same basic form, although details such as the density of stars across
the distribution, and their precise dimensions and orientation,
necessarily depend on the details of the model, the orbit, and the
potential in question.

We show by simulation that real disrupted clusters do indeed
form streams with tracks that are poorly predicted by the
progenitor orbit. However, we show that by utilizing a simple,
straight-line model of the action-space distribution, 
we are able to predict the real-space
stream track of a stream with perfect accuracy.

\subsubsection{Non-spherical systems}

In \secref{mech:sec:nonsph}, we extend the results of the previous
sections to systems with a Hamiltonian described by three actions,
allowing us to consider the stream-forming properties of non-spherical
systems.  As our example, we utilize oblate, axisymmetric \stackel\
models with asymptotically logarithmic density profiles \citep{de-z-2}.
We consider two such models: one highly flattened, and therefore
representative of a heavy galactic disk; and one that is only
slightly flattened, but is roughly consistent with the observed
Milky Way rotation curve from the Solar circle outwards.

In both potentials, we find that the ratio of the eigenvalues of
the map is very large; hence, disrupted clusters will always
form streams in these potentials. We further find that the
principal direction of the map is misaligned with the frequency
vector by $\near 1\deg$ when the flattening is only slight, and
by $\near 10\deg$ when the flattening is substantial. On this basis,
we generally expect streams not to be well represented by orbits
in these potentials, and we expect the representation to be worse
when the potential is flatter.

We performed an N-body simulation of a disrupting cluster on an
approximately planar orbit in the highly flattened \stackel\
potential.  We found that the angle-space misalignment between the
stream and the frequency vector is indeed large, and that the
resulting real-space track is very poorly represented by the
progenitor orbit. However, we again find that a simple straight-line
model of the action-space distribution predicts the corresponding
real-space stream track with superb accuracy.

We took the less flattened \stackel\ potential to be a
model of the Milky Way potential, and in this potential
we performed N-body simulations of streams that are superficially
similar to the observed Milky Way streams GD-1 \citep{gd1-discovery}
and the Orphan stream \citep{orphan-discovery}.
We found that, in both cases, the angle-space stream
and the frequency vector were almost perfectly aligned. This occurs
because of a property of spherical logarithmic potentials, which causes
the action-space image of the frequency vector to align closely with
the action-space distribution of a disrupted cluster.

In the case of GD-1, this results in a stream that is
almost perfectly represented by its progenitor orbit. Hence, we found
that the assumption made in \chapref{chap:galplx} that the GD-1 stream
perfectly delineates the orbits of its stars was a fortuitously fair
one, and that the analysis of the Galactic parallax of GD-1 therefore
needs no revision in light of this work.  However, this conclusion
comes with the caveat that the \stackel\ model used is not a
particularly realistic model of the Milky Way potential, and the
result should be confirmed with a better model.

In the case of the Orphan stream, we find that despite the stream
and the frequency vector being perfectly aligned in angle-space, the
real-space track of the stream is still not well represented by
the trajectory of the progenitor orbit. This occurs because
the variation in action down the stream is sufficient to
alter the curvature of the stream track near apsis, resulting
in the divergence of the stream track and the orbit. However, we find
once again that a simple model of the action-space distribution
allows us to predict this stream track with perfect accuracy.

Lastly, we found that in both our \stackel\ models, the action-space distribution
resulting from the disruption of clusters is directly comparable to that found in the
isochrone potential, confirming the generality of those observations.

\subsection{Discussion and future directions}

We have found that stream tracks cannot be used as reliable proxies
for orbits without detailed consideration of the mechanics of their
formation. One immediate consequence is to generally render as unreliable
those techniques that attempt to constrain the parameters of the Galactic
potential (e.g.~\citealp{newberg-orphan}, \citealp{willett}
and \citealp{koposov}) without an appropriate analysis of whether the
stream is actually well represented by its progenitor orbit.

Such a one-off analysis could be performed using N-body simulation, although
utilizing the techniques presented in this chapter would be quicker.
If the simulation confirms that the stream is well modelled by an
orbit, then one may proceed as before. However, if the stream is not well
modelled by an orbit, as will generally be the case, the technique
of fitting orbits can no longer be used.

In such circumstances, one might resort to N-body shooting methods,
such as \cite{johnston-nbody-fitting}, to compute stream tracks to
feed to an optimization algorithm.  However, the problem with
such methods is the sensitivity of the output to the detail of the
initial conditions. One must perform a multitude of
simulations, over a range of initial conditions, in order to reject a
single potential from further consideration. This requires both great
computational resources, and even then, it is impossible to consider
more than a tiny fraction of the $\near 10^7$ candidate orbits that we
found it necessary to consider in order to fully test the parameter
space and definitively exclude a potential (\chapref{chap:radvs}). The
conclusions drawn from such as an exercise could therefore only be weak,
and therefore unsatisfactory.

The results of this chapter present a possible alternative. We have found
that with simple models of the action-space distribution of a disrupted
cluster, such as can be readily obtained from a single N-body simulation,
we can reliably and accurately predict the track of a stream, even
when it diverges significantly from the trajectory of the progenitor
orbit. In principle, we can compute these tracks with no more computational
effort than it takes to integrate an orbit. Hence, the techniques for
fitting orbits, such as those presented in the earlier chapters of this
thesis, could be readily adapted to fit stream tracks instead.

To achieve this goal, the following hurdles need to be overcome:
\begin{enumerate}
\item We must achieve a detailed understanding of the action-space
  structure of a disrupted cluster, for any problem parameters of our
  choice.

  One approach would be to use N-body simulations to obtain the
  action-space distribution for a small number of cluster models on a
  set of possible orbits, in a given potential.  The resulting
  distributions could be used as a basis set, which would be
  interpolated and distorted to provide an estimate for the
  action-space structure for any given cluster on any chosen orbit.
  The required distortions for changes to cluster model and orbit
  parameters have already been touched upon in this chapter, although
  for general applicability, a complete quantitative theory of these
  distortions will be required.

  However achieved, a rigorous and systematic schema for efficiently
  modelling the action-space structure of clusters  must be developed if our
  insights are to be practically useful.

\item Application of the techniques of this chapter requires the
  ability to compute action-angle variables from conventional
  phase-space coordinates, and vice versa, with reasonable
  accuracy. In this chapter, we have restricted ourselves to those few
  potentials in which the transformation can be readily made. In
  general, one would like to work with more sophisticated potentials,
  for which no easy transformation between action-angle variables and
  phase-space coordinates can be made.

  Fortunately, the ``torus machine'' of \cite{mcmillan-torus}, which
  builds on the prior work of \cite{kaas-torus} and \cite{mcgill-torus},
  enables the actions, the frequencies and their derivatives to be
  accurately and quickly computed for regular orbits in realistic
  Galactic potentials. Combined with the torus machine, the techniques
  explored in this chapter could be extended to work with such models.

\item We have seen that, in certain circumstances, the precise
  trajectory of a stream can be sensitive to the variation in the
  actions of the stars along the stream. When this is the case, it is necessary
  to know the elapsed period since the first pericentre passage of a
  cluster, in order to predict the resulting stream track with maximum
  precision.  Further work is necessary to discover to what extent it
  is possible to estimate this elapsed period from observations of
  resulting streams themselves.

  Unfortunately, it is likely that when these certain circumstances
  prevail, efforts to correctly predict the entirety of the stream
  track will prove to be error prone. However, it should always be
  possible to accurately predict those parts of the track that are
  most likely to be erroneous. Work should therefore be done to
  provide a quick and robust method for detecting when these certain
  circumstances arise, and which parts of the predicted track are
  affected, such that appropriate levels of uncertainty can be
  attached to the erroneous regions.

\item Given success in items 1--3 above, we will be able to compute
  the track of a stream, given an initial condition, with similar
  computational expenditure as is currently required to compute an
  orbital trajectory from the same.

  It would then require only a small technical change to alter the
  many techniques that attempt to fit orbits to observations of
  streams, for instance those of \cite{willett}, in order to have them
  fit accurate stream tracks instead.

  Many of these techniques already have the power to diagnose the
  Galactic potential, based on their fitting of orbits.  Once
  converted to fit stream tracks instead, the existing code could be
  utilized directly to constrain the Galactic potential, with neither
  the assumption that streams follow orbits, nor the accompanying risk of
  erring should they not.

\end{enumerate}

A suggested programme of immediate further work is to tackle items
1--3 from this list in sequence.  The resulting stream-predicting
engine can then be coupled to existing orbit-fitting techniques, and
immediately applied to the Orphan stream data of \cite{newberg-orphan}
and the GD-1 data of \cite{koposov}.

% The techniques
% of \cite{mcmillan-torus} should be married to the work of this chapter,
% to enable stream-track predictions to be made in realistic potentials.
% Simultaneously, further theoretical study of the action-space
% distribution of disrupting clusters should take place, with the aim
% of producing simple and accurate models of these distributions,
% without constantly resorting to N-body simulations.

In the future, the techniques presented here may well be applicable to
the Sagittarius dwarf stream. In this chapter we have assumed that our
low-mass clusters do not affect the host potential.
This may not be true in the case of a heavy Sagittarius
progenitor.  Further study of the effect of a live host potential
on the mechanics of stream formation will be required for the
techniques to be reliably applicable to the Sagittarius dwarf stream.

%% file: concs/concs.tex
\chapter{Conclusions}
\label{chap:concs}

% N-body sims -> structure formation
% Merger fossil record -> chemistry

\section{Overview}

The automated Sloan Digital Sky Survey \citep[SDSS,][]{sdss} has
revealed a tremendous amount of substructure in the stellar halo of
the Milky Way Galaxy.  When appropriate cuts to the data are made, the
halo is revealed to be streaked with streams formed from the shattered
remains of galaxies, once captured by the Milky Way and now in the
process of being subsumed (\figref{intro:fig:fos};
\citealp{field-of-streams}). Careful analysis of the data from SDSS
and its follow-on extension programme \citep[SEGUE,][]{segue}
has uncovered large numbers of these streams \citep{odenkirchen-delineate,majewski-sag,yanny-stream,field-of-streams,orphan-discovery,
  grillmair-orphan,gd1-discovery,ngc5466,
  grillmair-2009,newberg-streams-2009}.

Tidal streams have been empirically noted to delineate the orbits
of their progenitors \citep{mcglynn-streams-are-orbits,johnston-nbody-fitting}.
Given this conjecture, the diagnostic power that a single stream provides
over the form of the Galactic potential is extraordinary \citep{binney08}.
Unfortunately, simulation results have latterly revealed that this conjecture
is untrue: tidal streams do not delineate individual orbits
(\citealp{choi-etal}; \chapref{chap:radvs}).

The subject of this thesis has been the study of these streams, and
how we can best unlock their diagnostic power, given real data from
streams that may not precisely delineate an orbit. In pursuit of this
goal, we have examined the following problems.

\subsection{Fitting orbits using radial velocity data}

The conventional approach to exploit the diagnostic power of streams
is as follows: guess a trial potential, and then integrate orbits from a
range of initial conditions in order to identify those orbits that
best match the data. In practice, this technique is difficult to
implement, because the combined space of initial conditions and trial
potentials is too vast to search effectively. The search rarely yields
an orbits that is a satisfactory fit to the data, and even if it does,
there is no guarantee that the correct orbit/potential combination has
been chosen.

One alternative are so-called geometrodynamical techniques, which utilize
measured velocity data in order to reduce the scope of the
orbit-fitting problem \citep{jin-reconstruction,binney08,
geometrodynamics}.
In \chapref{chap:radvs} we advance the work of \cite{binney08}
and \cite{jin-reconstruction} by adapting the geometrodynamical
reconstruction algorithm they present, such that it is able to handle
erroneous input data. It is not necessary to specify the
nature of the errors in order to proceed: they can be systematic or
statistical or a combination of both. In particular, it is no longer
necessary to assume that streams perfectly delineate orbits.
However, in all cases the errors must be contained within a specified boundary.

We find that we can successfully overcome the limitations the
prior methods face when handling real data, and that we can identify those orbits---should
any exist---that are consistent with a given potential and
the input data. We further find that, given sufficiently precise
input data, the ability shown by the \cite{binney08} algorithm
to diagnose the correct form of the potential is retained.

\subsection{Fitting orbits using proper motion data}

The key limitation on the work in \chapref{chap:radvs} is the lack of
availability of line-of-sight velocity data for the hundreds of
main-sequence stars that make up a stream.  It is unlikely that
sufficient 8-m class telescope time will be afforded, in the near
future, to remedy this deficit for more than a few example cases.
Conversely, the Pan-STARRS survey \citep[currently
commissioning]{pan-starrs}, the Gaia mission \citep{gaia} and the LSST
\citep{lsst} will soon produce catalogues of proper motions for
billions of main-sequence stars in the Milky Way.

In \chapref{chap:pms}, we complement the work of \cite{binney08}, by
creating a geometrodynamical algorithm to reconstruct orbits by
utilizing proper-motion measurements instead of radial-velocity
measurements. We showed that this algorithm retains the ability
of the radial-velocity algorithm to diagnose the Galactic potential.

\subsection{Galactic parallax}

In \chapref{chap:galplx} we explore a technique, which arises out of the
work of \chapref{chap:pms}, that permits the measurement of distances
to remote stars in streams given only the measured proper motions
of those stars, and no assumption being made about the form of the potential.

The technique utilizes an effect that we call Galactic parallax: the
apparent motion of stars in a direction other than along their stream
due to the reflex motion of the Sun.  Given knowledge of the velocity
of the Sun with respect to the Galactic centre, this effect enables
trigonometric distances to be calculated for stars far beyond the
range of conventional parallax.

We have examined in detail the practicality of this technique, and we
demonstrate its use by 
measuring the distance to the tidal stream GD-1.

\subsection{The mechanics of streams}

In \chapref{chap:mech}, we studied the mechanics of stream formation
from first principles. Our motivation for doing so was to investigate
under what circumstances streams could be relied upon to delineate
orbits.

We discovered that, in general, streams do not precisely delineate
orbits. The degree to which they do not is dependent upon a number
of factors: the shape of the potential, the orbit of the stream's
progenitor, and the size and shape of the progenitor itself.

We find that the real-space manifestation of this failure depends
upon the phase at which the stream is observed. Specifically,
if the stream is observed away from apsis, the stream will be
misaligned with its orbit, while if the stream is observed
close to apsis, the stream will display a different curvature
to the orbit.

We find that constraining the parameters of a the Galactic potential
by using a stream---while wrongly assuming that the stream
delineates an orbit---can cause large systematic errors in the
reported parameters.

However, we do find that, given a simple model of the phase-space
distribution of a stream that is informed by the results of
this chapter, we are able to predict the real-space tracks
of streams with high accuracy, even when those streams
are poorly represented by an orbit.

\section{This work in context}

\subsection{The dark matter distribution of the Galaxy}

The dark matter distribution in the Milky Way Galaxy is still very much
unknown, and devising techniques to effectively probe it is a key
challenge in galactic astrophysics. Attempts to directly detect
the annihilation of dark matter particles have failed thus far,
meaning that the only probe for the distribution of dark matter
in our Galaxy is its gravitational effect on the dynamics of baryonic matter.

The recently discovered wealth of substructure in the form of ``fossil
relics'' from merger events in the Galactic environment provides an
effective probe of the gravitational field of the Galaxy, and
therefore of the dark matter distribution as well. Many standard
techniques exist that attempt to harness streams to probe the
potential, and almost all involve fitting orbits to the streams.

The principles of stream formation that we have elucidated in this
thesis should now raise caution in any attempt to utilize streams as
environmental probes. In particular, the failure of streams to
precisely delineate orbits can cause serious systematic error when
attempting to constrain potential parameters.  It may be possible,
with further work, to repair these standard techniques to fit
stream-tracks, instead of orbits, to their target data. Otherwise, the
use of the standard techniques will have to be restricted to those
streams which can be demonstrably shown to be well represented by
orbits.

Meanwhile, the alternative reconstruction techniques that we have
developed may well prove key to constraining the dark
matter distribution. In particular, they can compensate for the failure
of streams to delineate orbits, and they have the power to show that
a given potential is completely inconsistent with a stream,
which would rule it out as a possible Galactic potential.

The only hurdle to the widespread application of
these techniques is the lack of appropriate velocity data for the
remote main-sequence stars that make up tidal streams.
The shortage of radial-velocity data for such 
stars in the Milky Way is unlikely to be remedied soon.
However, there is likely to be an explosion in the availability
of high-quality proper-motion data in the near future, as
several forthcoming astrometric projects come online.

Methods such as those presented in this thesis should then be immediately
applied to all known tidal streams for which the data become available.
This list would be quite long: GD-1, Pal 5, the Orphan stream, the
tails of the cluster NGC 5466 and the Sagittarius Dwarf stream.
Undoubtedly, more streams will be discovered because of the
arrival of the data itself, so this list is expected to grow.

The key to rapid success will be to work out, in the time before
these data arrive, how to best constrain the Galactic potential
given the combination of all possible sources of information:
proper-motion fitting of these streams, 
radial-velocity fitting of those streams for which the data is to hand,
and constraints on the Galactic potential from other observations.

This is an important challenge to which Galactic astrophysics should
commit itself over the next few years. If it does so, the scientific
reward could well be enormous.

\subsection{Distance estimation}

Distance estimation in our Galaxy is a difficult problem.
Trigonometric parallax produces accurate distance measurements
that can be used to calibrate other methods, but its range
is the most limited of all standard techniques. Meanwhile, distance estimation
based on the analysis of starlight has practically indefinite
range, but is subject to the effects of extinction and reddening caused
by intervening matter, and also to some of the unknown effects of
chemical make-up on stellar luminosity.

The technique of Galactic parallax, which we present, produces distance
estimates to remote stream stars that are as fundamental as
conventional trigonometric parallax, but with vastly superior range.
The technique is restricted, in that the effect can only be measured
for stars that are part of a stream. However, the technique can be
used to calibrate or independently check the calibration of other
techniques, such as photometric distance estimation, which have more
widespread applicability.

When high-quality proper motion data from next-generation
astrometric projects becomes available in the next few years,
this widespread application of this technique will be possible:
the data from Gaia and Pan-STARRS will put much of the Galaxy
in the range of trigonometric distance calculation for the
first time.

The only caveat to its use is the requirement to predict the
rest-frame direction of motion of a star from the track of its stream
on the sky.  In those cases where streams are well represented by orbits,
estimating this direction is trivial: the star moves tangent to the stream.
In the general case, where streams are not well represented by orbits,
Galactic parallax can still be computed, but the techniques expounded
in \chapref{chap:mech} of this thesis will need to be used to correct
the for the misalignment between the motion of the star and the tangent
to the stream.

\section{Future work}

Each chapter of this thesis contains its own discussion of the immediate directions
for extending the work within it. We include a short summary of those
possibilities below, in addition to some general observations.

\begin{enumerate}
\item The radial-velocity algorithm of \chapref{chap:radvs} should be immediately
applied to those streams for which we have data, namely, the Orphan stream
\citep{newberg-orphan} and the GD-1 stream \citep{koposov}. In addition,
work should be undertaken to improve the efficiency of the optimization
routine. Future work might also examine the routines for solving the reconstruction
equations, in order to see if the noise floor affecting the calculation
can be decreased further.
\item The proper-motion algorithm of \chapref{chap:pms} should be adapted
to the modification-and-search routines presented in \chapref{chap:radvs},
so that it can then be used with real data. The resulting procedure should
then be applied to the proper-motion data for the GD-1 stream \citep{koposov},
with the results compared to the radial-velocity analysis of the same.
\item The technique of Galactic parallax should be immediately applied
  to any stream for which sufficiently accurate proper-motion
  measurements can be found. Further, the technique is actually
  applicable to any galactic structure for which a proper motion can
  be measured, and which enables the rest-frame trajectory of its
  constituents to be predicted. It may be that some features of the
  Galaxy, such as spiral arms, allow the rest-frame motion of the
  stars to be predicted statistically: if a proper motion can also
  be measured for these features, then a Galactic parallax can also
  be computed.
\item It is important that the work in \chapref{chap:mech} on the mechanics
  of streams be quickly repeated in a realistic model for the Galaxy potential,
  so that we may assess the degree to which Milky Way streams are well-represented
  by orbits. This requires marriage to those techniques that allow action-angle
  variables to be computed in general potentials \citep{mcmillan-torus}. If we
  then discover that Milky Way streams do significantly deviate from orbits,
  attention should be paid to repairing existing potential-optimization routines
  by having them fit stream-tracks instead of orbits. 
\end{enumerate}

In 2012, the Gaia mission is expected to launch, and will
return full phase-space data for over one billion Milky Way stars.
Given the number of streams identified in the comparatively poor SDSS
data, we may expect many more discoveries to be made from Gaia data,
and many of the already-known streams will also become fully
characterized.

The Gaia era will represent a golden age in Galactic astronomy, and
will allow a Galactic model of unprecedented utility to be
constructed, once the right tools for the task are available.  In
order to achieve this goal without delay, in the few years until this
data-set becomes available, a substantial theoretical
effort should be directed towards understanding how to combine the
output of the methods presented here---and others---in a way that
places the Galactic potential under the strongest possible
constraints.

%% file: appendix_stackel/ap_stackel.tex
%% chapter-specific preamble stuff
\def\tdz85{Z85}

\chapter{Some results in \stackel\ potentials}
\label{appendix:stackel}

This appendix contains some useful results in \stackel\ potential
calculations. They are based on the work of \citet[herein
\tdz85]{de-z-1}, but to the best of our knowledge, most of them have
not been published, and so we include them here. Those that have been
published but are included nonetheless are required in order to
inform the later results. In order to maintain consistency, we adopt
the notation from \tdz85.  Furthermore, to remain concise, we will not
explain this notation, except where it departs from that presented in
\tdz85. Reference should be made to that work at all times.

\section{Computing the actions}

Equation~(118) from \tdz85 shows that,
\begin{equation}
J_\tau = {1 \over 2\pi} \oint p_\tau \, \d\tau.
\label{stack:eq:actionint}
\end{equation}
In asking a computer to perform this quadrature, we want to make the integrand as
flat as possible, with respect to the quadrature scheme that we are pursuing.
In this case, we only expect ill-conditioned behaviour near the endpoints, $(\tau_0, \tau_1)$.
We re-write \eqref{stack:eq:actionint}
\begin{equation}
J_\tau = {1 \over 2\pi} \oint \sqrt{(\tau - \tau_0)(\tau_1 - \tau)} \tilde{p}(\tau) \, \d\tau,
\label{stack:eq:aint2}
\end{equation}
where $\tilde{p}$ is a well-behaved function of $\tau$ that becomes
flat near the endpoints $(\tau_0, \tau_1)$.

Although the integrand of \eqref{stack:eq:aint2} is well-behaved everywhere,
we would like to remove explicit dependence upon radicals of the dummy
variable. This is because any numerical noise near the endpoints could
send the argument of these radicals negative, which would not compute.

We write
\begin{equation}
\bar{\tau} = {1\over2}(\tau_0 + \tau_1); \qquad \hat{\tau} = {1 \over 2}(\tau_1 - \tau_0);
\qquad \tau = \hat{\tau} \sin \theta + \bar{\tau}.
\label{stack:eq:coords}
\end{equation}
\Eqref{stack:eq:aint2} becomes,
\begin{equation}
J_\tau = {1 \over 2\pi} \oint \hat{\tau}^2 \cos^2 \! \theta \, \tilde{p}(\tau(\theta)) \, \d\theta,
\end{equation}
where we observe that
\begin{equation}
(\tau - \tau_0)(\tau_1 - \tau) = \hat{\tau}^2 \cos^2 \theta,
\label{stack:eq:transf}
\end{equation}
and the integral, which was over $\tau = (\tau_0, \tau_1)$,
is now over $\theta = (-\pi/2, \pi/2)$.

\section{Computing the frequencies}

Equations 125 and 126 in \tdz85 show that the frequencies are given by,
\begin{equation}
\Omega_\lambda = {1 \over \Delta} {\partial (J_\mu, J_\nu) \over
  \partial(I_2, I_3)}, \qquad
\Omega_\mu = {1 \over \Delta} {\partial (J_\nu, J_\lambda) \over
  \partial(I_2, I_3)}, \qquad
\Omega_\nu = {1 \over \Delta} {\partial (J_\lambda, J_\mu) \over
  \partial(I_2, I_3)},\label{stack:eq:frequencies}
\end{equation}
where we have defined the determinant,
\begin{equation}
\Delta = {\partial (J_\lambda, J_\mu, J_\nu) \over
  \partial (E, I_2, I_3)}.
\end{equation}
We must calculate explicitly the partial derivatives of the actions $J_\tau$ wrt to the
integrals $(E, I_2, I_3)$ by explicitly differentiating the equations \blankeqref{stack:eq:actionint}.
We find,
\begin{align}
{\partial J_\tau \over \partial E} &=
{1 \over 8 \pi} \oint {\d\tau \over
(\tau + \beta) p_\tau},\label{stack:eq:dJdH}\\ 
{\partial J_\tau \over \partial I_2} &=
-{1 \over 8 \pi} \oint {\d\tau \over
(\tau + \alpha)(\tau + \beta) p_\tau},\label{stack:eq:dJdI2}\\ 
{\partial J_\tau \over \partial I_3} &=
-{1 \over 8 \pi} \oint {\d\tau \over
(\tau + \gamma)(\tau + \beta) p_\tau},\label{stack:eq:dJdI3}
\end{align}
where we have noted that
\begin{align}
{\d p_\tau \over \d E} &= {1 \over 4 p_\tau (\tau + \beta)},\nonumber\\
{\d p_\tau \over \d I_2} &= -{1 \over 4 p_\tau (\tau + \beta)(\tau + \alpha)},\\
{\d p_\tau \over \d I_3} &= -{1 \over 4 p_\tau (\tau + \beta)(\tau + \gamma)}.\nonumber
\end{align}
The factors of $(\tau + \alpha)$, etc, in the denominators of these
expressions are well-behaved over the entire range of integration
$\tau = (\tau_0, \tau_1)$. The factors of $1/p_\tau$, however, give
integrable singularities at the ends of the range, since
$p_\tau^2 \propto (\tau - \tau_0)$ as $\tau \rightarrow \tau_0$.

We approach the problem using the same coordinate transformation as
above (equation~\ref{stack:eq:coords}). \Eqref{stack:eq:dJdH} becomes,
\begin{equation}
{\partial J_\tau \over \partial E} =
{1 \over 8 \pi} \oint {\hat{\tau} \cos \theta \d\theta \over
(\tau + \beta) \hat{\tau} \tilde{p_\tau} \cos \theta }
= 
{1 \over 8 \pi} \oint {\d\theta \over
(\tau + \beta) \tilde{p_\tau}}. \label{stack:eq:dJdHflat}
\end{equation}
This expression can now be accurately evaluated using standard quadrature
algorithms. Similarly, equations~\blankeqref{stack:eq:dJdI2} and~\blankeqref{stack:eq:dJdI3} are
written,
\begin{align}
{\partial J_\tau \over \partial I_2} &=
-{1 \over 8 \pi} \oint {\d\theta \over
(\tau + \alpha)(\tau + \beta) \tilde{p_\tau}},\\ 
{\partial J_\tau \over \partial I_3} &=
-{1 \over 8 \pi} \oint {\d\theta \over
(\tau + \gamma)(\tau + \beta) \tilde{p_\tau}}.
\end{align}

\section{Computing the derivatives of the frequencies}

For the purposes of computing the Hessian $\hessian$ in
\chapref{chap:mech}, it is necessary to be able to evaluate the
derivatives $\partial \Omega_i / \partial J_j$.

For a function $X(E,I_2,I_3)$ of the integrals $(E,I_2,I_3)$, the definition of the partial
derivative gives
\begin{align}
%\left.{\partial X \over \partial J_\tau}\right|_{J_\mu, J_\nu}
{\partial X \over \partial J_\tau}
&= {\partial X \over \partial E}{\partial E \over \partial J_\tau}
+ {\partial X \over \partial I_2}{\partial I_2 \over \partial J_\tau}
+ {\partial X \over \partial I_3}{\partial I_3 \over \partial J_\tau}\nonumber\\
&= {\partial X \over \partial E}\Omega_\tau
+ {\partial X \over \partial I_2}{\partial I_2 \over \partial J_\tau}
+ {\partial X \over \partial I_3}{\partial I_3 \over \partial J_\tau}.
\end{align}
The derivatives of the frequencies wrt the actions can thus be calculated in terms of the
derivatives of the frequencies wrt the integrals, and the derivatives of the
integrals wrt the actions.

\subsection{The derivatives of the integrals wrt the actions}

The derivatives of the integrals wrt the actions can be computed as follows. We may write
\begin{align}
{\partial I_2 \over \partial E}
&=
{\partial I_2 \over \partial J_\lambda}{\partial J_\lambda \over \partial E}
+
{\partial I_2 \over \partial J_\mu}{\partial J_\mu \over \partial E}
+
{\partial I_3 \over \partial J_\lambda}{\partial J_\nu \over \partial E} = 0,\nonumber\\
{\partial I_2 \over \partial I_2}
&=
{\partial I_2 \over \partial J_\lambda}{\partial J_\lambda \over \partial I_2}
+
{\partial I_2 \over \partial J_\mu}{\partial J_\mu \over \partial I_2}
+
{\partial I_3 \over \partial J_\lambda}{\partial J_\nu \over \partial I_2} = 1,\\
{\partial I_2 \over \partial I_3}
&=
{\partial I_2 \over \partial J_\lambda}{\partial J_\lambda \over \partial I_3}
+
{\partial I_2 \over \partial J_\mu}{\partial J_\mu \over \partial I_3}
+
{\partial I_3 \over \partial J_\lambda}{\partial J_\nu \over \partial I_3} = 0,\nonumber
\end{align}
The derivatives of $I_2$ wrt the actions are then given by the
cofactor expressions,
\begin{equation}
{\partial I_2 \over \partial J_\lambda} =
{ 1 \over \Delta }
{\partial (J_\nu, J_\mu) \over \partial (E, I_3)}; \qquad
{\partial I_2 \over \partial J_\mu} =
{ 1 \over \Delta }
{\partial (J_\lambda, J_\nu) \over \partial (E, I_3)}; \qquad
{\partial I_2 \over \partial J_\lambda} =
{ 1 \over \Delta }
{\partial (J_\mu, J_\lambda) \over \partial (E, I_3)}.
\end{equation}
Similarly, the derivatives of $I_3$ are given by the cofactor
expressions,
\begin{equation}
{\partial I_3 \over \partial J_\lambda} =
{ 1 \over \Delta }
{\partial (J_\mu, J_\nu) \over \partial (E, I_2)}; \qquad
{\partial I_3 \over \partial J_\mu} =
{ 1 \over \Delta }
{\partial (J_\nu, J_\lambda) \over \partial (E, I_2)}; \qquad
{\partial I_3 \over \partial J_\lambda} =
{ 1 \over \Delta }
{\partial (J_\lambda, J_\mu) \over \partial (E, I_2)}.\label{stack:eq:dintdj}
\end{equation}

\subsection{The derivatives of the frequencies wrt the integrals}

The derivatives of the frequencies wrt the integrals are
computed as follows.
Differentiating \eqref{stack:eq:frequencies} wrt some integral $Z$, we find
\begin{align}
{\partial \Omega_\lambda \over \partial Z}  = 
&-{1 \over \Delta^2}  {\partial \Delta \over \partial Z}
{\partial (J_\mu, J_\nu) \over \partial (I_2, I_3)}\nonumber\\
&+ {1 \over \Delta} \left\{
{\partial J_\nu \over \partial I_3}{\partial^2 J_\mu \over
\partial Z \partial I_2}
+ {\partial J_\mu \over \partial I_2}
{\partial^2 J_\nu \over \partial Z \partial I_3}
- {\partial J_\nu \over \partial I_2}
{\partial^2 J_\mu \over \partial I_3 \partial Z}
- {\partial J_\mu \over \partial I_3}
{\partial^2 J_\nu \over \partial Z \partial I_2}\right\},
\end{align}
and also
\begin{align}
{\partial \Omega_\mu \over \partial Z}  =  &
-{1 \over \Delta^2} {\partial \Delta \over \partial Z}
{\partial (J_\nu, J_\lambda) \over \partial (I_2, I_3)}\nonumber\\
&+ {1 \over \Delta} \left\{
{\partial J_\lambda \over \partial I_3}{\partial^2 J_\nu \over
\partial Z \partial I_2}
+ {\partial J_\nu \over \partial I_2}
{\partial^2 J_\lambda \over \partial Z \partial I_3}
- {\partial J_\lambda \over \partial I_2}
{\partial^2 J_\nu \over \partial I_3 \partial Z}
- {\partial J_\nu \over \partial I_3}
{\partial^2 J_\lambda \over \partial Z \partial I_2}\right\},
\end{align}
and finally
\begin{align}
{\partial \Omega_\nu \over \partial Z}  = &
-{1 \over \Delta^2} {\partial \Delta \over \partial Z}
{\partial (J_\lambda, J_\mu) \over \partial (I_2, I_3)}\nonumber\\
&+ {1 \over \Delta} \left\{
{\partial J_\mu \over \partial I_3}
{\partial^2 J_\lambda \over \partial Z \partial I_2}
+ {\partial J_\lambda \over \partial I_2}
{\partial^2 J_\mu \over \partial Z \partial I_3}
- {\partial J_\mu \over \partial I_2}
{\partial^2 J_\lambda \over \partial I_3 \partial Z}
- {\partial J_\lambda \over \partial I_3}
{\partial^2 J_\mu \over \partial Z \partial I_2}\right\}.
\end{align}
The second derivatives in these expressions are evaluated by
explicit differentiation of \eqref{stack:eq:dJdHflat}, etc., 
wrt $Z$.
We have to perform all calculations with $\theta$
as the dummy variable, since the integrand of \eqref{stack:eq:dJdH}, etc.,
is not defined at the limits of the integration.
We note that $\tau$ is now regarded as a function of $\theta$ and the
integrals, and its derivative must be considered when differentiating
any expression involving $\tau$ wrt an integral.
We find
\begin{align}
{\partial^2 J_\tau \over \partial E^2} &=
- {1 \over 8\pi} \oint \d\theta \left(
{1 \over (\tau + \beta) \tilde{p}_\tau^2}
{\partial \tilde{p}_\tau \over \partial E}
+ {1 \over (\tau + \beta)^2 \tilde{p}_\tau}
{\partial \tau \over \partial E}\right)
,\label{stack:eq:d2JdH2}\\
{\partial^2 J_\tau \over \partial E \partial I_2} &=
{1 \over 8\pi} \oint { \d\theta \over (\tau + \alpha)(\tau + \beta) \tilde{p}_\tau}
\left\{
{1 \over (\tau + \alpha)}
{\partial \tau \over \partial E} + 
{1 \over (\tau + \beta)}
{\partial \tau \over \partial E} + 
{1 \over \tilde{p}_\tau}
{\partial \tilde{p}_\tau \over \partial E}
\right\}
\nonumber\\
&=
-{1 \over 8\pi} \oint {\d\theta \over (\tau + \beta) \tilde{p}_\tau}
\left\{
{1 \over (\tau + \beta)}
{\partial \tau \over \partial I_2}
+
{1 \over \tilde{p}_\tau}
{\partial \tilde{p}_\tau \over \partial I_2}
\right\}
,\\
{\partial^2 J_\tau \over \partial E \partial I_3} &=
{1 \over 8\pi} \oint {\d\theta \over (\tau + \gamma)(\tau + \beta) \tilde{p}_\tau}
\left\{
{1 \over (\tau + \gamma)}
{\partial \tau \over \partial E} +
{1 \over (\tau + \beta)}
{\partial \tau \over \partial E} +
{1 \over \tilde{p}_\tau}
{\partial \tilde{p}_\tau \over \partial E}
\right\}
\nonumber\\
&= -{1 \over 8\pi} \oint {\d\theta \over (\tau + \beta) \tilde{p}_\tau}
\left\{
{1 \over (\tau + \beta)}
{\partial \tau \over \partial I_3} +
{1 \over \tilde{p}_\tau}
{\partial \tilde{p}_\tau \over \partial I_3}
\right\},\\
{\partial^2 J_\tau \over \partial I_2^2} &=
{1 \over 8\pi} \oint {\d\theta \over (\tau + \alpha)(\tau + \beta) \tilde{p}_\tau}
\left\{
{1 \over (\tau + \alpha)}
{\partial \tau \over \partial I_2}+
{1 \over (\tau + \beta)}
{\partial \tau \over \partial I_2}+
{1 \over \tilde{p}_\tau}
{\partial \tilde{p}_\tau \over \partial I_2}
\right\},\\
{\partial^2 J_\tau \over \partial I_2 \partial I_3} &=
{1 \over 8\pi} \oint {\d\theta \over (\tau + \alpha)(\tau + \beta) \tilde{p}_\tau}
\left\{
{1 \over (\tau + \alpha)}
{\partial \tau \over \partial I_3} +
{1 \over (\tau + \beta)}
{\partial \tau \over \partial I_3} +
{1 \over \tilde{p}_\tau}
{\partial \tilde{p}_\tau \over \partial I_3}
\right\}\nonumber\\
&={1 \over 8\pi} \oint {\d\theta \over (\tau + \gamma)(\tau + \beta) \tilde{p}_\tau}
\left\{
{1 \over (\tau + \gamma)}
{\partial \tau \over \partial I_2} +
{1 \over (\tau + \beta)}
{\partial \tau \over \partial I_2} +
{1 \over \tilde{p}_\tau}
{\partial \tilde{p}_\tau \over \partial I_2}
\right\},\\
{\partial^2 J_\tau \over \partial I_3^2} &=
{1 \over 8\pi} \oint {\d\theta \over (\tau + \gamma)(\tau + \beta) \tilde{p}_\tau}
\left\{
{1 \over (\tau + \gamma)}
{\partial \tau \over \partial I_3} +
{1 \over (\tau + \beta)}
{\partial \tau \over \partial I_3} +
{1 \over \tilde{p}_\tau}
{\partial \tilde{p}_\tau \over \partial I_3}
\right\},
\end{align}
where we understand that everything in the integrand is evaluated at
constant $\theta$. We further observe that
\begin{equation}
{1 \over (\tau_x + \alpha)(\tau_x + \gamma)}{\partial \tau_x \over \partial E} = 
-{1 \over (\tau_x + \gamma)}{\partial \tau_x \over \partial I_2} =
-{1 \over (\tau_x + \alpha)}{\partial \tau_x \over \partial I_3}.
\end{equation}
We now wish to evaluate ${\partial \tilde{p}_\tau / \partial Z}|_\theta$.
We observe from \eqref{stack:eq:transf} that
\begin{equation}
\tilde{p}^2 \hat{\tau}^2 \cos^2 \theta = p^2,
\end{equation}
Differentiating the above expression and rearranging, we find
\begin{align}
\left.{\partial \tilde{p} \over \partial Z}\right|_\theta = 
{1 \over 2 \tilde{p} \hat{\tau}^2 \cos^2 \theta}
\left.{\partial p^2 \over \partial Z}\right|_\theta
- {\tilde{p} \over \hat{\tau}}
\left.{\partial \hat{\tau} \over \partial Z}\right|_\theta.
\label{stack:eq:dptwiddle}
\end{align}
We note that
\begin{equation}
{\d p^2 \over \d Z} = \left.{\partial p^2 \over \partial Z}\right|_\tau
+ \left.{\partial \tau \over \partial Z}{\partial p^2 \over \partial \tau}\right|_Z,
\end{equation}
and that,
\begin{equation}
\left.{\partial p^2 \over \partial E}\right|_\tau 
= {1 \over 2(\tau + \beta)}; \qquad
\left.{\partial p^2 \over \partial I_2}\right|_\tau 
= -{1 \over 2(\tau + \alpha)(\tau+\beta)}; \qquad
\left.{\partial p^2 \over \partial I_3}\right|_\tau 
= -{1 \over 2(\tau + \beta)(\tau+\gamma)}.
\end{equation}
We also note that by differentiating \eqref{stack:eq:transf}, we find
\begin{equation}
\left.{\partial \hat{\tau} \over \partial Z}\right|_\theta
= {1 \over 2 \hat{\tau} \cos^2 \theta}\left\{
(\tau_1 - \tau)\left( \left.{\partial \tau \over \partial Z}\right|_\theta
- \left.{\partial \tau_0 \over \partial Z}\right|_\theta \right)
+ (\tau - \tau_0)\left( \left.{\partial \tau_1 \over \partial Z}\right|_\theta
- \left.{\partial \tau \over \partial Z}\right|_\theta \right) \right\},
\end{equation}
and that $\partial \tau / \partial Z|_\theta$ can be similarly obtained
as a function of $\partial \tau_0 / \partial Z|_\theta$
and $\partial \tau_1 / \partial Z|_\theta$ by differentiating
\eqref{stack:eq:coords}.
To evaluate \eqref{stack:eq:dptwiddle}, we now have to compute the
derivatives of the apses $\tau_x = (\tau_0, \tau_1) $ wrt the
integrals.  From the definition of $\tau_x$
\begin{equation}
p_\tau^2 (\tau_x) = 0 = 
{(\tau_x + \alpha)(\tau_x + \gamma)E- (\tau_x + \gamma)I_2 - (\tau_x + \alpha) I_3 + F(\tau_x)
\over 2(\tau_x + \alpha)(\tau_x + \beta)(\tau_x + \gamma)}.
\end{equation}
Differentiating the above expression, we find
\begin{equation}
{\partial \tau_x \over \partial E} \left(
(2\tau_x + \gamma + \alpha)E - I_2 - I_3 + 
\left.{\partial F \over \partial \tau}\right|_{\tau_x} \right)
= -(\tau_x + \alpha)(\tau_x + \gamma),\label{stack:eq:taux}
\end{equation}
for non-trivial values of $(\alpha, \beta, \gamma, \tau_x)$. Similarly, we
obtain expressions for the derivatives of the apses wrt the other integrals
\begin{align}
{\partial \tau_x \over \partial I_2} &\left(
(2\tau_x + \gamma + \alpha)E - I_2 - I_3 + 
\left.{\partial F \over \partial \tau}\right|_{\tau_x} \right)
= (\tau_x + \gamma),\\
{\partial \tau_x \over \partial I_3} &\left(
(2\tau_x + \gamma + \alpha)E - I_2 - I_3 + 
\left.{\partial F \over \partial \tau}\right|_{\tau_x} \right)
= (\tau_x + \alpha),
\end{align}
again, for non-trivial values of $(\alpha, \beta, \gamma, \tau_x)$.
We must now evaluate $\partial F / \partial \tau$.
We recall that,
\begin{equation}
F(\tau) = (\tau+\alpha)(\tau+\gamma) G(\tau),
\end{equation}
and therefore,
\begin{equation}
{\d F \over \d \tau} =
G(\tau)(\tau + \gamma)
+ G(\tau)(\tau + \alpha)
+ (\tau+\alpha)(\tau+\gamma){\d G \over \d \tau},
\end{equation}
which we can evaluate once we have chosen de Zeuuw's function $G(\tau)$.
We now have expressions for all terms, and can evaluate the
derivatives of the frequencies with respect to the actions.

\section{Computing the angles}

We start with the generating function for the oblate axisymmetric \stackel\
potential
\begin{equation}
S(\tau, E, I_2, I_3) = S_\lambda + S_\phi + S_\nu,
\end{equation}
where,
\begin{align}
S_\tau &= \int_{\tau_0}^\tau p_\tau \d \tau,\label{stack:eq:gentau}\\ 
S_\phi &= \int_{0}^\phi L_z \d \phi.\label{stack:eq:genphi}
\end{align}
The generalization to the triaxial \stackel\ potential is easy, since
$\phi \to \nu$, and we lose \eqref{stack:eq:genphi} and gain another
instance of \eqref{stack:eq:gentau} in its stead.
In all cases, the angles $\vartheta_i$ are given by,
\begin{equation}
\vartheta_i = {\partial S \over \partial J_i} = \sum_{j,k} {\partial S_j \over \partial I_k}
{\partial I_k \over \partial J_i},
\end{equation}
where the sum in $(j,k)$ is over each of $(\lambda,\phi,\nu)$,
and where $I_i$ denote the classical integrals $(E,I_2,I_3)$. 
We note that we have already calculated the 
$\partial I_i / \partial J_i$ in \eqref{stack:eq:dintdj}, etc.

In the case of $S_\phi$, we can see immediately that
\begin{equation}
{\partial S_\phi \over \partial I_i} =
\begin{cases}
{\phi \over \sqrt{2I_2}}& \text{if $J_i = L_z$},\\
0 & \text{otherwise.}
\end{cases}
\end{equation}
For $S_\tau$, we see that
\begin{equation}
{\partial S_\tau \over \partial I_i} =
\begin{cases}
{1 \over 4} \int_{\tau_0}^\tau {\d \tau \over p_\tau (\tau + \beta)} & \text{if $I_i = E$},\\
-{1 \over 4} \int_{\tau_0}^\tau {\d \tau \over p_\tau (\tau + \alpha)(\tau+\beta)}
& \text{if $I_i=I_2$},\\
-{1 \over 4} \int_{\tau_0}^\tau {\d \tau \over p_\tau (\tau + \gamma)(\tau+\beta)}
& \text{if $I_i=I_3$}.
\end{cases}
\end{equation}

%% file: thesis.bbl
\begin{thebibliography}{99}
\expandafter\ifx\csname natexlab\endcsname\relax\def\natexlab#1{#1}\fi

\bibitem[{{Ahmed}(2009)}]{direct-d-failed}
{Ahmed}, Z. 2009, arXiv:0912.3592

\bibitem[{{Aumer} \& {Binney}(2009)}]{ab09}
{Aumer}, M., \& {Binney}, J.~J. 2009, \mnras, 397, 1286

\bibitem[{{Belokurov} {et~al.}(2006){Belokurov}, {Zucker}, {Evans}, {Gilmore},
  {Vidrih}, {Bramich}, {Newberg}, {Wyse}, {Irwin}, {Fellhauer}, {Hewett},
  {Walton}, {Wilkinson}, {Cole}, {Yanny}, {Rockosi}, {Beers}, {Bell},
  {Brinkmann}, {Ivezi{\'c}}, \& {Lupton}}]{field-of-streams}
{Belokurov}, V., {et~al.} 2006, \apjl, 642, L137

\bibitem[{{Belokurov} {et~al.}(2007){Belokurov}, {Evans}, {Irwin},
  {Lynden-Bell}, {Yanny}, {Vidrih}, {Gilmore}, {Seabroke}, {Zucker},
  {Wilkinson}, {Hewett}, {Bramich}, {Fellhauer}, {Newberg}, {Wyse}, {Beers},
  {Bell}, {Barentine}, {Brinkmann}, {Cole}, {Pan}, \&
  {York}}]{orphan-discovery}
---. 2007, \apj, 658, 337

\bibitem[{Bethe(1972)}]{bethe-nobel}
Bethe, H. 1972, Energy Production in Stars, Nobel Lectures, Physics 1963-1970
  (Amsterdam: Elsevier)

\bibitem[{{Binney}(2008)}]{binney08}
{Binney}, J. 2008, \mnras, 386, L47

\bibitem[{{Binney} \& {Merrifield}(1998)}]{bm98}
{Binney}, J., \& {Merrifield}, M. 1998, {Galactic Astronomy} (Princeton:
  Princeton University Press)

\bibitem[{{Binney} \& {Tremaine}(2008)}]{bt08}
{Binney}, J., \& {Tremaine}, S. 2008, {Galactic Dynamics:~Second Edition}
  (Princeton: Princeton University Press)

\bibitem[{{Bullock} \& {Johnston}(2005)}]{bullock-johnston}
{Bullock}, J.~S., \& {Johnston}, K.~V. 2005, \apj, 635, 931

\bibitem[{{Bullock} {et~al.}(2001){Bullock}, {Kolatt}, {Sigad}, {Somerville},
  {Kravtsov}, {Klypin}, {Primack}, \& {Dekel}}]{bullock-etal}
{Bullock}, J.~S., {Kolatt}, T.~S., {Sigad}, Y., {Somerville}, R.~S.,
  {Kravtsov}, A.~V., {Klypin}, A.~A., {Primack}, J.~R., \& {Dekel}, A. 2001,
  \mnras, 321, 559

\bibitem[{{Bullock} {et~al.}(2000){Bullock}, {Kravtsov}, \&
  {Weinberg}}]{bullock-missing-clusters}
{Bullock}, J.~S., {Kravtsov}, A.~V., \& {Weinberg}, D.~H. 2000, \apj, 539, 517

\bibitem[{{Choi} {et~al.}(2007){Choi}, {Weinberg}, \& {Katz}}]{choi-etal}
{Choi}, J., {Weinberg}, M.~D., \& {Katz}, N. 2007, \mnras, 381, 987

\bibitem[{{Cox} \& {Loeb}(2008)}]{mw-m31-merger}
{Cox}, T.~J., \& {Loeb}, A. 2008, \mnras, 386, 461

\bibitem[{{de Zeeuw}(1985)}]{de-z-1}
{de Zeeuw}, T. 1985, \mnras, 216, 273

\bibitem[{{de Zeeuw} {et~al.}(1986){de Zeeuw}, {Peletier}, \& {Franx}}]{de-z-2}
{de Zeeuw}, T., {Peletier}, R., \& {Franx}, M. 1986, \mnras, 221, 1001

\bibitem[{{Dehnen} \& {Binney}(1998{\natexlab{a}})}]{db98-pot}
{Dehnen}, W., \& {Binney}, J. 1998{\natexlab{a}}, \mnras, 294, 429

\bibitem[{{Dehnen} \& {Binney}(1998{\natexlab{b}})}]{db98-kinematics}
{Dehnen}, W., \& {Binney}, J.~J. 1998{\natexlab{b}}, \mnras, 298, 387

\bibitem[{{Dehnen} {et~al.}(2004){Dehnen}, {Odenkirchen}, {Grebel}, \&
  {Rix}}]{dehnen-pal5}
{Dehnen}, W., {Odenkirchen}, M., {Grebel}, E.~K., \& {Rix}, H. 2004, \aj, 127,
  2753

\bibitem[{{Drimmel} \& {Spergel}(2001)}]{drimmel}
{Drimmel}, R., \& {Spergel}, D.~N. 2001, \apj, 556, 181

\bibitem[{{Dubinski} {et~al.}(1996){Dubinski}, {Mihos}, \&
  {Hernquist}}]{dubinksi-antennae-tails-probe}
{Dubinski}, J., {Mihos}, J.~C., \& {Hernquist}, L. 1996, \apj, 462, 576

\bibitem[{{Einstein}(1915)}]{gr}
{Einstein}, A. 1915, Sitzungsberichte der K{\"o}niglich Preu{\ss}ischen
  Akademie der Wissenschaften (Berlin), 844

\bibitem[{{Eyre}(2010)}]{galplx}
{Eyre}, A. 2010, \mnras, 403, 1999

\bibitem[{{Eyre} \& {Binney}(2009{\natexlab{a}})}]{eb09b}
{Eyre}, A., \& {Binney}, J. 2009{\natexlab{a}}, \mnras, 399, L160

\bibitem[{{Eyre} \& {Binney}(2009{\natexlab{b}})}]{eb09a}
---. 2009{\natexlab{b}}, \mnras, 400, 548

\bibitem[{{Faber} \& {Gallagher}(1979)}]{faber-tidal-tails}
{Faber}, S.~M., \& {Gallagher}, J.~S. 1979, \araa, 17, 135

\bibitem[{{Fellhauer} {et~al.}(2007{\natexlab{a}}){Fellhauer}, {Evans},
  {Belokurov}, {Wilkinson}, \& {Gilmore}}]{fellhauer-ngc5466}
{Fellhauer}, M., {Evans}, N.~W., {Belokurov}, V., {Wilkinson}, M.~I., \&
  {Gilmore}, G. 2007{\natexlab{a}}, \mnras, 380, 749

\bibitem[{{Fellhauer} {et~al.}(2007{\natexlab{b}}){Fellhauer}, {Evans},
  {Belokurov}, {Zucker}, {Yanny}, {Wilkinson}, {Gilmore}, {Irwin}, {Bramich},
  {Vidrih}, {Hewett}, \& {Beers}}]{fellhauer-orphan}
{Fellhauer}, M., {et~al.} 2007{\natexlab{b}}, \mnras, 375, 1171

\bibitem[{Gamow(1970)}]{gamow-autobio}
Gamow, G. 1970, My World Line: An Informal Autobiography (New York: Viking
  Press)

\bibitem[{{Ghez} {et~al.}(2005){Ghez}, {Salim}, {Hornstein}, {Tanner}, {Lu},
  {Morris}, {Becklin}, \& {Duch{\^e}ne}}]{s-stars}
{Ghez}, A.~M., {Salim}, S., {Hornstein}, S.~D., {Tanner}, A., {Lu}, J.~R.,
  {Morris}, M., {Becklin}, E.~E., \& {Duch{\^e}ne}, G. 2005, \apj, 620, 744

\bibitem[{{Gillessen} {et~al.}(2009){Gillessen}, {Eisenhauer}, {Trippe},
  {Alexander}, {Genzel}, {Martins}, \& {Ott}}]{gillessen}
{Gillessen}, S., {Eisenhauer}, F., {Trippe}, S., {Alexander}, T., {Genzel}, R.,
  {Martins}, F., \& {Ott}, T. 2009, \apj, 692, 1075

\bibitem[{{Grillmair}(2006{\natexlab{a}})}]{grillmair-orphan}
{Grillmair}, C.~J. 2006{\natexlab{a}}, \apjl, 645, L37

\bibitem[{{Grillmair}(2006{\natexlab{b}})}]{anticentre-discovery}
---. 2006{\natexlab{b}}, \apjl, 651, L29

\bibitem[{{Grillmair}(2009)}]{grillmair-2009}
---. 2009, \apj, 693, 1118

\bibitem[{{Grillmair} {et~al.}(2008){Grillmair}, {Carlin}, \&
  {Majewski}}]{grillmair-anticentre-radvs-pms}
{Grillmair}, C.~J., {Carlin}, J.~L., \& {Majewski}, S.~R. 2008, \apjl, 689,
  L117

\bibitem[{{Grillmair} \& {Dionatos}(2006)}]{gd1-discovery}
{Grillmair}, C.~J., \& {Dionatos}, O. 2006, \apjl, 643, L17

\bibitem[{{Grillmair} \& {Johnson}(2006)}]{ngc5466}
{Grillmair}, C.~J., \& {Johnson}, R. 2006, \apjl, 639, L17

\bibitem[{{Helmi}(2008)}]{helmi-review-halo}
{Helmi}, A. 2008, \aapr, 15, 145

\bibitem[{{Helmi} \& {White}(1999)}]{helmi-white-halo}
{Helmi}, A., \& {White}, S.~D.~M. 1999, \mnras, 307, 495

\bibitem[{{Helmi} {et~al.}(1999){Helmi}, {White}, {de Zeeuw}, \&
  {Zhao}}]{helmi-white-nature}
{Helmi}, A., {White}, S.~D.~M., {de Zeeuw}, P.~T., \& {Zhao}, H. 1999, \nat,
  402, 53

\bibitem[{{Hubble}(1929)}]{hubble-expansion}
{Hubble}, E. 1929, Proceedings of the National Academy of Science, 15, 168

\bibitem[{{Hubble}(1926)}]{hubble-extra}
{Hubble}, E.~P. 1926, \apj, 64, 321

\bibitem[{{Ibata} {et~al.}(1995){Ibata}, {Gilmore}, \& {Irwin}}]{ibata-sag}
{Ibata}, R.~A., {Gilmore}, G., \& {Irwin}, M.~J. 1995, \mnras, 277, 781

\bibitem[{{Ivezi{\'c}} {et~al.}(2008){Ivezi{\'c}}, {Monet}, {Bond},
  {Juri{\'c}}, {Sesar}, {Munn}, {Lupton}, {Gunn}, {Knapp}, {Tyson}, {Pinto}, \&
  {Cook}}]{ivezic}
{Ivezi{\'c}}, {\v Z}., {et~al.} 2008, in IAU Symposium, ed. {W.~J.~Jin,
  I.~Platais, \& M.~A.~C.~Perryman}, Vol. 248, 537--543

\bibitem[{{Jin} \& {Lynden-Bell}(2007)}]{jin-reconstruction}
{Jin}, S., \& {Lynden-Bell}, D. 2007, \mnras, 378, L64

\bibitem[{{Jin} \& {Lynden-Bell}(2008)}]{geometrodynamics}
---. 2008, \mnras, 383, 1686

\bibitem[{{Johnston} {et~al.}(1996){Johnston}, {Hernquist}, \&
  {Bolte}}]{johnston-delineate}
{Johnston}, K.~V., {Hernquist}, L., \& {Bolte}, M. 1996, \apj, 465, 278

\bibitem[{{Johnston} {et~al.}(2005){Johnston}, {Law}, \&
  {Majewski}}]{johnston-nbody-fitting}
{Johnston}, K.~V., {Law}, D.~R., \& {Majewski}, S.~R. 2005, \apj, 619, 800

\bibitem[{{Johnston} {et~al.}(1995){Johnston}, {Spergel}, \&
  {Hernquist}}]{johnston-spergel-hernquist-sag}
{Johnston}, K.~V., {Spergel}, D.~N., \& {Hernquist}, L. 1995, \apj, 451, 598

\bibitem[{{Juri{\'c}} {et~al.}(2008){Juri{\'c}}, {Ivezi{\'c}}, {Brooks},
  {Lupton}, {Schlegel}, {Finkbeiner}, {Padmanabhan}, {Bond}, {Sesar},
  {Rockosi}, {Knapp}, {Gunn}, {Sumi}, {Schneider}, {Barentine}, {Brewington},
  {Brinkmann}, {Fukugita}, {Harvanek}, {Kleinman}, {Krzesinski}, {Long},
  {Neilsen}, {Nitta}, {Snedden}, \& {York}}]{juric}
{Juri{\'c}}, M., {et~al.} 2008, \apj, 673, 864

\bibitem[{{Kaasalainen} \& {Binney}(1994)}]{kaas-torus}
{Kaasalainen}, M., \& {Binney}, J. 1994, \mnras, 268, 1033

\bibitem[{{Kaiser} {et~al.}(2002){Kaiser}, {Aussel}, {Burke}, {Boesgaard},
  {Chambers}, {Chun}, {Heasley}, {Hodapp}, {Hunt}, {Jedicke}, {Jewitt},
  {Kudritzki}, {Luppino}, {Maberry}, {Magnier}, {Monet}, {Onaka}, {Pickles},
  {Rhoads}, {Simon}, {Szalay}, {Szapudi}, {Tholen}, {Tonry}, {Waterson}, \&
  {Wick}}]{pan-starrs}
{Kaiser}, N., {et~al.} 2002, in SPIE Conference Series, ed. {J.~A.~Tyson \&
  S.~Wolff}, Vol. 4836, 154--164

\bibitem[{{King}(1966)}]{kingmodel}
{King}, I.~R. 1966, \aj, 71, 64

\bibitem[{{Klypin} {et~al.}(1999){Klypin}, {Kravtsov}, {Valenzuela}, \&
  {Prada}}]{klypin-missing-satellites}
{Klypin}, A., {Kravtsov}, A.~V., {Valenzuela}, O., \& {Prada}, F. 1999, \apj,
  522, 82

\bibitem[{{Koposov} {et~al.}(2010){Koposov}, {Rix}, \& {Hogg}}]{koposov}
{Koposov}, S.~E., {Rix}, H., \& {Hogg}, D.~W. 2010, \apj, 712, 260

\bibitem[{{Laughlin}(2009)}]{laughlin-stability}
{Laughlin}, G. 2009, \nat, 459, 781

\bibitem[{{Law} {et~al.}(2005){Law}, {Johnston}, \& {Majewski}}]{law-modelling}
{Law}, D.~R., {Johnston}, K.~V., \& {Majewski}, S.~R. 2005, \apj, 619, 807

\bibitem[{{Londrillo} {et~al.}(2003){Londrillo}, {Nipoti}, \& {Ciotti}}]{fvfps}
{Londrillo}, P., {Nipoti}, C., \& {Ciotti}, L. 2003, Memorie della Societa
  Astronomica Italiana Supplement, 1, 18

\bibitem[{{Magnier} {et~al.}(2008){Magnier}, {Liu}, {Monet}, \&
  {Chambers}}]{pan-starrs-3pi}
{Magnier}, E.~A., {Liu}, M., {Monet}, D.~G., \& {Chambers}, K.~C. 2008, in IAU
  Symposium, ed. {W.~J.~Jin, I.~Platais, \& M.~A.~C.~Perryman}, Vol. 248,
  553--559

\bibitem[{{Majewski} {et~al.}(2003){Majewski}, {Skrutskie}, {Weinberg}, \&
  {Ostheimer}}]{majewski-sag}
{Majewski}, S.~R., {Skrutskie}, M.~F., {Weinberg}, M.~D., \& {Ostheimer}, J.~C.
  2003, \apj, 599, 1082

\bibitem[{{Mathewson} {et~al.}(1974){Mathewson}, {Cleary}, \&
  {Murray}}]{mag-stream}
{Mathewson}, D.~S., {Cleary}, M.~N., \& {Murray}, J.~D. 1974, \apj, 190, 291

\bibitem[{{McGill} \& {Binney}(1990)}]{mcgill-torus}
{McGill}, C., \& {Binney}, J. 1990, \mnras, 244, 634

\bibitem[{{McGlynn}(1990)}]{mcglynn-streams-are-orbits}
{McGlynn}, T.~A. 1990, \apj, 348, 515

\bibitem[{{McMillan} \& {Binney}(2008)}]{mcmillan-torus}
{McMillan}, P.~J., \& {Binney}, J.~J. 2008, \mnras, 390, 429

\bibitem[{{McMillan} \& {Binney}(2010)}]{mb09}
---. 2010, \mnras, 402, 934

\bibitem[{{Miyamoto} \& {Nagai}(1975)}]{miyamoto}
{Miyamoto}, M., \& {Nagai}, R. 1975, \pasj, 27, 533

\bibitem[{{Montuori} {et~al.}(2007){Montuori}, {Capuzzo-Dolcetta}, {Di Matteo},
  {Lepinette}, \& {Miocchi}}]{montuori-nbody-streams}
{Montuori}, M., {Capuzzo-Dolcetta}, R., {Di Matteo}, P., {Lepinette}, A., \&
  {Miocchi}, P. 2007, \apj, 659, 1212

\bibitem[{{Moore} \& {Davis}(1994)}]{lmc-ram-pressure}
{Moore}, B., \& {Davis}, M. 1994, \mnras, 270, 209

\bibitem[{{Moore} {et~al.}(2006){Moore}, {Diemand}, {Madau}, {Zemp}, \&
  {Stadel}}]{moore-2006-missing-satellites}
{Moore}, B., {Diemand}, J., {Madau}, P., {Zemp}, M., \& {Stadel}, J. 2006,
  \mnras, 368, 563

\bibitem[{{Moore} {et~al.}(1998){Moore}, {Governato}, {Quinn}, {Stadel}, \&
  {Lake}}]{moore-etal}
{Moore}, B., {Governato}, F., {Quinn}, T., {Stadel}, J., \& {Lake}, G. 1998,
  \apjl, 499, L5+

\bibitem[{{Munn} {et~al.}(2004){Munn}, {Monet}, {Levine}, {Canzian}, {Pier},
  {Harris}, {Lupton}, {Ivezi{\'c}}, {Hindsley}, {Hennessy}, {Schneider}, \&
  {Brinkmann}}]{munn-etal}
{Munn}, J.~A., {et~al.} 2004, \aj, 127, 3034

\bibitem[{{Navarro} {et~al.}(1997){Navarro}, {Frenk}, \& {White}}]{nfw}
{Navarro}, J.~F., {Frenk}, C.~S., \& {White}, S.~D.~M. 1997, \apj, 490, 493

\bibitem[{{Newberg} {et~al.}(2010){Newberg}, {Willett}, {Yanny}, \&
  {Xu}}]{newberg-orphan}
{Newberg}, H.~J., {Willett}, B.~A., {Yanny}, B., \& {Xu}, Y. 2010, \apj, 711,
  32

\bibitem[{{Newberg} {et~al.}(2009){Newberg}, {Yanny}, \&
  {Willett}}]{newberg-streams-2009}
{Newberg}, H.~J., {Yanny}, B., \& {Willett}, B.~A. 2009, \apjl, 700, L61

\bibitem[{{Odenkirchen} {et~al.}(2009){Odenkirchen}, {Grebel}, {Kayser}, {Rix},
  \& {Dehnen}}]{oden-2009}
{Odenkirchen}, M., {Grebel}, E.~K., {Kayser}, A., {Rix}, H., \& {Dehnen}, W.
  2009, \aj, 137, 3378

\bibitem[{{Odenkirchen} {et~al.}(2003){Odenkirchen}, {Grebel}, {Dehnen}, {Rix},
  {Yanny}, {Newberg}, {Rockosi}, {Mart{\'{\i}}nez-Delgado}, {Brinkmann}, \&
  {Pier}}]{odenkirchen-delineate}
{Odenkirchen}, M., {et~al.} 2003, \aj, 126, 2385

\bibitem[{{Ostriker} \& {Hausman}(1977)}]{ostriker-cannibalism}
{Ostriker}, J.~P., \& {Hausman}, M.~A. 1977, \apjl, 217, L125

\bibitem[{{Perryman} {et~al.}(2001){Perryman}, {de Boer}, {Gilmore}, {H{\o}g},
  {Lattanzi}, {Lindegren}, {Luri}, {Mignard}, {Pace}, \& {de Zeeuw}}]{gaia}
{Perryman}, M.~A.~C., {et~al.} 2001, \aap, 369, 339

\bibitem[{{Press} {et~al.}(2002){Press}, {Teukolsky}, {Vetterling}, \&
  {Flannery}}]{press-etal}
{Press}, W.~H., {Teukolsky}, S.~A., {Vetterling}, W.~T., \& {Flannery}, B.~P.
  2002, {Numerical Recipes in C++. The Art of Scientific Computing: Second
  Edition} (Cambridge: Cambridge University Press)

\bibitem[{{Reid} \& {Brunthaler}(2004)}]{reid-brunthaler}
{Reid}, M.~J., \& {Brunthaler}, A. 2004, \apj, 616, 872

\bibitem[{{Rix} {et~al.}(1997){Rix}, {de Zeeuw}, {Cretton}, {van der Marel}, \&
  {Carollo}}]{rix-ellipticals}
{Rix}, H., {de Zeeuw}, P.~T., {Cretton}, N., {van der Marel}, R.~P., \&
  {Carollo}, C.~M. 1997, \apj, 488, 702

\bibitem[{{Sales} {et~al.}(2008){Sales}, {Helmi}, {Starkenburg}, {Morrison},
  {Engle}, {Harding}, {Mateo}, {Olszewski}, \& {Sivarani}}]{sales-orphan}
{Sales}, L.~V., {et~al.} 2008, \mnras, 389, 1391

\bibitem[{{Schlegel} {et~al.}(1998){Schlegel}, {Finkbeiner}, \&
  {Davis}}]{reddening1}
{Schlegel}, D.~J., {Finkbeiner}, D.~P., \& {Davis}, M. 1998, \apj, 500, 525

\bibitem[{{Skrutskie} {et~al.}(2006){Skrutskie}, {Cutri}, {Stiening},
  {Weinberg}, {Schneider}, {Carpenter}, {Beichman}, {Capps}, {Chester},
  {Elias}, {Huchra}, {Liebert}, {Lonsdale}, {Monet}, {Price}, {Seitzer},
  {Jarrett}, {Kirkpatrick}, {Gizis}, {Howard}, {Evans}, {Fowler}, {Fullmer},
  {Hurt}, {Light}, {Kopan}, {Marsh}, {McCallon}, {Tam}, {Van Dyk}, \&
  {Wheelock}}]{2mass}
{Skrutskie}, M.~F., {et~al.} 2006, \aj, 131, 1163

\bibitem[{{Smith} {et~al.}(2007){Smith}, {Ruchti}, {Helmi}, {Wyse},
  {Fulbright}, {Freeman}, {Navarro}, {Seabroke}, {Steinmetz}, {Williams},
  {Bienaym{\'e}}, {Binney}, {Bland-Hawthorn}, {Dehnen}, {Gibson}, {Gilmore},
  {Grebel}, {Munari}, {Parker}, {Scholz}, {Siebert}, {Watson}, \&
  {Zwitter}}]{rave-escape-v}
{Smith}, M.~C., {et~al.} 2007, \mnras, 379, 755

\bibitem[{{Soares}(2001)}]{soares-hubble}
{Soares}, D.~S.~L. 2001, \jrasc, 95, 10

\bibitem[{{Spitzer}(1987)}]{spitzer87}
{Spitzer}, L. 1987, {Dynamical Evolution of Globular Clusters} (Princeton:
  Princeton University Press)

\bibitem[{{Toomre} \& {Toomre}(1972)}]{toomre2tails}
{Toomre}, A., \& {Toomre}, J. 1972, \apj, 178, 623

\bibitem[{{Tremaine}(1999)}]{tremaine99}
{Tremaine}, S. 1999, \mnras, 307, 877

\bibitem[{{Tyson}(2002)}]{lsst}
{Tyson}, J.~A. 2002, in SPIE Conference Series, ed. {J.~A.~Tyson \& S.~Wolff},
  Vol. 4836, 10--20

\bibitem[{{van Leeuwen}(2007)}]{newhipparcos}
{van Leeuwen}, F. 2007, {Hipparcos, the New Reduction of the Raw Data},
  Astrophysics and Space Science Library (Dordrecht: Springer)

\bibitem[{{Velazquez} \& {White}(1995)}]{velazquez-tail-prediction}
{Velazquez}, H., \& {White}, S.~D.~M. 1995, \mnras, 275, L23

\bibitem[{{Vergely} {et~al.}(1998){Vergely}, {Ferrero}, {Egret}, \&
  {Koeppen}}]{vergely}
{Vergely}, J., {Ferrero}, R.~F., {Egret}, D., \& {Koeppen}, J. 1998, \aap, 340,
  543

\bibitem[{{White} \& {Rees}(1978)}]{white-rees-merging}
{White}, S.~D.~M., \& {Rees}, M.~J. 1978, \mnras, 183, 341

\bibitem[{{Willett} {et~al.}(2009){Willett}, {Newberg}, {Zhang}, {Yanny}, \&
  {Beers}}]{willett}
{Willett}, B.~A., {Newberg}, H.~J., {Zhang}, H., {Yanny}, B., \& {Beers}, T.~C.
  2009, \apj, 697, 207

\bibitem[{Wright(1750)}]{thom-wright}
Wright, T. 1750, An Original Theory or New Hypothesis of the Universe (London)

\bibitem[{{Yanny} {et~al.}(2003){Yanny}, {Newberg}, {Grebel}, {Kent},
  {Odenkirchen}, {Rockosi}, {Schlegel}, {Subbarao}, {Brinkmann}, {Fukugita},
  {Ivezic}, {Lamb}, {Schneider}, \& {York}}]{yanny-stream}
{Yanny}, B., {et~al.} 2003, \apj, 588, 824

\bibitem[{{Yanny} {et~al.}(2009){Yanny}, {Rockosi}, {Newberg}, {Knapp},
  {Adelman-McCarthy}, {Alcorn}, {Allam}, {Allende Prieto}, {An}, {Anderson},
  {Anderson}, {Bailer-Jones}, {Bastian}, {Beers}, {Bell}, {Belokurov},
  {Bizyaev}, {Blythe}, {Bochanski}, {Boroski}, {Brinchmann}, {Brinkmann},
  {Brewington}, {Carey}, {Cudworth}, {Evans}, {Evans}, {Gates}, {G{\"a}nsicke},
  {Gillespie}, {Gilmore}, {Gomez-Moran}, {Grebel}, {Greenwell}, {Gunn},
  {Jordan}, {Jordan}, {Harding}, {Harris}, {Hendry}, {Holder}, {Ivans},
  {Ivezi{\v c}}, {Jester}, {Johnson}, {Kent}, {Kleinman}, {Kniazev},
  {Krzesinski}, {Kron}, {Kuropatkin}, {Lebedeva}, {Lee}, {Leger}, {L{\'e}pine},
  {Levine}, {Lin}, {Long}, {Loomis}, {Lupton}, {Malanushenko}, {Malanushenko},
  {Margon}, {Martinez-Delgado}, {McGehee}, {Monet}, {Morrison}, {Munn},
  {Neilsen}, {Nitta}, {Norris}, {Oravetz}, {Owen}, {Padmanabhan}, {Pan},
  {Peterson}, {Pier}, {Platson}, {Fiorentin}, {Richards}, {Rix}, {Schlegel},
  {Schneider}, {Schreiber}, {Schwope}, {Sibley}, {Simmons}, {Snedden}, {Smith},
  {Stark}, {Stauffer}, {Steinmetz}, {Stoughton}, {Subba Rao}, {Szalay},
  {Szkody}, {Thakar}, {Thirupathi}, {Tucker}, {Uomoto}, {Vanden Berk},
  {Vidrih}, {Wadadekar}, {Watters}, {Wilhelm}, {Wyse}, {Yarger}, \&
  {Zucker}}]{segue}
---. 2009, \aj, 137, 4377

\bibitem[{{York} {et~al.}(2000){York}, {Adelman}, {Anderson}, {Anderson},
  {Annis}, {Bahcall}, {Bakken}, {Barkhouser}, {Bastian}, {Berman}, {Boroski},
  {Bracker}, {Briegel}, {Briggs}, {Brinkmann}, {Brunner}, {Burles}, {Carey},
  {Carr}, {Castander}, {Chen}, {Colestock}, {Connolly}, {Crocker}, {Csabai},
  {Czarapata}, {Davis}, {Doi}, {Dombeck}, {Eisenstein}, {Ellman}, {Elms},
  {Evans}, {Fan}, {Federwitz}, {Fiscelli}, {Friedman}, {Frieman}, {Fukugita},
  {Gillespie}, {Gunn}, {Gurbani}, {de Haas}, {Haldeman}, {Harris}, {Hayes},
  {Heckman}, {Hennessy}, {Hindsley}, {Holm}, {Holmgren}, {Huang}, {Hull},
  {Husby}, {Ichikawa}, {Ichikawa}, {Ivezi{\'c}}, {Kent}, {Kim}, {Kinney},
  {Klaene}, {Kleinman}, {Kleinman}, {Knapp}, {Korienek}, {Kron}, {Kunszt},
  {Lamb}, {Lee}, {Leger}, {Limmongkol}, {Lindenmeyer}, {Long}, {Loomis},
  {Loveday}, {Lucinio}, {Lupton}, {MacKinnon}, {Mannery}, {Mantsch}, {Margon},
  {McGehee}, {McKay}, {Meiksin}, {Merelli}, {Monet}, {Munn}, {Narayanan},
  {Nash}, {Neilsen}, {Neswold}, {Newberg}, {Nichol}, {Nicinski}, {Nonino},
  {Okada}, {Okamura}, {Ostriker}, {Owen}, {Pauls}, {Peoples}, {Peterson},
  {Petravick}, {Pier}, {Pope}, {Pordes}, {Prosapio}, {Rechenmacher}, {Quinn},
  {Richards}, {Richmond}, {Rivetta}, {Rockosi}, {Ruthmansdorfer}, {Sandford},
  {Schlegel}, {Schneider}, {Sekiguchi}, {Sergey}, {Shimasaku}, {Siegmund},
  {Smee}, {Smith}, {Snedden}, {Stone}, {Stoughton}, {Strauss}, {Stubbs},
  {SubbaRao}, {Szalay}, {Szapudi}, {Szokoly}, {Thakar}, {Tremonti}, {Tucker},
  {Uomoto}, {Vanden Berk}, {Vogeley}, {Waddell}, {Wang}, {Watanabe},
  {Weinberg}, {Yanny}, \& {Yasuda}}]{sdss}
{York}, D.~G., {et~al.} 2000, \aj, 120, 1579

\bibitem[{{Zwicky}(1933)}]{zwicky33}
{Zwicky}, F. 1933, Helvetica Physica Acta, 6, 110

\end{thebibliography}
